\documentclass[final,1p]{article}
\usepackage{graphicx}
\usepackage{amsmath}
\usepackage{amssymb}
\usepackage{amsthm}
\usepackage{float}
\floatstyle{boxed}
\restylefloat{figure}
\begin{document}
\title{{\large Stochastic mechano-chemical kinetics of molecular motors:\\
a multidisciplinary enterprise from a physicist's perspective}}
\author{Debashish Chowdhury\\
Department of Physics, Indian Institute of Technology, Kanpur 208016, India}
\maketitle
\begin{abstract}
A molecular motor is made of either a single macromolecule or a 
macromolecular complex. Just like their macroscopic counterparts, 
molecular motors ``transduce'' input energy into mechanical work. 
All the nano-motors considered here operate under isothermal 
conditions far from equilibrium. Moreover, one of the possible 
mechanisms of energy transduction, called Brownian ratchet, does 
not even have any macroscopic counterpart. But, molecular motor 
is not synonymous with Brownian ratchet; a large number of molecular 
motors execute a noisy power stroke, rather than operating as 
Brownian ratchet.  
We review not only the structural design and stochastic kinetics 
of individual single motors, but also their coordination, cooperation 
and competition as well as the assembly of multi-module motors in 
various intracellular kinetic processes. Although all the motors 
considered here execute mechanical movements, efficiency and power 
output are not necessarily good measures of performance of some 
motors. Among the intracellular nano-motors, we consider the porters, 
sliders and rowers, pistons and hooks, exporters, importers, packers 
and movers as well as those that also synthesize, manipulate and 
degrade ``macromolecules of life''. We review mostly the quantitative 
models for the kinetics of these motors. We also describe several of 
those motor-driven intracellular stochastic processes for which quantitative 
models are yet to be developed. In part I, we discuss mainly the methodology 
and the generic models of various important classes of molecular motors. 
In part II, we review many specific examples emphasizing the 
unity of the basic mechanisms as well as diversity of operations 
arising from the differences in their detailed structure and kinetics. 
Multi-disciplinary research is presented here from the perspective of 
physicists.\\
{\bf Keywords:}
Motor protein, enzyme, ATP, ion-motive force, myosin, kinesin, dynein, 
microtubule, F-actin, helicase, translocase, polymerase, ribosome, 
ATP synthase, bacterial flagellar motor. \\
\\
This is an author-created, un-copyedited final version of the article 
published in {\bf Physics Reports} (\copyright Elsevier).\\ 
Submitted to arXiv with the permission of the publisher (Elsevier).
Elsevier is not responsible for any errors or omissions in this version of the manuscript or any version derived from it.
\end{abstract}


\newpage
\begin{verbatim}

1. Introduction
2. Why should physicists study molecular motors? 
Part I: General concepts, essential techniques, generic models and results
3. Motoring on a "landscape": conformation and structure
4. Molecular motors and fuels: classification, catalogue and 
   some basic concepts
  4.1. Classification of molecular machines 
    4.1.1. Cytoskeletal motors and filaments 
    4.1.2. Machines for synthesis, manipulation and degradation of 
           macromolecules of life 
    4.1.3. Rotary motors 
  4.2. Fuels for molecular motors 
    4.2.1. Chemical fuel generates generalized chemical force 
    4.2.2. Electro-chemical gradient of ions generates ion-motive force 
    4.2.3. Some uncommon energy sources for powering mechanical work 
    4.2.4. Manufacturing energy currency from external energy supply 
  4.3. Some basic concepts  
    4.3.1. Directionality, processivity and duty ratio  
    4.3.2. Force–velocity relation and stall force   
    4.3.3. Mechano-chemical coupling: slippage and futile cycles 
5.￼Experimental methods for molecular motors: 
   ensemble-averaged and single-molecule techniques   
6. Chemical physics of enzymatic activities of molecular motors: 
   concepts and techniques   
  6.1. Enzymatic reaction in a cell: 
       special features and levels of theoretical description   
  6.2. Enzyme as a chemo-chemical cyclic machine: 
       free energy transduction  
  6.3. Enzymatic activities of molecular motors  
    6.3.1. Average rate of enzymatic reaction: Michaelis–Menten equation  
    6.3.2. Specificity amplification by energy dissipation: 
           kinetic proofreading  
    6.3.3. Effect of external force on enzymatic reactions 
           catalyzed  by motors  
    6.3.4. Effects of multiple ligand-binding sites: 
           spatial cooperativity and allosterism in molecular motors   
    6.3.5. ATPase rate and velocity of motors: evidence for tight coupling?   
  6.4. Sources of fluctuations in enzymatic reactions and their effects   
    6.4.1. Fluctuations caused by low-concentration of reactants 
    6.4.2. Fluctuations caused by conformational kinetics of the enzyme: 
           ‘‘dynamicdisorder’’   
  6.5. Substrate specificity and specificity amplification   
    6.5.1. Role of conformational kinetics in selecting specific substrate  
    6.5.2. Temporal cooperativity in enzymes: hysteretic, mnemonic enzymes 
           and energy relay  
7. Thermodynamics of energy transduction: equilibrium and beyond  
  7.1. Phenomenological linear response theory for molecular motors: 
       modes of operation           
8. Modeling stochastic chemo-mechanical kinetics: 
   continuous landscapes vs. discrete networks  
  8.1. Fully atomistic model, limitations of MD and normal mode analysis  
  8.2. Coarse-grained model, elastic networks and normal mode analysis   
  8.3. Stochastic mechano-chemical model: wandering on landscapes  
    8.3.1. Motor kinetics as wandering in a time-independent 
           mechano-chemical free-energy landscape   
    8.3.2. Motor kinetics as wandering in the time-dependent mechanical 
           (real-space)free-energy landscape  
  8.4. Markov model: motor kinetics as a jump process in a network of 
       fully discrete mechano-chemical states  
9. Solving the forward problem by stochastic process modeling: 
   from model to data    
  9.1. Average speed and load–velocity relation  
  9.2. Beyond average: dwell time distribution (DTD) 
    9.2.1. A matrix-based formalism for the DTD   
    9.2.2. Extracting kinetic information from DTD    
10. Solving the inverse problem by probabilistic reverse engineering: 
    from data to model   
  10.1. Frequentist versus Bayesian approach   
    10.1.1. Maximum-likelihood estimate    
    10.1.2. Bayesian estimate  
  10.2. Hidden Markov models 
    10.2.1. HMM: formulation for a generic model of molecular motor   
11. Motoring along filamentous tracks: generic models of porters  
  11.1. Phenomenological linear response theory and modes of operation  
  11.2. A generic model of a motor: kinetics on a discrete 
        mechano-chemical network   
  11.3. 2-headed motor: generic models of hand-over-hand and inchworm 
        stepping patterns  
12. Sliders and rowers: generic models of filament alignment, bundling 
    and contractility   
13. Nano-pistons, nano-hooks and nano-springs: generic models  
  13.1. Push of polymerization: generic model of a nano-piston  
    13.1.1. Phenomenological linear response theory for chemo-mechanical  
            nano-piston: modes of operation  
    13.1.2. Stochastic kinetics of chemo-mechanical nano-piston  
  13.2. Pull of de-polymerization: generic model of a nano-hook  
14. Exporters and importers of macromolecules: generic models  
15. Motoring along templates: generic models of template-directed 
    polymerization    
  15.1. Common features of template-directed polymerization   
  15.2. A generic minimal model of the kinetics of elongation by 
        a single machine  
  15.3. A generic minimal model of simultaneous polymerization by 
        many machines   
16. Rotary motors: generic models  
Part II: Kinetic models of specific motors  
17. Cargo transport by cytoskeletal motors: specific examples of porters   
  17.1. Processivei dimeric myosins   
    17.1.1. Myosin-V: a plus-end directed processive dimeric motor   
    17.1.2. Myosin-VI: a minus-end directed processive dimeric motor   
    17.1.3. Myosin-XI: the fastest plus-end directed myosin   
  17.2. Processive dimeric kinesin  
    17.2.1. Plus-end directed homo-dimeric porters: 
            members of kinesin-1 family   
    17.2.2. Plus-end directed hetero-trimeric porters: 
            members of kinesin-2 family   
  17.3. Single-headed myosins and kinesin  
    17.3.1. Single-headed kinesin-3 family    
    17.3.2. Single-headed myosin-IX family  
  17.4. Processive dimeric dynein  
  17.5. Collective transport by porters  
    17.5.1. Collective transport of a ‘‘hard’’ cargo: 
            load-sharing, tug-of-war and bidirectional movements 
    17.5.2. Many cargoes on a single track: molecular motor traffic jam  
    17.5.3. Trip to the tip: intracellular transport in eukaryotic cells 
            with long tips 
    17.5.4. Fluid membrane-enclosed soft cargo pulled by many motors: 
            extraction of nanotubes  
  17.6. Collective transport of filaments by motors: 
        non-processivity and bistability 
  17.7. Section summary 
18. Filament depolymerization by cytoskeletal motors: 
    specific examples of chippers   
  18.1. Section summary   
19. Filament crossbridging by cytoskeletal motors: 
    specific examples of sliders and rowers  
  19.1. Acto-myosin crossbridge and muscle contraction 
  19.2. Sliding of acto-myosin bundle in non-muscle cells: stress fibers  
  19.3. Sliding MTs by axonemal dynein and beating of flagellai  
  19.4. Sliding MTs by dynein and platelet production  
  19.5. Sliding MTs by kinesin-5  
  19.6. Section summary   
20. Push/pull by polymerizing/depolymerizing cytoskeletal filaments: 
    specific examples of nano-pistons and nano-hooks  
  20.1.￼Force generated by polymerizing microtubules in eukaryotes  
  20.2. Force generated by polymerizing actin: dynamic cell protrusions 
        and motility  
    20.2.1. Force generation and cell protrusion by actin polymerization  
  20.3. Cell polarization: roles of cytoskeletal filaments and motors  
  20.4. Section summary  
21. Mitotic spindle: a self-organized machinery for eukaryotic chromosome 
    segregation  
  21.1.￼Mitotic spindle: inventory of force generators and list of stages  
    21.1.1. Mitotic spindle: key components and force generators  
    21.1.2. Mitosis: successive stages of chromosomal ballet   
  21.2. Spindle morphogenesis  
    21.2.1. Centrosome-directed astral pathway: ‘‘search-and-capture’’ 
            as a first-passage time problem  
    21.2.2. Chromosome-directed anastral pathways via sliding and sorting 
            of MTs  
    21.2.3. Amphitelic attachments: a determinant of fidelity of segregation   
    21.2.4. Chromosomal congression driven by poleward and anti-poleward 
            forces  
    21.2.5. Positioning and orienting spindle: role of MT-cortex coupling  
  21.3. Pull to the poles 
    21.3.1. Kinetochore pulling by MT filaments: Brownian ratchet or 
            power stroke?  
    21.3.2. Chromosome oscillation  
    21.3.3. Force–exit time relation: a first passage problem  
    21.3.4. Chromosome segregation in the anaphase: separated sisters 
            transported to opposite poles 
  21.4. Section summary 
22. Macromolecule translocation through nano-pore by membrane-associated 
    motors: specific exporters, importers and packers 
  22.1. Properties of macromolecule, membrane and medium that affect 
        translocation  
  22.2. Export and import of proteins 
    22.2.1. Bacterial protein secretion machineries  
    22.2.2. Machines for protein translocation across membranes of 
            organelles in eukaryotic cells   
  22.3. Export and import of macromolecules across eukaryotic nuclear envelope 
  22.4. Export/import of DNA and RNA across membranes  
    22.4.1. Export/import of DNA across bacterial cell membranes  
    22.4.2. Machines for injection of viral DNA into host: phage DNA 
            transduction as example 
  22.5. Machine-driven packaging of viral genome  
    22.5.1. Energetics of packaged genome in capsids   
    22.5.2. Structure and mechanism of viral genome packaging motor  
  22.6. ATP-binding cassette (ABC) transporters: two-cylinder ATP-driven 
        engines of cellular cleaning pumps 
  22.7. ection summary
23. Motoring into nano-cage for degradation: specific examples of nano-scale 
    mincers of macromolecules  
  23.1. Exosome: a RNA degrading machine  
  23.2. Proteasome: a protein degrading machine 
  23.3. Section summary 
24. Polymerases motoring along DNA and RNA templates: template-directed 
    polymerization of DNA and RNA 
  24.1. Transcription by RNAP: a DdRP 
    24.1.1. Effects of RNAP–RNAP collision and RNAP traffic congestion 
    24.1.2. Primae: a unique DdRP￼￼
  24.2. Replication by DNAP: a DdDP  
    24.2.1. Coordination of elongation and error correction by a 
            single DdDP: speed and fidelity 
    24.2.2. Replisome: coordination of machines within a machine  
    24.2.3. Coordination of two replisomes at a single fork   
    24.2.4. Traffic rules for replication forks and TECs: DNAP–DNAP and 
            DNAP–RNAP collisions  
    24.2.5. Initiation and termination of replication: where, which, how 
            and when? 
    24.2.6. Genome-wide replication: analogy with nucleation, growth and 
            coalescence
25. Ribosome motor translating mRNA track: template-directed polymerization 
    of proteins 
  25.1. Composition and structure of a single ribosome and accessory devices   
    25.1.1. Molecular composition and structural design of a ribosome 
  25.2. Polypeptide elongation by a single ribosome: speed versus fidelity 
    25.2.1. Selection of amino-acid: two steps and kinetic proofreading   
    25.2.2. Peptide bond formation: peptidyl transfer 
    25.2.3. Translocation: two steps of a Brownian ratchet? 
    25.2.4. Dwell time distribution and average speed of ribosome 
  25.3. Initiation and termination of translation: ribosome recycling   
  25.4. Translational error from sources other than wrong selection   
  25.5. Polysome: traffic-like collective phenomena   
    25.5.1. Experimental studies: polysome profile and ribosome profile   
    25.5.2. Modeling polysome: spatio-temporal organization of ribosomes  
    25.5.3. Effects of sequence inhomogeneity: codon bias  
  25.6. Summary of sections on machines and mechanisms for template-directed 
        polymerization   
26. Helicase motors: unzipping of DNA and RNA 
  26.1. Non-hexameric helicases: monomeric and dimeric  
  26.2. Hexameric helicases   
  26.3. Section summary   
27. Rotary motors I: ATP synthase (F0F1-motor) and similar motors  
  27.1.￼Rotary motor F0F1-ATPase  
    27.1.1. F0 motor: Brownian ratchet mechanism of energy transduction 
            from PMF during ATP synthesis   
    27.1.2. F1 motor: power stroke mechanism in reverse mode powered by 
            ATP hydrolysis   
    27.1.3. F0 –F1 coupling  
  27.2. Rotary motors similar to F0F1-ATPase  
    27.2.1. Rotary motor V0 V1 -ATPase: a ‘‘gear’’ mechanism?   
    27.2.2. Rotary motor A0A1-ATPase   
28.￼Rotary motors II: flagellar motor of bacteria  
  28.1. Summary of the sections on rotary motors  
29. Some other motors   
30. Summary and outlook 
    Acknowledgments   
    Appendix A. Eukaryotic and prokaryotic cells: differences in the 
                internal organization of the micro-factories   
    A.1. Model eukaryotes and prokaryotes   
    Appendix B. Molecules of a cell: motor components and raw materials   
    B.1. Natural DNA, RNA, and proteins   
 ￼￼￼Appendix C. Information transfer in biology: replication, gene expression 
               and central dogma   
    Appendix D. Cytoplasmic and internal membranes of a cell  
    Appendix E. Internal compartments of a cell   
    Appendix F. Viruses, bacteriophages and plasmids: hijackers or 
                poor parasites?  
    F.1. Baltimore classification of viruses according to their genome 
    Appendix G. Organization of packaged genome: from virus and prokaryotes 
                to eukaryotes  
    Appendix H. Experimental methods: introduction to the working principles 
    H.1. FRET: tool for monitoring conformational kinetics 
    H.2. Optical microscopy: diffraction-limited and beyond 
         H.2.1. Diffraction-limited microscopy  
         H.2.2. Sub-diffraction microscopy (or, super-resolution nanoscopy) 
    H.3. Single-molecule imaging and single-molecule manipulation 
    H.4. Determination of structure: X-ray crystallography and 
         electron microscopy 
    Appendix I. Modeling of chemical reactions 
    I.1. Deterministic non-spatial models of chemical reactions: 
         rate equations for bulk systems  
         I.1.1. Thermodynamic equilibrium, transient kinetics and 
                non-equilibrium steady states 
    I.2. Stochastic non-spatial models of reaction kinetics 
         I.2.1. Chemical master equation  
         I.2.2. Chemical Langevin and Fokker–Planck equations  
    I.3. Enzymatic reactions: regulation by physical and chemical means  
    Appendix J. Elastic stiffness of polymers   
    J.1. Freely jointed chain model and entropic elasticity   
    J.2. Worm-like chain and its relation with freely jointed chain: 
         persistence length 
    Appendix K. Cytoskeleton: beams, struts and cables  
    K.1. Cytoskeleton of eukaryotic cells  
    Appendix L. Kinetics of nucleation, polymerization and depolymerization 
                of polar filaments: treadmilling and dynamic instability  
    References  
\end{verbatim}
\newpage
\section{\bf Introduction}
\label{sec-introduction}

All living systems are made of cells. A single cell itself can be an
{\it uni-cellular} organism whereas {\it multi-cellular} organisms
consist of different types of cells that communicate and interact with
each other, and perform specialized functions. Cells are not only basic
{\it structural units} but also basic {\it functional units} of life
\cite{albertsbook,lodishbook}.

\noindent$\bullet${\bf Cell is an ``open system'': homeostasis of the ``milieux interieur''}

The typical size of a cell can very between approximately 1 micron to 
10 microns. 
As the name suggests, a cell is a small compartment that is bounded by
a membrane and is filled with an inhomogeneous concentrated aqueous
medium containing wide varieties of chemicals. However, a cell is not
a bag of ``passive'' mixture of chemicals. It is an open system that
not only exchanges materials with its external environment, but also
continues the opposite activities of breaking down and synthesis of its
own molecular constituents \cite{bertalanffy50}. 
In spite of the non-vanishing flux of matter and energy, the internal 
environment maintains homeostasis (i.e., non-equilibrium stationary 
state or dynamic equilibrium), a concept which originated in the works 
of Claude Bernard and William B. Cannon \cite{cannon1932,recordati04}.

\noindent$\bullet${\bf Molecular motors: ``nano-machines'' in a ``micro-factory''}

In this review, we view a cell as a ``micro-factory'' \cite{alberts98} 
where operation of the participating nano-machines are well coordinated 
in space and time. An intracellular nano-machine is either a single 
macromolecule or a macromolecular complex  
\cite{alberts98,pollard92,mavroidis04,baumgaertner05,goodsell09,cozzarelli06a,rittie08,stark10,carlier10a,frank11a,roux11b,rossmann12a}
Just like their macroscopic counterparts, molecular machines have an 
``engine'', an input and an output.

All the great thinkers from Aristotle to Descartes and Leibnitz compared 
the whole organism with a machine, the organs being the coordinated parts 
of that machine. Cell was unknown; even micro-organisms became visible 
only after the invention of the optical microscope in the seventeenth 
century. Marcelo Malpighi, father of microscopic anatomy,  speculated in 
the 17th century about the existence molecular machines in living systems. 
He wrote (as quoted in english by Marco Piccolino \cite{picco00}) that 
the organized bodies of animals and plants been constructed with  `` very 
large number of machines''. He went on to characterize these as ``extremely 
minute parts so shaped and situated, such as to form a marvelous organ''. 
Unfortunately, the molecular machines were invisible not only to the naked 
eye, but even under the optical microscopes available in his time. In fact, 
individual molecular machines could be ``caught in the act'' only in the 
last quarter of the 20th century. A strong impact across disciplinary 
boundaries was made by the influential paper of Bruce Alberts 
\cite{alberts98}, then the president of the National Academy of Sciences 
(USA). He wrote that ``{\it the entire cell can be viewed as a factory 
that contains an elaborate network of interlocking assembly lines, each of 
which is composed of a set of large protein machines}'' \cite{alberts98}. 

If the output of the machine involves 
mechanical movement, the machine is usually referred to as a {\it motor}
\cite{howard01a,vale99,vale00,schliwa03a,hackney04,squire05,duke02b,fisher07a,wang08,goel08,howard08,chowdhury09,hwang09,veigel11}.
$$\mathop{\rightarrow}^{\rm Input~ energy}~~~{\framebox{Machine}}~~~\mathop{\rightarrow}^{\rm Output~ energy}$$
$$\mathop{\rightarrow}^{\rm Input~ energy}~~~{\framebox{Motor}}~~~\mathop{\rightarrow}^{\rm Mechanical~ output}$$
However, we'll use the terms machine and motor interchangeably in 
this review.

The processes driven by molecular motors include not only intracellular
motor transport (as the name might suggest), but also manipulation,
polymerization and degradation of the bio-molecules 
\cite{cozzarelli06a,rittie08,stark10}.
Molecular motors also drive complex processes like cell motility, mitosis
(cell division) and morphogenesis (development of an entire organism). 
In this review we study the roles of molecular motors in several 
``vectorial'' processes 
\cite{cui06}
where molecules move, {\it on the average}, in a ``directed'' manner
\cite{jencks80,krupka98,mitchell91,harold05}.

\noindent$\bullet${\bf Beyond inventory; structure, energetics and kinetics}

To gain insight into the functions of the molecular motors, it is not 
enough to prepare just an {\it inventory of their parts} or a {\it 
catalogue of their structural design} \cite{aitchison03}. 
In between two successive mechanical steps, a motor transits through a 
number of chemical states; typical chemical transitions being attachment 
to and detachment from the track, binding to a fuel molecule, breakdown 
of the fuel molecule and releasing the resulting product molecules, etc. 
Besides, a single motor can be capable of performing several different 
functions. Therefore, to understand the mechanisms of molecular motors 
one has to study their {\it dynamics} during various physico-chemical 
processes in which they are involved \cite{shrager03}.
A comprehensive overview of the operational mechanisms of molecular 
motors would ultimately emerge from a thorough study of the correlation 
between their {\it structural design}, {\it energetics} and {\it stochastic 
kinetics}. However, the major emphasis of this review, which is written 
from the perspective of statistical physicists, are the {\it energetics} 
and {\it stochastic kinetics} of molecular motors. Nevertheless, the 
prototypical structural designs of motors are sketched at the beginning 
of our discussion of different types of motors.

\noindent$\bullet${\bf Top-down, bottom-up, proximate, ultimate causation: design optimization by tinkering:} 

To describe the operation of a motor, we may need to use objects at 
different levels of biological organization, starting from a single 
molecule to giant supra-molecular aggregates to the entire cell. 
Therefore, explanation of the operational mechanism of a motor may  
raise questions of ``top-down'' and ``bottom-up'' causation 
\cite{ellis12}.
However, in this review, we'll not address such philosophical questions.

Explaining the operational mechanism of a motor requires finding the 
causes of the observed phenomena associated with its operation.  
Cause and effect can be correlated in biology at different temporal scales 
\cite{mayr61,laland11,mazzocchi10} 
Explaining the observed motor-driven ``vectorial processes'' in terms of 
the present-day structure and kinetics of the corresponding motor(s) 
exposes what Ernst Mayr \cite{mayr61} would identify as the ``proximate'' 
cause of these phenomena. However, explaining the present-day structure 
and kinetics of the motors in terms of the evolutionary tinkerings in its 
design over millions of years reveals what, in Ernst Mayr's terminology, 
would qualify as the  ``ultimate'' cause \cite{mayr61}.
I believe that the response of a motor to input and its assembly-disassembly 
can provide us clues as the ``proximate'' cause of the vectorial processes  
driven by these motors. In contrast, the evolutionary tinkering of 
their design thereby, possibly, leading to the adaptive alteration of 
their function fall in the category of ``ultimate'' cause of the observed 
features of vectorial processes that they drive 
\cite{tinbergen63,purushotham10}. 

Unlike man-made macroscopic motors, molecular motors are products of 
Nature's {\it evolutionary design} over billions of years by tinkering 
\cite{jacob77}. 
``Nothing in biology makes sense except in the light of evolution'' 
\cite{dobzhansky73}. 
In fact, cell has been compared to an ``archeological excavation site'' 
\cite{gyorgyi72}, the oldest modules of functional devices are the analogs 
of the most ancient layer of the exposed site of excavation.
Does evolution tend to optimize the design of the molecular motors 
\cite{parker90}?
However, in this review we'll restrict our discussions mostly to proximate 
cause at the level of single molecule and supra-molecular assemblies. 
Occasionally, we'll mention the names of the evolutionary ancestors of 
some of the motors.

\noindent$\bullet${\bf Wet lab and dry lab: complementary approaches of experiment, theory and computer simulation}

Laboratory experiment, theoretical analysis and computer simulations 
are the three complementary approaches of investigation in physical 
sciences. 

\centerline{{\framebox{Theory}}}

\centerline{$\swarrow$$\nearrow$~~~~~~~~~~$\nwarrow$$\searrow$}

\centerline{{\framebox{Laboratory experiment}}~ $\longleftrightarrow$ ~{\framebox{Computer simulations}}}
In biological sciences the divide between the ``wet labs'' (where 
experiments are performed) and ``dry labs'' (where theoretical or 
computational biologists work) is gradually falling apart and 
the two communities are meeting at a ``moist'' zone \cite{penders08}.  

Although both theory and experiment are needed to make progress, in 
this article we critically review mainly the theoretical understanding 
of the mechanisms of molecular motors. 
Theory provides {\it understanding} and {\it insight}. These allow
us to formulate hypotheses, systematically organize and interpret 
the empirical observations, recognize the importance of the various 
ingredients. Theory also makes it possible to generalize from 
observations and to create a framework for addressing the next level 
of question and to make predictions which can be tested by carrying 
out new experiments.
However, as far as possible,
we have tried to strike a balance between theory, experiments and
computer simulations.

\noindent$\bullet${\bf Modeling: deterministic and stochastic, forward and inverse, top-down and bottom-up}

Theorization requires a model of the system. A theoretical model is
an abstract representation of the real system which helps in
understanding the real system
\cite{barnes10,epstein08,lander10,bailey98,beard05,mogilner06a,laubichler07,krakauer11}.
This representation can be pictorial (for example, in terms of cartoons
or graphs) or symbolical (e.g., a mathematical model). Qualitative
predictions may be adequate for understanding some complex phenomena
or for ruling out some plausible scenarios. But, a desirable feature
of any theoretical model is that it should make quantitative predictions
\cite{mogilner06a,oates09,bailey98,beard05,cohen04,lazebnik02}. 
Models can be formulated at several levels of biological organization 
\cite{hunter05,andrews06,schnell07,yaliraki07}, but it should be possible to derive a higher level 
model by integrating details of a lower level model. 
Results for a given model can be obtained analytically by mathematical 
manipulations. But, most often even approximate analytical treatment of 
realistic models becomes extremely difficult. Results are then obtained by 
numerical computation \cite{noble02,fisher07b}.  
The model can be individual-based or population-based. It can be deterministic 
or stochastic 
\cite{mcquarrie67,erdi89,turner04,gillespie05,gillespie07,gadgil08,sun08,andrews09,wilkinson09,ullah10,chen10,qian12}

The ``forward problem'' of process modeling \cite{voit02} starts with a 
model that is formulated on the basis of {\it apriori} hypotheses which 
are, essentially, educated guess as to the mechano-chemical kinetics of 
the motor.  Standard theoretical treatments of the model yields data on 
various aspects on the modeled motor; this approach is expressed below 
schematically. \\
\centerline{{\framebox{Theoretical model}} ~~$\rightarrow$~~ {\framebox{Experimental data}} }
Consistency between theoretical prediction and experimental data validates
the model. However, any inconsistency between the two indicates a need to
modify the model.

The ``inverse problem'' of inferring the model from empirical data 
has to be based on the theory of probability. Such ``statistical 
inference'' \cite{balding11} can be drawn by
following methods developed by statisticians over the last one century.
This inverse problem is expressed below schematically.\\
\centerline{{\framebox{Theoretical model}} ~~$\leftarrow$~~ {\framebox{Experimental data}} }

Inferring the complete network of mechano-chemical states and kinetic
scheme of a molecular motor from its observed properties is reminiscent
of inferring the operational mechanism of a given functioning macroscopic
motor by ``reverse engineering'' \cite{chikofsky90,csete02}. 
It would be desirable to follow Platt's \cite{platt64} principle of
``strong inference'' \cite{beard09} which is an extension of Chamberlin's
\cite{chamberlin1890} ``method of multiple working hypothesis'' 
\cite{beard05}.
The relative scores of the competing models (and the corresponding 
underlying hypotheses) would be a true reflection of their merits.
Both the directions of investigations, i.e. the forward problem and 
the inverse problem are equally important and complementary to each 
other \cite{kell03}.

Although, because of the usual perspective of statistical physicists, 
most of the theories reviewed here are based on modeling the kinetic 
processes, we also explain and review statistical modeling of the 
experimental data on molecular motors. In the concluding section, 
we shall summarize our assessment of the achievements and limitations 
of both these approaches to theoretical studies of molecular motors.

\noindent$\bullet${\bf From motor molecules to functional modules: systems biology of molecular motors?}

In a living cell, several motors cooperate or coordinate with each
other thereby forming ``{\it functional modules}'' \cite{hartwell99}.
Some functional modules consist of a single assembly whereas the
components of other functional modules are dispersed spatially 
\cite{bolker00}. 
Thus, each module may be viewed as a ``network'' of motors and 
the physiology of an organism may be regarded as a network of networks  
\cite{han08,barabasi04}.
Modularity can also increase the robustness of motor \cite{kitano02,kitano04}.
The nature of the forward-inverse problems and the bottom-up, top-down 
modeling strategies needed for integrating motors at different levels 
have some similarities, at least in spirit, with those followed in 
systems biology 
\cite{kitano02,wingren06,bruggeman06,noble10,omalley12}
and in the more ambitious physisome project \cite{hunter03}.
However, we'll consider only a couple of modules, formed by the integration 
of motors \cite{hofmann06}, in the part II of this review.

\noindent$\bullet${\bf A multidisciplinary enterprise from a physicist's perspective}

({\it Something there is that doesn't love a wall, 
That wants it down.} - Robert Frost, in:{\it Mending Wall}.) 

As a system of scientific investigation, molecular motors are of current 
interest in several disciplines, e.g., biophysics, biochemistry, molecular 
cell biology, nanotechnology, etc. Therefore, many papers cited in this 
review appeared in journals that are not part of the usual list of core 
journals in physics. Bold and adventurous readers who do not mind browsing 
journals of other disciplines may find a treasure house of phenomena 
related to molecular motors that are begging for modeling and explanation. 

\noindent$\bullet${\bf Organization of this review: parts I and II, appendices}

I am fully aware of the challenges of reviewing a multi-disciplinary 
research topic like molecular motors 
\cite{editorial11}.
As far as possible, I have tried to ``provide fresh scientific insight'' 
by carrying out a ``novel synthesis'' of the results scattered in the 
primary literature of several different disciplines. However, the 
presentation has been made from the perspective of a statistical physicist.
The review is divided into two parts.
In part I we develop the general conceptual foundation and the essential 
technical framework that are essential for understanding the basic physical 
principles which govern the operation of molecular motors. In this part  
the applications of the formalism are restricted to only simple generic 
models of molecular motors. However, the motivation for these models can 
be appreciated by browsing the catalogue of the real molecular motors 
that we present in the beginning of the part I in the form of tables and 
a brief description. 

``The world of life can be studied from two points of view- that of its 
unity and that of its diversity'' \cite{dobzhansky64}.
Therefore, in part II we review more detailed models of specific motors 
and the corresponding kinetic processes. 
From the catalogues provided in part I, readers may pick and choose 
motors of their interest and find the corresponding details in the part  
II.
The results summarized in module I emphasize the generic features of
molecular motors while the distinct features of different types of
motors are presented in module II.

Not all the readers of this review are expected to be familiar with the 
biological pre-requisites. Therefore, very brief summary of some of 
the biological facts that are essential for appreciating molecular motors 
and their functions are presented in the appendices.

\section{\bf Why should physicists study molecular motors?} 
\label{sec-whymotors}

Biomolecular motors operate in a domain where the appropriate units of
length, time, force and energy are, {\it nano-meter}, {\it milli-second},
{\it pico-Newton} and $k_BT$, respectively ($k_B$ being the Boltzmann
constant and $T$ is the absolute temperature). 
Aren't the operational mechanism of molecular motors similar to their 
macroscopic counterparts except, perhaps, the difference of scale? NO.
In spite of the striking similarities, it is the differences between
molecular motors and their macroscopic counterparts that makes the
studies of these systems so interesting from the perspective of
physicists.

\noindent $\bullet${\bf Nature of the dominant forces: viscous drag and Brownian 
force}

Force is one of the most fundamental quantities in physics. The forces 
which dominate the dynamics of molecular motors have negligible 
effect on macroscopic motors. 
Consider a solid object of linear size $L$ moving through a fluid of 
density $\rho$ at a speed $v$. The Reynolds number $Re$ is a dimensionless 
number that measures the ratio of the inertial and viscous forces acting 
on the object. On the basis of elementary arguments one can derive 
\cite{phillips09} 
\begin{equation}
Re = \rho L v/\eta = L v/\nu
\label{eq-Reynolds} 
\end{equation}
where $\eta$ is the viscosity and $\nu = \eta/\rho$ is the kinematic 
viscosity of the fluid. At room temperature, for water 
$\nu = 10^{-6} m^{2}/s$. Therefore for a fish \cite{berg93,phillips09} 
of length $L = 0.1$m moving at a speed of 1 m/s, $Re = 10^5$. 
In sharp contrast, for a globular protein \cite{howard01a} 
of radius $L = 10 nm$ moving at the same speed of $v = 1$ m/s in the 
same medium, $Re = 10^{-2}$; it would be even smaller at slower speeds. 
For a human swimmer, a Reynold's number of $10^{-2}$ would arise if 
(s)he tried to swim, for example, in honey! 
Thus, the dynamics of molecular motors is expected to be dominated by 
hydrodynamics at low Reynold's number \cite{purcell77,brody96}. 

The dominant forces acting on a typical Brownian particle are listed 
below.\\
\centerline{{\framebox{Forces on a Brownian particle}}}

\centerline{$\swarrow$~~~~~~~~~~~$\downarrow$~~~~~~~~~~~~$\searrow$}

\centerline{{\framebox{Conservative}}~~{\framebox{Dissipative}}~~{\framebox{Random}}}

\centerline{$\downarrow$~~~~~~~~~~~$\downarrow$~~~~~~~~~~~~$\downarrow$}

\centerline{{\framebox{electrostatic force}}~~{\framebox{viscous drag}}~~{\framebox{thermal force}}}

Already in the first half of the twentieth century D'Arcy Thompson, 
a pioneer in bio-mechanics, realized the importance of {\it viscous 
drag} and {\it Brownian forces} in this domain. He pointed out that 
in the microscopic world of cells, ``{\it gravitation is forgotten}'' 
\cite{darcy} (i.e., inertia is negligible), and ``{\it the viscosity 
of the liquid}'' and the ``{\it molecular shocks of the Brownian 
movement}'' as well as the ``{\it electric charges of the ionized
medium}'', have the strongest influence. 
Thus, the kinetics of molecular motors are dominated by fluctuations
and irreversibilities; besides, these exhibit some counter-intuitive
phenomena which are characteristics of hydrodynamics at low Reynold's 
number.

\noindent $\bullet${\bf Energy transduction: isothermal engine far from equilibrium}

Molecular motors are made of {\it soft matter} whereas macroscopic
motors are normally made of hard matter to withstand wear and tear.
Nature seems to exploit the high deformability of the ``active'' 
soft material \cite{mackintosh10},  
of which a molecular motor is made, for its biological function. 
The special characteristics which make the energy transduction 
by molecular motors interesting from the perspective of physics 
are as follows \cite{seifert12}: 
(i) these motors are {\it isothermal}, in contrast to the heat 
engines of the macroscopic motors \cite{julicher06}, 
(ii) the cycle times of these cyclic motors are finite and the 
power output is non-zero; the formalisms of neither equilibrium 
thermodynamics nor endo-reversible thermodynamics \cite{berry00a} 
are applicable for reasons that we'll explain later. 
(iii) Molecular motors operate, in general, under conditions far 
from thermodynamic equilibrium and, therefore, the formalism of 
non-equilibrium thermodynamics \cite{katchalsky67} for 
coupled mechano-chemical processes is also not applicable. 
(iv) The energy released by the a single ``fuel'' molecule is about
$10^{-21} J$. Interestingly, the mean thermal energy $k_BT$ associated
with a molecule at a temperature of the order of $T \sim 100 K$, is
also $k_B T \sim 10^{-21} J$. Moreover, equating this thermal energy
with the work done by the thermal force $F_t$ in causing a displacement
of $1 nm$ we get $F_t \sim 1 pN$. This is comparable to the elastic
force experienced by a typical motor protein when stretched by $1$nm.
Thus, a motor protein that gets bombarded from all sides by random
thermal forces is similar to a tiny creature getting bombarded randomly 
from all sides by hailstones! Therefore, the position of its center of 
mass as well as the positions of its atomic constituents with respect 
to its center of mass fluctuate. Furthermore, because of the low 
concentrations of the other molecular species involved in its operation, 
fluctuations of the cycle time is also another unavoidable intrinsic 
features of the kinetics of molecular motors. Consequently, in contrast 
to the deterministic dynamics of the macroscopic motors, the dynamics of 
molecular motors is stochastic (i.e., probabilistic). Therefore, one has 
to use the more sophisticated toolbox of stochastic processes and 
non-equilibrium statistical mechanics for theoretical treatment of 
molecular motors.

Noise \cite{bialek} need not be a nuisance for a motor \cite{samoilov06}; 
instead, a motor can move forward by gainfully exploiting this noise.  
A noise-driven mechanism of molecular motor transport, which does not 
have any counterpart in the macroscopic world of man-made motors, is 
closely related to fundamental questions on the foundations of statistical 
physics.

\noindent $\bullet${\bf Spatial symmetry breaking: polar track, directed motility, asymmetric cells}

Molecular motors and their respective filamentous tracks not only 
exhibit intrinsic {\it spatial} asymmetries in their key properties, 
but are also responsible for the {\it spatial} organization, including 
the spatial asymmetries of the emergent patterns, that are observed 
at the subcellular as well as cellular levels of organization \cite{paluch06}. 

The cause and effects of broken symmetry of molecular motors can be 
examined in the broader context of the fundamental principles of 
symmetry breaking in physics and biology 
(see the articles in the special ``Perspectives on Symmetry
Breaking in Biology'' \cite{cshperspectives}).
For macroscopic systems in thermodynamic equilibrium, symmetry
breaking is explained in terms of the form of the free energy.
However, since living cells are far from thermodynamic equilibrium, the 
theory of  symmetry breaking in those systems cannot be based on 
thermodynamic free energy. As we'll see repeatedly, kinetics cannot be 
ignored in the study of symmetry breaking in living systems.

\noindent $\bullet${\bf Directed motility of a single motor on a polar track}

Energy is a scalar quantity whereas velocity is a vector. How does
consumption of energy give rise to a non-zero average velocity of
a molecular motor? Moreover, a directed movement that a motor
exhibits on the average, requires breaking the forward-backward
symmetry on its track. What are the possible {\it cause} and
{\it effects} of this broken symmetry at the molecular level?

As far as the {\it cause} of this asymmetry is concerned, the
asymmetry of the tracks alone cannot explain the ``directed''
movement of the motors, because on the same track members of
different families of motors can, on the average, move in opposite
directions. Obviously, the structural design of the motors and
their interactions with the respective tracks also play crucial
roles in determining their direction of motion along a track.

\noindent $\bullet$ {\bf Coordination, cooperation and competition of motors: intra-cellular self-organization}

Collective dynamics of molecular motors can be viewed at several 
different levels \cite{vermeulen02}: 
(i) {\it coordination} of the different subunits of a single motor; 
(ii) {\it cooperation} and {\it competition} of a few motors in 
moving either a single cargo (if they walk on a immobilized filamentous 
track) or a single filament (if the motors are immobilized and the 
filament can move); (iii) traffic of a large population of motors 
on a fibrous network of many filaments; (iv) {\it integration} of 
different types of motors \cite{mallik04b} and other energy-transducing force generators 
within a single modular machinery that performs a specific function.

The size, shape, location and number of intracellular compartments 
\cite{marshall04,marshall08a,chan10,marshall11,chan12}
as well as modular intracellular machineries are self-organized 
\cite{kirschner00,karsenti08,keller08,keller09,kurakin09,kurakin10}, 
rather than self-assembled. Dissipation takes place in ``self-organization''
and distinguishes it from ``self-assembly'' \cite{misteli01,halley08}; 
the latter corresponds to the minimum of thermodynamic free energy 
whereas self-organized system does not attain thermodynamic equilibrium. 
The coordination, cooperation and/or competition of the ``directed'' 
movements of the individual motors on their respective tracks and the 
push / pull of the other force generators are necessary for the 
intracellular self-organization process \cite{kirschner00,karsenti08}.

\noindent $\bullet${\bf Cell motility, morphogenesis and development: pattern formation}

The broken symmetry at the molecular level, e.g., asymmetric growth 
kinetics of the polar filaments and the ``directed'' movement of 
molecular motors, generate forces required for the motility of a 
cell as a whole. Moreover, the interplay of the kinetics of motors and 
the filaments play crucially important roles in cell morphogenesis as 
well as in the development of an organism, both of which are essentially 
pattern formation phenomena. 

\newpage 

\centerline{\Huge {Part I:}}
\centerline{\Huge {General concepts, essential techniques,}}
\centerline{\Huge {generic models and results}}

\vspace{3cm} 

{\bf 
\noindent ``{\it Science is nothing without generalisations. Detached and ill-assorted facts are only raw material, and in the absence of a theoretical solvent, have but little nutritive value. }''- Lord Rayleigh, Presidential address (1884), British Association for the Advancement of Science.

}

\newpage

\section{\bf Motoring on a ``landscape'': conformation and structure}
\label{sec-landscapes}

The terms ``conformation'' and ``structure'' are used extensively to 
describe the kinetics of molecular motors. The main aim of this section 
is to clarify the subtle differences between these two concepts.

The energy landscape of a chemical reaction is a graphical way of showing
how the energy of the reacting system depends on the degrees of freedom
of the system which include the positions (and orientations) of all the
atoms of the reactant and product molecules.
For any single event of the occurrence of the reaction, the trajectory
in this landscape does not necessarily proceed along the bottom of the
valley, but occasionally also makes excursions up the walls of the
valley. However, when averaged over large number of such trajectories,
the reaction process can be described as an effective route in this
landscape that corresponds to the lowest energy from the entrance to
the exit over a saddle point. This average route in the multidimensional
energy landscape is called the {\it reaction coordinate} which we'll
denote by the symbol $\xi$. Moving along this pathway alters the
coordinates of all the atoms involved in the reaction; therefore, this
reaction coordinate is actually a composite coordinate. The magnitude
of this reaction coordinate expresses how far the reaction has progressed.
Often the energy of the system is plotted against the reaction coordinate;
the reactants and the products correspond to two local minima separated
by a maximum which corresponds to the saddle point on the multi-dimensional
energy landscape. The state of maximum energy along the reaction coordinate,
is called the {\it transition state}. The energy landscape can be
surveyed by a detailed quantum chemical calculation \cite{wales03}.

The abundant materials available to nature for designing and manufacturing 
molecular motors in living cells were proteins and nucleic acids both of 
which are linear polymers. The individual {\it monomeric residues} that 
form proteins and nucleic acids are {\it amino acids} and {\it nucleotides}, 
respectively. Although the monomeric subunits are covalently bonded along 
the linear chain, the secondary and tertiary structures (and, therefore, 
the three-dimensional shape) of proteins are determined by much weaker 
non-covalent bonds (e.g., hydrogen bonds, Van der Waals interactions, etc.) 
between these chains. Since the strengths of these non-covalent bonds are 
comparable to the thermal energy $k_BT$, the high-order structures 
exhibit significant thermal fluctuations  even when such ``soft'' materials 
are neither subjected to any external force nor participate in any chemical 
reaction. Thus, proteins are dynamic intrinsically \cite{wildman07}.

According to our convention, a {\it conformational} state (or, simply, 
conformation) of a protein is given by the coordinates of all the 
constituent atoms.  
{\it In thermodynamic equilibrium}, a protein persistently goes through
a large number of conformational states which are typically within
$k_BT$ of the conformation that has the lowest free-energy. If the
fluctuations in the positions of the atoms are not too large, we can
regard the different conformations as small deviations about a state
which is the time-average of these conformations. Such a {\it
time-averaged conformational state} is called a {\it structural state}.

We now explain the relations between conformations and structure more 
quantitatively \cite{howard01a}.
Suppose a protein can exist in any of the $N$ different conformational
states each with the corresponding potential energy $U_{i}$ ($i=1,..,N$).
Since the conformations are assumed to be canonically distributed in
thermodynamic equilibrium, the probability of finding the protein in
the $i$-th conformation is
\begin{equation}
p_{i} = \frac{exp(-\beta U_{i})}{Z}
\end{equation}
where  the partition function $Z$ is given by
\begin{equation}
Z = \sum_{i=1}^{N} exp(-\beta U_{i}).
\end{equation}

For simplicity, suppose the conformational states segregate into two
ensembles where the first is associated with the structural state
${\cal E}_1$ while the second ensemble is associated with the
structural state ${\cal E}_2$. For example, ${\cal E}_1$ and ${\cal E}_2$
may correspond to the ``pre-stroke'' and ``post-stroke'' states of a
motor protein.  Suppose the first ensemble consists of the
$n$ conformational states with energies $U_{1}, U_{2},...,U_{n}$
while the remaining $N-n$ conformational states with energies
$U_{n+1}, U_{n+2},...,U_{N}$  belong the second ensemble. Then,
the probability of finding the protein in the {\it structural} state
${\cal E}_{1}$ is given by
\begin{equation}
P_{1} = \sum_{i=1}^{n} p_{i} = Z_{1}/Z
\end{equation}
where
\begin{equation}
Z_{1} = \sum_{i=1}^{n} exp(-\beta U_{i}).
\end{equation}
is the restricted partition function. Similarly, the probability of finding 
the protein in the {\it structural} state ${\cal E}_{2}$ is given by
\begin{equation}
P_{2} = \sum_{i=n+1}^{N} p_{i} = Z_{2}/Z
\end{equation}
where
\begin{equation}
Z_{2} = \sum_{i=n+1}^{N} exp(-\beta U_{i}).
\end{equation}
Thus, $P_{2}/P_{1} = Z_{2}/Z_{1}$. But, 
$Z_{1} = exp(-\beta F_{1})$,  $Z_{2} = exp(-\beta F_{2})$ 
where $F_1$ and $F_2$ are the free energies of the 
structures ${\cal E}_1$ and ${\cal E}_2$, respectively. Hence,
\begin{equation}
P_{2}/P_{1} = exp(-\beta \Delta F)
\label{eq-strratio}
\end{equation}
where
\begin{equation}
\Delta F = F_{2} - F_{1}
\end{equation}
Thus, the probability of finding a protein in a conformational state
with energy $U_{c}$ is proportional to $exp(-\beta U_{c})$ whereas that
of finding the protein in a structural state with free energy $F_{s}$
is proportional to $exp(-\beta F_{s})$.
Most of the biological processes, in which molecular motors participate,
take place at constant temperature and constant pressure. Therefore,
the most appropriate thermodynamic potential (i.e., free energy) is the
Gibbs free energy $G = U - TS + PV$. Therefore, any change $\Delta G$
of the Gibbs free energy can be expressed as the sum of the contributions
from the changes in the enthalpy $H$ and entropy $S$:
$\Delta G = \Delta H - T \Delta S$.

\begin{figure}[ht]
\begin{center}
\includegraphics[angle=-90,width=0.55\columnwidth]{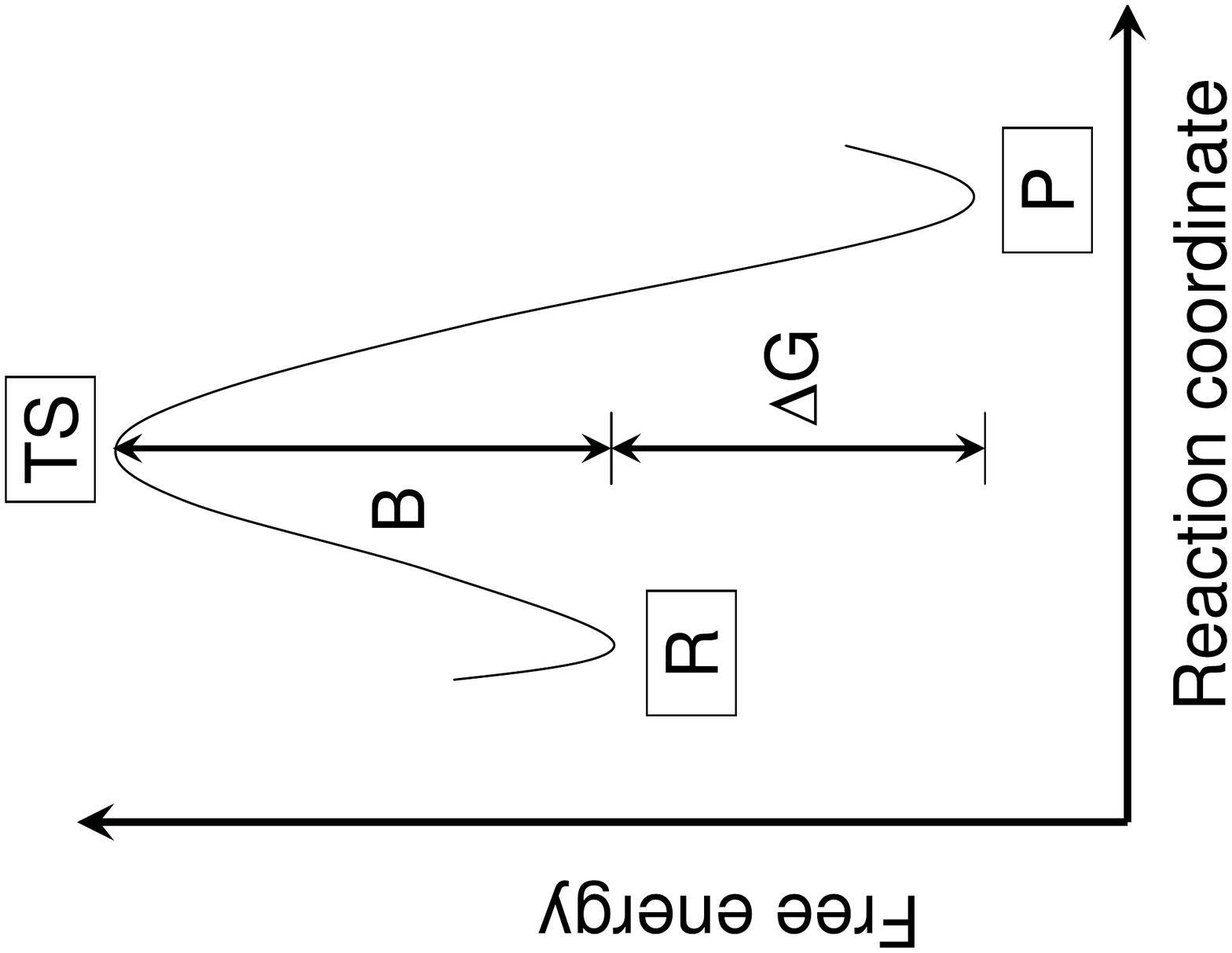}
\end{center}
\caption{The free energy of a chemically reacting system is plotted 
schematically against the reaction coordinate. The symbols R and P 
correspond to the reactant(s) and the product(s), respectively, while 
TS represents the transition state. The free energy barrier for this 
reaction is denoted by B while $\Delta G$ is the free energy 
difference between the reactant(s) and the product(s).
}
\label{fig-kramers}
\end{figure}

For reactions involving two small molecules, for example, the dimension
and complexity of the energy landscape are still small enough and the 
description of the dynamics in this landscape is useful. However, for
reactions catalyzed by enzymes (i.e., proteins), manyfold increase in
the dimension and complexity of the landscape makes its use cumbersome,
if not impractical. However, even for such reactions, a simpler landscape
can be constructed by averaging over the fast degrees of freedom which
are not important for describing the mechanism of the reaction that
takes place on much longer time scales \cite{wales03,wales06a,wales06b}.
Such an averaging over a subset of the degrees of freedom yields a
``free energy'' that still depends on the remaining degrees of freedom;
such a ``free energy'' landscape may be viewed as a projection of the
energy landscape onto a much lower-dimensional space. Usually the
reaction coordinate is one of the coordinates which span the
low-dimensional ``free energy landscape''.
The cross-section of the free energy landscape along the reaction
coordinate is usually plotted as shown in Fig.\ref{fig-kramers}
where the deeper of the two local minima corresponds to the products
while the other local minimum corresponds to the reactants.
The landscape picture and reaction coordinate diagrams are used also to 
describe the thermodynamics and kinetics of molecular motors 
\cite{whitford12,scholey13}.

\section{\bf Molecular motors and fuels: classification, catalogue and some basic concepts}
\label{sec-catalogue}

\subsection{\bf Classification of molecular machines}

Molecular machines can be classified in many different ways depending on
the characteristic property used for classification.
From the perspective of (mechanical-) engineers, the biomolecular machines
can be classified according to their similarities with their macroscopic
counterparts.  {\it Cyclic machines} operate in repetitive cycles and are
very similar to the cyclic engines which run our cars. In contrast, some
other  molecular machines  are {\it one-shot} machines that exhaust an
internal source of free energy in a single round.
The most common type of cyclic machines that we'll consider here
are {\it motors} 
\cite{howard01a,schliwa03a,hackney04,squire05,duke02b,fisher07a,wang08,goel08,howard08,veigel11} 
and pumps. 
In this review we focus exclusively on motors. 
So far as the intracellular transport system system is
concerned, its components are as follows: \\

\centerline{{\framebox{Intracellular motor transport system}} = {\framebox{motor}} + {\framebox{fuel}} + {\framebox{external regulation \& control}}   }

\centerline{{\framebox{motor}} = {\framebox{engine}} + {\framebox{transmission system (gear, clutch,etc.)}}  }

Therefore, for understanding the intracellular motor transport system,
it is not enough to understand the individual motors in isolation.
One also needs to pay attention to the regulation of motor transport
\cite{wolkenhauer05}. However, a detailed discussion of the mechanisms 
of regulation of molecular motors is beyond the scope of this review.

During its life time, a cell goes through a sequence of different phases
before it gets divided into two daughter cells thereby completing one 
{\it cell cycle}. In this review we discuss the energetics and kinetics 
of molecular motors and motor assemblies which drive key processes during 
the successive phases of cell cycle.

\subsubsection{\bf Cytoskeletal motors and filaments}
\label{sec-catalogue1}

The cytoskeleton of a cell is the analogue of the human skeleton
\cite{howard01a}. However, it not only provides mechanical strength
to the cell, but its filamentous proteins also form the networks of
``highways'' (or, ``tracks'') on which cytoskeletal motor proteins
\cite{howard01a,schliwa03a} can move.  Filamentous actin (F-actin)
and microtubules (MT) which serve as tracks are ``polar'' in the
sense that the structure and kinetics of the two ends of each
filament are dissimilar.

\begin{table}
\begin{tabular}{|c|c|c|c|} \hline
Motor superfamily & Filamentous track & Minimum step size & Appendix, Section \\\hline
Myosin & F-actin & 36 nm & \ref{app-cytoskeleton}, \ref{sec-specificporters} \\ \hline
Kinesin & Microtubule & 8 nm & \ref{app-cytoskeleton},\ref{sec-specificporters} \\ \hline
Dynein & Microtubule & 8 nm & \ref{app-cytoskeleton},\ref{sec-specificporters} \\ \hline
\end{tabular}
\caption{Superfamilies of motor proteins, corresponding tracks, minimum 
step size and section number in module II where further details can be 
found.
 }
\label{table-cytomot}
\end{table}

The superfamilies of cytoskeletal motors and the corresponding filamentous
tracks are listed in table \ref{table-cytomot}. These are {\it linear} 
molecular motors because they move along special filamentous linear tracks, 
performing mechanical work, while consuming some form of (free-) energy 
input. These are analogs of trains which move on railway tracks. 
Every superfamily can be further divided into families. Members of every
family move always in a particular direction on its track; for example,
kinesin-1 and cytoplasmic dynein move towards + and - end of MT,
respectively. Similarly, myosin-V and myosin-VI move towards the + and
- ends of  F-actin, respectively.

For their operation, each motor must have a track-binding site and
another site that binds and ``burns'' a ``fuel'' molecule (usually 
hydrolyzes a molecule of Adenosine triphosphate, abbreviated ATP).  
Both these sites are located, for example, in the  {\it head}
domain of myosins and kinesins. 
The motor-binding sites on the tracks are equispaced; the actual
step size of a motor can be, in principle, an integral multiple of
the minimum step size which is the separation between two neighboring
motor-binding sites on the corresponding track.

\noindent$\bullet${\bf Porters: intracellular cargo transport} 

Some linear motors are {\it cargo transporters}. Such a motor ``walks'' `
for a significant distance on its track carrying the cargo. For obvious 
reasons, such motors are referred to as {\it porters} \cite{leibler93}. 
The distinct possible stepping patterns of the motor proteins will 
be discussed in section \ref{sec-genericporters}.

\noindent$\bullet${\bf Depolymerases: chipping of filamentous tracks}

A MT {\it depolymerase} is a kinesin motor that chips away its own track
from one end \cite{howard07}. Members of the kinesin-13 family can reach
either end of the MT diffusively (without ATP hydrolysis) and, then,
start chipping the track from the end where it reaches. In contrast,
members of the kinesin-8 family walk towards the plus end of the MT track
hydrolyzing ATP and after reaching that end starts chipping it from there. 
Chipping by both families of depolymerase kinesins are energized by ATP
hydrolysis.

\noindent$\bullet${\bf Sliders and rowers: motor-filament crossbridge in motility and contractility} 

\begin{table}
\begin{tabular}{|c|c|c|c|} \hline
Motor & Sliding filaments  & Function (example) & Section \\\hline
Myosin & ``Thin filaments'' of muscle fibers & Muscle contraction & \ref{sec-muscle}\\ \hline
Myosin & ``Stress fibers'' of non-muscle cells & Cell contraction & \ref{sec-stressfiber} \\ \hline
Myosin & Cytokinetic ``contractile ring'' in eukaryotes & Cell division & \ref{sec-miscellaneous} \\ \hline
Kinesin & Interpolar microtubules in mitotic spindle & Mitosis & \ref{sec-spindle} \\ \hline
Dynein & Microtubules of axoneme & Beating of eukaryotic flagella & \ref{sec-flagellabeat} \\ \hline
Dynein & Microtubules of megakaryocytes & Blood platelet formation & \ref{sec-platelet} \\ \hline
\end{tabular}
\caption{Few example of cytoskeletal rowers and sliders as well as their biological functions.
 }
\label{table-rowslide}
\end{table}

Some motors are capable of sliding two different filaments with respect
to each other by stepping simultaneously on these two filaments
\cite{zemel09}. Some {\it sliders} work in groups and each detaches from
the filament after every single stroke; these are often referred to as
{\it rowers} because of the analogy with rowing with oars 
\cite{leibler93,lecarpentier01}.
The oars of rowers come in contact with water for a very brief period,
giving a stroke and then comes out of water, completing one cycle.
Similarly, ``rower'' molecular motors also remain attached to their
track for a small fraction of their ATPase cycle, i.e., the duty
ratio of these nonprocessive motors is usually small. However, the
collective stroke of a very large number of such tiny motor molecules
can generate forces large enough to slide filaments over a significant 
distance.
{\it Contractility}, rather than motility, at the subcellular and cellular
level are driven by the sliders and rowers.
Some examples of this category are listed in the table \ref{table-rowslide}. 

\begin{table}
\begin{tabular}{|c|c|c|c|} \hline
Polymer  & mode of force generation & Function (example) & Section \\\hline
MT & polymerization & organizing cell interior & \ref{sec-specificpistonhookspring}\\\hline
F-actin & polymerization & cell motility & \ref{sec-specificpistonhookspring}  \\\hline
FtsZ & polymerization & bacterial cytokinesis & \ref{sec-specificpistonhookspring}  \\\hline
MSP & polymerization & motility of nematode sperm cells & \ref{sec-specificpistonhookspring} \\\hline
Type-IV pili & polymerization  & bacterial motility & \ref{sec-specificpistonhookspring}  \\\hline
MT & de-polymerization & Eukaryotic chromosome segregation & \ref{sec-specificpistonhookspring} \\\hline
spasmin & spring-like & vorticellid spasmoneme &  \\\hline
Coiled actin & spring-like & egg fertilization by sperm cells  &  \\\hline
\end{tabular}
\caption{Force generation by polymerizing/depolymerizing, coiling/uncoiling filaments: pistons, hooks and springs.
 }
\label{table-piston}
\end{table}

\noindent$\bullet${\bf Cytoskeletal polymerizing/depolymerizing filaments: pistons, hooks and springs}

Motor proteins are not the only force generators in a cell. In fact, no 
homolog of motpr proteins have been found so far in prokaryotic cells. 
Dynamic filamentous proteins also generate forces. 
Elongation of a filamentous biopolymer that presses against a light object
(e.g., a membrane) can result in a ``push'' \cite{mogilner03a}.
Similarly, a depolymerizing tubular filament can ``pull'' a light
ring-like object by inserting its hook-like outwardly curled
depolymerizing tip into the ring \cite{mcintosh10}. 
The interplay of the pushing and pulling forces dominate the dynamic 
organization of the cell interior \cite{tolic08}.
A flexible filament, upon compression by input energy, can store energy
that can perform mechanical work when the filament springs back to its
original relaxed shape \cite{mahadevan00}. Some typical examples are
given in table \ref{table-piston}.

The architecture of the diverse MT-based intracellular superstructures 
are determined by a combined operation of the MT-based motor proteins 
and other non-motor MT-associated proteins (MAPs)
\cite{mandelkow95,amos05,maiato04,sedbrook04,hamada07,morrison07,wade09,subramanian12}. 
Similarly, actin-based motor proteins and the non-motor actin-related 
proteins (ARPs) 
\cite{remedios03,muthugapattai04,goley06,carlier98,cooper00,staiger06,pollard07,evangelista03,zigmond04,watanabe04,baum05b,kerkhoff06,anton07}
determine the overall architecture of the actin-based intracellular 
superstructures. 
Some of the superstructures self-organized in an {\it in-vitro} 
motor-filament system, in the absence of MAPs and ARPs, will be 
mentioned in section \ref{sec-mitosis}.
Microtubule plus-end tracking proteins (+TIPs)
\cite{akhmanova08,xiang06,wu06,carvalho03,schuyler01b}
are special MAPs that accumulate
at the plus end of microtubules; depolymerase motors proteins that 
target the plus-end of MT filaments are also +TIPs.

\subsubsection{\bf Machines for synthesis, manipulation and degradation of macromolecules of life}
\label{sec-catalogue2}

\noindent$\bullet${\bf Membrane-associated motors for translocation of macromolecules across membranes} 

In many situations, the motor remains immobile and pulls a macromolecule;
the latter are often called {\it translocase}. Some translocases
{\it export} (or, {\it import}) either a protein \cite{schatz96} or
a nucleic acid strand \cite{burton10,stewart10} across the plasma membrane
of the cell or, in case of eukaryotes, across internal membranes. A list
is provided in table \ref{table-memtranslocase}.

The genome of many viruses are packaged into a pre-fabricated empty
container, called {\it viral capsid}, by a powerful motor attached to
the entrance of the capsid \cite{guo07}.

\begin{table}
\begin{tabular}{|c|c|c|} \hline
Membrane & Polymer & Section \\\hline
Nuclear envelope & RNA/Protein & \ref{sec-specificexim}  \\\hline
Membrane of endoplasmic reticulum & Protein & \ref{sec-specificexim} \\\hline
Membranes of mitochondria/chloroplasts & Protein & \ref{sec-specificexim} \\\hline
Membrane of peroxisome & Protein & \ref{sec-specificexim}  \\\hline
\end{tabular}
\caption{Membrane-bound translocases.
 }
\label{table-memtranslocase}
\end{table}

\noindent$\bullet${\bf Machines for degrading macromolecules of life}

Restriction-modification (RM) enzyme defend bacterial hosts against
bacteriophage infection by cleaving the phage genome while the DNA
of the host bacteria are not cleaved \cite{pingoud05}.
{\it Exosome} and {\it proteasome} are nano-cages into which RNA 
and proteins are translocated and shredded into smaller fragments 
\cite{lorentzen06}. Similarly, there are machines 
for degrading polysachharides, e.g., cellulosome (a cellulose 
degrading machine) \cite{bayer04}, starch degrading enzymes 
\cite{smith05a}, chitinase (chitin degrading enzyme) \cite{chuan06}, 
etc.  These machines are listed in table \ref{table-degradosomes}.

\begin{table}
\begin{tabular}{|c|c|c|} \hline
Polymer  & Examples of Machines & Section/reference \\\hline
DNA (polynucleotide) & RM enzyme & \cite{pingoud05} \\\hline
RNA (polynucleotide) & Exosome & \ref{sec-specificdegraders}  \\\hline
Protein (polypeptide) & Proteasome & \ref{sec-specificdegraders}  \\\hline
Cellulose (polysachharide) & Cellulosome & \cite{bayer04}  \\\hline
Starch (polysachharide) & Starch degrad. enzyme & \cite{smith05a} \\\hline
Chitin (polysachharide) & Chitinase & \cite{chuan06}  \\\hline
\end{tabular}
\caption{Machines for degradation of macromolecules of life.
 }
\label{table-degradosomes}
\end{table}

\noindent$\bullet${\bf Machines for template-dictated polymerization}

Two classes of biopolymers, namely, polynucleotides and polypeptides
perform wide range of important functions in a living cell. DNA and RNA
are examples of polynucleotides while proteins are polypeptides.
Both polynucleotides and polypeptides are made from a limited number of
different species of monomeric building blocks, namely, nucleotides and
amino acids,respectively. The sequence of the monomeric subunits to be used for
synthesis of each of these are dictated by that of the corresponding template.
These polymers are elongated, step-by-step, during their birth by
successive addition of monomers, one at a time. The template itself
also serves as the track for the polymerizer machine that
takes chemical energy as input to polymerize the biopolymer as well as
for its own forward movement. Therefore, these machines are also referred
to as motors.

Depending on the nature of the template and product nucleic acid strands,
polymerases can be classified as DNA-dependent DNA polymerase (DdDP),
DNA-dependent RNA polymerase (DdRP), etc.  as listed in the table
\ref{table-polymerase}.
\begin{table}
\begin{tabular}{|c|c|c|c|c|} \hline
Machine & Template & Product & Function & Section \\ \hline
DdRP & DNA & RNA & Transcription & \ref{sec-specificpoly} \\ \hline
DdDP & DNA & DNA & DNA replication & \ref{sec-specificpoly} \\ \hline
RdRP & RNA & RNA & RNA replication & \ref{sec-specificpoly} \\ \hline
RdDP & RNA & DNA & Reverse transcription & \ref{sec-specificpoly} \\ \hline
Ribosome & mRNA & Protein & Translation & \ref{sec-specificribo} \\ \hline
\end{tabular}
\caption{Types of polymerizing machines, the templates they use
and the corresponding product of polymerization.
 }
\label{table-polymerase}
\end{table}

\noindent$\bullet${\bf Unwrappers, unzippers and untanglers of DNA: chromatin remodellers, Helicases and topoiomerases}

In an eukaryotic cell DNA is packaged in a hierarchical structure
called {\it chromatin}. In order to use a single strand of the
DNA as a template for transcription or replication, it has to be
unpackaged either locally or globally. ATP-dependent chromatin
remodelers \cite{clapier09}
are motors that perform this unpackaging. However,
only one of the strands of the unpackaged duplex DNA serves as a
template; the duplex DNA is {\it unzipped} by a DNA helicase motor
\cite{pyle08}.
Similarly, a RNA helicase motor unwinds a RNA secondary structure.
During DNA replication, a helicase moves ahead of the polymerase,
like a mine sweeper, unzipping the duplex DNA and dislodging other
DNA-bound proteins. However, the transcriptional and translational
machineries do not need assistance of any helicase because these
are capable of unzipping DNA and unwinding RNA, respectively, on
their own.  

In order to control and modulate the DNA topology, a cell uses a 
class of machines designed specifically for this purpose. These 
machines, called topoisomerase, can untangle DNA by passing one 
DNA through a transient cut in another \cite{vos11}.

\noindent$\bullet${\bf Quality control: a delicate balance in an 
unreliable factory}

The molecular machines that synthesize the macromolecules in a cell 
are far from perfect. Therefore, template-directed polymerization 
is an error-prone process. Any defective protein is likely to misfold 
and, therefore, would be unsuitable for its biological function. 
Misincorporation of a nucleotide during the polymerization of a mRNA 
would produce an erroneous template for protein synthesis. Error in 
DNA replication would produce defective genome for the daughter cells. 
In order to maintain macromolecular integrity, each cell has a 
``quality-control system'' \cite{rorth08}. In the context of molecular 
machines for synthesis and degradation of macromolecular machines, the 
following questions are of fundamental interest: (i) does the quality 
control system detect the perfect product or the defective product? 
(ii) Does this detection take place during the ongoing polymerization 
process (e.g, immediately after committing an error) or after the 
product molecule is released by the machinery at the end of synthesis 
of the complete product?
(iii) Is the detection mechanism based on the principles of equilibrium 
thermodynamics or kinetics? 
(iv) Once an error is detected, is the error corrected or is the 
defective product degraded? 
(v) What are the possible short-term and long-term consequences of 
an error if the error escapes detection or/and correction/degradation 
process of the quality control system?\\
Although in this review we focus exclusively on the machines and 
mechanisms that ensure high quality of the macromolecular products, 
the quality-control system of a normal eukaryotic cell acts on  
multiple levels- molecular, organellar as well as cellular levels  
\cite{rorth08}. 

\subsubsection{\bf Rotary motors}

Rotary molecular motors \cite{pilizota09} (see table \ref{table-rotarymotors}) 
are, at least superficially, very similar to the motor of a hair dryer. 
Two rotary motors have been studied most extensively. (i) A rotary motor 
embedded in the membrane of bacteria drive the bacterial flagella 
which, the bacteria use for their swimming in aqueous media.
(ii) A rotary motor, called ATP synthase is embedded on the membrane of 
mitochondria, the powerhouses of a cell. A {\bf synthase} drives a 
chemical reaction, typically the synthesis of some product; the ATP 
synthase produces ATP, the ``energy currency'' of the cell, from ADP.

\begin{table}
\begin{tabular}{|c|c|c|} \hline
Motor & Function & Section \\ \hline
ATP-synthase & Synthesis of ATP & \ref{sec-specificrotary1} \\ \hline
Bacterial flagellar motor & Rotating bacterial flagella & \ref{sec-specificrotary2} \\ \hline
\end{tabular}
\caption{Two major rotary motors.
 }
\label{table-rotarymotors}
\end{table}

\subsection{\bf Fuels for molecular motors}

Energy sources available not only explain the differences in the 
``lifestyles'' of prokaryotes and eukaryotes \cite{lane11} 
but also provides an alternative perspective on the fundamental 
question of the origin of life \cite{deamer97}. It is thermodynamics 
and kinetics which ultimately decided the allowed processes that led 
to the emergence of life from inanimate matter. Throughout the 
subsequent evolution of life, energy has fuelled the machineries in 
living systems \cite{wallace10}. Therefore, this aspect of the 
investigations on molecular machines is intimately related to the 
subject of bioenergetics \cite{gyorgyi57,gyorgyi72,bligh87}.

Several polymerase motors are capable of extracting the required input 
energy directly from the substrates that they use for polymerizing 
a macromolecule. On the other hand, some motors that degrade nucleic 
acids are powered by the free energy released by the degrading nucleic 
acid strand. However, motors that use filamentous polymers as track 
use a separate fuel molecule; in most cases the fuel molecule is 
adenosine tri-phosphate (ATP). Contributions to the input energy for 
a motor come from (a) the binding of ATP, (b) hydrolysis of the bound 
ATP molecule, as well as (c) release of the products of hydrolysis. 

\subsubsection{\bf Chemical fuel generates generalized chemical force}

Before considering any specific chemical reaction that provides the 
input chemical energy for a specific motor, let us keep the discussion 
as general as possible. We consider the reaction 
\begin{equation}
{\cal C}_{1} \mathop{\rightleftharpoons}^{k_f}_{k_r} {\cal C}_{2}
\label{eq-unifuel}
\end{equation}
where higher energy compound $C_1$ gets converted to the lower energy 
compound $C_2$ spontaneously. The forward and reverse fluxes are given 
by $J_f = k_f [C_1]$ and $J_r = k_r [C_2]$, respectively.
In thermodynamic equilibrium of this system, 
\begin{eqnarray}
\frac{k_{f}}{k_{r}} = \frac{[{C}_{2}]_{eq}}{[{C}_{1}]_{eq}} = P_{2}/P_{1} = exp(-\beta \Delta G^{0}), 
\label{eq-massaction}
\end{eqnarray}
i.e., 
\begin{equation}
\Delta G^{0} + k_B T ln\biggl(\frac{[{C}_{2}]_{eq}}{[{C}_{1}]_{eq}}\biggr) = 0 
\label{eq-zeroX}
\end{equation}
where $\Delta G^{0} = G_2 - G_1$, and hence $J_f = J_r$.
What happens if the concentrations of $C_1$ and $C_2$ deviate slightly 
from the equilibrium concentrations? The populations of the 
two molecular species keep changing by conversion from one species 
to the other till the new concentrations again satisfy the equilibrium 
condition (\ref{eq-massaction}). What drives the system towards 
equilibrium and which way does this proceed- forward or reverse?

In order to address the question posed at the end of the last paragraph, 
suppose, there are $n_1$ molecules of ${C}_{1}$ (each of free energy 
$G_1$) and $n_2$ molecules of ${C}_{2}$ (each of free energy $G_2$). 
The corresponding free energy of the entire system is given by 
\begin{eqnarray} 
G_{i} = n_1 G_1 + n_2 G_2 + (n_1+n_2) k_B T \biggl[\biggl(\frac{n_1}{n_1+n_2}\biggr) ln\biggl(\frac{n_1}{n_1+n_2}\biggr)+\biggl(\frac{n_2}{n_1+n_2}\biggr) ln\biggl(\frac{n_2}{n_1+n_2}\biggr)\biggr]
\label{eq-Gini}
\end{eqnarray}
If one molecule of $C_1$ now gets converted to one molecule of $C_2$ 
by the reaction (\ref{eq-unifuel}), the new free energy of the 
system can be obtained from (\ref{eq-Gini}) by replacing $n_1$ and 
$n_2$ by $n_1-1$ and $n_2+1$, respectively. Let us denote the 
corresponding change in the free energy of the entire system by 
$\Delta G$. When $n_1$ and $n_2$ are sufficiently large, it is 
straightforward to show that 
\begin{equation}
\Delta G \simeq \Delta G^{0} + k_B T ln\biggl(\frac{[{C}_{2}]}{[{C}_{1}]}\biggr)
\label{eq-deltaG}
\end{equation}
Comparing eqn.(\ref{eq-deltaG}) with eqn.(\ref{eq-zeroX}), we find 
that $\Delta G$ vanishes in equilibrium. Moreover, $\Delta G^{0}$ 
indicates merely the direction of spontaneous conversion of a molecule 
with high free energy into a molecule of low free energy. But, when 
the concentrations of the molecules deviate from equilibrium, it is 
$X=\Delta G$ that drives the chemical system towards equilibrium. 
Furthermore, change in the free energy caused by the conversion of one 
molecule is identical to the change in the chemical potential $\Delta \mu$. 
Therefore, we define $\Delta \mu = \Delta G$ as the ``generalized chemical 
force'' $X$.

\noindent {\bf Example 1: ATP hydrolysis vs. ATP synthesis}

\begin{figure}[htbp]
\begin{center}
\includegraphics[angle=-90,width=0.45\columnwidth]{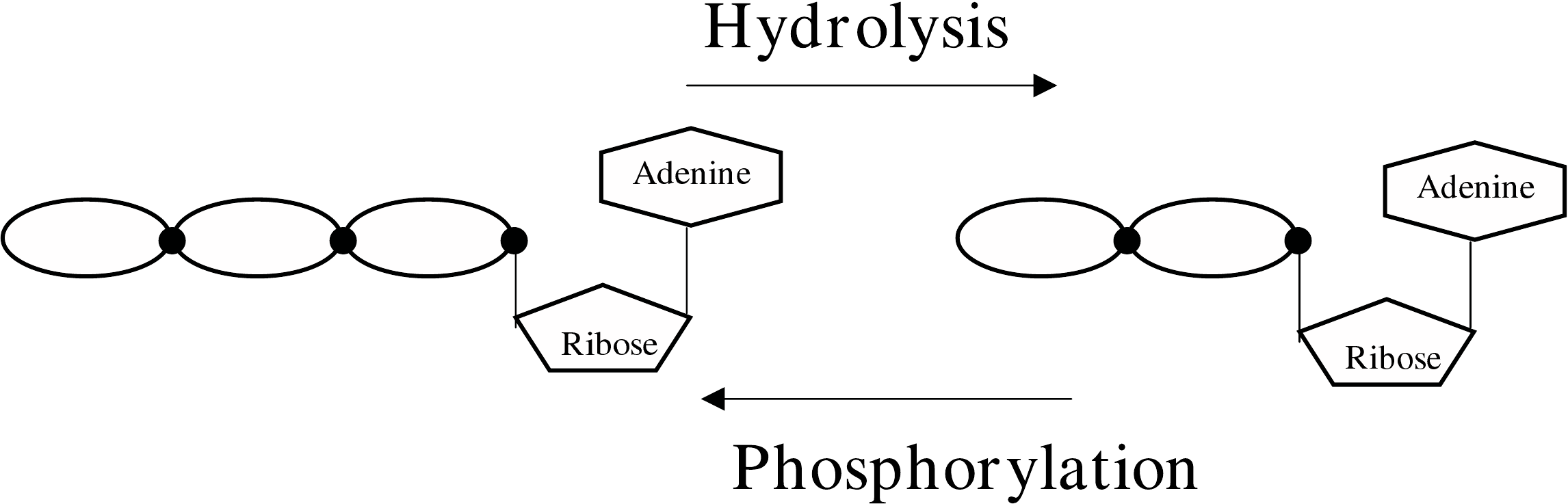}
\end{center}
\caption{Schematic representation of hydrolysis of adenosine 
triphosphate (ATP) to adenosine diphosphate (ADP).
}
\label{fig-atphydrolysis}
\end{figure}

The most common way of supplying energy to a natural nano-motor is
to utilize the chemical energy (or, more appropriately, free energy)
released by a chemical reaction. Most of the motors use the so-called
``high-energy compounds''- particularly, nucleoside triphosphates (NTPs)-
as an energy source to generate the mechanical energy required
for their directed movement. However, the term ``high-energy compound'', 
although widely used colloquially, is confusing. Here ``high-energy'' or 
``energy-rich'' merely means that the free energy change $\Delta G^{0}$ 
associated with the chemical reaction, that the compound undergoes to 
supply input (free-)energy for the motor, is {\it strongly negative} 
\cite{peschek11}.  The most common chemical reaction is the
{\it hydrolysis} of ATP to ADP $ ATP \rightarrow ADP + P_i$ (see 
fig.\ref{fig-atphydrolysis}).
ATP analogoues \cite{bagshaw01} are very useful substitutes for normal
ATP for exploring the role of ATP in the operational mechanism of a
molecular motor.

Some other high-energy compounds can also supply input energy; one
typical example being the hydrolysis of Guanosine Triphosphate (GTP)
to Guanosine Diphosphate (GDP). Inorganic pyrophosphate (PP$_{i}$), 
which forms naturally during the hydrolysis of ATP into Adenosine 
mono-phosphate (AMP) by the reaction ATP $\to$ AMP + PP$_i$, is also  
used as fuel in some living systems \cite{serrano07a,rea93}. 
Interestingly, PP$_i$ is a member of the family of inorganic 
polyphosphates \cite{kornberg99,achbergerova11} which are believed 
to be an ancient energy source in living systems.

For the reaction (see fig.\ref{fig-atphydrolysis})
\begin{equation}
ATP  \rightleftharpoons  ADP + P_i,
\end{equation}
\begin{equation}
X = \Delta G = \Delta G_0 - k_B T ~~ln \frac{[ATP]_c}{[ADP]_c [P_i]_c}
\label{eq-deltagATP}
\end{equation}
where $\Delta G_0 = - 54 \times 10^{-21} J$.

A curiosity in the choice of phosphates as the energy currency of the 
cell: why did Nature choose phosphates? This question can be answered 
only by examining its relative stability and its enhanced rate of 
hydrolysis by the enzymes as compared to alternative compounds which 
might have been available to Nature during the course of evolution. 
It has been argued \cite{westheimer87}
that phosphates were the best choice for Nature but, perhaps, not
for the present-day organic chemists.

If a cyclic machine runs on a specific chemical fuel then the spent
fuel must be removed as waste products and fresh fuel must be supplied
to the machine. Fortunately, normal cells have machineries for
recycling waste products to manufacture fresh fuel, e.g., synthesizing
ATP from ADP. This raises an important question: since ATP is a
higher-energy compound than ADP, how are the ATP-synthesizing machines
driven to perform this energetically ``uphill'' task? Fortunately,
chemical fuel is not the only means by which input energy can be
supplied to intracellular molecular machines; ATP synthesis is driven 
by ion-motive force (IMF) that we discuss in the next subsubsection.

\subsubsection{\bf Electro-chemical gradient of ions generates ion-motive force}

During Darwinian evolution, cells seem to have selected only two atomic
species of ions for the electro-chemical gradient- hydrogen ion $H^{+}$
(which is is essentially a single proton) and sodium ion $Na^{+}$.
The evolutionary advantages of these two ionic species over all other
possible candidates and the sequence in which these might have been
selected in the course of evolution are still debated but will not be
discussed here
\cite{skulachev88,skulachev92,konings94,konings06,speelmans95,mulkidjanian08,mulkidjanian09}.

An electro-chemical gradient of protons across the membrane of a cell or 
that of an organelles of an eukaryotic cell generates the proton-motive force 
(PMF). 
The strength of the PMF is generally expressed in terms of the free energy 
$\Delta G$ required to create it. Suppose $V$ denotes the electric potential 
and $[H]$ is the concentration of the protons (hydrogen ions). Traditionally, 
in the literature on active transport across membranes \cite{bligh87,inesi94} 
$\Delta G$ is expressed as
\begin{equation}
\Delta G = RT ln\frac{[H]_{in}}{[H]_{out}} +  F(V_{in}-V_{out}).
\label{eq-elecgrad}
\end{equation}
where the subscripts $in$ and $out$ refers to inside and outside of the
membrane-bound compartment, $F$ is the Faraday constant and $R$ is the 
gas constant ($R = N_{A} k_B$ where $N_A$ is the Avogadro number). 
The first and second terms on the right hand
side of (\ref{eq-elecgrad}) correspond to the concentration (chemical)
gradient and electrical potential gradient, respectively. Since
$pH=log_{10}(1/[H])$ and since  $ln_{e} x = 2.303 log_{10} x$ equation
(\ref{eq-elecgrad}) can also be recast in terms of the pH values on the
two sides of the membrane. A similar expression describes the sodium-motive
force (SMF) generated by the electro-chemical gradient of sodium ions 
\cite{dimroth97}. 

\subsubsection{\bf Some uncommon energy sources for powering mechanical work}

The spring-like action of spasmoneme is powered neither by any NTP nor 
by any IMF. Instead, binding of Ca$^{2+}$ ions causes contraction of 
the spring thereby storing elastic energy that is later released when 
the spring rapidly extends to its full length because of the unbinding 
of the calcium ions 
\cite{mahadevan00}. 
Similarly, the switching of a forisome from a spindle-like elongated 
shape to a balloon-like swollen plug is energized also by the binding of 
calcium ions 
\cite{knoblauch04,pickard06,tuteja10,tuteja10b}. 
In living plants movements can be caused by the variation of internal 
pressure (also called turgor) of cells that arise from uptake or loss 
of water \cite{martone10}.
However, we'll not discuss these mechanisms of force generation in this 
review. 

For designing artificial nanomotors, light is often the preferred 
choice as the input energy.
The advantages of using light, instead of chemical reaction, as the
input energy for a molecular motor are as follows: (i) light can be
switched on and off easily and rapidly, (ii) usually, no waste
product, which would require disposal or recycling, is generated.

\subsubsection{\bf Manufacturing energy currency from external energy supply}

A cell gets its energy from external sources. It has special machines
to convert the input energy into some ``energy currency''. For example,
chemical energy supplied by the food we consume is converted into an
electro-chemical potential $\Delta \mu$ that not only can be used to
synthesize ATP, but can also directly run some other machines.
In plants similar proton-motive forces are
generated by machines which are driven by the input sunlight.

Thus, study of molecular machines deals with two complementary
aspects of bioenergetics: (a) conversion of energy input from the
external sources into the energy currency of the cell, and (b)
utilization of the energy currency to drive various other
active processes \cite{peschek11}.

ATP was discovered by Lohmann and, independently, by Fiske and Subbarow 
\cite{maruyama91,simoni02,langen08}. 
As we will discuss in section \ref{sec-specificrotary1}, synthesis of 
ATP from ADP is driven by a PMF (or, SMF). But, the mechanism of 
generating the PMF (and, SMF) from metabolic energy was discovered 
by Peter Mitchell \cite{orgel99,pasternak93,prebble01,prebble02a,prebble02b,weber06a}. 
For the history of the discoveries of the reaction chains that convert 
other forms of input energy into the standard energy currencies of the 
cell, see ref.\cite{edsall74,gest02}).

\subsection{\bf Some basic concepts}

\subsubsection{\bf Directionality, processivity and duty ratio}

All the members of a distinct family of motor protein moves in a specific 
direction on its track which is a polar filament, i.e, whose two ends are 
not equivalent. One of the key features of the kinetics of molecular
motors is their ability to attach to and detach from the corresponding
track. A motor is said to be attached to a track if at least one
of its domains remains bound to the corresponding track.

One can define processivity in three different ways:\\
(i) Average number of {\it chemical cycles} in between attachment and
the next detachment from the filament;\\
(ii) {\it attachment lifetime}, i.e., the average time in between an
attachment and the next detachment of the motor from the filament;\\
(iii) {\it mean distance} spanned by the motor on the filament in
a single run.\\
The first definition is intrinsic to the process arising from the
{\it mechano-chemical} coupling. But, it is extremely difficult to
measure experimentally. The other two quantities, on the other
hand, are accessible to experimental measurements.

Leibler and Huse \cite{leibler93} presented a unified scenario for the
function of the cytoskeletal motor proteins and argued that the different
processivities of the motors arise from the different rate limiting
processes in their mechano-chemical cycle. The details of their
arguments will be examined in part II.
To translocate processively, a motor may utilize one of the three
following strategies:\\
{\bf Strategy I}: the motor may have more than one track-binding domain
(oligomeric structure can give rise to such a possibility quite
naturally). Most of the cytoskeletal motors, like conventional two-headed
kinesin, use such a strategy. One of the track-binding sites remains
bound to the track while the other searches for its next binding site. \\
{\bf Strategy II}: A motor may possess non-motor extra domains or some
accessory protein(s) bound to it which can bind to the track even when
none of the motor domains of the motor is directly attached to the track.\\
{\bf Strategy III}: it can use a ``clamp-like'' device to remain attached
to the track; opening of the clamp will be required before the motor
detaches from the track. Many motors utilize this strategy for moving
along the corresponding nucleic acid tracks.

During one cycle, suppose a motor spends an average time $\tau_b$
bound (attached) to the filament, and the remaining time $\tau_{u}$
unbound (detached) from the filament. 
Clearly, the period during which it exerts its {\it working} stroke
is $\tau_{b}$ and its {\it recovery} stroke takes time $\tau_{u}$.
The {\it duty ratio}, $r$, is defined as the fraction of the time
that each head spends in its attached phase, i.e.,
\begin{equation}
r = \tau_{b}/(\tau_{b} + \tau_{u})
\end{equation}

\subsubsection{\bf Force-velocity relation and stall force}

An external force that opposes the natural directed movement of a motor 
is called a {\it load} force. As the strength of the load force 
is increased, the average velocity $V$ of the motor decreases. 
The force-velocity relation $V(F)$ is one of the most important 
characteristics of a molecular motor; its status in the studies 
of molecular motors is comparable to that of the I-V characteristics 
of a device in the studies of semiconductors.  
The minimum load force $F_{s}$ which just stalls the motor is called
the {\it stall force} and it is the true measure of the maximum force
that a motor can generate.

The force-velocity relation can have the general form \cite{kunwar10} 
\begin{equation}
V(F) = V(0)[1-(F/F_{s})^{\alpha}]
\end{equation}
where $V(F)$ is the average velocity of the motor in the presence of 
a load force $F$ and $F_{s}$ is the stall force. Both the unloaded 
velocity $V(0)$ and the stall force $F_{s}$ are important measurable 
characteristics of a motor. The magnitude of $\alpha$ determines the 
curvature of the plot. 
For example, $\alpha=1$ corresponds to a linear force-velocity relation. 
In contrast, {\it sub-linear} and {\it super-linear} force-velocity 
relations, which arise for $\alpha < 1$ and $\alpha > 1$, respectively, 
appear convex-up and concave-up when plotted graphically. 

What happens to the motor if it is subjected to a load force that is
stronger than the stall force? Clearly, there are three possibilities: \\
(a) the motor may simply detach from the track;
(b) the motor may walk backward (i.e., in a direction that is opposite
to its natural direction of motion in the absence of any load force),
but this motion is driven by the load force alone because the motor
no longer hydrolyzes any ``fuel'' molecule;
(c) the motor walks backward, but is hydrolyzes ``fuel'' molecules
exactly the same way as it does while moving forward for load forces
$F < F_{s}$. Can a motor synthesize, instead of hydrolyzing, ATP 
while walking under the action of load force $F > F_{s}$ opposite to 
its natural direction of motion? We shall see in part II that, different 
families of motors exercise different options among (a), (b) and (c) above.
The load force need not be directed exactly parallel to the filament. 
Consequences of vectorial loading of molecular motors have also been 
investigated \cite{fisher05,kim05}

\subsubsection{\bf Mechano-chemical Coupling: slippage and futile cycles}

A molecular motor has to coordinate its three cycles:
(a) enzymatic cycle in which it hydrolyzes one molecule of the
``fuel'' (ATP or GTP);
(ii) cycle of attachment to- and detachment from the track; and
(iii) stepping cycle in which it moves forward or backward on the
track by one mechanical step.

In this context, some of the fundamental questions on the nature and
strength of the mechano-chemical coupling are as follows:\\
(i) how many cycles of hydrolysis of ATP (or GTP) occurs during a
single cycle of mechanical stepping of the motor? \\
(ii) Is ``slippage'' between the chemical cycle of ATP (or GTP)
hydrolysis and the mechanical cycle of stepping possible? In other
words, is it possible that hydrolysis of fuel turns out to be ``futile''
in the sense that it does not lead to any stepping of the motor?
For such motors, the output is loosely coupled to the input; the output
work extracted from the same amount of input energy (e.g., hydrolysis
of a single ATP molecule) fluctuates from one cycle to another
\cite{oosawa00}.

This is in sharp contrast the output of macroscopic motors are usually tightly coupled to the
corresponding input; the chemical energy is converted into mechanical
work via a strictly scheduled sequence of stages where in each stage
there is one-to-one correspondence between the movements of the parts
of the motor and the work done. We define the strength of the coupling by
\begin{equation}
\kappa = \frac{({\rm {average~ velocity~ of~ motor}})}{({\rm {average~ rate~ of~ reaction}}) \times {\ell}}
\label{eq-mccoupling}
\end{equation}
where ${\ell}$ is the step size. Note that $\kappa$ is the probability
that the motor takes a mechanical step in space per chemical reaction.
Tight coupling corresponds to $\kappa = 1$ whereas all $\kappa < 1$
if the coupling is loose. Moreover, $\kappa > 1$ if the motor can
take more than one mechanical step per cycle of chemical reaction.

\section{\bf Experimental methods for molecular motors: ensemble-averaged and single-molecule techniques}

Most of the traditional experimental techniques of biophysics and
biochemistry relied on collection of data for a large collection of
molecules and thereby getting their ensemble-averaged properties. The
amplification of the signals caused by the presence of large number
of such molecules makes it easier to detect and collect the data.
However, there are practical limitations of the bulk measurements in the
specific context of understanding the operational mechanisms of cyclic
molecular machines because it is practically impossible to synchronize
their cycles. That's why single-molecule techniques are required.
The single molecule of interest also acts like a {\it reporter} of the
local ``nano-environment'' because its own properties are influenced by
those of the molecules in its immediate surroundings \cite{moerner03}.

Thus, experimental techniques for probing the operational mechanisms of
molecular motors can be divided broadly into two groups \cite{tinoco11}:


\centerline{\framebox{Experimental techniques}}

\centerline{~~~~$\swarrow$ ~~ $\searrow$}

\centerline{{\framebox{Ensemble-averaged}} ~~{\framebox{Single-molecule}}}


The advantages of single-molecule techniques are as follows:
(a) Single molecule imaging exposes the inhomogeneity and disorder
in a sample even when the dynamic inhomogeneities average out over
longer period of time, (b) enable the observer to monitor a molecule
as it moves in a complex fluid medium, (c) probe the kinetics of
the molecule and reveal even the rare pathways which would not be
detected in the ensemble-averaged measurements over bulk systems,
while the single-molecule techniques of manipulation, in addition,
yield (d) quantitative measures of forces, distances and velocities.
The basic principles of some of the most useful techniques are 
summarized in appendix \ref{app-expttech}.

\section{\bf Chemical physics of enzymatic activities of molecular motors: concepts and techniques}
\label{sec-chemphys}

The input (free-) energy of chemo-mechanical molecular motors is derived 
from {\it chemical reactions}. Therefore, in order to understand the 
mechanisms of bio-molecular motors, it is necessary to understand not 
only how these move in response to the mechanical forces but also how 
these are affected by generalized ``chemical forces''. 
Enzymes and ribozymes constitute two classes of biological catalysts; 
enzymes are proteins whereas ribozymes are RNA molecules. As we'll 
illustrate in parts II, most of the molecular motors considered here 
are either enzymes or consist of ribozymes. Therefore, it is desirable 
to have some background knowledge in the theory of enzymatic reactions 
before embarking on a study of bio-molecular motors. 

For any motor that doesn't step backwards, the position of its center 
of mass advances in the forward direction by one step at a time. 
Similarly, if a single enzyme molecule catalyzes a chemical reaction 
that is practically irreversible, the population of the product molecules 
increases by one in each enzymatic cycle. 
In recent years this formal analogy between mechanical stepping and 
enzymatic reaction has enriched the fields of biophysics and chemical 
biology by exchange of novel ideas. Molecular motors have the unique 
distinction of involving both these phenomena in its core mechanism 
of operation.
The main aim of this section is to provide a brief summary of the 
essential concepts and techniques for studying enzymatic reactions, 
particularly in the context of molecular motors.

\begin{figure}[htbp]
\begin{center}
\includegraphics[angle=-90,width=0.45\columnwidth]{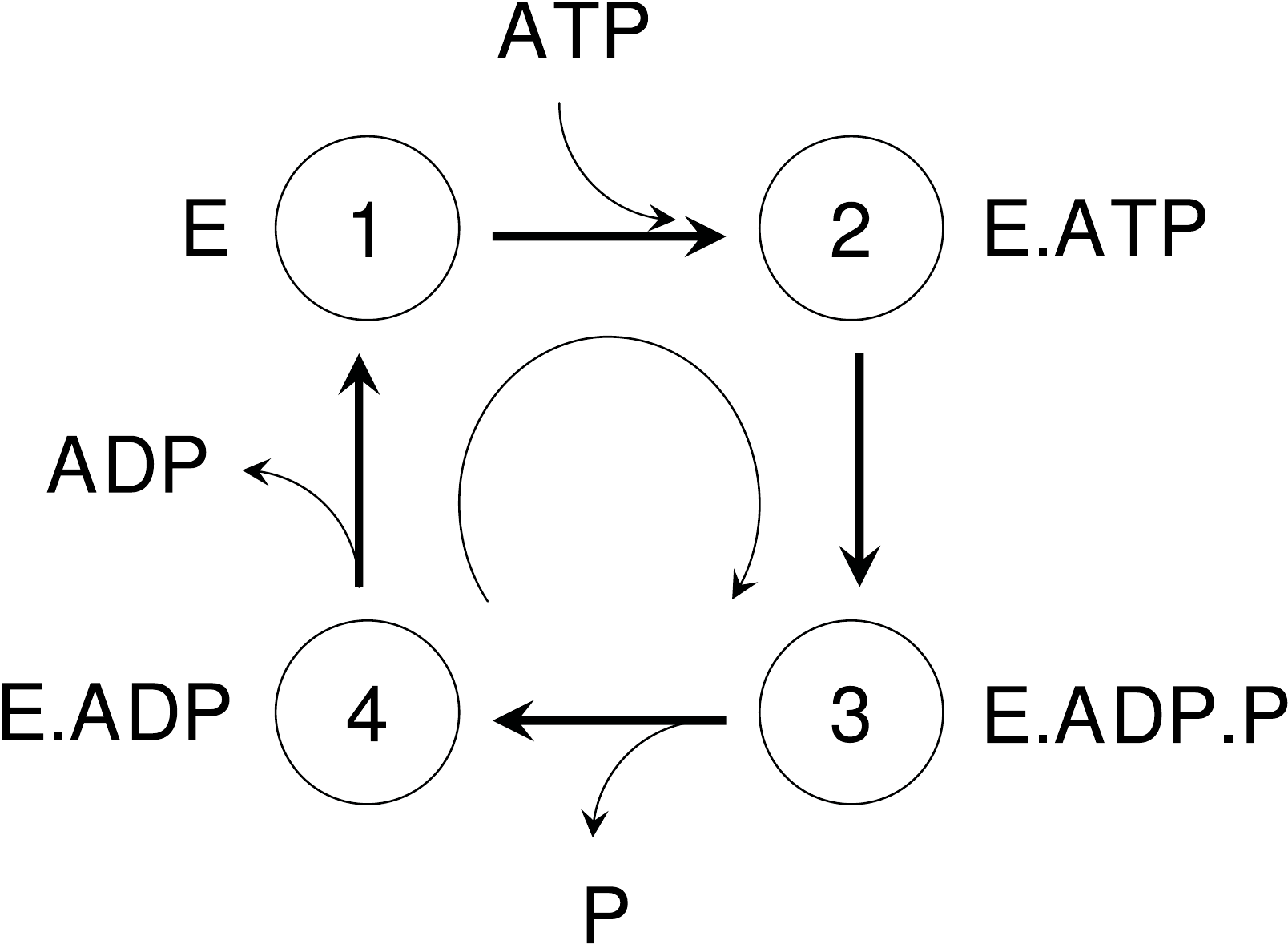}
\end{center}
\caption{A biochemical cycle, consisting of four states, of a 
typical ATPase enzyme. Binding of ATP with the enzyme E 
leads to the formation the complex E.ATP. Hydrolysis of ATP, 
catalyzed by E, produces ADP and P. The enzyme returns to its 
original state, and is ready for the next cycle, after releasing 
sequentially the products of hydrolysis, namely P and ADP.
}  
\label{fig-enzyme}
\end{figure}

Two classes of enzymes that are most relevant in the context of
molecular motors are the (i) ATPases (which hydrolyze ATP), and
(ii) GTPases (which hydrolyze GTP) 
\cite{bourne95}. For obvious reasons, proposals
have been made to include these NTPases in one single class and
to name the class an ``energases'' \cite{purich01} although this
proposal has been criticised \cite{purich01}.

Even for a given single motor domain, a large number of chemical states
are involved in each enzymatic cycle. In principle, there are, many
{\it pathways} for the hydrolysis of ATP, i.e., there are several
different sequences of states that defines a complete hydrolysis cycle.
Although, all these pathways are allowed, some paths are more likely
than others. The most likely path is identified as the {\it hydrolysis
cycle}.  Let us consider an ATPase,
an enzyme that hydrolyzes ATP (see Fig.\ref{fig-enzyme}). Under normal
conditions, the spontaneous rate of hydrolysis of ATP extremely low.
However, ATPases are enzymes which specifically speed up this reaction.

\subsection{\bf Enzymatic reaction in a cell: special features and levels of theoretical description}

All chemical reactions are intrinsically {\it reversible} and have the
general form {\sl Reactants} $\rightleftharpoons$ {\sl Products}. However,
if the rate of the reverse reaction is very small compared to that of
the forward reaction, or if the products are continuously removed from
the reaction chamber as soon as thes are formed, the reaction becomes,
effectively, {\it irreversible} and takes the form
{\sl Reactants} $\rightarrow$ {\sl Products}.
Chemical kinetics is a framework for studying how fast the amounts of
reactants and products change with time.

\noindent $\bullet${\bf Special features of enzymatic reactions in-vivo}

How do chemical reactions within cells differ from those occuring
{\it in-vitro}? (i) First, because of the dense crowd of molecules
in a solution, the reactions take place in the presence of a
``background'' that occupies a large fraction of the cell itself.
Consequently, even if this background does not participate actively
in the reaction, it can (a) shift the equilibrium concentrations of 
the reactants and the products, and (b) change the reaction rates 
\cite{minton06,zhou08,schnell04}. 
For example, the reactant and/or product molecules may be adsorbed 
reversibly and non-specifically on a nearby fiber or membrane that 
is a constituent of this ``background''; such adsorption can influence 
the course of the reaction. A concrete example is that of a molecular 
motor that hydrolyzes ATP; the rate of ATP hydrolysis depends on 
whether or not the motor is interacting with a filamentous track. 
(ii) Second, interior of a cell is so inhomogeneous that the rate of 
the same biochemical reaction may vary significantly depending on the 
location of the reaction. 
(iii) Third, for many reactions inside a cell, the population of the 
reactants can be so low that rate of the reaction may fluctuate strongly 
from one instant to another, even at the same spatial location. For 
example, the duration of the ATPase cycle of a molecular motor is a 
fluctuating quantity.

\noindent $\bullet${\bf Levels of description in theories of chemical reactions}

\centerline{{\framebox{Rate equations: deterministic ODEs; no spatial fluctuation- well-stirred approximation}}}

\centerline{$\downarrow$}

\centerline{{\framebox{Reaction-diffusion equations: deterministic PDEs; spatial variations captured}}}

\centerline{$\downarrow$}

\centerline{{\framebox{Chemical master- or chemical Fokker-Planck/Langevin equations: stochastic description}}}

\centerline{$\downarrow$}

\centerline{{\framebox{Quantum-mechanical theories (quantum chemistry) }}}

Modelling the electronic processes through which chemical bonds are
made and broken would require a quantum mechanical formalism. However,
our interest in this article is restricted to phenomena which occur
on {\it length scales} longer than the spatial extent of the molecules
and on {\it time scales} longer than those of electron dynamics. The
effects of the electronic degrees of freedom get averaged out on the
length and time scales of our interest. Therefore, we do not present
here the quantum mechanical formalisms of chemical reaction kinetics.

Theory of chemical reactions can be developed at several different 
levels depending on the purpose of the investigation
\cite{andrews06,andrews09,grima08}.
Moreover, at a given level, the equations can be formulated at least
in two different ways: (i) equations which govern the time evolution
of the {\it populations} of the molecular species involved in the
reaction, (ii) equations which describe the motion of {\it individual}
molecules
\cite{tolle06}.
Furthermore, equations for chemical kinetics are often developed 
ignoring the possibility of spatial variations. However, spatial 
variations in the populations of the reactants and products can be 
taken into account, for example, by replacing the ordinary 
differential equations by partial differential equations. Finally, 
depending on the physical situation and the level of description,
the equations of chemical kinetics can be either deterministic or
stochastic. (A brief technical summary of these alternative 
formulations of chemical reaction kinetics is presented in appendix 
\ref{app-chemreaction}).

\subsection{\bf Enzyme as a chemo-chemical cyclic machine: free energy transduction}

A general cyclic reaction can be written as
\begin{equation}
{\cal E}_{1} {\rightleftharpoons} {\cal E}_{2} {\rightleftharpoons} ...... {\rightleftharpoons} {\cal E}_{n} {\rightleftharpoons} {\cal E}_{1}
\end{equation}
In this section we show how cyclic chemical reaction can be exploited
to design a chemo-chemical machine for which both input and output
are chemical energies \cite{hillbook,kamp88}. Such machines are chemical
analogues of simple mechano-mechanical machines like a simple lever.

In order to motivate the design of a chemo-chemical machine, consider
a reaction
\begin{equation}
A \rightarrow C
\label{eq-fav}
\end{equation}
which is {\it strongly favored} as the corresponding change of free
energy $\Delta G = G_C - G_A \ll 0$. On the other hand, the reaction
\begin{equation}
B \rightarrow D
\label{eq-unfav}
\end{equation}
is {\it weakly unfavored} as the corresponding
$\Delta G = G_D - G_B > 0$.  So, given an opportunity, $A$
molecules will spontaneously transform to $C$ whereas $D$ molecules
will spontaneously transform into $B$. Is it possible to utilize the
large change of free energy of the first reaction (\ref{eq-fav}) to
drive the second reaction (\ref{eq-unfav})? If this is possible, this
would be an example of ``free energy transduction'' and system would
operate as a chemo-chemical machine. Some of the free energy released
in the reaction (\ref{eq-fav}) is, then, used to pay the free energy
cost required to drive the unfavorable reaction (\ref{eq-unfav}).

On many occasions it is hard to see how the two reaction would couple
together to transduce the free energy on their own. On the other hand,
free energy transduction is quite common in living cells; in these
processes, usually, a large protein molecule or a macromolecular
complex plays the role of a ``broker'' or a ``middleman''. In fact,
most of the molecular motors we consider here fall in this category
of ``brokers''.

\noindent $\bullet${\bf Example of a chemo-chemical machine}

\begin{figure}[htbp]
\begin{center}
\includegraphics[angle=-90,width=0.45\columnwidth]{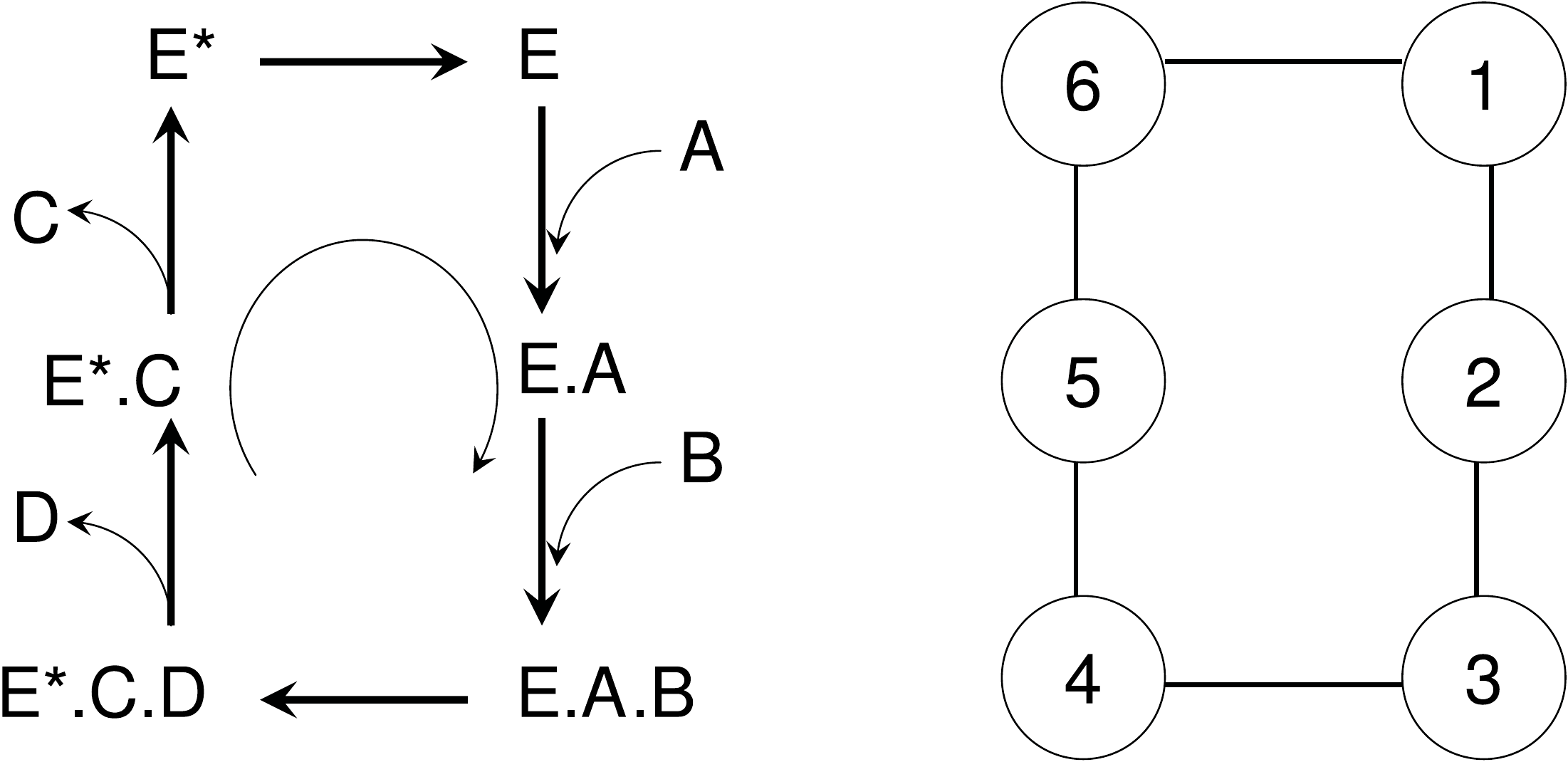}
\end{center}
\caption{The kinetic states and transitions of a chemo-chemical 
machine that drives the free-energetically unfavorable reaction 
(\ref{eq-unfav}) by coupling it to a highly favorable reaction 
(\ref{eq-fav}). The six distinct states on the left panel 
are labelled by the integers 1-6 on the right panel. E and 
E$^{*}$ are two distinct ligand-free conformational states of 
the same enzyme. The straight arrows denote transitions whereas 
the curved arrows indicate binding of substrates and release of 
products. The semicirclar arrow shows the overall direction of 
the enzymatic process.
}
\label{fig-cycle1}
\end{figure}

To illustrate the mechanism let us consider a hypothetical (but, in
principle, possible) model shown in fig.\ref{fig-cycle1} where
$E$ is the enzyme. Note that $E$ exists in this model in two different
conformational states, denoted by $E$ and $E*$, which are interconvertible.
There is one binding site for $A$ and another for $B$ on the same
conformation $E$ of the enzyme. On the other hand, in the conformational
state $E*$ of the enzyme, these binding sites are accessible only to
the molecules $C$ and $D$. Therefore, once $A$ and $B$ bind to their
respective binding sites on $E$, the enzyme makes a transition to the
state $E*$ forcing $A$ and $B$ to make the corresponding transitions
to $C$ and $D$, respectively.

\begin{figure}[htbp]
\begin{center}
\includegraphics[angle=-90,width=0.45\columnwidth]{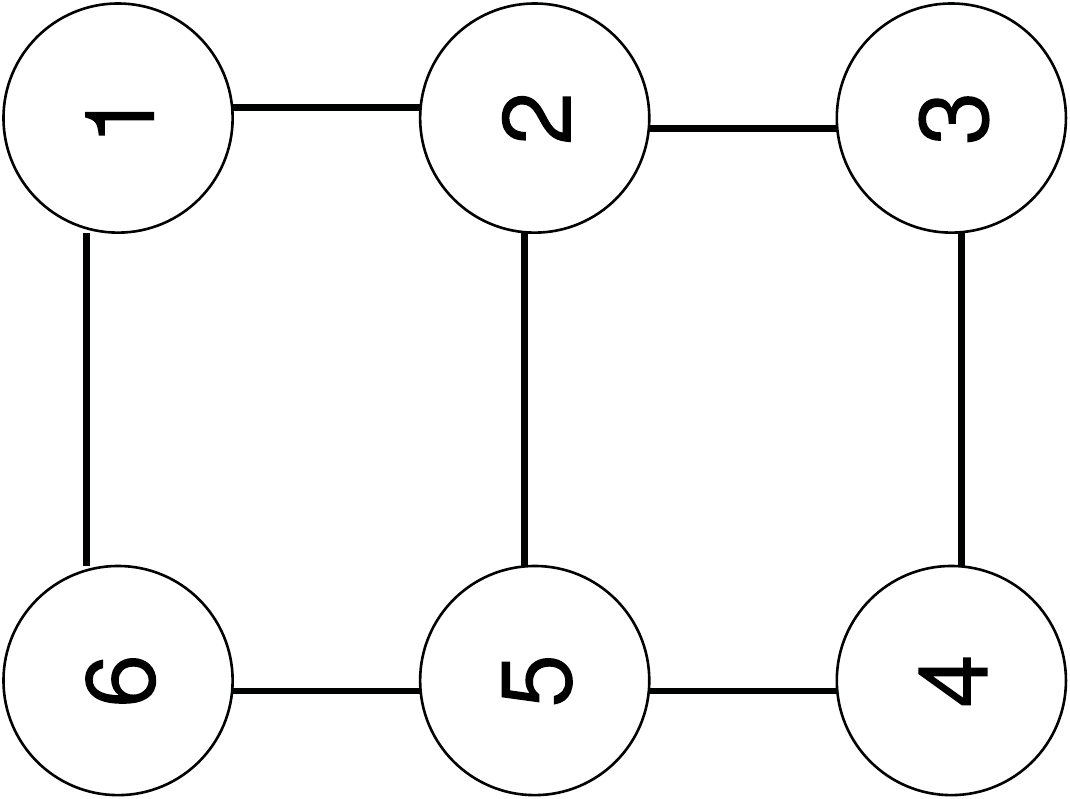}
\end{center}
\caption{A generalization of the cycle shown in fig.\ref{fig-cycle1} 
by allowing direct transition between $EA$ and $E*C$.
}
\label{fig-cycle2}
\end{figure}

Thus, in this model, the enzyme exists in six states numbered by the
sequence of integers shown in fig.\ref{fig-cycle1}. If one enzyme
completes one cycle in the clockwise (CW) diretion, the net effect is
to convert one $A$ molecule and one $B$ molecule into one $C$ molecule
and one $D$ molecule; the enzyme itself is not altered by the complete
cycle.

With the use of only one cycle, as shown in fig.\ref{fig-cycle1},
there is {\it tight} coupling between the two reactions (\ref{eq-fav})
and (\ref{eq-unfav}), i.e., the stoichiometry is exactly one-to-one:
each complete cycle coverts exactly one $A$ and one $B$ into exactly
one $C$ and one $D$ molecule.

\begin{figure}[htbp]
\begin{center}
\includegraphics[angle=90,width=0.45\columnwidth]{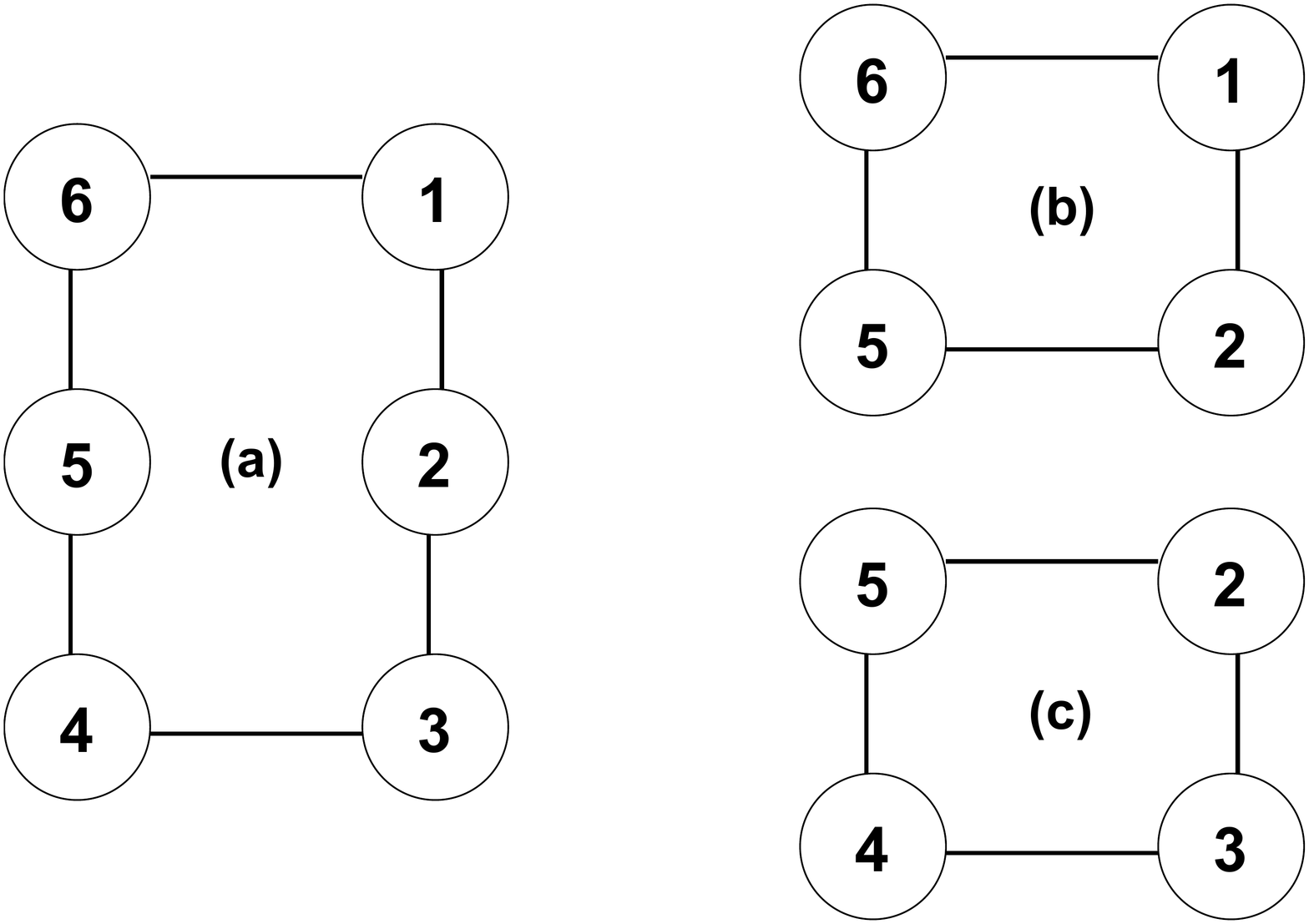}
\end{center}
\caption{Three elementary cycles that are possible in the kinetic model 
shown in fig.\ref{fig-cycle2}.
}
\label{fig-cycle3}
\end{figure}

Fig.\ref{fig-cycle2} is a generalization of the model where
possible transitions between $EA$ and $E*C$ are now also included.
This small extension has non-trivial consequences as we shall explain
below. Note that now there are three possible cycles as shown in
the figure \ref{fig-cycle3}. The possible directions are chosen
arbitrarily in the CW direction in all three cycles. As explained
above, cycle (a) transduces free energy. The cycle (b) runs
spontaneously; but, from the point of view of free energy transduction,
this cycle does not contribute and it simply dissipates some of the
free energy of $A$. The cycle (c), which runs opposite to the direction
of spontaneous progress of the reaction, is an wasteful cycle from
the perspective of free energetics. Only the cycle (a) transduces
free energy. However, if all the cycles (a), (b) and (c) occur, the
cycles (b) and (c) spoil the exact stoichiometry thereby reducing the
overall efficiency of the free energy transduction. More precisely,
if the transitions between $EA$ and $E*C$ occur, the tight coupling
of the model is lost because of the ``slippage'' caused by the
cycles (b) and (c) converting the model into a ``weak-coupling'' model.
Thus, for free energy transduction, the kinetic diagram must have
at least one cycle that involves both free energy supply and free-energy
demanding transitions.

An appropriate measure of the efficiency of the free energy transduction
in any chemo-chemical machine is given by
\begin{equation}
\eta_{ch} = \frac{(\Delta G)_{out}}{-(\Delta G)_{in}}
\end{equation}
For the abstract chemo-chemical machine designed above,
\begin{equation}
\eta_{ch} = \frac{(\Delta G)_{B \to D}}{-(\Delta G)_{A \to C}}
\end{equation}

\subsection{\bf Enzymatic activities of molecular motors}

Although, for appreciating the mechanisms of molecular motors, one has 
to be familiar with the enzymatic activities of only the motors, we 
present the basic principles from a much broader perspective 
\cite{jencks75,kraut03,kamerlin10}. 
Implications of the rate of the enzymatic activities of a motor for 
its mechanical movements will be examined repeatedly in this review 
for the generic models as well as for the models of specific motors.

\subsubsection{\bf Average rate of enzymatic reaction: Michaelis-Menten equation}

\begin{figure}[htbp]
\begin{center}
\includegraphics[width=0.45\columnwidth]{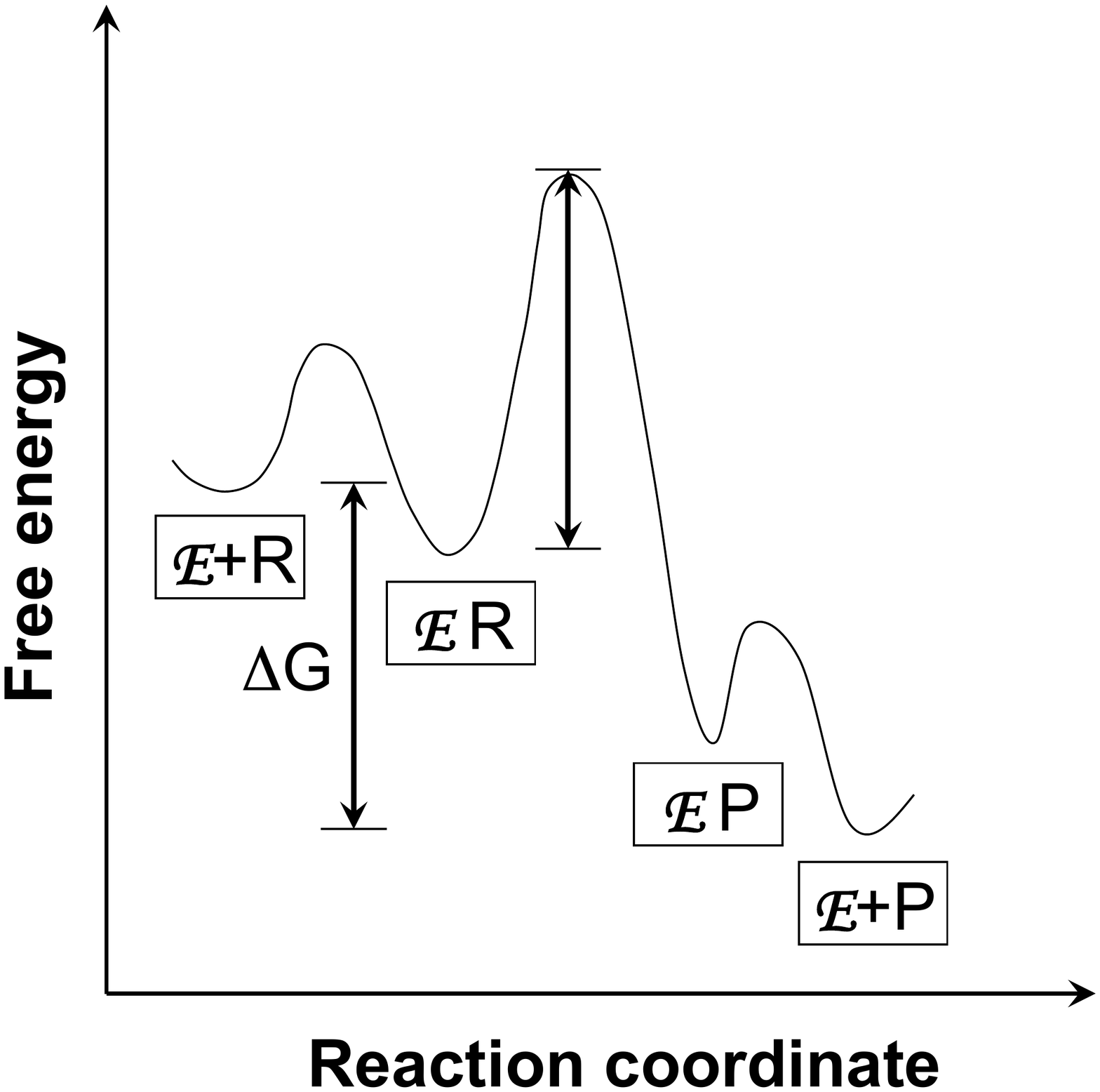}
\end{center}
\caption{Detailed counterpart of the fig.\ref{fig-kramers}. 
The formation of the enzyme-substrate complex $ER$, by the association 
of the enzyme $E$ and substrate $R$, as well as that of the free 
enzyme $E$ and product $P$, by dissociation of the complex $EP$, 
are shown explicitly.
}
\label{fig-multimin}
\end{figure}

It has been felt for a long time that the scenario depicted in
fig.\ref{fig-kramers} is an oversimplified description of chemical 
reactions, particularly those which are catalyzed by enzymes. 
A specific reactant molecule, after diffusing in the medium, comes 
sufficiently close to the enzyme ${\cal E}$ to bind reversibly
forming an enzyme-reactant complex ${\cal E}R$. Then, ${\cal E}R$
converts to the enzyme-product complex ${\cal E}P$ catalytically
and thereafter ${\cal E}P$ dissociates whereby the product $P$
is released; the free enzyme ${\cal E}$ is available again for the
next cycle. This scheme can be represented as follows:
\begin{eqnarray}
{\cal E} + R &\rightleftharpoons& {\cal E}R \rightleftharpoons {\cal E}P \rightleftharpoons {\cal E} + P \nonumber \\
\label{eq-MMscheme0}
\end{eqnarray}
For such a simple scheme, the counterpart of the fig.\ref{fig-kramers}
would be as shown in fig.\ref{fig-multimin} which exhibits more maxima 
and minima than those in Fig.\ref{fig-kramers}. 

\vspace{1cm}

In a more general situation, the conformation of the enzyme ${\cal E^{*}}$ 
immediately after releasing the bound ligands may not be identical to its 
original relaxed ligand-free conformation ${\cal E}$. Such situations can 
be captured by generalizing the scheme (\ref{eq-MMscheme0}) to 
\begin{eqnarray}
{\cal E} + R &\rightleftharpoons& {\cal E}R \rightleftharpoons {\cal E^{*}}P \rightleftharpoons {\cal E}^{*} + P \nonumber \\
{\cal E}^{*} &\rightarrow& {\cal E}
\label{eq-MMscheme}
\end{eqnarray}
If the rate of conversion ${\cal E}^{*} \rightarrow {\cal E}$ is
sufficiently rapid, ${\cal E}^{*}$ can be approximated by ${\cal E}$ 
and the scheme (\ref{eq-MMscheme}) would reduce to the form 
(\ref{eq-MMscheme0}).
Often a simpler reaction scheme of the type \cite{dixon79book}
\begin{eqnarray}
{\cal E} + R \mathop{\rightleftharpoons}^{k_{1}}_{k_{-1}} &I_{1}& \mathop{\rightarrow}^{k_{2}} {\cal E}^{*} + P \nonumber \\
{\cal E}^{*} \mathop{\rightarrow}^{\delta} &{\cal E}&
\label{eq-MMscheme1}
\end{eqnarray}
is adeqate where the symbol $I_1$ represents an intermediate molecular 
complex and it is assumed to yield ${\cal E}^{*}$ and $P$ irreversibly. 
For the sake of simplicity,
we have assumed only a single intermediate state $I_{1}$, the
treatment can be easily extended if more than one intermediate
states are involved in the reaction, for example,
\begin{eqnarray}
{\cal E} + R \mathop{\rightleftharpoons}^{k_{1}}_{k_{-1}} &I_{1}& \mathop{\rightarrow}^{k_{2}} I_{2} \mathop{\rightarrow}^{k_{3}}~....~\mathop{\rightarrow}^{k_{n}} I_{n} \mathop{\rightarrow}^{k_{n+1}}  {\cal E}^{*} + P \nonumber \\
{\cal E}^{*} \mathop{\rightarrow}^{\delta} &{\cal E}&
\label{eq-MMnint}
\end{eqnarray}

\noindent$\bullet${\bf A derivation of MM equation under steady-state assumption}

For the enzymatic reaction 
\begin{equation}
{\cal E} + R \mathop{\rightleftharpoons}^{k_{1}}_{k_{-1}} I_{1} \mathop{\rightleftharpoons}^{k_{2}}_{k_{-2}} {\cal E} + P,  
\label{eq-MMreverse}
\end{equation}
which does not distinguish between ${\cal E}$ and ${\cal E}^{*}$, 
the rate equations are 
\begin{equation}
\frac{d[R]}{dt} = - k_1 [{\cal E}] [R] + k_{-1} [I_1]
\label{ezeq-1}
\end{equation}
\begin{equation}
\frac{d[I_1]}{dt} =  k_1 [{\cal E}] [R] - (k_{-1}+k_2) [I_1]
\label{ezeq-2}
\end{equation}
\begin{equation}
\frac{d[P]}{dt} = k_2 [I_1] - k_{-2} [{\cal E}] [P]
\label{ezeq-3}
\end{equation}
Moreover, as the total amount of enzyme $[{\cal E}]_{0}$ is, by
definition, conserved, we must have
\begin{equation}
[{\cal E}] + [I_1] = [{\cal E}]_0 = {\rm constant}.
\label{ezeq-4}
\end{equation}
We now make two simplifying assumptions.\\
{\it Assumption 1:} $k_{-2} \simeq 0$, i.e., the second step is
practically irreversible; this condition can be implemented by
removing the products from the reaction chamber as soon as these
are released by the enzyme in each enzymatic cycle.
Then, the equation (\ref{ezeq-3}) simplifies to
\begin{equation}
\frac{d[P]}{dt} = k_2 [I_1]
\label{ezeq-5}
\end{equation}
Eliminating $[{\cal E}]$ from (\ref{ezeq-2}) and (\ref{ezeq-4}) we get
\begin{equation}
\frac{d[I_1]}{dt} =  k_1 ([{\cal E}]_0 - [I_1]) [R] - (k_{-1}+k_2) [I_1] = k_1 [{\cal E}]_0 [R] - (k_{-1} + k_2 + k_1 [R]) [I_1]
\label{ezeq-6}
\end{equation}
{\it Assumption 2: (steady-state approximation)} for the intermediate
complex $I_{1}$, i.e., $d[I_1]/dt = 0$. This situation arises if,
for example, $[R] >> [{\cal E}]_0$, i.e., the reactants are in
large excess, compared to the total initial amount of enzyme. We
can now envisage a situation where, during a very brief initial
period, the intermediate complex $I_{1}$ is formed and soon its
concentration attains a steady (i.e., time-independent) value.
For all successive times, the rate of conversion of $I_{1}$ into the
product $P$ and free enzyme can exactly balance the rate of formation
of $I_1$ thereby maintaining the steady concentration of $I_{1}$
\begin{equation}
[I_1] = \frac{k_1 [{\cal E}]_0 [R]}{(k_{-1}+k_2+k_1[R])}.
\end{equation}
Moreover, {\it assuming} $[R]$ to be practically constant (because
there is so much excess of R), $[I_1]$, indeed, reaches the
above mentioned steady state with a relaxation time
\begin{equation}
\tau = \frac{1}{k_{-1} + k_2 + k_1 [R]}
\end{equation}
starting from $[I_1](t=0) = 0$.

Under these assumptions, the speed of the reaction is
\begin{equation}
V = \frac{d[P]}{dt} = k_2 [I_1] = \frac{k_1 k_2 [{\cal E}]_0 [R]}{k_{-1}+k_2+k_1[R]}
\end{equation}
which is conventionally expressed in the form
\begin{equation}
V = \frac{k_2 [{\cal E}]_0 [R]}{K_M + [R]}
\label{eq-michaelis}
\end{equation}
where the so-called {\it Michaelis constant}
\begin{equation}
K_M = \frac{k_{-1}+k_2}{k_1} = \biggl(\frac{[{\cal E}][R]}{[I_{1}]}\biggr)_{ss}
\label{eq-michconstant}
\end{equation}
is the ratio of the total rates of reactions {\it out of} $I_1$ and
that {\it into} $I_1$.

We'll now explore the physical meaning and significance of the Michaelis
constant $K_M$. Writing $V = k_2 [{\cal E}]_0/[1 + (K_M/[R])]$, we find that
$V \rightarrow V_{max} = k_2 [{\cal E}]_0$ as $[R] \rightarrow \infty$, where
$V_{max}$ is the maximum possible reaction rate. Therefore, the equation
(\ref{eq-michaelis}) can be recast as
\begin{equation}
\frac{1}{V} = \frac{K_M}{V_{max}}\biggl(\frac{1}{[R]}\biggr) + \frac{1}{V_{max}}
\label{eq-michael2}
\end{equation}
From (\ref{eq-michael2}) we see that for $K_M = [R]$, $V = V_{max}/2$,
i.e., $K_M$ is the reactant concentration at which the reaction rate
is half of its maximum possible value.

\noindent$\bullet${\bf A derivation of MM equation under quasi-equilibrium approximation}

In order to get further insight into the MM equation, let us make a
``quasi-equilibrium'' (i.e., near equilibrium) approximation where the
first step
$${\cal E} + R \mathop{\rightleftharpoons}^{k_1}_{k_{-1}} I_{1}$$
is assumed to attain equilibrium whereas the rate $k_{2}$ of the
second step
$$I_{1} \mathop{\rightarrow}^{k_{2}} {\cal E} + P$$
is assumed to be very small (infinitesimal).
Then, for the first step (note that from the consideration of free
energetics, $I_{1}$ is the reactant and ${\cal E}$ and $R$ are the
products),
\begin{equation}
K^{1}_{eq} = \biggl(\frac{[{\cal E}][R]}{[I_{1}]}\biggr)_{eq} = \biggl(\frac{([{\cal E}_{0}]-[I_{1}])[R]}{[I_{1}]}\biggr)_{eq} = \frac{k_{-1}}{k_1}
\label{eq-MMeq}
\end{equation}
No product formation is possible if this equilibrium is strictly
maintained. However, suppose the deviation from equilibrium is
extremely small so that the equation (\ref{eq-MMeq}) still holds
approximately. Then the rate of product formation is given by
\begin{equation}
V = d[P]/dt = \frac{k_2 [{\cal E}]_{0} [R]}{K^{1}_{eq}+[R]}
\label{eq-MMeq2}
\end{equation}
which, formally, appears similar to (\ref{eq-michaelis}) except that
$K^{1}_{eq} = \frac{k_{-1}}{k_1} = \biggl(\frac{[{\cal E}][R]}{[I_{1}]}\biggr)_{eq}$ replaces
$K_M = \frac{k_{-1}+k_2}{k_1} = \biggl(\frac{[{\cal E}][R]}{[I_{1}]}\biggr)_{ss}$;
the difference between the two can be made as small as one wishes
by reducing $k_2$ accordingly.

\noindent$\bullet${\bf Analysis of experimental data and testing validity of MM scheme}

Two important parameters which characterize the MM equation are
$V_{max}$ and $K_{M}$. Several different methods of curve plotting
has been followed in the literature to extract these two parameters
from the experimental data.
(i) According to the equation (\ref{eq-michael2}), which is also called
{\it Lineweaver-Burk} equation, plotting experimentally measured values
of $1/V$ against $1/[R]$, one should get a straight line with slope
$K_M/V_{max}$ and intercept $1/V_{max}$ from which both $V_{max}$ and
$K_M$ can be extracted. \\
(ii) Alternatively, equation (\ref{eq-michael2}) can be recast as
\begin{equation}
\frac{V}{[R]} = - \frac{V}{K_{M}} + \frac{V_{max}}{K_{M}} ~{\rm (Eadie-Hofstee ~plot)}
\end{equation}
Therefore, plotting $V/[R]$ against $V$, one can get $K_{M}$ and
$V_{max}$ using the slope and intercept of the straight line.\\
(iii) A third alternative is to use the form
\begin{equation}
\frac{[R]}{V} = - \frac{[R]}{V_{max}} + \frac{K_M}{V_{max}} ~{\rm (Hanes ~plot)} 
\end{equation}
of the same equation (\ref{eq-michael2}) to extract $V_{max}$ and
$K_{M}$ from the slope and intercept of the straight line obtained
by plotting $[R]/V$ against $[R]$.
Critical analysis of the available experimental data 
indicate that the rates of a large class of enzymatic reactions do
not follow the MM equation \cite{hill77}.
For a critical evaluation of the assumptions made in deriving the MM
equation and the reliability of the methods of estimating $V_{max}$
and $K_M$ from the experimental data using the above scenario, see
ref.\cite{schnell03a,chen10}.
The validity of the steady-state assumption and the possibility of
extending the domain of its validity have been examined critically
over the last few decades (see, for example,
ref.\cite{segel88,borghans96,tzafriri03,tzafriri05,tzafriri07,pedersen07,gunawardena12}).

\subsubsection{\bf Specificity amplification by energy dissipation: kinetic proofreading} 
\label{sec-kinproofreading}

Enzymes are specific in the sense that every reaction is catalyzed by a 
specific enzyme. Machines for template-directed polymerization, that also 
qualify as molecular motors (reviewed in sections \ref{sec-genericpolyribo}, 
\ref{sec-specificpoly}, \ref{sec-specificribo}), select monomeric subunits 
of the growing polymer at every step as directed by the corresponding 
template. The fidelity of the polymerization process depends, at least 
in part, on the accuracy of this selection of the substrate that is then 
enzymatically bonded to the growing polymer. 
In this section we discuss a particular mechanism of specificity
amplification, called {\it kinetic proofreading} \cite{hopfield74,ninio75}, 
that enhances accuracy beyond what would be normally allowed from purely 
thermodynamic considerations.

Suppose an enzyme E catalyzes specifically both the reactions
$R_c \to P_c$ and $R_w \to P_w$. Now consider a situation where, $P_c$
is the desired product of the reaction catalyzed by E because $P_c$
is needed for some specific biological function. However, both $R_c$
and $R_w$ are present so that the enzyme molecules can catalyze both
the reactions thereby producing both the correct product $P_c$ and
the wrong product $P_w$. The lowest free energy of the enzyme-reactant
complex ${\cal E}-R_c$ along the reaction pathway is expected to be
lower than that of the ${\cal E}-R_w$ complex by an amount $\Delta G$.
Therefore, the smallest ratio of the populations of the wrong and
correct products is expected to be $\phi_{0} = exp[-\Delta G/(k_BT)]$.
If the two reactants are very similar, $\Delta G$ may not be sufficiently
large to keep $\phi_0$ below a certain pre-determined tolerance level
of error. 

{\it Kinetic proofreading} is a kinetic mechanism designed for specificity 
amplification, e.g., for decreasing the fraction of population of the 
erroneous product to $\phi_0^2$ (or, in general, to $\phi_0^n$ with $n > 2$).
Let us assume that the catalytic reactions with both the correct and 
incorrect substrates follow the Michaelis-Menten scheme \ref{eq-MMscheme}. 
For simplicity, we also assume that all the rate constants, except 
$k_{-1}$, are identical for both the substrate species, i.e., the 
substrate discrimination arises only from the differences between 
$k_{-1}^{(c)}$ and $k_{-1}^{(w)}$. For simplicity, we also present 
the arguments under the quasi-equilibrium approximation although the 
general conclusions are valid also for the more realistic steady-state 
approximation.
Then, the average rates of the corresponding reactions, in the 
quasi-equilibrium approximation, are 
\begin{eqnarray} 
V_{c} &=& \frac{k_{2} [{\cal E}]_{0}[R]_{c}}{K_{eq}^{(c)}+[R]_{c}} \nonumber \\
V_{w} &=& \frac{k_{2} [{\cal E}]_{0}[R]_{w}}{K_{eq}^{(w)}+[R]_{w}} 
\end{eqnarray}
where $K_{eq}^{(c)} = (k_{-1}^{(c)}/k_{1})$ and 
$K_{eq}^{(w)} = (k_{-1}^{(w)}/k_{1})$. 
We define $f_0 = V_{w}/V_{c}$ to be the ratio of the rates of formation 
of the wrong and correct products. For the same initial concentrations 
of the two substrates, i.e., $[R]_{c} = [R]_{w}$, we get 
$f_{0} \simeq k_{-1}^{(c)}/k_{-1}^{(w)} = K_{eq}^{(c)}/K_{eq}^{(w)} = exp(-\Delta G)$.
So, with pure MM-scheme of the enzymatic reaction the substrate 
discrimination is limited by the free energy different between the two.

Next, let us consider the kinetic scheme shown below
\begin{eqnarray}
{\cal E}+R \rightleftharpoons I_{1} \rightarrow &I_{2}& \rightarrow {\cal E}+P \nonumber \\
&\downarrow& \nonumber \\
{\cal E}&+&R
\label{eq-MMlikeproof}
\end{eqnarray}
which is an extension of the MM scheme; in this extended version an 
extra
intermediate state $I_2$ and a branched path from $I_2$ have been
added. This scheme is one of the simplest possible implementations
of kinetic proofreading \cite{hopfield74,ninio75}.
In this assuming that the rate of the transition 
$I_{2} \rightarrow {\cal E}+P$ to be extremely small 
compared to that for the transition along the branched pathway 
$I_{2} \rightarrow {\cal E} + R$, one gets $f \simeq f_{0}^{2}$ 
where $f$ is the ratio of the rates of formation of wrong and 
correct products according to the scheme (\ref{eq-MMlikeproof}). 

Kinetic proofreading amplifies substrate specificity beyond what 
is allowed purely on the basis of equilibrium thermodynamics. 
The two features are {\it essential} for kinetic proofreading are 
as follows \cite{yarus92a,yarus92b,burgess93}: 
(i) a strongly forward driven step that results in a high-energy
intermediate complex, and\\
(ii) one or more branched pathways along which dissociation of
the enzyme-reactant complex, and rejection of the reactant, can 
take place before the complex gets an opportunity to make the final
transition to yield the product.

Many other mechanisms of specificity amplification have been proposed. 
One of these, based on ``energy relay'' will be discussed later in 
this section. Another kinetic proofreading scheme \cite{lindo11} is 
based on ``{\it inter-molecular frustration}''.

\subsubsection{\bf Effect of external force on enzymatic reactions catalyzed by motors}

External force affect not only mechanical movements over significant 
distances, but also alter the rates of chemical reactions in each 
cycle of a molecular motor. The effects of force on enzymatic reactions 
catalyzed by motor proteins have been investigated extensively 
\cite{khan97a,bustamante04}, 
particularly after single-molecule techniques were developed 
\cite{tinoco02,tinoco04,tinoco06,cebollada10}.  

The free energy landscape is altered by an external force; it affects
not only the equilibrium populations of the various structural states,
but also the rates of transitions among these states.
\cite{khan97a,bustamante04,tinoco02,tinoco06,cebollada10}
For the purpose of explaining these phenomena, let us again consider
the reaction (\ref{eq-simplest}). Suppose an external force $F$ is
applied on the protein and the force is directed from ${\cal E}_{1}$
to ${\cal E}_{2}$. Then
\begin{equation}
\Delta G = \Delta G^{0} - F (\Delta x)
\end{equation}
where $\Delta G^{0}$ is the free energy difference between ${\cal E}_{2}$
and ${\cal E}_{1}$ in the absence of the external force $F$. Obviously,
in equilibrium,
\begin{eqnarray}\frac{k_{f}(F)}{k_{r}(F)} = \frac{[{\cal E}_{2}]_{eq}}{[{\cal E}_{1}]_{eq}} = exp(-\beta \Delta G) = K_{eq}^{0} exp(\beta F \Delta x),
\label{eq-fdep}
\end{eqnarray}
i.e., the structural state ${\cal E}_{2}$ is more probable than the
state ${\cal E}_{1}$.

The equation (\ref{eq-fdep}) implies that we can write the individual
rate constants for the forward and reverse transitions as
\cite{khan97a,bustamante04,tinoco02,tinoco06}
\begin{equation}
k_{f}(F) = k_{f}(0) e^{\theta \beta F(\Delta x)}
\label{eq-fdepf}
\end{equation} 
and
\begin{equation}
k_{r}(f) = k_{r}(0) e^{-(1-\theta) \beta f(\Delta x)}
\label{eq-fdepr}
\end{equation} 
where $\theta$ is a fraction of the distance $\Delta x$ and determines 
how the external load is shared by the forward and reverse transitions.
The forms of
$F$-dependence assumed in (\ref{eq-fdepf}) and (\ref{eq-fdepr}) is
used routinely for molecular motors while deriving their force-velocity
relations which are among the fundamental characteristics of each
family of motors.\\

\subsubsection{\bf Effects of multiple ligand-binding sites: spatial cooperativity and allosterism in molecular motors}
\label{sec-allostery}

The term ``cooperativity'' is used to describe wide variety of biochemical
phenomena. Cooperative interactions involving proteins can take place
at various levels of organization \cite{whitty08}:
(a) {\it intra-molecular} interaction between different regions of the
same protein (e.g., in a monomeric enzyme),
(b) {\it inter-molecular} interaction between the different protein
molecules of an oligomeric single enzyme,
(c) {\it inter-enzyme} interactions in a multi-enzyme complex, etc.

In the context of enzymes and motors, the term ``cooperativity'' refers 
to a process in which one event affects another event of similar type 
(e.g., binding of a ligand) by means of intra-molecular (in a single
protein) or inter-molecular (in a multi-protein macromolecular complex)
communication. For example, several types of motors have more than one 
binding sites for ATP. Almost all motors have multiple binding sites 
also for other ligands. Linear motors must also have a binding site for 
attaching to the track. The emergent properties of such motors are 
results of the  ``cooperative'' effects.

Cooperativity in enzymatic kinetics has been studied
extensively over several decades
\cite{rabin67,frieden71,neet80a,neet80b,neet95,ricard87,acerenza97}.
Quantitative measure of cooperativity can be defined both in terms of
thermodynamic equilibrium and kinetics. The increase (or decrease) of
binding of one ligand following that of another can be quantified in
terms of free energies of binding. 
If the binding of the first ligand helps (inhibits) the binds of
the second, the cooperativity is called {\it positive} ({\it negative}) 
If the two ligands are of the same type, the cooperativity is
{\it homotropic} whereas {\it heterotropic} cooperativity involves
two different types of ligands. Both {\it homotropic} and {\it
heterotropic} cooperativity can be either {\it positive} or {\it
negative}.

In the context of enzymatic reaction kinetics, usually non-Michaelis-Menten 
behavior of an enzyme is identified as the signature of cooperativity. 
However, more objective quantitative measures of the extent of 
cooperativity have been used in the literature \cite{neet95}.
What is the reason for identifying MM kinetics as a non-cooperative 
phenomenon? Note that for $N$ independent trials of a biased coin, 
for which head and tail occur with probabilities $p$ and $q$, 
respectively, the expected number of heads is $Np/(p+q)$. 
Comparing this form with the average rate $k_2[{\cal E}]_0[R]/(k_M+[R])$ 
of MM reaction, we conclude that the form (\ref{eq-michaelis}) is a 
signature of non-cooperativity.

\begin{figure}[htbp]
\begin{center}
\includegraphics[angle=-90,width=0.45\columnwidth]{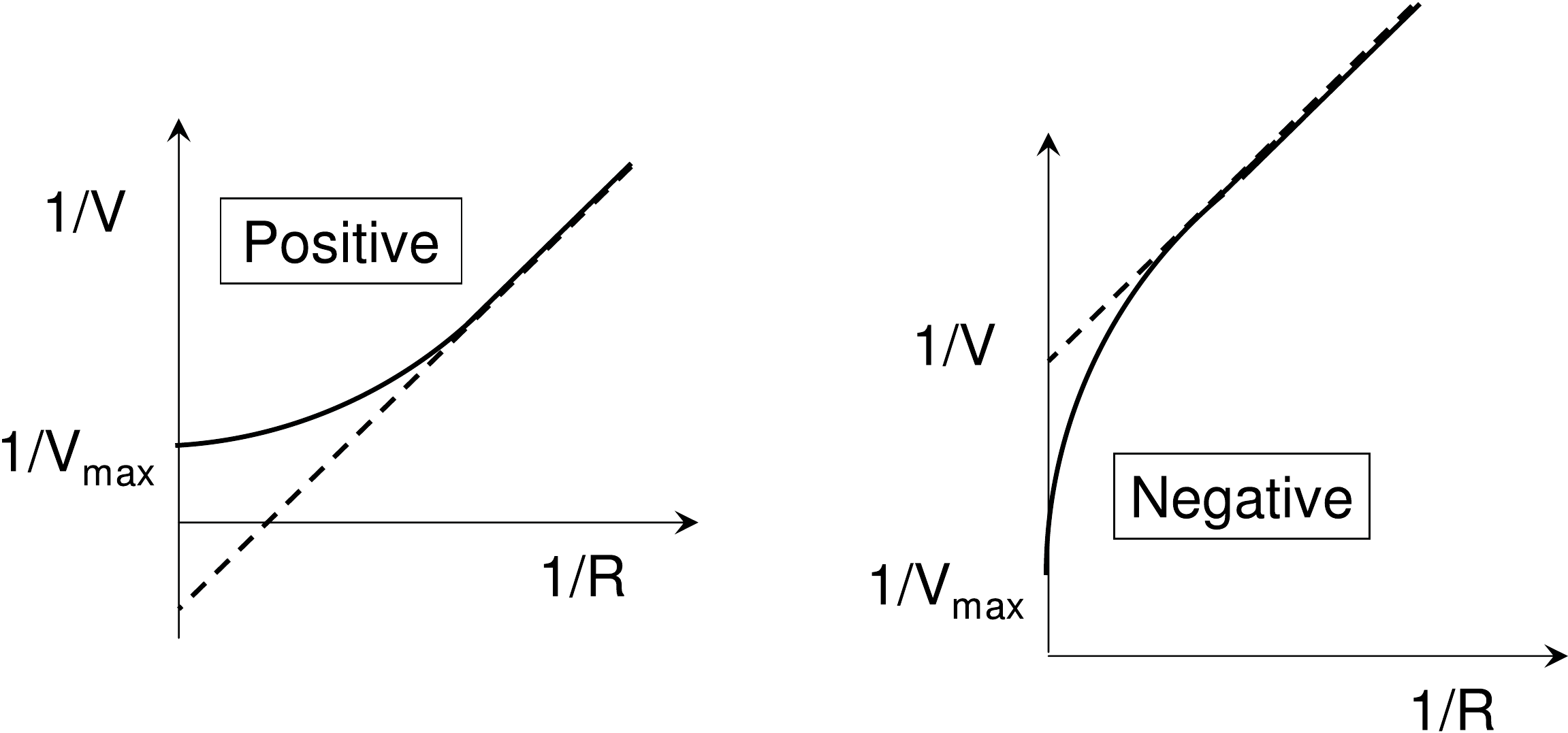}
\end{center}
\caption{Graphical sketch of the deviations from the MM form 
(\ref{eq-michaelis}), which indicate positive and negative 
cooperativities in enzymatic kinetics depending on the sign of 
the curvature. 
}
\label{fig-coop1}
\end{figure}

When plotted graphically, the crucial difference between the MM-type 
equation $y=x/(K+x)$ and the Hill-type equation $y=x^{n}/(K^{n}+x^{n})$ 
is that the curvature of the former is negative for all $x \geq 0$ whereas, 
for all $n > 1$ that of the latter changes sign from positive to negative 
with gradual increase of $x$. A Hill-type form would be an indicator of 
cooperativity; $n>1$ and $n<1$ indicate positive and negative 
cooperativities, respectively (see fig.\ref{fig-coop1}).

The interaction between the ligands is not direct. Instead, conformational
changes in the enzyme following the binding of one ligand influences the
binding of another ligand to the same enzyme. Past history of occupation of
a ligand-binding site can propagate {\it temporally}, and affect the binding
of ligands to the same enzyme in future, if conformation of the enzyme does
not relax to the original conformation of free enzyme before the next round
of ligand-binding \cite{neet80a,neet80b,neet95}. We'll consider temporal 
cooperativity in the context of molecular motors later in this section. 
Alternatively, information on the occupational status of one binding site 
can be transmitted {\it spatially} to the other binding site(s).

For an enzyme with at least two binding sites the phenomenon of {\it
spatial} cooperativity leads to the interesting cooperative phenomenon
of allosterism \cite{cui08a}. Allostery is a mechanism for regulation of the
structure, dynamics and function of an enzyme by the binding of
another molecule, called effector, which can be a small molecule (a
ligand) or another macromolecule
\cite{monod65,koshland66,changeux98,changeux05,changeux10,changeux12,hilser12,koshland02}.
The three defining characteristics of allostery are \cite{fenton08}:
(i) the effector is chemically distinct from the substrate,
(ii) the binding site for the effector is spatially well separated
from that of the substrate, and
(iii) binding of an effector molecule affects at least one of the
functional properties of the enzyme; the functional property could
be either (a) the binding affinity for its specific substrate or
(b) the rate of the reaction it catalyzes.

The models of allosteric control were originally analyzed within the
framework of equilibrium thermodynamics \cite{monod65,koshland66}.
Over the last decade, a more general mathematical formulation of
this phenomenon has been reported \cite{duke99a,duke01,bray04,mochrie10}.
Unlike deterministic picture of the thermodynamic formulation, this
statistical mechanical theory allows spontaneous fluctuations and
introduces the concept of {\it conformational spread (CS)}. 
The CS model postulates that each subunit of the enzyme can be in 
either an active or an inactive conformation and can make rapid 
transitions between these states. In this model, an individual 
subunit can also bind a ligand present in the surrounding solution. 
The probability of a subunit being active or inactive depends on 
(i) whether or not it is bound to a ligand, and (ii) the conformational 
state of its neighbors.  This model
may be regarded as an extension of the Ising model which is one one of
the simplest models in equilibrium statistical mechanics \cite{chowdhury00a}.
The properties of this CS model have been derived using the formal
techniques which are widely used for analyzing the Ising model.
The CS model reduces to the two pioneering models \cite{monod65,koshland66}
of allosteric control in two different special limits.

\begin{figure}[htbp]
\begin{center}
\includegraphics[angle=-90,width=0.55\columnwidth]{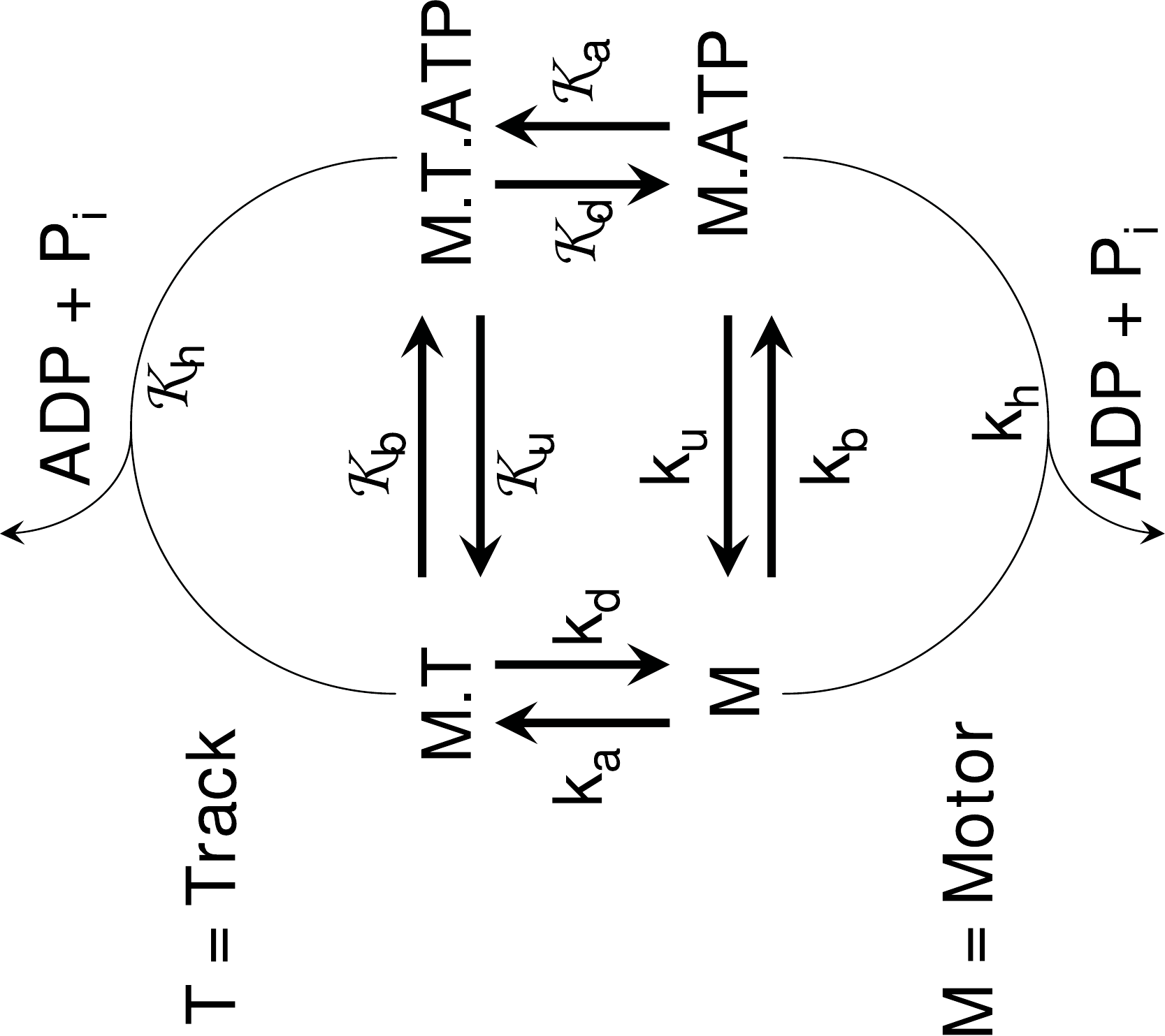}
\end{center}
\caption{A schematic representation of the generic scenario of
hydrolysis of ATP by a motor enzyme in the presence of the 
corresponding cytoskeletal filament.
}
\label{fig-atphydro}
\end{figure}

Allosterism is not restricted only to proteins; allosteric ribozymes
are also receiving attention in recent years \cite{fastrez09}.
A motor protein has separate sites for binding the fuel molecule
and the track. Therefore, the mechano-chemical cycle of a motor can
be analyzed from the perspective of allosterism
\cite{goldsmith96,bray04,vologodskii06}.
The cycles of molecular motors can be represented as a sequence of
allosteric transitions which are caused by the binding or release
of fuel molecules (e.g., ATP) and the spent fuel (e.g., ADP and
$P_{i}$) as well as attachment and detachment of the filamentous
track.
Typical generic cycles in the absence and in the presence of
the corresponding track are shown in fig.\ref{fig-atphydro}. 
Presence of track has very significant effects on the ATP binding 
and hydrolysis. 
There are some common features of the enzymatic cycle of cytoskeletal
motors, in spite of some crucial differences \cite{hackney96}.

\subsubsection{\bf ATPase rate and velocity of motors: evidence for tight coupling?}

If ATP hydrolysis fuels the mechanical movement of a molecular motor, 
is there a direct relation between this ATPase rate and the average 
velocity of the motor on its track? For example, is the ATP-dependence 
of the average velocity governed by a MM-like equation when the average 
rate of ATP hydrolysis, by the same motor, follows MM-equation? If so, 
does this MM-like form survive even when the motor is subjected to a 
load force $F$?  If it does, then the MM-like form of the force-dependent 
average velocity $V(F)$ of a motor would be 
\cite{fisher01,lattanzi01,zhang12b}
\begin{equation}
V(F) = \frac{\kappa {\ell} V_{max}(F) [ATP]}{K_{M}(F)+[ATP]}
\label{eq-FdepV}
\end{equation}
where $\kappa$ is the strength of mechano-chemical coupling (defined by 
eqn.(\ref{eq-mccoupling})) and ${\ell}$ is the step size of the motor. 
The $F$-dependence of both the characteristic parameters $V_{max}$ and 
$K_{M}$ are mentioned explicitly in (\ref{eq-FdepV}). Since, apriori, 
the mechano-chemical coupling of molecular motors is expected to be 
weak, one curiosity is to find out whether any motor displays 
tight-coupling (i.e., $\kappa=1$). 
All these fundamental questions have been addressed in the last decades 
by many careful experiments and sophisticated analysis of the data;  
in part II we'll find answers to these questions in the context of specific 
molecular motors.

\subsection{\bf Sources of fluctuations in enzymatic reactions and their effects}

Apart from drawing input energy from enzymatic reactions, some motors 
also catalyze other types of chemical reactions. For example, a DdDP, 
whose main function is template directed polymerization of a DNA molecule, 
also catalyzes DNA cleavage for error correction. Therefore, understanding 
the causes and consequences of the fluctuations in enzymatic reactions 
is required for getting a broader picture of the performance of some 
motors.

To my knowledge, a stochastic treatment of the MM scheme of enzymatic 
reactions was published already in 1962 by Bartholamay \cite{bartholamay62}. 
In this pioneering work, he made clear distinction between ``two types of 
irreproducibilities'': those arising from experimental noise and those 
caused by intrinsic fluctuations. He also emphasized that ``even in the 
total absence of experimental irregularities a concentration time course 
has an independent existence as a statistical entity'' \cite{bartholamay62}.
With remarkable clarity, Bartholomay \cite{bartholamay62} identified the 
sources of these fluctuations to be the ``Brownian-like motions of the 
reactant molecules'', the ``random intermolecular collisions'', and the 
accompanying intramolecular (i.e., conformational) transitions.

\noindent $\bullet$ {\bf Sources of fluctuations in enzymatic reactions}

Let us now summarize the sources of fluctuations in enzymatic reactions.
There are at least three different sources which might contribute
to the fluctuations in the turnover times \cite{dan07}: \\
(i) intrinsic stochasticity arising from the reservoir that provides
the thermal energy required for barrier crossing;\\
(ii) low concentration of the reactants makes the arrival of the
substrate molecules to the enzyme stochastic; and\\
(iii) Conformational fluctuations of the enzyme can introduce novel
features which are absent when the catalyst is a rigid molecule.

\subsubsection{\bf Fluctuations caused by low-concentration of reactants} 

\noindent$\bullet${\bf Micro-macro correspondence: MM equation in thermodynamic limit}

\begin{figure}[htbp]
\begin{center}
\includegraphics[angle=-90,width=0.65\columnwidth]{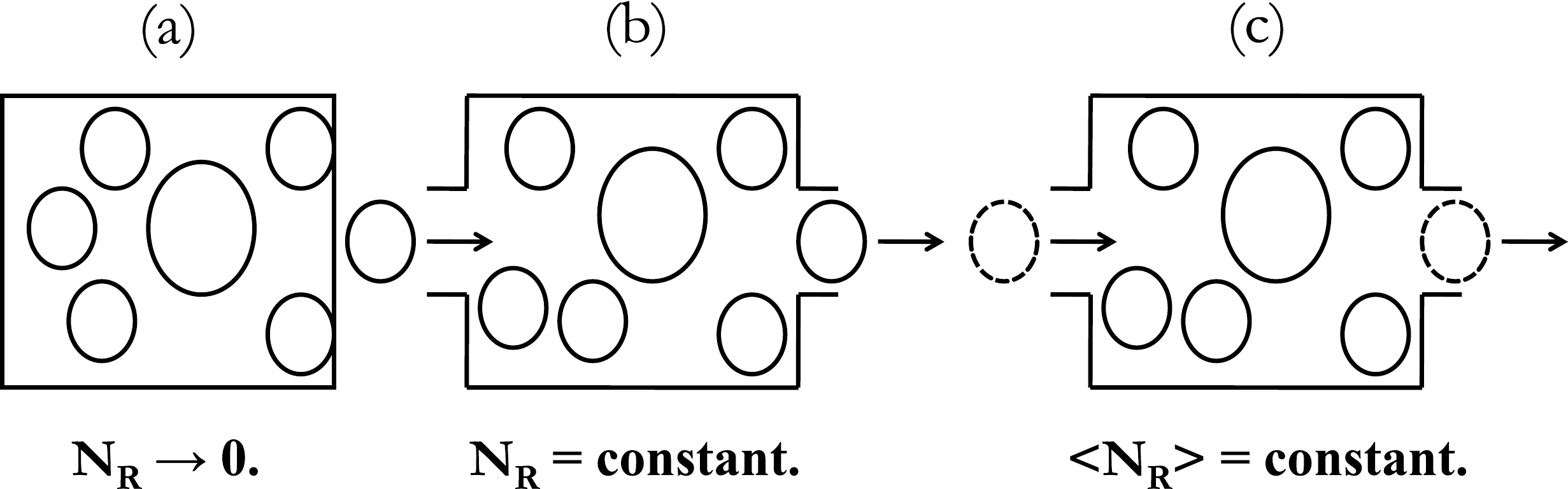}
\end{center}
\caption{Schematic illustration of the three distinct scenarios
considered in ref.\cite{stefanini05}. The large ellipse represents
the single enzyme while the smaller ellipses represent the
substrate molecules; $N_{R}$ is the number of substrate molecules
in the reaction volume. (a) The reaction volume is perfectly {\it
isolated} and, therefore $N_{R} \to 0$ as $t \to \infty$. (b) A
counter keeps track of the number of molecules and the depletion
of substrate population is {\it exactly} compensated by fresh
addition so as to maintain $N_{R}$=constant. $V([{\cal R}])$
satisfy MM equation exactly. (c) A counter ensures that the
substrate concentration remains constant only on the average.
Deviation from MM equation, observed in this case, are caused by
the fluctuations in $N_{R}$; the smaller the number of molecules,
the stronger are the fluctuations. Therefore, the deviation from
the MM equation decreases with increasing $<N_{R}>$ and vanishes
in the limit $<N_{R}> \to \infty$ (figure adapted from ref.
\cite{stefanini05}).
}
\label{fig-stefanini}
\end{figure}

Let us consider an enzymatic reaction which is given by the scheme
\begin{eqnarray}
{\cal E} + R \mathop{\rightleftharpoons}^{k_{1}}_{k_{-1}} &I_{1}& \mathop{\rightarrow}^{k_{2}} {\cal E} + P
\end{eqnarray}
Suppose there is a single enzyme molecule and $N_{R}$ substrate molecules
in a reaction volume. If $N_{R}$ can be maintained strictly constant at
all times by inserting a substrate molecule whenever one gets converted
to product, the MM equation describes the average rate of the reaction.
But, if the substrate concentration is allowed to fluctuate around a
constant mean, i.e., $<N_{R}>$=constant, deviation of the average rate of
the reaction from the corresponding MM equation is found \cite{stefanini05}.
Not surprisingly, the average rate of the reaction approaches the MM
equation in the ``thermodynamic limit'' $<N_{R}> \to \infty$
(see fig.\ref{fig-stefanini}).

\noindent$\bullet${\bf Single-molecule enzymology: turnover time and average rate}

Single-molecule enzymology 
\cite{claessen10}
is relevant also for understanding single-motor mechanism.
The time needed to complete one catalytic cycle of an enzyme is called
its {\it turnover time}. The inverse of the mean turnover time gives
the average rate of the reaction. Each turnover consists of the
following stages:

\centerline{{\framebox{Substrate selection}}$\rightarrow${\framebox{product formation}}$\rightarrow${\framebox{release of product \& enzyme relaxation}}  }

Therefore, the turnover time should be the sum of the times taken for
each of these stages.

Let us compare and contrast various types of assays that can be used
for enzymology \cite{bagshaw03,bagshaw06}.
Suppose the numbers of enzymes and reactant molecules are denoted by
$N_{e}$ and $N_{r}$, respectively.
In biochemical experiments on enzymatic reactions using bulk samples,
usually, $N_{r} \gg N_{e} \gg 1$; for example reactants can be in
micro-molar range while the enzyme may be in nano-molar range.
This scenario can be described as {\it ``multiple enzyme, multiple
turnover''} because each of the enzyme molecule goes through multiple
rounds of enzymatic cycle. In general, the cycles of different enzyme
molecules are not synchronized and the population of the product
molecules increases smoothly, and linearly, with time.

However, if $N_{r} = N_{e} \gg 1$, and if the reaction is not too
fast, each enzyme molecule will catalyze the reaction only once;
this scenario may be described as {\it ``multiple enzyme, single
turnover''}. In this case, the population of the product molecules
increases smoothly within a period of time and, then, saturates as
all the reactants are exhausted and fully converted to the product.

In contrast, if $N_{r} \gg N_{e} = 1$, the enzyme goes through many
rounds of enzymatic cycle in this {\it ``single enzyme, multiple
turnover''} scenario. The population of the product molecules
increases by unity in each cycle of the enzyme and the turnover time
fluctuates randomly.

We now show that, in spite of the fluctuations of the turnover times 
\cite{yang11}
the average rate may still satisfy MM equation 
\cite{kou05,min05a,english06,min06a,qian02b,xue06}. 
Consider the MM reaction scheme \cite{kou05}
\begin{eqnarray}
E + R \mathop{\rightleftharpoons}^{k_{1}}_{k_{-1}} &I_{1}& \mathop{\rightarrow}^{k_{2}} E^{*} + P \nonumber \\
E^{*} \mathop{\rightarrow}^{\delta} E
\label{eq-singMMscheme}
\end{eqnarray}
where $R$ is the reactant and $P$ is the product of the reaction
catalyzed by $E$ while $E^{*}$ is the same enzyme in an excited
state. In the limit $\delta \to \infty$ the reaction scheme
(\ref{eq-singMMscheme}) reduces to the original reaction scheme
(\ref{eq-MMscheme}) for which we derived the MM equation
(\ref{eq-michael2}) using reaction rate equation approach.
\begin{equation}
\frac{dP_{E}(t)}{dt} = - k_1^{0} P_{E}(t) + k_{-1} P_{I1}(t) 
\label{singezeq-1}
\end{equation}
\begin{equation}
\frac{dP_{I1(t)}}{dt} =  k_1^{0} P_{E}(t) - (k_{-1}+k_2) P_{I1}(t) 
\label{singezeq-2}
\end{equation}
\begin{equation}
\frac{dP_{E^*}(t)}{dt} = k_2 P_{I1}(t)
\label{singezeq-3}
\end{equation}
where
\begin{equation}
k_{1}^{0} = k_{1} [R]
\label{singezeq-4}
\end{equation}
We {\it assume} that $R$ is independent of time $t$; this is a
good approximation because the reaction is driven by one single
enzyme molecule whereas the initial amount of the reactant is
sufficiently large.
Note that $f(t) \Delta t =$ probability that one reaction has
been completed in the time interval between $t$ and $t+\Delta t$
$=$ probability that the enzyme molecule was {\it not} in the
state $E^*$ upto time $t$ and is in the state $E^*$ between
$t$ and $t+\Delta t =$ probability that at time $t$ the enzyme
molecule was in state $I_{1}$ and made a transition to $E^{*}$
in the next time interval $\Delta t$ $= k_2 P_{I1}(t) \Delta t$.
Thus,
\begin{equation}
f(t) = k_2 P_{I1}(t)
\label{eq-ftdiffeqn}
\end{equation}
Moreover, as the total amount of enzyme is, by definition, conserved,
we must have
\begin{equation}
P_{E}(t) + P_{E^*}(t) + P_{I1}(t) = 1
\label{singezeq-5}
\end{equation}
We solve the equations (\ref{singezeq-1})-(\ref{singezeq-3}),
with the constraint (\ref{singezeq-5}), using the initial conditions
\begin{equation}
P_{E}(0) = 1, ~  P_{I1}(0) = 0 = P_{E^*}(0)
\label{singezeq-6}
\end{equation}
Hence,
\begin{equation}
f(t) = \biggl(\frac{k_1 k_2 [R]}{2 B}\biggr)\biggl\{e^{-(A-B)t} - e^{-(A+B)t}\biggr\} 
\label{eq-ftsingMM}
\end{equation}
where
\begin{equation}
A = \frac{(k_{1}[R]+k_{-1}+k_{2})}{2}
\label{eq-singMMA}
\end{equation}
\begin{equation}
B = \biggl(\frac{(k_{1}[R]+k_{-1}+k_{2})^{2}}{4}-k_{1} k_{2}[R]\biggr)^{1/2}
\label{eq-singMMB}
\end{equation}

\noindent$\bullet${\bf Distribution of turnover times: generic features of first and second moments}

Substituting (\ref{eq-ftsingMM}) into the definition
\begin{equation}
\langle t \rangle = \int_{0}^{\infty} t f(t) dt,
\end{equation}
of the mean turnover time $\langle t \rangle$ and relating it with
the average rate $V$ of the reaction by $V = 1/\langle t \rangle$,
we recover the MM equation (\ref{eq-michael2}) for $V$.

Next, let us begin with the oversimplified linear enzymatic reaction scheme, 
with $N$ distinct kinetic states, where all the transitions (i) are 
completely irreversible, and (ii) take place with the same rate $\omega$. 
For this scheme the distribution of the turnover times is the 
Gamma-distribution 
\begin{equation}
f(t) = \frac{\omega^{N} t^{N-1} e^{-\omega t}}{\Gamma(N)} 
\end{equation}
where $\Gamma(N)$ is the gamma function.
Interestingly, for the Gamma-distribution, the randomness parameter 
\cite{schnitzer95} (also called the Fano factor \cite{fano47})  
\begin{equation} 
r = (<\tau^2>-<\tau>^2)^{1/2}/<\tau> 
\label{eq-fano} 
\end{equation} 
is exactly given by $r=1/N$. Therefore, 
\begin{equation} 
n_{min} = <\tau>^{2}/(<\tau^2>-<\tau>^2) 
\label{eq-nmin} 
\end{equation} 
provides a strict lower bound on the number of kinetic states 
\cite{moffitt10a}.

For a very general class of kinetic schemes, which can be interpreted 
either as that of an enzymatic reaction or as that of a molecular motor, 
Moffitt et al.\cite{moffitt10a} derived a very general expression for 
$n_{min}$ that has a status similar to the MM-expression for $<\tau>$. 
Their derivation was based on the assumptions that (i) the scheme is 
a linear chain without branching or parallel pathways, (ii) the last 
step of the transitions is irreversible, and (iii) $<\tau>$ obeys the 
MM equation. Under these assumptions, they derived \cite{moffitt10a} 
\begin{equation}
n_{min} = \frac{N_L N_S \biggl(1+\frac{[S]}{K_M}\biggr)^2}{N_S+2\alpha\biggl(\frac{[S]}{K_M}\biggr)+N_L\biggl(\frac{[S]}{K_M}\biggr)^2}
\label{eq-nminS}
\end{equation}
which involves, in addition to the Michaelis constant $K_M$, three 
dimensionless parameters $N_L$, $N_S$ and $\alpha$. In spite of the 
difference in the details of the kinetic schemes, all of which satisfy 
the assumptions made in the derivation of eqn.(\ref{eq-nminS}), the 
corresponding $n_{min}$ can be expressed in terms of $K_M, N_L, N_S$ 
and $\alpha$ exactly as in (\ref{eq-nminS}). However, the actual 
functional dependence of these parameters on the rate constants depends 
on the details of the kinetic scheme. 
Note that $N_L = lim_{[S]/K_M \to 0} n_{min}$ and
$N_S = lim_{[S]/K_M \to \infty} n_{min}$. Moreover, occurrence of a 
maximum or minimum in $n_{min}$ at some intermediate concentration 
of the substrate depends on the magnitude of $\alpha$ as compared to 
those of $N_L$ and $N_S$. 

For reactions that are more complex than MM scheme, a perturbative 
technique has been developed by de Ronde et al.\cite{ronde09}. 
Bel et al.\cite{bel10} and Munsky et al.\cite{munsky09} calculated 
the distribution of the completion times of a specific class of models 
for kinetic proofreading process.

\subsubsection{\bf Fluctuations caused by conformational kinetics of the enzyme: ``dynamic disorder''} 

While dealing with fluctuations of enzymatic reactions, so far we have 
not paid any attention to the conformational kinetics of the enzymes. 
In general, the conformational dynamics of proteins \cite{wildman07} 
span a wide range of length and time scales- from a fraction of nanometer 
to tens of nanometers and from femtoseconds to seconds, or even longer 
\cite{karplus00}.
What makes the study of this dynamics so challenging is the coupling 
between the motions that occur on a hierarchy of time scales which, 
in turn, is a consequence of a hierarchy of energy barriers.  
Conformational fluctuations of a enzyme gives rise to {\it temporal} 
fluctuations in the reaction rates of an enzyme molecule; this randomness 
is called ``{\it dynamic disorder}'' \cite{zwanzig90,reichman06}.

\noindent$\bullet${\bf Irreversible decay as an example:}

In order to get an intuitive analytical understanding of the effects
of interconversion of motor (or, enzyme) conformations, one can begin, 
alternatively, with a discrete formulation of the reaction \cite{xie02} 
\begin{eqnarray}
&{\cal E}_{1}& \mathop{\rightarrow}^{k_1} \nonumber \\
k &\upharpoonleft \downharpoonright& k \nonumber \\
&{\cal E}_{2}& \mathop{\rightarrow}^{k_2} \nonumber \\ 
\label{eq-xie}
\end{eqnarray}
where ${\cal E}_{1}$ and ${\cal E}_{2}$ are two distinct conformations
of the same motor (or, enzyme). The reaction considered here could be, for example,
the decay of the fluorescent state to a non-fluorescent state. The rate
of the decay, however, is assumed to depend on the conformation of
the fluorescent state, i.e., in general $k_1 \neq k_2$.
Initially, the protein can be either in ${\cal E}_{1}$ or in
${\cal E}_{2}$ with the probabilities $C$ and $1-C$, respectively.
Let $P_{1}(t)$ and $P_{2}(t)$ denote the probabilities of
finding the protein in the conformations ${\cal E}_{1}$ and
${\cal E}_{2}$, respectively, at any arbitrary time $t$.

First consider the special case where no interconversion of the
conformations ${\cal E}_{1}$ and ${\cal E}_{2}$ is allowed on the
time scale of the decay.  Solving the corresponding master equations
\begin{eqnarray}
\frac{dP_{1}(t)}{dt} &=& - k_{1} P_{1}(t) \nonumber \\
\frac{dP_{2}(t)}{dt} &=& - k_{2} P_{2}(t)
\label{eq-f2uncoup}
\end{eqnarray}
under the initial conditions $P_{1}(0)=C$ and $P_{2}(0)=1-C$,
we get $P_{1}(t) = C exp(-k_{1}t)$ and $P_{2}(t) = (1-C) exp(-k_{2}t)$.
Hence, the survival probability
\begin{equation}
P(t) = P_{1}(t) + P_{2}(t) = [C exp(-k_{1}t) + (1-C) exp(-k_{2}t)]
\label{eq-ftcoupled}
\end{equation}

Now let us allow reversible interconversion of the conformations
${\cal E}_{1}$ and ${\cal E}_{2}$ with the same rates $k$ for the
forward and backward transitions, as shown in eqn(\ref{eq-xie}) 
\cite{xie02,gopich06}.
The equations (\ref{eq-f2uncoup}) are modified to
\begin{eqnarray}
\frac{dP_{1}(t)}{dt} &=& k P_{2} - (k+k_{1}) P_{1}(t) \nonumber \\
\frac{dP_{2}(t)}{dt} &=& k P_{1} - (k+k_{2}) P_{2}(t)
\label{eq-f2coup}
\end{eqnarray}
In this case $f(t)$ is still a sum of two terms each of which
decays exponentially with $t$ but, in contrast to the decay rates
$k_1$ and $k_2$ in (\ref{eq-ftcoupled}), the decay rates of the
two exponentials are
$k_{\pm} = [(k_1+k_2+2k)\pm\sqrt{(k_1-k_2)^2+4k^2}]/2$.
In the limit $k \gg k_{1}$ and $k \gg k_{2}$, $k_{+} \simeq 2k$
and $k_{-} \simeq (k_1+k_2)/2$.

In this and the next few subsections, we explore the roles of the 
conformational kinetics of enzymes \cite{goodey08} 
(i) on the turnover times; 
(ii) in generating temporal {\it correlations}, if any, between
the times taken for its catalytic cycles in a multiple turnover 
and enzymatic {\it hysteresis},
(iii) in the {\it selection} of specific substrates and specificity 
amplification.
Conformational kinetics of enzymes have been studied following two 
alternative mathematical approaches. One of these visualizes the 
kinetics as wandering in an energy landscape whereas the other 
represents kinetics in terms of jumps on a discrete network of states. 
This classification of the approaches is summarized below emphasizing 
that the two are related to each other.
\\
\centerline{{\framebox{Mathematical formalisms for conformational kinetics of enzymatic reactions}}}

\centerline{$\swarrow$~~~~~~~~~~$\searrow$}

\centerline{{\framebox{Wandering in an energy landscape}}~ $\longleftrightarrow$
 ~{\framebox{jumping around on a discrete network}}}

\noindent$\bullet${\bf Conformational dynamics as wandering in an energy landscape}

Let us portray the reaction on a two-dimensional landscape where the
two axes quantify the ``reaction coordinate'' and ``conformational
coordinate'' while the height at each point is a measure of the
corresponding free energy \cite{kruse07a,atkins11} 
(see fig.\ref{fig-atkins11})
This free energy diagram resembles a model of a mountain range
and is obtained by averaging over all the other degrees of freedom
which correspond to faster dynamics.  Each local minimum in the
two-dimensional free energy landscape corresponds to a distinct
conformational state. Fluctuations of length scales much shorter
than inter-minima separation and on time scales much shorter
than the time required for hopping from one local minimum to a
neighboring one manifest as vibrations around the corresponding
local minimum \cite{cartling85}. Usually, the barriers separating
the successive minima along the conformational coordinate are
relatively low and, therefore, can be overcome by thermal activation
on relatively short time scales. In contrast, the barriers to be
crossed along the reaction coordinate are usually much higher and,
therefore, the reaction proceeds at a slower rate.

As discussed earlier, in the older picture, a catalytic cycle consists
of a sequence of intermediate enzyme-substrate or enzyme-product
complexes along the reaction coordinate. In the current scenario,
the free enzyme, as well as these intermediate complexes, exist as
an ensemble of conformations along the conformational coordinate
\cite{goodey08}.
Thus, what appears as an effectively unique transition state in
fig.\ref{fig-kramers} turns out to be the ``transition state {\it
ensemble}'' \cite{ma00} on this two-dimensional landscape.
This ensemble of transition states forms a plane, which resembles
a stretch of high ``mountain peaks'' and runs perpendicular to the
reaction coordinate, bisects the diagram. Reactants (and the
enzyme-reactant complexes) are on one side of this plane while the
products (and enzyme-product complexes) are on the other side
\cite{kruse07a,benkovic08}.

If the conformational dynamics are much faster than the reaction, then
for a given value of the reaction coordinate, an ensemble-average
over the conformational coordinate yields projection of the free-energy
landscape onto the reaction coordinate.
However, for many enzymatic reactions, barriers in both the directions
are of comparable height. For such reactions,  multiple pathways on
this two-dimensional landscape are available for the reaction to occur.

Let us first consider two extreme limiting cases.
(a) First consider the scenario where the ``mountain peak range'' running
perpendicular to the reaction coordinate are much higher than small
``hills'' on the two sides of this range. In this limit, because of
the fast conformational transitions, the enzyme-reactant complex explores
all possible conformations before its conversion to enzyme-product
complex. Consequently, the one-dimensional free energy profile obtained
by the projection of the two-dimensional free energy landscape onto the
reaction coordinate provides an adequate description of the reaction.
Classical treatment of enzymatic reactions in terms of the MM scheme
(or similar scenarios) is sufficient for quantitative estimation of the
average rate of the reaction.
(b) In the opposite limit, where conformational transitions are much
slower than the rates of interconversion of the intermediate complexes,
each complex remains essentially ``frozen'' in a particular conformation,
before its conversion to the next complex along the reaction coordinate.
Different enzymes may remain frozen in different conformations during
individual enzymatic cycle. In such a situation, the molecule-to-molecule
random variation of the reaction rate in a {\it population} of the same
species of molecules is called ``{\it static disorder}''.

More interesting phenomena are expected in the intermediate situations
where the rates of transitions along the conformational coordinate are
comparable to those along the reaction coordinate. In this case, the
random {\it temporal} fluctuations in the reaction rates of an enzyme
molecule is called ``{\it dynamic disorder}'' \cite{zwanzig90}.

\begin{figure}[htbp]
\begin{center}
{\bf Figure NOT displayed for copyright reasons}.
\end{center}
\caption{(a) A discrete ``catalytic network'' formed by the conformational 
states of the enzyme and enzyme-substrate complex. (b) A schematic 
continuum representation of the conformational states on a energy landscape 
where the two mutually perpendicular directions on the planar ``land'' 
correspond to the reaction coordinate and conformational coordinate, 
respectively. 
Reprinted from Biochemistry    
(ref.\cite{atkins11}), 
with permission from American Chemical Society \copyright (2011). 
}
\label{fig-atkins11}
\end{figure}

\noindent$\bullet${\bf Conformational dynamics as a jump process on discrete network}

Let us consider the reaction
\begin{eqnarray}
R+{\cal E} \mathop{\rightleftharpoons} I_{1} \mathop{\rightleftharpoons} I_{2} \mathop{\rightleftharpoons}...\mathop{\rightleftharpoons} I_{m} \mathop{\rightleftharpoons} {\cal E}+P
\end{eqnarray}
catalyzed by the enzyme $E$ in a bulk biochemical reactor. The $m$
distinct intermediate complexes $I_{1}, I_{2},...I_{m}$ can form
on the pathway leading to the product $P$, starting from the reactant
$R$. However, because of conformational fluctuations, each complex
may exist in $n$ different conformations. For the sake of simplicity
and for the convenience of sketching this intuitive picture, we assume
that the number of conformational states corresponding to all the
intermediate complexes is the same. We use the symbol $I_{\mu}^{c}$
to denote the $c$-th conformation of the $\mu$-th intermediate
complex as shown in eqn.\ref{map-catnetwork} \cite{kou05,kou08}.
Thus, the conformational states form of a ``{\it catalytic network}''
\cite{yon98,kruse07a,benkovic08,hammes02,ma10} 
(see fig.\ref{fig-atkins11}).
From the perspective biochemical reaction networks, each horizontal
row can be interpreted as a reaction channel whereas different channels
interconvert because of the conformational dynamics. Depending on
the rates of the individual transitions on a specific realization
of the catalytic network, some pathways may dominate over others.

\begin{eqnarray}
R+&{\cal E}^{(1)}& \mathop{\rightleftharpoons} ~I_{1}^{(1)}~ \mathop{\rightleftharpoons}  ~I_{2}^{(1)}~ \mathop{\rightleftharpoons}...\mathop{\rightleftharpoons} ~I_{m}^{(1)}~ \mathop{\rightleftharpoons} {\cal E}^{(1)}+P \nonumber \\
&\downarrow \uparrow&~~~~\downarrow \uparrow~~~~~~\downarrow \uparrow~~~~~~~~~~~~\downarrow \uparrow~~~~~~\downarrow \uparrow \nonumber \\
R+&{\cal E}^{(2)}& \mathop{\rightleftharpoons} ~I_{1}^{(2)}~ \mathop{\rightleftharpoons}  ~I_{2}^{(2)}~ \mathop{\rightleftharpoons}...\mathop{\rightleftharpoons} ~I_{m}^{(2)}~ \mathop{\rightleftharpoons} {\cal E}^{(2)}+P \nonumber \\
&\downarrow \uparrow&~~~~\downarrow \uparrow~~~~~~\downarrow \uparrow~~~~~~~~~~~~\downarrow \uparrow~~~~~~\downarrow \uparrow \nonumber \\
...&.....&...................................................................\nonumber \\
&\downarrow \uparrow&~~~~\downarrow \uparrow~~~~~~\downarrow \uparrow~~~~~~~~~~~~\downarrow \uparrow~~~~~~\downarrow \uparrow \nonumber \\
R+&{\cal E}^{(c)}& \mathop{\rightleftharpoons} ~I_{1}^{(c)}~ \mathop{\rightleftharpoons}  ~I_{2}^{(c)}~ \mathop{\rightleftharpoons}...\mathop{\rightleftharpoons} ~I_{m}^{(c)}~ \mathop{\rightleftharpoons} {\cal E}^{(c)}+P \nonumber \\
&\downarrow \uparrow&~~~~\downarrow \uparrow~~~~~~\downarrow \uparrow~~~~~~~~~~~~\downarrow \uparrow~~~~~~\downarrow \uparrow \nonumber \\
...&.....&...................................................................\nonumber \\
&\downarrow \uparrow&~~~~\downarrow \uparrow~~~~~~\downarrow \uparrow~~~~~~~~~~~~\downarrow \uparrow~~~~~~\downarrow \uparrow \nonumber \\
R+&{\cal E}^{(n)}& \mathop{\rightleftharpoons} ~I_{1}^{(n)}~ \mathop{\rightleftharpoons}  ~I_{2}^{(n)}~ \mathop{\rightleftharpoons}...\mathop{\rightleftharpoons} ~I_{m}^{(n)}~ \mathop{\rightleftharpoons} {\cal E}^{(n)}+P \nonumber \\
\label{map-catnetwork}
\end{eqnarray}


\noindent$\bullet${\bf Enzymatic reactions with dynamic disorder: GLE-based approach}

Information on the structure of the landscape or the topology of the 
network, which are required for the above mentioned approaches to the 
conformational kinetics of the enzymes, are usually not available. 
Therefore, there is a need for an alternative approach.

Kramers modelled the reaction as the Brownian motion of a fictitious 
particle- the rate at which the particle permanently escaped a metastable 
potential minimum was identified with the average rate of the corresponding 
reaction \cite{hanggi90}. The motion of this fictitious Brownian particle 
is described in terms of a Langevin (or, equivalent Fokker-Planck) equation. 
In order to capture the memory effects arising from dynamic disorder,  
an extended version of Kramers theory, based on the generalized Langevin 
equation (GLE) with a power-law memory kernel, has been adopted 
\cite{kou04,min05b,min06b,chaudhury06a,chaudhury06b}.

In the overdamped limit, the GLE for the particle subjected to an external 
$U(x)$ is given by  
\begin{equation}
- \zeta \int_{0}^{t} dt' K(t-t') v(t') - dU(x)/dx + \eta(t) = 0 
\label{eq-GLE}
\end{equation}
where $\zeta$ is a measure of dissipation and $\eta$ is a random force. 
The memory kernel $K$ is related to the noise through the fluctuation-
dissipation relation $\zeta k_B T K(|t-t'|) = <\eta(t) \eta(t')>$. 
The success of this approach depends on the appropriate choice of the 
form of the kernel $K$. A power-law kernel $K(t-t') \sim |t-t'|^{-1/2}$ 
can account for the experimental observations \cite{min06b}. 
It can be argued \cite{kou04,min05b} that a more general form for the 
memory kernel $K$ would be 
$K(|t-t'|) = 2 H (2H-1) |t-t'|^{2H-2}$ where $H$ ($1/2 \leq H \leq 1$) 
is a measure of the degree of temporal correlation in the noise.

What is the physical original of the power-law kernel? In the original 
treatment of reaction rate, Kramers assumed a clear separation of the 
times scales: the short time scales of fluctuations in the bath and 
much longer time scale of the reaction that requires hopping over 
barrier assisted by these fluctuations. This scenario may be valid for 
reactions involving small molecules. But, for enzymatic reactions such 
a clear separation of the time scales is not possible. Even if Kramers' 
assumption of infinitely fast relaxation of bath (i.e., white noise)  
is replaced by colored noise with finite relaxation time, the corresponding  
GLE \cite{grote80} cannot account for the memory effects that arise from 
the dynamic disorder in enzymatic reactions. The power-law kernel, which  
corresponds to fractional Gaussian noise \cite{kou04,chaudhury06b}, is 
as essential as the GLE to account for the observed memory effects.

\noindent$\bullet${\bf Enzymatic reactions with dynamic disorder: FP-based approach}

Let us assume that the number of conformations is so large that
these can be described by a continuous variable $q$.
The kinetics of the conformations along the $q$ coordinate includes 
a diffusion term and a drift term as well as a term that represents
a simple reaction which could be, for example, the decay of the
fluorescent state of the protein to a non-fluorescent state.
The Smoluchowski equation can be written as \cite{agmon83}
\begin{equation}
-\partial P(q,t)/\partial t = \partial J(q,t)/\partial q + k(q) P(q,t),
\end{equation}
where the probability flux $J$ is given by
\begin{equation}J(q,t) = - D\biggl(\frac{\partial}{\partial q} + \frac{1}{k_BT} \frac{\partial V
}{\partial q}\biggr) P(q,t)
\end{equation}
If no reaction takes place, i.e., $k(q) = 0$, this Smoluchowski equation
provides an equilibrium solution $P^{eq}(q) = exp[-V(q)/(k_BT)]/Z$
where $Z = \int exp[-V(q)/(k_BT)] dq$ is the partition function.
In the opposite special situation where no diffusion takes place, i.e.,
$D=0$, the probability $P(q,t)$ decays exponentially with time $t$ as
\begin{equation}
P(q,t) = P(q,0) exp[-k(q)t] ~~D \to 0,
\end{equation}
purely because of the reaction at a fixed $q$.
But, if $k \neq 0$ and $D \to \infty$, one can show that
\cite{agmon83}
\begin{equation}
P(q,t) = P(q,0) exp[-<k>_{eq}t], ~~D \to \infty
\end{equation}
where $<k>_{eq} = \int k(q) P^{eq}(q) dq$. Thus, in both the extreme
cases $D \to 0$ and $D \to \infty$, $P(q,t)$ decays with a single
exponential.

The main difficulty with this formulation of the problem is that the 
general case can be solved only numerically if $V(q)$ has a nontrivial 
$q$-dependence. Since any numerical solution requires discretization 
in any case, it is more convenient to reformulate the problem as a 
discrete jump process along $q$ coordinate from one potential minimum 
to a neighboring one and describe it in terms of a master equation:
\begin{equation}
\partial P_{j}(t)/\partial t = P_{j-1} W(j-1 \to j) + P_{j+1} W(j+1 \to j) - P_{j}[W(j \to j-1) + W(j \to j+1)] - k_{j} P_{j}
\end{equation}
where the integer index $j$ labels the successive discrete positions
along the coordinate $q$. As a simple illustrative example, one can
consider a two-state system, i.e., a system which has only two
conformational states, labelled by $j=1,2$ along $q$ coordinate.
In this case, for the initial condition $P_{1}(0) = P_{2}(0) = 1/2$,
the ``survival probability'' $Q(t) = \int P(q,t) dq$ is given by
\cite{agmon83}
\begin{eqnarray}
Q(t) = [exp(-k_{1}t) + exp(-k_{2}t)]/2, ~~D \to 0, \nonumber \\
Q(t) = exp[-(k_{1}t + -k_{2}t)/2], ~~D \to \infty.
\label{eq-surviveQ}
\end{eqnarray}
The forms of $Q(t)$ in equation (\ref{eq-surviveQ}) obtained in the two 
special cases $D \to 0$ and $D \to \infty$ are identical to the 
survival probabilities calculated earlier earlier for the model 
(\ref{eq-xie}) in the special limits $k \to 0$ and $k \to \infty$.

\begin{figure}[htbp]
\begin{center}
\includegraphics[angle=-90,width=0.45\columnwidth]{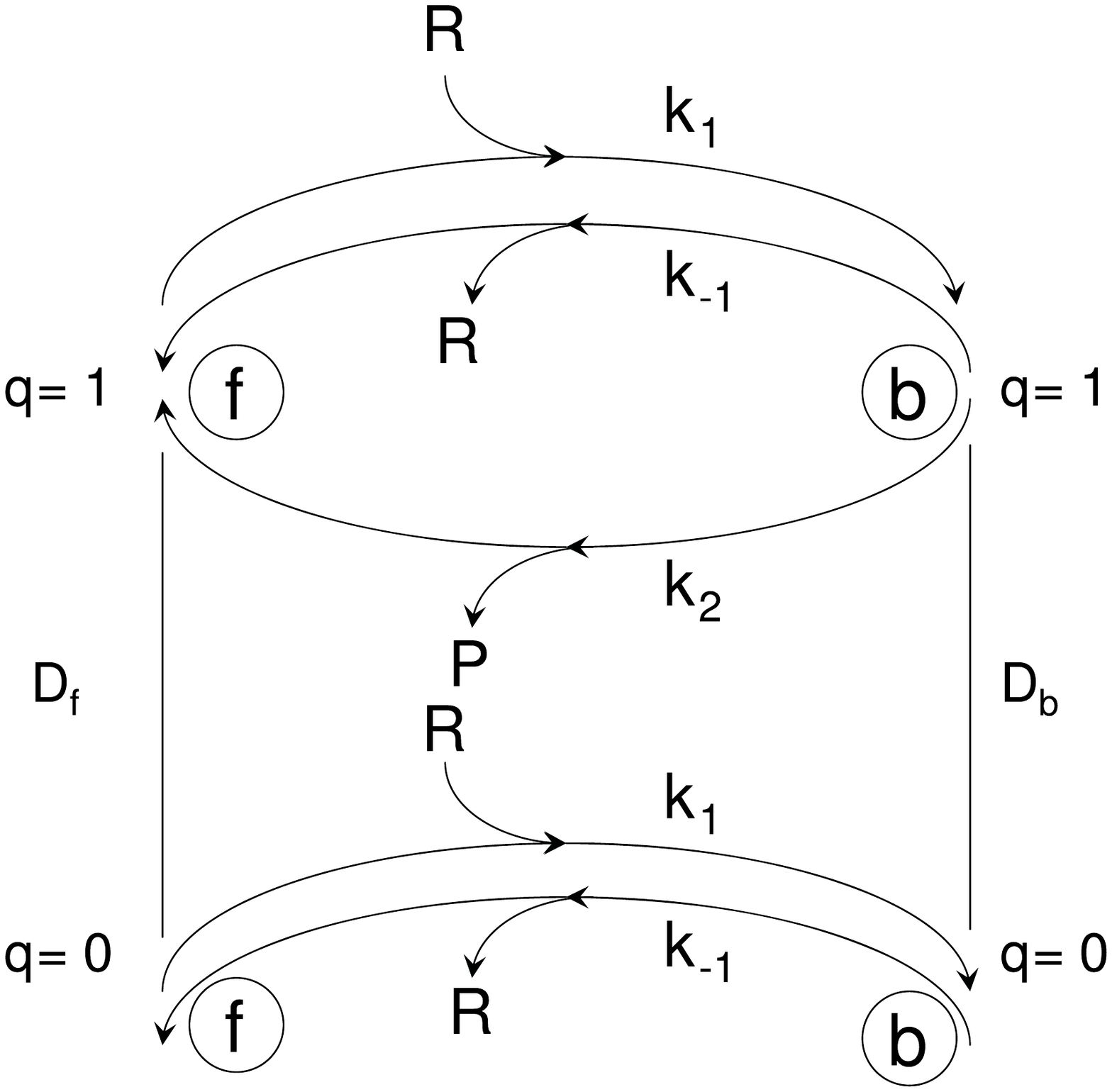}
\end{center}
\caption{Agmon's model \cite{agmon85} of Michaelis-Menten scheme for 
enzymatic reactions with purely diffusive conformational fluctuations. 
See the text for explanations.}
\label{fig-agmon}
\end{figure}

A generalization of the MM scheme was proposed many years ago \cite{agmon85}
to capture the effects of conformational variations. In this model
(see Fig.\ref{fig-agmon})
the enzyme can exist in two discrete forms: ``free'' and ``bound''.
The conformations are described by a continuous variable $q$ and
its range is $0 \leq q \leq 1$. The value $q=0$ represents the
``inactive'' conformation in which it can bind reversibly with the
reactant but cannot catalyze the reaction; the reactant can unbind.
$q=1$ represents the ``active'' form which can catalyze the reaction.
Suppose $P_f$ and $P_b$ denote the probabilities of finding the
enzyme in the free and bound states, respectively. {\it Assuming} that
the variations of the conformations is purely {\it unbiased diffusive}
motion along the coordinate $q$,
\begin{eqnarray}
\partial P_f/\partial t = D_f (\partial^2P_f/\partial q^2) \nonumber \\
\partial P_b/\partial t = D_b (\partial^2P_b/\partial q^2)
\end{eqnarray}
where the corresponding diffusion constants $D_f$ and $D_b$ are
measures of the rapidity of conformational dynamics. Moreover,
{\it assumption of steady state} along $q$ yields the equations
\begin{eqnarray}
\partial^2P_f/\partial q^2 = 0 ~{\rm and}~ \partial^2P_b/\partial q^2 = 0.
\end{eqnarray}
while the {\it assumption of steady state} along the reaction coordinate
$\xi$ provides the boundary conditions
\begin{eqnarray}
- D_f [\partial P_f/\partial q](q=0) = k_{-1} P_b(q=0) - k_{1} P_f(q=0) [R] = D_b [\partial P_b/\partial q](q=0)
\end{eqnarray}
and
\begin{eqnarray}
- D_f [\partial P_f/\partial q](q=1) = - k_{2} P_b(q=1) = D_b [\partial P_b/\partial q](q=1)
\end{eqnarray}
The rate of the reaction, under the steady-state conditions, is given
by \cite{agmon85}
\begin{eqnarray}
V = \frac{k_{2} [R]}{K^{eff}_M + \{1+(k_{2}/2)(D_f^{-1}+D_b^{-1})\}[R]}
\end{eqnarray}
where
\begin{equation}
K^{eff}_M = \frac{k_{-1}+k_{2} + (k_{1}k_{2}/D_b)}{k_{1}}
\end{equation}
In the special limit $D_f \to \infty$ and $D_b \to \infty$, in which
the effects of fast conformational dynamics gets averaged out, we
recover the original MM expression (\ref{eq-michaelis}) for the
average rate of the reaction, along with the form (\ref{eq-michconstant})
of the Michaelis constant $K_M$.

Note that in the Agmon model \cite{agmon85} the two kinetic steps of the
MM reaction scheme are incorporated as boundary conditions of the
diffusion equation that describes conformational dynamics in a
direction perpendicular to the reaction coordinate.

A more realistic description of the conformational transitions as
``diffusive'' motion along the coordinate $q$ should incorporate
the fact that the potential energy exhibits many ``wells'' separated
by small barriers. In principle, one should formulate a FP-like
equation for the probability density $P(q,\xi,t)$ in the two-dimensional
space spanned by the conformational coordinate $q$ and the reaction
coordinate $\xi$ \cite{xing07}. In the special situations where the
chemical reaction proceeds slowly and conformational transitions are
faster, $P(q,\xi,t)$ can be simplified to the form $P_{i}(q,t)$ where
the discrete index $i$ labels the distinct chemical states along $\xi$
\cite{xue06}. One can then develop a hybrid equation where a FP-like
part describe diffusive motion along continuous coordinate $q$ and
a master equation-like part accounts for the discrete jumps along
the discretized reaction coordinate. Those transitions which involve
simultaneous change of $q$ and the reaction coordinate $\xi$ cannot
be captured by this model \cite{qian09}. But, such mixed transitions
which couple reaction with conformational transition(s), have important
implications \cite{min08,min09}.

The MM-like form holds under all the following three conditions \cite{min06a}:\\
(i) {\it quasi-static} condition when the conformational fluctuations
of the free enzyme as well as the enzyme bound to reactant or product
are much slower than the other steps, e.g., the substrate-binding,
catalytic step of the reaction, and release of the product. \\
(ii) {\it quasi-equilibrium} condition when the reactant dissociation
is much faster than all the other steps, e.g., catalytic conversion
of the reactant(s) to product(s), irrespective of the amplitude or
the time scales of the conformational fluctuations.\\
(iii) {\it conformational equilibrium} condition when the rate constants
for the steps of the reaction depend on the same way on the conformational
coordinate $q$, i.e., $k_2(q)/k_{1}(q) = c$ independent of $q$.

\noindent$\bullet${\bf Enzymatic reactions with dynamic disorder: master equation-based approach}

{\bf Example 1: a model with 2 conformational states}

In order to get an indication of the possible complexities arising from
such dynamic disorder let us consider the MM reaction
(\ref{eq-singMMscheme}) in the special limiting situation
$k_{-1} \ll k_{1}$ so that the both the steps of the two-step reactionare irreversible, i.e.,
\begin{eqnarray}
E + R \mathop{\rightarrow}^{k_{1}} &I_{1}& \mathop{\rightarrow}^{k_{2}} E^{*} + P, \nonumber \\
E^{*} \mathop{\rightarrow}^{\delta} E
\label{eq-singMMirrev}
\end{eqnarray}
Now suppose $k_{2}$ is given by the Arrhenius equation
$k_{2} = k_{0} exp[-E_{a}/(k_BT)]$ where $E_{a}$ is the activation
barrier. Dynamic disorder is incorporated in this model in the
following way \cite{floyd10}:
for each round of the reaction catalyzed by the same individual single
enzyme, the magnitude of the barrier $E_{a}$ is obtained by drawing a
normally distributed random variable. The width $w(k_{2})$ of the
distribution of $k_{2}$ is, thus, determined by that of $E_{a}$. In
order to make both the steps rate limiting, the mean
$\langle k_{2}\rangle$ was kept fixed at a value that is identical
to the numerical value of $k_{1}$ which is non-random. From numerical
simulation of this model \cite{floyd10}, it was observed that
as the width of the distribution $w(k_{2})$ increases, $f(t)$ not
only becomes wider, but also approaches a {\it single} exponential.
Moreover $1/r$ also approaches unity with the increase of the width
of $w(k_{2})$. Although both these observations are mutually consistent,
these contradict the expectation that $r=2$ and $f(t)$ should be a
sum of two exponentials because this reaction involves two equally
rate-limiting steps. This simple example demonstrates that, because
of dynamic disorder, a multi-step reaction with $N$ intermediate
steps may appear to involve fewer steps.

{\bf Example 2: general model with n conformational states: a catalytic network} 

Some of the experimental observations in single-molecule enzymology 
cannot be explained by the simple stochastic formulation of the 
kinetics of the MM reaction without incorporating the effects of 
dynamic disorder. Extension of that stochastic model into a stochastic 
reaction network model \cite{kou05,kou08} can account for the 
experimental observations.

Thus, the apparent memory reflected in the correlation function does not
result from any intrinsic memory of the enzyme- it does not remember its
past. The apparent memory effect is caused by the inability of the
experimental set up to detect individual conformational states of the
enzyme and enzyme-substrate complexes. A single-enzyme experiment does
not observe the conformational states directly. Instead, it goes through
a relatively ``dark'' period interrupted by a fluorescence pulse followed
by another ``dark'' period.

Why was these memory effects not picked up in bulk measurements? The
answer is that such slow turnovers were not tracked in bulk measurements.
A unified description of fluctuating enzyme kinetics has been presented 
by Min et al.\cite{min10}. The phase diagram of this model depicts 
distinct behaviors of the enzymatic kinetics (i.e., distinct kinetic 
phases) in different parameter regimes.

\subsection{\bf{Substrate specificity and specificity amplification}}

Although we have already discussed one of the mechanisms of specificity 
amplification, namely kinetic proofreading, so far we have not explained 
the mechanism of substrate specificity itself. In this section, we 
summarize the current understanding of how specificity arises. We also 
discuss an alternative mechanism of specificity amplification, called 
energy relay, that is closely related to the phenomenon of temporal 
cooperativity.

\subsubsection{\bf Role of conformational kinetics in selecting specific substrate}

Investigations on the role of protein fluctuations in enzyme kinetics 
has a long history \cite{welch82}.

\noindent $\bullet${\bf Substrate specificity: from lock-and-key to induced fit}

Substrate specificity is a concrete example of a general phenomenon, called
``molecular recognition'', which plays important roles not only in
catalysis, but also, for example, in (a) immune response 
\cite{goldstein04,goldstein08},
(b) signal transduction \cite{swain02}, etc.

According to the oldest hypothesis of ``lock-and-key'' mechanism,
originally proposed by Emil Fischer, the specificity
arises from the complementarity of the shape of the substrate and that
of the catalytic pocket of the enzyme. However, this picture cannot
explain why the same enzyme does not catalyze all those smaller
substrates which would fit into the active site that is specifically
complimentary to a much larger substrate. It also fails to account for
the observed fact that some enzymes are highly selective whereas others
can catalyze several substrates of quite different shapes.

Later the rigid lock-and-key picture was replaced by the ``induced-fit
mechanism'' \cite{koshland94,johnson08}
according to which the substrate, upon binding to the enzyme, induces
conformational changes so as to fit it. In other words, lock-and-key
fitting is like fitting the pieces of a jigsaw puzzle whereas the
induced fit is like the fitting of a hand in a glove.

\noindent $\bullet${\bf substrate specificity: induced fit versus conformation selection}

\begin{figure}[htbp]
\begin{center}
\includegraphics[angle=-90,width=0.45\columnwidth]{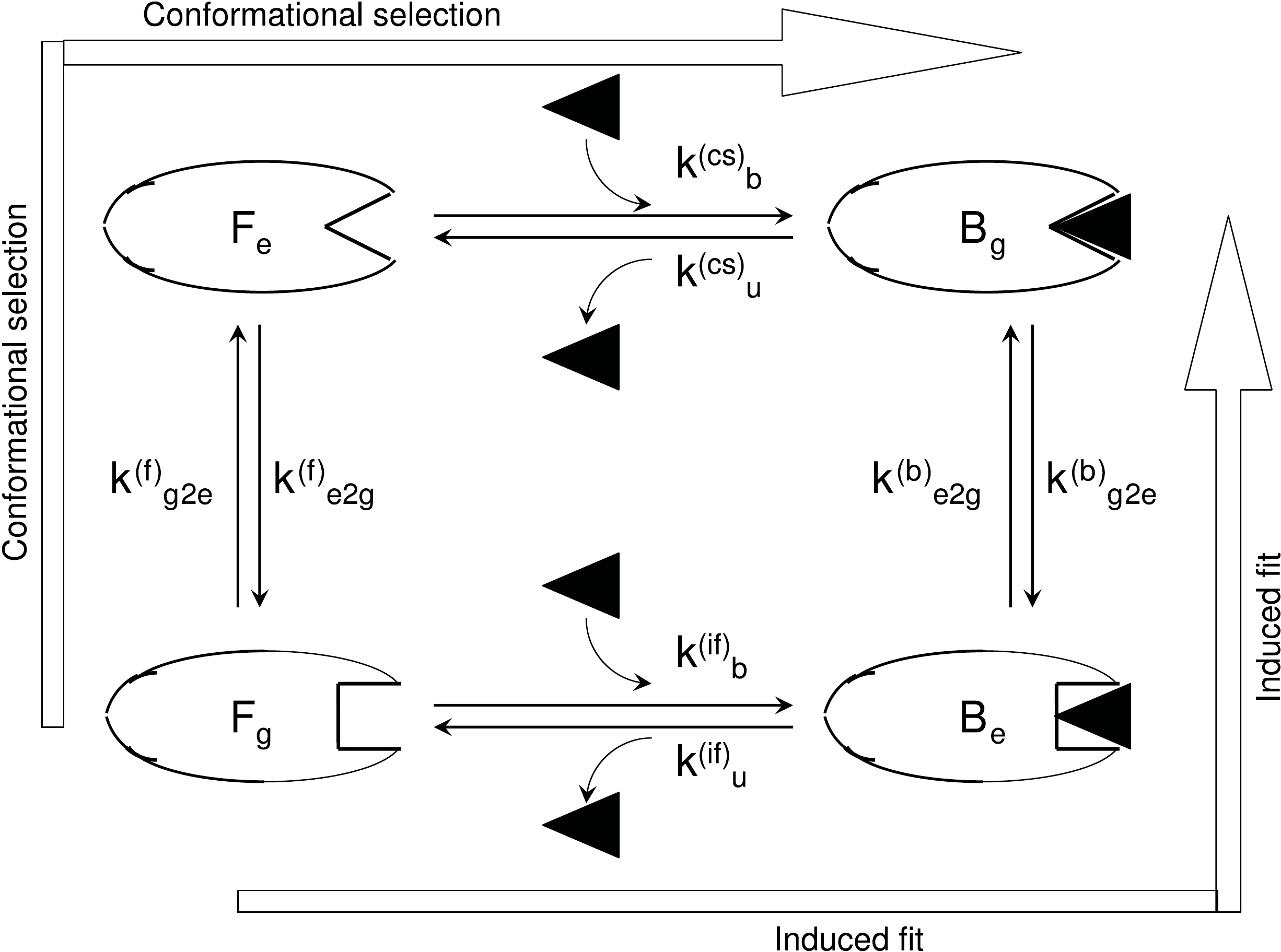}
\end{center}
\caption{Induced fit versus conformational selection mechanism of 
substrate specificity of enzymes.
}
\label{fig-indfitconfsel}
\end{figure}

In more recent times, an alternative scenario has been proposed. In
this ``conformation selection'' scenario,
\cite{tsai99,kumar00,bosshard01,giraldo06,boehr09,weikl09,okazaki08,hammes09,zhou10,bernado10,vertessy10,nussinov12}
populations of the enzyme pre-exist in different conformations because
of thermal fluctuations; a substrate merely ``selects'' the one that
fits it best. The difference between the induced-fit mechanism and the
conformation selection mechanism can be elucidated in terms of
(i) a dynamic landscape \cite{tsai99,kumar00,boehr09,okazaki08} and
(ii) kinetic pathways through discrete states \cite{weikl09,hammes09}
(see fig.\ref{fig-indfitconfsel}).

Mechanism of substrate specificity of an enzyme can be viewed as a
process of ``{\it conformational proofreading}'' \cite{savir07}.
Specificity is a manifestation of an enzyme's ability to discrminate
between competing substrates. In spite of the similarities, there are
also important differences between the concepts of conformational
proofreading and kinetic proofreading. For example, a coupling of an
enzymatic reaction to ATP hydrolysis inserts a {\it temporal delay}
in the main pathway. The couterpart of this in conformational
proofreading is a {\it spatial mismatch} \cite{savir07}. Moreover,
recall that kinetic proofreading drives the reaction away from
equilibrium by the coupling to the energy-consuming reaction (e.g.,
ATP hydrolysis). In contrast, conformational proofreading requires
only a quasi-equilibrium scenario.

\subsubsection{\bf Temporal cooperativity in enzymes: hysteretic, mnemonic enzymes and energy relay}

Temporal cooperativity can occur even in those enzymes which are
{\it monomeric} and has only a single binding site which is the
catalytic site of the enzyme. Cooperativity in such enzymes arise
from slow conformational dynamics. Concepts of hysteretic and
mnemonic enzymes as well as the related concept of enzyme memory,
which also arise from slow conformational dynamics, were formalized
in the nineteen sixties and seventies
\cite{rabin67,frieden70,ainslie72,ricard74,ricard77,buc77}
to account for some kinetic phenomena in biochemical experiments.
In the recent years these concepts are again at the focus of attention
because of the feasibility of single-enzyme experiments.

\begin{figure}[htbp]
\begin{center}
\includegraphics[angle=-90,width=0.45\columnwidth]{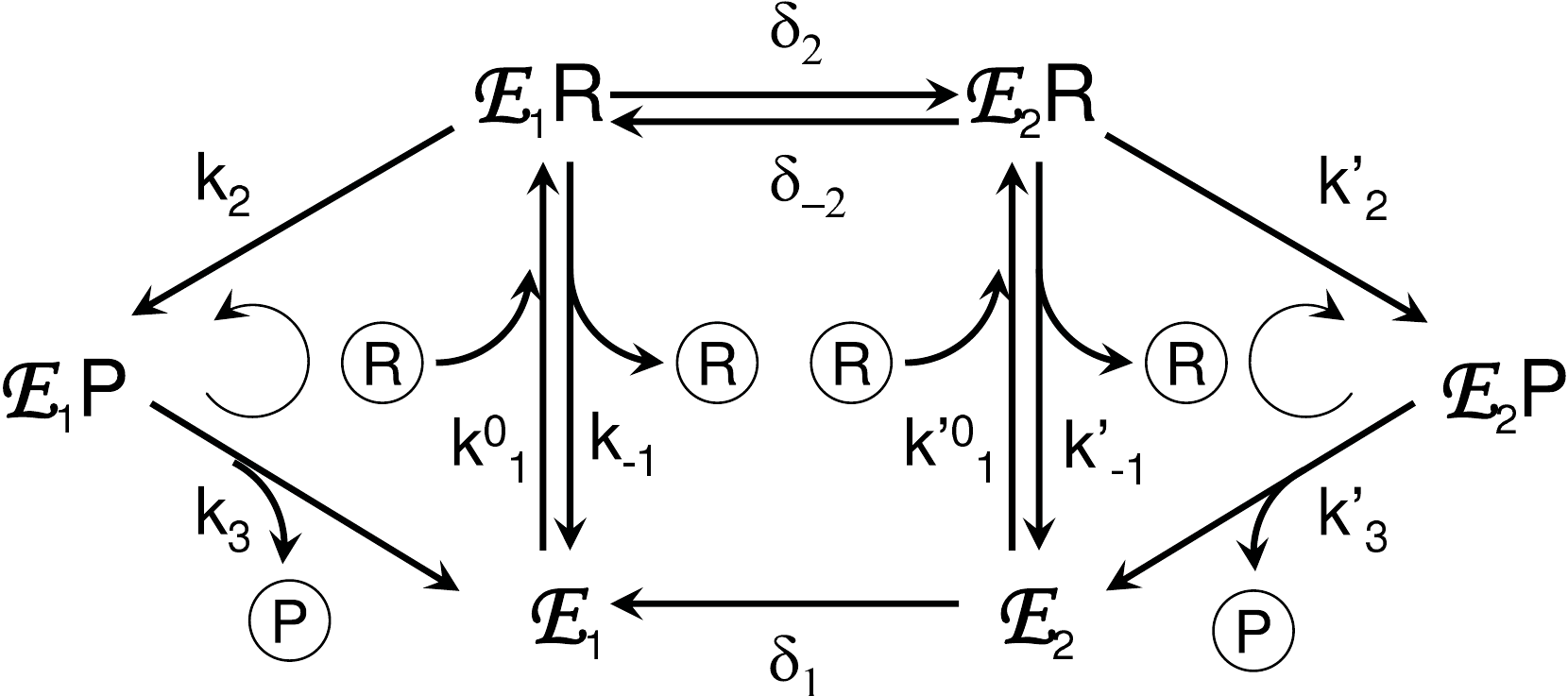}
\end{center}
\caption{Temporal cooperativity of enzymes (see the text for explanations).
}
\label{fig-tempcoop}
\end{figure}

Consider an enzyme which can exist in two different conformations
${\cal E}_{1}$ and ${\cal E}_{2}$ in the ligand-free state. 
As emphasized by the symmetrical figure \ref{fig-tempcoop}, 
if the interconversions 
${\cal E}_{2} \to {\cal E}_{1}$ and 
${\cal E}_{2}R \leftrightharpoons {\cal E}_{1}R$ 
are extremely slow compared to the other transitions, the enzymatic 
reaction will proceed either through the pathway 
${\cal E}_{1}+R \to {\cal E}_{1}R \to {\cal E}_{1}P \to {\cal E}_{1}+P$ 
or through the pathway 
${\cal E}_{2}+R \to {\cal E}_{2}R \to {\cal E}_{2}P \to {\cal E}_{2}+P$; 
the actual pathway followed by a given enzyme would depend on whether 
it was in ${\cal E}_{1}$ or ${\cal E}_{2}$ initially.

However, if interconversions are possible the kinetics can be quite complex. 
Suppose, ${\cal E}_{1}$ is thermodynamically more stable than ${\cal E}_{2}$. 
Therefore, let it be initially in  the conformation ${\cal E}_{1}$. Then, 
following the binding of a reactant $R$ to ${\cal E}_{1}$, a complex 
${\cal E}_{1}R$ is formed. This complex can lead to the formation of the 
product $P$ either by direct transition to ${\cal E}_{1}P$ or indirectly by 
getting converted to the complex ${\cal E}_{2}R$ which, in turn, produces $P$ 
following the step ${\cal E}_{2}R \to {\cal E}_{2}P$. Releasing the product, 
the free enzyme is recovered in the conformation ${\cal E}_{1}$ and 
${\cal E}_{2}$ in these alternative pathways. The possibilities of the 
interconversion ${\cal E}_{2} \to {\cal E}_{1}$ and the transition 
${\cal E}_{2} \rightleftharpoons {\cal E}_{2}R$ add further complexities 
to the kinetics.

For simplicity, let us assume $k_{2} = k_{3} \simeq 0$; in this special 
case, multiple turnovers by ${\cal E}_{1}$ is possible only if both 
$\delta_{1}$ and $\delta_{2}$ are nonzero. In addition, let us make the 
following {\it assumptions}: (i) among the sequence of steps 
starting with the combination of ${\cal E}_{1}$ and $R$ the rate 
limiting (i.e., the slowest) step is ${\cal E}_{1}R \to {\cal E}_{2}R$. 
Therefore, in the pathway 
${\cal E}_{1}+R \to {\cal E}_{1}R \to {\cal E}_{2}P \to {\cal E}_{2}+P$  
i.e., $\delta_2 << k'_{2}$.
(ii) $\delta_{-2} << k'_{2}$, so that the conversion 
${\cal E}_{2}R \to {\cal E}_{1}R$ is highly unlikely.
(iii) Note that $k'_{1} = k'^{0}_{1} [R]$. Therefore, if the numerical 
values of $\delta_{1}$ and $k'^{0}_{1}$ are such that 
$k'_{1} << \delta_{1}$ for small $[R]$ and $k'_{1} >> \delta_{1}$ at 
sufficiently large $[R]$, then direct conversion 
${\cal E}_{2} \to {\cal E}_{1}$ is almost certain at low $[R]$ and 
practically impossible at high $[R]$. 
(iv) Suppose, $k^{0}_{1} << k'^{0}_{1}$ so that the reactant is more 
likely to bind ${\cal E}_{2}$ than ${\cal E}_{1}$  \cite{rabin67,qian10}.

How does cooperativity arise in such a system? Note that the conformation 
${\cal E}_{2}$ is a byproduct of the first turnover of the enzyme that 
was initially in ${\cal E}_{1}$. Subsequently, ${\cal E}_{2}$ can convert 
substrate molecules into products bypassing the slow step (corresponding 
to $\delta_{2}$) in the reaction that starts with the enzyme in conformation 
${\cal E}_{1}$.  At low substrate
concentration, the free enzyme gets enough time to relax from the
conformation ${\cal E}_{2}$ to ${\cal E}_{1}$ before the encounter
with a substrate molecule. Thus, at sufficiently low concentrations,
the substrate molecules are converted to product by the route
${\cal E}_{1}+R \to {\cal E}_{2}+P$ whereas at sufficiently high
concentration, the more frequent route is 
${\cal E}_{2}+R \to {\cal E}_{2}+P$; a crossover from the first 
(slower) to the second (faster) route takes place at some intermediate 
concentration of substrates.

The effective $S$-shape of the resultant curve also explains the Hill-like, 
rather than MM-like, behavior of the kinetics observed in such systems.
Hopfield \cite{hopfield80,qian10} proposed a formal scheme for 
specificity amplification based on the concept of ``{\it energy relay}'' 
that exploits temporal cooperativity of enzymes.

\section{\bf Thermodynamics of energy transduction: equilibrium and beyond}

\noindent $\bullet${\bf Defining efficiency: from Carnot to Stokes}

The performance of {\it macroscopic} motors are characterized by a
combination of its efficiency, power output, maximum force or torque
that it can generate. Just like the performance of their macroscopic
counterparts with finite cycle time, that of molecular motors
\cite{schmiedl08} have also been characterized in terms of efficiency
at maximum power, rather than maximum efficiency. However, the
efficiency of molecular motors can be defined in several different
ways \cite{linke05}.

The efficiency of a motor, with finite cycle time, is generally
defined by
\begin{equation}
\eta = {\cal P}_{out}/{\cal P}_{in}
\label{eq-eff}
\end{equation}
where ${\cal P}_{in}$ and ${\cal P}_{out}$ are the input and output
powers, respectively. The usual definition of {\it thermodynamic
efficiency} $\eta_{T}$ is based on the assumption that, like its
macroscopic counterpart, a molecular motor has an output power
\cite{parmeggiani99}
\begin{equation}
{\cal P}_{out} = - F_{ext} V.
\label{eq-thermoout}
\end{equation}
where $F_{ext}$ is the externally applied opposing (load) force.
Although this definition is unambiguous, it is unsatisfactory for
practical use in characterizing the performance of molecular
motors. As explained earlier, a molecular motor has to work against
the omnipresent viscous drag in the intra-celluar medium even when
no other force opposes its movement (i.e., even if $F_{ext}=0$).

A generalized efficiency $\eta_{G}$ is also represented by the
same expression (\ref{eq-eff}) where, instead of (\ref{eq-thermoout}),
the output power is assumed to be \cite{derenyi99a}
\begin{equation}
{\cal P}_{out} = F_{ext} V + \gamma V^2.
\label{eq-genout}
\end{equation}
where the viscous drag force has been assumed to have the usual form 
$-\gamma v$. This definition treats the load force and viscous drag on 
equal footing.

In contrast, the ``Stokes efficiency'' $\eta_{S}$ for a molecular motor
driven by a chemical reaction is defined as \cite{wang02b}
\begin{equation}
\eta_{S} = \frac{\gamma V^2}{(\Delta G)\langle r \rangle + F_{ext} V}
\end{equation}
where $\langle r \rangle$ is the average rate of the chemical reaction
and $\Delta G$ is the chemical free energy consumed in each reaction
cycle. This efficiency is named after Stokes because the viscous drag
is calculated from Stokes law.

Power output is one of 
the standard measures of performance of a motor. Power output itself 
gets contributions from both force and velocity, the two key features 
of the motors whose output is mechanical work. However, higher output 
of larger motors may arise from a trivial dependence on its volume 
(or weight); in such cases larger power output may merely reflect 
contributions of larger number of force generators. Therefore, the 
{\it specific} power output, i.e., maximum power output per unit volume 
(or, weight) of the engine, is an {\it intrinsic} characteristic that 
should be used to compare the performance of motors irrespective of the 
difference in their volume (or, weight) \cite{nicklas84}.

\subsection{\bf Phenomenological linear response theory for molecular motors: modes of operation}

As we explained earlier with the example of a chemo-chemical machine, 
energy transduction by a molecular machine involves a direct coupling 
between a process favored by free energetics (i.e., a {\it spontaneous} 
process) and a unfavored process \cite{harvey94}.
In this subsection we present a general treatment of the {\it
phenomenological} linear response theory for such coupled processes
within the framework of thermodynamics of irreversible processes
\cite{katchalsky67,demirel04}.
For introducing the key concepts of of this formalism, we follow
mostly the classic papers \cite{kedem65,essig68,stucki80a,stucki80b}
and a few recent works \cite{hernandez08}.

For simplicity, as well as for the realistic situation of most motors, 
we consider only two coupled processes. The generalized currents 
$J_{\mu}$ ($\mu=1,2$) are assumed to be related to the two generalized 
forces $X_{\mu}$ ($\mu=12$) by 
\begin{eqnarray}
\left(
\begin{array}{c}
J_{1} \\
J_{2}
\end{array}
\right) =
\begin{pmatrix}
L_{11} & L_{12} \\
L_{21} & L_{22}
\end{pmatrix}
\left(
\begin{array}{c}
X_{1} \\
X_{2}
\end{array}
\right)
\label{eq-2by2}
\end{eqnarray}
$L_{12} = L_{21}$ is the well known Onsager reciprocity relation. 

Inverting the equations (\ref{eq-2by2}), the linear response relations
can also be expressed as
\begin{eqnarray}
\left(
\begin{array}{c}
X_{1} \\
X_{2}
\end{array}
\right) =
\begin{pmatrix}
R_{11} & R_{12} \\
R_{21} & R_{22}
\end{pmatrix}
\left(
\begin{array}{c}
J_{1} \\
J_{2}
\end{array}
\right)
\label{eq-2by2inv}
\end{eqnarray}
where $R = L^{-1}$. Thus, the elements of the matrix ${\underbar R}$
are related to those of the matrix ${\underbar L}$ by
\begin{eqnarray}
R_{11} &=& \frac{L_{22}}{L_{11}L_{22}-L_{12}^{2}} \nonumber \\
R_{22} &=& \frac{L_{11}}{L_{11}L_{22}-L_{12}^{2}} \nonumber \\
R_{12} &=& \frac{-L_{12}}{L_{11}L_{22}-L_{12}^{2}} = R_{21}.
\end{eqnarray}

The rate of (internal) entropy production $d_{i}\sigma/dt$ is
$d_{i}\sigma/dt = \sum_{k} X_{k} J_{k}$
which, in the case of two coupled processes of the type (\ref{eq-2by2})
takes the simple form 
\begin{equation}
d_{i}\sigma/dt = J_{1} X_{1} + J_{2} X_{2}
\label{eq-entrorate1}
\end{equation}

From the second law of thermodynamics, $d_{i}\sigma/dt > 0$ in
any irreversible process. Consequently, if one of the two terms
on the right hand side of (\ref{eq-entrorate1}) is negative,
the other term has to more than compensate it so as to make the
sum positive. Suppose $J_{2} X_{2} > 0$ and $J_1 X_1 < 0$ while 
$J_{1} X_{1} + J_{2} X_{2} > 0$. In this case the process 2 is 
a ``natural'' (or, spontaneous) process. In contrast, the process 
1  ``unnatural'' for which the flux $J_{1}$ flows against the 
corresponding generalized force $X_{1}$. In this case, the 
``natural'' process 2 plays the role of an energy source that 
drives the ``unnatural'' process 1 by coupling with it. 

Substituting the expressions (\ref{eq-2by2}) for $J_{1}$ and $J_{2}$
into the equation (\ref{eq-entrorate1}) we get
\begin{equation}
d_{i}\sigma/dt = L_{11} X_{1}^{2} + (L_{12}+L_{21})X_{1}X_{2} + L_{22} X_{2}^{2} > 0.
\label{eq-quadratic}
\end{equation}
Since either $X_{1}$ or $X_{2}$ can be switched off, and we must still
have $d_{i}\sigma/dt > 0$, $L_{11} > 0$, $L_{22} > 0$. Moreover, the
positive definite property of the quadratic form on the right hand side
of (\ref{eq-quadratic}) is guaranteed if and only if the determinant
of the matrix $\underbar{L}$ is non-negative, i.e.,
$(L_{11} L_{22} - L_{12} L_{21}) \geq 0$ which, in turn, implies
\begin{equation}
L_{11} L_{22} \geq L_{12}^{2}
\label{eq-Lconstraint}
\end{equation}
because of the Onsager reciprocity relation.
Using the relations between the elements of the matrix ${\bf L}$
and ${\bf R}$, it is straightforward to verify that
$R_{11} > 0$, $R_{22} > 0$ and $R_{11} R_{22} \geq R_{12}^{2}$.

It is often
convenient \cite{kedem65} to define the force-ratio $x$ and flux-ratio
$j$ by
\begin{equation}
x = X_{1}/X_{2}, ~~j=J_{1}/J_{2}.
\end{equation}
For later convenience, we introduce the notation
\begin{equation}
Zx = \tilde{x} ~{\rm and}~ j/Z = \tilde{j}
\end{equation}
where the ``{\it phenomenological stoichiometry}'' $Z$ is defined by
\begin{equation}
Z = (L_{11}/L_{22})^{1/2} = (R_{22}/R_{11})^{1/2}.
\end{equation}
Note that $Z$ is not to be confused with stoichiometry of chemical
reactions. 

The {\it degree of coupling} $q$ between the two processes is defined as
$q = L_{12}/\sqrt{(L_{11} L_{22})} = R_{12}/\sqrt{(R_{11} R_{22})}$. Note
that $0 \leq |q| \leq 1$ (more precisely, $-1 \leq q \leq 1$). 
The linear response relations (\ref{eq-2by2}) can be recast as a
single equation in terms of the dimensionless force $\tilde{x}$ and
dimensionless flux $\tilde{j}$, using the dimensionless parameter
$q$ as follows \cite{kedem65}:
\begin{equation}
\tilde{j} = \frac{q + \tilde{x}}{1+q\tilde{x}}.
\end{equation}
In the special limit $X_{1} \to 0$, and $X_{2} \neq 0$, i.e.,
$\tilde{x} \to 0$, $\tilde{j} \to q$. In the opposite limit
$X_{2} \to 0$ and $X_{1} \neq 0$, i.e., $\tilde{x} \to \pm \infty$,
$\tilde{j} \to 1/q$.  In the special limit where the two processes 1 and
2 become uncoupled from each other, $\tilde{j} = \tilde{x}$.
In the opposite limit $q^{2} = 1$, i.e., complete coupling of the
two processes, $\tilde{j} = 1$, irrespective of the value of
$\tilde{x}$. In fact, in the limit $q^{2} \to 1$, the determinant
of the $2 \times 2$ matrix $\underbar{L}$ vanishes; consequently,
in this limit the two equations (\ref{eq-2by2}) are not independent
of each other.

The {\it efficiency} of transduction is defined by
\begin{equation}
\eta = - \frac{J_1 X_1}{J_2 X_2}
\end{equation}
Within the linear response formalism the efficiency can be expressed
in terms of $\tilde{x}$ and $q$ as
\begin{equation}
\eta(\tilde{x},q) = - \frac{\tilde{x}+q}{\tilde{x}^{-1}+q}
\end{equation}
and in terms of $\tilde{j}$ and $q$ as
\begin{equation}
\eta(\tilde{j},q) = - \frac{\tilde{j}-q}{\tilde{j}^{-1}-q}
\end{equation}
For a given $q$, the maximum of efficiency $\eta(\tilde{x},q)$ is
attained at $\tilde{x}_{max} = - q/(1+\sqrt{1-q^{2}})$ and
the corresponding value of the efficiency is
$\eta_{max}(q) = q^{2}/(1+\sqrt{1-q^{2}})^{2}$

The {\it output power} ${\cal P}$ is defined as ${\cal P} = - J_1 X_1$.
In terms of $\tilde{x}$,
${\cal P} = - \tilde{x}(\tilde{x}+q)L_{22} X_{2}^{2}$. 
The efficiency at maximum power output is given by
$(\eta)_{{\cal P}_{max}} = q^{2}/(4-2q^{2})$. Thus,
in the limit $q=0$, both $\eta_{max}$ and $\eta_{{\cal P}_{max}}$
vanish. But, in the opposite limit $q=1$, $\eta_{max} = 1$, whereas
$(\eta)_{{\cal P}_{max}} = 1/2$.

\noindent$\bullet${\bf Modes of operation of a nano-motor}

\begin{figure}[htbp]
\begin{center}
\includegraphics[angle=-90,width=0.45\columnwidth]{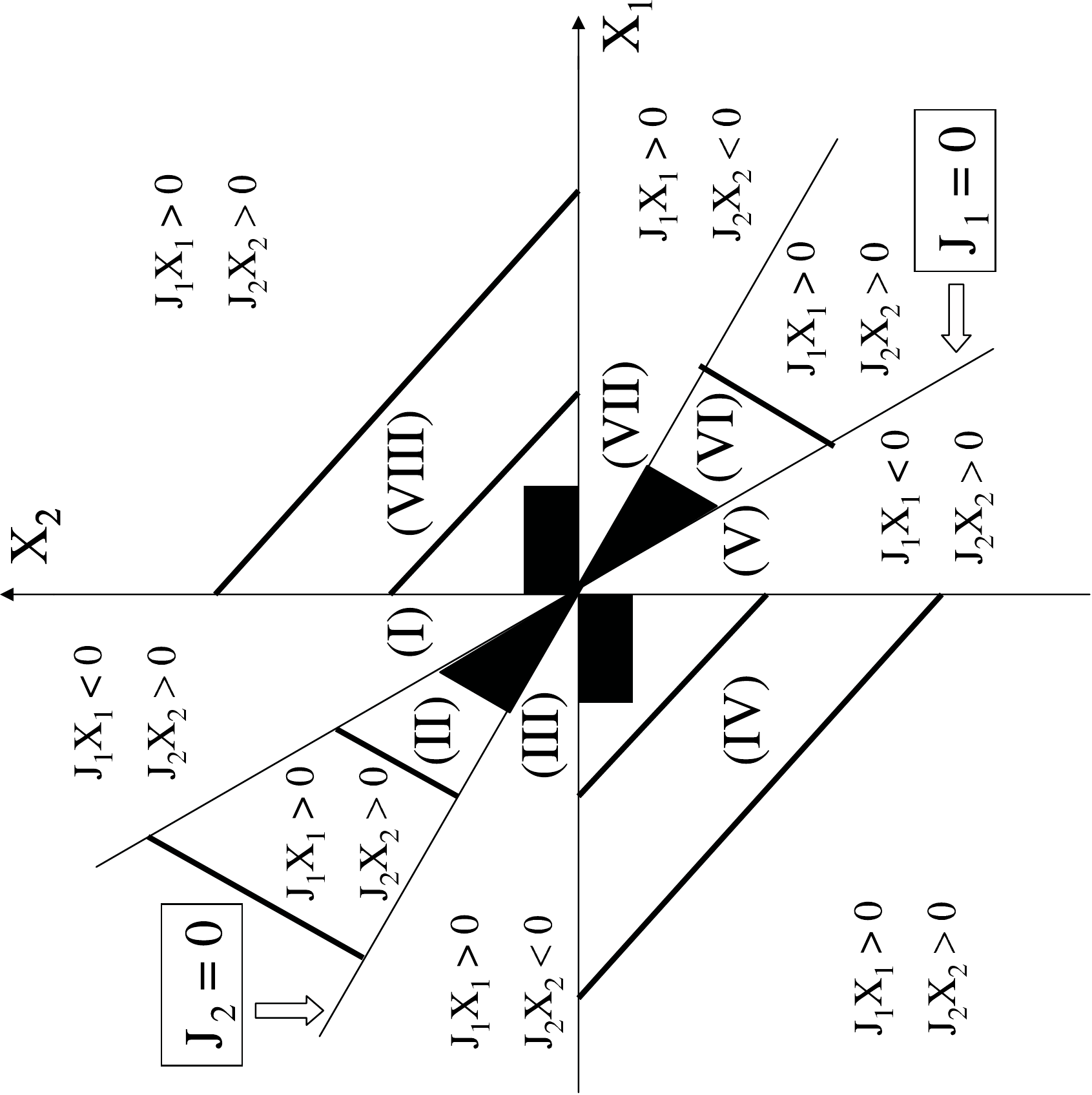}
\end{center}
\caption{Various modes of operation in the general coupled transport
model of a energy converter in the 2d plane spanned by the two
generalized forces $X_{1}$ and $X_{2}$. 
The sectors II, IV, VI 
and VIII are crossed out and marked in black to emphasize the fact that 
in these four sectors, no energy transduction takes place. 
}
\label{fig-lrtmodes}
\end{figure}

Let us identify the different modes of operation of the molecular
motors on the $X_1-X_2$ diagram.
The lines $J_{1} = 0$ and $J_{2} = 0$ are given by
$X_{2} = - (L_{11}/L_{12}) X_{1}$ and $X_{2} = - (L_{21}/L_{22}) X_{1}$,
respectively. The slope of the line $J_{1}=0$ is higher than that
of the line $J_{2}=0$, i.e., $L_{11}/L_{12} > L_{21}/L_{22}$
because of the general condition that $L_{11}L_{22}> L_{12}^2$.
The line $J_{1}=0$ divides the plane into two halves in one of
which $J_{1}$ is positive while in the other $J_{1}$ is negative.
Similarly $J_{2}$ also changes sign on crossing the line $J_{2}=0$.
In the regions II, IV, VI and VIII both the processes 1 and 2 are
spontaneous and, consequently, energy is merely dissipated.
In contrast, in the regions labelled by I, III, V and VII energy
transduction take place; in I and V, the motor the process 2
drives the process 1 whereas in III and VII the process 1 drives
the process 2.

\section{\bf Modeling stochastic chemo-mechanical kinetics: continuous landscapes vs. discrete networks} 
\label{sec-math}

In general for modeling molecular motors four key choices need to be made 
\cite{christen07}: 
(i) choice of the {\it degrees of freedom}, or {\it dynamical variables}, 
consistent with the intended level of spatio-temporal resolution, 
(ii) choice of the {\it form of the interactions} between the variables, 
(iii) choice of the {\it dynamical equations} depending on the nature of 
the dynamical variables, and 
(iv) choice of the {\it methods of solution} suitable for calculating the 
quantities of interest under the given initial and/or boundary conditions.

The hierarchy of the different levels of description used so far in
modeling mechanics of molecular motors is shown in table 
\ref{tab-levelmech} 
(MD $\equiv$ molecular dynamics, NMA $\equiv$ normal mode analysis).

\begin{table}
\centerline{{\framebox{Markov model: master equation for jump processes on ``network'' of states}}}

\centerline{$\uparrow$}

\centerline{{\framebox{Mechano-chemical model: Langevin/FP equation for BD on a ``landscape'' }}}

\centerline{$\uparrow$}

\centerline{{\framebox{Coarse-grained model: NMA for elastic ``bead-spring networks''}}}

\centerline{$\uparrow$}

\centerline{{\framebox{Atomistic model: Classical MD; NMA of collective dynamics}}}
\caption{Hierarchy of the levels of description in modeling kinetics of 
molecular motors.
}
\label{tab-levelmech}
\end{table}

In the following subsections, we mention a few alternative formalisms that 
model molecular motors at different levels of spatio-temporal resolution. 
We also explore the possible relations between them. Moreover, whereever 
possible, we mention a few alternative formalisms at the same level of 
spatio-temporal resolution and explain their relative advantages and 
disadvantages. In most of the approaches that we particularly emphasize 
below, we combine the fundamental principles of (stochastic) chemical 
kinetics with those of structural (Brownian) dynamics to formulate the 
general theoretical framework of {\it mechano-chemistry}  or {\it 
chemo-mechanics}. The generic models of molecular motors as well as those 
for specific examples, which we review in the subsequent sections, are 
based on these formalisms.

\subsection{\bf Fully atomistic model, limitations of MD and normal mode analysis}

Since we are interested neither in the making or breaking of covalent 
bonds nor in the very fast processes governed by quantum dynamics, 
solving time-dependent Schr\"odinger equation is not required. 
Therefore, in principle, classical molecular dynamins would be ideally 
suited to model the mechano-chemical dynamics of molecular motors. 
Unfortunately, in practice, the relevant time scales for the kinetics 
of molecular motors are too long to be accessed by MD simulation of fully 
atomistic models with the currently available computational resources. 
But, conceptually dividing these processes into shorter sub-processes, 
it has been possible to study the sub-processes independently by 
carrying out MD simulations of the corresponding atomistic models 
\cite{lee09a,gumbart11}. 

However, some insight into the conformational dynamics of the motor 
can be gained by carying out the standard normal mode analysis (NMA) 
\cite{cui06b}
of a fully atomistic model. The key idea behind NMA is the diagonalization 
of the Hessian matrix whose elements are the second derivatives of 
the potential energy in the harmonic appoximation. Obviously, starting 
with the fully atomistic force fields, the minimum energy configuration 
has to be found before studying the collective dynamics about such a 
configuration. The spectrum of the normal modes of these collective 
dynamics can be obtained numerically provided the $3N \times 3N$ 
Hessian matrix can be diagonalized using an efficient algorithm where 
$N$ is the number of atoms. In practice, huge computational resources 
are required for the energy minimization and the Hessian diagonalization 
for such fully atomistic models because $N$ is quite large for all 
molecular motors. 

Because of the technical difficulties in structural measurements based 
on X-ray crystallography, high-resolution atomic structures of many 
motors are yet to be determined. Instead, structural information at 
lower resolution are often available from other probes, e.g., cryo-electron 
microscopy. Therefore, it is desirable to follow modeling strategy for 
which the low-resolution strutural information and limited computational 
resources are adequate to study the key dynamical processes theoretically.  
We discuss such modeling strategies in the next few subsections.

\subsection{\bf Coarse-grained model, elastic networks and normal mode analysis}

If one is interested in developing a ``{\it structural}'' model that 
captures the intra-motor movements and conformational changes of the 
motor in each cycle, then the most convenient approximation would be a 
coarse-grained description \cite{levy06,cranford10a,cranford10b,tozzini05,voth08}. 
Usually, a group of atoms is clustered together to be represented as a 
``site'', or ``point particle'', of the coarse-grained model \cite{voth08}. 
These ``sites'' are assumed to interact through an appropriate effective 
potentials. Such a coarse-grained model can be formulated, for example,
by assigning a ``point mass'' (a bead) to each amino acid and postulating 
that these beads are connected by harmonic springs \cite{tirion96}. 
A more detailed model can be developed by assigning more than one bead per 
amino acid 
\cite{cranford10a}.

Insight into the actual kinetics of the motor can be gained by carrying 
normal mode analysis (NMA) which yields the spectrum of collective modes 
of the elastic network 
\cite{tama06,hayward08,miyashita08,miyashita11,bahar10a,bahar10b,bahar10c,dykeman10}. 
A class of arbitrary deviations of the network from equilibrium, caused by 
small displacement of the individual bead positions, can be expressed as a 
linear combination of the normal modes \cite{dykeman10}. 

The advantages and the limitations of this approach has been reviewed 
\cite{ma05,sanejouand12}. 
From a theorists perspective, one of the limitations of the coarse-grained 
models is that even the minimal version is too complex to be treated 
analytically; only numerical results can be obtained.
Even with the numerical techniques, in the absence of additional experimental 
information, it is very difficult to identify unambiguously which of the 
normal modes is functionally relevant for the given motor. 
Moreover, NMA considers small excursions from equilibrium whereas the 
deviations are quite large in many biological processes 
\cite{dykeman10}.
Nonlinear elastic effects, which are important for situations far from 
equilibrium, can be treated within the general framework of conformational 
relaxation of an elastic network \cite{togashi07}. 
Obviously, NMA is not a suitable technique for studying those dynamical 
features of the system that involve low degree of collectivity 
\cite{yang07}.

Coarse-grained approach cannot resolve chemical details, e.g., ATP-binding 
and hydrolysis. The effect of ATP binding is mimicked by establishment 
of new elastic links between the ATP-binding region of the motor and 
the coarse-grained domain where the ATP-binding site is located. Release 
of ADP and P$_i$ is captured by the breaking of these elastic links. 
The free energy change caused by the hydrolysis is implicitly incorporated 
by the relaxation of the elastic strain following elastic link formation 
and breaking. Clearly, such descriptions of ATP binding and hydrolysis 
are not suitable for elucidating how the ATPase activity of a motor is 
coupled to its mechanical movement.

\subsection{\bf Stochastic mechano-chemical model: wandering on landscapes} 

The collective oscillations are expected to be strongly damped by the 
surrounding aqueous medium. Interactions of the motor with this aqueous 
medium, including the effects of thermal fluctuations, can be captured 
within the framework of coarse-grained models discussed in the preceeding 
subsection   provided appropriate effective interactions between the 
elastic network and a coarse-grained representation of the aqueous medium 
can be prescribed 
\cite{cressman08}. 
In this subsection, we discuss an alternative approach that captures the 
effects of the aqueous medium indirectly in way that is standard practice 
in non-equilibrium statistical mechanics. Moreover, the coarse-grained 
models of this type involve far fewer dynamical variables and are capable 
to capturing non-collective dynamical features.

\subsubsection{\bf Motor kinetics as wandering in a time-independent mechano-chemical free-energy landscape} 

This formulation is useful for an intuitive physical explanation of the 
coupled mechano-chemical kinetics of molecular motors \cite{keller00}. 

Being a protein or a macromolecular complex, a motor has a large number 
of degrees of freedom. At the microscopic level, a conformation of a 
motor is described by specifying the positions of all the constituent 
atoms in the 3-dimensional space. In the coarse-grained description 
that we discuss in this subsubsection, we retain only a few variables 
$X_1,X_2,...,X_N$
that are required for describing the most important dynamical processes 
of the motor on the relatively long relevant time scales. 
The remaining degrees of freedom $Y_1,Y_2,...,Y_r$ are assumed to 
equilibrate so rapidly that they are treated as part of the reservoir 
that also includes the degrees of freedom associated with the surrounding 
aqueous medium. 

Suppose $V(X_1,X_2,...,X_N;Y_1,Y_2,...,Y_r)$ is the full potential that 
depends on all the microscopic degrees of freedom. Based on the assumption 
mentioned above, the free energy of the motor (also called the {\it potential 
of mean force}) in the reduced $N$-dimensional state space spanned by 
$X_1,X_2,...X_N$ is obtained from 
\begin{equation}
U(X_1,X_2,...,X_N) = - k_B T ln \biggl[\int \int...\int exp\biggl(-\frac{V(X_1,X_2,...,X_N;Y_1,Y_2,...,Y_r)}{k_B T}\biggr) dY_1 dY_2...dY_r \biggr]
\label{eq-integral}
\end{equation}  
The potential $U$ gets contribution from (i) intra-motor interactions 
and interaction of the motor with its track, (ii) interaction of 
the motor with the fuel molecule, and (iii) interactions of the 
motor, track and fuel molecules with the molecules of the surrounding 
aqueous medium \cite{bustamante01,wang07a}. 

Let us now divide the dynamical variables $X_1,X_2,...,X_N$ into two 
clases: ``mechanical'' variables $x_1,x_2,...,x_n$ and ``chemical'' 
variables $\sigma_1,\sigma_2,...,\sigma_m$, where $n+m=N$. 
At least one of the $N$ variables must be ``mechanical'' variable 
that gives the position of the motor. For a porter on a linear track, 
the position is its actual location on the track. In case of a rotary 
motor, the position variable is actually an angle. Thus the ``mechanical 
velocity'' in this case would be either the linear or the angular 
velocity of the motor in real space. Additional mechanical variables 
may be used to denote, for example, the angle between two subunits of 
the motor, angles made by each of the subunits with the track, etc. 
Similarly, at least one of the $N$ variables must be a ``chemical'' 
variable that accounts for the progress of the chemical reaction which 
supplies the chemical input energy of the motor. In this case, the 
``chemical velocity'' corresponds to the rate of the chemical reaction.
For a motor driven by an electro-chemical gradient the chemical variable, 
in principle, can be redefined accordingly.

\begin{figure}[htbp]
\begin{center}
\includegraphics[angle=90,width=0.45\columnwidth]{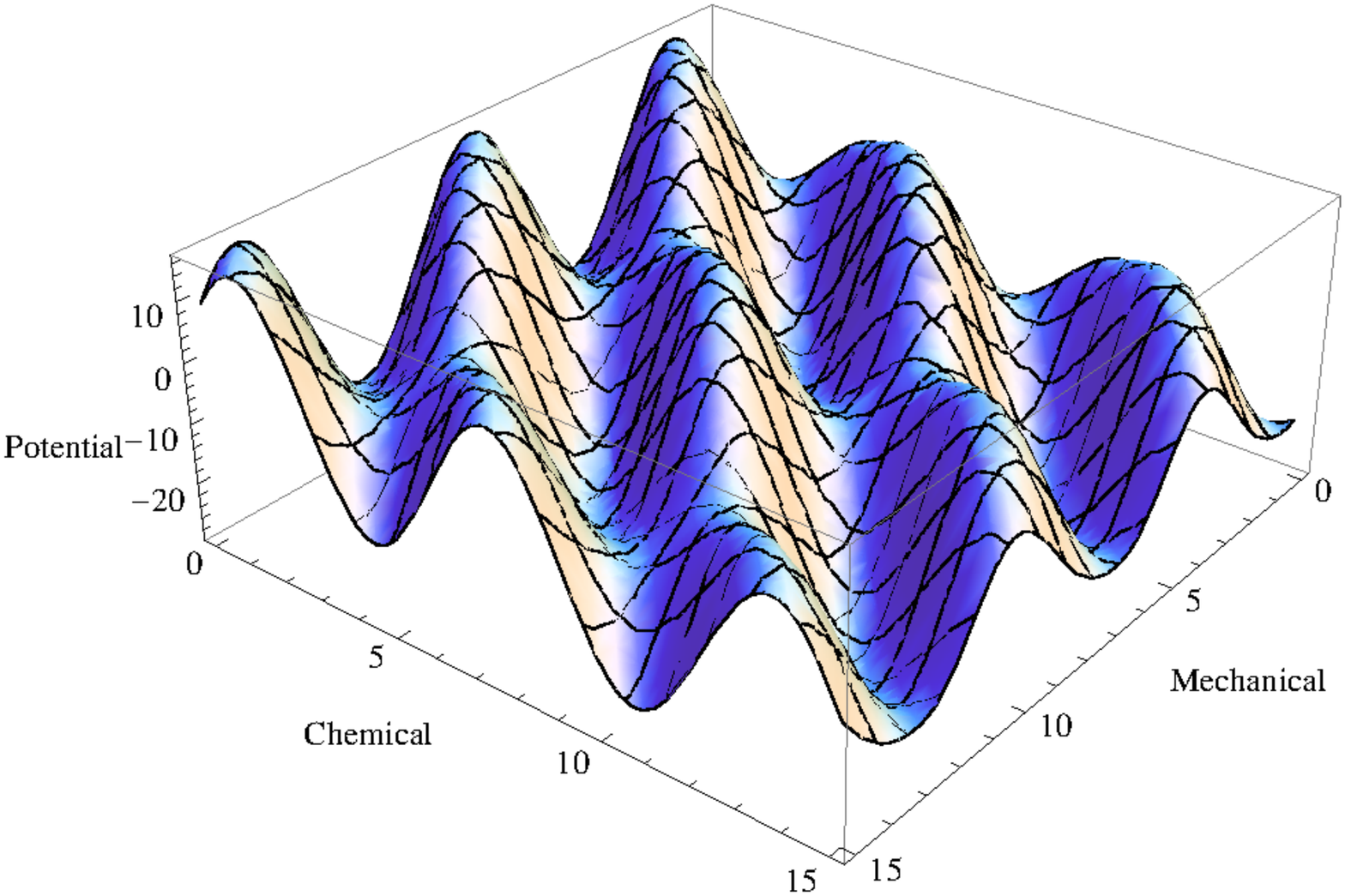}
\end{center}
\caption{A sketch of a landscape where the height denotes the potential 
of mean force (free eenrgy) of a hypothetical molecular motor that is 
described by a single ``mechanical'' variable and a single ``chemical'' 
variable (adapted from ref.\cite{keller00}; courtesy Ajeet K. Sharma)
}
\label{fig-2dlandscape}
\end{figure}

\begin{figure}[htbp]
\begin{center}
(a)\\ 
\includegraphics[width=0.55\columnwidth]{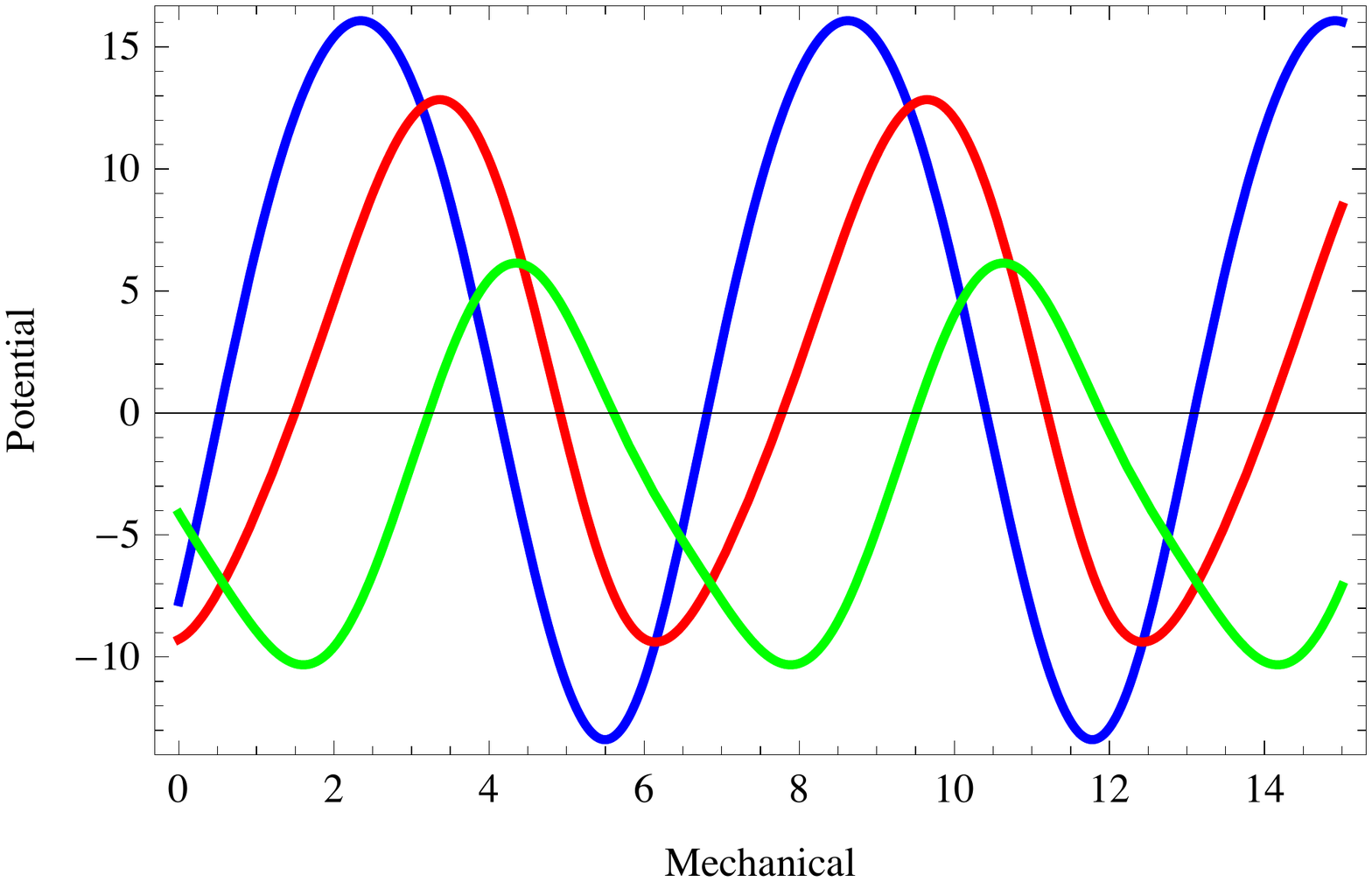}\\
(b) \\
\includegraphics[width=0.55\columnwidth]{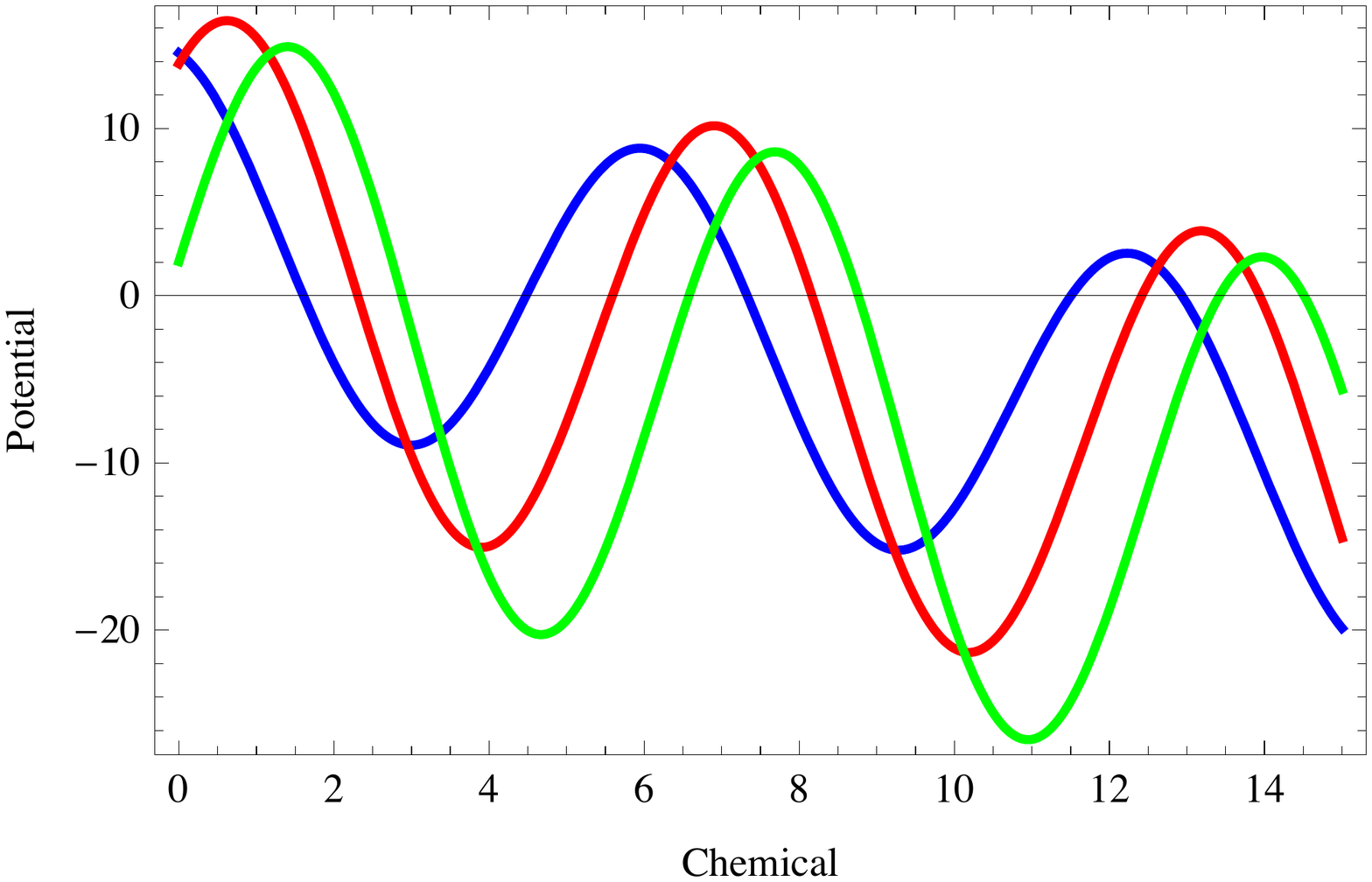}
\end{center}
\caption{Cross sections of the landscape shown in fig.\ref{fig-2dlandscape}
parallel to (a) the mechanical coordinate (i.e., for a few constant values 
of the chemical variable), and (b) the chemical variable (i.e., for a few 
constant value of the mechanical variable) (courtesy Ajeet K. Sharma).
}
\label{fig-1dlandscape}
\end{figure}

For the minimal case with a single mechanical variable and a single 
chemical variable the state space is essentially a 2-dimensional 
``land'', as shown schematically in fig\ref{fig-2dlandscape}. 
Suppose $x_1 \equiv x$ and $\sigma_1 \equiv \sigma$ denote the 
mechanical and chemical variables, respectively.
Then, the potential $U(x,\sigma)$ can be represented by the ``height'' 
at each point on the ``land''. 
Since a track has equi-spaced binding sites for the motor, a cross-section 
of this landscape parallel to $x$ (i.e., for $\sigma$=constant) is periodic 
(see fig.\ref{fig-1dlandscape}); 
all the local minima are equally deep and coincide with the location of 
the motor-binding sites on the track. 
On the other hand, as shown in fig.\ref{fig-1dlandscape}, a cross-section 
of this landscape parallel to $\sigma$ (i.e., for $x$=constant) looks 
like a typical free energy diagram for a chemical reaction plotted against 
the reaction coordinate- two local minima, that correspond to the reactants 
and products, are separated by a free energy barrier. Since a molecular 
motor is a cyclic machine, the local minima parallel to $\mu$ also exhibit 
periodicity, except that the profile is tilted forward along the chemical 
direction, i.e., 
\begin{eqnarray}
U(x+{\ell},\sigma) &=& U(x,\sigma) \nonumber \\
U(x,\sigma+\delta) &=& U(x,\sigma) - |\Delta G|
\end{eqnarray}
where $\Delta G$ is a constant and $\delta$ is the periodicity along $\sigma$.
Because of the forward tilt of the profile along $\sigma$ the bottom of the 
successive minima are deeper by $|\Delta G|$ which accounts for the 
lowering of free energy caused, for example, by ATP hydrolysis.

Using this scenario, Magnasco \cite{magnasco94} argued how a coupling 
between the mechanical and chemical cycles in this space would give 
rise a chemically-driven mechanical motor. This would be the 
chemo-mechanical analog of the chemo-chemical machine of the kind that 
we discussed in section \ref{sec-chemphys}.

It is well known that for a classical system coupled to a reservoir, 
the deterministic time evolution of the system as well as the 
constituents of the reservoir is governed by a set of coupled Newton's 
equations which exhibit the time-reversal symmetry.  
However, when the degrees of freedom associated with the reservoir 
are projected out, the dynamics of the system appears stochastic 
and irreversible; such time evolution of the system is described by 
a Langevin equation \cite{langevineqnbook}. Therefore, in the minimal 
case of a 2-dimensional free-energy landscape, the Langevin equations 
are of the form 
\begin{eqnarray}
\gamma_m (dx/dt) &=& - (\partial U/\partial x) + F_m + \eta_m \nonumber \\
\gamma_c (d\sigma/dt) &=& - (\partial U/\partial \sigma) + F_c + \eta_c  
\end{eqnarray}
where $\gamma_m$ and $\gamma_c$ are the phenomenological damping 
coefficients for the mechanical and chemical variables, respectively; 
$F_m$ and $F_c$ are the corresponding external generalized forces, 
$\eta_m$ and $\eta_c$ being the corresponding generalized Brownian 
(noise) forces. Although the degrees of freedom associated with 
the reservoir do not appear as dynamical variables, their effects on 
the system enter the Langevin equation through the viscous damping 
term and random force term. Most often, for simplicity, the random 
generalized forces are assumed to be Gaussian white noise.
The inertial terms have been neglected; 
this assumption is well justified for nano-motors whose motions are, 
in general, overdamped on the time scales relevant for motor movements. 

An alternative, but equivalent, formulation is based on the Fokker-Planck 
equation \cite{FPeqnbook} which, in the general case of $N$-dimensional 
mechano-chemical state space, has the form 
\begin{eqnarray}
\frac{\partial {\cal P}(X_1,X_2,...,X_N,t)}{\partial t} = \sum_{i=1}^{N} \biggl[\frac{k_T}{\gamma_i} \frac{\partial^2 {\cal P}}{\partial X_i^2} - \frac{1}{\gamma_i} \frac{\partial}{\partial X_i} \biggl\{\biggl(-\frac{\partial U}{\partial X_i}+F_i\biggr){\cal P} \biggr\}\biggr] 
\label{eq-FPkeller}
\end{eqnarray}
where ${\cal P}(X_1,X_2,...,X_N,t)$ is the probability that the motor protein is 
in the state $X_1,X_2,...,X_N$ at time $t$. The two terms on the right hand 
side of eq.(\ref{eq-FPkeller}) account for the diffusion and drift of the 
motor in the free-energy landscape.

\subsubsection{\bf Motor kinetics as wandering in the time-dependent mechanical (real-space) free-energy landscape}
\label{sec-timedeplandscape}

In this subsection we consider those special situations where the chemical 
states of the motor are {\it long lived} and change in {\it fast discrete 
jumps}. Consequently, the mechanical variables of the motor can continue 
to change without alteration in its chemical states, except during chemical 
transitions when the mechanical variables remains frozen and at least one 
of the chemical variables changes abruptly. Because of this clear separation 
of the time scales of variation of the mechanical and chemical variables, 
no mixed mechano-chemical transition is allowed in this scenario.

In this formulation we assume that each of the mechanical variables is 
continuous whereas all the chemical variables are discrete. We now use 
the symbols $\vec{x} \equiv (x_1,x_2,...,x_n)$ and 
$\vec{\sigma} \equiv (\sigma_1,\sigma_2,...,\sigma_m)$ 
to denote the mechanical and chemical variables. 
As before, the free energy of the system is given by $U(\vec{x},\vec{\sigma})$.
Although, in principle, $U(\vec{x},\vec{\sigma})$ can be derived from 
the microscopic potential using eq.(\ref{eq-integral}), in practice, 
its explicit form is most often postulated based on physical arguments.

The minimal model in this case also requires $m=1=n$ where the continuous 
variable $x$ denotes the position of the motor while the discrete variable 
$\sigma$ ($\sigma=1,2,...,\mu$) labels the $\mu$ chemical states. 
For example, if $\mu=4$, the four distinct values of $\sigma$ may correspond 
to the following chemical states of the motor:   
(a) ligand-free state, (b) ATP-bound state, (c) ADP-P$_i$-bound state, 
and (d) ADP-bound state. The free energy $U(x,\sigma)$ is plotted as a 
function of $x$ for a given discrete value of $\sigma$; in general, 
different profiles correspond to different values of $\sigma$. A sequence 
of transitions of the chemical state is accompanied by the corresponding 
sequential change of the profile.

For the simplicity of explanation of this modeling strategy, let us 
consider the minimal case $m=1=n$. In this case, the actual 
$2$-dimensional potential landscape can be replaced by a sequence of $\mu$ 
one-dimensional potentials where $\mu$ is the number of discrete values 
allowed for the chemical state variable $\sigma$.

Because of the discrete nature of the chemical state variables, the 
eq.(\ref{eq-FPkeller}) is replaced by the equation 
\begin{eqnarray}
\frac{\partial {\cal P}(\vec{x},\vec{\sigma},t)}{\partial t} &=& \sum_{i=1}^{n} \biggl[\frac{k_T}{\gamma_i} \frac{\partial^2 {\cal P}}{\partial x_i^2} - \frac{1}{\gamma_i} \frac{\partial}{\partial x_i} \biggl\{\biggl(-\frac{\partial U}{\partial x_i}+F_i\biggr){\cal P} \biggr\}\biggr] \nonumber \\
&+& \sum_{j} {\cal P}(\vec{x},\sigma_1,...,\sigma_j',..,\sigma_m) W_{\sigma_j',\{\sigma\}_j \to \sigma_j,\{\sigma\}_j}(\vec{x}) \nonumber \\
&-& \sum_{j} {\cal P}(\vec{x},\sigma_1,...,\sigma_j,..,\sigma_m) W_{\sigma_j,\{\sigma\}_j \to \sigma_j',\{\sigma\}_j}(\vec{x}) \nonumber \\
\label{eq-hybrid}
\end{eqnarray}
where $W_{\sigma_j,\{\sigma\}_j \to \sigma_j',\{\sigma\}_j}(\vec{x})$ is the 
transition probability per unit time for the transition from $\sigma_j$ 
to $\sigma_j'$ while the mechanical variables $\vec{x}$ remain 
frozen at the current instantaneous values; the symbol $\{\sigma\}_j$ 
denotes values of all the chemical variables except the $i$-th 
chemical variable. The condition of detailed balance imposes restrictions 
on the choice of these transition probabilities.
Note that there is no term in this equation which would correspond to 
a mixed mechano-chemical transition.

\noindent$\bullet${\bf Brownian ratchet}

\begin{figure}[htbp]
\begin{center}
\includegraphics[angle=90,width=0.65\columnwidth]{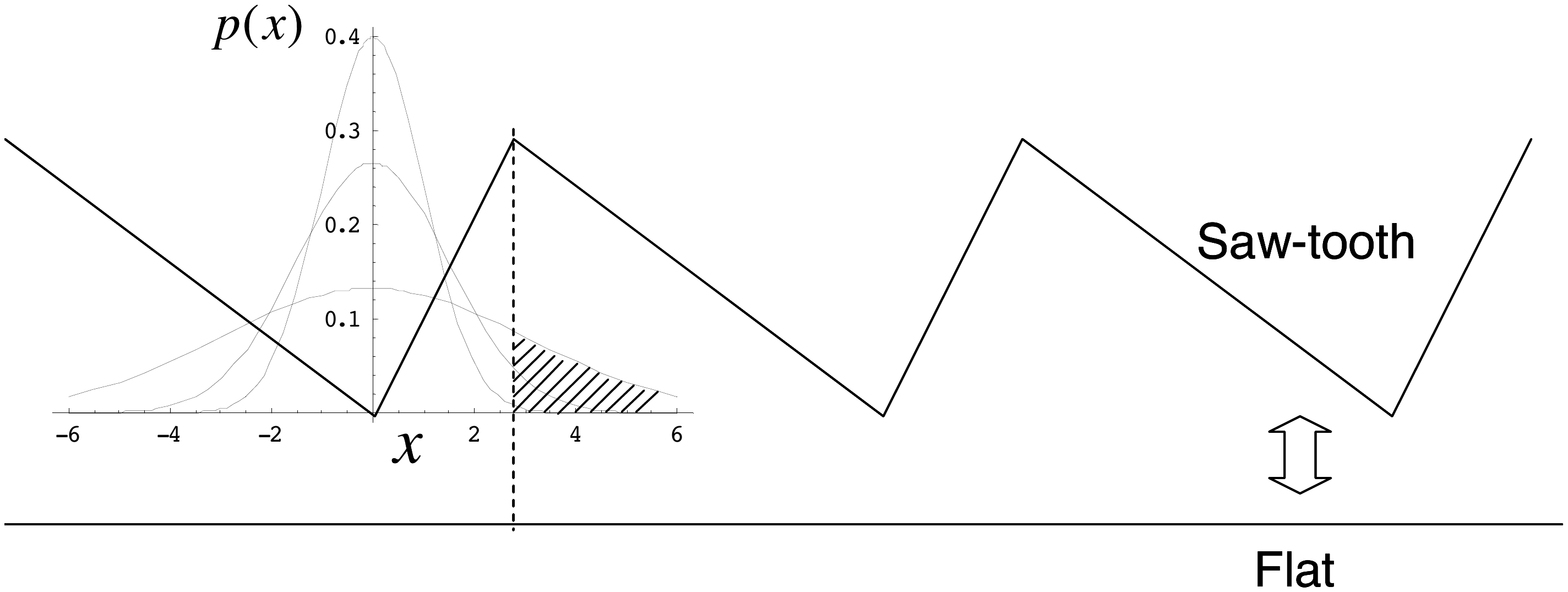}
\end{center}
\caption{Two forms of the time-dependent potential used for implementing 
the Brownian ratchet mechanism. The spreading of the Gaussian profile 
of the probability density with time is shown in the inset.
(Reprinted from Physical Review E  (ref.\cite{greulich07})). Copyright 
2007, American Physical Society. 
} 
\label{fig-sawtooth}
\end{figure}

Let us assume the special values $m=1=n$. Moreover, suppose $\mu=2$, i.e., 
the chemical variable $\sigma$ is allowed to take one of the only two 
allowed values so that only two different profiles of the free-energy 
landscape are possible. Let one of these be the flat form $U(x)=0$ for all 
$x$ whereas the other be the sawtooth form shown in fig.\ref{fig-sawtooth}. 
Note the two key features of the sawtooth: (i) it is periodic, (ii) within 
each period, it has a spatial asymmetry.

Suppose, initially the potential has the sawtooth shape and the motor 
is located at a point that corresponds to a minimum of the potential. 
As long as the potential remains unchanged, in spite of its spatial 
asymmetry, the average force on the motor (averaged over a single spatial 
period) 
$<f> = [\int F(x) dx]/{\ell_s} = 0$ because $F = - \partial U/\partial x$ 
and $U(x) = U(x+{\ell_s})$. Thus, even if the motor position initially 
does not coincide with a minimum of the potential, given enough time, 
it will settle to the location of the nearest minimum of the free energy 
because of the energy dissipation by viscous drag against it. From then 
onwards, its position will exhibit only small amplitude thermal fluctuations 
around the potential minimum provided the thermal energy is much smaller 
compared to the height of the walls separating the successive wells of 
the potential profile. 
In addition, it will execute thermally-assisted occasional jumps to the 
neighboring wells symmetrically in both the forward and backward 
directions that, in the absence of any other processes, would result in 
only diffusion.

Let the chemical state $\sigma$ of the motor make a transition to its 
other allowed state so that the potential profile makes a transition 
from its sawtooth form to its flat form. Immediately, the free particle 
begins to execute a Brownian motion and the corresponding Gaussian profile
of the probability distribution begins to spread with the passage of
time.  If the potential is again switched on before the Gaussian profile
gets enough time for spreading beyond the original well, the particle
will return to its original initial position. But, if the period
during which the potential remains off is sufficiently long, so that
the Gaussian probability distribution has a non-vanishing tail
overlapping with the neighboring well on the right side of the
original well, then there is a small non-vanishing probability that
the particle will move forward towards right by one period when the
potential is switched on. 
Furthermore, the Gaussian profile may have a non-vanishing overlap 
with the well to the left of the original well that gives rise to 
the possibility of also moving backward in a cycle. But, because of 
the asymmetric shape of each period of the sawtooth, the overlap with 
the well on the right is larger than that with the well on the left; 
consequently, on the average, the particle would move to the right.

Thus, stochastic switching of the chemical variable $\sigma$ back and 
forth between its two allowed discrete values can give rise to a 
``directed'' movement of the motor, on the average, because of the 
asymmetric shape of each of the periods of the periodic sawtooth 
shaped potential profile. The resulting probability current is given by 
\cite{ajdari92,astumian94} 
\begin{equation}
J = J_{+} - J_{-}
\end{equation} 
with 
\begin{eqnarray}
J_{+} &=& \biggl(\frac{\omega_{f}}{4}\biggr) ~erf(\frac{\alpha}{2}\sqrt{\omega_{f}}) \nonumber \\
J_{-} &=& \biggl(\frac{\omega_{f}}{4}\biggr) ~erf(\frac{1-\alpha}{2}\sqrt{\omega_{f}})
\end{eqnarray}
where $erf(x)$ is the error function, the parameter $0 < \alpha < 1$ 
is a measure of the spatial asymmetry of the sawtooth potential 
(symmetric sawtooth corresponds to $\alpha=1/2$) and $\omega_{f}$ is 
the rate of flipping of the time-dependent potential. 
 
Note that in this mechanism, the particle moves forward not because of 
any force imposed on it but because of its Brownian motion. The system
is, however, not in equilibrium because energy is pumped into it during 
every period in switching the potential between the two forms. In other 
words, the system works as a rectifier where the Brownian motion, in 
principle, could have given rise to both forward and backward movements 
of the particle in the multiples of $\ell$, but the backward motion of 
the particle is suppressed by a combination of (a) the time dependence 
and (b) spatial asymmetry of the potential. In fact, the direction of 
motion of the particle can be reversed by reversing the asymmetry in each 
period of the potential. The mechanism of directional movement discussed 
above is called a Brownian ratchet
\cite{julicher97a,astumian97,astumian01,astumian02a,astumian02b,astumian07,reimann02a}
The concept of Brownian ratchet was popularized by Feynman through his
lectures \cite{feynmanbook} although, historically, it was introduced by
Smoluchowski. 

For a class of porters, for example, the switching of the chemical state 
is caused by ATP hydrolysis. The asymmetric periodic potential (whose 
shape is, at least qualitatively, somewhat similar to the sawtooth form) 
arises from the effect of the polarity of the filament on the motor-track 
interaction \cite{astumian96,astumian00z}

The rectification of the noise required for Brownian ratchet mechanism 
can also be achieved by, for example, the binding of a ligand that 
stabilizes conformations in the ``forward'' direction \cite{howard06a}. 
This process has strong similarity with the conformational selection 
mechanism for substrate specificity of enzymes that we have discussed 
in section \ref{sec-chemphys}. 
The main difference between this mechanism of Brownian ratchet and 
the power stroke has been explained by Howard (see fig.1 of ref. 
\cite{howard06a}) with a  simple illustration.

\noindent$\bullet${\bf Efficiency of Brownian motors}

Suppose, $W$ is the work done against a load force $F$. The input
energy $E_{in}$ for a Brownian motor can be calculated from
\begin{equation}
E_{in} = <\int_{0}^{T_{on}} \frac{dU(x(t))}{dx} dx(t) >
\end{equation}
assuming a specific $x$- and $t$-dependence of the potential $U$,
where the angular bracket $<.>$ denotes average over many ratchet
cycles. Hence the efficiency of a Brownian motor can be obtained
from the definition
\begin{equation}
\eta = \frac{W}{E_{in}} = \frac{F~v}{\dot{E_{in}}}
\end{equation}

The directional movement of Brownian motors arises from the rectification of random thermal noise. For
such motors, an alternative measure of performance is the {\it
rectification efficiency} \cite{suzuki03}.

\subsection{\bf Markov model: motor kinetics as a jump process in a network of fully discrete mechano-chemical states}

In this subsection we simplify the continuum landscape-based scenario 
developed in the preceeding subsection to formulate a fully discrete 
scheme making well-motivated approximations \cite{keller00,lan12}.  
With each local minimum of the free-energy landscape we associate a 
discrete state. The probability $P_i$ of finding the system in the 
$i$-th discrete state is given by \cite{keller00}
\begin{equation}
P_i(t) = \int_{i-th ~zone} {\cal P}(\vec{X},t) d\vec{X} 
\end{equation}
where $i$-th zone is the immediate surrounding of the $i$-th local minimum 
of the free energy landscape. Thus, the continuum of states $\vec{X}$ is 
replaced by set of discrete states identified by the above procedure. 
Moreover, instead of the probability densities ${\cal P}(\vec{X})$ 
defined on the free energy landscape we now deal with the probabilities $P_i$. 

The local minima in the free-energy landscape are separated by 
low-energy passes so that thermal fluctuations occasionally cause 
the system to leave the neighbourhood of one local minimum and arrive 
at that of a neighboring one; such wanderings on the free energy 
landscape are identified as transitions from one discrete state to 
another in the fully discrete formulation. The corresponding rate 
constants (i.e., the probabilities of transition per unit time) 
can also be obtained from an analysis of the probability fluxes on the 
continuous landscape \cite{keller00,lan12}. Obviously, the rate 
constants depend on the shape of the free energy landscape; the 
dependence of the rate constants on the external force arise from 
that of the landscape shape on the external force.

The discrete state space of this formulation can be regarded as a 
network \cite{lipowsky00a,lipowsky00b,lipowsky03,jaster03}. 
Just like the continuum formulation, the minimal model must have one 
mechanical variable and a chemical variable both of which are discrete.
Let $P_{\mu}(i,t)$ be the probability of finding the motor at the
discrete position labelled by $i$ and in the ``chemical''
state $\mu$ at time $t$. Then, the master equation for $P_{\mu}(i,t)$
is given by
\begin{eqnarray}
\frac{\partial P_{\mu}(i,t)}{\partial t}&=&[\sum_{j\neq i} P_{\mu}(j,t) k_{\mu}(j \to i) - \sum_{j \neq i} P_{\mu}(i,t) k_{\mu}(i \to j)] \nonumber \\ 
&+& [\sum_{\mu'} P_{\mu'}(i,t) W_{\mu'\to \mu}(i) - \sum_{\mu'} P_{\mu}(i,t) W_{\mu\to \mu'}(i)] \nonumber \\
&+& [\sum_{j\neq i} \sum_{\mu'} P_{\mu'}(j,t) \omega_{\mu'\to \mu}(j \to i) \nonumber \\
&-& \sum_{j\neq i} \sum_{\mu'} P_{\mu}(i,t) \omega_{\mu\to \mu'}(i \to j)]
\label{eq-fullmaster}
\end{eqnarray}
where the terms enclosed by the three different brackets $[.]$ correspond
to the purely mechanical, purely chemical and mechano-chemical transitions,
respectively.
For obvious reasons, these equations are sometimes referred to as the 
stochastic rate equations.  

As a concrete example, which will be used also for on several other 
occasions later in this review, consider a 2-state motor that, at 
any site $j$, can exist in one of the only two possible chemical 
states labelled by the symbols $1_j$ and $2_j$. We assume the 
mechano-chemical cycle of this motor to be 
\begin{equation}
1_{j} \mathop{\rightleftharpoons}^{\omega_{1}}_{\omega_{-1}} 2_{j} \mathop{\rightarrow}^{\omega_{2}} 1_{j+1}
\label{eq-MMlikemot}
\end{equation}
where the rates of the allowed transitions are shown explicitly 
above or below the corresponding arrow. Note that the transition 
$1_j \rightleftharpoons 2_{j}$ is purely chemical whereas the 
transition $2_{j} \rightarrow 1_{j+1}$ is a mixed mechano-chemical 
transition. The corresponding master equations are given by 
\begin{eqnarray}
\frac{dP_1(i)}{dt} = \omega_{2} P_2(i-1) + \omega_{-1} P_{2}(i) - \omega_{1} P_{1}(i) \nonumber \\
\frac{dP_2(i)}{dt} = \omega_{1} P_1(i) - \omega_{-1} P_{2}(i) - \omega_{2} P_{2}(i) \nonumber \\
\end{eqnarray} 
We'll see some implications of these equations in several specific 
contexts later in this article.

\noindent$\bullet${\bf Microscopic reversibility and balance conditions for mechano-chemical kinetics: cycles, and flux}

The principle of microscopic reversibility \cite{blackmond09} 
has important implications in the free energy transduction by 
molecular motors \cite{astumian12}. 

On a discrete mechano-chemical state space, each state is denoted by 
a {\it vertex} and the direct transition from one state (denoted by, 
say, the vertex $i$) to another (denoted by, say, the vertex $j$) is 
represented by a directed {\it edge} $|ij>$. The opposite transition 
from $j$ to $i$ is denoted by the directed edge $|ji>$.  
A {\it transition flux} can be defined along any edge of this diagram. 
The forward transition flux $J_{|ij>}$ from the vertex $i$ to the 
vertex $j$ is given by $W_{ji} P_{i}$ while the reverse flux, i.e., 
transition flux $J_{|ji>}$ from $j$ to $i$ is given by $W_{ij} P_{j}$. 
Therefore, the net transition flux in the direction {\it from} the 
vertex $i$ {\it to} the vertex $j$ is given by 
$J_{ji} = W_{ji} P_{i} - W_{ij} P_{j}$. 

A {\it cycle} in the kinetic diagram consists of at least three 
vertices. Each cycle $C_{\mu}$ can be decomposed into two directed 
cycles (or, {\it dicycles}) \cite{lipowsky08b} $C_{\mu}^{d}$ where 
$d=\pm$ corresponds the clockwise and counter-clockwise cycles. 
The net cycle flux $J(C_{\mu})$ in the cycle $C_{\mu}$, in the CW 
direction, is given by $J(C_{\mu}) = J(C_{\mu}^{+}) - J(C_{\mu}^{-})$. 

For each arbitrary dicycle $C_{\mu}^{d}$, let us define the {\it 
dicycle ratio} 
\begin{equation}
{\cal R}(C_{\mu}^{d}) = \Pi_{<ij>\epsilon C_{\mu}^{+d}} W_{ji}/\Pi_{<ij>\epsilon C_{\mu}^{-d}} W_{ij} = \Pi_{<ij>}^{\mu,d} (W_{ji}/W_{ij}).
\end{equation}
where the superscript $\mu,d$ on the product sign denote a product 
evaluated over the directed edges of the cycle. So, by definition, 
${\cal R}(C_{\mu}^{-}) = 1/{\cal R}(C_{\mu}^{+})$.

It has been proved rigorously \cite{vankampenbook} that, for detailed 
balance to hold, the necessary and sufficient condition to be 
satisfied by the transition probabilities is   
\begin{equation}
{\cal R}(C_{\mu}^{d}) = 1 ~~{\rm for ~all ~dicycles}~~ C_{\mu}^{d}
\end{equation}

For a non-equilibrium steady state (NESS), one can define 
\cite{lipowsky08b} the {\it dicycle frequency} 
$\Omega^{ss}(C_{\mu}^{d})$ which is the number of dicycles $C_{\mu}^{d}$ 
completed per unit time in the NESS of the system. Then, 
\begin{equation}
\Omega^{ss}(C_{\mu}^{+})/\Omega^{ss}(C_{\mu}^{-}) = \Pi_{<ij>}^{\mu,d} (W_{ji}/W_{ij}) = {\cal R}(C_{\mu}^{d})
\end{equation}
Clearly, ${\cal R}(C_{\mu}^{d}) \neq 1$, in general, for any NESS.
There are close relations between this network formalism of energy 
transduction by molecular motors and a master equation based general 
network theory for microscopic dynamics \cite{schnackenberg76}  
which is the microscopic counterpart of Kirchoff's macroscopic theory 
of electrical networks.

Detailed balance is believed to be a property of systems in equilibrium 
whereas the conditions under which molecular motors operate are far 
from equilibrium. Does it imply that detailed balance breaks down for  
molecular motors?  The correct answer this subtle question needs a 
careful analysis \cite{wang05,astumian05,thomas01a}.

If one naively {\it assumes} that the entire system returns to 
its original initial state at the end of each cycle one would 
get the erroneous result that the detailed balance breaks down. 
But, strictly speaking, the free energy of the full system gets 
lowered by $|\delta G|$ (e.g., because of the hydrolysis of ATP) 
in each cycle although the cyclic machine itself comes back to 
the same state. When the latter fact is incorporated correctly 
in the analysis \cite{wang05,astumian05,thomas01a}, 
one finds that detailed balance is not violated by molecular 
motors. This should not sound surprising- the transition rates 
``do not know'' whether or not the system has been driven out of 
equilibrium by pumping energy into it.

\section{\bf Solving forward problem by stochastic process modeling: from model to data}
\label{sec-forwardproblem}

\begin{figure}[htbp]
\begin{center}
(a)\\
\includegraphics[angle=-90,width=0.35\columnwidth]{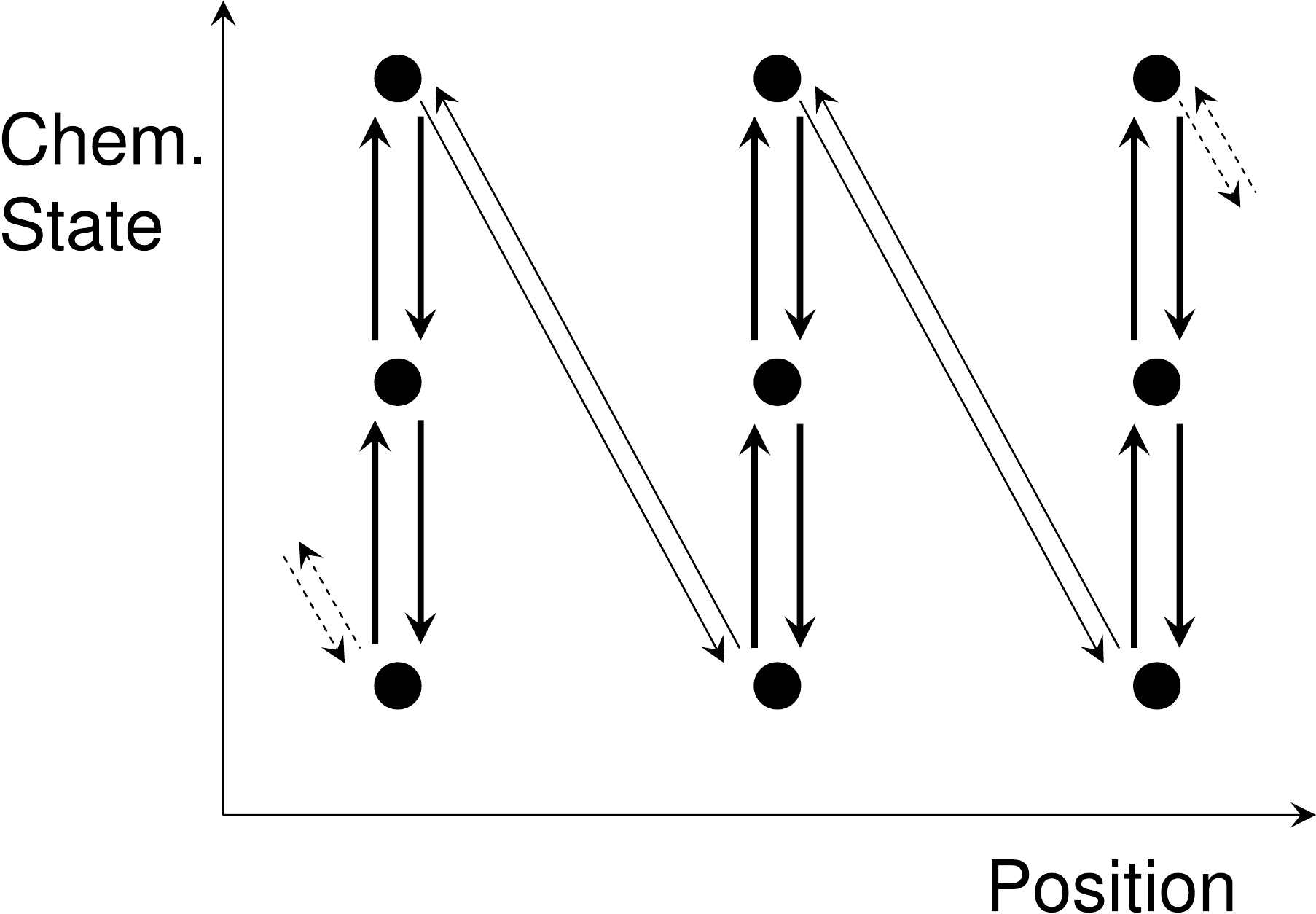}\\
(b)\\
\includegraphics[angle=-90,width=0.35\columnwidth]{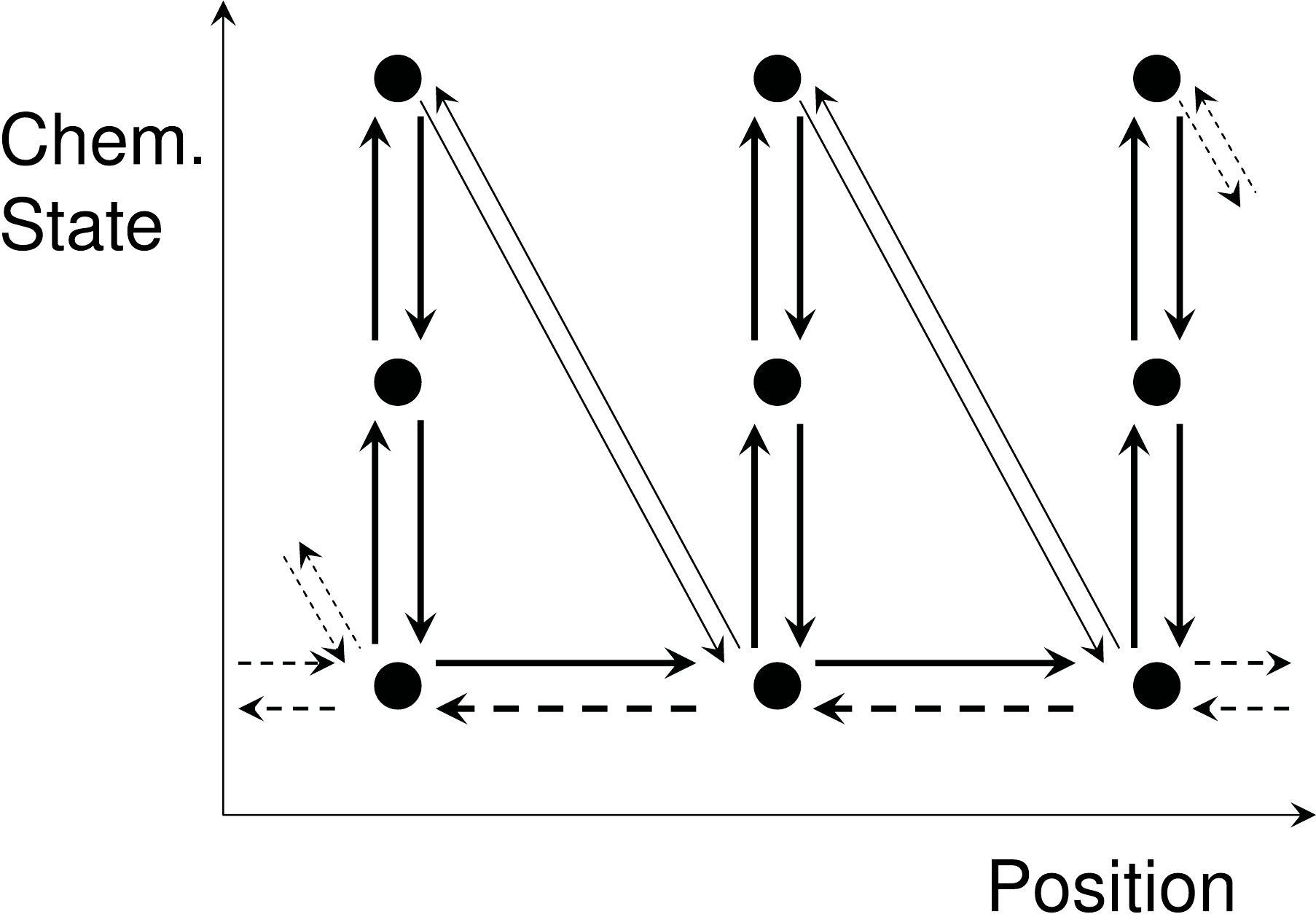}\\
(c)\\
\includegraphics[angle=-90,width=0.35\columnwidth]{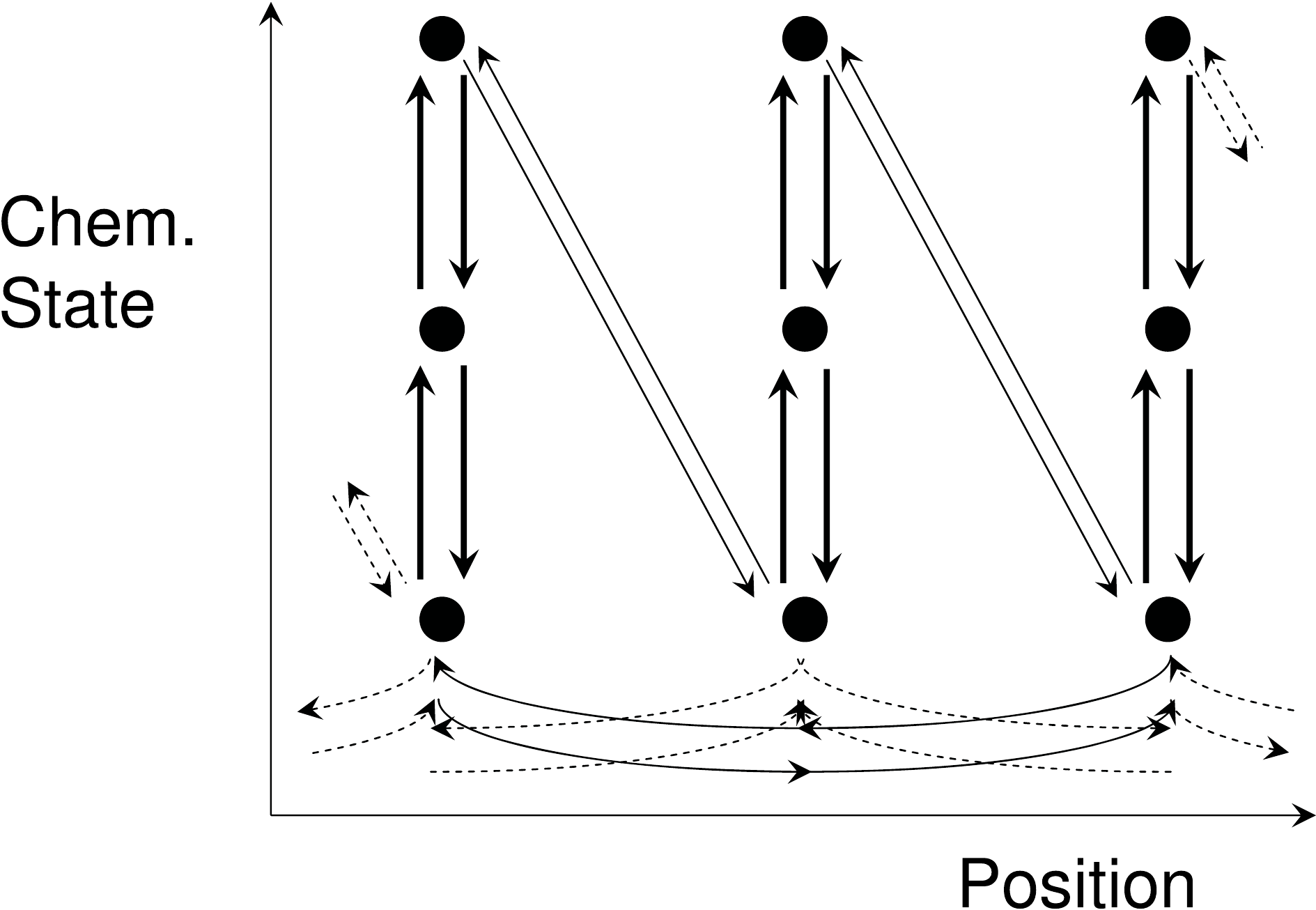}
\end{center}
\caption{Three examples of different types of networks of 
discrete mechano-chemical states. The bullets represent the 
distinct states and the arrows denote the allowed transitions 
between the two corresponding states. The scheme is (a) is 
unbrached whereas that in (b) has branched pathways connecting 
the same pair of states. The mechanical step size is unique 
in both (a) and (b) whereas steps of more than one size are 
possible in (c). (Adapted from fig.1 of ref.\cite{chemla08}).
}
\label{fig-DTD}
\end{figure}

\subsection{\bf Average speed and load-velocity relation }

For simplicity, let us consider the kinetic scheme shown in fig. 
(\ref{fig-DTD}(a)). In terms of the Fourier transform 
\begin{equation}
\bar{P}_{\mu}(k,t) = \sum_{j=-\infty}^{\infty} P_{\mu}(x_{j},t) e^{-ikx_j} 
\label{eq-FourierP}
\end{equation}
of $P_{\mu}(x_{j},t)$, the master equations can be written as a 
matrix equation 
\begin{equation}
\frac{\partial {\bf \bar{P}}(k,t)}{\partial t} = {\bf W}(k) {\bf \bar{P}}(k,t) 
\label{eq-masterq}
\end{equation} 
where ${\bf \bar{P}}$ is a column vector of $M$ components $P_{\mu}(k,t)$ 
($\mu=1,...M$) and ${\bf W}(k)$ is the transition matrix in the $k$-space 
(i.e., Fourier space). Other than the rate constants, ${\bf W}(k)$ also 
involves the $k$-dependent factors 
\begin{equation}
\rho_{+}(k) = e^{-ik\ell} ~~ {\rm and}~~ \rho_{-}(k) = e^{ik\ell} 
\end{equation}
Summing over the hidden chemical states, we get the {\it position 
probability density}
\begin{equation}
\bar{P}(k,t) = \sum_{\mu=1}^M \bar{P}_{\mu}(k,t) = \sum_{\mu=1}^M \sum_{j=-\infty}^{\infty} P_{\mu}(x_j,t) e^{- i k x_j}. 
\label{eq-barPk}
\end{equation}
Taking derivatives of both sides of (\ref{eq-barPk}) with respect to 
$k$ we get \cite{koza99,koza00} 
\begin{eqnarray} 
i \biggl(\frac{\partial \bar{P}(k,t)}{\partial k}\biggr)_{k=0} &=& <x(t)> \nonumber \\
- \biggl(\frac{\partial^2 \bar{P}}{\partial k^2}\biggr)_{k=0} + \biggl(\frac{\partial P}{\partial k}\biggr)^2_{k=0} &=& <x^2(t)>-<x(t)>^2 \nonumber \\ 
\end{eqnarray}
where $<x(t)> = \sum_{j} x_{j} \sum_{\mu} P_{\mu}(x_{j},t)$. 
Evaluating $\bar{P}(k,t)$, in principle, the stationary 
drift velocity (i.e., asymptotic mean velocity) $V$ and the 
corresponsding diffusion constant $D$ can be obtained from 
\begin{eqnarray} 
&&V = lim_{t \to \infty} i \frac{\partial}{\partial t}\biggl[\biggl( \frac{\partial \bar{P}(k,t)}{\partial k}\biggr)_{k=0}\biggr] \nonumber \\
&&D = lim_{t \to \infty} \frac{1}{2} \frac{\partial}{\partial t}\biggl[ \biggl(-\frac{\partial^2\bar{P}(k,t)}{\partial k^2}\biggr)_{k=0} + \biggl(\frac{\partial\bar{P}(k,t)}{\partial k}\biggr)^2_{k=0}\biggr] \nonumber \\
\end{eqnarray}

It may be tempting to attempt a direct utilization of the general form 
\begin{equation}
\bar{P}(k,t) = \sum_{\mu} B_{\mu} e^{\lambda_{\mu}(k) t} 
\label{eq-gensol}
\end{equation}
where the coefficients $B_{\mu}$ are fixed by the initial conditions and 
$\lambda_{\mu}(k)$ are the eigenvalues of ${\bf W}(k)$. 
However, for the practical implementation of this method analytically 
the main hurdle would be to get all the eigenvalues of ${\bf W}(k)$. 
Fortunately, only the smallest eigenvalue $\lambda_{min}$, which 
dominates $\bar{P}(k)$ 
in the limit $t \to \infty$, is required for evaluating $V$ and $D$ 
\cite{koza99,koza00}: 
\begin{eqnarray} 
V &=& i \biggl(\frac{\partial \lambda_{min}(k,t)}{\partial k}\biggr)_{k=0} \nonumber \\
D &=& - \frac{1}{2} \biggl(\frac{\partial^2\lambda_{min}(k,t)}{\partial k^2}\biggr)_{k=0} 
\label{eq-lambda}
\end{eqnarray}
Even the forms (\ref{eq-lambda}) are not convenient for evaluating 
$V$ and $D$. Most convenient approach is based on the characteristic 
polynomial $Q(k)$ associated with the matrix ${\bf W}(k)$, i.e., 
$Q(k;\lambda) = det[\lambda {\bf I} - {\bf W}(k)]$. Therefore, 
$\lambda_{min}(k)$ is a root of the polynomial $Q(k;\lambda)$, i.e., 
solution of the equation
\begin{equation}
Q(k;\lambda) = \sum_{\mu=0}^M C_{\mu}(k) [\lambda(k)]^{\mu} = 0.
\label{eq-poly}
\end{equation}
Hence \cite{koza99,koza00} 
\begin{eqnarray}
V &=& - i \frac{C_{0}'}{C_{1}(0)} \nonumber \\
D &=& \frac{C_{0}'' - 2i C_{1}'(0) V - 2 C_{2}(0) V^{2}}{2 C_{1}(0)} 
\label{eq-Fourier}
\end{eqnarray}
where $C_{\mu}' = [\partial C_{\mu}(k)/\partial k]_{k=0}$. 

For a postulated kinetic scheme, ${\bf W}$ is given. Then the 
expressions (\ref{eq-Fourier}) are adequate for analytical 
derivation of $V$ and $D$ for the given model. However, in order to 
calculate the distributions of the dwell times of the motors, it is 
more convenient to work with the Fourier-Laplace transform, rather 
than the Fourier transform, of the probability densities. Therefore, 
we now derive alternative expressions for $V$ and $D$ in terms of 
the full Fourier-Laplace transform of the probability density. 

Taking Laplace transform of (\ref{eq-FourierP}) with respect to time 
\begin{equation}
\tilde{P}_{\mu}(k,s) = \int_{0}^{\infty} \bar{P}_{\mu}(k,t) e^{-s t}, 
\end{equation}
and inverting the matrix form of the master equation leads to the 
solution 
\begin{equation}
\tilde{{\bf P}}(k,s) = {\bf R}(k,s)^{-1} {\bf P}(0) 
\label{eq-Pflspace}
\end{equation}
where 
\begin{equation}
{\bf R}(k,s) = s{\bf I} - {\bf W}(k) 
\label{eq-Rmatrix}
\end{equation}
and ${\bf P}(0)$ is the column vector of initial probability densities. 

Applying the defnition of inverse of a matrix to the matrix ${\textbf{R}}$ 
expressed in eqn.(\ref{eq-Rmatrix}), the eqn.(\ref{eq-Pflspace}) takes the 
form  
\begin{equation}
\tilde{P}(k,s)=\dfrac{\sum\limits_{\mu,\nu=1}^M Co_{\nu,\mu}P_{\nu}(0)}{\lvert{\textbf{R}}(k,s)\rvert}
\label{eq-tprob}
\end{equation}
where $Co_{\nu,\mu}$ are the cofactors of the {\textbf{R}}(k,s).

Now the determinant of $\textbf{R}(k,s)$, i.e., the characteristic 
polynomial has the general form 
which can be expressed as \cite{chemla08}
\begin{equation}
|{\bf R}(k,s)| = \sum_{\mu=0}^M c_{\mu}(k) s^{\mu} 
\label{eq-poly2}
\end{equation}
where, for reasons that will be clear as we proceed, the coefficients 
for only the three lowest order terms will be relevant for calculating 
the quantities of our interest. 

Equation (\ref{eq-poly2}) is formally similar to (\ref{eq-poly}). 
As expected, we get \cite{chemla08} 
\begin{eqnarray}
V &=& - i \frac{c_{0}'(0)}{c_{1}(0)} \nonumber \\
D &=& \frac{c_{0}''(0) - 2i c_{1}'(0) V - 2 c_{2}(0) V^{2}}{2 c_{1}(0)} 
\label{eq-FourierLaplace}
\end{eqnarray}

\subsection{\bf Beyond average: dwell time distribution (DTD)}

Two motors with identical average velocities may exhibits widely 
different types of fluctuations. 
Suppose the successive mechanical steps are taken by the motor at 
times $t_{1}, t_{2},...,t_{n-1}, t_{n}, t_{n+1},...$. Then, the 
time of dwell before the $k$-th step is defined by 
$\tau_{k} = t_{k}-t_{k-1}$. In between successive steps, the motor 
may visit several ``chemical'' states and each state may be visited 
more than once. But, the purely chemical transitions would not be 
visible in a mechanical experimental set up that records only its 
position. The number of visits to a given state and the duration of 
stay in a state in a given visit are random quantities.

In order to appreciate the origin of the fluctuations in the dwell times, 
let us consider the simple $N$-step kinetics: 
\begin{equation}
M_{1} \rightleftharpoons M_{2} \rightleftharpoons M_{3} ...\rightleftharpoons M_{j} \rightleftharpoons ...\rightleftharpoons M_{N}
\label{eq-cycle}
\end{equation}
Suppose $t_{\mu,\nu}$ is the duration of stay of the motor in the 
$\mu$-th state during its $\nu$-th visit to this state. If $\tau$ 
is the dwell time, then 
\begin{equation}
\tau = \sum_{\mu=1}^{N} \sum_{\nu=1}^{n_{\mu}} t_{\mu,\nu}
\end{equation}
where $n_{\mu}$ is the number of visits to the $\mu$-th state.
It is straightforward to check that 
\begin{eqnarray} 
<\tau> &=& \sum_{\mu=1}^{N} <n_{\mu}><t_{\mu}> \nonumber \\
\end{eqnarray}
where $<n_{\mu}>$ is the average number of visits to the $\mu$-th 
state and $<t_{\mu}>$ is the average time of dwell in the $\mu$-th 
state is a single visit to it. More interestingly \cite{moffitt10b}, 
\begin{eqnarray} 
&&<\tau^2>-<\tau>^2 = \sum_{\mu=1}^{N} (<t_{\mu}^2> - <t_{\mu}>^2) <n_{\mu}> \nonumber \\ 
&+& \sum_{\mu=1}^{N} (<n_{\mu}^2> - <n_{\mu}>^2) <t_{\mu}>^2 \nonumber \\
&+& 2 \sum_{\mu > \nu} (<n_{\mu} n_{\nu}>-<n_{\mu}><n_{\nu}>)<t_{\mu}><t_{\nu}> \nonumber \\
\label{eq-correl}
\end{eqnarray}
The first and second terms on the right hand side of (\ref{eq-correl}) 
capture, respectively, the fluctuations in the lifetimes of the 
individual states and that in the number of visits to a kinetic state. 
Note that the number of visits to a particular state depends on the 
number of visits to the neighboring states on the kinetic diagram; 
this interstate correlation is captured by the third term on the 
right hand side of (\ref{eq-correl}).

Several different analytical and numerical techniques have been 
developed for calculation of the dwell time distribution 
\cite{shaevitz05,liao07,linden07,chemla08}. 
Since the dwell time 
is essentially a first passage time \cite{rednerbook}, an absorbing 
boundary method \cite{liao07} has been used.

\noindent{\it An example: DTD for an irreversible motor with linear chain of ststes}

As an example, we consider again the simple scheme (\ref{eq-MMlikemot}). 
In this case, the DTD is 
\begin{equation}
f(t) = \biggl( \frac{\omega_{1}\omega_{2}}{\omega_{-}-\omega_{+}}\biggr)(e^{-\omega_{+}t} - e^{-\omega_{-}t}) 
\label{eq-coupledft}
\end{equation}
where 
\begin{equation} 
\omega_{\pm} = \frac{\omega_{1}+\omega_{-1}+\omega_{2}}{2}\pm\biggl[\sqrt{\frac{(\omega_{1}+\omega_{-1}+\omega_{2})^2}{4}-\omega_{1}\omega_{2}}\biggr]
\label{eq-omegapm}
\end{equation}
In the special case $\omega_{-1} = 0$, $\omega_{+} = \omega_{1}$ and 
$\omega_{-} = \omega_{2}$
and, hence, 
\begin{equation}
f(t) = \biggl( \frac{\omega_{1}\omega_{2}}{\omega_{2}-\omega_{1}}\biggr)(e^{-\omega_{1}t} - e^{-\omega_{2}t}) 
\label{eq-uncoupledft}
\end{equation} 
Similar sum of exponentials for DTD have been derived also for machines  
with more complex mechano-chemical kinetics (see, for example, 
refs.\cite{garai09a,tripathi09,sharma11a}).

\subsubsection{\bf A matrix-based formalism for the DTD}

The matrix-based formalism that we have discussed above \cite{chemla08}, 
for the calculation of $V$ and $D$, has been extended to formulate 
general prescriptions for calculating the dwell time distribution 
$\psi(t)$. Let the Laplace transform of the DTD be denoted by 
$\tilde{\psi}(s)$. 
The general strategy, developed by Chemla et al.\cite{chemla08}, 
for the calculation of $\psi(t)$ is based three main steps: 
(i) obtaining $\tilde{P}(k,s)$ by solving the master equation for 
$P(x_j,t)$ in the Fourier-Laplace space; 
(ii) deriving a relation between $\tilde{P}(k,s)$ and $\tilde{\psi}(s)$ 
and using it, together with the solution $\tilde{P}(k,s)$ obtained in 
step (i), to get $\tilde{\psi}(s)$; and 
(iii) obtaining either $\psi(t)$ from $\tilde{\psi}(s)$ by inverse 
Laplace transform or, extracting at least the first few moments of 
$\psi(t)$ from $\tilde{\psi}(s)$.

\noindent$\bullet${\it DTD for a motor that never steps backward}

We now illustrate the method with the simple example 
\begin{equation}
1_{j} \mathop{\rightleftharpoons}^{k_1}_{k_{-1}} 2_{j} \mathop{\rightleftharpoons}^{k_2}_{k_{-2}} ....M_{j-1} \mathop{\rightleftharpoons}^{k_{(j-1)}}_{k_{-(j-1)}}M_{j} \mathop{\rightarrow}^{k_M} 1_{j+1}
\label{eq-Mirrev}
\end{equation}
where the integer subscripts $j$ and $j+1$ label the discrete positions 
$x_j$ and $x_{j+1}$ of the motor on its track. The total number of 
``chemical'' states of a motor at each position is $M$. Note that the 
``chemical'' transitions are reversible, but the ``mechanical step'' is 
irreversible; the latter rules out any possibility of backward stepping 
of the motor. 

Suppose the initial condition is $P_{\mu}(0) = \delta_{\mu 1}$, i.e., 
the motor is certainly in the chemical state 1 at $t=0$. For the 
kinetic scheme (\ref{eq-Mirrev}) under this initial condition, 
eqn.(\ref{eq-tprob}) becomes 
\begin{equation}
\tilde{P}(k,s) = \frac{1}{s} \frac{|\textbf{R}(0,s)|}{|\textbf{R}(k,s)|}  = \frac{s^{M-1} + ...+ c_{2} s + c_{1}}{s^{M}+...+c_{2}s^{2}+c_{1}s+c_{0}(k)}
\label{eq-Pks}
\end{equation}
Moreover, in this case, the relation between $\tilde{P}(k,s)$ and 
$\tilde{\psi}(s)$ is \cite{chemla08} 
\begin{equation}
\tilde{P}(k,s) = \frac{1-\tilde{\psi}(s)}{s[1-\rho_{+}(k)\tilde{\psi}(s)]}
\label{eq-PvsPsi}
\end{equation}
Equating the right hand sides of the eqns.(\ref{eq-Pks}) and (\ref{eq-PvsPsi}), 
we get 
\begin{equation}
\tilde{\psi}(s) = \frac{|\textbf{R}(k,s)|-|\textbf{R}(0,s)|}{|\textbf{R}(k,s)|-\rho_{+}(k)|\textbf{R}(0,s)|} 
\end{equation}

\noindent$\bullet${\it DTD for a motor that steps both forward and backward}

If $n_{+}$ and $n_{-}$ are the numbers of forward and backward steps, 
respectively, then for large $n = n_{+} + n_{-}$, the corresponding 
{\it step splitting} probabilities are $\Pi_{+} = n_{+}/n$ and 
$\Pi_{-} = n_{-}/n$. The dwell times {\it before} a forward step and 
before a backward step can be measured separately. Hence, the 
{\it prior dwell times} $\tau_{+}^{\leftarrow}$ and $\tau_{-}^{\leftarrow}$ 
can be obtained by restricting the summations in 
\begin{equation}
\tau_{\pm}^{\leftarrow} = \frac{1}{n_{\pm}} \sum^{\pm} \tau_{k}
\end{equation}
to just forward (+) or just backward (-) steps, respectively. In 
terms of splitting probabilities and prior dwell times, the 
mean dwell time $\langle \tau \rangle$ can be expressed as 
\begin{equation}
\langle \tau \rangle = \Pi_{+} \tau_{+}^{\leftarrow} + \Pi_{-} \tau_{-}^{\leftarrow}
\end{equation}

Compared to the prior dwell times, more detailed information on 
the stepping statistics is contained in the four {\it conditional 
dwell times}, which are defined as follows: 
\begin{eqnarray} 
\tau_{++} &=& {\rm dwell ~time ~between ~a ~+ ~step ~followed ~by ~a ~+ ~step} \nonumber \\
\tau_{+-} &=& {\rm dwell ~time ~between ~a ~+ ~step ~followed ~by ~a ~- ~step} \nonumber \\
\tau_{-+} &=& {\rm dwell ~time ~between ~a ~- ~step ~followed ~by ~a ~+ ~step} \nonumber \\
\tau_{--} &=& {\rm dwell ~time ~between ~a ~- ~step ~followed ~by ~a ~- ~step} \nonumber \\
\end{eqnarray} 
It is helpful to introduce {\it pairwise step probabilities} 
$\Pi_{++}, \Pi_{+-}, \Pi_{-+}, \Pi_{--}$. Note that 
$\Pi_{++}+\Pi_{+-} = 1$, and $\Pi_{-+} + \Pi_{--} = 1$. 
Neglecting finite time corrections, 
\begin{eqnarray} 
\Pi_{++} = n_{++}/(n_{++}+n_{+-}),~~\Pi_{+-} = n_{+-}/(n_{++}+n_{+-})\nonumber \\
\Pi_{-+} = n_{-+}/(n_{-+}+n_{--}),~~\Pi_{--} = n_{--}/(n_{-+}+n_{--})\nonumber \\
\end{eqnarray} 
Hence 
\begin{eqnarray}
\tau_{+}^{\leftarrow} &=& \Pi_{++} \tau_{++} + \Pi_{+-} \tau_{-+} \nonumber \\
\tau_{-}^{\leftarrow} &=& \Pi_{-+} \tau_{+-} + \Pi_{--} \tau_{--} 
\end{eqnarray} 
Defining the {\it post dwell times} $\tau_{\pm}^{\rightarrow}$ in 
a fashion similar to that used for defining the {\it prior dwell 
times}, we get 
\begin{eqnarray}
\tau_{+}^{\rightarrow} &=& \Pi_{++} \tau_{++} + \Pi_{+-} \tau_{+-} \nonumber \\
\tau_{-}^{\rightarrow} &=& \Pi_{-+} \tau_{-+} + \Pi_{--} \tau_{--} 
\end{eqnarray} 
and 
\begin{equation}
\langle \tau \rangle = \Pi_{+} \tau_{+}^{\rightarrow} + \Pi_{-} \tau_{-}^{\rightarrow}
\end{equation}

For motors which can step both forward and backward, more relevant  
information on the kinetics of a motor are contained in the 
{\it conditional} dwell time distributions \cite{tsygankov07,chemla08}. 
We illustrate the concepts and the matrix-based formalisms for such 
motors with the simple linear chain of states where all the transitions, 
including the mechanical transition, are reversible.
\begin{equation}
1_{j} \mathop{\rightleftharpoons}^{k_1}_{k_{-1}} 2_{j} \mathop{\rightleftharpoons}^{k_2}_{k_{-2}} ....M_{j-1} \mathop{\rightleftharpoons}^{k_{(j-1)}}_{k_{-(j-1)}}M_{j} \mathop{\rightleftharpoons}^{k_M}_{k_{-M}} 1_{j+1}
\label{eq-Mrev}
\end{equation}
We define the {\it conditional branching probabilities} $p_{\pm\pm}$ 
as the probability of taking a forward (+) or backward (-) step, 
given the previous step being forward (+) or bakward (-). Similarly, 
instead of one single DTD $\psi(t)$, we now have four {\it conditional 
dwell time distributions} (cDTD) $\psi_{\pm\pm}(t)$. For convenience 
of calculation, we define the $2 \times 2$ matrix 
\begin{equation}
\mathbf{\psi(s)}=
\begin{bmatrix}
      p_{++}\psi_{++}(s) & p_{+-}\psi_{+-}(s) \\
      p_{-+}\psi_{-+}(s) & p_{--}\psi_{--}(s)\\
\end{bmatrix}
\end{equation}
the diagonal matrix 
\begin{equation}
\mathbf{\rho(q)}=
\begin{bmatrix}
    \rho_{+}(q) & 0 \\
     0 & \rho_{-}(q)
\end{bmatrix}
\end{equation}
and the column vector 
\begin{equation}
\mathbf{\Psi(s)}=\dfrac{1}{s}
\begin{bmatrix}
    1-p_{++}\psi_{++}(s)-p_{+-}\psi_{+-} \\
    1-p_{-+}\psi_{-+}(s)-p_{--}\psi_{--}(s)\\
\end{bmatrix}
\end{equation}

In this case the relation between $\tilde{P}(k,s)$ and the cDTDs is 
\cite{chemla08} 
\begin{equation}
\tilde{P}(q
,s)=  \textbf{p}_{0}^{T}( \textbf{I}- \mathbf{\psi}(s) \mathbf{\rho}(q))^{-1} \mathbf{\Psi}(s)
\label{eq-m20}
\end{equation}
where $\textbf{p}_{0}$ is the vector of initial conditions. For 
example, $\textbf{p}_{0}^{T} = (1 0)$ corresponds to the given  
condition that the motor has taken the initial step in the forward 
(+) direction.

Extracting the cDTDs exploiting the relation (\ref{eq-m20}) and the 
solution for $\tilde{P}(k,s)$ is more complicated than the procedure 
we followed in the case of a single DTD. Let us begin with the case 
$\textbf{p}_{0}^{T} = (1 0)$ (i.e., given initial forward stepping) 
and the initial condition $P_{\mu}(0) = \delta_{\mu 1}$. For this 
case \cite{chemla08} 
\begin{equation}
\left.\dfrac{1}{s\tilde{P}_{+}(q,s)}\right|_{\{\rho_{-}(q)=0\}}=\dfrac{1-\rho_{+}(q)p_{++}\tilde{\psi}_{++}(s)}{1-p_{++}\tilde{\psi}_{++}(s)-p_{+-}\tilde{\psi}_{+-}(s)}
\label{eq-m25}
\end{equation}
Equation (\ref{eq-m25}) can be re expressed as
\begin{equation}
\left.\dfrac{1}{s\tilde{P}_{+}(q,s)}\right|_{\{\rho_{-}(q)=0\}}=a_0+a_+\rho_+(q)
\label{m26}
\end{equation}
where
\begin{eqnarray}
a_0 &=& \dfrac{1}{1-p_{++}\tilde{\psi}_{++}(s)-p_{+-}\tilde{\psi}_{+-}(s)} \nonumber \\
a_+ &=& - \dfrac{p_{++}\tilde{\psi}_{++}(s)}{1-p_{++}\tilde{\psi}_{++}(s)-p_{+-}\tilde{\psi}_{+-}(s)}
\label{a0a+}
\end{eqnarray}
Hence,
\begin{equation}
p_{++}\tilde{\psi}_{++}(s)=-\frac{a_{+}}{a_{0}}
\label{m27}
\end{equation}
and
\begin{equation}
p_{+-}\tilde{\psi}_{+-}(s)=\frac{a_{0}+a_{+}-1}{a_{0}}
\label{m27b}
\end{equation}
Next we need to obtain 
$\left.\dfrac{1}{s\tilde{P}_{+}(q,s)}\right|_{\{\rho_{-}(q)=0\}}$
directly from (\ref{eq-tprob}) and, by comparing it with eqn.(\ref{m26}),
find out the expressions for $a_{0}$ and $a_{+}$; substituting these
expressions for $a_{0}$ and $a_{+}$ into equations (\ref{m27}) and
(\ref{m27b}) we get
$p_{++}\tilde{\psi}_{++}(s)$ and $p_{+-}\tilde{\psi}_{+-}(s)$, respectively.

Similarly, for the case $\textbf{p}_{0}^{T} = (0 1)$ (i.e., given 
initial backward stepping) and the initial condition 
$P_{\mu}(0) = \delta_{\mu M}$, one can derive the relation between 
$\left.\dfrac{1}{s\tilde{P}_{-}(q,s)}\right|_{\{\rho_{+}(q)=0\}}$ 
and $p_{-+}\tilde{\psi}_{-+}(s)$ and $p_{--}\tilde{\psi}_{--}(s)$.
Obtaining the solution 
$\left.\dfrac{1}{s\tilde{P}_{-}(q,s)}\right|_{\{\rho_{+}(q)=0\}}$
directly from (\ref{eq-tprob}) and, by utilizing its relation with 
$p_{-+}\tilde{\psi}_{-+}(s)$ and $p_{--}\tilde{\psi}_{--}(s)$ we 
can get the cDTDs $p_{-\pm}\tilde{\psi}_{-\pm}(s)$ in the Laplace 
space.

\subsubsection{\bf Extracting kinetic information from DTD}

It is possible to establish on general grounds that, for a motor 
with $N$ mechano-chemical kinetic states like (\ref{eq-cycle}), 
the DTD is a sum of $N$ exponentials of the form \cite{moffitt10b}
\begin{equation}
f(t) = \sum_{j=1}^{N} C_{j} e^{-\omega_{j}t}
\label{eq-sumexpo1}
\end{equation}
where $N-1$  of the $N$ coefficients $C_{j}$ ($1 \leq j \leq N$) 
are independent of each other because of the constraint imposed by 
the normalization condition for the distribution $f(t)$. Also note 
that the prefactors $C_{j}$ can be both positive or negative while 
$\omega_{j} > 0$ for all $j$. 

Recall that for the Gamma distribution, the randomness parameter $r=1/N$. 
Can the experimentally measured DTD 
be used to determine the number of states $N$? Unfortunately, 
for real motors, (i) not each step of a cycle is fully irreversible, 
(ii) the rate constants for different steps are not necessarily identical, 
(iii) branched pathways are quite common. Consequently, $1/r$ may 
provide just a bound on the rough estimate of $N$. 

Can one use the general form (\ref{eq-sumexpo1}) of DTD to extract 
all the rate constants for the kinetic model by fitting it with the 
experimentally measured DTD? The answer is: NO.
First, even if a good estimate of $N$ is available, the number of 
parameters that can be extracted by fitting the experimental DTD 
data to (\ref{eq-sumexpo1}) is $2N-1$ ($n$ values of $\omega_{j}$ 
and $N-1$ values of $C_{j}$). 
On the other hand, the number of possible rate constants may be much 
higher \cite{moffitt10b}. For example, if transitions from every kinetic 
state to every other kinetic state is allowed, the total number of rate 
constants would be $N(N-1)$. In other words, in general, the kinetic 
rate constants are underspecified by the DTD. 
Second, as the expression (\ref{eq-coupledft}) for the DTD of the 
example (\ref{eq-MMlikemot}) shows explicitly, the $\omega$'s that appear 
in the exponentials may be combinations of the rate constants for the 
distinct transitions in the kinetic model. It is practically impossible 
to extract the individual rate constants from the estimated $\omega$'s  
unless any explicit relation like (\ref{eq-omegapm}) between the estimated 
$\omega$'s and actual rate constants is {\it apriori} available.

The systematic method for extracting all the kinetic informations from 
experimental data are described in the next section where the utility 
of the DTD will be shown again.

\section{\bf Solving inverse problem by probabilistic reverse engineering: from data to model} 
\label{sec-inverseproblem}

A discrete kinetic model of a molecular motor can be regarded as a 
network where each node represents a distinct mechano-chemical state. 
The directed links denote the allowed transitions. Therefore, such a 
model is unambiguously specified in terms of the following parameters: 
(i) the total {\it number} $N$ of the nodes, 
(ii) the $N \times N$ matrix whose elements are the {\it rates} of 
the transitions among these states; a vanishing rate indicates a 
forbidden direct transition. 

In the preceding sections we handled the ``forward problem'' by starting 
with a model assuming the structure of the network and the transition 
rates. In this section we discuss the inverse problem for molecular 
motors after introducing the methodology. In most real situations the 
numerical values of the rate constants of the kinetic model are not known 
apriori. In principle, these can be extracted by analyzing the experimental 
data in the light of the model.

\subsection{\bf Frequentist versus Bayesian approach}

Suppose, $\vec{m}$ be a column vector whose $M$ components are the $M$
parameters of the model, i.e., the transpose of $\vec{m}$ is
$\vec{m}^{T} = (m_1,m_2,...,m_M)$
Let the data obtained in $N$ observations of this model are represented
by the $N$-component column vector $\vec{d}$ whose transpose is
$\vec{d}^{T} = (d_1,d_2,...,d_N)$. Our ``inverse problem'' is to infer 
information on $\vec{m}$ from the observed $\vec{d}$.
The philosophy underlying the frequentist approach, i.e., approaches
based on maximum-likelihood (ML) estimation and the Bayesian approach 
for extracting these information are different in spirit, as we explain 
in the next two subsubsections \cite{cowan07}.

For simplicity, let us assume that a device has only two possible
distinct states denoted by ${\cal E}_{1}$ and ${\cal E}_{2}$.
\begin{equation}
{\cal E}_{1} \mathop{\rightleftharpoons}^{k_f}_{k_r} {\cal E}_{2}
\label{eq-stateseq}
\end{equation}
Let us imagine that we are given the actual sequence of the states,
over the time interval $0 \leq t \leq T$, generated by the Markovian
kinetics of the device. But, the magnitudes of the rate constants
$k_f$ and $k_r$ are not supplied. We'll now formulate a method,
based on ML analysis \cite{andrec03} to extract the numerical
values of the parameters $k_f$ and $k_r$.

Suppose $t_{j}^{(1)}$ and $t_{j}^{(2)}$ denote the time interval of
the $j$-th residence of the device in states ${\cal E}_{1}$ and
${\cal E}_{2}$, respectively. Moreover, suppose that the device
makes $N_1$ and $N_2$ visits to the states ${\cal E}_{1}$ and
${\cal E}_{2}$, respectively, and $N = N_1+N_2$ is the total
number of states in the sequence. Therefore, total time of dwell
in the two states are
$T_{1} = \sum_{j=1}^{N_1} t_{j}^{(1)}$ and
$T_{2} = \sum_{j=1}^{N_2} t_{j}^{(2)}$
where $T_{1}+T_{2} = T$.

Since the dwell times are exponentially distributed for a Poisson
process, the likelihood of any state trajectory $S$ is the 
conditional probability density 
\begin{eqnarray}
P(S|\underline{k_f,k_r}) &=& \biggl(\Pi_{j=1}^{N_1} k_f e^{-k_f t_j^{(1)}}\biggr) \nonumber \\
&&\biggl(\Pi_{j=1}^{N_2} k_r e^{-k_r t_j^{(2)}}\biggr) \nonumber \\
&=& \biggl(k_f^{N_1} e^{-k_f T_1}\biggr)\biggl(k_r^{N_2} e^{-k_r T_2}\biggr) 
\label{eq-likeli}
\end{eqnarray}

\subsubsection{\bf Maximum-likelihood estimate}

ML approach \cite{myung03} is based on finding the estimates
of the set of model parameters that corresponds to the maximum of
the likelihood $P(\vec{d}|\underline{\vec{m}})$ for a fixed set of
data $\vec{d}$.
For the kinetic scheme shown in equation (\ref{eq-stateseq}), the
the ML estimates of $k_f$ and $k_r$ are obtained by using (\ref{eq-likeli}) 
in
$d[ln P(S|\underline{k_f,k_r})]/dk_f=0 = d[ln P(S|\underline{k_f,k_r})]/dk_r$.
It is straightforward to see \cite{andrec03} that these estimates are
$k_f=N_{1}/T_{1}$ and $k_r=N_{2}/T_{2}$.

\subsubsection{\bf Bayesian estimate}

For drawing statistical inference regarding a kinetic model, the
Bayesian approach has gained increasing popularity in recent
years \cite{raeside71a,raeside71b,ulrych01,scales01,eddy04a,kou05c}.
The areas of research where this has been applied successfully 
include various biological processes in, for example, genetics 
\cite{schoemaker99,beaumont04}, biochemistry \cite{golightly11},
cognitive sciences \cite{kruschke10}, ecology \cite{ellison04}, etc. 

In the Bayesian method there is no logical distinction between the
model parameters and the experimental data; in fact, both are
regarded as random. The only distinction between these two types
of random variables is that the data are observed variables whereas
the model parameters are unobserved. The problem is to estimate the 
{\it probability distribution} of the model parameters from the 
distributions of the observed data.

The Bayes theorem states that
\begin{equation}
P(\vec{m}|\underline{\vec{d}}) = \frac{P(\vec{d}|\underline{\vec{m}}) P(\vec{m})}{P(\vec{d})}
\label{eq-bayes1}
\end{equation}
where $P(\vec{d})$ can be expressed as
\begin{equation}
P(\vec{d}) = \int P(\vec{d}|\underline{\vec{m}}) P(\vec{m}) d\vec{m}
\label{eq-bayes2}
\end{equation}
The {\it likelihood} $P(\vec{d}|\underline{\vec{m}})$ is the conditional
probability for the observed data, given a set of particular values of
the model parameters, that is predicted by the kinetic model.  However, 
implementation of this scheme also requires $P(\vec{m})$ as input.
In Bayesian terminology $P(\vec{m})$ is called the {\it prior} because
this probability is {\it assumed} {\it apriori} by the analyzer
{\it before} the outcomes of the experiments have been analyzed.
In contrast, the left hand side of equation (\ref{eq-bayes1}) gives the
{\it posterior} probability, i.e., after analyzing the data.

Thus, an experimenter learns from the Bayesian analysis of the data.
Such a learning begins with an input in the form of a prior probability;
the choice of the prior can be based on physical intuition, or general
arguments based, for example, on symmetries. Prior choice can become
simple if some experience have been gained from previous measurements.
Often an uniform distribution of the model parameter(s) is assumed over 
its allowed range if no additional information is available to bias its
choice. To summarize, Bayesian analysis needs not just the likelihood
$P(\vec{d}|\underline{\vec{m}})$ but also the prior $P(\vec{m})$.

For the kinetic scheme shown in equation (\ref{eq-stateseq}), the Bayes'
theorem (\ref{eq-bayes1}) takes the form
\begin{eqnarray}
P(k_f,k_r|\underline{S}) &=& \frac{P(S|\underline{k_f,k_r}) P(k_f,k_r)}{P(S)} \nonumber \\
&=& \frac{P(S|\underline{k_f,k_r}) P(k_f,k_r)}{\sum_{k_f',k_r'} P(S|\underline{k_f',k_r'}) P(k_f',k_r')}
\label{eq-bayes3}
\end{eqnarray}
We now assume a uniform prior, i.e., constant for positive $k_f$ and $k_r$,
but zero otherwise. Then, $P(k_f,k_r|\underline{S})$ is proportional to
the likelihood function $P(S|\underline{k_f,k_r})$ (within a normalization
factor). Normalizing, we get
\begin{equation}
P(k_f,k_r|\underline{S}) =
\biggl[\frac{T_{1}^{N_1+1}}{\Gamma(N_1+1)}k_f^{N_1} e^{-k_f T_1}\biggr]\biggl[\frac{T_{2}^{N_2+1}}{\Gamma(N_2+1)}k_r^{N_2} e^{-k_r T_2}\biggr]
\label{eq-bayes4}
\end{equation}
The mean of $k_f$ obtained from the posterior distribution is 
$(N_1+1)/T_1$ whereas the corresponding ML estimate is $N_1/T_1$. 
Similarly, the mean obtained from the posterior distribution and 
the ML estimate of $k_r$ are obtained by replacing the subscripts 
$1$ by subscripts $2$. Moreover, the variance of $k_f$ and $k_r$ 
calculated from the posterior distribution are $(N_1+1)/T_1^2$ and
$(N_2+1)/T_2^2$, respectively.

\subsection{\bf Hidden Markov Models}

The actual sequence of states of the motor, generated by the underlying
Markovian kinetics, is not directly visible. For example, a sequence of
states that differ ``chemically'' but not mechanically do not appear as
distinct on the recording of the position of the motor in a single motor 
experiment.
This problem is similar to an older problem in cell biology: ion-channel
kinetics \cite{ball92,hawkes04}. 
Current passes through the channel only when it is in the ``open'' state.
However, if the channel has more than one distinct closed states, the
recordings of the current reveals neither the actual closed state in
which the channel was nor the transitions between those closed states when
no current was recorded.

Hidden Markov Model (HMM) \cite{eddy04b,rabiner89,krogh98,talaga07,vogl10}
has been applied to analyze FRET trajectories \cite{mckinney06,lee09b}, 
stepping recordings of molecular motors \cite{mullner10,syed10}, and 
actomyosin contractile system \cite{smith01b}
to extract kinetic information.

For a pedagogical presentation of the main ideas behind HMM, we start
with the kinetic scheme shown in (\ref{eq-bayes1}) and
a simple, albeit unrealistic, situation and then by gradually
adding more and more realistic features, explain the main concepts in
a transparent manner \cite{andrec03}. First, suppose that the actual 
sequence of states (trajectory) itself is visible; this case can be 
analyzed either by the ML-analysis of by Bayesian approach both of 
which we have presented above. We now relax the strong assumption about 
the trajectory and proceed as below.

\noindent$\bullet${\bf If state trajectory is hidden and visible trajectory is noise-free}

The sequence of states of the device is, as before, generated by a 
Markov processes which is {\it hidden}. Suppose the device emits 
photons from time to time that are detected by appropriate 
photo-detectors. For simplicity, we assume just two detection 
channels labelled by $1$ and $2$. For the time being, we also 
assume a perfect one-to-one correspondence between the state of the 
light emitting device and the channel that detects the photon.
If the channel $1$ ($2$) clicks then the light emitting device was
in the state ${\cal E}_{1}$ (${\cal E}_{2}$) at the time of emission.
The interval $\Delta t_{j} = t_{j+1}-t_{j}$ between the arrival of
the $j$-th and $j+1$-th photons ($1 \leq j \leq N$) is random.

Thus, from the photo-detectors we get a visible sequence of the 
channel index (a sequence made of a binary alphabet) which we call 
``noiseless photon trajectory'' \cite{andrec03}. 
The sequence of states in the noiseless photo trajectory is also 
another Markov chain that is conventionally referred to as  the 
``random telegraph process''. Note that the photon detected by 
channel 1 (or, channel 2) can take place at any instant during the 
dwell of the device in state 1 (or, state 2). Therefore, the sequence 
of states in the noiseless photon trajectory does not reveal the 
actual instants of transition from one state of the device to another.

Since the noiseless phton trajectory corresponds to a random telegraph 
process, the transition probabilities for this process are 
\begin{eqnarray}
P({\cal E}_{1}|\underline{{\cal E}_{1};k_f,k_r,\Delta t_{j}}) &=& \frac{k_r}{k_f+k_r} + \frac{k_f}{k_f+k_r} e^{-(k_f+k_r)\Delta t_j} \nonumber \\
P({\cal E}_{1}|\underline{{\cal E}_{2};k_f,k_r,\Delta t_{j}}) &=& \frac{k_r}{k_f+k_r} [1 - e^{-(k_f+k_r)\Delta t_j}] \nonumber \\
P({\cal E}_{2}|\underline{{\cal E}_{1};k_f,k_r,\Delta t_{j}}) &=& \frac{k_f}{k_f+k_r} [1 - e^{-(k_f+k_r)\Delta t_j}] \nonumber \\
P({\cal E}_{2}|\underline{{\cal E}_{2};k_f,k_r,\Delta t_{j}}) &=& \frac{k_f}{k_f+k_r} + \frac{k_r}{k_f+k_r} e^{-(k_f+k_r)\Delta t_j} \nonumber \\
\label{eq-condprob}
\end{eqnarray}
where $P({\cal E}_{\mu}|\underline{{\cal E}_{\nu};k_f,k_r,\Delta t_{j}})$
is the conditional probability that state of the device is ${\cal E}_{\mu}$
given that it was in the state ${\cal E}_{\nu}$ at a time $\Delta t$ earlier.

The likelihood of the visible data sequence $\{V\}$ is now given by
\begin{equation}
P(\{V\}|k_f,k_r) = P(V_1|k_f,k_r) \Pi_{j=1}^{N-1} P(V_{j+1}|V_{j};k_f,k_r,\Delta t_j)
\label{eq-nonoise}
\end{equation}
where the first factor on the right hand side is the initial probability
(usually assumed to be the equilibrium probability). The transition
probabilities on the right hand side of equation (\ref{eq-nonoise}) are
the conditional probabilities given in equation (\ref{eq-condprob}).
Unlike the previous simpler case, where the state sequence itself was
visible, no analytical closed-form solution is possible in this case.
Nevertheless, analysis can be carried out numerically.

\noindent$\bullet${\bf If state trajectory is hidden and visible trajectory is noisy}

In the preceeding case of a noise-free photon trajectory, we assumed that
from the channel index we could get perfect knowledge of the state of the
emitting device. More precisely, the conditional probabilities were
\begin{eqnarray}
P(1|\underline{{\cal E}_{1}}) &=& 1\nonumber \\
P(1|\underline{{\cal E}_{2}}) &=& 0 \nonumber \\
P(2|\underline{{\cal E}_{1}}) &=& 0 \nonumber \\
P(2|\underline{{\cal E}_{2}}) &=& 1 \nonumber \\
\label{eq-emissionprob1}
\end{eqnarray}

However, in reality, background noise is unavoidable. Therefore, if a 
photon is detected by the channel $1$, it could have been emitted by 
the device in its state ${\cal E}_{1}$ (i.e., it is, indeed, a signal 
photon) or it could have come from the background (i.e., it is a noise 
photon). Suppose $p_{s}$ is the probability that the detected photon 
is really a signal that has come from the emitting device. Suppose 
$p_{b1}$ is the probability of arrival of a background photon in the 
channel 1. The probability that a background photon arrives in channel 
2 is $1-p_{b1}$. Then \cite{andrec03},
\begin{eqnarray}
E(1|\underline{{\cal E}_{1}}) &=& p_s + (1-p_s) p_{b1} \nonumber \\
E(2|\underline{{\cal E}_{1}}) &=& 1- P(1|\underline{{\cal E}_{1}}) = (1-p_s) (1-p_{b1}) \nonumber \\
E(1|\underline{{\cal E}_{2}}) &=& (1-p_s) p_{b1} \nonumber \\
E(2|\underline{{\cal E}_{2}}) &=& 1 - P(1|\underline{{\cal E}_{2}}) = p_s + (1-p_s) (1-p_{b}) \nonumber \\
\label{eq-emissionprob2}
\end{eqnarray}
Thus, in this case, the relation between the states of the hidden and 
visible states is not one-to-one, but one-to-many.Therefore, given a 
hidden state of the device, a set of ``emission probabilities'' 
determine the probability of each possible observable state; these 
are listed in equations (\ref{eq-emissionprob2}) for the device 
(\ref{eq-stateseq}).

\subsubsection{\bf HMM: formulation for a generic model of molecular motor}

On the basis of the simple example of a 2-state system presented above, 
we conclude that, for data analysis based on a HMM four key ingredients 
have to be specified:\\
(i) the alphabet of the ``visible'' sequence $\{\mu\}$ ($1 \leq \mu \leq N$),
i.e., $N$ possible distinct visible states;
(ii) the alphabet of the ``hidden'' Markov sequence $\{j\}$
($1\leq j \leq M$), i.e., $M$ allowed distinct hidden states,
(iii) the hidden-to-hidden {\it transition} probabilities $W(j\to k)$, and
(iv) hidden-to-visible {\it emission} probabilities $E(j \to \mu)$.
In addition to the transition probabilities and emission probabilities,
which are the parameters of the model, the HMM also needs the initial
state of the hidden variable(s) as input parameters.

\centerline{Visible: {\framebox{V$_0$}}~~~~~~~~~{\framebox{V$_1$}}~~~~~...~~~~~{\framebox{V$_t$}}~~~~~~~~~ {\framebox{V$_T$}} }
~~~~~~~~~~~~~~~~~~~~~~~~~~~~~$\Uparrow$~~~~~~~~~~~~~$\Uparrow$~~~~~~~~~~~~~~~~$\Uparrow$~~~~~~~~~~~~~$\Uparrow$\\
\centerline{Hidden: {\framebox{H$_0$}}~~~$\rightarrow$~~~~{\framebox{H$_1$}}~$\rightarrow$~...$\rightarrow$~{\framebox{H$_t$}}~~~$\rightarrow$~~~~{\framebox{H$_T$}} }

Suppose $P(\{V\}|\underline{HMM,\{\lambda\}})$ denotes the probability that an
HMM with parameters $\{\lambda\}$ generates a visible sequence $\{V\}$.
Then,
\begin{equation}
P(\{V\}|\underline{HMM,\{\lambda\}}) = \sum_{\{H\}} P(\{V\}|\underline{\{H\};\{\lambda\}}) P(\{H\}|\underline{\{\lambda}\})
\label{eq-HMM1}
\end{equation}
where $P(\{H\}|\underline{\{\lambda\}})$ is the conditional probability 
that the HMM generates a hidden sequence $\{H\}$ for the given parameters 
$\{\lambda\}$ and $P(\{V\}|\underline{\{H\};\{\lambda\}})$ is the 
conditional probability that, given the hidden sequence $\{H\}$ (for 
parameters $\{\lambda\}$) the visible sequence $\{V\}$ would be obtained.

Once $P(\{V\}|\underline{HMM,\{\lambda\}})$ is computed, the parameter set
$\{\lambda\}$ are varied to maximize
$P(\{V\}|\underline{HMM,\{\lambda\}})$ (for the convenience of numerical
computation, often ln$P(\{V\}|\underline{HMM,\{\lambda\}})$ is maximized.
The total number of possible hidden sequences of length $T$ is $T^{MN}$.
In order to carry out the summation in equation (\ref{eq-HMM1}) one has 
to enumerate all possible hidden sequences and the corresponding 
probabilities of occurrences. A successful implementation of the HMM 
requires use of an efficient numerical algorithm; the {\it Viterbi 
algorithm} \cite{forney73,viterbi06} is one such candidate.

In case of a molecular motor, the ``chemical states'' are not visible
in a single molecule experiment. Moreover, even its mechanical state
that is ``visible'' in the recordings may not be its true position
because of (a) measurement noise, and (b) steps missed by the detector.
Let us denote the ``visible'' sequence by the recorded positions
$\{Y\}$ whereas the hidden sequence is the composite mechano-chemical
states $\{X,C\}$ where $X$ and $C$ denote the true position and chemical
state, respectively. The transition probabilities are denoted by
$W(X_{t-1},C_{t-1} \to X_t,C_t)$
One possible choice for the emission probabilities $E$ is \cite{mullner10}
\begin{equation}
E(X_t \to Y_t) = \sqrt{\frac{1}{(2\pi\sigma_t^2)}}exp[-\frac{(Y_t-X_t)^2}{(2\sigma_t^2)}]
\end{equation}
In this case,
\begin{equation}
P(\{Y\}|\underline{HMM,\{\lambda\}}) = \sum_{\{X,C\}} P(\{Y\}|\underline{\{X,C\};\{\lambda\}}) P(\{X,C\}|\underline{\{\lambda}\})
\label{eq-HMM2}
\end{equation}
where 
\begin{equation}
P(\{X,C\}|\underline{\{\lambda}\}) = P_{X_0,C_0} W(X_0,C_0 \to X_1,C_1) W(X_1,C_1 \to X_2,C_2).....W(X_{T-1},C_{T-1} \to X_T,C_T)
\end{equation}
and
\begin{equation}
P(\{Y\}|\underline{\{X,C\};\{\lambda\}}) = E(X_0 \to Y_0) E(X_1 \to Y_1)...E(X_t \to Y_t)...E(X_T \to Y_T)
\end{equation}

The usual strategy \cite{mckinney06,mullner10} consists of the following 
steps: Step I: {\it Initialization} of the parameter values for iteration.
Step II: {\it Iterative re-estimation} of parameters for {\it maximum likelihood}:
the parameters 
$\{W(X_{t-1},C_{t-1} \to X_t,C_t)\}$, $\{E(X_t \to Y_t)\}$ and $P_{X_0,C_0}$
are re-estimated iteratively till $P(\{Y\}|\underline{HMM,\{\lambda\}})$
saturates to a maximum.

Step III: Construction of {\it ``idealized'' trace}: using the final
estimation of the model parameters, the position of the motor as a
function of time is reconstructed; naturally, this trace is noise-free.

Step IV: Extraction of the {\it distributions of step sizes and dwell
times}: the distributions of the steps sizes and dwell times are obtained 
by constructing the distributions of the 
vertical and horizontal segments, respectively, of the idealized trace. 
These distributions can be compared with the corresponding theoretical 
predictions.

The strategy developed above turns out to be very successful in extracting 
the parameters for a well-defined model. However,
in the case of specific molecular motors, not only the rate constants 
but also the number of states and the overall architecture of the 
mechano-chemical network as well as the kinetic scheme postulated by 
the model may be uncertain. In that case, the experimental data should 
be utilized to ``select'' the most appropriate model from among the 
plausible ones. In fact, more than one model, based on different 
hypotheses, may appear to be consistent with the same set of experimental 
data within a level of accuracy. The experimental data can be exploited 
at least to ``rank'' the models in the order of their success in 
accounting for the same data set. Unfortunately, as we'll summarize 
later in this review, very little effort has been made so far in this 
direction for inferring and ranking models of molecular motors based on 
empirical data.

\section{\bf Motoring along filamentous tracks: generic models of porters}
\label{sec-genericporters}
 
So far as intracellular transport is concerned, the two basic mechanisms 
are: (i) passive diffusion, and (ii) active transport driven by molecular 
motors. Stochastic models of these two processes have been reviewed very 
recently from the perspective of applied mathematicians \cite{bresloff13}. 
Here we focus on the molecular motors and motor-driven active processes 
from the perspective of statistical physicists.
The generic models ignore the details of the composition and structure 
of the track as well as those of the architectural design of the motors.

\subsection{\bf Phenomenological linear response theory and modes of operation}

We identify the external load force $F_{ext}$ opposing the movement of 
the motor and the chemical potential difference 
$\Delta \mu = \mu_{ATP} - \mu_{ADP+P}$ as the two generalized forces 
$X_{1}$ and $X_{2}$. The corresponding generalized fluxes $J_{1}$ and 
$J_{2}$ are, respectively, the average spatial velocity $<V>$ of the 
motor and the average rate $<r>$ of ATP hydrolysis, measured in terms 
of the average number of ATP molecules hydrolyzed per unit time 
\cite{parmeggiani99}.
The modes of operation on the $F_{ext}-\Delta \mu$ plane is identical 
to the generic ones sketched on the $X_{1}-X_{2}$ plane in 
fig.\ref{fig-lrtmodes}.

\begin{figure}[htbp]
\begin{center}
\includegraphics[angle=-90,width=0.45\columnwidth]{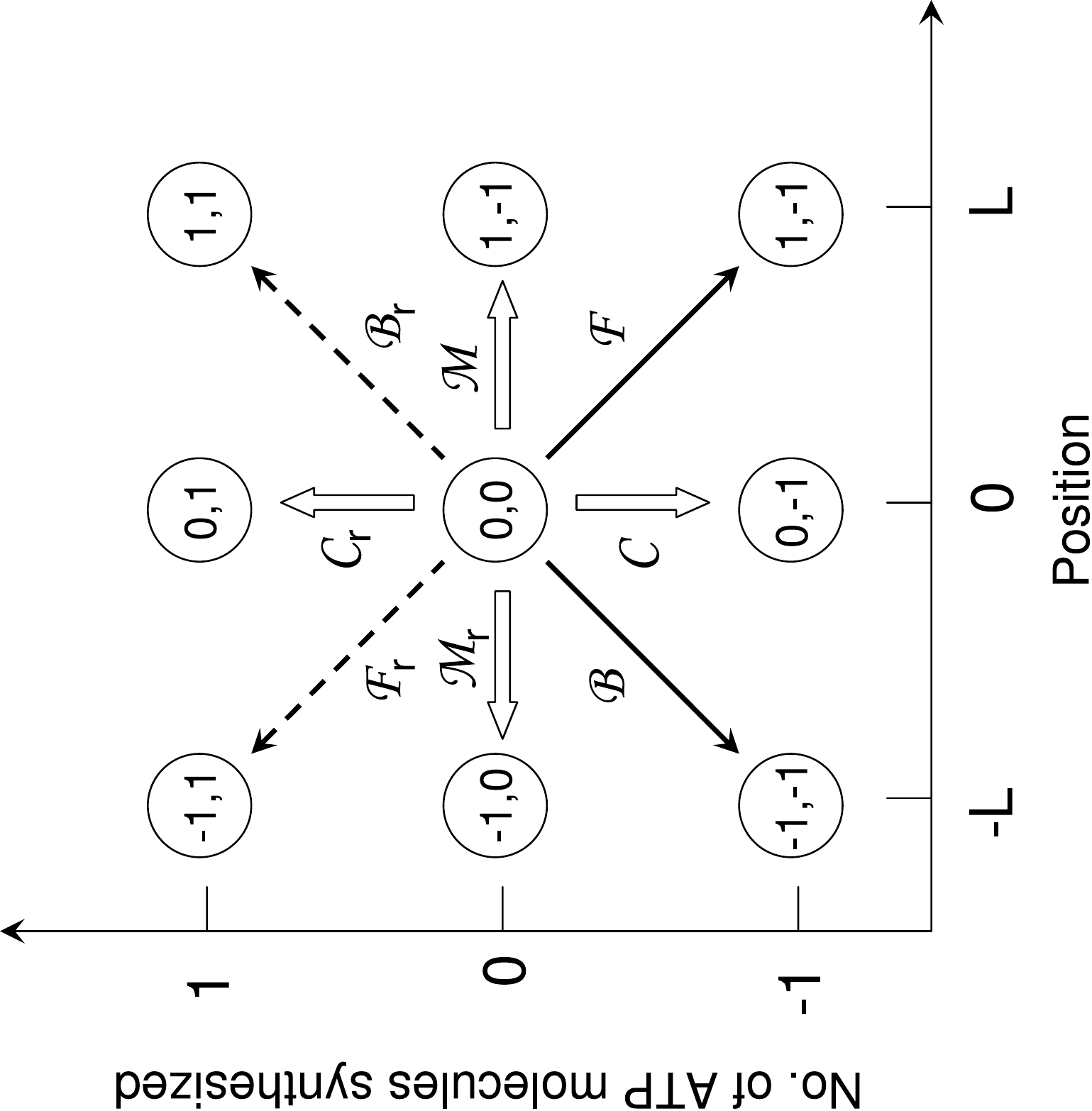}
\end{center}
\caption{Possible changes in the position and the number of ATP
molecules in a cycle (adapted from ref.\cite{astumian10}; see text 
for details).
}
\label{fig-coupledvr}
\end{figure}

\vspace{4cm} 

\begin{figure}[htbp]
\begin{center}
{\bf Figure NOT displayed for copyright reasons}.
\end{center}
\caption{Modes of operation of the molecular motors and the corresponding 
regimes on the $F-\Delta \mu$ space 
Reprinted from Biophysical Journal  
(ref.\cite{astumian10}), 
with permission from Elsevier \copyright (2010) [Biophysical Society]. 
}
\label{fig-modesNLR}
\end{figure}

The above scenario is deterministic and holds only in the linear response 
regime. In order to go beyond \cite{astumian08,astumian10}, 
let us consider the concrete case of cytoskeletal motors. A motor can step
forward (${\cal F}$) or backward (${\cal B}$) both by hydrolyzing
one molecule of ATP (see fig.\ref{fig-coupledvr}). The corresponding
reverse processes (${\cal F}_r$ and ${\cal B}_r$) {\it synthesize}
one molecule of ATP. The process ${\cal M}$ and the corresponding
reverse process ${\cal M}_r$ are purely mechanical processes which
do not change the number of ATP molecules. Similarly, ${\cal C}$
and ${\cal C}_r$ are purely chemical processes which do not change
the position of the motor. Let the instantaneous state of the
system at any arbitrary instant of time be denoted by the position
of the motor and the number of ATP molecules (see fig.\ref{fig-coupledvr}). 
The symbols $W({\cal F}), W({\cal F}_r), W({\cal B}), W({\cal B}_r), 
W({\cal M}), W({\cal M}_r), W({\cal C}), W({\cal C}_r)$ denote the 
probabilities of the respective processes defined above. 
These probabilities must satisfy the relations \cite{astumian08,astumian10}
\begin{eqnarray}
\frac{W({\cal F}_r)}{W({\cal F})} &=& exp\biggl(\frac{-\Delta \mu - F L}{k_BT}\biggr) \nonumber \\
\frac{W({\cal B}_r)}{W({\cal B})} &=& exp\biggl(\frac{-\Delta \mu + F L}{k_BT}\biggr) \nonumber \\
\frac{W({\cal M}_r)}{W({\cal M})} &=& exp\biggl(\frac{- F L}{k_BT}\biggr) \nonumber \\
\frac{W({\cal C}_r)}{W({\cal C})} &=& exp\biggl(\frac{- \Delta \mu}{k_BT}\biggr)
\label{eq-wratio}
\end{eqnarray}
where $L$ is the step size of the motor.
If $\tau_{cyc}$ is the time of a cycle, the velocity $v$ of the motor
is then given by
\begin{equation}
v = [L W({\cal F}) - L W({\cal F}_r) - L W({\cal B}) + L W({\cal B}_r) + L W({\cal M}) - L W({\cal M}_r)]/\tau_{cyc}
\end{equation}
Hence, rearragning the terms, and using (\ref{eq-wratio}), we get \cite{astumian10}
\begin{equation}
v = \biggl[\biggl(1-e^{(-\Delta \mu - F L)/(k_BT)}\biggr)
- \biggl(1-e^{(-\Delta \mu + F L)/(k_BT)}\biggr)\frac{W({\cal B})}{W({\cal F})}
+ \biggl(1-e^{(-F L)/(k_BT)}\biggr)\frac{W({\cal M})}{W({\cal F})}\biggr]\frac{LW({\cal F})}{\tau_{cyc}}
\label{eq-NLrelV}
\end{equation}
Similarly, the reaction rate $r$ is given by \cite{astumian10}
\begin{equation}
r = \biggl[\biggl(1-e^{(-\Delta \mu - F L)/(k_BT)}\biggr)
+ \biggl(1-e^{(-\Delta \mu + F L)/(k_BT)}\biggr)\frac{W({\cal B})}{W({\cal F})}
+ \biggl(1-e^{(-\Delta \mu)/(k_BT)}\biggr)\frac{W({\cal C})}{W({\cal F})}\biggr]\frac{W({\cal F})}{\tau_{cyc}}
\label{eq-NLrelR}
\end{equation}
Using the equations (\ref{eq-NLrelV}) and (\ref{eq-NLrelR}), instead of 
linear response relations between $v,r$ and $F,\Delta \mu$, the modes 
of operation of the molecular motor in this scenario can be 
analyzed. The resulting modes and the corresponding sectors on the 
$F-\Delta \mu$ space are shown in fig.\ref{fig-modesNLR}.
A similar analysis was reported also by Liepelt and Lipowsky \cite{liepelt09}.

However, it has been argued \cite{lipowsky08b} that,
for motors at a fixed temperature $T$ and driven by ATP hydrolysis,
the 2-dimensional space spanned by the load force $F$ and the
chemical potential diference $\Delta \mu = \mu_{ATP}-\mu_{ADP}-\mu_{P}$
is only a sub-space of the full 4-dimensional space spanned by
$F$, $\mu_{ATP}$, $\mu_{ADP}$ and $\mu_{P}$.

\begin{figure}[htbp]
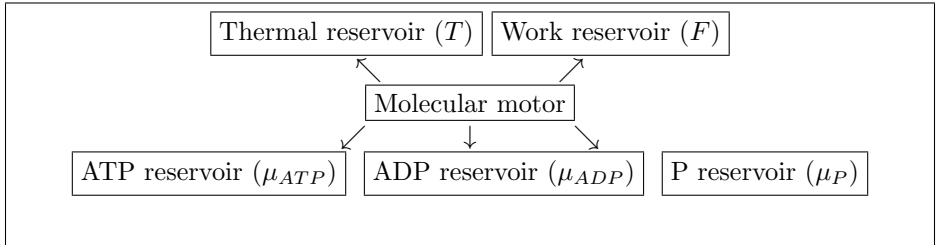

\begin{center}
\centerline{{\framebox{Thermal reservoir ($T$)}}~{\framebox{Work reservoir ($F$)}}}

\centerline{$\nwarrow$~~~~~~~~~~~~~~~~~~~~$\nearrow$}

\centerline{{\framebox{Molecular motor}}}

\centerline{$\swarrow$~~~~~~~~~~~$\downarrow$~~~~~~~~~~~$\searrow$}

\centerline{{\framebox{ATP reservoir ($\mu_{ATP}$)}}~~{\framebox{ADP reservoir ($\mu_{ADP}$)}}~~{\framebox{P reservoir ($\mu_{P}$)}}}
\caption{The reservoirs with which a molecular motor, fuelled by ATP 
hydrolysis, can exchange matter and energy.  }
\end{center}
\end{figure}

\subsection{\bf A generic model of a motor: kinetics on a discrete mechano-chemical network}

\begin{figure}[htbp]
\begin{center}
\includegraphics[angle=-90,width=0.85\columnwidth]{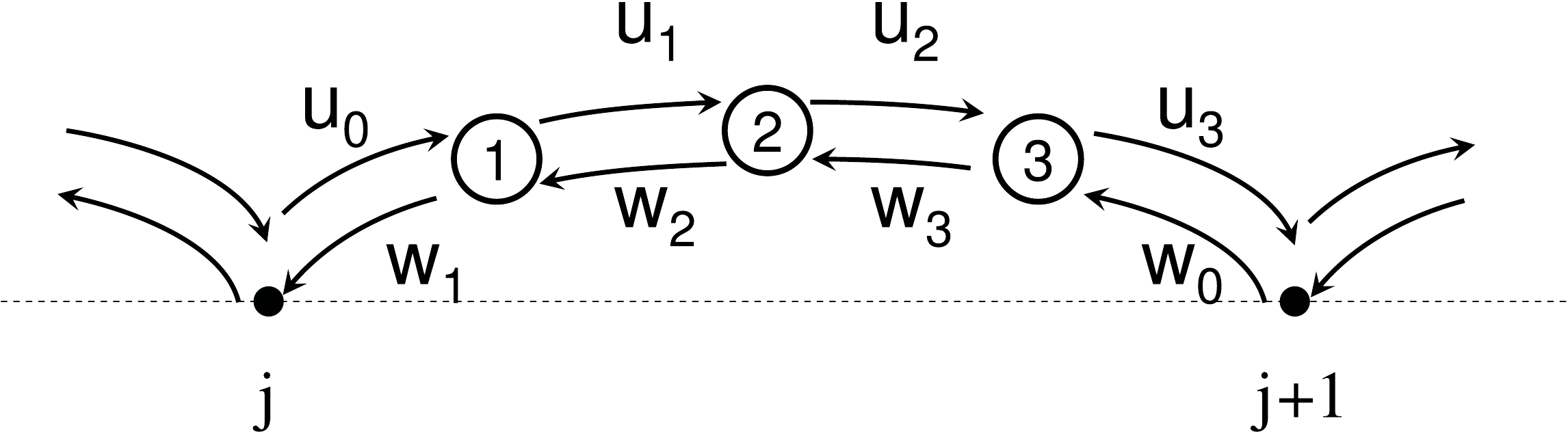}
\end{center}
\caption{An unbranched mechano-chemical cycle of the molecular motor 
with $M = 4$. The horizontal dashed line
shows the lattice which represents the track; $j$ and $j+1$
represent two successive binding sites of the motor. The circles
labelled by integers denote different ``chemical'' states in
between $j$ and $j+1$. (Adapted from fig.7 of ref.\cite{garai09a}).
}
\label{fig-fishkolo}
\end{figure}

To my knowledge, a generic 3-state stochastic model, with unbranched 
cyclic kinetics, was proposed first by Qian \cite{qian97,qian00} for 
the mechano-chemistry of molecular motors. 
As an illustrative example, let us consider the unbranched mechano-chemical 
cycle \cite{fisher07a} with $M=4$, as shown in fig.\ref{fig-fishkolo}. 
This special value of $M$ is motivated by the typical example of a 
kinesin motor for which the four essential steps in each cycle are as  
follows: 
(i) a substrate-binding step (e.g., binding of an ATP molecule), 
(ii) a chemical reaction step (e.g., hydrolysis of ATP), 
(iii) a product-release step (e.g., release of ADP), and 
(iv) a mechanical step (e.g., power stroke). 

Suppose, The forward transitions take place at rates $u_j$ whereas the 
backward transitions occur with the rates $w_j$.  The average
velocity $V$ of the motor is given by \cite{fisher07a}
\begin{equation}
 V=\frac{1}{R_{M}}\biggl[1-\prod^{{M}-1}_{j=0}\biggl(\frac{w_j}{u_j}\biggr)\biggr] 
\label{eq-kolov}
\end{equation}
where
\begin{equation}
R_{M}=\sum^{{M}-1}_{j=0}r_j   
\label{eq-fv3}
\end{equation} 
with
\begin{equation}
r_j=\biggl(\frac{1}{u_j}\biggr)\biggl[1+\sum^{{M}-1}_{k=1}\prod^{k}_{i=1}\biggl(\frac{w_{j+i}}{u_{j+i}}\biggr)\biggr] 
\label{eq-fv4}
\end{equation}
while $D$ is given by
\begin{equation}
 D=\biggl[ \frac{(VS_M+dU_M)}{R^2_M}-\frac{(M+2)V}{2} \biggr]\frac{d}{M}
\end{equation}
where \begin{equation}
 S_M=\sum^{M-1}_{j=0}s_j\sum^{M-1}_{k=0}(k+1)r_{k+j+1}
\end{equation}
and
\begin{equation}
 U_M=\sum^{M-1}_{j=0}u_jr_js_j
\end{equation}
while,
\begin{equation}
 s_j=\frac{1}{u_j}\biggl(1+\sum_{k=1}^{M-1}\prod^{k}_{i=1}\frac{w_{j+1-i}}{u_{j-i}}\biggr).
\end{equation}
For various extensions of this scheme see ref.\cite{fisher07a}. 

In the simpler case shown in (\ref{eq-MMlikemot}), 
where $M=2$, and the second step is purely irreversible,  
using the step size ${\ell}$ explicitly (to make the 
dimensions of the expressions explicitly clear), we get 
\begin{eqnarray}
V &=& {\ell} \biggl[\frac{\omega_{1} \omega_{2}}{\omega_{1}+\omega_{-1}+\omega_{2}}\biggr] \nonumber \\
D &=& \frac{\ell^2}{2} \biggl[\frac{(\omega_{1}\omega_{2})-2 (V/\ell)^2}{\omega_{1}+\omega_{-1}+\omega_{2}}\biggr] 
\label{eq-2strir}
\end{eqnarray} 
Note that if, in addition, $\omega_{-1}$ vanishes, i.e., if both the 
steps are fully irreversible, then 
$d/V = \omega_{1}^{-1} + \omega_{2}^{-1}$, 
i.e., the average time taken to move forward by one site is the sum 
of the mean residence time in the two steps of the cycle.

\subsection{\bf 2-headed motor: generic models of hand-over-hand and inchworm stepping patterns} 
\label{subsec-HoHInch}
 
\begin{figure}[htbp]
\begin{center}
\includegraphics[angle=90,width=0.65\columnwidth]{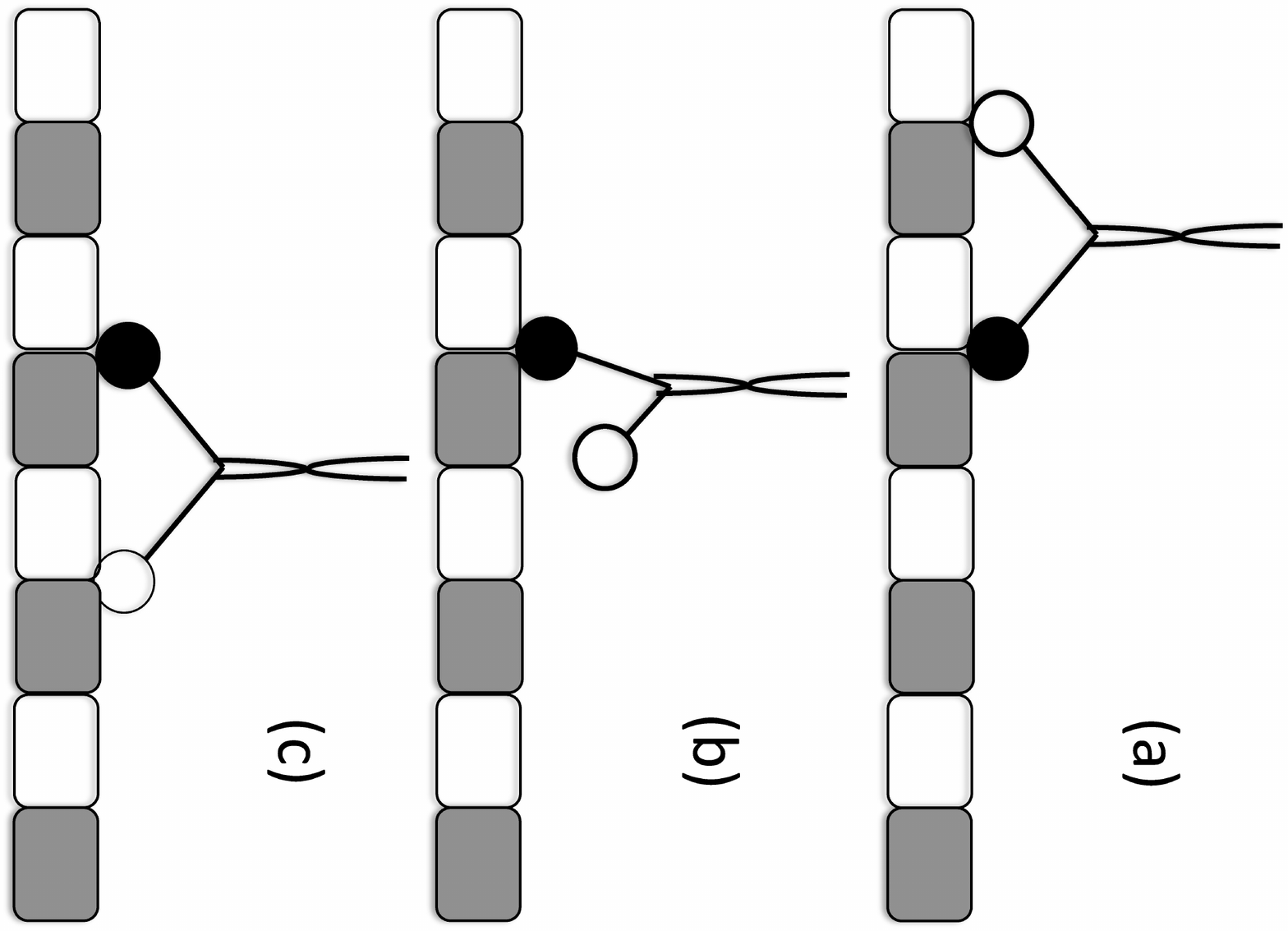}
\end{center}
\caption{A schematic representation of the hand-over-hand stepping 
pattern of a 2-headed motor. Each disc denotes a ``head''; one is 
filled and the other is not filled just to label the heads distinctly. 
}
\label{fig-HoH}
\end{figure}

\begin{figure}[htbp]
\begin{center}
\includegraphics[angle=90,width=0.65\columnwidth]{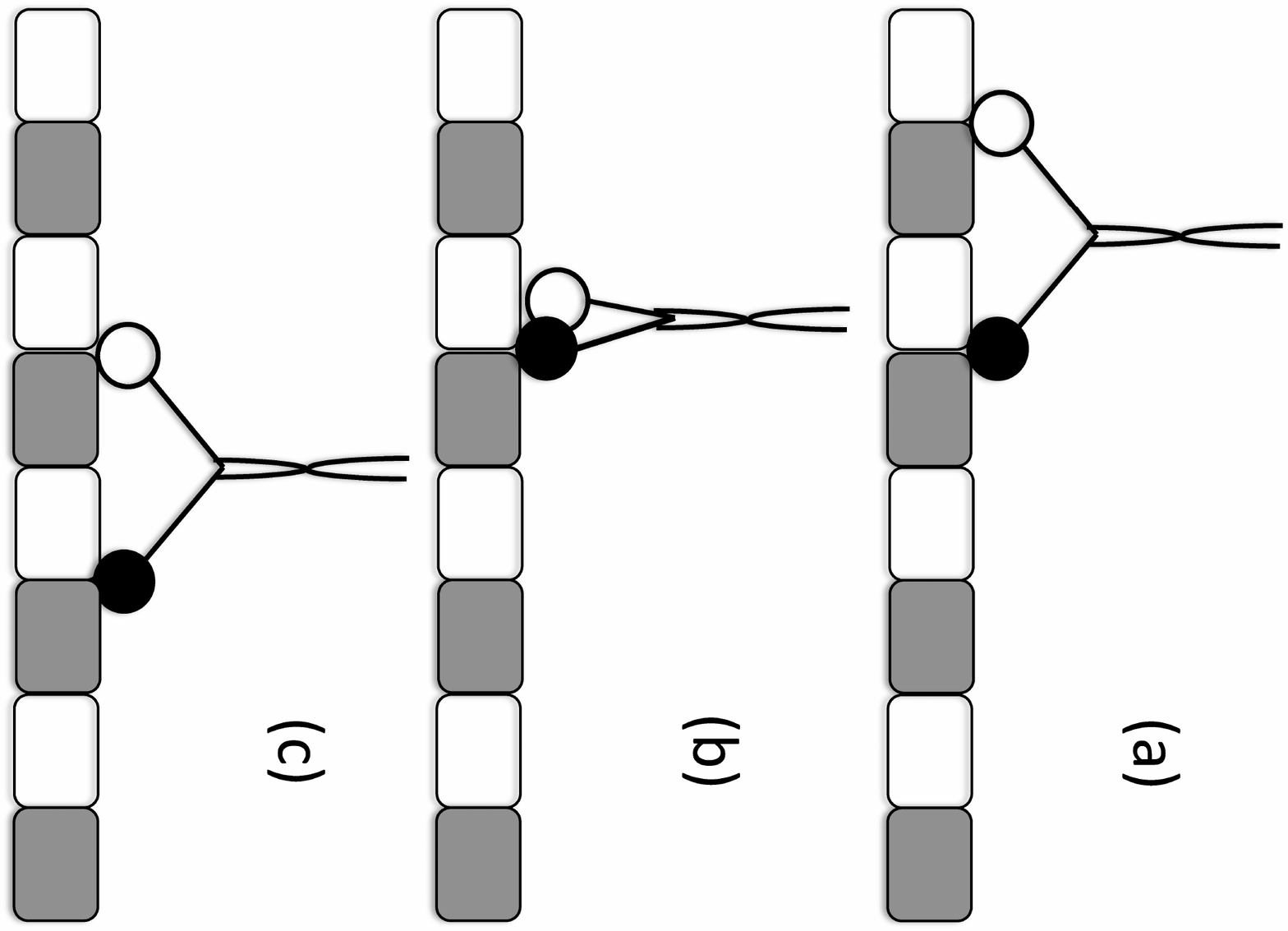}
\end{center}
\caption{A schematic representation of the inchworm stepping 
pattern of a 2-headed motor. Each disc denotes a ``head''; one is 
filled and the other is not filled just to label the heads distinctly. 
}
\label{fig-Inchworm}
\end{figure}

The stepping pattern of a 2-headed motor can be either ``hand-over-hand'' 
or ``inchworm'' (see figs.\ref{fig-HoH} and \ref{fig-Inchworm}). 
In the HoH stepping pattern, each head alternates between leading and 
lagging position as the motor steps forward. In contrast, in the inchworm 
pattern, the leading head always leads while the lagging head always lags.

One of the approaches for modeling 2-headed motors is based on Brownian 
ratchets. In this approach one begins with two identical heads each of 
which, at least for part of each cycle, is subjected to a periodic 
potential which represents its interaction with the periodic track. 
Then the two heads are coupled by an elastic spring to construct a 
2-headed motor. One writes Langevin equations for each head which 
are coupled because of the spring. In some models (see, for example, 
Dan et al.\cite{dan03}) the potential seen by each head is 
time-dependent so that each head is effectively a flashing ratchet, 
e.g., potential switching alternately between a sawtooth ($V_{1}(x)$) 
and a flat form ($V_{2}(x)$). The potential felt by the two heads are 
out of phase so that when one feels $V_{1}(x)$, the other feels 
$V_{2}(x)$ and vice-versa (see ref.\cite{munarriz08} for a slightly 
different formulation in terms of two sawtooth potentials that 
are shifted with respect to each other by half the spatial period). 
In an alternative formulation Derenyi and Vicsek \cite{derenyi96} 
assumed the potential to be time-independent whereas the relaxed length 
of the spring was assumed to alternate between 0 and 8 nm in each cycle. 
Mogilner et al.\cite{mogilner98} modelled a generic 2-headed motor 
where the conformational changes induced by ATP binding and/or hydrolysis 
gives rise to asymmetric internal velocity fluctuations. They showed 
that (noisy) directed motion of the motor is a consequence of the 
rectification of these velocity fluctuations by ``protein friction'' 
\cite{tawada91}.

Kumar et al.\cite{kumar08} modeled the inchworm stepping of 2-headed 
motors by a generic Brownian ``active elastic dimer''. In this model 
the two heads, whose positions are given by the coordinates are 
$x_1$ and $x_2$, are coupled by a elastic spring. The damping coefficient 
of the heads are assumed to depend on the relative coordinate 
$x = x_1-x_2$, i.e., on the elastic strain. 
The Langevin equation for the individual monomers include not only a 
strain-dependent damping, but also two noise terms one of which is 
thermal noise while the other is a non-equilibrium (or, active) noise. 
The model displays a counterintuitive reversal of the velocity of the 
center of mass with the variation of characteristics of damping. The 
key ingredient is the strain-dependent damping. 
As argued by the authors \cite{kumar08}, the stretch-dependent 
damping in this model is ``encoded'' in the fact that the ATP-bound 
(free) first head encounters a higher (lower) barrier against motion 
than that faced by the second head.

Next we study a generic model for the hand-over-hand stepping 
pattern of a 2-headed motor. As before, the track is represented by a 
linear chain with equispaced binding sites that are labelled by the 
integer index $j$ ($-\infty \leq j \leq \infty$). 

We first sketch the strategy adopted by Kolomeisky and Phillips 
\cite{kolomeisky05a}  
who extended the Fisher-Kolomeisky model \cite{fisher01,fisher07a} 
described above. 
There are $N$ discrete biochemical states on the pathway in between 
two successive sites $j$ and $j+1$. The separation between the 
successive binding sites is ${\ell}$. In the HoH stepping pattern, 
the two heads are assumed to move alternately: the trailing head, 
labelled as $1$, goes through transitions among the $N$ intermediate 
states while the other head, labelled as $2$, remains anchored to its 
own position; in this process the head $1$ becomes the leading head 
and the head $2$ becomes the trailing head. Next, head $1$ remains 
static while head $2$ executes its transitions among its own $N$ 
intermediate states to regain its leading position ahead of head $1$. 
Thus, in a single cycle each head moves ahead by a distance $2{\ell}$ 
and, the center of mass moves forward by ${\ell}$. Consequently, for 
the purpose of quantitative calculations, the overall translocation 
of the motor can be represented as a motion of two particles on two 
parallel periodic lattices; the distance between two neighboring 
sites on each lattice is $2{\ell}$.  In this case, the average 
velocity and the diffusion constant are given by the expressions 
(\ref{eq-2strir}). 

An alternative approach was developed earlier by Peskin and Oster 
\cite{peskin95}.
The two heads of the motor are connected at a hinge. The angle made by 
the two heads at the hinge can increase up to a maximum that corresponds 
to the separation ${\ell}$ between the two heads. 
For simplicity, we assume that each head of the 2-headed motor can exist 
in one of the two allowed states which are designated as ``strongly'' 
attached state (labelled by index 1) and ``weakly'' attached (labelled 
by index 0) \cite{peskin95}.  The transition $1 \to 0$ corresponds to 
{\it detachment} of the head from the track whereas the next transition 
$0 \to 1$ corresponds to its re-attachment. 

When both the heads are attached to the track, the rates of detachment 
of the front head and that of the back head are denoted by the symbols 
$\beta_{f}$ and $\beta_{b}$, respectively. If only one head is attached 
to the track, $\alpha$ is the rate of reattachment of the unattached 
head. The unattached head attaches in front of the already attached 
head with probability $p$ (the unattached head attaches behind the 
already attached head with probability $1-p$). 
Since we are interested in the average velocity of the motors during a 
single run, both the heads can be detached only before, and after, each 
run.  We now make the following assumptions \cite{peskin95}: 
(I) $\beta_{b} > \beta_{f}$, and (II) $p > (1/2)$. 

\begin{figure}[htbp]
\begin{center}
\includegraphics[angle=-90,width=0.85\columnwidth]{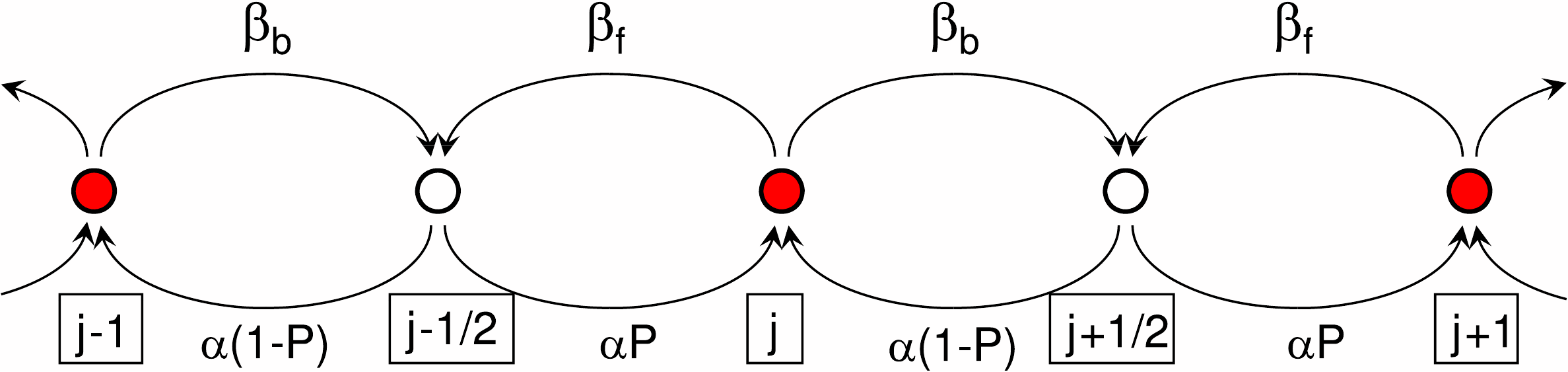}
\end{center}
\caption{A Markov chain representation of the movement of a 2-headed 
motor in the generic model developed by Peskin and Oster
(adapted from ref.\cite{peskin95}).
}
\label{fig-peskinKIN}
\end{figure}

The state of the motor, as a whole, is specified by specifying the 
corresponding states of the two heads. In principle, there are three 
state, namely (1,1), (1,0), (0,1) because, for reasons explained 
above, the state (0,0) is not allowed. We simply the description 
even further by noting that the states (1,0) and (0,1) describe the 
same state. Thus, the motor has two states, namely, (1,1) in which 
both the heads are attached and (1,0)=(0,1) in which only one head is 
attached while the other is detached. According our convention, the 
motor position $x_m$ is obtained by the following rule: if both the 
heads are attached then $x_m$ coincides with the mid-point which is 
also the position of the hinge. However, if only one head is attached, 
then $x_m$ is identified as the position of the bound head. Thus, $x_m$ 
can change by $\pm({\ell}/2)$. Denoting the positions of the states 
(1,1) by integer index $j$ and those of the states (1,0)=(0,1) by 
half-integer indices $j+1/2$, we can represent the hand-over-hand 
stepping pattern of the motor by the Markov chain shown in the 
fig.\ref{fig-peskinKIN}. 

\vspace{5cm}

The probabilities that the motor position is $j$ and $j+1/2$ are 
denotes by $P_{j}$ and $P_{j+1/2}$, respectively. The corresponding 
master equations are 
\begin{eqnarray} 
\frac{dP_{j}}{dt} = \alpha p P_{j-1/2} + \alpha (1-p) P_{j+1/2} - (\beta_b+\beta_f) P_{j} \nonumber \\
\frac{dP_{j+1/2}}{dt} = \beta_b P_{j} + \beta_f P_{j+1} - \alpha P_{j+1/2} 
\label{eq-peskinKIN}
\end{eqnarray}
We define the $k$-th moments 
$M_{k} = \sum_{j=-\infty}^{\infty} j^{k} P_{j}$ and   
$N_{k} = \sum_{j=-\infty}^{\infty} (j+1/2)^{k} P_{j+1/2}$. It follows 
from (\ref{eq-peskinKIN}), in the steady state, 
\begin{eqnarray} 
M_{0} = \frac{\alpha}{\alpha+\beta_b+\beta_f} \nonumber \\
N_{0} = \frac{\beta_b+\beta_f}{\alpha+\beta_b+\beta_f} 
\label{eq-M0N0}
\end{eqnarray}
and the average velocity 
\begin{eqnarray} 
<V> = {\ell} \frac{d(M_1+N_1)}{dt} = \frac{\alpha \ell}{2(\alpha+\gamma)}\biggl[\delta + 2 \biggl(p-\frac{1}{2}\biggr)\gamma\biggr]
\label{eq-VavpeskinKIN}
\end{eqnarray}
where $\delta = \beta_b - \beta_f$ and $\gamma = \beta_b + \beta_f$. 
Note that the asymmetries of both detachment and reattachment inherent  
in the assumptions (I) and (II) give rise to the two positive terms 
within the square bracket of eqn.(\ref{eq-VavpeskinKIN}). This generic 
model of 2-headed motor assumes these asymmetries; we'll explain the 
physical origin of such asymmetries of real two-headed motors when we 
study specific examples in part II of this review.

\noindent$\bullet${\bf Traffic-like collective movement of motors: totally asymmetric simple exclusion process}

So far we have discussed mostly a single isolated porter walking on its 
track. Now we present a generic model of traffic-like collective 
movements of many porters simultaneously on the same filamentous track. 
The model developed by Aghababaie et al.\cite{aghababaie99} for this 
purpose is based on an abstract formulation of Brownian ratchet.  
However, most of the subsequent works have been based on the  {\it 
asymmetric simple exclusion process} (ASEP) \cite{gunter01}. 

An ASEP is a simple particle-hopping model where ``particles'' can hop, 
with some probability per unit time, from one lattice site to a neighboring 
site if, and only if, the target site is not already occupied by another 
``particle''. Thus, simultaneous occupation of any site by more than one 
particle is ruled out in this model; this fact is expressed by the term 
``simple exclusion''. The terms ``asymmetric'' expresses the fact that 
the particles have a preferred direction of motion. If the particles 
move only in one direction, and never in the opposite direction, the 
model is further specialized to {\it totally asymmetric simple exclusion 
process} (TASEP). In the general case, a particle hops forward with a 
rate $q$ if the target site is empty. So far as the kinetics is concerned, 
random sequential updating of the states of the system is implemented in 
discrete time steps. The number of particles passing through a given site 
per unit time is defined as the {\it flux}; in the steady state, because of 
the equation of continuity, the flux $J$ is independent of the lattice site.

\begin{figure}[htbp]
\begin{center}
\includegraphics[angle=90,width=0.65\columnwidth]{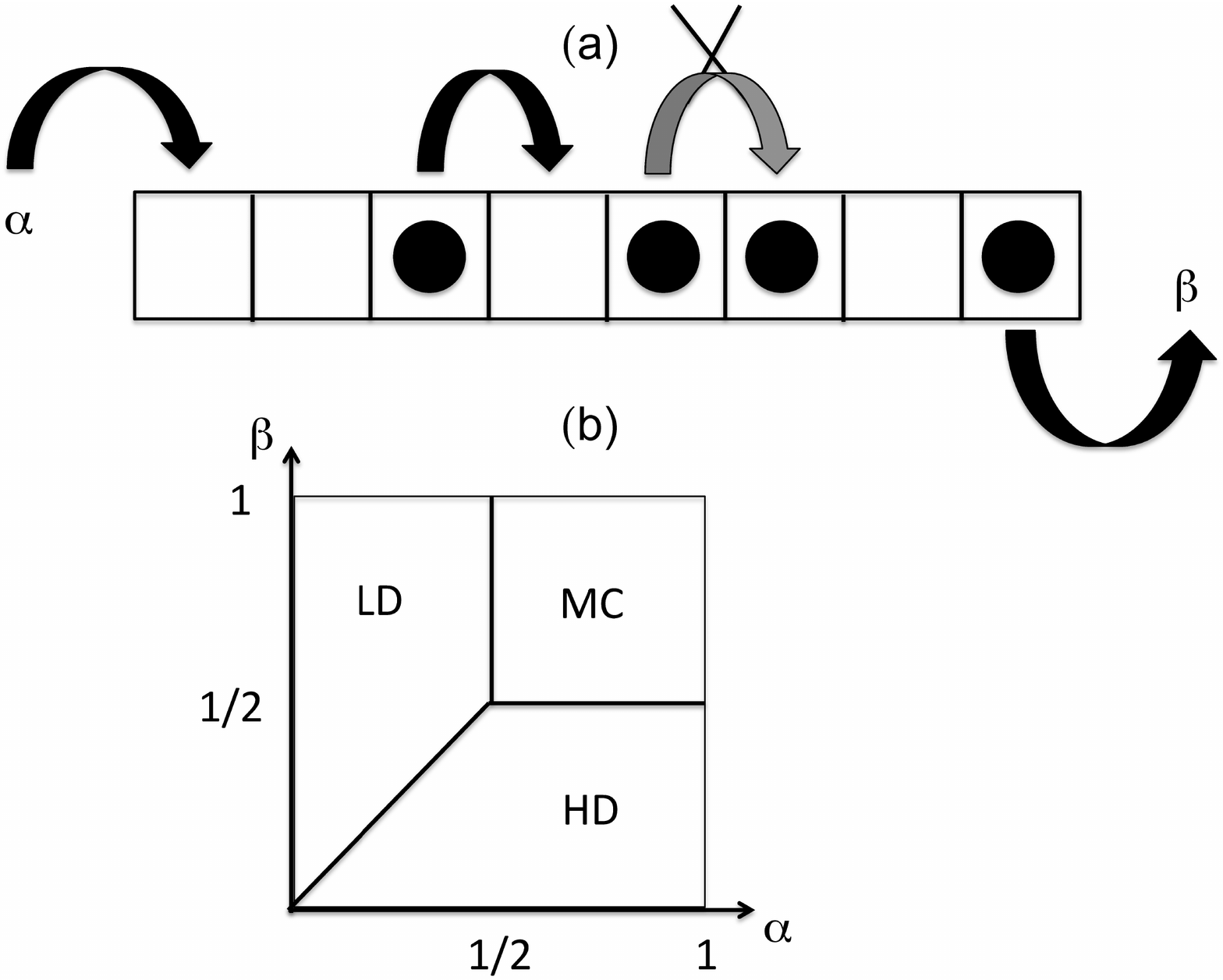}
\end{center}
\caption{(a) The totally asymmetric simple exclusion process (TASEP)
under open boundary condition (OBC), and (b) the corresponding phase 
diagram (see the text for the details).
}
\label{fig-TASEPobc}
\end{figure}

For a finite lattice boundary conditions have to be imposed. Although 
periodic boundary conditions are imposed often for the simplicity of 
calculation or as an intermediate step, the open boundary conditions 
capture the molecular motor traffic more realistically. In the latter 
case, particles enter from one end at a rate $\alpha$ and exit from the 
opposite end at the rate $\beta$. 
At any instant of time, the number density $\rho$ of the particles is 
defined by $\rho=N/L$ where $N$ is the total number of particles 
distributed over the lattice that consists of a total of $L$ sites. 
Obviously, under the periodic boundary conditions, $\rho$ is independent 
of time and the steady-state flux $J$ depends on $\rho$ and $q$. 
The plot of $J$ against $\rho$ is called the {\it fundamental diagram}.
As expected, at sufficiently low number density of the particles, the 
flux increases with $\rho$, although the rate of increase decreases with 
increasing $\rho$ because of the stronger hindrance effects. After 
attaining a maximum at a certain density $\rho_{m}$, the flux keeps 
decreasing with further increase of $\rho$, eventually vanishing at 
$\rho=1$. In contrast, under open boundary conditions, the number density 
$\rho$ fluctuates. This version of TASEP serves as the prototype for the 
non-equilibrium systems that exhibit {\it boundary-induced phase transitions} 
\cite{krug91}. 
On a plane spanned by the parameters $\alpha$ and $\beta$, the system 
exhibits an interesting phase diagram \cite{gunter01}. 
The three phases are identified as (see fig.\ref{fig-TASEPobc}) 
the (i) low-density (LD) phase, 
(ii) high-density (HD) phase, and (iii) maximal current (MC) phase; 
the rate limiting process for the three phases correspond to the rate 
constants $\alpha$, $\beta$ and $q$ respectively.

TASEP and its various extensions have been used extensively for modeling 
vehicular traffic and many similar systems 
\cite{chowdhury00,mahnke05,schadschneider11}. 
In the context of molecular motors, the ``particle'' represents a motor 
while the lattice represents its track, the lattice sites being the 
binding sites for the motor. A one-dimensional TASEP is adequate for 
generic models of molecular motor traffic \cite{chowdhury05a}. 
In the generic models of molecular motor traffic, one should also 
include the possibility of attachment of a motor at any vacant binding 
site and detachment from any occupied site of the lattice. Almost all 
the models of molecular motor traffic reported in the physics literature 
\cite{lipo1,lipo2,lipo3,lipo4,lipo5,lipo6,lipo7,frey,santen1,santen2,popkov1} 
are essentially extensions of TASEP that incorporates, in addition, a 
Langmuir-like kinetics of adsorption and desorption of the particles.
Analyzing their model, Parmeggiani et al.\cite{frey} 
demonstrated a novel phase of the system in which the low- and high-desity 
regions, separated by a domain wall, co-exist \cite{santen1,santen2}. 
This spatial organization can be interpreted as a traffic jam of molecular 
motors.

\section{\bf Sliders and rowers: generic models of filament alignment, bundling and contractility}
\label{sec-genericslidersrowers}

Flexible parallel filaments are known to form bundles in the presence of 
passive cross-linking molecules with two adhesive end groups 
\cite{kierfeld05}. 
Effects of cross-linking by active cross-linkers can be more dramatic. 
For example, molecular motors can align two filaments that are initially 
not parallel to each other; such ``zipping'' of two polar filaments is 
a cooperative effect of multiple motors  
\cite{uhde04,ziebert09}.
Alignment of more than two such filaments by the collective effort of 
multiple motors can lead to the formation of filament bundles. 
Suppose $\phi$ denotes the angle between the two polar filaments. Then, 
in the kinetic model of motor-induced alignments of these two filaments 
\cite{ziebert09}, 
the time-evolution of $\phi$ is described by a torque balance equation. 
Alignment of the two filaments involves not only rotation about the 
point of intersection of the two filaments, but also movement of the 
intersection point itself. Moreover, these two aspects of the dynamics 
are driven collectively by cross-linking molecular motors. Therefore, 
in addition to the equation for $\phi$, a separate equation accounts for 
the movement of the point of intersection (crossing point) of the two 
filaments. Furthermore, these two equations are coupled to the appropriate 
equations for the stepping kinetics of the motors that describe their 
diffusion, drift as well as their attachment to and detachment from the 
filaments. The mean time needed for such motor-driven alignment of two 
polar filaments is at least an order of magnitude faster than that 
required for alignment by passive cross-linkers.

Continuum models for the generic situation of relative sliding of 
active filaments by active linkers were developed by Kruse, 
J\"ulicher and collaborators 
\cite{kruse01,kruse02,kruse03a,kruse03b}. 
Suppose $c^{+}(x,t)$ and $c^{-}(x,t)$ denote the number densities 
(i.e., number per unit length) of the filaments whose plus ends 
are oriented in the $+X$ and $-X$ directions, respectively. 
The total number density $c(x,t)$ and the polarization density 
$p(x,t)$ of the filaments are given by $c(x,t) = c^{+}(x,t)+c^{-}(x,t)$ 
and $p(x,t) = c^{+}(x,t)-c^{-}(x,t)$, respectively. If all the 
filaments are oriented in the same direction, $p = \pm c$. 
The dynamics of the system is described by the two equations 
of continuity, with the respective source terms, $s$ and $s_p$ 
\cite{kruse03b} 
\begin{eqnarray}
\frac{\partial c}{\partial t} + \frac{\partial j}{\partial x} &=& s \nonumber \\
\frac{\partial p}{\partial t} + \frac{\partial j_p}{\partial x} &=& s_p
\end{eqnarray}
where the currents $j$ and $j_p$ arise from the change of relative 
positions and orientations of the filaments induced by the active 
cross linkers. These currents include, in addition to drift, also 
diffusion currents caused by the fluctuations in $c$ and $p$. 
In the absence of polymerization and depolymerization of the filaments, 
$s = 0 = s_p$. This model has been extended by Peter et al.\cite{peter08} 
incorporating effects of its coupling to a visco-elastic network.

\begin{figure}[htbp]
\begin{center}
\includegraphics[angle=90,width=0.65\columnwidth]{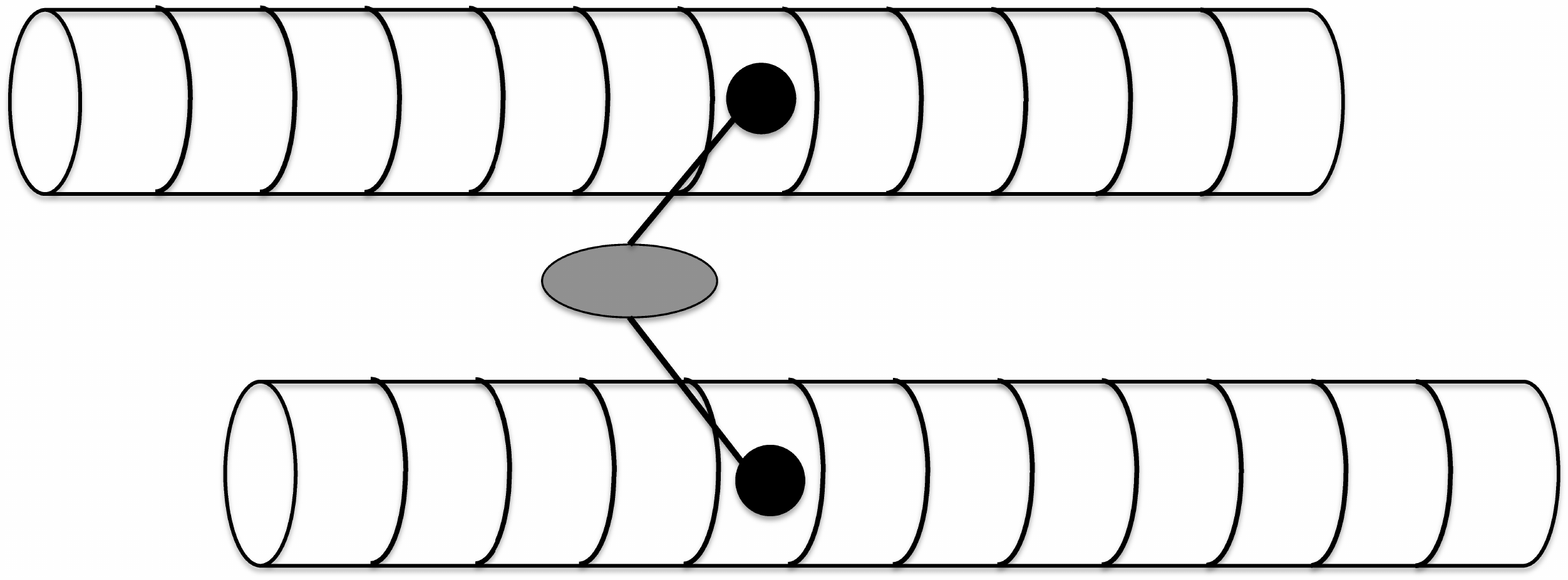}
\end{center}
\caption{Schematic illustration of the discrete model proposed by 
Kruse and Sekimoto \cite{kruse02} for motor-induced sliding of two 
polar cytoskeletal filaments. The cylinders represent the two polar 
filaments. The black discs represent the two head domains of a motor 
that are capable of attaching to the two motor-binding sites on the 
two filaments provided the two binding sites are closest neighbors 
of each other. Stepping of the motor on the filaments gives rise to 
the sliding of one filament with respect to the other.
(Adapted from ref.\cite{chowdhury05a}). 
}
\label{fig-kruseseki}
\end{figure}

Kruse and Sekimoto \cite{kruse02} proposed a discrete model for 
motor-induced relative sliding of two filamentary motor tracks 
(see Fig.\ref{fig-kruseseki}). Each 
of the two-headed motors is assumed to consist of two particles that 
are connected to a common neck and are capable of binding with two
filaments provided the two corresponding binding sites are closest 
neighbours. Each particle can move forward following a TASEP-like 
rule and every step of this type causes sliding of the two filaments 
by one single unit. The average relative velocity of the filaments
was found to be a non-monotonic function of the concentration of the 
motors \cite{kruse02}.

\begin{figure}[htbp]
\begin{center}
\includegraphics[angle=90,width=0.65\columnwidth]{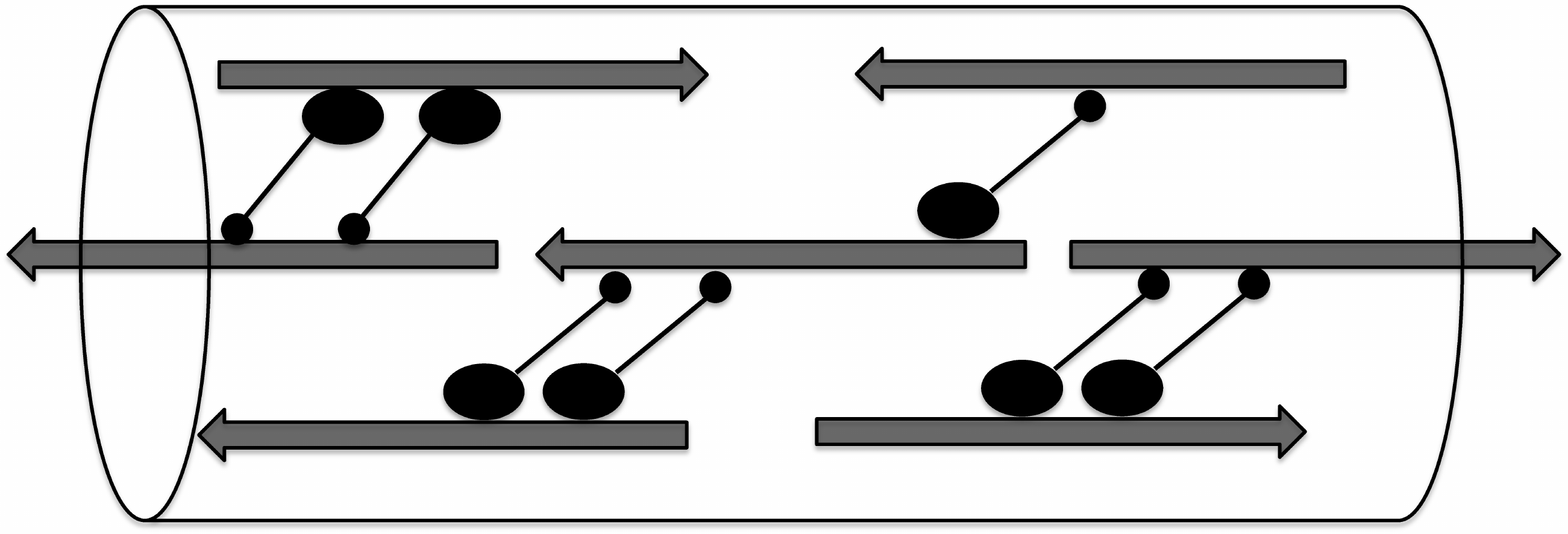}
\end{center}
\caption{A schematic depiction of the MT bundle crossbridged by 
sliders in the generic model developed by Zemel and Mogilner 
\cite{zemel09}. The arrow of each filament indicates the direction of the gliding of the motor on that filament. The cargo-binding domain at the end 
a motor is marked with a small disc while the motor domain is 
represented by an ellipse. (Adapted from ref.\cite{zemel09}). 
}
\label{fig-zemelmogil}
\end{figure}

Zemel and Mogilner \cite{zemel09} studied a generic discrete model of 
motor-driven sliding and bundling of filaments by computer simulations 
(see fig.\ref{fig-zemelmogil}). 
In this model the equation of motion of the $j$-th filament is assumed 
to have the form $\gamma \vec{V}_{j} = \vec{F}_j$ ($j=1,..,N$), where 
the actual form of the force $F$ depends on the force-velocity relations 
postulated for the filament-sliding motors. Suppose $M$ is the total 
number of potential crossbridges between pairs of filaments in the 
bundle. Then, $f_{b} = M_{b}/M$ is the fraction of active overlaps 
in the bundle if $M_{b}$ is the actual number of crossbridged formed 
by the sliding motors. $f_{b}$ and the number density $\rho_{b}$ of the 
motors (i.e., number of motors per unit length) bound to the filaments 
are the two important model parameters that, in turn, depend on the 
concentration of the motors. The model can simulate wide varieties of 
situations. For example, in general, a fraction $f_{L}$ of the filaments 
can have left-polarity while the remaining fraction $f_{R}=1-f_{L}$ 
has right-polarity. Both unipolar and bipolar sliding motors were 
modeled. Unipolar motors, which have motor domain at one end and the 
cargo-binding domain at the other, mimic kinesin-1 or cytoplasmic dynein. 
In contrast, bipolar motors have motor domains at both ends and mimic,  
for example, kinesin-5. Moreover, the results depend on the choice of 
the boundary conditions; both period and open boundary conditions were 
imposed in different sets of simulations. Although, for simplicity, 
it was formulated as a one-dimensional model, its simulation provided 
interesting insight into physics of filament sorting and pattern 
formation \cite{zemel09}.

The Brownian ratchet mechanism of free energy transduction by molecular 
motors for sliding of filaments have been discussed in the literature 
for many years \cite{vale90,oosawa95,oosawa08}. 
A Brownian ratchet model for the kinetics of ``cross-bridge'' of a 
single motor and a filament  was developed by Cordova et al. 
\cite{cordova92}. It can also account for $N$ ($>1$) cross-bridges 
formed by multiple motors simultaneously with the same filament. 
Physically motivated forms of the probabilities of attachment and 
detachment of the motors (i.e., probabilities for the formation 
and break-up of cross-bridges) were postulated. 
The sliding of the filament by the intact cross-bridges is taken 
into account by equating the velocity of the fiber with the 
collective velocity of the cross-bridges. The overdamped Langevin 
equation describing the one-dimensional dynamics of the filament 
includes not only the viscous drag and random Brownian force, but 
also the elastic restoring force of the stretched cross-bridge 
and the external load force. The force-velocity relation for this 
model can be obtained by solving this equation numerically.

\section{\bf Nano-pistons, nano-hooks and nano-springs: generic models}
\label{sec-genericpistonhookspring}

So far the cytoskeletal filaments have played a secondary role as tracks
for the respective motors whose mechanisms of energy transduction and
force generation were of primary interest to us. In this section we 
review the generic mechanisms of force generation by polymerizing 
and depolymerizing filaments which work, effectively, as nano-pistons 
and nano-hooks, respectively. Since the elastic stiffness of these 
filaments are crucial for force generation, a brief introduction to 
their elastic properties, particularly their stretching and bending 
stiffnesses, is given in the appendix \ref{appendix-elasticpol}.

Operation as nano-piston and nano-hook are not the only possible modes 
of motor-independent force generation by filamentous polymers. 
Spring-like actions of filamentous structures are known to drive fast 
motility of some biological systems \cite{mahadevan00}. 
One well known example of such biological spring is the vorticellid 
spasmoneme whose major protein component is spasmin 
\cite{mahadevan11,misra10}.
The sperm cell of the horse-shoe crab {\it Limulus polyphemus} also
utilizes the spring-like action of a coiled bundle, which consists
mainly of actin filaments, to penetrate into an egg for its fertilization 
\cite{shin03,shin07}. 
The cytoskeleton of the red blood cell has also been proposed to be a 
form of active nano-spring \cite{gov07}. 
However, in this review we'll neither review these mechanisms nor the 
mechanisms of force generation by a jet of oozing gel through a nozzle
\cite{wolgemuth02,wolgemuth05c,wolgemuth05d,jeon05,mignot07,kaiser08,zhang12c}
and that by shrinking gels resulting from crosslinking and bundling of 
filaments \cite{sun10}. The effects of the distinct structural and 
kinetic features of the different types of polymerizing filaments on 
the force generation will be discussed in section 
\ref{sec-specificpistonhookspring}.

\subsection{\bf Push of polymerization: generic model of a nano-piston}

\subsubsection{\bf Phenomenological linear response theory for chemo-mechanical nano-piston: modes of operation}

\begin{figure}[htbp]
\begin{center}
\includegraphics[angle=-90,width=0.45\columnwidth]{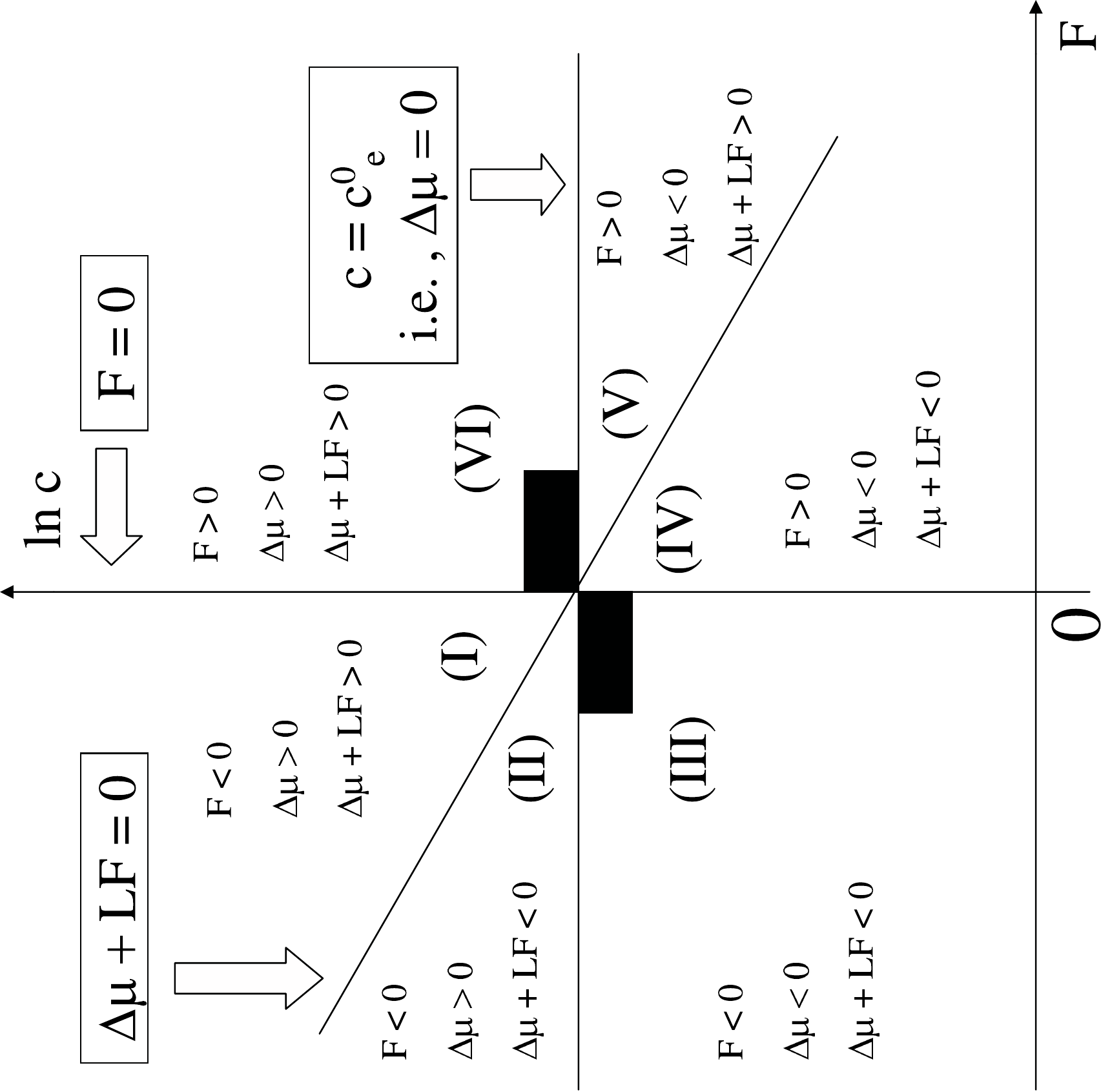}
\end{center}
\caption{Various modes of operation of a polymerizing-depolymerizing 
filament, in the linear response theory, depicted schematically in  
the 2d plane spanned by the two corresponding generalized forces.
Adapted from ref.\cite{hill81a} 
}
\label{fig-lrtpiston}
\end{figure}

Suppose $c$ denotes the concentration of the monomeric subunits of the 
filamentous polymer in solution and $F$ denotes the load force. Then, 
$F$ and $ln c$ can be treated as the two relevant ``generalized forces'' 
for a phenomenological linear response theory for nano-pistons 
\cite{hill81a}. 
The possible modes of operation of the nano-piston and the corresponding 
parameters regimes on the 2d plane spanned by these two generalized forces 
\cite{hill81a} are shown schematically in fig.\ref{fig-lrtpiston}.

\subsubsection{\bf Stochastic kinetics of chemo-mechanical nano-piston}

Suppose the rates of attachment (on-rate) and detachment (off-rate) 
of the $\alpha-\beta$ tubulin dimers to the MT are denoted by 
$k_{on}(0)$ and $k_{off}(0)$, respectively, in the absence of load 
force. Let $\Delta G$ be the free energy difference between the on 
and off states. Obviously, $k_{on}/k_{off} = exp(\Delta G/k_BT)$. 
However, in the presence of a load force opposing the filament growth, 
we get
\begin{equation}
\frac{k_{on}(F)}{k_{off}(F)} = e^{(\Delta G-F{\ell})/k_BT} = \frac{k_{on}(0)}{k_{off}(0)} e^{-F{\ell}/k_BT}
\end{equation}
based on purely thermodynamic arguments (see, for example, 
ref.\cite{dogterom02}) where ${\ell}$ is the monomer length.
Hence, the force-velocity relation for the nano-piston is given by 
\begin{equation}
V(F) = {\ell}[k_{on} e^{-F{\ell}\delta/k_BT} - k_{off} e^{-F{\ell}(1-\delta)/k_BT}]
\end{equation}
where the parameter $0 < \delta < 1$ accounts for the load distribution 
\cite{kolomeisky01}. Therefore, the corresponding stall force is 
\cite{peskin93a}
\begin{equation}
F_{s} = k_BT~ln\biggl(\frac{k_{on}(F)}{k_{off}(F)}\biggr)  = \Delta G/{\ell}. 
\end{equation} 
This final result is not surprising because the entire chemical free 
energy input is spent in stalling the nano-piston.

A statistical mechanical treatment that takes into account fluctuations, 
was formulated by Peskin et al.\cite{peskin93a} in terms of a Fokker-Planck 
equation. Let us put the origin of the coordinate system on the tip of 
the polymerizing filament. We represent the position of the membrane 
by a ``particle''; thus, the position of the particle ($x$) is also the 
gap between the tip of the MT and the membrane. We define the density 
$c(x,t)$ such that $\int_{x_1}^{x_2} c(x,t) dx$ is the number of filaments 
(in an ensemble) that has $x$ value lying between $x_1$ and $x_2$ at time 
$t$. The FP equations for $c(x,t)$ are \cite{peskin93a} 
\begin{eqnarray}
\frac{\partial c(x,t)}{\partial t} &=& D \biggl(\frac{\partial^2 c(x,t)}{\partial x^2} \biggr) + \biggl(\frac{DF}{K_BT}\biggr)\biggl(\frac{\partial c(x,t)}{\partial x} \biggr)\nonumber \\ 
&+& \alpha[c(x+{\ell},t)-H(x-{\ell})c(x,t)] \nonumber \\
&+& \beta[H(x-{\ell})x(x-{\ell},t)-c(x,t)] \nonumber \\
\end{eqnarray}
where $\alpha = k_{on} \times$ the concentration of $\alpha-\beta$ dimers 
in the solution, $\beta = k_{off}$ and $H(x-{\ell})$ is the Heaviside step 
function. In the steady-state the force-velocity relation is obtained from 
\begin{equation}
V(F) = \biggl[\frac{\alpha\int_{\ell}^{\infty} c(x) dx - \beta \int_{0}^{\infty} c(x) dx}{\int_{0}^{\infty} c(x) dx}\biggr]{\ell} 
\end{equation}

\subsection{\bf Pull of de-polymerization: generic model of a nano-hook}

The force generated by a depolymerizing filament was originally 
developed in the context of chromosome segregation. Therefore, we 
defer a detailed discussion to the section \ref{sec-mitosis}. 
In this subsection we study force generation by depolymerizing 
filaments in a similar in-vitro experiment \cite{peskin95b}. 
In this experiment a micron-size bead diffuses on a filament that 
is simultaneously depolymerizing. 

The filament is represented by a one-dimensional lattice of lattice 
constant $\ell$. The right end of the lattice corresponds to the 
depolymerizing tip. Therefore, the lattice shortens, from the right to 
left, by a length ${\ell}$, at the rate $\beta$ per unit time. 
Unless located on the tip of the filament, the bead hops towards right 
and left with rates $\gamma_+$ and $\gamma_-$, respectively, per unit 
time. When the bead is located exactly on the tip of the filament and hops 
to the left, it has a probability $p$ of ``peeling off'' the terminal 
subunit from the filament. Solving the kinetic equations in the 
steady-state, one gets the force-velocity relation \cite{peskin95} 
\begin{equation}
V(F) = \gamma e^{-f/2}\biggl[\frac{p\gamma(e^{f/2}-e^{-f/2})+\beta}{\gamma(e^{f/2}-pe^{-f/2})+\beta}\biggr]{\ell}
\end{equation} 
where $\gamma=\gamma_+=\gamma_-$ and $f = F{\ell}/k_BT$.

\section{\bf Exporters and importers of macromolecules: generic models}
\label{sec-genericexim}

The cell membrane separates the interior of the cell from its surroundings.
It is essential for maintaining the integrity of the cell. At the same
time, the cell cannot survive without exchange of matter and energy with
its surroundings. Therefore, the cell membrane must be capable of
performing a remarkable task: on the one hand, it must allow export/import
of molecules across itself that are necessary for sustaining the life of
the cell and, on the other, it should prevent all those transport processes
which can threaten the survival of the cell. The same properties are also
shared by internal membranes of eukaryotic cells which maintain the
integrity of various compartments that perform specialized functions. 

Macromolecules to be translocated across the pore may be {\it hydrophobic} 
or may be electrically charged. Therefore, it is not surprising if it 
encounters an energy barrier while trying to translocate across the pore. 
However, what makes macromolecule translocation even more interesting from 
statistical physics perspective is that the macromolecule also encounters 
an {\it entropic} barrier \cite{muthukumar07}. 
The number of allowed conformations of the
macromolecular chain, and hence its entropy, is drastically reduced when 
it translocates across a narrow pore. Therefore, in general, the barrier 
encountered by the translocating macromolecular chain is a {\it free 
energy} barrier. Naturally, in order to overcome the free energy barrier 
against its passage across a membrane a macromolecule, in general, 
requires energy supply which is provided, most often, by the action of 
the corresponding translocation motor \cite{palmen94,palmen97}.

In this section we review the generic models of ``translocators'' 
\cite{baumgartner08}, i.e., motors that translocate linear polymers 
across a narrow passage on a surface. 
The process of macromolecule translocation can
be divided into two steps. Step I: The tip of the macromolecule just
enters the pore; step II: the entire length of the chain crosses the pore.
The first process is analogous to putting the tip of a thread through the
hole of a needle whereas the second is the analogue of pulling a length
$L$ of that thread through the same hole after successful insertion of
the tip. We focus exclusively on the latter aspect of the phenomenon. 

For simplicity, let us model the translocating linear polymer as a {\it rigid 
rod} on which there are equispaced binding sites for a class of large 
molecules called {\it chaperonins} \cite{simon92}. 
The separation between the successive chaperonin binding sites is $\ell$. 
We model the membrane as a flat thin rigid wall with a ``pore'' in it. 
Initially, the entire length of the polymer is on 
one side of the wall and is oriented perpendicular to the surface of the 
wall with one of its tips located just on the ``pore''. The  
size of the pore is just enough for the rod to pass through it. We designate 
this side of the wall as the initial side and the opposite side as the target 
side of the wall. 

The model can be formulated as a one dimensional diffusion of the rod 
along an axis, passing through the pore, that is perpendicular to the wall. 
As soon as a binding site crosses the wall and enters the target side, a 
chaperonin binds with this site {\it irreversibly}. Since the chaperonin 
cannot pass through the pore, the rod essentially works as a Brownian 
ratchet. It is then straightforward to see that the average velocity of 
translocation of the rod would be $<V> = 2 D/{\ell}$ where $D$ is the 
diffusion constant. Note that in the absence of chaperonin binding a 
rigid rod of total length $L$ would take a mean time $t_d = L^2/(2D)$ 
to cross the pore by pure diffusive motion. When the chaperonin binding 
takes place irreversibly, it takes an average time $t_r = t_d/M$ where 
$M = L/{\ell}$ is the total number of chaperonin-binding sites.

Next, let us consider a slightly more general scenario. Suppose a 
binding site does not get occupied by a chaperonin as soon as it enters the 
target side of the wall and that the chaperonin binding is not totally 
irreversible. Instead both binding and unbinding of chaperonins continue 
at the rates $\omega_{a}$ and $\omega_{d}$, respectively. In this case 
the average velocity of translocation is \cite{simon92,peskin93a}
\begin{equation}
<V> = \biggl(\frac{2D}{\ell}\biggr)\biggl(\frac{\omega_{a}}{\omega_{a}+2\omega_{d}}\biggr)
\label{eq-SPO}
\end{equation}
Thus, in this case, the average speed of translocation depends on the 
ratio of the rates of binding and unbinding of the chaperonins. 
Note that if $\omega_{d}=0$ the eqn.(\ref{eq-SPO}) reduces to 
$<V> = 2D/{\ell}$, the result quoted above in case of irreversible 
binding of the chaperonins.
A load force $F_{load}$ can be taken into account and the stall force 
$F_{stall}$ turns out to be \cite{peskin93a} 
\begin{equation}
F_{stall} = \frac{k_BT}{\ell}~ ln\biggl(1+\frac{\omega_a}{\omega_d}\biggr)
\end{equation}

The treatment summarized above captures the rectification of thermal 
fluctuations by the chaperonins on the translocating polymer. However, 
this did not take into account the details of the allowed configurations 
of the chaperonins. By incorporating these details into an extended model, 
Zandi et al.\cite{zandi03} showed the existence of an {\it entropic} force, 
also called ``Langmuir force'', which speeds up translocation beyond the 
value implied by equation (\ref{eq-SPO}). 

Ambjornsson and Metzler \cite{ambor04} carried out a detailed systematic 
analysis to clarify different possible regimes of chain translocation 
in the presence of chaperonins. These regimes can be distinguished by 
the relative magnitudes of three different time scales in the problem. 
$\tau_{d}$ is the time required for the chain to diffuse a distance 
${\ell}$ whereas $\tau_{occ}$ and $\tau_{unocc}$ are the durations 
for which a binding site remains occupied and unoccupied, respectively.
In deriving this analytical expression, Simon et al.\cite{simon92} 
assumed that the binding-unbinding kinetics is very fast compared to the 
diffusion of the rod, so that the bound cheparonins achieved equilibrium 
practically instantaneously as soon as a new cheperonin-binding site 
entered te target side of the wall. This approximation, however, need 
not be valid in many real situations \cite{elston00b}. 
The original model of Simon et al.\cite{simon92} was extended by Sung and Park 
\cite{sung96} by incorporating the effects of the flexibility of the polymer 
chain. The native conformations of the translocating polymers get altered 
significantly thereby losing entropy and erecting a free energy barrier 
against translocation. 

We shall see in section \ref{sec-specificexim} how the generic models 
summarized above are extended to capture the distinct characteristic 
features of the different specific cases.

\section{\bf Motoring along templates: generic models of template-directed polymerization}
\label{sec-genericpolyribo}

Biological information is chemically encoded in the sequence of the
species of the monomeric subunits of a class of linear polymers that
play crucial roles in sustaining and propagating ``life''. Nature also
designed wonderful machineries for polymerizing such macromolecules,
step by step adding one monomer at each step, using another existing
biopolymer as the corresponding template. Compared to the enzymatic 
activities of other enzymes, that of the machines of template-directed 
polymerization is quite unique. Since the sequence on the template is, 
in general, heterogeneous, the substrate selected for incorporation 
as monomers to the nascent polymers belong to different molecular 
species (i.e., 4 possible NTPs in case of polynucleotide polymerase or 
20 possible amino acids in case of ribosome). Yet, the same machine 
catalyzes the incorporation of these different at the respective 
positions on its template. Obviously, the template plays a more active 
role than merely specifying the nature of the substrate; it must also 
cooperate with the machine to perform its catalytic function with such 
diverse species of substrates.

In this section we summarize
the recent progress in understanding the common generic features of
the structural design of these machines and stochastic kinetics of the
polymerization processes. Later, in sections \ref{sec-specificpoly},
and \ref{sec-specificribo},
we consider specific examples of such machines and the unique distinct
features of their structural and kinetic properties.

\subsection{\bf Common features of template-directed polymerization}

In spite of the differences between their constituent monomers as well
as in their primary, secondary and tertiary structures, nucleic acids
and proteins share some common features in the birth and maturation:\\
(i) Both nucleic acids and proteins are made from a limited number of
different species of monomeric building blocks. \\
(ii) The sequence of the monomeric subunits to be used for synthesis
are directed by the corresponding template. \\
(iii) These polymers are elongated, step-by-step, during their birth by
successive addition of monomers, one at a time. \\
(iv) Synthesis of each chain (polynucleotide and polypeptide) begins
and ends when the machine encounters well-defined start and stop signals
on the template strand.\\
(v) The free energy released by each event of the phosphate ester hydrolysis,
that elongates the polynucleotide by one subunit, serves as the input
energy for driving the mechanical movements of the corresponding
polymerase by one step on its track. Moreover, as we'll discuss in
detail later, GTP molecules are hydrolyzed during the process of
polymerization of polypeptides. Therefore, the machines for template-
directed polymerization are also regarded as molecular motors; these
use the template itself also as the track for their translocation. \\
(vi) The main stages in the process of template-directed polymerization
are common:

(a) {\it initiation}: The start signal is chemically encoded on the
template. This stage is completed when the machinery gets stabilized
against dissociation from the template. \\
(b) {\it elongation}: During this stage, the nascent product polymer 
gets elongated by the addition of monomers.  \\
(c) {\it termination}: Normally, the process of synthesis is terminated,
and the newly polymerized full length product molecule is released,
when the machine encounters the {\it terminator} (or, stop) sequence
on the template. Throughout our discussion we'll focus mostly on the
process of {\it elongation}.\\
(vii) The primary product of the synthesis, namely, polynucleotide or
polypeptide, often requires ``processing'' whereby the modified
product matures into functional nucleic acid or protein, respectively.\\

\noindent$\bullet${\bf Fidelity of template-directed polymerization: proofreading and editing}

For template-directed polymerization, the selection of the correct
molecular species of subunit requires a mechanism of ``molecular
recognition''. However, if this mechanism is not perfect, errors can
occur.
The typical proability of the errors in the final product is 
\cite{sharma12bprl} about $1$
(i) in $10^{3}$ polymerized amino acids, in case of protein synthesis,
(ii) in $10^4$ polymerized nucleotides in case of mRNA syntheis and
(iii) in $10^9$ polymerized nucleotides in case of replication of DNA.
Purely thermodynamic discrimination of different species of nucleotide
monomers cannot account for such high fidelity of polymerization.
Therefore, a normal living cell has mechanisms of ``proofreading''
and ``editing'' so as to correct errors.

\noindent$\bullet${\bf A polymerase is a ``tape-copying'' Turing machine}

Polymerization of a nucleic acid strand by a polymerase can be analyzed
from the perspective of information processing \cite{mooney98}.
From this perspective, a polymerase displays many similarities with 
the Turing machine, an idealized computing device that was introduced 
by Alan Turing in 1936 \cite{turing36}. A Turing machine reads input 
information from a digital tape and produces an output by using its 
rules that are based on digital logic. A polymerase also reads the 
input information from the template and the output of its ``computation'' 
is another digital tape. Therefore, a polymerase is analogous to a 
``tape-copying Turing machine'' that would polymeize its output tape, 
instead of writing its output on a pre-syntesized tape \cite{bennett82}. 
However, in contrast to the digital logic of a Turing machine, logical 
decisions of a polymerase are based on equilibria of various conformations 
and competing rates of transitions among these conformations. These 
``decisions'' of a polymerase are regulated by intrinsic as well as 
extrinsic input informations \cite{mooney98}. Dissipationless 
computation by a Turing machine would correspond to an error-free 
polymerization by a polymerase which is possible only at a vanishingly 
small speed, i.e., in the reversible limit \cite{bennett79}.

\subsection{\bf A generic minimal model of the kinetics of elongation by a single machine}

A minimal model of the mechano-chemical kinetics during the
template-directed polymerization must incorporate at least two
facts: (i) substrate selection as well as the possibility of
substrate rejection, and (ii) mechanical stepping of the machine
on the template should accompany elongation of the nascent polymer.
To keep the model as simple as possible, we represent the template
as a one-dimensional lattice of total length $L$ where the sites
are labelled by the integer index $j$ ($1 \leq j \leq L$).
We also assume sequence-homogeneity of the template and, therefore,
the rate constants of the kinetic processes are independent of the
index $j$ of the lattice site.

At an arbitrary instant of time during the elongation, the nascent
polymer of length $n$ is denoted by the symbol $P_{n}$. The machine
is denoted by the symbol $M$. A minimal generic kinetic scheme would
be as follows:
\begin{eqnarray}
M_{j}+P_{n}+S \rightleftharpoons I_{1} \rightarrow \mathop{I_{2}}_{\downarrow} \rightarrow M_{j+1}+P_{n+1} \nonumber \\
\label{eq-genericchem}
\end{eqnarray}
The scheme (\ref{eq-genericchem}) consists of three sub-steps.
The forward stepping of the machine can take place either in the
first substep, i.e., with the arrival of the substrate, or in the
third step, i.e., with the completion of elongation of the nascent
polymer by one monomer. In the special situations where the 
branched pathway is traversed rarely, the average speed of the motor 
would be identical to the average speed of the corresponding elongation
reaction which would obey the MM equation. We do not need to elaborate 
these calculation further here.

\subsection{\bf A generic minimal model of simultaneous polymerization by many machines}

\begin{figure}[htbp]
\begin{center}
\includegraphics[angle=90,width=0.45\columnwidth]{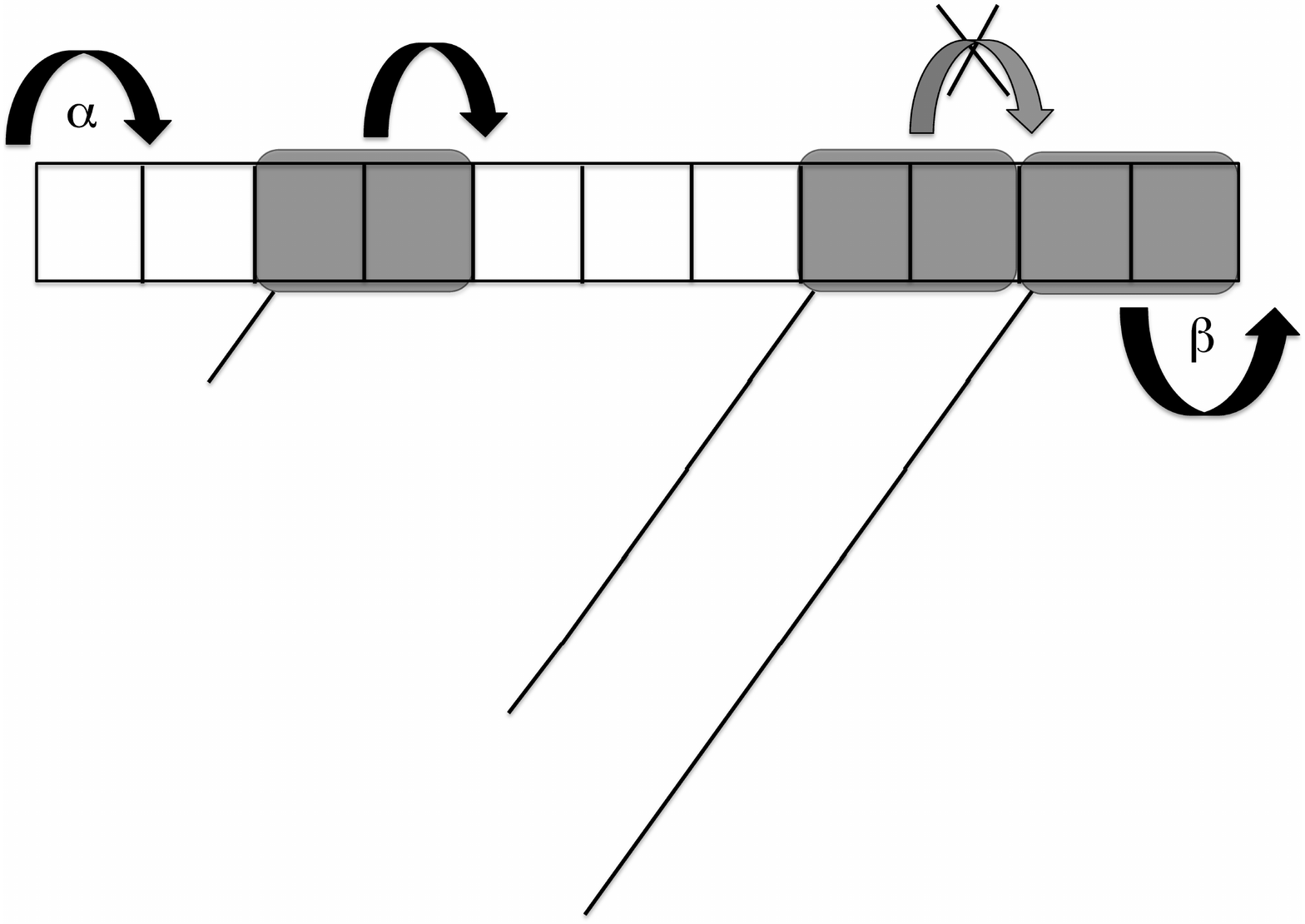}
\end{center}
\caption{Traffic-like situation that arises when several machines for 
template-directed polymerization move like motors simultaneously on 
the same template strand. Each motor polymerizes a distinct copy of 
the same polymer. Each unfilled rectangle depicts a subunit of the 
template and corresponds to a specific subunit of the elongating polymer. 
Each motor is represented by a grey rectangle that covers $\ell$ 
($\ell=2$ in this figure) subunits on the template.
}
\label{fig-TASEPtemPOL}
\end{figure}

In a living cell most often the machines for template-directed
polymerization do not work in isolation. A template serves as the
track simultaneously for several machines. Therefore, discovering
the ``traffic-rules'' for these machines is essential for
understanding the collective effects. In this section, however,
we consider only a special type of collective phenomena which
addresses the following question:
when many machines move on the same track in the same direction,
(as shown schematically in fig.\ref{fig-TASEPtemPOL})
can a traffic-jam like situation arise? What are the causes and
consequences of such traffic jams?

In order to take into account the steric hindrance of one machine
against another, we represent each by a rigid rod of length ${\ell}$
in the unit of the lattice constant. Therefore, each machine can
cover ${\ell}$ lattice sites simultaneously, but moves forward
by one lattice site at a time. No lattice site can be covered by
more than one machine at a time. The individual mechano-chemistry
of each machine gives rise to an effective rate $q$ of ``hopping''
forward. But, a given machine can step forward if, and only if,
the target site is not already covered by another machine.  For
convenience, the lattice is assumed to have length $L+{\ell}-1$
of which only the first ${\ell}$ sites constitute the template.
Moreover, a new machine can initiate polymerization by occupying
the first ${\ell}$ sites of the lattice, with a rate $\alpha$,
provide all these ${\ell}$ sites are not covered by any other
machine. Similarly, if the last ${\ell}$ sites are occupied by
a machine, it terminates the polymerization by ``hopping out''
of the system with a rate $\beta$. Thus, the traffic-like
collective movement of many machines simultaneously on a single
template is a TASEP of hard rods \cite{chou11}. Indeed, to my 
knowledge, this was the first application of TASEP to a biological 
system \cite{macdonald68,macdonald69}.

\section{\bf Rotary motors: generic models} 
\label{sec-genericrotary}

Some of the most important rotary motors in a living cell are driven by 
ion-motive force (IMF). The motor is embedded, at least partly, in a 
membrane. The concentrations of the relevant ion (normally, either 
hydrogen ion H$^+$, i.e., a proton, or sodium ion Na$^+$) that drives 
this rotary motor, is higher on one side of the membrane than on the 
opposite side. A phenomenological theory, along the same lines as the 
linear response theory developed earlier for linear motors 
\cite{parmeggiani99}, has been attempted \cite{pikin01}. However, 
the choice of the generalized forces and the conjugate generalized 
currents is more subtle in this case because of the full vectorial 
character of the non-chemical variables.

Berry and Berg \cite{berry99} developed the simplest kinetic model for 
IMF-driven rotary motors. It does not involve any explicit structural 
consideration. The three kinetic states are labelled by the letters 
E, A and B. In the kinetic scheme shown below
\begin{equation}
E \mathop{\rightleftharpoons}^{k_{f1}}_{k_{b1}} A \mathop{\rightleftharpoons}^{k_{f2}}_{k_{b2}} B \mathop{\rightleftharpoons}^{k_{f3}}_{k_{b3}} E 
\label{eq-BerryBerg}
\end{equation}
the forward transitions $E \to A$ and $B \to E$ indicate, respectively, 
entry of an ion from the high-concentration side and its exit to the 
low-concentration side of the membrane. All the other events, including 
rotation of the motor, which take place during the transit of the ion 
inside the motor are incorporated in the single transition $A \to B$. 
Although the dominant pathway would be $E \to A \to B \to E$, several 
other possibilities also arise because of the reversibility of each of 
the transitions. The rate constants are assumed to satisfy the detailed 
balance conditions appropriately and depend on both the IMF and the work 
done against the load torque. The three master equations satisfied by the 
probabilities $P_E, P_A, P_B$ for the states E, A, B, respectively, can 
be solved in the steady state and the angular velocity $\omega$ of the 
motor can be obtained from $\omega = (k_{f2} P_A - k_{b2} P_B) \phi$ 
where $\phi$ is the step size, i.e., angle by which the motor rotates in 
one cycle.

\begin{figure}[htbp]
\begin{center}
\includegraphics[angle=90,width=0.45\columnwidth]{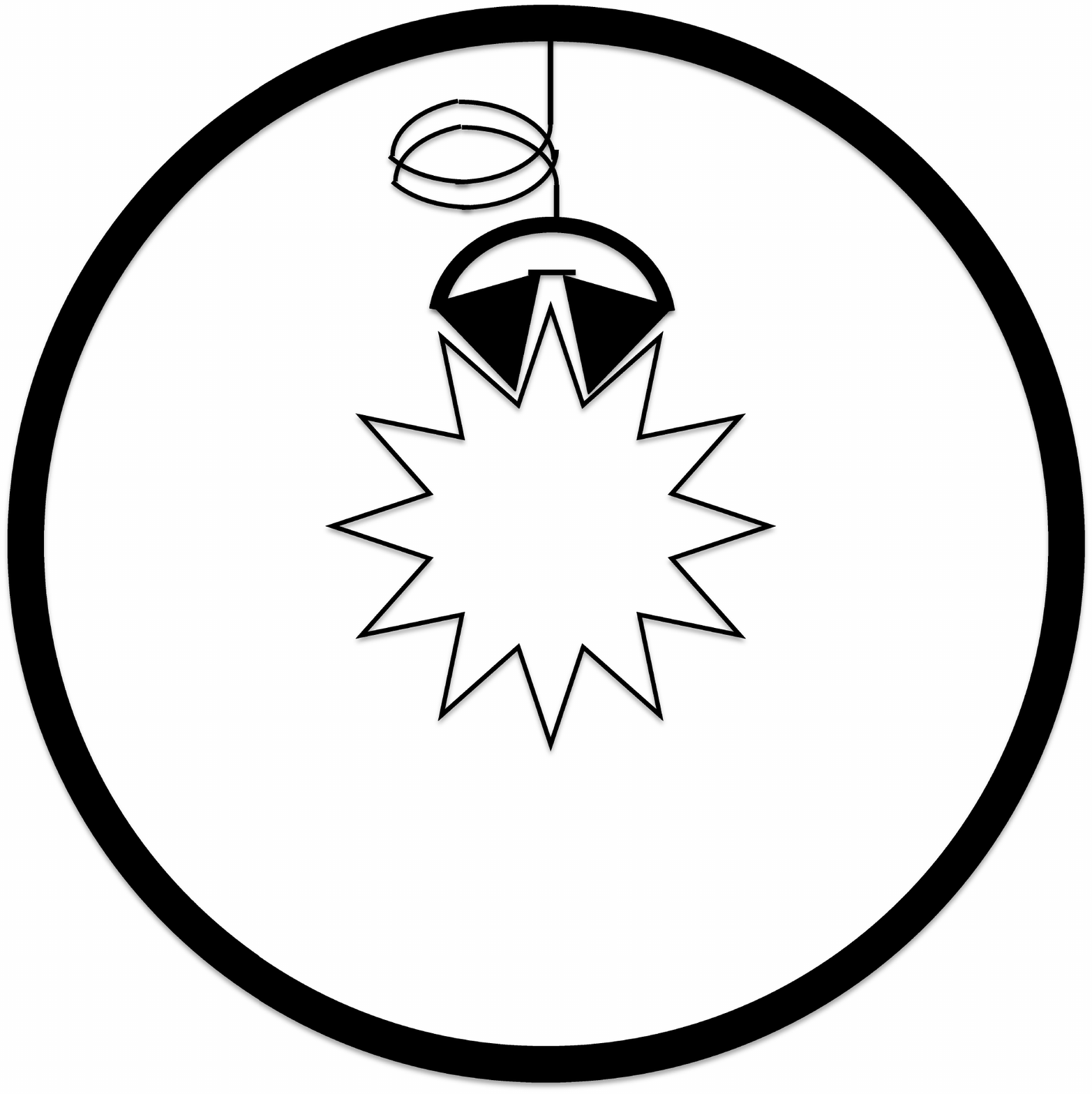}
\end{center}
\caption{A schematic representation of a generic model of a rotary motor 
based on {\it stator-rotor cross-bridge}. The black outer circle is the 
stator while the 12-teeth star represents the rotor. The crossbridge is 
established when the assembly consisting of two inverted triangles 
extends from the stator and attaches with the rotor. The elastic strain 
in the crossbridge is captured by the extension or compression of the 
spring. The affinity of this element for the rotor depends on the ionic 
charges on the rotot units. (This figure is inspired by fig.2 of 
ref.\cite{khan88}).
}
\label{fig-RotGen}
\end{figure}

In order to explain the mechanism of energy transduction some essential 
structural features, which are shared by wide varieties of rotary motors, 
should be incorporated in the model without sacrificing its generic 
character.  At least two different classes of such generic models for 
the IMF-driven rotary motors can be conceived \cite{blair95}. \\
{\it Stator-rotor ``cross-bridge'' models:} 
As emphasized by the schematic depiction in fig.\ref{fig-RotGen},
these models have similarities 
with ``motor-filament'' crossbridge system that we have discussed in section 
\ref{sec-genericslidersrowers}. In these generic models the ions are the 
analogs of ATP, while the stator and the rotor are the analogs of a motor 
and the filament, respectively. In such a generic scenario, the binding and 
release of the ions to the ``flexible'' stator not only alter its affinity 
for the rotor (and hence stator-rotor binding), but also trigger 
conformational changes (power stroke) that generate the torque needed 
for the angular displacement of the rotor.\\
{\it Non-contact stator-rotor interaction models:} These models 
involve neither stator-rotor binding nor significant conformational 
changes of the stator. Instead, rotation is caused by the combined effects 
of (i) the constraints imposed by the hydrophobicity of the membrane, 
in which the motor is embedded, and (ii) electrostatic interactions of 
the various charged components of the motor with the ions transiting 
through it. In some of the models, each ion hops from the stator onto 
the rotor and gets a ride for a part of its journey through the motor 
before being offloaded. But, in the other models an ion remains confined 
within the stator region throughout its transit. Since some concrete 
examples of this mechanism will be discussed later in this review in 
sections \ref{sec-specificrotary1} and \ref{sec-specificrotary2}, we 
do not discuss it further here.
\\

\newpage 

\centerline{\Huge {Part II:}}
\centerline{\Huge {Kinetic models}}
\centerline{\Huge {of}}
\centerline{\Huge {specific motors}} 
\vspace{1cm}
{\it ``The beauty of Nature lies in detail: the message, in generality.
Optimal appreciation demands both''}- {\bf Stephen J. Gould, in ref. \cite{gould89}}. 

\newpage

\section{\bf Cargo transport by cytoskeletal motors: specific examples of porters} 
\label{sec-specificporters}

In the generic models of porters that we discussed in section 
\ref{sec-genericporters}, the track was assumed to provide only 
equidistant binding sites for the motor. However, in reality, 
a MT plays a much more subtle role which have started emerging 
from the X-ray studies of the motor-MT complexes \cite{kikkawa08}. 
Moreover, the generic models neither take into account the fact 
that MT consists of 13 protofilaments and F-actin is effectively 
a double helix (see appendix \ref{app-cytoskeleton}); 
while ``lane changing'' on the former cannot be  ruled out 
\cite{brunnbauer12}, 
tilting and twirling on the latter seems possible \cite{warshaw12}. 
Motors, particularly kinesins, also play key roles as regulators 
of the dynamics and organization of the MT filaments \cite{daire11}.
Furthermore, most of the motors normally consist of more than one 
domain or subunit whose coordination is essential for the motility 
and other specific operations of the motor \cite{goodsell00}. 
Besides, the motor molecules are not always in the transport 
competent ``active'' state; the mechanisms of autoinhibition and 
activation by binding to cargo and MT tracks shed light on their 
regulation and control \cite{verhey09}. 
Although head-to-tail interaction is one way of such regulation 
\cite{verhey09}, tail-independent mechanism has also been reported 
\cite{xu12}.
Modeling the determinants of the directionality and processivity of the 
different families of the same superfamily \cite{endow99} 
cannot be done satisfactorily without incorporating the specific 
distinct features of the different families of motors. 

In this section we focus on the operational mechanism of a single motor
taking into account its architectural design, the energetics of its interaction
with the track and that with the fuel molecule, its mechano-chemical 
kinetic pathways as well as the mechanisms of their regulation and control
\cite{howard01a,cross00a,spudich94,howard97,vale99,vale00a,vale03a,goldstein01,schliwa03b,tyreman03,mallik06,amos08,reilein01,chowdhury08a}.
Intracellular motor-driven transport 
\cite{cross01}
is a crucially important process not only for the maintenance of the 
cell, but also for its morphogenesis. In fact, porters supply the raw 
materials needed for the formation of long tips of many specialized 
cells, e.g., neurons in animal cells, cilia in algae, fungal hyphae, 
etc. In this section, and a few following sections, we review not only 
the distinct features of single molecules of specific families of 
molecular motors but also coordination, cooperation and competition of 
motors of the same family and different families in intracellular transport.

\noindent$\bullet${\bf Common features of architectural designs and mechano-chemical kinetics}

All the porters share common functional features- these generate force 
and walk along a filamentous track. Interestingly, their functional 
commonality is related to some common features of their structural 
design and mechano-chemical kinetics. 

All the cytoskeletal motor proteins have a {\it head} domain (or, 
motor domain) that contains a site for ATP binding (and hydrolysis); 
the ATPase site serves, effectively, as the ``engine'' of the motor 
where the chemical fuel is ``burnt''. The track-binding sites of myosin 
and kinesin are also located on the head domain, but are distinct from 
the ATP-binding site. In contrast, the track-binding sites of dynein 
are located on the ``antennae''-like extensions of the head domains.  
The identity of the ligand bound to 
the ATPase site (i.e., whether ATP of products of its hydrolysis) 
regulates the affinity of the motor for its track. The porters walk on 
their ``heads'' carrying cargo that are bound to their tail domains 
located at the distal end of the stalk.

The head
domain of the kinesins is the smallest (about 350 amino acids), that of
myosins is of intermediate size (about 800 amino acids) whereas the head
of dyneins is very large (more than 4000 amino acids) \cite{marx05}.
The ``identity card'' for members of a superfamily is the sequence
of amino acids in the motor domain. The members of a given superfamily
exhibit a very high level of ``sequence homology'' in their motor domain.
But, the amino acid sequence as well as the size, etc. of the other
domains differ widely from one member to another of the same superfamily.
The tail domain exhibits much more diversity than the head domain
because of the necessity that the same motor should be able to recognize
(and pick up) wide varieties of cargoes.

According to the widely accepted nomenclature, myosins 
\cite{hartman11,hartman12} 
are classified into families bearing numerical (roman) suffixes (I, II, 
..., etc.) \cite{berg01}. According to the latest standardized nomenclature of kinesins 
\cite{endow10}
the name of each family begins with the word ``kinesin'' followed by an 
arabic number ($1$, $2$, etc.) 
\cite{lawrence04}. Moreover,
large subfamilies are assigned an additional letter ($A$, $B$, etc.)
appended to the familty name. For example, kinesin-14A and kinesin-14B
refer to two distinct subfamilies both of which belong to the family
kinesin-14. Myosins and kinesins have a common ancestor called G protein 
\cite{vale96y,kull98}.  Dyneins can be broadly divided into {\it two} 
major classes: (i) cytoplasmic dynein, and (ii) axonemal dynein. 

In an automobile, the site that processes the chemical fuel must be 
linked through ``intermediate components'' to the site that ultimately 
generates the motion. In the automobile, the breakdown of the chemical 
fuel in the engine is coupled to the stroking of a piston, which in 
turn is linked through the crankshaft and transmission to the turning 
of the wheels. Similarly, a motor molecule has to ``sense'' a 
nucleotide-dependent small conformational change resulting from ATP 
hydrolysis and amplify it to generate the power stroke. 
The actual moving element of a myosin and kinesin is a lever-like 
component. In case of myosin, this component is called a lever arm 
whereas the neck-linker of a kinesin serves essentially a similar role. 

The typical duty ratios of kinesins and cytoplasmic dynein
are at least $1/2$ whereas that of conventional myosin can be
as small as $0.01$. Unlike conventional myosin-II, which has a very
low duty ratio ($ \leq 0.05$), unconventional myosin-V and myosins-VI
have quite high ($0.7-0.8$) duty ratios. Myosin-X has moderate duty
ratio.

\noindent$\bullet${\bf {\it In-vitro} Motility assays}

There are two geometries used for {\it in-vitro} motility assays
\cite{scholey93}:
(i) the {\it gliding} assay and (ii) the {\it bead}
assay.  In the gliding assay, the motors themselves are fixed to a
substrate and the filaments are observed (under an optical microscope)
as they glide along the motor-coated surface. In the bead assay, the
filaments are fixed to a substrate. Small plastic or glass beads, whose
diameters are typically of the order of $1 \mu m$, are coated with the
motors. These motors move along the fixed filaments carrying the bead
as their cargo. The movements of the beads are recorded optically.

\subsection{\bf Processive dimeric myosins} 
\label{sec-proc2myosins}

For the historic reason that the first myosin discovered (muscle myosin 
that belongs to the family myosin-II) turned out to be non-processive, 
those discovered later \cite{korn04} were called ``unconventional''  
\cite{kieke03,mooseker08,mezgueldi08,sweeney10a}. 
In this subsection, we highlight the main distinct features of two 
families of unconventional myosin motors that are both processive and 
serve as intracellular porters.

\subsubsection{\bf Myosin-V: a plus-end directed processive dimeric motor}

Myosin-V is a dimeric ``unconventional'' myosin that moves processively
towards the plus end of the actin track
\cite{mehta01,vale03b,veigel02,sweeney04,sellers08,trybus08,delacruz09,hammer12}.
Some of the key features of the mechano-chemical kinetics of myosin-V are 
as follows \cite{rief00}: (i) ATP-binding triggers fast detachment of the 
head from the F-actin track, (ii) hydrolysis of ATP takes place quite 
rapidly, (iii) phosphate release is also fast, (iv) the ADP-bound head has 
high affinity for the F-actin track. It is desirable that a minimal model 
of myosin-V should account for all these facts. 

Quantitative theoretical models of myosin-V has been developed by
several groups from different perspectives and at different levels
(see refs.\cite{mehta01,vilfan09} for reviews). 
Kolomeisky and Fisher \cite{kolomeisky03} adapted their generic model, 
which we sketched in section \ref{sec-genericporters}, for explaining 
several aspects of the experimental data on the kinetics of myosin-V. 
The kinetics of the motor was described by an unbranched pathway. 
Skau et al.\cite{skau06} introduced a more elaborated 7-state kinetic 
model that allows the possibility of branching of the pathway thereby 
giving rise to a {\it futile cycle} in addition to the main productive 
cycle of myosin-V. This kinetic scheme is closely related a qualitative 
model, developed earlier by Rief et al. \cite{rief00}, except that 
two additional features have been incorporated: (i) futile cycle, 
and (ii) detachment from the track. The mechanical stepping was assumed 
to take place in two sub-steps: a working stroke  of size 
$d_w \simeq 25$nm and a diffusive excursion through $d_D \simeq 11$nm. 
Both the average velocity and the run length were calculated as 
functions of ATP concentration and load force. 

In the discrete kinetic model studied by Wu et al. \cite{wu07}, the 
rear head can exist, at a time, in one of the following 4-distinct 
states:\\ 
$E$: the ligand-binding site is empty and the head is attached to the 
F-actin track;\\ 
$D$: the ligand-binding site is occupied by ADP and the head is attached 
to the F-actin track;\\ 
$T$: the ligand-binding site is occupied by ATP and the head is attached 
to the F-actin track;\\ 
$T'$: the ligand-binding site is occupied by ATP and the head is detached 
from the F-actin track. \\
On the other hand, the leading (front) head is assumed to be in one of 
the only two states, namely,
$D^{s}$ and $D^{w}$ in which the head is occupied by ADP and is bound to 
the F-actin track strongly and weakly, respectively. 
Thus, in this model, the homo-dimeric motor has 8 distinct kinetic states; 
a 2-letter code is used to describe these states where the letters on the 
left and right represents the states of the rear and front heads, 
respectively. It is also assumed that a {\it power stroke} is exerted 
when the leading head is in the state $D^{s}$ and that when the trailing 
head overtakes the leading head and re-attaches with the F-actin track 
it must be in the state $D^{w}$. Moreover, at the end of the power 
stroke, when the erstwhile leading head becomes the trailing head, the 
superscript $s$ is dropped from the label $D^{s}$ symbolizing it 
because the ADP-bound state of the trailing head is unique.  
Thus, both the transitions $TD^{s} \to DD^{w}$ and $T'D^{s} \to DD^{w}$ 
correspond to power strokes, albeit of different sizes (see ref.
\cite{wu07} for further details of the differences).  

The three kinetic pathways in this model are \cite{wu07}  

$$DD^{w} \rightleftharpoons DD^{s} \rightleftharpoons ED^{s} \mathop{\rightleftharpoons}^{k'_{2}} TD^{s} \rightleftharpoons \mathop{DD^{w}}^{{\uparrow}~{kt2}} $$

$$DD^{w} \rightleftharpoons ED^{w} \rightleftharpoons ED^{s} \mathop{\rightleftharpoons}^{k'_{2}} TD^{s} \rightleftharpoons \mathop{DD^{w}}^{{\uparrow}~{kt2}} $$

$$DD^{w} \rightleftharpoons ED^{w} \mathop{\rightleftharpoons}^{k_{2}} TD^{w} \rightleftharpoons \mathop{T' D^{w}}^{{\uparrow}~{kt1}}  \rightleftharpoons T'D^{s} \rightleftharpoons \mathop{DD^{w}}^{{\uparrow}~{kt2}}$$

Wu et al.'s model \cite{wu07} is an extension of an earlier model proposed 
by Uemura et al.  \cite{uemura04}. It is also related to another kinetic 
model suggested independently by Baker et al.  \cite{baker04}. One of the 
main quantities of interest calculated by Wu et al. \cite{wu07} is the 
steady-state average velocity  
\begin{equation}
V = \biggl(k_{2}' P_{ED^s} [ATP] + k_{2} P_{ED^w} [ATP]\biggr){\ell}_{step}
\end{equation}
where we have chosen the ATP-dependent transitions for writing the 
expression.

Another interesting feature of this model is that termination of the 
walk can take place from the states $DD^{w}$ and $T'D^{w}$. Since the 
overall rate of termination is 
\begin{equation}
k_{term} = k_{t1} P_{T'D^w} + k_{t2} P_{DD^w} 
\end{equation}
the average run length ${\ell}_{run}$ is obtained from 
\begin{equation}
{\ell}_{run} = V/k_{term}.
\end{equation}
Wu et al.\cite{wu07} examined the trends of variation of $V$ and 
${\ell}_{run}$ with the concentrations of ATP and ADP.

\vspace{5cm}

\begin{figure}[htbp]
\begin{center}
{\bf Figure NOT displayed for copyright reasons}.
\end{center}
\caption{Mechano-chemical network of 9 states for myosin-V before 
approximating with fewer states in the Bierbaum-Lipowsky model 
\cite{bierbaum11}. See the text for details.
Reprinted from Biophysical Journal  
(ref.\cite{bierbaum11}), 
with permission from Elsevier \copyright (2011) [Biophysical Society]. 
}
\label{fig-bierbaum1}
\end{figure}

For modeling the mechano-chemical kinetics of myosin-V, Bierbaum and 
Lipowsky \cite{bierbaum11} adapted the Lipowsky-Liepelt model 
\cite{liepelt07a,liepelt07b,liepelt09}, which was developed earlier 
in terms of a network of mechano-chemical states, and applied 
successfully to dimeric kinesin. 
In this model each head can be in one of the three distinct states: 
ATP-bound (T), ADP-bound (D) and empty (E). Thus, in principle, 
the 2-headed myosin-V should have 9 distinct states 
(see fig.\ref{fig-bierbaum1}). 
However, identifying the dominant pathways, Bierbaum and Lipowsky 
\cite{bierbaum11} reduced the number of distinct states to 6, namely, 
EE, ET, ED, DD, DT, TD. Moreover, the only 6 chemical transitions 
and 2 mechanical transitions (which swap the two motor heads) were 
allowed in this reduced description (see fig.\ref{fig-bierbaum2}).

\vspace{4cm}

\begin{figure}[htbp]
\begin{center}
{\bf Figure NOT displayed for copyright reasons}.
\end{center}
\caption{Reduced network of 6 mechano-chemical states in the Bierbaum-Lipowsky 
kinetic model for myosin-V \cite{bierbaum11}. See the text for details.
Reprinted from Biophysical Journal  
(ref.\cite{bierbaum11}), 
with permission from Elsevier \copyright (2011) [Biophysical Society]. 
}
\label{fig-bierbaum2}
\end{figure}

Denoting the six distinct states at position $j$ by the integer indices 
1-6 and the corresponding states at $j+1$ by the primed indices 1'-6', 
the mechanical transitions were represented by $5 \rightleftharpoons 5'$ 
and $3 \rightleftharpoons 4'$ (see fig.\ref{fig-bierbaum2}).  
The average velocity of the motor was calculated from 
$V = [(k_{55'} P_{5} - k_{5'5} P_{5'}) + (k_{34'} P_{3} - k_{4'3} P_{4'})]{\ell}$, 
where ${\ell}=36$nm is the step size. Moreover, assuming that detachment 
of the motor is most likely from the state $DD$ (labelled by the index 1), 
and with rate $\omega_{u}$, the average run length ${\ell}_{run}$ was 
calculated from ${\ell}_{run} = V/(\omega_{u} P_{1})$. 
Force-velocity relation was also computed by incorporating the 
force-dependence of the rate constants through the usual exponential 
factor.

Purely kinetic models of the type discussed above assume, rather than 
explain, the processive hand-over-hand stepping pattern of myosin-V.  
Lan and Sun \cite{lan05a,lan06} developed a coarse-grained model of 
myosin-V which captures the essential features of its structural  
design and energetics. The state of each motor domain is described 
\cite{lan06} in terms of two mechanical variables $\theta_i,\phi_i$ 
and a chemical variable $\mu_i$ ($i = 1,2$). The mechanical variables 
$\theta_1,\theta_2$ and $\phi_1,\phi_2$ are the angles shown in the 
fig\ref{fig-lanmyo}. Moreover, each motor domain can exist in one of the $N$ 
distinct chemical states, i.e., $\mu_i$ ($i=1,2$) can take one of the 
$N$ distinct allowed values.\\

\begin{figure}[htbp]
\begin{center}
{\bf Figure NOT displayed for copyright reasons}.
\end{center}
\caption{A schematic description of the mechano-chemical model for 
myosin-V developed by Lan and Sun \cite{lan05a}. 
Reprinted from Biophysical Journal  
(ref.\cite{lan06}), 
with permission from Elsevier \copyright (2006) [Biophysical Society]. 
}
\label{fig-lanmyo}
\end{figure}

In this model the dynamics of the mechanical variables can be formulated
in terms of the Langevin equation, or equivalent Fokker-Planck equation, 
in the overdamped limit. Assuming a physically motivated form of the 
energy $E(\theta_1,\theta_2,\mu_1,\mu_2,f)$, the corresponding torques 
were obtained by evaluating the appropriate derivatives. For example, 
$\tau_{\theta_i} = \partial E(\theta_1,\theta_2,\mu_1,\mu_2,f)/\partial \theta_i$ 
is the torque in the $\theta_i$-direction. These torques were then 
used in the stochastic equation of motion (e.g., Fokker-Planck equation) 
for the mechanical variables. A master equation is ideally suited to 
describe the stochastic kinetics of the discrete chemical states.

Lan and Sun \cite{lan06} split the elastic energy of the homo-dimeric 
myosin as
\begin{equation}
E(\theta_1,\phi_1,\theta_2,\phi_2,\mu_1,\mu_2,F) = E_0(\theta_1,\phi_1,\mu_1)+ E_0(\theta_2,\phi_2,\mu_2) + E_1(\theta_1,\phi_1,\theta_2,\phi_2,z,F) 
\end{equation}
where $z$ is the separation between the two heads of the myosin 
and $F$ is the externally applied load force.
For the elastic energy of the single myosin head labelled by $i$ 
($i=1,2$), they assumed the explicit form  
$E_0(\theta_i,\phi_i,\mu_i) = (1/2) k(\mu_i)[\theta_i-\theta_0(\mu_i)]^2 + (1/2) k'\phi_i^2 + c(\mu_i)$ 
where $\theta_o(\mu_i)$ is a $\mu_i$-dependent preferred angle and 
the $\mu_i$-dependent constant $c(\mu_i)$ accounts for the differences 
in the free energies of different chemical states even when 
$\theta_i = \theta_0$. 
The term $E_{1}(\theta_1,\phi_1,\theta_2,\phi_2,z,F)$ is the elastic 
energy of the chain domain that links the two motor domains. Note 
that $E_{1}$ is independent of $\mu_1$ and $\mu_2$ because this 
elastic energy is independent of the chemical states of the motor 
domains. When both the motors are bound to the actin filament, 
$E_{1}$ depends on $z$, the separation between the binding sites. 
A physically motivated form of the interaction energy 
$E_1(\theta_1,\phi_1,\theta_2,\phi_2,z,F)$ was also assumed. 

Models developed by Vilfan \cite{vilfan05a} and Xie et al.\cite{xie06a} 
are similar, in spirit, to those developed by Lan and Sun \cite{lan06}. 
However, the details are not identical. Besides the trajectories and 
step size distributions, the force-velocity relation was the main 
quantity of interest.  
Following an approach similar to that followed by Lan and Sun \cite{lan06} 
and by Vilfan \cite{vilfan05a}, Craig and Linke \cite{craig09} have 
formulated a model that provides further insight into the role of 
strain-induced gating in coordinating its two heads that is essential 
for its  processivity.
Few specific steps in the mechano-chemical kinetics of myosin-V have 
been elucidated by carrying out normal mode analysis of structural 
models \cite{cecchini08,zheng09a,duttmann12}. Some of these structures 
are based on proteins data bank while others are coarse-grained elastic 
networks.

\subsubsection{\bf Myosin-VI: a minus-end directed processive dimeric motor}

An unloaded myosin-VI walks towards the pointed end (i.e., minus end) 
of an actin filament whereas an unloaded myosin-V walks towards the 
barbed end (i.e., the plus end). For several years, the step size  
(36 nm) of myosin-VI was believed to be much larger than what would be 
expected on the basis of the prevalent interpretation of its structure 
at that time. In recent years this puzzle has been solved in terms of a 
swing of the lever arm by 180$^0$ and special structural features of 
its tail domain \cite{spudich10,sweeney10b}.  
To our knowledge, only the Lan-Sun model for myosin-VI \cite{lan06}, 
which was adapted from their earlier model for myosin-V, incorporates 
some structural features of this motor. However, in view of the new 
interpretations of the structures of the lever arms and tail domains 
\cite{spudich10,sweeney10b}, the theoretical model needs refinements. 

\subsubsection{\bf Myosin-XI: the fastest plus-end directed myosin}
\label{sec-myosinXI}

Myosin-XI is, perhaps, the fastest among the myosins \cite{tominaga12}. 
Just like myosin-V, it is a plus-end directed motor. But, its processivity 
and duty ratio are much lower than those of myosin-V motors \cite{ito07}. 
Although the relevant thermodynamic and kinetic parameters that 
characterize myosin-XI have already been extracted from experimental data 
\cite{tominaga12,ito07}, no Markov model for this kinetics have been 
reported so far. It would also be interesting to develop a coarse-grained 
model or, at least, a mechano-chemical model model of myosin-XI that would 
correlate its fast kinetics with the dynamics of its structural components. 

\subsection{\bf Processive dimeric kinesin}

The structures and mechanisms of all the kinesin families have been 
summarized in several reviews \cite{vale97,ray06,hirokawa09a,endow10}.

\subsubsection{Plus-end directed homo-dimeric porters: members of kinesin-1 family}

\noindent$\bullet${Key structural features of kinesin-1}

If the walk of a kinesin-1 is dominated by power stroke, then a mechanism 
for translating chemical changes (ATP hydrolysis) into mechanical movements 
must exist in kinesin just like that in myosin. 

As we explain below, a nucleotide-dependent small conformational 
change of kinesin is ``amplified'' to generate its power stroke 
\cite{vale99,vale00,vale00c}.
A ``sensor-element'' in kinesin senses the main enzymatic transitions 
and then relays this information to the track-binding interface and 
the ``mechanical element'' which is responsible for the mechanical 
movement. This pathway operates in reverse as well, because track-binding 
or strain on the ``mechanical element'' can affect rates of enzymatic 
reaction.

\noindent {\it The nucleotide-binding site}: 

In order to change conformations between ATP- and ADP-bound states, 
motor proteins must sense the presence or absence of a single phosphate 
group. From the structural studies the ``$\gamma$-phosphate sensor'' 
was identified by comparing structures with and without bound ATP analogs. 
The sensor consists of two loops called {\it switch I} and {\it switch II}.
Very similar switch loops also operate in myosin as well as in
G-proteins indicating that switch loops are, from evolutionary
point of view, ancient and existed even before the appearance of
molecular motors.

\noindent {\it Piston-like motion of a helix: a relay element}

Small movement of the ``$\gamma$-phosphate sensor'' are transmitted to
distant regions of the motor protein using a long helix that is
connected to the switch-II loop at one terminus. Since the helix is
long, but practically incompressible, it works like a piston. This
helix is a key structural element in the communication pathway
linking the catalytic site, the track-binding site and the mechanical
element; the mechanical element for a kinesin is the ``neck-linker''
which we describe below.

\noindent {\it Neck-linker: the mechanical element} 

The neck-linker (NL) of a kinesin is a region adjacent to its ``catalytic 
core''. The NL consists of about 15 amino acids. Since it is connected
to a ``coiled-coil'' dimerization domain its motion in one head of
kinesin gets conveyed to the other head by a mechanism mechanism 
that we'll discuss below.

The mechanism of the coordination of the two heads has been at the focus 
of intense experimental investigation over the last decade 
\cite{woehlke00,sablin04,asbury05,yildiz05,cross04a,carter06,guydosh07b,block07,gennerich09}. 
This coordination is now viewed as a ``gating'' phenomenon. The main idea 
behind ``gating'' is that one of the two heads has to wait till the other 
head opens a ``gate'' that allows the waiting head to resume the steps of 
its own mechano-chemical cycle. In principle, the gate can be operated in 
at least two different ways \cite{gennerich09}: (i) {\it polymer gate}: 
by the attachment or detachment of the gate operator head to the MT track, 
or (ii) {\it nucleotide gate}: by the binding or hydrolysis of ATP or 
unbinding of ADP (and/or P$_{i}$) from the gate operator head. Moreover, 
there are two possible choices for the gated head \cite{guydosh06,block07}: 
it could be the front head or the rear head. Although several plausible 
mechanisms of gating have been postulated 
\cite{guydosh06,block07,gennerich09,sindelar11}, 
the range of their validity remain controversial.  

The forward bias for re-attachment of the unattached head seems to arise 
from the free energetics and conformational kinetics of the NL.  
The key feature of the neck linker is that it can ``dock'' on a head, i.e., 
it can bind to the head aligning itself towards the forward direction of 
motion. ATP-binding with the forward head triggers ``zippering'' and  
``docking'' of the NL on it. This docking, in turn, drives the detachment 
of the rear head from the track and powering its forward swing towards 
the next binding site on the track ahead of the bound head. Thus, the NL 
is believed to be involved in both ``gating'' and forward ``biasing''.
The effects of length, charge and structure of the NL on the speed and 
processivity of kinesin-1 have been investigated experimentally by 
Shastry and Hancock \cite{shastry10}. Subsequently, they have also 
studied the effects of NL length on the processivities of some other 
families of kinesins \cite{shastry11} that we discuss later in this review. 

\noindent$\bullet${\bf Detailed kinetic model of kinesin-1}

Fisher and Kolomeisky \cite{fisher01} adapted their generic model,
which we discussed in section \ref{sec-genericporters}, for explaining
several aspects of the experimental data on the kinetics of kinesin-1.

The generic models of 2-headed motors that we discussed earlier either 
directly or indirectly {\it assumed} a processive hand-over-hand 
stepping pattern rather than explaining how this pattern emerges from 
underlying molecular interactions and kinetic processes. 
More specifically, the Peskin-Oster model \cite{peskin95} assumed 
(i) a ``gating'' mechanism, whereby the front head waits in its 
MT-attached state while the rear head detaches from the MT and 
searches for a nearby binding site for its re-attachment; and 
(ii) a ``biasing'' mechanism whereby re-attachment of unattached head 
in front of the attached head is more probable than that of the 
unattached head behind the attached head. For any model developed 
specifically for 2-headed kinesin, it would be desirable to incorporate 
the molecular mechanisms of ``gating'' and ``biasing'' explicitly within 
its kinetic scheme. 

In contrast to the models that assume tight coordination between the 
two heads in a hand-over-hand stepping pattern, Xie et al.\cite{xie06b} 
proposed a model in which the two heads are {\it partially coordinated}. 
Because of such partial coordination both backward stepping and futile 
ATP hydrolysis are possible in this model. Based on their recent 
experiments, Clancy et al.\cite{clancy11} have proposed a 5-state kinetic 
model (see fig.\ref{fig-clancy}) that incorporates not only both forward 
and backward steppings, but also futile cycles.

\begin{figure}[htbp]
\begin{center}
{\bf Figure NOT displayed for copyright reasons}.
\end{center}
\caption{Model for kinesin-1 developed by Clancy et al.\cite{clancy11}. 
(a) The five distinct states are labelled by the integers 1,2,...,5. 
Each of the two heads is coded by one particular color (red and blue).   
(b) The allowed transitions and the corresponding rate constants are 
shown. The forward, backward and futile pathways are shaded by yellow, 
orange and light green colors.
Reprinted from Nature Structural \& Molecular Biology  
(ref.\cite{clancy11}), 
with permission from Mcmillan Publishers Ltd. \copyright (2011). 
}
\label{fig-clancy}
\end{figure}

Following the steps prescribed by Chemla et al.  \cite{chemla08}, which 
we have discussed in section \ref{sec-forwardproblem}, 
Clancy et al.  \cite{clancy11} also calculated the average velocity 
and randomness parameter analytically for this 5-state kinetic model.
A model for the hand-over-hand stepping pattern of dimeric kinesin 
was developed by Shao and Gao \cite{shao06} by formulating the 
equations of motion of the two heads in terms of Langevin equations. 
A network model for the same motor has been developed by Liepelt 
and Lipowsky \cite{liepelt07a,liepelt07b,lipowsky08b} 
in which the kinetics is formulated in terms of master equations.

In some simplified models the NL is not incorporated explicitly, only its 
effect is captured in a simplified manner. For example, Mogilner et al. 
\cite{mogilner01a} extended the Peskin-Oster model \cite{peskin95} 
by assigning 3 possible states to each motor head, namely, the 
zippered state, unzippered state and the strained state.  
Wilson \cite{wilson09} models the effect of NL through a ``switch''; 
activation of the switch mimics the zippering of the NL whereas 
unzippering of the NL is indicated by the deactivation of the switch. 
(See also ref.\cite{thomas02,xie08a}).

Derenyi and coworkers \cite{czovek08,czovek11} have developed a theoretical 
framework that incorporates the effects of the docking and undocking of the 
NL in terms of a freely-jointed-chain (FJC) model for the NL. Each head is 
assumed to have 6 possible states: in addition to the detached state, there 
are five possible attached states, namely, ATP-containing NL-undocked state 
(T), ATP-containing NL-docked state (T$^*$), ADP-containing NL-undocked 
state (D), ADP-containing NL-docked state (D$^*$), nucleotide-free 
NL-undocked state (0).  The $6 \times 6$ distinct states of the dimeric 
kinesin were plotted on a two-dimensional state space each axis of which 
depicts all the mechano-chemical states of one of the two heads. The rate 
constants for the transitions among the states in this state-space of the 
dimeric kinesin were derived from (i) the force-free rate constants for a 
monomeric kinesin head, and (ii) properties of the FJC model of the NL. 
Imposing some criteria for extraction of the parameters in an optimization 
procedure, Czovek et al.\cite{czovek11} observed that over a narrow range 
of the parameters, the model could account for observed data. 
FJC model is not the only way in which the effects of the NL can be 
incorporated. Alternative formulations that treat the NL either as an 
elastic spring or as a worm-like chain have also been developed \cite{kutys10}.

Most of the models that incorporate NL are quantified in terms of either 
master equation or equivalent Brownian dynamics. However, a clearer picture 
may emerge if an model with structural details could be simulated. Data 
obtained from simulations of some structural models, performed under 
restricted conditions \cite{hyeon07} indicate the important regulatory 
roles of the elastic strain in the NL. 
Recent Brownian dynamics simulations incorporating electrical charges of 
the amino acids indicate enhancement of the forward bias of kinesin motors 
by their electrostatic interactions with the MT track \cite{grant11}.

The free energy difference between the docked and undocked conformations 
of the NL is about 5 pN nm \cite{rice03}. One of the controversial issues 
is how such a small free energy can drive a forward stepping of the motor 
by 8 nm against a load force as large as 5 pN 
\cite{mather06,hwang08,kawaguchi08}. 
Carrying out a Brownian dynamics 
simulation of a coarse-grained model with hydrodynamics interactions, 
Zhang and Thirumalai \cite{zhang12} demonstrated that, in this model, 
the distance of 16 nm covered by the trailing head of a kinesin-1 in a 
single step involves three major stages. In the first stage, the NL docks 
driving translocation of the trailing head by about 5-6 nm.  In the second 
stage, the trailing head moves ahead by another 6-8 nm by anisotropic 
translational diffusion. Finally, in the third stage, an optimal interaction 
of the trailing head with the MT and its eventual binding completes its 
forward movement by $\sim$16 nm in a single step. The importance of 
(biased) Brownian motion resulting in part of the stepping of kinesin-1 
has been known for many years \cite{kinosita07}. However, interpreting the 
entire stepping process in terms of a pure Brownian ratchet-and-pawl device 
\cite{wang07z,fan08,zheng09b} 
is an interesting, but controversial, idea.

On those occasions when a cargo-carrying kinesin-1 has only one of its 
heads attached to the track, thermal fluctuation can lead to its detachment 
from the track rapidly followed by a re-attachment. However, the 
re-attachment need not take place at its original location because of its 
possible displacement during the detached state resulting from relaxation 
of its elastic strain. Such a process of detachment and rapid re-attachment 
would manifest as ``hopping'' of the motor \cite{imafuku11}. The effects 
of hopping on the processivity and force-velocity relation has been 
examined theoretically using a 4-state kinetic model \cite{imafuku11}.  

\noindent$\bullet${\bf Limping gait of dimeric kinesin}

The fact that the steps of a heterodimeric kinesin can alternate between 
a fast and a slow one \cite{kaseda03} may not sound very surprising. 
However, what is even more surprising is that a similar stepping pattern 
was observed even for a homo-dimeric kinesin with genetically shortened 
stalk \cite{asbury03}. The alternating short and long dwell times 
is such that, for sufficiently short stalks, the longer dwell time could 
be about an order of magnitude longer than the shorter one. 
In analogy with limping gait of macroscopic bipeds, the pattern of 
alternating fast and slow steps of the artificially constructed kinesins 
were also called ``limping'' \cite{block07}. By performing a series of 
careful experiments Fehr et al.\cite{fehr09} ruled out several mechanisms 
speculated earlier for explaining limping of kinesin. The experimental 
observations were found to be consistent with a kinetic model proposed 
by Xie et al.\cite{xie09x}. In this model the differences in the two 
successive steps arise from the different {\it vertical} forces acting on 
the kinesin head in the two steps. Structural origin of the limping has 
been established by FRET measurements implicating the direct interaction 
of the neck linker with the MT track for the asymmetry \cite{martin10}. 

The possible consequences of vectorial loading of a motor were analyzed  
by Fisher and Kim \cite{fisher05a,kim05} from theoretical considerations 
assuming, however, perfect symmetry between the successive steps. 
This analysis was subsequently extended by Zhang and Fisher \cite{zhang11a} 
in the light of the insights gained from experimental investigations of 
limping \cite{fehr09}. Zhang and Fisher \cite{zhang11a} introduced the 
concept of {\it limping factor}, a quantitative measure of the limping. 
Suppose, $t_{j}$ is the dwell time at the $j$-th step. Then, 
\begin{equation}
T_{o}(n) = \sum_{j=1}^{n} t_{2j-1} ~{\rm and}~ T_{e}(n) = \sum_{j=1}^{n} t_{2j}
\end{equation}
are the total dwell times at the odd and even steps, respectively.
Zhang and Fisher \cite{zhang11a} defined the limping factor by 
\begin{equation}
L_{n} = max\biggl(\frac{T_{o}(n)}{T_{e}},\frac{T_{e}(n)}{T_{o}}\biggr)
\end{equation}
The {\it intrinsic limping factor} is defined by the limiting value 
\begin{equation}
L_{in} = lim_{n\to\infty} <L_{n}>
\end{equation}
where the angular brackets denote average over many runs (i.e., many 
independent sequences of 2$n$ steps). This quantity was calculated 
analytically for a stochastic kinetic model of kinesin that captures 
the asymmetry of the stepping rates. A more detailed mechano-chemical 
model was developed, and treated numerically, by Shao and Gao \cite{shao07}. 

\subsubsection{\bf Plus-end directed hetero-trimeric porters: members of kinesin-2 family}

Trimeric members of the kinesin-2 family consists of two different motors 
and a non-motor subunit. These are plus-end directed processive motors.
KIF3A and KIF3B, together with kinesin-associated protein 3 (KAP3),  
form a hetero-trimeric complex \cite{zhang04}. 
Another example is KLP11/KLP20 which makes a hetero-trimer by 
associating wih the kinesin-associated protein 1 (KAP1) 
\cite{brunnbauer10} 
These motors can be collectively represented as $M_1,M_2$ where 
$M_1$ and $M_2$ are the two different types of motors. A more 
elaborate notation would be $H_1T_1,H_2T_2$ where $H$ and $T$ 
denote the head and stalk-tail domains of each motor separately. 
Two artificial homo-dimeric constructs $H_1T_1,H_1T_1$ and $H_2T_2,H_2T_2$ 
as well as a doubly-heterogeneous construct $H_1,T_2,H_2T_1$ have been 
used in experiments \cite{pan10} for a comparative study to understand 
the distinct roles of the different head and tails domains in the  
speed, processivity and force-generation of the wild-type motors 
\cite{zhang04,brunnbauer10,pan10}. 
The effects of the length of the NL on the processivity of kinesin-2 
motors have been reported by shastry and Hancock \cite{shastry10}. 

Das and Kolomeisky \cite{das08} extended the generic model developed 
by Fisher-Kolomeisky to incorporate the distinct features of the two 
motors. Corresponding to a 2-state model for homo-dimeric motor 
$M_1,M_1$ and another distinct 2-state model for the homo-dimeric 
motor $M_2,M_2$, each with step size $\ell=8$nm, they formulated a 
4-state model, with an effective step size $\ell=16$nm, for the 
hetero-dimeric motor $M_1,M_2$.

\subsection{\bf Single-headed myosins and kinesin} 

\subsubsection{\bf Single-headed kinesin-3 family} 

In the initial investigations, kinesin KIF1A, a member of kinesin-3 family, 
was an enigma. It was found to be practically as processive as a kinesin-1. 
But, it is a single-headed motor. Therefore, its processivity cannot be 
explained by any mechanism similar to the coordinated hand-over-hand 
stepping pattern of the 2-headed motors. 
Another puzzling feature of KIF1A is its step-size distribution. It can 
step both forward and backward although forward steps are taken more 
often that backward steps. Moreover, in both the directions, its step 
size is not restricted to only 8 nm; steps size up to $\pm$32 nm, in the 
integral multiples of $\pm$8 nm, have also been observed. 

In the mutants of KIF1A constructed by Hirokawa, Okada and collaborators 
the number of charged amino acid residues in the so-called K-loop of the 
motor was varied systematically to investigate the effects on its 
processivity. Moreover, the effect of removing the charged E-hook of the 
tubulins was also explored. On the basis of these experiments and other 
complementary structural studies, it was established 
(see ref.\cite{hirokawa09b} for a review) 
that The processivity of KIF1A arises from the electrostatic attraction 
between the oppositely charged K-loop of KIF1A and E-hook of the tubulins. 
Experimental data indicated that the mechano-chemical cycle of a KIF1A 
can be divided roughly into two different parts: in one of these the motor 
is strongly attached to the MT track whereas in the other it is weakly 
tethered to the MT while executing an, effectively, one-dimensional 
Brownian motion. The overall mechanism of energy transduction by KIF1A is 
a physical realization of the abstract Brownian ratchet mechanism that we 
discussed in section \ref{sec-chemphys}. 
This mechanism is captured by the 2-state stochastic kinetic model of 
KIF1A developed by Nishinari et al. 
\cite{nishinari05,greulich07,chowdhury08b,garai11} 
(from now onwards, referred to as the NOSC model). All the kinetic 
parameters of the NOSC model were extracted from the data collected 
from single molecule experiments. The dwell time distribution of the 
single-headed KIF1A has been calculated by Garai and Chowdhury 
\cite{garai11} using this NOSC model. Furthermore, the model can account 
also for the effects of steric interactions between the motors at higher 
concentrations when traffic congestion takes place on the MT track. 
The phase diagram of the NOSC model was plotted in a space spanned by 
experimentally accessible parameters \cite{nishinari05,greulich07}. 
In principle, a single-headed motor like KIF1A {\it in-vitro} can change 
its ``lane'' by shifting to a neighboring protofilament on the same MT 
at the end of a single mechano-chemical cycle. The consequences of such 
``lane changing'' on the KIF1A traffic has been explored theoretically 
\cite{chowdhury08b}.

Xie and collaborators \cite{xie07kif1a} have extended the generic Brownian 
ratchet type models of molecular motors to develop a stochastic kinetic 
model specifically for KIF1A. The kinetics is formulated in terms of the 
Langevin equations that needs the shape of the potential(s) as input. 
In the spirit of Brownian ratchet models discussed in section 
\ref{sec-timedeplandscape}, Xie et al.\cite{xie07kif1a} assumed sawtooth 
and its variants at different stages of the cycle depending on the nature 
of the bound ligand. 

\subsubsection{\bf Single-headed myosin-IX family}

Myosin-IX is a plus-end-directed motor \cite{liao10}.
Just like KIF1A, myosin-IX family members also appear to move processively 
along actin filaments in spite of being single-headed motor. 
The mechano-chemical kinetic model proposed by Xie \cite{xie10myo9} 
for this single-headed motor is very similar to the model he developed 
earlier for single-headed kinesins \cite{xie07kif1a}

\subsection{\bf Processive dimeric dynein}

In this section we consider only cytoplasmic dynein 
\cite{king11,hook06,hook10,allan11} 
because it functions as a porter in a cell (see Gibbons \cite{gibbons11} 
for a history of the discovery of dynein). Dynein is a minus-end directed 
porter. Surprisingly, the cytoplasmic dynein alone 
is capable of minus-end directed cargo transportation whereas for the 
plus-end directed transportation several different families of kinesin 
participate. The dynein can transport diverse cargoes because of its 
regulation by many different molecular adaptors \cite{kardon09,vallee12}. 

Each dynein operates as a homo-dimer. The nomenclature for dynein has 
been standardized \cite{pfister05}. Its large head consists of 6 domains 
arranged in the form a ring that is a typical characteristics of the 
members of the AAA+ superfamily of ATPases 
\cite{vale00aaa,numata08,ogura01,hanson05,erzberger06,white07,tucker07,snider08}. 
However, compared to the other members of the AAA+ superfamily, the 
ring-like head of a dynein has an unusual structure and enzymatic function 
\cite{numata08}. 
First, four of the six modules contain sites capable of binding ATP whereas 
the remaining 2 modules are believed to play only regulatory roles 
(see fig.\ref{fig-dyneinEXPT}). 

\vspace{4cm}

\begin{figure}[htbp]
\begin{center}
{\bf Figure NOT displayed for copyright reasons}.
\end{center}
\caption{Schematic representation of cytoplasmic dynein.
Reprinted from Nature Cell Biology  
(ref.\cite{vallee12}), 
with permission from Mcmillan Publishers Ltd. \copyright (2012). 
}
\label{fig-dyneinEXPT}
\end{figure}

Secondly, from this ring-shaped head a stalk and a tail protrude. 
Unlike kinesin and myosin, the head of a dynein does not bind directly 
to its MT track. Instead, the small globular tip of a 12- to 15-nm long 
stalk, that projects approximately radially outward from the ring-shaped 
head, binds with the MT. Since the distance between an ATPase site and 
the corresponding MT-binding site of a dynein is much longer than those 
of myosin and kinesin, the mechanism of communication between the ring-like 
head and the globular tip of the stalk remains controversial 
\cite{carter10}. 
In fact, questions on the intra-molecular communication in dynein can be 
posed in three categories: \cite{houdusse09,carter10,cho12}: 
(i) communication between the modules of the ring-like head,  
(ii) communication between the ring and the linker, 
(iii) communication between the ring and stalk. 
 
The head is connected to the cargo-binding tail by a linker that is 
believed to functions as a mechanical lever in the force generation 
process \cite{koonce04,sakakibara11}. 
An alternative mechanism, in which the ring works like a winch, has also 
been suggested 
\cite{burgess04,sakakibara11}.  
Dynein seems to have a ``gear'' that controls its step size in response 
to load force \cite{mallik04a}. 
In the absence of any load force the step size is predominantly $4{\ell}$ 
where ${\ell} \simeq 8$nm. When subjected to a sufficiently low load force, 
the most probable step size first decreases to $3{\ell}$ and, on further 
increase of the load, it decreases to $2{\ell}$. Finally, it attains  
the smallest step size ${\ell}$ at even higher load.   
The stepping pattern of dynein is also quite unusual and appears very 
different from the standard hand-over-hand pattern followed by processive 
dimeric kinesins and myosins \cite{qiu12}.

Dynactin \cite{schroer04} is a multisubunit protein complex that can 
bind to dynein. Dynein alone is not as processive as kinesin-1. However, 
dynactin \cite{gross03a,dell03,holleran98}. 
is believed to enhance the processivity of dynein by acting like a tether. 
A cargo may be hauled by a mixture of active motors and passive 
tethers. In the simplest situation where one motor and one tether 
transport a cargo, the passive tether can either supress the rate of 
detachment of the active motor from the track or increase the rate of 
its re-attachment to the track thereby enhancing the effective processivity 
of the motor \cite{posta09}. 
 
Singh et al. \cite{singh05} carried out a Monte Carlo (MC) simulation of 
a model of a {\it single head} of a dynein motor.  The different AAA 
domains are labelled by integer indices 1-6 (see fig.\ref{fig-singh}). 
The label C marks a non-AAA domain. ATP hydrolysis occurs primarily at 
domain 1. The direction of the power stroke, caused by the response of 
ADP from domain 1 is indicated by the thick curved arrow. The domains 
2, 3, and 4 are assumed to play only regulatory roles.

\begin{figure}[htbp]
\begin{center}
{\bf Figure NOT displayed for copyright reasons}.
\end{center}
\caption{A sketch of the main components of the head domain of a single 
dynein motor.
Reprinted from Proceedings of the National Academy of Sciences  
(ref.\cite{singh05}),
with permission from  National Academy of Sciences, U.S.A. \copyright (2005).
}
\label{fig-singh}
\end{figure}

The main assumptions are as follows: 

{\it Assumption 1}: Four distinct step sizes: Singh et al.\cite{singh05} 
proposed that if no ATPs are bound at the secondary sites, the motor 
attempts to take a 32-nm step; if one ATP is bound at a secondary site, 
dynein tries to take a 24-nm step; if two ATPs are bound at secondary 
sites the motor attempts a 16-nm step; and if all secondary sites are 
occupied the attempted step size is 8-nm. However, because of thermal 
noise, the actual step size could exhibit a distribution that has peaks 
at each of the four step sizes listed.\\
{\it Assumption 2}: ATP-binding affinities: in the absence of any load 
force, the binding affinities were assumed to be  ordered as follows: 
$K^{1}(F=0) > K^{2}(F=0) > K^{3}(F=0) > K^{4}(F=0)$. This assumption was 
implemented in the MC simulation by assigning numerical values for the 
on- and off-rates that satisfy the conditions 
$k^{1}_{on} = k^{2}_{on} > k^{3}_{on} > k^{4}_{on}$ and $k^{1}_{off} < k^{2}_{off} = k^{3}_{off} = k^{4}_{off}$, 
where the superscript refers to the number of bound ATP molecules. 
If $n-1$ ATP molecules are already bound to the head, then the 
probability of binding the nth ATP to a secondary site is 
$P^{n} = k^{n}_{on} [ATP] ~\Delta t$, and the probability of the nth ATP 
unbinding from a secondary site is $P^{n}_{off} = k^{n}_{off} ~\Delta t$ 
where the choice for the elementary time step was 
$\Delta t = 2 \times 10^{-4}$ s.\\
{\it Assumption 3}: Load-dependence of on/off rates at secondary sites: 
It was assumed that $k^{j}_{on} =  k^{j}_{on}(F=0) exp(F{\ell}/k_BT)$ 
where $j=2,3,4$ is the number of bound ATP molecules and ${\ell}$ is an 
adjustable length. The off-rates were treated as effectively 
load-independent parameters.\\
{\it Assumption 4}: Load-dependence of hydrolysis rate: The rate of ATP 
hydrolysis by the domain 1 was assumed to decrease exponentially with 
increasing load force $F$. \\
{\it Assumption 5}: Regulators' effect of ATP hydrolysis: The rate of ATP 
hydrolysis was assumed to be enhanced by a multiplicative factor if at least 
one of the secondary sites is occupied by ATP. \\
{\it Assumption 6}: The hydrolysis of ATP was assumed to be reversible 
before the actual release of ADP and P$_{i}$. \\ 
From the simulation data, Singh et al.\cite{singh05} obtained the 
step-size distribution. They also computed the force-velocity relation 
and studied the ATP-dependence of (a) the stall force, (b) average velocity. 

An alternative approach was followed by Gao \cite{gao06} to account for 
the same phenomena (including the load-dependent step size) that 
Singh et al.\cite{singh05} tried to explain with their model. 
Gao introduced two coordinates: a physical coordinate $x$ 
and a chemical coordinate $\alpha$. The coordinate $x$ denotes the position 
of the motor along the MT track. In contrast, $\alpha$ represents the 
``conformational changes'' that control the chemical processes like ATP 
binding and hydrolysis as well as the release of ADP and P$_{i}$. Since 
both $x$ and $\alpha$ were assumed to vary continuously, two separate  
overdamped Langevin equations were written for the time evolution of these 
two variables. The forces entering into these two equations were derived 
from a potential profile (a landscape) $V(x,\alpha;i)$ which depends on 
the chemical state $i$ (i.e., whether bound to ATP or ADP) at that instant 
of time. The actual forms of the rates of chemical transition $k_{ij}$ 
depend on $x$ and $\alpha$. Using this model, Gao \cite{gao06} studied the 
ATP-dependence of step sizes and stall force as well as the variation of 
average velocity and rate of ATP hydrolysis on the load force and ATP 
concentration. The experimentally observed gear-like function of dynein 
\cite{mallik04a} is explained in this model to be a consequence of the 
loose chemo-mechanical coupling \cite{gao06}.

\vspace{2cm}

\begin{figure}[htbp]
\begin{center}
{\bf Figure NOT displayed for copyright reasons}.
\end{center}
\caption{The structural model of dynein developed by Tsygankov et al. 
\cite{tsygankov11b}. (A) and (B) represent the pre- and post-stroke 
conformations while (C) depicts the transition from the pre-stroke 
to post-stroke conformation. The five angles used to describe the 
shape of dynein are shown in the schematic diagram in (D). Few 
representative snapshots in (E) show the relative positions of the 
heads during a forward step of the dynein motor. 
Reprinted from Biophysical Journal  
(ref.\cite{tsygankov11b}), 
with permission from Elsevier \copyright (2011) [Biophysical Society]. 
}
\label{fig-tsygankovdyn}
\end{figure}

Tsygankov et al.\cite{tsygankov11b} combined the kinetic model that they 
developed earlier for dynein’s ATP hydrolysis cycle \cite{tsygankov09}  
with a coarse-grained  structural model to formulate a full mechanochemical 
model for a hand-over-hand stepping model of a homodimeric dynein. Based 
on the assumption that a dynein does not step sideways to any other 
protofilament, the model   model describes its mechanical dynamics in a 
two-dimensional space (see fig.\ref{fig-tsygankovdyn}). 
The AAA+ ring of each head is modeled as a circle 
of a fixed radius. Moreover, both the stalk and the tail are assumed to 
bendable, but inextensible. Furthermore, the {\it tail} is assumed to 
emerge {\it tangentially} from the AAA+ ring whereas the stalk is assumed 
to emerge perpendicularly from the same ring. Thus, the structure of {\it 
each head} is described by 6 variables. $(X_i,Y_i)$ ($i=1,2$) are the 
cartesian coordinates of the center of the ring and several angular 
variables. The angle $\phi$ describes the rotation of the ring around 
its center. The angles $\alpha$ and $\beta$, which characterize the 
curvature of the stalk and tail, respectively. The angle $\gamma$, which 
is a quantitative measure of the relative position of the stalk and the 
tail, takes two distinct mean values in the pre-stroke and post-stroke 
configurations. The angle between the stalk and the MT track is denoted 
by $\delta$ whereas the relative orientations of the heads at the point 
of tail junction is denoted by $\xi$ . The variables describing the 
structural features evolved following Langevin equations. The 6 discrete 
biochemical states of each head  were denoted by 
$T,T^{*},D,D^{*},DP,\Phi^{*}$ where T and D correspond to ATP-bound and 
ADP-bound states, DP describes the state bound to both ADP and P$_{i}$ 
while $\Phi$ indicates empty site. The asterisk indicates post-stroke 
state conformations. These biochemical states were updated according to 
the corresponding master equations. 
A multi-scale modeling approach was followed by Serohijos et al. 
\cite{serohijos09} for studying dynein at different levels of 
spatio-temporal resolution. Zheng \cite{zheng12} carried out a 
normal mode analysis of a coarse-grained elastic network model 
of dynein establishing the key role of a closed AAA3-AAA4 interface 
in the mechano-chemical coupling in dynein.

\subsection{\bf Collective transport by porters} 
\label{sec-collectiveporters}

In living cells, common cargoes are vesicles, organelles, etc.
However, a dielectric bead is often used as a cargo while performing
experiments {\it in-vitro}. At least some motors are known to associate 
with other accessory proteins and/or macromolecular complexes that 
serve effectively as adaptors which can alter the intrinsic properties 
of the motors. The molecular link between the cargo and the cargo-binding 
domain of the motor is normally elastic. In most of the theoretical 
treatments \cite{elston00c,elston00d,chen01} 
this link is modelled as a harmonic spring. 
In principle, the cargo would be free to move in three dimensions 
although the motor transporting it would walk on a linear filamentous 
track. However, for simplicity of modeling, the cargo may also be 
restricted to move only in one dimension that is parallel to the 
track of the motor. However, the position of the center of mass of 
the cargo is a continuous variable while the motor moves forward only 
in discrete steps.

One of the common features of these cargoes is that these are much 
bigger than the individual motors that haul them. Therefore, the 
cargo can mediate interaction between the motors giving rise to 
collective effects. In this subsection we review the progress in 
understanding some of the effects of cooperation and competition 
among the porters involved in intracellular transport 
\cite{chowdhury06,constantinou10}. 
The type of the collective phenomenon depends on the nature of the 
cargo, viz., whether the cargo if ``hard'' or ``soft''; we define the 
``hardness" and ``softness'' in the appropriate contexts below. 
In principle, the collective properties of the motors \cite{guerin10}
are likely to depend on the following single-motor properties: 
(i) the processivity of individual motors, and the force-dependence of 
the detachment (and, attachment) rates, 
(ii) single-motor velocity and, more generally, the force-velocity relation, 
(iii) rule for load-sharing by the motors.

\subsubsection{\bf Collective transport of a ``hard'' cargo: load-sharing, tug-of-war and bidirectional movements}

When a single ``hard'' cargo is hauled by more than one motor of the same 
type (i.e., either all kinesin or all dynein) along a single MT, it does 
not deform in shape. In contrast, a ``soft'' cargo would get elongated 
when pulled by the motors. A real intracellular cargo is never rigid, 
but the softness may vary from one cargo to another. The motility and 
shape change of the cargo by the collective effect of the teams of 
porters is a relatively new area of investigation \cite{lipowsky10,berger11}.

\noindent$\bullet${\bf Co-directional motors: load-sharing and cooperation}

How do the average velocity and the run length of 
such a cargo scale with the number of motors? Note that the total number 
of motors engaged in pulling the motor is not constant, but keeps 
fluctuating with time because of the detachment and reattachment of the 
individual motors. Suppose $N$ is the maximal number of motors that can 
engage simultaneously with a single hard cargo. If a load force is applied 
{\it against the cargo}, how does the {\it collective force-velocity 
relation} vary with the variation of the parameter $N$? This issue remains 
controversial in spite of many experimental investigations over the last 
decade. Moreover, the dependence of multi-motor cargo transport on the 
single-motor velocity has just begun to receive attention \cite{xu12b}.

A kinetic model for this phenomenon was developed by Klumpp and Lipowsky 
\cite{klumpp05}. 
In the model the load on the cargo is assumed to be shared equally by the 
motors ({\it mean field approximation}). 
A master equation is written for $P_n$, the probability that the cargo 
is bound to the filamentous track by $n$ ($0 \leq n \leq N$) motors. 
This theory predicted an increase in the run length of the cargo 
with increasing maximal number of motors $N$. Assuming a linear 
force-velocity relation for each single motor, this theory  
also predicted a nonlinear force-velocity relation for all $N > 1$, 
where the stall force is an increasing function of $N$.
Experimental data have been analyzed within the framework of this 
theory \cite{beeg08,fallesen11}. 
The effect of cargo-mediated effective assisting or resisting force on 
the motors was treated in a transparent intuitive manner by Wang and Li 
\cite{wang09b}.  
The Klumpp-Lipowsky model \cite{klumpp05} was extended by Korn et al. 
\cite{korn09} by allowing unequal sharing of the load by the motors. 
Stochastic sharing of the load by the $n$ bound motors take place in 
the computational model studied by Kunwar et al. \cite{kunwar08}. 
In this model the motors were treated as special floppy linkages/springs. 
Later studies of collective transport by Kunwar and Mogilner \cite{kunwar10} 
using a combination of the nonlinear force-velocity relation and the 
stochasticity gives rise a collective force-velocity curve that is 
almost linear provided at least three motors carry the load. 
Kunwar and Mogilner's computational model \cite{kunwar10} was also used 
by McKenney et al.\cite{mckenney10} to investigate collective transport 
by dynein motors.
The dependence of the collective force-velocity relation on the nature 
of the force-velocity relation of the individual single motors deserves 
further systematic thorough investigation. Kunwar and Mogilner's 
\cite{kunwar10} has been extended by Bouzat and Falo \cite{bouzat11}.

Most of the theoretical works in the early stage of investigations were 
based on essentially one-dimensional models. However, more recently, 
some of the effects of the organization of the motors on the surface of 
a real three-dimensional cargo has been studied by computer simulations 
\cite{erickson11}. 

What happens when a mixed population of co-directional fast-moving and 
slow-moving motors share the same track simultaneously \cite{larson09}? 
One possibility is that the queues of motors may form behind the 
slow-moving ones, a well known phenomenon in vehicular traffic 
\cite{chowdhury00,schadschneider11}. 
An alternative possibility is that the faster-moving motors can increase 
the rate of dissociation of the slow-moving motors from the track. 
Indeed, as the relative fraction of the fast-moving motors is increased, 
a sharp transition from slow cooperative transport to fast cooperative 
transport is observed \cite{larson09}. In a mixed population of proteins, 
one species may be an active motor while the other may serve as a 
passive tether \cite{posta09}. In this case, the enhanced processivity 
of the cargo arises from either the suppression of the detachment or 
enhancement of re-attachment of the active motors \cite{posta09}. 

\noindent$\bullet${\bf Opposing motors: tug-of-war and bidirectional movements}

It is well known that some motors reverse the direction of motion by 
switching over from one track to another which are oriented in 
anti-parallel fashion. In contrast to these types of reversal of
direction of motion, we consider in this section those reversals where
the cargo executes a bidirectional motion on the same MT track because 
of a ``tug-of-war'' between kinesins and dyneins 
\cite{gross04,welte04,welte08,verhey11a,bryantseva12}. 
Tug-of-war is not restricted only to kinesins and dyneins that move along 
MT tracks. Similar phenomena have been observed also during cargo transport 
by myosin V and myosin VI both of which walk along filamentous actin 
\cite{ali11}.

At least three possible mechanisms of bidirectional transport have been
postulated. (i) One possibility is that either only + end directed motors
or only - end directed motors are attached to the cargo at any given
instant of time. Reversal of the direction of movement of the cargo
is observed when the attached motors are replaced by motors of opposite
polarity. (ii) The second possible mechanism is the closest to the
real life ``tug-of-war''; the competition between the motors of opposite
polarity, which are simultaneously attached to the same cargo and tend
to walk on the same filament, generates a net displacement in a direction
that is decided by the stronger side. (iii) The third mechanism is
based on the concept of regulation; although motors of opposite polarity
are simultaneously attached to the cargo, only one type of motors are
activated at a time for walking on the track. In this mechanism, the
reversal of the cargo movement is caused by the regulator when it
disengages one type of motor and engages motors of the opposite polarity.
For experimentalists, it is a challenge not only to identify the
regulator, if such a regulator exists, but also to identify the mechanism
used by the regulator to act as a switch for causing the reversal of
cargo movement. 

Carrying measurements in live cells, Soppina et al.\cite{soppina09} 
demonstrated a tug-of-war between a single kinesin and 4 to 8 dyneins. 
They also speculated on how the cells might exploit the competition  
between a ``strong and tenacious'' kinesin against 4 to 8 ``weak and 
detachment-prone'' dyneins.  

Muller et al.\cite{mueller08} extended the formalism developed by Klumpp 
and Lipowsky \cite{klumpp05} for one type of motors by incorporating 
two oppositely moving motors. The stochastic kinetics of the system 
is now described by the master equation for $P(n_{+},n_{-},t)$, the 
probability that at time $t$ the cargo is attached with the track by 
$n_{+}$ plus-end directed and $n_{-}$ minus-end directed motors, 
respectively. Two linear force-velocity relations, with different sets 
of parameters, were assumed for the single motors of the two types of 
motors. A mathematical analysis of the steady-states of this model has 
been carried out by Zhang \cite{zhang09} in a special limit in which 
the numbers of motors of both species are infinite; this limit is not  
realistic. 

The current status of the models seem to be far from satisfactory. For 
example, a simple stochastic model of load sharing would predict more 
pauses of the cargo when more motors are involved; However, this is not 
supported by the experimental observations \cite{kunwar11}. 
It is possible that some regulatory mechanisms, which are not incorporated 
in the current models, have significant influence on the collective 
bidirectional transport of a cargo.

\noindent$\bullet${Force-dissociation kinetics in collective motor transport}

Force-dissociation kinetics determines how fast individual motors would 
detach from the track under load. This kinetics plays an important role 
in the collective transport of a single hard cargo by many motors. 
In most of the theoretical works on collective motor transport the 
dissociation rate was assumed to be an exponentially increasing function 
of the applied load force. However, recent experimental results indicate 
that this  assumption is not always valid \cite{kunwar11}.

\noindent$\bullet${Multi-motor hauling of single cargo across MT-MT, AF-AF and MT-AF crossings}

What happens to a cargo hauled by multiple motors at a crossing of 
two filamentous tracks? Several possible situations can be conceived 
and some of these have already been investigated by experiments {\it 
in-vitro} \cite{gross07,holzbaur10}; we list a few of these below 
in table \ref{table-crossing}.

\begin{table}
\begin{tabular}{|c|c|} \hline
Intersection & motors \\ \hline
MT-MT & all kinesins \\ \hline
MT-MT & all dyneins \\ \hline
MT-MT & kinesins and dyneins \\ \hline
AF-AF & all myosin-V \\ \hline
AF-AF & all myosin-VI \\ \hline
AF-AF & myosin-V and myosin-VI \\ \hline
MT-AF & kinesins/dyneins and myosin-V/myosin-VI \\ \hline
\end{tabular}
\caption{Various types of crossings of filamentous tracks and motors that 
appriach the crossing hauling a single cargo.
}
\label{table-crossing}
\end{table}


\subsubsection{\bf Many cargoes on a single track: Molecular motor traffic jam}

As the cargoes are always
much bigger than the motors (in-vitro as well as in-vivo), direct
steric interactions of the cargoes become significant when several
cargoes are carried by sufficiently dense population of motors along
the same track. Such situations are reminiscent of vehicular traffic
where mutual hindrance of the vehicles cause traffic jam at sufficiently
high densities. In analogy with vehicular traffic, we shall refer to
the collective movement of molecular motors along a filamentary track
as ``molecular motor traffic''; we shall explore the possibility of
molecular motor traffic jam and its possible functional implications.

Most of the minimal theoretical models of interacting molecular motors
utilize the similarities between molecular motor traffic on MT and
vehicular traffic on highways both of which can be modeled by
appropriate extensions of the totally asymmetric simple exclusion
process. In such models the motor is represented by a ``self-propelled''
particle and its dynamics is formulated as an appropriate extension of
the dynamics of the totally asymmetric simple exclusion process.
In such models, in addition to forward ``hopping'' from one binding
site to the next, the motor particle is also allowed to detach from
the track. Moreover, attachment of a motor particle to an empty site
is also allowed.

In reality, a molecular motor is an enzyme that hydrolyses ATP and
its mechanical movement is coupled to its enzymatic cycle. In some
recent works on cytoskeletal motor traffic, the essential features
of the enzymatic cycle of the individual motors have been captured.
Ciandrini et al.\cite{ciandrini10} developed a model where the 
extent of details incorporated falls in between TASEP-type model 
(which do not incorporate any mechano-chemistry) and those that 
incorporate lot of those details. Nishinari et al. 
\cite{nishinari05,greulich07}
extended the TASEP by incorporating the minimal details of the 
mechano-chemical cycles of individual KIF1A motors to predict their 
collective spatio-temporal organization, specifically jamming of motors.

\subsubsection{\bf Trip to the tip: Intracellular transport in eukaryotic cells with long tips}
\label{sec-IFT}

\noindent$\bullet${\bf Motor transport in fungal hyphae}

Fungal hyphae already attracted the attention of Marcello Malpighi 
in the seventeenth century (as evident from the graphical illustration 
reproduced from Malpighi's original in ref.\cite{money08}). 
Most of the works on the growth of fungal hyphae focussed mainly on the 
biomechanics of the cell wall to predict the shape of the growing tip 
\cite{keijzer08,riquelme11}.
Microtubules and cytoskeletal motors are known to play several important 
roles in hyphal growth in filamentous fungi 
\cite{steinberg07a,steinberg07b,steinberg07c}. 
One of the unique feature of the growth of fungal hyphae is the 
existence of ``spitzenk\"orper'' which is believed to play a distinct 
role as a vesicle supply center.  
Complementary investigations on the transport of materials required for 
this growth by the cytoskeletal motors began in more recent years 
\cite{sugden07,sugden09}. 
These models are extensions of the TASEP that incorporate the elongation 
of hyphae by allowing the lattice to elongate according to an appropriate 
dynamic rule. This extended model is called the {\it dynamically extending 
exclusion process} (DEEP) \cite{sugden09}.
This model has been extended further to model bacterial flagellar growth 
\cite{schmitt11}. 
Interesting regulatory roles of dynein has been discovered in the 
bidirectional transport of cargoes along fungal hyphae and quantified  
mathematically in terms of first passage time \cite{schuster11}. 
The queueing of the motors near the tip of a microtubule in a fungal 
hyphae has also been modeled by an extended version of TASEP 
\cite{ashwin10}.

\noindent$\bullet${\bf Motor transport in plant root hair and pollen tube}

Pollen tube and root hair are long extensions in plants. The transport 
of organelles here is dominated by acto-myosin system 
\cite{cai09,cai10}.
A consequence of this organelles movements is that it gives rise to 
cytoplasmic streaming \cite{shimmen07,lubicz10}.
The role of this streaming in mixing up the interior of the cell is 
similar to that of the molecular motors in many other situations 
\cite{brangwynne08}.

\noindent$\bullet${\bf Intraflagellar transport in algae}

Flagella and cilia are important organelles in many eukaryotic cells. 
Eukaryotic flagella and cilia, e.g., those of green algae {\it 
Chlamydomonas reinhardtii}, and long extensions of some apical 
cells in brown algae, e.g., those of {\it Sphacelariales}  
are cell appendages that are shaped as a long ``tip''.  
The history of research on these cell appendages over the last 150 
years have been explored recently by Bloodgood \cite{bloodgood09}.
In this subsubsection we review only intraflagellar transport (IFT) 
\cite{hao09,rosenbaum99,rosenbaum02,cole03,scholey03,scholey08,blacque08,pederson08,emmer10}, 
a phenomenon that is driven by molecular motors, that has a much 
shorter history \cite{kozminski12}. 
The machines and mechanisms of the undulatory motion of cilia and flagella 
are reviewed in section \ref{sec-flagellabeat}.
A theoretical model of IFT has been developed by Bressloff \cite{bresloff06}.
In this model ``particles'' reach the tip of the cilium by hopping and 
elongate it by one unit.

One of the special features of cilia is that although a cilium is not 
an organelle in the strictest sense it is also not a continuous extension 
of the cytoplasm. The ciliary compartment is separated from the cytoplasmic 
compartment by a ``diffusion barrier'' \cite{nachury10,hu11a}. 
Very recent experiments \cite{kee12} have established a close relation 
between the molecular components of this barrier and those on the nuclear 
pore complex \cite{obado12} which we'll discuss later in this review. 
Based on this similarity, Kee et al. \cite{kee12} have proposed the 
existence of ``ciliary pore complex''.  The kinetics of 
crossing this barrier diffusively has not been included in any 
theoretical model of IFT explicitly. 

\noindent$\bullet${\bf Axonal transport}

\begin{figure}[htbp]
\begin{center}
\includegraphics[angle=90,width=0.65\columnwidth]{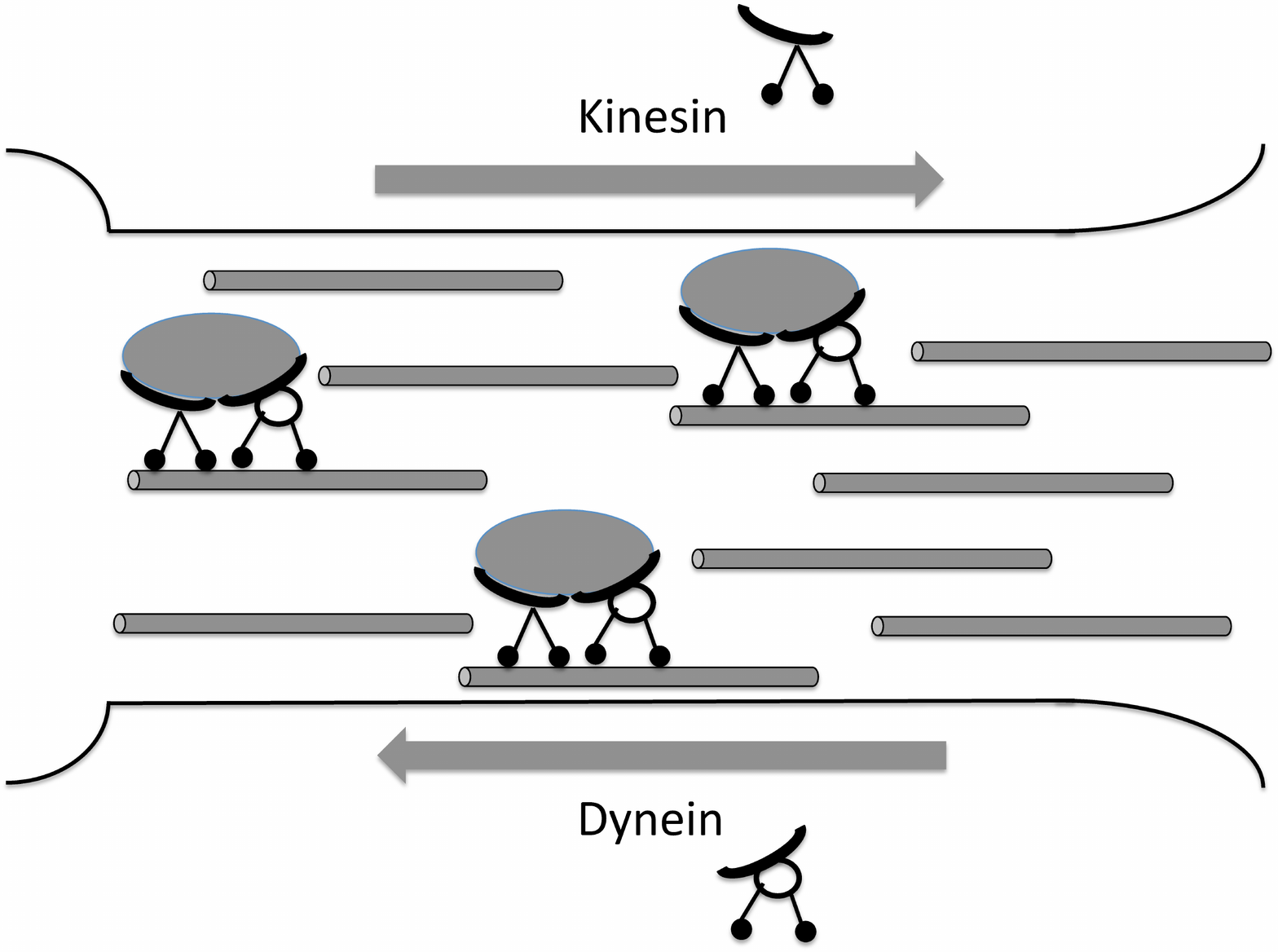}
\end{center}
\caption{A schematic depiction of axonal transport; only the axon 
of the neuron is shown. The cell body and dendrites (on the left 
of the figure), and the synaptic junction between the axon with 
the dendrite of another neuron (on the right of the figure) are 
not shown explicitly.
}
\label{fig-axonaltr}
\end{figure}

In a human body, the axon can be as long as a meter whereas the
corresponding cell body is only about 10 microns in length. Almost 
all the proteins needed to maintain the synapses are synthesized in the
cell body. How are these proteins transported to the synapse along
the long axon \cite{goldstein00,stokin06}? 
The problem is even more challenging in animals like elephant and 
giraffe which have even longer axons. A bundle of parallel MTs 
usually run along the axons and dendrites of a neuron, (see appendix 
for a brief description of the cytoskeleton of a neuron).  
These MTs form the track for the motorized transport of intracellular 
cargoes, e.g., vesicles, organelles, etc., along axons and dendrites 
\cite{conde09,baas11}.
Movement of the cargo in a direction away from the cell body is called 
{\it anterograde} whereas that in the reverse direction is called 
{\it retrograde}; therefore, both kinesins and dyneins are involved 
\cite{hirokawa05,hirokawa08,hirokawa10} (see Fig.\ref{fig-axonaltr}).

Every cargo in an axon spends a fraction of its journey time actually 
moving and even during those periods its movements are bidirectional.  
Two distinct patterns of cargo movement in axonal transport have been 
named ``fast'' and ``slow'' transport \cite{brown09,miller08}. 
Fast transport occurs at an average velocity of several microns per 
second, i.e., equivalently, several hundreds of millimeters per day. 
In contrast, the slow axonal transport takes place at the rate of a 
maximum of tens of nanometers per second, i.e., about few millimeters 
per day \cite{brown09}. 
This difference may be caused by the differences in the fraction of 
time they spend moving although the underlying mechanism of motor 
transport may be the same \cite{brown03}. 
For historical accounts of research on fast axonal transport, see 
ref.\cite{dahlstrom10}.

To our knowledge, the earliest quantitative model of axonal transport was 
developed by Blum and Reed \cite{blum89} at a time when the nature of 
the key components was not clear. Subsequently, Brown \cite{brown00,brown03} 
proposed his hypothesis of ``stop-and-go'' traffic in axon. In order 
to test the basic idea of this hypothesis a quantitative model was 
developed by Brown et al.\cite{brown05}. In this model it was assumed 
that (i) neurofilaments are cargo which can switch between two ``relatively 
persistent directional states'', namely anterograde and retrograde, 
and (ii) in either state the neurofilaments can move or pause. Although 
the movements in both directions can be rapid, the overall transport 
is slow because the filament dwells most of the time in paused states. 
A slightly different dynamical model was studied by Cracium et al. 
\cite{cracium05}. In this 5-state model, the neurofilament can also be 
bound to a anterograde or retrograde motor that pauses while off-track. 
This model can be extended to a 6-state version by distinguishing between 
the two off-track (paused) states: neurofilament attached to an 
anterograde motor and neurofilament attached to a retrograde motor 
\cite{jung09}.  Possible effects of cooperativity of the motors on 
axonal transport have been explored \cite{mitchell09,mitchell12} by 
extending the model developed by Cracium et al.\cite{cracium05}. 
When looked at from a broader perspective, axonal transport is essentially 
a concrete physical realization of cooperation and competition of a group 
of kinesins and dyneins where the anterograde and retrograde transport 
observed in a specific case in an emergent phenomenon.

\noindent$\bullet${\bf Effects of defect and disorder on cytoskeletal transport}

We have already come across the possibility of randomness of one
kind in motor traffic: the different properties of different types
of motors in a mixture can be captured by randomizing the motor
properties. In other words, the randomness of the motor properties
have already been considered.
But, so far we have implicitly regarded the track for the cytoskeletal
motors to be a perfectly periodic array of motor-binding sites.
However, in reality, a motor can encounter defects or disorder
on its path.

\subsubsection{\bf Fluid membrane-enclosed soft cargo pulled by many motors: extraction of nanotubes}

In many eukaryotic cells ER tubules are pulled out of membrane reservoirs 
by molecular motors that walk collectively on MT tracks \cite{hu11b}.
Molecular motors can extract membrane tubes also from vesicles \cite{leduc10}. 
Because of the liquid-like nature of the membranes of the vesicles, 
the motors bound to the vesicle surface do not experience any resistance 
from the membrane except at the leading edge. Consequently, all the 
motors, except those at the leading edge, move freely along the membrane 
surface walking along a filamentous track. In contrast, the motors at the 
leading edge, because of the load force exerted by the membrane move at a 
slower rate. The queueing up of the faster motors behind the slower ones 
is analogous to queueing of vehicles in traffic \cite{evans96,ktitarev97}. 
Thus, the assumption of equal sharing of external load, that is often 
used for theoretical calculations on multi-motor hauling of hard cargo, 
does not hold in this case. Moreover, the motors pulling the same 
membrane tube are not strongly coupled to each other.  

The motors fail to extract a nanotube if their number density is smaller 
than a critical value; at densities above this critical value, motors 
cooperatively pull a long thin tube out of the vesicle and the tube 
elongates at a steady average velocity \cite{derenyi07}. 
Using the formalisms of ASEP, Campas et al.\cite{campas06a} derived an 
analytical expression for the velocity $V_{N}$ of a cluster of $N$ motors 
denoting the forward and backward hopping rates by $p$ and $q$, respectively:
\begin{equation}
V_{N} = p\frac{[1-e^{F{\ell}/k_BT}(q/p)^N][1-(q/p)]}{e^{\delta F{\ell}/k_BT}[1-(q/p)]+e^{F{\ell}/k_BT}[(q/p)-(q/p)^N]}
\end{equation}
where $0 < \delta < 1$ is a dimensionless parameter characterizing the 
position of the energy barrier between two neighboring lattice sites. 
Once tubulation takes place, TASEP-type models predict that the motors 
can exhibit varieties of dynamical phenomena \cite{tailleur09}, e.g., 
shocks and inverse shocks, re-entrant phase transitions, etc.  
The phenomenon of membrane pulling by multiple motors has been formulated 
mathematically also in terms of Brownian ratchets whose kinetics are 
governed by appropriate Langevin equations \cite{orlandi10}.

It has been argued that, even collectively, motors cannot generate 
strong enough force to extract membrane nanotubes if all of them move 
along a single protofilament; motors must be using several protofilaments 
simultaneously when they successfully extend a membrane tube \cite{campas08}.
The nature of the dynamics of the tube, however, depends on the
extent of processivity of the motors \cite{shaklee08}. 
For example, Ncd is a nonprocessive motor. Ncd can extract nanotubes from 
vesicles, but exhibits a {\it bidirectional switching}; the tube 
alternates between forward and backward movements with variable speeds 
\cite{shaklee08}. 
Such richer dynamical behavior of tube extraction by nonprocessive 
motors is in sharp contrast with the monotonic tube growth observed 
when pulled by processive motors. In fact, this bidirectional motion 
for collective pulling by nonprocessive motors resembles the 
bidirectional motion of a filamentous backbone rigidly connected to 
multiple nonprocessive motors (which we'll discuss in section 
\ref{sec-badoual}).

\subsection{\bf Collective transport of filaments by motors: non-processivity and bistability}
\label{sec-badoual}

Consider a group of identical motors bound to an elastic backbone 
\cite{julicher95,julicher97b,badoual02,shu04,mouri08,vilfan98,vilfan99,vilfan00}. 
Two classes of such coupled motors (motors coupled to a single
backbone) have been considered. In the first, the individual
motors are modeled as Brownian ratchets whereas in the second the
individual motors exert a power stroke. In the first case, it has
been demonstrated that even if each individual motor is non-processive,
such a system of elastically coupled motors can exhibit bidirectional
processive motion. In this mode of movement, the motors move
collectively on a filamentary track in a processive manner in one
direction for a period of time and, then, spontaneously reverse its
direction of motion. Such spontaneous oscillations can account for
the dynamics of axonemes, which are core constituents of eukaryotic
cilia, as well as oscillatory motions of flight muscles of many insects.
In the second case, several different forms of strain-dependent rates
of detachment of the motors from the track have been considered.

The works cited above establish that bidirectional motion does not 
necessarily require pulling a cargo or a filament by antagonistic 
motors. A group of identical motors, each of which lacks directionality 
of its average motion, can give rise to a bidirectional movement of 
the polar filament \cite{badoual02}. More recent investigations have 
revealed that a group of identically directed motors can give rise 
to bidirectional movement of a bundle that consist of filaments with 
alternating polarities (and, therefore, apolar, on the average) 
\cite{gilboa09,gur10,farago11,leduc10b}.

\subsection{Section summary} 

In this section we have reviewed the kinetics of the members of several 
different specific families of porters as well as processes driven by 
single and multiple porters. Over the last couple of decades many 
research groups have investigated the effects of the following: 
(i) structural designs of the motors and the corresponding tracks, 
(ii) the conformational dynamics of the motor in each cycle and the 
(de-)polymerization kinetics of the track, 
(iii) the nature of the cargo and motor-cargo coupling, 
(iv) manner in which the load is shared by multiple motors, and 
regulation of oppositely-directed motors, engaged with the same cargo, 
and (vi) motor-traffic on the same track. 

In section \ref{sec-whymotors} we listed several different levels at 
which collective dynamics of molecular motors can be viewed. In the 
specific context of porters, models have been developed to study the 
coordination between (a) subunits of a single motor, (b) members of 
the same family of porters, (c) members of different families of the 
myosin superfamily that move in opposite directions on the same F-actin 
track, (c) members of kinesin and dynein superfamilies that move in 
opposite directions on the same MT track. We have also presented a 
brief review of intracellular transport in eukaryotic cells, particularly 
some of those with long tips. 

However, the understanding of the intracellular transport {\it in-vivo} 
cannot be complete without integrating the MT-based transport system with 
the F-actin-based transport system. Although some {\it in-vitro} experiments 
have provided initial insights, to my knowledge, no quantitative kinetic 
model of this integrated transport system has been reported so far.  


\section{\bf Filament depolymerization by cytoskeletal motors: specific examples of chippers}
\label{sec-specificchippers}

Depolymerases \cite{howard07}, which chip way from the tips of MT tracks, 
form two families of kinesins \cite{howard03a,wordeman05}, namely 
kinesin-8 and kinesin-13. Kip3p and MCAK are the two extensively studied 
members of these two families \cite{helenius06,varga06}. 
The depolymerase activity of MCAK has received most of the attention 
so far \cite{helenius06}. Very little attention has been paid to its 
role as force generator \cite{oguchi11}.

Klein et al.\cite{klein05} theoretically investigated the depolymerization 
of MT by MCAK using a one-dimensional model. The origin of the 
coordinate system ($x=0$) is permanently located at the depolymerizing 
tip of the MT, i.e., the description is based on a frame of reference 
that moves with respect to lab-fixed system. Denoting the concentration 
of MCAK in the bulk solution by $c$, their MT-binding rate was assumed 
to be $\omega_{a}c/\rho_{max}$ where $\rho_{max}$ is the maximal density 
of MCAK for which binding sites on the MT saturate. $\omega_{d}$ is 
the rate of detachment of the MCAK motors from the MT. The density 
profile $\rho(x,t)$ of the MCAK along the MT satisfies the equation 
of continuity 
\begin{equation}
\frac{\partial \rho}{\partial t} + \frac{\partial J}{\partial x} = 
\omega_{a} c \biggl(1 - \frac{\rho}{\rho_{max}}\biggr) - \omega_{d} \rho
\end{equation}
where the two terms on the right hand side are the source- and sink-
terms, respectively. The current density $J$ is given by the sum of 
diffusion and drift current, i.e., 
\begin{equation}
J = - D (\partial \rho/\partial x) - V \rho, 
\end{equation}
where $D$ is is the diffusion coefficient of the MCAK motors along the 
MT and $ V = V_{0} + V_{d}$ is the sum of the average velocity of the 
motor with respect to the filament and $V_{d} \geq 0$ is the velocity 
of MT depolymerization. For MCAK $V_{0}$ can be neglected. Moreover, 
if $\alpha-\beta$ dimeric subunits 
have length ${\ell}$ and are removed from the MT at a rate $\Omega$, 
then, $V_{d} = \Omega {\ell}/N$ where $N$ is the total number of 
protofilaments of a MT. $\Omega$ is expected to depend on the density 
of the MCAK at the depolymerizing tip of the MT; a phenomenological 
form of $\Omega(\rho_{0}$ is assumed. So far as the boundary conditions 
are concerned, $\rho \to \rho_{\infty}$ as $x \to \infty$ where 
\begin{equation}
\rho_{\infty} = \frac{\omega_{a} c \rho_{max}}{\omega_{a}c + \omega_{d} \rho_{max}} 
\end{equation} 
is the equilibrium density (Langmuir formula) for motors with attachment-
detachment kinetics. Similarly, another physically motivated boundary 
condition is specified for the boundary at $x=0$. 
Klein et al.\cite{klein05} showed that, depending on the values of the 
set of model parameters, MCAK motors can either accumulate at the 
depolymerizing end of the MT or their population there gets depleted. 
The dynamical accumulation of the MCAK, caused by the capturing of 
the motors bound along the MT filament by the retracting MT end, is a 
collective phenomenon. Occurrence of this phenomenon requires sufficiently 
high processivity of the MCAK, i.e., high probability that the 
depolymerizer MCAK remains attached to the MT carrying our several 
rounds of subunit removal before, ultimately detaching from the MT.  

In the corresponding discrete stochastic model, motors are represented 
by particles and the MT is represented by a one-dimensional chain 
of equispaced discrete binding sites. No more than one particle can 
occupy a site simultaneously; the occupation variable $n_{j}$ can 
take only one of the two allowed values $n_{j} = 0$ (empty) and 
$n_{j}=1$ (occupied). From the later equation for the occupational 
probabilities, one gets the rate equation for the average occupation 
numbers $<n_{j}>$: 
\begin{eqnarray}
\frac{d<n_{j}>}{dt} &=& \omega_{h} (<n_{j+1}>+<n_{j-1}>-2<n_{j}>) \nonumber \\
&+& \omega_{a}c<1-n_{j}> - \omega_{d}<n_{j}> + <\Omega n_{1}(n_{j+1}-n_{j})>
\end{eqnarray} 
where the last term essentially relabels the binding sites if the 
subunit at the at the MT tip is lost by depolymerization. A mean-field 
treatment of this model confirms the dynamic accumulation of MCAK at 
the depolymerizing tip of the MT provided $\Omega = \Omega^{0}(1-P_{p}n_{2})$ 
when $n_{1}+1$ and the processivility $P_{p}$ is sufficiently high. 

Since Klein et al.\cite{klein05} considered a semi-infinite MT, their 
model could not be used for studying the effects of MCAK on the 
distribution of he lengths of the MTs. These effects of depolymerases 
were calculated by Govindan et al.\cite{govindan08}. One of the key 
points is that the typical ``residence time'' of a single depolymerase 
after its adsorption on the MT is $\tau_r \sim 1/\omega_{d}$ on a 
sufficiently long MT before desorbing. Therefore, only those MCAK 
motors which bind to the MT within a ``trapping zone'' of length 
${\ell}_{trap} \sim \sqrt{D \tau{r}}$ from the depolymerizing tip of 
the MT get adsorbed at the tip. Those MCAK motors which bind to the 
MT at a distance larger than ${\ell}_{trap}$ from the MT tip get 
detached before getting trapped by the MT tip. Once trapped, a MCAK 
can escape from the MT tip only during its depolymerization activity 
by accompanying a $\alpha-\beta$ subunit chipped from the MT tip. 

One of the limitations of the model developed by Govindan et al. 
\cite{govindan08} is that it does not take into account the steric 
exclusion of the motors on the MT track, even at high concentrations 
of the motors. This model was extended by Hough et al.\cite{hough09} 
incorporating the effects of steric exclusion by a prescription that 
was used earlier by Parmeggiani et al.\cite{frey} for modeling 
molecular motor traffic. Very recently jamming of MCAK motors on the 
MT track has been observed experimentally \cite{leduc12}.  

\subsection{Section summary}

In my opinion, the consequences of MT depolymerization by the depolymerase 
motors have been the main focus of attention of theoretical models so far. 
The causes of the filament depolymerization induced by these depolymerase 
motors has received very little attention. The difference in the modes of 
translocation of the members of kinesin-8 and kinesin-13 families also 
needs a clear explanation in terms of the differences in their structure 
and conformational dynamics.

\section{\bf Filament crossbridging by cytoskeletal motors: specific examples of sliders and rowers}
\label{sec-specificslidersrowers}

In this section we discuss a few specific examples of sliders and rowers 
that are responsible for the contractility and shape changes of many cells 
and subcellular structures. As generators of contractile forces, acto-myosin 
is ubiquitous in living systems. Various modes of actomyosin contraction 
and the corresponding spatial and temporal patterns of force generation 
have been reviewed \cite{martin10b,levayer12} in the context of cell 
division, cell motility and morphognesis.

\subsection{\bf Acto-myosin crossbridge and muscle contraction}
\label{sec-muscle}

There are some chemical differences between the muscles of vertebrates
and invertebrates (e.g, flight muscles of insects). Muscle cells of
vertebrates can be broadly classified into ``striated'' and smooth
(``non-striated'') types. Vertebrate striated muscle cells can be
further divided into two categories- skeletal and cardiac. Although
skeletal muscles of vertebrates (e.g., those of frog and rabbit)
were used in most of the early investigations on the mechanism of
muscle contraction, the cardiac muscle has been getting attention in
recent years because of its implications in cardiac disease control.

Each muscle fiber is actually an enormous multinucleated cell
produced by the fusion of many mononucleated precursor cells
during development whose nuclei are retained in the adult
muscle cell. The diameter of muscle cells is typically $10-100$
$\mu$m and the length can range from less than a millimeter to
a centimeter. Each of these cells is enclosed by a plasma membrane.
The nuclei are squeezed to the peripheral region just beneath
the plasma membrane.

About $80$ percent of the cytoplasm of a skeletal muscle fiber
(i.e., muscle cell) is occupied by cylindrical rods of protein
and are known as myofibrils. Many myofibrils, each about $1 \mu$m
in diameter, are contained within the cross section of a single
muscle cell. The muscle cells also contain mitochondria sandwiched
between the myofibrils.

Myofibrils are the structures that are responsible for muscle
contraction. The most distinctive feature of myofibrils is their
banded appearance; the dark bands correspond to higher density
of protein. The spatial periodicity of the alternating light and
dark bands is 2.3-2.6 $\mu$m in the resting state of a muscle;
the entire repeating structure, from one $Z$-disc to the next,
is known as sarcomere.

The banded appearance of the sarcomere is produced by hundreds of
protein filaments bundled together in a highly ordered fashion.
The two main types of filament are:\\
(i) thick filaments, about 15 $nm$ in diameter, are made mostly
of myosin; \\
(ii) thin filaments, about 7 $nm$ in diameter, consist mostly of
actin.\\
Both these types of filaments contain also other types of proteins
which help to hold them in correct steric arrangement and regulate
the process of contraction.

Arrays of thin and thick filaments overlap in the sarcomere in a
manner similar to that of two stiff bristle brushes.
Myosin molecules are arranged in such a way on the thick filament
that their heads point away from the mid-zone towards either end
of the filament. The thick filaments come within about 13 nm of the
adjacent thin filament which is close enough for the formation of
{\it cross-bridges} between the myosin heads belonging to the thick
filament and actin molecules constituting the thin filaments.

Research on molecular machines was focussed almost exclusively on the 
mechanism of muscle contraction during the first half of the 20th 
century and it was dominated by Archibald Hill and Otto Meyerhof  
\cite{avhill1927,needham71} 
who shared the Nobel prize in Physiology (or medicine) in 1922. 
We'll not pursue the historical developments in muscle research; 
interestedreaders are referred to refs.
\cite{needham71,spudich01,gyorgyi58,gyorgyi04,ahuxley82,ahuxley88,hhuxley90,ahuxley00b,hhuxley00,hhuxley04,hhuxley08,weber02,cooke04,holmes04,sellers04,offer06}.

In two landmark papers published in 1954, A.F. Huxley and Niedergerke
\cite{ahuxley54} and, independently, H. E. Huxley and Hanson 
\cite{hhuxley54} proposed the {\it sliding filament hypothesis} of
muscle contraction 
\cite{bagshaw92}. 
According to this hypothesis, it is the sliding
of the thick and thin filaments past each other, rather than folding
of individual proteins, that leads to the contraction of the muscle.
This theory was formulated clearly and quantitatively in another
classic paper of A.F. Huxley in 1957. The essential assumptions of 
this model are as follows \cite{ahuxley00a} (see fig.\ref{fig-actomyocross}): 

\begin{figure}[htbp]
\begin{center}
\includegraphics[angle=-90,width=0.45\columnwidth]{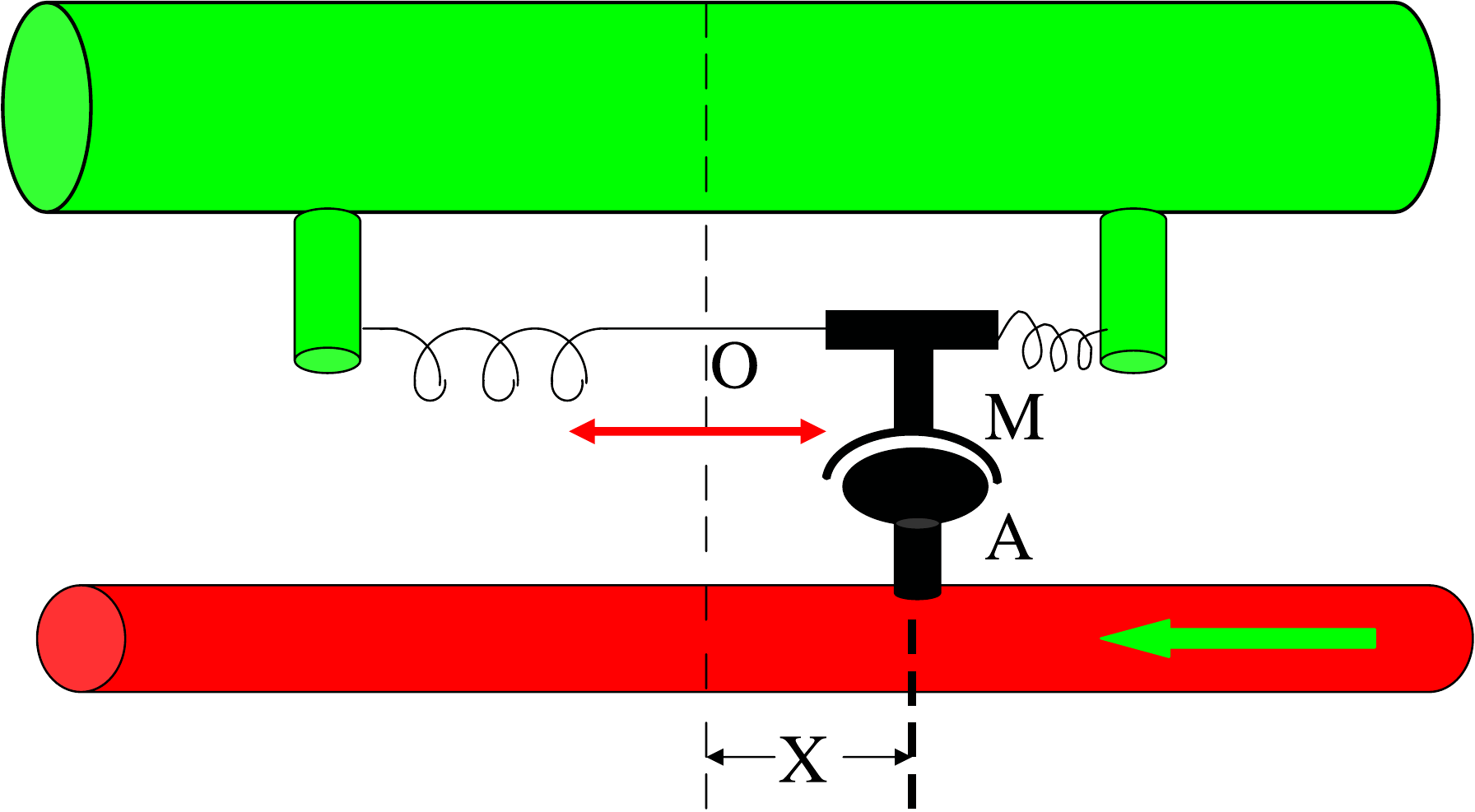}
\end{center}
\caption{Cross-bridge model proposed by Andrew Huxley in 1957 (adapted 
from ref.\cite{ahuxley57}).
}
\label{fig-actomyocross}
\end{figure}

(i) Each myosin has a binding site $M$ for the actin filament of the same 
cross-bridge. While unbound, the myosin executes a one-dimensional 
Brownian motion parallel to the actin filament and can bind to the 
filament with a rate constant $f(x)$ where $x$ is the extension of 
its elastic element. \\
(ii) A myosin head bound to the actin filament can detach from the 
actin filament with a rate $g(x)$. \\
(iii) $f(x)$ is moderate within a certain range of $x$ provided 
$x >0$; however, $f(x) = 0$ for $x \leq 0$. In contrast, for all 
$x > 0$, $g(x) < f(x)$ whereas $g(x)$ is a large constant for all 
$x < 0$. 

Let $P(x,t)$ be the probability that a motor at position $x$ 
(i.e., with strain $x$) is attached to the corresponding binding 
site on the actin filament. Then, 
\begin{equation}
\frac{\partial P(x,t)}{\partial t} + \frac{\partial P}{\partial x} \frac{dx}{dt} = (1-P) f - P g 
\end{equation}
If a steady state exists then $dx/dt = V$ and we have a simpler equation 
\begin{equation}
- V \frac{\partial P(x)}{\partial x} = (1-P)f - P g 
\end{equation}
The average force generated can be computed from 
\begin{equation}
F = N \int kx P(x) dx 
\end{equation}
where $N$ is the total number of myosin motors. In principle, the 
force-velocity relation can be obtained by first evaluating the 
steady-state probability $P(x)$.

Does this scheme correspond to a power stroke or a Brownian ratchet? 
Although non-power stroke mechanism for muscle contraction has appeared 
in the literature in many apparently different versions (see, for example, 
refs.\cite{mitsui88}), only a few interpreted the energy transduction 
mechanism in Andrew Huxley's theory as a Brownian ratchet 
\cite{vale90,cordova92}. 
The functional forms of the strain-dependent rates of attachment and 
detachment of the myosin motors to the actin filaments that Huxley 
assumed \cite{ahuxley57} is responsible for the Brownian ratchet-like 
mechanism. In several later papers other authors assumed more 
complicated functional forms of these rates (see, for example, 
ref.\cite{smith87}) to overcome some of the limitations of the 
original formulation.

In the original version of the sliding filament model, developed in the
nineteen fifties, it was generally assumed that the cross bridges moved
back and forth along the backbone of the thick filaments remaining
firmly attached to it laterally. However, later X-ray studies
demonstrated that the filament separation could vary without apparently
interfering with the actin-myosin interactions. On the basis of this
observation, in 1969, H.E. Huxley proposed the myosin ``lever arm''
hypothesis \cite{hhuxley69}. This model was developed further and formulated quantitatively
by A.F. Huxley and Simmons \cite{ahuxley71} in 1971 (to appreciate the 
status of the theory at that time see Huxley's review of 1974 
\cite{ahuxley74}).

Many subsequent extensions of the Huxley-Simmons model either incorporate 
larger number of chemical states of larger number of pathways. For example, 
a 4-state mechano-chemical kinetic model for muscle contraction, developed 
by Eisenberg and Hill \cite{eisenberg78}, 
was essentially an extension of the Huxley-Simmons model \cite{ahuxley71}. 
In the extended version \cite{eisenberg78} that Eisenberg et al. also 
compared with experimental data \cite{eisenberg80a}, the energetics of the 
elastic strain and the changes in the chemical states are coupled (see 
ref.\cite{eisenberg80b} for furher details of this approach). 
The importance of more than one pathways in the mechano-chemical cycle 
of muscle myosins was emphasized by Piazzesi et al.\cite{piazzesi95}.  
In recent years many powerful experimental techniques have provided deeper 
insight into the acto-myosin dynamics\cite{thompson01,thomas09,koubassova11}.  

One of the relatively recent theoretical works on muscle contraction 
incorporates the cooperativity of the rowers through strain-dependent 
chemical kinetic processes \cite{duke99b,duke00a,duke02b}. 
The collective force-velocity relation of the rowers depends on the 
fraction of the bound heads $r = f/(f+g)$. If $r$ is small, the load-free 
sliding velocity s large because of the successive members of a 
``relay teams'' cooperate with each other.  However, the force generated 
is small because only a few motors form cross-bridges at a time. 
In contrast, in the opposite limit of large $r$, larger number of 
cross-bridges gives rise to stronger force; however, hindrance caused 
by the crowding of cross-bridges leads to low velocity of sliding. 
Many numerical as well as a few analytical treatments of the theories 
of various aspects of muscle contraction have been reported over the 
last decade 
\cite{vilfan03,lan05c,woo05,vilfan01}

\subsection{\bf Sliding of acto-myosin bunldle in non-muscle cells: stress fibers}
\label{sec-stressfiber}

In non-muscle cells actomyosin bundles form stress fibers where filament 
sliding driven by myosin motors has strong similarity with actomyosin 
system in muscle cells \cite{pellegrin07,tojkander12}. 
These actin bundles are involved in cell adhesion, contractility as well 
as in motility \cite{naumanen08}.
The stress fibers are essential components also in machineries involved 
in mechano-transduction. 
The structure and mechanism of operation of stress fibers in motile 
cells are different from those in non-motile (but contractile) cells 
\cite{deguchi09}.  

The roles of chemical signaling in the alignment of stress fibers 
during cell adhesion has been modeled mathematically by Scholey et al. 
\cite{scholey05d}. However, this model does not deal with the 
actomyosin crossbridges explicitly.
A continuum model for the kinetics of cell contractility was developed 
by Deshpande et al. \cite{deshpande06} and implemented computationally 
using the finite element method. This model includes 
(i) a simple form of time-dependence of an activation signal that triggers 
formation of stress fibers, 
(ii) an equation for the kinetics of the stress fibers where the 
signal-dependent recruitment of actin and myosin competes against 
their tension-dependent dissociation; and 
(iii) phenomenological equations that relate the bundle contraction 
(or extension) rate to the tension thereby accounting for the 
acto-myosin crossbridge dynamics. This model and its numerical 
implementation are based on fairly standard strategies of modeling 
and simulation in engineering for elastic continua. 

Most of the models mentioned above did not incorporate the details of the 
signaling pathways. An attempt to capture at least some of these details 
was made by Besser and Schwarz \cite{besser07}. They modelled stress fiber 
contraction by combining somewhat detailed biochemical signaling processes 
with the mechanics of the contractile fibers. Each sarcomeric unit of the 
stress fiber has been modelled by extending the Kelvin-Voigt model for a 
viscoelastic material. A Kelvin-Voigt unit consists of a spring and a 
dashpot joined in parallel. The effects of the the myosins, which slide 
the actin fibers, is captured by adding, also in parallel, a ``contractile 
module'' that generates a contraction force $F_{m}$. For simplicity, a 
linear force-velocity relation was assumed for the motor-generated contractile 
force. The one-dimensional model of the stress fiber is a chain of coupled 
series of sarcomeric units. In the continuum limit of this chain the equation 
for the displacement variable $u(x,t)$ satisfies a partial differential 
equation in which mixed derivatives of $u(x,t)$ also appear \cite{besser07}: 
\begin{eqnarray} 
\biggl[\frac{\partial}{\partial x} \eta_{e}(x,t) \frac{\partial}{\partial x} \frac{\partial}{\partial t} + \frac{\partial}{\partial x} k(x) \frac{\partial}{\partial x}\biggr] u(x,t) = = -\frac{1}{a} \frac{\partial}{\partial x} F_{stall}(x,t)
\end{eqnarray} 
where the spring constant $k(x)$ need not necessarily depend on $x$, but 
the spatially-varying effective viscosity $\eta_{e}(x,t)$ is enhanced 
by the motor activity. The effective stall force $F_{stall}(x,t)$ is 
assumed to depend linearly on the active fraction $n(x,t)$ of the myosin 
heads so that $F_{stall}(x,t) = F_{max} n(x,t)$ attains its maximum value 
$F_{max}$ only if all the myosins within the cross-section at $x$ are active.  
The myosins get activated by a biochemical signaling pathway. The fraction 
$n(xt)$ is determined by solving the corresponding system of 
reaction-diffusion equations that describes the biochemical siganling 
events. One of the main predictions of this theoretical model is a 
heterogeneous contraction of the stress fibers; this result is consistent 
with experiments \cite{besser07}.

The models mentioned above address mostly the questions on collective 
behavior of the stress fibers. A model has been developed by Stachowiak 
and O'Shaughnessy \cite{stachowiak08} to describe the kinetics of 
individual stress fibers that explicitly accounts for the actomyosin 
crossbridges (see fig.\ref{fig-stressfiber}).

\begin{figure}[htbp]
\begin{center}
{\bf Figure NOT displayed for copyright reasons}.
\end{center}
\caption{A schematic representation of the stress fiber model developed 
by Stachowiak and O'Shaughnessy \cite{stachowiak08}. (a) Along the axis 
of the stress fiber, myosin-containing regions (blue) alternate with 
$\alpha$-actinin-containing regions (green). Each end of the fiber is 
connected to a focal adhesion. (b) A sarcometic unit of the stress fiber 
consists of actin (grey), myosin (blue) and titin (orange). The force 
$f_{myo}$ is generated by the myosin motors whereas the force $f_{titin}$ 
is associated with the springlike action of the passive protein titin. 
Te fiber tension $T$ is exerted by the two neighboring sarcomeres on the 
two sides while the force $p$ resist the overlap of actin filaments at 
their pointed ends (near the center of the sarcomere).
Reprinted from New Journal of Physics  
(ref.\cite{stachowiak08}),
with permission from Institute of Physics \copyright (2008).
}
\label{fig-stressfiber}
\end{figure}

In this one dimensional model regions containing myosin motors alternate 
with regions containing $\alpha$-actinin. Each end of the stress fiber 
is connected to a transmembrane protein complex, called focal adhesion, 
that is anchored to the extracellular matrix. Actin and myosin (and 
titin) form a sarcomeric structure that resembles the sarcomere of 
muscles cells. Suppose $x$ denotes the length of the sarcomere. 
Let $z$ be the extent of overlap of the actin filaments at their 
pointed ends (see fig.\ref{fig-stressfiber}). The rates of polymerization 
of the actin filaments at the barbed ends and that of their depolymerization 
at the pointed end are denoted by $V_{+}$ and $V_{-}$, respectively. 
It is assumed that $V_{+}$ is constant whereas increasing overlap of 
the pointed ends of the actin filaments increases the depolymerization 
rate $V_{-}$.

Assuming, as usual, the validity of overdamped approximation, the force 
balance equation leads to the following equation for the sarcomere length 
variation with time: 
\begin{equation} 
\gamma [(dx/dt)-V_{+}] = \underbrace{- k_{t} x}_\text{elastic force of titin } + \underbrace{k_{o} z}_\text{elastic force of overlap} - \underbrace{F_{s}}_\text{myosin stall force} + \underbrace{T}_\text{tension} 
\end{equation} 
where $\gamma$ is the phenomenological drag coefficient. 
Moreover, the equation 
\begin{equation}
(dx/dt) + (dz/dt) = V_{+} - V_{-}^{0}e^{k_{o}z/F_{*}}
\end{equation}
imposes the length constraint, where $F_{*}$ is a characteristic force.
Study of the 
kinetics of relaxation gets simplified by a separation of two different 
relevant times scales; usually, actin overlap and polymerization relax in 
seconds whereas the sarcomere length relaxation requires minutes 
\cite{stachowiak08}.  
The dimensionless parameter $r = \gamma V_{+}/F_{*}$ is good measure of 
the actin turnover rate. The experimental data analyzed by Stachowiak and 
O'Shaughnessy for real stress fibers correspond to $r \ll 1$. In this 
limit, they \cite{stachowiak08} predict that the relaxation time 
$\tau_{sarc}$ for the sarcomere length is $\tau_{sarc} = F_{*}/(k_{t} V_{+})$. 

Contact of salmonella bacteria with a host cell can activate the formation
of a contractile acto-myosin machinery that resembles stress fibers.
\cite{haenisch12}.
Contraction of this machinery generates sufficiently strong force that
pulls the bacterium inward thereby driving the invagination of the host
cell.
The sarcomeric organization is not essential for the contractility of
actomyosin crossbrige. Contractile actomyosin bundles, without sarcomeric
organization, can arise from buckling of actin filaments
\cite{lenz12a,lenz12b}.

\subsection{\bf Sliding MTs by axonemal dynein and beating of flagella}
\label{sec-flagellabeat}

The molecular composition, structure and dynamics of eukaryotic flagella
are totally different from those of bacterial flagella. Moreover,
structurally, eukaryotic flagella and cilia are qualitatively similar
cell appendages; their quantitative differences lie in their size and
distribution on the cell. Therefore, in the past suggestions have been
made(see, for example, ref.\cite{margulis80}) that eukaryotic flagella
and cilia should be called ``undulopodia'' because of their common
undulatory movements.

In section \ref{sec-IFT} we have already reviewed intraflagellar transport
(IFT).
In this subsubsection we consider only the physical processes driven by
the cytoskeletal filaments and the motors which lead to the beating
of the flagella. How the various patterns of these beatings in a
fluid medium propels the eukaryotic cell is a problem of fluid dynamics
and will not be discussed in this article. Fir historical developments
in the research on the machinery causing the beating of cilia and
flagella, see for example, refs.
\cite{gray29,sleigh62,gibbons81,bloodgood10,satir95}.

A cilium (or eukaryotic flagellum) has a very special organization of
MTs and axonemal dyneins.
\cite{ginger08,lindemann10,fisch11}
The core of the machinery that drive ciliary beating is the {\it axoneme}.
It consists of parallel doublet of MTs. Normally 9 such ``outer doublets''
are arranged so as to form the outer surface of a cylinder. Inside this
cylinder, usually a pair of ``singlet'' microtubules runs along the axis
and there are spokes that radially extend towards each outer doublet.
Let us label the doublets by integer $j$ (j=1,...,N) where $j$ increases
in the clockwise direction when viewed from the basal end of the axoneme.
Rows of axonemal dynein form ``crossbridges'' between successive doublet,
i.e., doublet $j$ with the doublet $j+1$ (j=1,...,N). Driven by ATP
hydrolysis, each row of dynein slide the two doublets $j$ and $j+1$ with
respect to each other. This sliding gets converted to a bending of the
cilium because of their anchoring at the basal body and other linkages.
In spite of these general features, wide variations have also been
observed in the structures of cilia and flagella \cite{mencarelli08}.

Many investigators have made important contributions in the theoretical
modeling of flagellar and ciliary beating
\cite{lindemann10}
There are some superficial similarities between muscle contraction and
flagellar beating- both are driven by sliding of filaments by molecular
motors \cite{brokaw75}. A sliding filament model for flagellar beating
was suggested by Brokaw \cite{brokaw72}.

Beating requires a ``switching'' phenomenon. Two different types of
switching can be envisaged: (i) switching at {\it temporal} ``switch
points'', and (ii) switching at {\it spatial} ``switch points''
\cite{brokaw09}.
In order to complete the model, the mechanism of the switching has to 
be incorporated. Brokaw \cite{brokaw72,brokaw75} 
proposed a {\it curvature control} model based on the hypothesis that 
when the flagellum bends up to a critical curvature, it triggers the 
inactivation (switching OFF) of one set of dyneins and activation 
(switching ON) of the opposing set of dyneins. 

Lindemann \cite{lindemann94a,lindemann94b,lindemann95,lindemann04,lindemann07} 
suggested a {\it geometric clutch} hypothesis that, at first sight, may 
appear an alternative mechanism of switching. In this scenario,  
in the intact straight axoneme the dyneins are far enough from their 
respective binding sites so that practically no crossbridge is found. 
In contrast, when a flagellum bends the stretching of the nexin links, 
that hold the outer doubles in a ring-like geometry, generates a force 
transverse to the bend (t-force). This t-force squeezes some MT doublets 
close enough that dyneins can now form crossbridges between them. However, 
the dyneins now generate a torque that pushes the doublets apart thereby 
disengaging the active dyneins (inactivation of the crossbridges that 
they formed) and engaging their opposing set of dyneins (activation of 
another set of crossbridges). Because of their obvious analogy with 
clutches, this switching mechanism is called geometric clutch hypothesis.
 
Alternative hypotheses for switching are based on either control exerted 
through the central-pair spoke \cite{omoto80,omoto99}, or coordinated 
with the dynein crossbridge cycle \cite{riedelkruse07}. 
The role of the central pair of MTs is not fully understood, particularly 
because some cilia do not possess these central MTs \cite{mitchell04}.

The special 9+2 or 9+0 design of the axoneme and dynein-driven sliding 
of the axonemal MTs cause the  beating of cilia and flagella. However, 
for similar beating of a bundle of MTs much fewer molecular components 
and simpler design seems to be adequate \cite{sanchez11}. In their 
{\it in-vitro} studies, Sanchez et al. \cite{sanchez11}, used only the 
following main components: (i) a cluster of biotin-labelled kinesin motors 
by binding with multimeric streptavidin, (ii) taxol-stabilized MTs, and 
(iii) polythylene glycol (PEG). The bundling of MTs is induced by PEG 
and relative sliding of the MTs in a bundle are driven by the artificially 
constructed multimeric kinesin. The active MT bundles are attached to a fixed 
boundary that serves as the counterpart of the basal body to which the 
axonemes are attached in real eukaryotic cilia and flagella. In spite of 
several crucial differences in the components and design, the beat patterns 
of the active MT bundles are very similar to those of cilia and flagella 
\cite{sanchez11}. The generic beating of the active filamentous bundles has 
been predicted theoretically \cite{camelet99,camelet00} (see also 
ref.\cite{hilfinger08,hilfinger09}).

\subsection{\bf Sliding MTs by dynein and platelet production}
\label{sec-platelet}

{\it Megakaryocytes} are precursor cells that reside primarily in the 
bone marrow. Remodeling of each megakaryocyte through a complex series 
of processes leads to the formation of thousands of {\it platelets} 
that are released into the bloodstream 
\cite{patel05,hartwig06,italiano07,thon10a}. 
The sequence of these processes begins with the formation of a long 
protrusion of the megakaryocyte that serves as the site of organization 
of a {\it proplatelet}, the precursor of a platelet. These protrusions  
elongate, become thinner, and branch out to form tubular projections. 
Alignment of many microtubules within the proplatelet leads to the 
formation of a bundle just under the cell cortex. These bundles loop 
around forming buds at the tips of the proplatelets. 

Although the MT bundle keeps growing in length by ongoing MT 
polymerization, the proplatelet enlargement is not driven by piston-like 
action of the polymerizing MTs. In fact, the plus ends of the MTs are 
dispersed throughout the cortex of the proplatelet and not all the MTs 
are oriented parallelly in the bundle \cite{hartwig06,italiano07,thon10a}. 
It is the sliding of the MTs relative to each other by dynein motors that 
is believed to be responsible for the growth of the proplatelets. 
Thus, platelet formation can be divided roughly into three phases: 
(i) emergence of the protrusion of a megakaryocyte thereby initiating 
the formation of a proplatelet, 
(ii) elongation, thinning and branching of the proplatelet, and 
(iii) release of the platelets from the tips of the proplatelets. 
The general principles of cell protrusions driven by cytoskeletal 
filaments and associated motors will be discussed in the next section. 

In this section we briefly mention the key ingredients of a 
computational model that has been developed very recently \cite{thon12} 
for primarily the stage (ii) of the process, namely the emergence of 
the shape of the proplatelet. This model is based on the assumption 
that the shape of the proplatelet is determined by a balance of the 
forces acting on the MT bundle that runs along its periphery. 
The MT bundle is modeled as node-spring loop where each node interacts 
with two adjacent neighbors on its two sides. Both stretching and bending 
result in restoring forces. The extension of the MT bundle is implemented 
by gradual elongation of the rest lengths of all the springs in the 
loop; the plausible microscopic physical origins of these elongations 
(e.g., MT sliding and polymerization) are, however, not incorporated 
explicitly. Bundling proteins, that ``zipper'' the MT bundles on the 
opposite sides of a narrow corridor of the barbell-shaped proplatelet, 
are mimicked by transient elastic bonds. The compressive force exerted 
by the cell cortex is captured by an effective pressure $P$. Using this 
model, Thon et al. \cite{thon12} demonstrated the effects of the initial 
perimeter of the proplatelet and the number of MTs in the bundle on the 
transition from spherical to barbell shaped proplatelet.

\subsection{\bf Sliding MTs by kinesin-5}
\label{sec-slidebyEg5}

Kinesin-5 is homotetrameric in the sense that it has four identical motor 
domains with a pair of motor domains at each end of a rod-like stalk. 
Most of the {\it in-vitro} experiments on kinesin-5 have been performed 
with Eg5, a member of this family. It is a plus-end directed motor. 
Each kinesin-5 motor can crosslink a pair of MTs such that the two pairs 
of its heads, located at the two ends of the stalk, walk on the two 
different MTs that are crosslinked by it \cite{kaseda09}.  

From {\it in-vitro} experiments with polarity-labelled MTs, Kapitein et al. 
\cite{kapitein05} established that parallel MTs crosslinked by Eg5 remain 
practically static whereas an Eg5 crosslinker slides two mutually 
antiparallel MTs. In the latter case the sliding velocity $2V$ of the 
crosslinked antiparallel MTs arises from the fact that one pair of its 
heads walk on one of the crosslinked MTs with a velocity $V$ while the 
other pair of heads walk with a velocity $-V$ on the oppositely oriented MT.  
Eg5 gets activated and its directional motility gets triggered only 
when it crosslinks two MTs \cite{kapitein08}. This is very similar to 
the activation of kinesin-1 by cargo binding \cite{verhey09};  
one of the two MTS cross-linked by the kinesin-5 can be viewed as the 
cargo for kinesin-5 while the other is regarded as a track \cite{kapitein08}.

The structural transitions and chemical steps in the ATPase cycle of 
individual Eg5 motors have been monitored simultaneously \cite{jun10}. 
Monitoring the domain movements of Eg5, using FRET as the probe, Rosenfeld 
et al. \cite{rosenfeld05} proposed a kinetic scheme for the ATPase cycle  
of individual heads (motor domains) of Eg5.

So far as the mechanism of force generation and stepping pattern is 
concerned, a comparison of the members of kinesin-5 and kinesin-1 
families has been reported \cite{valentine06a}. For example, at a given 
ATP concentration, Eg5 is much slower than kinesin-1 although the 
ATP-dependence of both follow the same Michaelis-Menten equation 
\cite{valentine06a}. 
In order to understand the mechanisms of Eg5 and compare it with 
conventional dimeric kinesin-1, dimeric Eg5 (essentially, a truncated 
`half Eg5'') have been constructed genetically. At first sight, the 
mechano-chemical kinetics of such a dimeric Eg5 might be expected to be 
similar to that of a kinesin-1. But, biochemical experiments as well as 
single-molecule manipulations have revealed that the dimeric Eg5 is 
(i) slower, (ii) less processive, and (iii) less sensitive to load force 
than dimeric kinesin-1 
\cite{krzysiak06,krzysiak08,valentine06b,valentine07,valentine09} 
These differences may be consequences of the differences in the stiffness 
and docking/undocking of their neck linkers \cite{rosenfeld05}. 
These structural differences may be the results of their evolutionary 
adaptation for distinct functional roles-  full length tetrameric Eg5 
motors usually work as MT sliders in small groups whereas dimeric 
kinesin-1 work as lonely porters \cite{valentine06b}. 
A 5-state kinetic model has been proposed for the stepping of dimeric 
Eg5 during a processive run \cite{valentine09}.

In order to understand the relevance of the coordination of the two heads 
of dimeric Eg5 constructs, Kaseda et al. \cite{kaseda08} engineered an 
even further truncated Eg5 construct that has only a single head. 
Based on their experimental observations on this single-headed Eg5, 
they claimed that MT sliding driven by full tetrameric Eg5 is 
very similar to sliding of actin filaments by myosin-II in muscles. 
For any Eg5, they claim, only one of the two heads interacting with a 
MT generates force while the other is redundant; coordinated hand-over-hand 
processive walking of the two heads of Eg5 are rare events \cite{kaseda08}.
These claims are not consistent with the results obtained from single 
molecule experiments on dimeric Eg5. Plausible reasons for these 
discrepancies have been listed by Kaseda et al.\cite{kaseda09}.

\subsection{Section summary}

In this section we have reviewed several different motor-filament 
crossbridge system that display striking similarities of motor-induced 
sliding of cytoskeletal filaments. 

Interestingly, specific power output of muscles and eukaryotic flagella
are comparable. For both, large number of motors collectively slide
cytoskeletal filaments although the force producers are quite different.
In contrast, the specific power output of the cytokinetic furrow is few
orders of magnitude lower than that of muscle in spite of the fact that
both are acto-myosin systems. This difference may be a consequence of
the widely different density of the myosin motors in the two systems.

\section{\bf Push / pull by polymerizing / depolymerizing cytoskeletal filaments: specific examples of nano-pistons and nano-hooks}
\label{sec-specificpistonhookspring}

Earlier in section \ref{sec-genericpistonhookspring}, we have 
discussed only a few generic models that account for the pushing and 
pulling forces generated by nano-pistons and nano-hooks, respectively. 
In this section we discuss more explicit models for the force 
generation by cytoskeletal filaments in eukaryotes as well as that 
by their prokaryotic homologs by taking into account some of the key 
specific features of their structure and dynamics. 
The polymerizing filaments ``polarize'' cells, form dynamic cell 
``protrusions'' and drive the engines of motility of single cells 
as well as collective migration of a group of cells.

Unicellular microorganisms have developed diverse molecular mechanisms
of locomotion. The actual mechanism used by a specific type of organism
depends on the nature of the environment in the natural habitat of the
organisms. If a micro-organism lives in a bulk fluid, it's natural mode 
of motility is {\it swimming}. In contrast, if a micro-organism lives in 
a thin fluid film close to a solid surface (i.e., in a wet surface),
{\it gliding} should be its mechanism of movement 
\cite{spormann99,heintzelman06,mauriello10,mcbride01,nan11}. 
Of course, some micro-organisms may be capable of utilizing both these 
modes of motility.

Unicellular eukaryotes, like free-living protozoa, move primarily for
food. In multicellular prokaryotes, cell locomotion is essential in
development. Moreover, leukocytes move to offer immune response.
Furthermore, fibroblasts, which are normally stationary, move during
wound healing. Swimming, gliding and crawling are some of the most
common modes of motility of eukaryotic cells.

One of the fundamental questions on cell motility is the molecular
mechanisms involved in the generation of required forces. Broadly
speaking, three different mechanisms have been postulated and their
possibility in specific contexts have been explored: (i) Force
generated by polymerization of cytoskeletal protein filaments
(actin and microtubules), (ii) Force generated by cytoskeletal motors
by their interactions with filamentous tracks, and (iii) forces of
osmotic of hydrostatic origin.

Since our aim here is limited to a discussion of the mechanisms of 
force generation by the cytoskeletal filaments and their prokaryotic 
analogs, and since a detailed discussion of cell motility 
is beyond the scope of this review, we'll explain these phenomena 
only briefly and provide relevant references to the literature in the 
appropriate contexts.

\subsection{\bf Force generated by polymerizing microtubules in eukaryotes}

Pushing force generation by polymerizing MTs has been investigated 
experimentally for the last one and half decades (for reviews, see 
ref. \cite{kalisch11,oster04,theriot00,mogilner99b,dogterom07,laan10,carlier10b,erickson10,nudleman04,tao98}).
A normal MT consists of 13 protofilaments. If the tips of all the 
protofilaments always touched the same obstacle, one could replace 
the MT by a single rigid rod. However, in reality, only a fraction 
of these protofilament may touch the obstacle at a time and this 
fraction may fluctuate because of the stochastic kinetics of the 
polymerization process. Thus, the load force is shared by only 
those protofilaments that touch the membrane. At any instant, those 
load-bearing protofilaments are too close to the obstacle to polymerize 
because the gap in between the obstacle and their tips are not wide 
enough to accomodate a $\alpha-\beta$ tubulin dimer. However, by 
supporting a larger share of the load, these protofilaments ``subsidize'' 
the growth of those neighboring filaments whose tip is farther from 
the obstacle. The Brownian ratchet model \cite{peskin93a} of force 
generation by filament polymerization, which we sketched in section 
\ref{sec-genericpistonhookspring}, was appropriately modified by 
Mogilner and Oster to incorporate this ``subsidy effect''  \cite{mogilner99b}.

\begin{figure}[htbp]
\includegraphics[angle=90,width=0.8\columnwidth]{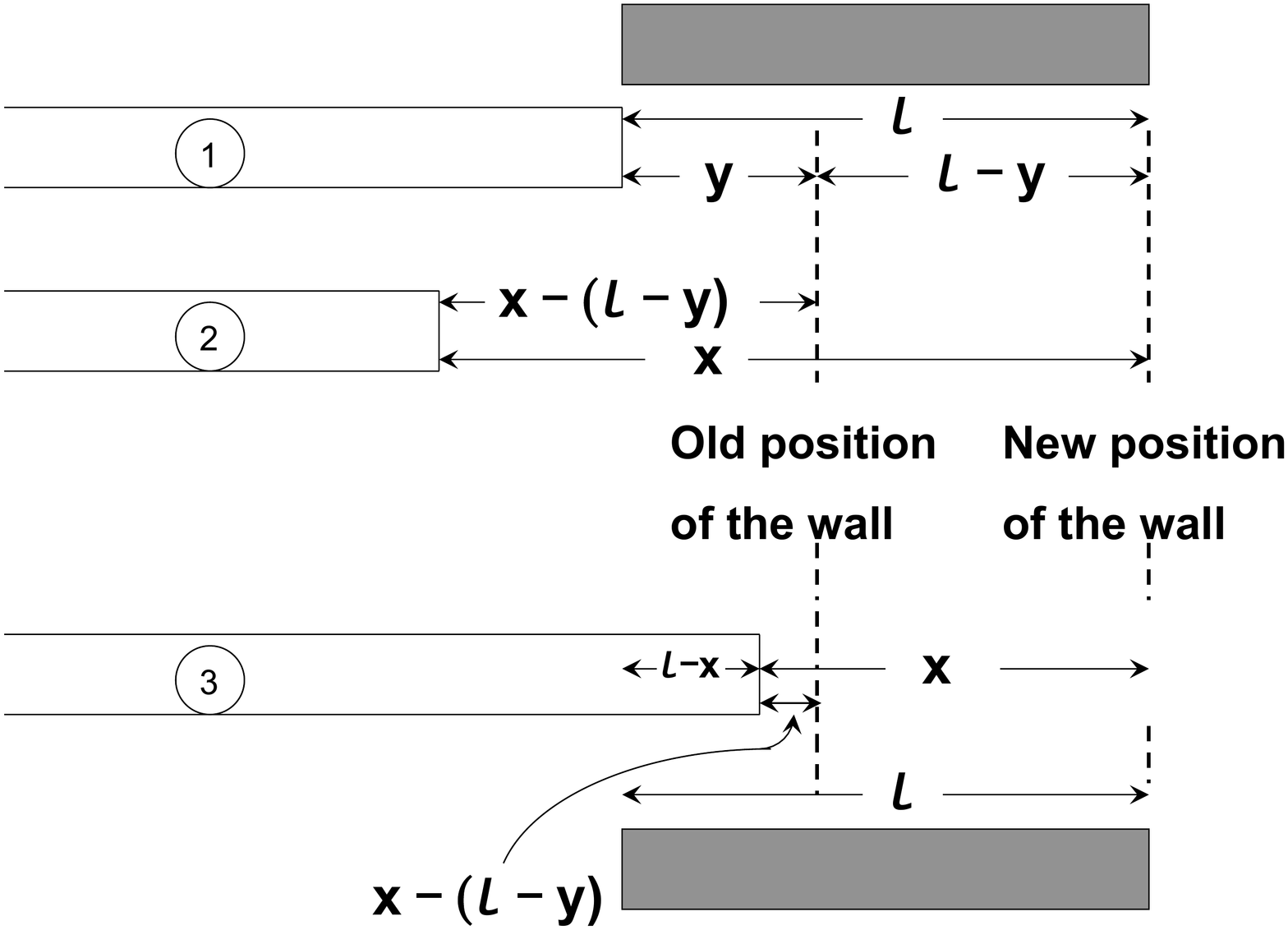}
\caption{The ``subsidy effect'' in MT polymerization where ${\ell}$ 
denotes the length of a $\alpha-\beta$ tubulin dimer. The protofilament 
1 was initially at a distance $y < {\ell}$ from the leading tip; 
attachment of a $\alpha-\beta$ tubulin dimer to its tip causes it to 
become the leading tip and, thus, the leading tip advances by a distance 
${\ell}-y$. Although the protofilaments 2 and 3 have neither elongated 
nor shortened, the position of their tips with respect to that of the 
leading tip has changed because of the elongation of the protofilament 1. 
The protofilament 2 now finds its tip at a new distance $x > {\ell}$ from 
the new leading tip whereas its tip was originally at a distance 
$x-({\ell}-y)$ from the old leading tip. In contrast, the protofilament 3 
now finds its tip at a new distance $x < {\ell}$ from the new leading tip 
whereas its tip was originally at a much shorter distance $x - ({\ell}-y)$.   
(adapted from ref.\cite{mogilner99b}).
}
\label{fig-subsidy}
\end{figure}

Mogilner and Oster \cite{mogilner99b} assumed that the longest 
protofilament can support the MT against the membrane. In this 
one-dimensional model the origin of the coordinate system is 
placed at the tip of the longest protofilament (see fig.\ref{fig-subsidy}). 
Suppose, $N(x,t)$ represents the number of protofilament tips at a 
distance $x$ from the tip of the leading protofilament at time $t$, 
where $x=\{({\ell}/13)j\}$, with $j=0,1,...$ labelling the 
protofilaments and ${\ell}=8$nm is the length of a $\alpha-\beta$ 
tubulin dimer. The protofilaments within a distance ${\ell}$ from 
the membrane are identified as the ``working protofilaments'' 
\cite{mogilner99b}. If the tip of a protofilament is at a distance 
$y$ from the tip of the leading protofilament just before a 
$\alpha-\beta$ tubulin dimer assembles on its tip, then the leading 
tip is advance by a distance ${\ell}-y$ (see fig.\ref{fig-subsidy}).  

In order to get the continuum limit $N(x,t)$ is scaled to 
the density $n(x,t) = N(x,t)/({\ell}/13)$ with the constraint 
$N = \int n(x,t) dx$ arising from the conservation of the total 
number of tips. $k_{on}$ and $k_{off}$ denote the rates of attachment 
and detachment of tubulin dimers at a MT tip far from the wall. 
In the presence of a load force $F$, the rate of growth is altered 
from $k_{on}$ to $\kappa_{on}(F,y)=k_{on} exp[F(y-{\ell})/(k_BT)]$.  
The rate equations for $n(x,t)$ are \cite{mogilner99b} 
\begin{eqnarray} 
\frac{\partial n(x,t)}{\partial t} &=& k_{on} n(x+{\ell}) + k_{off} n(x-{\ell}) 
- (k_{on}+k_{off}) n(x) \nonumber \\
&+& \underbrace{\int_{0}^{\ell} \kappa_{on}(F,y) n(y) n(x+y-{\ell}) dy}_\text{Gain of tips at x because of polymerization of tips at y} \nonumber \\  
&-& \underbrace{n(x) \int_{0}^{\ell} \kappa_{on}(F,y) n(y) dy}_\text{Loss of tips at x because of polymerization of tips at y}, ~{\rm for}~ x \geq {\ell} \nonumber \\ 
\end{eqnarray}  
(as depicted by the protofilament 2 in fig.\ref{fig-subsidy}), and 
\begin{eqnarray} 
\frac{\partial n(x,t)}{\partial t} &=& k_{on} n(x+{\ell})  
- k_{off} n(x) \nonumber \\
&+& \underbrace{\int_{{\ell}-x}^{\ell} \kappa_{on}(F,y) n(y) n(x+y-{\ell}) dy}_\text{Gain of tips at x because of polymerization of tips at y} \nonumber \\  
&-& \underbrace{n(x) \int_{0}^{\ell} \kappa_{on}(F,y) n(y) dy}_\text{Loss of tips at x because of polymerization of tips at y}, ~{\rm for}~ 0 \leq x < {\ell} \nonumber \\ 
\end{eqnarray}  
(as depicted by the protofilament 3 in fig.\ref{fig-subsidy}).  
The average velocity of the membrane protrusion can be calculated from 
\begin{equation}
V(F) = - k_{off} {\ell} + \int_{0}^{\ell} ({\ell}-x) \kappa_{on}(F,x) n(x) dx
\end{equation} 
Mogilner and Oster \cite{mogilner99b} obtained not only the force-velocity 
relation (relation between the average velocity of MT growth as a function 
of the load force), but also the steady-state distribution of the tips of 
the protofilaments for a few distinct strengths of the load force. 
Predictions of both the Brownian ratchet theory of pushing force 
generated by polymerizing MTs, both with and without proper accounting 
of subsidy effect have been the subjects of many experimental studies 
\cite{dogterom07,kalisch11}.

\subsection{\bf Force generated by polymerizing actin: dynamic cell protrusions and motility}

Proper biological function of a cell requires appropriate spatio-temporal 
organization. For spatial organization within the cell, the number, size, 
shape and the internal environment of the organelles need to self-organize.

\subsubsection{\bf Force generation and cell protrusion by actin polymerization}

Actin filaments are much more flexible than MT filaments. Moreover, 
these not only have a double-helical structure but also form a 
branched network. The Brownian ratchet model \cite{peskin93a} was 
extended to the ``elastic Brownian ratchet'' \cite{mogilner96a} by 
incorporating the interplay of bending elasticity and thermal 
fluctuations in the model. The space in between the obstacle and 
the filament can be created by the thermally induced fluctuations 
of the actin filaments rather than that of the obstacle.
In many real situations, a subpopulation of the actin filaments 
remain attached to the obstacle while the remaining population 
are detached from it. The tethered filaments hold the obstacle 
while the free filaments can push it by polymerization. By capturing 
these two distinct populations separately in the model, the elastic 
ratchet model was extended by Mogilner and Oster to the ``tethered 
ratchet model'' \cite{mogilner03b}.

One distinct feature of the actin filaments is that new branches 
can nucleate on existing filaments thereby creating a branched 
network. Such branching  has been modeled by Carlsson 
\cite{carlsson01,carlsson03}.  
A mesoscopic model for force generation by actin polymerization was 
developed by Gerbal et al.\cite{gerbal00}. In this model, the 
polymerized network of the actin filaments is described as a gel 
\cite{sykes10}
and treated as a continuous elastic medium that is anchored to 
the surface of the obstacle. Growth of the actin filaments is 
captured as addition of new layers on the gel causing compression 
of the previously formed layers. The release of the stored elastic 
energy relaxes the gel and pushes the barrier. Thus, the free energy 
of polymerization does not directly push the barrier; it is first 
stored as elastic energy which, in turn, performs the mechanical work.

Forces generated by actin polymerization \cite{zhu06} not only gives rise to 
dynamic cell protrusions at the leading edge of a cell 
(see table \ref{table-dynprotru})
but also plays the central role in actin-based cell motility  
\cite{condeelis93,chhabra07,insall09,ridley11,revenu04,small10,faix09,gupton07,mattila08,weber06b,ayala06,buccione04,murphy11,derosier00,pollard02,cramer10,huber08,lee08b,lee09c}. 
as well as in the collective migration of a group of cells 
\cite{ilina09,friedl09,rorth09,rorth11,weijer09,aman10}. 
Actin-based cell protrusions also play important roles in growth and 
morphogenesis of non-motile organisms like filamentous fungi 
\cite{berepiki11}.
Crawling of animal cell results from a coordinated cycle of three key 
processes: (i) formation of cell protrusions in the forward direction, 
(ii) adhesion of the cell to a solid substrate, and (iii) retraction 
from the rear. However, in this review, we restrict our discussions 
only to the role of force generators in creating cell protrusions. 

The actin-based protrusions of eukaryotic cells 
\cite{condeelis93,chhabra07,insall09,ridley11,revenu04,small10,faix09,gupton07,mattila08,weber06b,ayala06,buccione04,murphy11,derosier00,pollard02,cramer10,huber08}. 
can be broadly divided into two categories based on the nature of the 
actin networks: 
(i) branched arrays, and (ii) actin bundles \cite{nambiar10}. 
The actin networks of lamellipodia of crawling cells and the invadopodia 
of cancer cells are common examples of branched arrays. In contrast, 
the crosslinked parallel filaments in filopodia, microvili and sterocilia 
are examples of actin bundles. 
Filopodia protruding from a lamellipodium is not uncommon.
Other protrusions of the eukaryotic cell include
pseudopodia, ruffles, microvilli, invadopodia, etc.  Although the key 
role in the formation of these protrusions is played by actin, the 
myosin motors also participate in the process, particularly in 
maintaining the polarity \cite{nambiar10}.

\begin{table}
\begin{tabular}{|c|c|c|} \hline
Protrusion & Width  & Duration \\\hline
Lamellipodium & 0.1-0.2 $\mu$m & Minutes \\ \hline
Filopodium & 0.1-0.3 $\mu$m & Minutes \\ \hline
Podosome & 0.5-2.0 $\mu$m & Minutes \\ \hline
Invadopodia & 0.5-2.0 $\mu$m & Hours \\ \hline
\end{tabular}
\caption{Width and duration of dynamic cell protrusions (adapted from .
ref.\cite{murphy11} }
\label{table-dynprotru}
\end{table}

Two different models for the formation of the actin-bundles
of filopodia have been proposed. In the ``convergent elongation model'',
the filopodial actin filaments are assumed to originate from the
lamellipodial actin network. But, in the ``de novo filament nucleation
model'', the filopodial actin filaments are assumed to nucleate
separately in the filopodia. 

A common feature of the actin networks in lamellipodia and filopodia is 
that the fast growing (barbed) ends of the actin filaments are oriented 
towards the membrane which gets pushed by the piston-like action of the 
polymerizing actin filaments. This piston-like pushing by polymerizing 
actin is very similar to piston-like action of polymerizing microtubules, 
which we discussed earlier, except that actin can form branched structures 
whereas microtubules do not.  
Moreover, actin-capping proteins \cite{cooper08} can block the barbed ends 
of the actin filaments thereby increasing the concentration of monomeric 
actin which can be channelled towards the faster growth of non-capped 
actin filaments \cite{clainche08}. But, processive cappers like formin 
protect the barbed ends from capping proteins thereby enabling the 
formation of long actin filaments \cite{watanabe04}. The role of the 
elastic energy of the formin-capped barbed end in the diversity of the 
actin polymerization rates have been studied theoretically \cite{shemesh07}. 
From the perspective of transport, long-distance movement of a processive 
capping protein at the tip of a polymerizing actin filament can be viewed 
as a molecular motor powered by actin polymerization \cite{watanabe04}.
Theoretical models developed for the dynamic cell protrusions have been 
reviewed extensively
\cite{mogilner96a,mogilner96b,mogilner03b,carlsson00,carlsson01,gholami08,burroughs05,mogilner02,grimm03,keren08,mogilner05a,mogilner06b,atilgan05,atilgan06,murphy11}.

\noindent$\bullet${\bf Force generated by depolymerizing MSP in nematodes}

The uterus of a nematode female is normally packed densely with eggs. 
Therefore, it would be extremely difficult for a nematode sperm to 
swim under these conditions. Perhaps, that is the reason why, instead 
of swimming like in other eukaryotes, the sperm cell of nematodes crawl 
\cite{singaravelu11}. 
However, a more interesting feature of nematode sperm motility is the 
fact that it does not possess actin! Instead, a protein, called major 
sperm protein (MSP), acts like actin \cite{theriot96,roberts00} forming dynamic 
filaments which drive the motility of the cell \cite{italiano01}. 
However, in contrast to actin filaments, the individual filaments of 
MSP have no polarity. Therefore, these cannot serve as tracks for any 
motor proteins (or their analogs). 

There are patches on the individual filaments of MSP which cross-link 
filaments into ``bundles''. Ideas similar to ``tethered ratchet'', 
developed originally for actin filaments, could be adapted to explain 
the mechanism of cell protrusion by MSP filament bundles in the leading 
edge of the nematode sperm 
\cite{bottino02,wolgemuth05b,mogilner03c,demekhin09}. 
However, in the absence of myosin and analogous motor proteins, an 
altogether different mechanism had to be invoked for the observed 
retraction of the rear of the nematode sperm \cite{shimabukaro11}. 
This mechanism of force generation by {\it depolymerizing} MSP bundles 
is a physical realization of the generic mechanism outlined \cite{sun10} 
in section \ref{sec-genericpistonhookspring}. 
A pH gradient was postulated to regulate MSP assembly at the leading 
edge and disassembly at the rear. More recently, biochemical kinetics 
that regulates the biophysical processes of force generation has been 
incorporated in a model \cite{stajic09}. Besides, an alternative plausible 
scenario of MSP assembly has been proposed \cite{dickinson07} to explain 
the motility of nematode sperm cells. 
 
\noindent$\bullet${\bf Force generation by depolymerization of type-IV pili in bacteria}

Myxobacteria have two different types of engines at their two poles. 
One of these assembles type IV pili, a class of dynamic appendages, 
whose retraction propels the cell forward and the corresponding 
mode of motility is called twitching
\cite{mattick02,merz02,whitchurch06,proft09,wall99,kaiser10}. 
A typical type IV pilus is 6 nm thick and can 
extend up to about 5 $\mu$m from the surface of the cell. In rod-shaped 
bacteria, these appendages are normally located at the cell poles. 
A pilus is a polymeric helical filament consisting of pilin subunits. 
\cite{mattick02,merz02,pelicic08,allen12,craig08,kaiser00,kaiser08,burrows05}

Suppose, polymerization of a pilus is energy-consuming whereas the 
depolymerization is a spontaneous process. PilT, an ATPase, could 
catalyze the removal of a stabilizing cap at the base of the pilus 
thereby triggering its depolymerization from the base. This process 
of retraction of the pilus exerts pulling force on the surface to 
which its distal tip is tethered. 
Alternatively, suppose the polymerization of a pilus is spontaneous. 
Then, PilT can peel off subunits from its base in an ATP-dependent 
manner causing its retraction. In both these alternative scenario, 
the energy consumed does not directly pull the pilus towards the 
membrane. Instead, energy input assures its depolymerization at the 
base. Therefore, these have been called Brownian ratchet mechanism 
and facilitated ratchet mechanism, respectively \cite{merz02}. 
In the power stroke mechanism, PilT walks like a ATP-fuelled motor 
towards the distal end of the pilus and its stepping forces the 
pilus directly towards its base \cite{merz02}.

\noindent$\bullet${\bf Force generation by actin comets for motility of bacterial pathogens}

A classic example of actin-based cell motility is that of the 
intracellular bacterial pathogens, like {\it Listeria monocyteogenes}, 
that are propelled by ``actin comets'' 
\cite{theriot95,ireton97,dramsi98,cameron00,goldberg01,merz03,cossart04,cossart08,lambrechts08,haglund11}
(for an historical account of its discovery, see ref\cite{portnoy12}). 
In this case a comet-like tail of polymerizing actin filaments push
the pathogen in the host cell. Unlike, cell crawling, which is also
driven by actin-polymerization, neither adhesion to a solid substrate
nor retraction of the rear of the cell is required.

\subsection{\bf Cell polarization: roles of cytoskeletal filaments and motors}

Most of the living cells get ``polarized'' \cite{rafelski08}. 
{\it Polarization} of a cell is defined as a ``redistribution of multiple 
proteins and lipids in the cell'' \cite{jilkine11}. 
Two essential properties of cell polarity are \cite{li08a}: 
(i) asymmetric distribution of mobile molecular species between two opposite 
poles of the cell; and 
(ii) the oriented organization of intrinsically polar cytoskeletal filaments 
(e.g., microtubules and actin filaments) along the axis of polarity. 
Polarization leads to the formation of cell protrusions which are used by 
a wide variety of cells for their motility. 
The ``universal'' features of polarized cells (i.e., features shared by most 
of the polarized cells) have been listed recently \cite{jilkine11}. 
Establishment, as well as the subsequent maintenance, of cell polarity 
depend, at least partly, on the dynamic assembly of the cytoskeletal 
filaments and the motor-driven transport that they support 
\cite{nelson03c,li08a,mullins10,siegrist07}. 

Establishment of polarization of a cell may be viewed as a symmetry 
breaking phenomenon \cite{li10a}. 
In biology, symmetry breaking can take place at several different 
levels of organization; at each level, the asymmetry can be attributed 
to underlying asymmetries at a lower level of organization. For example, 
the polarity of the polar cytoskeletal filaments arise from the 
asymmetry in the structure of their subunits and in the kinetics of 
their polymerization. Similarly, the polarity of the cytoskeletal 
filaments plays key roles in generating polarity of the cell which, 
in turn, leads to the asymmetries at the levels of tissue and the 
organism \cite{li10a}. 

The process of establishing polarity of a cell can get assisted by 
either {\it internal} processes or by its interaction with the {\it 
external} environment or by a combination of the two. In the absence 
of any external cause, symmetry of a cell can be broken by amplification 
of the spontaneous internal fluctuations \cite{li10a,soldner03}. 
Alternatively, a mechanical stress generated by the extracellular matrix 
or a bio-chemical signal sent by the surrounding aqueous medium (or 
a combination of chemo-mechanical interactions of the cell with its 
environment) can polarize it \cite{gucht09}.
Whether a cell can polarize spontaneously or only in response to an 
external asymmetric signal may depend crucially on the geometry of the 
organization of the cytoseletal filaments \cite{hawkins09}. 
The cell membrane is likely to play am important role in this process. 
For example, phase segregation of membrane proteins can give rise to 
cell protrusions if these proteins assist force generation by promoting 
actin polymerization \cite{gov06a,gov06b,veksler09,kabaso11}.   
Different alternative approaches to mathematical modeling of cell 
polarization have been reviewed and compared in recent years 
\cite{onsum09,jilkine11,mogilner12}. 

Once a cell becomes polarized, how long does the polarity persist? 
The answer to this question depends on the type of the cell. 
Some cells utilize this polarity for their motility either to chase their 
enemies or to search for a mate. For such cells, ability to respond to 
temporal variations in external stimuli and corresponding adaptation 
requires changes in the polarization pattern. Transient cell protrusions 
can lead to the formation of a nano-tubes 
\cite{davis08,gerdes07,gurke08,hurtig10} connecting two cells for 
their communications \cite{rorth03}.
In contrast, for some cells, like neuron, polarity needs to be maintained 
stably throughout the life time of the cell. Therefore, establishment of 
the polarized structure of the cell is a part of its morphogenesis. 
Interestingly, a single cell exhibits all the hallmarks of development 
of an entire multicellular organism, viz., anterior-posterior asymmetry, 
dorsal-ventral asymmetry as well as the formation of the overall pattern 
\cite{marshall11}.  

\subsection{Section summary}

In this section we have reviewed force generation by polymerizing and 
depolymerizing cytoskeletal filaments. A single MT is a stiff linear 
nano-tube; its polymerization involves an interesting cooperativity 
in the load-sharing by the leading protofilaments. In contrast, actin 
can form either bundles or branched network of filaments; cooperativity 
of the polymerization-depolymerization kinetics of these filaments 
dominate the emerging morphology and motility of a cell.

\section{\bf Mitotic spindle: a self-organized machinery for eukaryotic chromosome segregation}
\label{sec-mitosis}

The asters and vortices observed in {\it in-vitro} mixtures of tubulins 
and motors, 
\cite{sharp00a,nedelec97,nedelec01,surrey01,nedelec03,karsenti06,karsenti08,vorobjev01,cytrynbaum04}
have been theoretically demonstrated to be the emergent patterns 
\cite{bassetti00,lee01a,kim03,sankararaman04,ziebert05,aranson05,aranson06}.  
Such patterns are generic \cite{kruse04,zumdieck05} and the 
corresponding acto-myosin system forms not only asters, but also rings, 
and various typesof networks 
\cite{backouche06,smith07}.
In this section we review a self-organized machine called mitotic spindle; 
it may be regarded as a system of two interacting asters, and it generates  
forces that drive chromosome segregation in eukaryotic cells.

In eukaryotic cells 
chromosome segregation is preceded by the replication and condensation 
of chromosomes which lead to the formation of sister chromatids. 
The complex process whereby the sister chromatids in eukaryotic cells are 
finally segregated is called {\it mitosis} 
\cite{mitchison01,rieder03,scholey01,inoue08b,scholey03a,aist99,bloom08,dumont11,mcintosh12}  
and is carried out by the mitotic spindle  
\cite{karsenti01,gadde04,smith04,kwon04a,bouck08,bloom10a,nicklas88a,inoue95,walczak08,scholey10a,mogilner10,duncan11,zhang11b}.
A similar machinery, called the {\it meiotic spindle}, runs the 
related process of {\it meiosis}. 
The evolution of the main ideas on the mitotic machinery over almost 
one and a quarter century are well documented \cite{gourret95,mogilner06c}. 
An excellent review \cite{mcintosh12} of mitosis has appeared very 
recently.

\subsection{\bf Mitotic spindle: inventory of force generators and list of stages}
\label{sec-spindle}

\subsubsection{\bf Mitotic spindle: key components and force generators}

The spindle shape in all eukaryotic cells share some common features 
irrespective of the pathway that leads to its assembly. When the spindle 
assembly is completed, it takes its characteristic fusiform shape. 
It looks like an ellipsoid made of fibers which are actually bundles 
of MT filaments. The minus ends of the MTs are focussed into the two 
``poles'' located at the opposite ends while the plus ends of the MTs 
extend towards the spindle ``equator''. 

Just before chromosome segregation begins, the sister chromatids remain 
attached with each other, mostly in a typical ``X''-shaped structure. 
The region where the two sister chromatids are closest to each other 
(the intersection of the two arms of ``X''-shaped structure) is called 
{\it centromere}. At the centromere region of each sister chromatid a 
protein complex of a specialized composition and architectural design 
is located; this complex, called {\it kinetochore} \cite{rieder98,maiato04b}, plays a crucial role 
in mitosis. We'll describe its structure and function in further detail 
later in this section.

MTs in the spindle can be classified into two main categories on the 
basis of the interacting partners of their plus ends- (i) {\it kinetochore} 
MTs (kMT), and (ii) non-kinetochore MTs. The plus end of the kMTs 
attach with the kinetochores; the MT-kinetochore coupling is essential 
for chromosome segregation. Non-kinetochore MTs can be further 
subdivided into two major classes- 
(a) {\it astral} MT (aMT), and (b) {\it interpolar} MT (ipMT). 
There is another set of MTs that interacts with the chromosome arms. 
The aMTs radiate from the poles towards the cell cortex; 
they are believed to play important roles in the positioning of the 
spindle as a whole and in marking the plane for subsequent assembly of 
the cleavage furrow during cytokinesis. Two sets of ipMTs, roughly equal 
in number, originate from the opposite poles; these antiparallel MTs 
overlap and interact (possibly crosslinked by MAPs or MT-based motors) 
at the equatorial plane thereby linking the two halves of the spindle 
mechanically. 

The key force generators of a mitotic spindle are \cite{mcintosh02} 
(i) cytoskeletal filaments, and 
(ii) cytoskeletal molecular motors 
\cite{scholey03b,sharp00b,hildebrandt00,brunet01,gatlin10,wordeman10}. 
The major kinetic processes involved in the force generation are  
(a) polymerization and depolymerization of the cytoskeletal filaments, 
caused by dynamics instability, predominantly of the MTs, that result 
in pushing and pulling forces 
(b) depolymerization of MTs by depolymerizers (the ``shredders''), 
generating pulling forces \cite{mcclung10}; 
(c) relative sliding of the filaments by crosslinking ``rower'' and 
``slider'' molecular motors \cite{peterman09,ferenz10,hentrich10}, (d) transport of molecular cargoes by 
``porter'' motors, (e) stretching of the chromosomes or bending of 
cytoskeletal filaments that generate spring-like elastic forces. 
Although MT, MT-based motors and MAPS are the main structural and functional 
components of mitotic spindle and mitosis \cite{glotzer09}, 
actin and myosin are also suspected to play some roles \cite{sandquist11}.
We have reviewed the kinetic models of all these processes separately 
in the preceding sections. It is the integration of so many processes 
in cell division that poses the main conceptual challenge to theoretical 
modelers.

Recall that a single molecular motor generates a force that is of the 
order of pico-Newtons and leads to mechanical movements over a maximum 
of tens of nanometers. In contrast, during cell division coordination 
of a large number of force generators takes place thereby exerting 
forces as large as a few nano-Newtons and causing movements over microns 
to tens of microns. 
A satisfactory theoretical model must show how such large forces and 
long distance movements emerge from the cooperation and/or competition 
between the molecular components of the machinery of cell division.

\subsubsection{\bf Mitosis: successive stages of chromosomal ballet}

The mitotic phase of cell cycle is collectively designated as the M phase.
The M phase is subdivided into a sequence of several phases of shorter 
duration.
Assembly of the mitotic spindle and its coupling with the chromosomes 
begins in the {\it prometaphase} and the appropriate positioning of 
the chromosomes in the equatorial plane gets completed in the 
{\it metaphase}. Chromosome segregation takes places in two stages 
of anaphase. During the first part of anaphase, called {\it anaphase A}, 
the sister chromatids are pulled apart towards the poles of the spindle 
while the pole-pole separation remains practically unchanged. However, 
in the second part of anaphase, called {\it anaphase B}, pole-pole 
separation keeps increasing simultaneously with the poleward movement 
of the chromosomes. In the next phase, called {\it telophase}, the spindle 
is disassembled while two separate nuclei of the two daughter cells 
form around the segregated chromatin. We'll review the kinetics of 
the mitotic machinery mainly during the period that covers approximately 
the prometaphase, metaphase and anaphase.

\subsection{\bf Spindle morphogenesis} 

Spindle morphogenesis \cite{duncan11} involves at least three {\it positioning} tasks: 
(i) positioning of the spindle as a whole in the parent cell, 
(ii) positioning of the chromosomes within the spindle in the 
equatorial plane, and 
(iii) positioning of the poles within the spindle at a certain distance 
from the chromosomes. Such positioning requires a subtle interplay 
of several force generators that are parts of the complex machinery. 
Positioning of poles is, however, different from the other two in 
one respect; it establishes a spatial scale, namely, the pole-to-pole 
separation which need not be 
determined by the cell size alone \cite{marshall04,chan10,marshall11}.  
Studying the pole-to-pole separation in terms of the forces generated 
by the various force-generating components of the mitotic machinery has 
received attention of both experimentalists and theorists.  

Spindle morphogenesis also involves correct orientations: 
(a) correct orientation of the major axis of ellipsoidal symmetry 
which decides the directions in which the two sister chromatids are 
pulled apart in the anaphase; 
(b) orientation of the sister chromatids with respect to the MT 
filaments. 

There are more than one pathways for spindle formation, the major 
pathways being (i) centrosome-directed astral pathway and (ii) 
chromosome-directed anastral pathway; each pathway consists of many 
steps \cite{compton00}. Computer simulations \cite{paul09} indicate 
that nature may use a combination of these two pathways to speed up 
spindle self-organization. In this subsection we review the roles of 
the various components, particularly the force-generators, in the 
kinetics of spindle morphogenesis.

\subsubsection{\bf Centrosome-directed astral pathway: ``search-and-capture'' as a first-passage time problem} 

In one of the common pathways, MTs nucleate at the centrosomes located 
at the poles and grow towards the equator by polymerization. The growing 
microtubules explore (or ``search'') the three-dimensional nuclear/cellular 
space until they are captured by one of the sister kinetochores on a 
sister chromatid.  Such a chromosome is called ``monooriented'' because it is 
attached to only a single pole of the spindle. Subsequently, when the 
other sister kinetochore is captured by another MT approaching from the 
opposite pole, the ``bioriented'' chromosome is said to be correctly 
aligned. Correct alignment of all the sister chromatids is a pre-requisite 
for proper segregation of the chromosomes. 

The ``search-and-capture'' mechanism was originally proposed by Kirschner 
and Mitchison \cite{kirschner86}. 
A growing MT may have difficulty finding a kinetochore because of the 
small size of the latter and also because it may not be located in the 
direction of growth of the MT. However, because of dynamic instability, 
a futile growth of a MT in a wrong direction can be corrected. The MTs 
randomly explore the space and once a MT makes a chance encounter with a 
kinetochore the contact gets stabilized whereas those MTs that do not 
make a successful contact with a kinetochore would soon depolymerize. 

For simplicity of an elementary calculation, based on heuristic 
arguments, let us ignore the possibility of rescue. Suppose $d$ is 
the distance of the target kinetochore from the MT nucleation site. 
Suppose, $p$ is the probability of a successful search. Let $t_s$ and 
$t_u$ be the durations of typical successful and unsuccessful searches, 
respectively. Then, the search time would be 
\begin{equation}
T_{search} = p t_s + p (1-p)(t_s+t_u) + p(1-p)^{2}(t_s+2t_u) ...= t_s + \biggl(\frac{1-p}{p}\biggr) t_u
\label{eq-holy}
\end{equation}
where $p$ can be expressed as a product $p = p_{d} p_{r}$ where 
$p_{d}$ is the probability of growing in the direction of the target 
and $p_r$ is the probability of reaching a distance $d$ before 
depolymerizing completely.

Since $p_d$ is 
proportional to the solid  angle subtended by the target kinetochore,  
$p_{d} = \pi r_{kt}^2/(4 \pi d^2) = r_{kt}^2/(4 d^2)$.
where $r_{kt}$ is the effective radius of a kinetochore. Thus, $p_{d}$  
is expected to be small enough to satisfy the condition $p \ll 1$. 
In this limit, 
$T_{search} \simeq t_{u}/p$. So, we need to estimate $t_u$ and $p$. 

Suppose, $V_{g}$ and $V_{s}$ are the velocities of the MT in its 
growing and shrinking phases, respectively.
In the simple Hill model \cite{hill84a,hill84b} (or, the corresponding 
Dogterom-Leibler continuum version \cite{dogterom93}), 
$p_{r} \simeq exp(-d/<L>)$ where $<L> \simeq V_{g}/f_{cat}$ 
is the average length of a MT in the steady state; $V_{g}$ and 
$f_{cat}$ being the rates of growth and catastrophe, respectively. 
Moreover, $t_{u} = <L>/V$, where both the growth and shrinkage 
rates (assumed to be approximately equal) are represented by a single 
symbol $V$. So, finally, according to this approximate analysis 
\cite{holy94}, 
\begin{equation}
T_{search} \simeq \biggl(\frac{<L>}{V}\biggr)\biggl(\frac{4d^2}{r_{kt}^2}\biggr) e^{d/<L>} 
\label{eq-holyleibler}
\end{equation}
This mechanism can be efficient provided the frequencies of catastrophe 
and rescue satisfy the following two conditions \cite{holy94}:\\
(i) the rescue frequency should not be large;\\ 
(ii) the catastrophe frequency should be such that the resulting average 
length of a MT is equal to the mean separation between the centrosome 
and the kinetochore. \\
Because of the condition (i) MTs would not waste time searching 
repeatedly in a ``wrong'' direction. The condition (ii) ensures that 
a MT neither suffers premature catastrophe while growing in the ``right'' 
direction nor waste time continuing its growth in a ``wrong'' direction. 
It is also obvious that the average time needed to capture one 
kinetochore can be reduced by increasing the number of MTs. Similarly, 
for a given number of MTs, longer time will be required to capture all 
the kinetochores. 

A more systematic, and somewhat more general, calculation of the search 
time was reported by Wollman et al.\cite{wollman05}. Then, for a system 
consisting of a total of $N_m$ MTs and $N_k$ kinetochores, the average 
time needed for an  ``unbiased'' search-and-capture is \cite{wollman05} 
\begin{eqnarray}
<T_{search}^{N_m,N_k}> 
\simeq <T^{1,1}_{search}> \frac{ln~N_k}{N_{m}}
\end{eqnarray} 
where the expression 
\begin{equation}
<T_{search}^{1,1}> = \biggl(\frac{V_{g}+V_{s}}{V_{s}f_{cat}}\biggr)\biggl(\frac{4d^2}{r_{kt}^2}\biggr)exp(d f_{cat}/V_{g})
\label{eq-T11}
\end{equation}
differs slightly from the heuristically derived expression 
(\ref{eq-holyleibler}). 
Thus, the search time is inversely proportional to the total number 
of MTs and proportional to the logarithm of the total number of 
kinetochores. Wollman et al.\cite{wollman05} argued that this simple 
model of ``unbiased'' search-and-capture can be made at least $10$ times 
faster by biasing the search process. 

The model used above \cite{wollman05} suffers from three limitations: 
(i) By assuming that the entire spindle space is 
available for search by the MTs it overestimates the search efficiency 
because, in reality, the chromosome arms occupy a significant region 
of this space and hinder search;  
(ii) It is based on the astral pathway for spindle assembly and fails 
if anastral pathway dominates;    
(iii) It's main aim is to investigate the {\it speed} of the 
search-and-capture process without paying attention to the accuracy of 
the resulting assembly.

The average time needed for capture is a mean first-passage time. A 
systematic calculation, that is more rigorous than the previous works, 
was carried out only for the special case for $M=1$ by Gopalakrishnan 
and Govindan \cite{gopalakrishnan11}.
A search cone for a MT nucleation site is defined by the corresponding 
solid angle $\Delta \Omega$. If a target kinetochore at a distance 
$d$ from the nucleation site falls within this cone and has a cross 
sectional area $a$, then the probability that nucleation takes place 
in the ``correct'' direction is $p = a/(d^2 \Delta \Omega)$. For 
simplicity, they assumed that within a search cone the cell boundary is 
at the same distance $R$ from the center. Let $\Phi(d,T)$ denote 
the conditional first passage time density (CFPD) for a freshly 
nucleated microtubule to reach a target at a distance $d$, for the 
first time, without ever shrinking to vanishing length in between. 
$Q_{X}(T)$ is the CFPD for shrinking to vanishing length after a 
life time $T$ without ever reaching a distance $X$ from the nucleation 
site in between. Similarly, $\Psi(T)$ is the CFPD for a freshly 
nucleated MT to disappear after a time interval $T$, following its 
encounter with the cell boundary at least once in between.  Let the 
symbols $t_{c}$ and $t_{w}$ denote that mean time spent in searching in 
the correct and wrong directions, respectively. Moreover, suppose, $\nu$ 
is the nucleation rate and
$t_{\nu}$ is the time in between successive nucleations, i.e., the 
mean time between the disappearance of a MT and re-nucleation of the next 
MT at the same site. According to this analysis \cite{gopalakrishnan11}, 
\begin{eqnarray}
<T_{search}^{N_m,N_k}> = N_{s} [pt_{c} + (1-p) t_{w} + t_{\nu}] = T_{c} + \frac{1-p}{p} T_{w} + \frac{1}{p} T_{\nu}
\end{eqnarray}
where $N_{s} = 1/[p\tilde{\Phi}(0)]$ is the mean number of unsuccessful 
search events before each successful event, 
$T_{c} = - [\tilde{\Phi}'(d,0) + \tilde{Q}'(d,0)]/[\tilde{\Phi}(d,0)]$, 
$T_{w} = - [\tilde{Q}'(R,0) + \tilde{\Psi}'(0)]/[\tilde{\Phi}(d,0)]$, 
$\nu T_{\nu} = 1/[\tilde{\Phi}(d,0)]$, 
are written in terms of the Laplace transforms $\tilde{\Phi}(X,s)$ and 
$\tilde{Q}(X,s)$ and $\tilde{\Psi}(s)$ whose exact expressions were 
obtained explicitly. The results reported by Gopalakrishnan et al. 
\cite{gopalakrishnan11} has been re-derived by Mulder \cite{mulder12} 
following an alternative mathematical approach. 

\noindent$\bullet${\bf From asters to spindle}

There are some general principles of spindle formation that are shared 
by both the astral and anastral pathways \cite{wadsworth11}; these are (a) nucleation 
and growth of MTs, (b) formation of well defined poles and equator, 
(c) attachment of the chromosomes to MTs of the spindle. What makes the 
two pathways different is the sequence of these events. In the astral 
pathway, nucleation and growth of the MTs from the poles first forms 
two asters which then interact with each other as well as with the 
chromosomes to position the sister chromatids on a plane that forms the 
spindle equator. To explain the role of molecular motors in this pathway, 
Nedelec \cite{nedelec02} (see also ref.\cite{channels08}) developed an 
in-silico model in which two asters exist initially in the presence of 
a both plus-end directed and minus-end directed motors. Computer 
simulations of this model demonstrated the formation of a spindle pattern 
by motor-driven fusion of the two asters. Computer simulation of a 
model that includes static MT-end crosslinking (or, MT-bundling) proteins 
and molecular motors to explore the focussing of the minus ends of the 
MTs into spindle poles \cite{chakravarty04,janson07}.

\subsubsection{\bf Chromosome-directed anastral pathways via sliding and sorting of MTs} 

There exists another pathway for spindle formation in which the MTs 
nucleate around chromosomes, instead of nucleating at centrosomes 
\cite{mazia84,merdes97,bannigan08}.
If each of the chromosomes could organize their respective MTs into 
separate mini spindles, a human cell could assemble 46 mini spindles 
simultaneously \cite{oconnell07}. But, instead, a cell deploys an 
elaborate team of motors which sort the microtubules into an 
antiparallel array to generate one single bipolar spindle where all the sister 
chromatids are aligned correctly at the equatorial plane. A spindle 
assembled by this anastral pathway appears very similar to astral  
spindles except that no centrosome exists at the spindle poles. 
Thus, the sequence of the main events along this pathway differ 
from those along the astral pathway although the basic processes 
are the same, i.e., both involve (i) nucleation and growth of MTs, 
(ii) formation of well defined poles and equator, and (iii) attachment 
of the chromosomes to the MTs. 

Interestingly, motor-driven sliding and sorting of MTs nucleated on 
just chromatin-coated beads can lead to the formation of spindle 
patterns {\it in-vitro} in spite of the absence of both centrosomes 
and kinetochores \cite{gadde04,dinarina09}. This has also been supported by 
the corresponding theoretical model developed by Schaffner and Jose 
\cite{schaffner06,dinarina09}.   

In the ``slide-and-cluster'' model \cite{burbank07} the MTs, after 
nucleation near the chromosomes, {\it slide} and {\it cluster}. 
Occasionally, because of the dynamic instability, MTs can be lost 
if it is not rescued after a catastrophe. Under suitable conditions, 
which we explain below, a spindle may emerge from such a slide-and-cluster 
kinetics.  For simplicity, let us assume that the motion of the MTs, 
nucleated near the chromosomes, are controlled by only two types of 
motors: sliding motors (e.g., Eg5) and clustering motors (e.g., dynein) 
(see ref.\cite{hallen09} for another slide-and-cluster model that 
differs in some respect from the slide-and-cluster model of Burbank 
et al. \cite{burbank07}).

A sliding motor that crosslinks two antiparallel MTs walks towards
the {\it plus} end of each thereby pushing the minus ends of the two
MTs away from each other. On the other hand, a sliding motor resists
the relative sliding of two parallel MTs. In contrast, a clustering
motor moves towards the {\it minus} ends of the two MTs that it 
links; upon reaching the minus end of one of the two MTs it stops
moving on that MT while it continues moving on the other. When such
a clustering motor crosslinks two parallel MTs it moves their minus
ends close together; the {\it minus} ends of a group of parallel MTs
are thus clustered by clustering motors. Thus, on the average, half
of the MTs get their minus ends focussed at a point in one half of
the spindle while those of the remaining MTs get focussed in the
other half of the spindle thereby forming the two opposite poles.
Near the chromosomes, the two species of motors {\it cooperate}
because both pull the MTs outward. But, they {\it compete} in the
regions away from the chromosomes. The pole-to-pole separation
stabilizes to a steady value when the opposing velocities imparted
by the two species balance each other.

As a concrete example, let us quantify these intuitive ideas to develop
a one-dimensional mathematical model for anastral spindle morphogenesis .
Suppose, $F_s$ and $F_c$ represent the stall forces of the sliding
and clustering motors, respectively. The average number of sliding
motors per unit length of MT is denoted by $c_s$. Similarly, the
average number of clustering motors per minus-end crosslink is
denoted by $n_c$. Suppose, $V_s$ and $V_c$ are the zero-load velocities
for single head of a sliding motor and a clustering motor, respectively.

Making the simplifying assumption that the viscous drag and the random
thermal force on a MT are negligibly small \cite{burbank07}, the net
force on a MT which has its {\it minus} end at $x$ is given by
\cite{burbank07}
\begin{equation}
F_{net}(x) = F_{sl,par} (x) + F_{sl,apar}(x) + F_{cl,par}(x)
\label{eq-Fburbank}
\end{equation}
where the first and second terms on the right hand side of the equation
(\ref{eq-Fburbank}) are the forces arising from its sliding against
parallel and an antiparallel MTs, respectively whereas the third term
represents the force experienced by it because of motors clustering it
with other parallel MTs.

Suppose, $\rho(x)$ and $\tilde{\rho}(x)$ are the number densities of the
{\it minus} ends of {\it right} moving and {\it left} moving MTs,
respectively.
A sliding motor that crosslinks two parallel MTs is assumed to exert
a ``drag'' force that is proportional to their relative velocity. For
simplicity, it is also assumed that all the MTs have the same length
$L$ \cite{burbank07}. If the {\it minus} ends of the two crosslinked
parallel MTs are at $x$ and $y$ (and $|y-x| < L$), then,
\begin{equation}
F_{sl,par}(x) = \int_{x-L}^{x+L} F_{s} c_{s}(L-|y-x|) \biggl[\frac{v(y)-v(x)}{2V_s}\biggr] \rho(y) ~dy
\label{eq-Fslpar}
\end{equation}
Similarly, assuming that a sliding motor crosslinking two antiparallel
MTs pushes them apart with a force that vanishes when the relative
velocity is $2V_s$, one gets \cite{burbank07}
\begin{equation}
F_{sl,apar}(x) = \int_{x-2L}^{x} F_{s} c_{s}(L-|L+z-x|) \biggl[1 + \frac{\tilde{v}(z)-v(x)}{2V_s}\biggr] \tilde{\rho}(z) ~dz
\label{eq-Fslapar}
\end{equation}
which exploits the fact that antiparallel MTs with minus ends at
$x$ and $z$ will overlap if $0 < |x-z| < 2L$. Note that force
experienced by a MT because of its crosslinking with another parallel
MT by a clustering motor depends on which of the two is farther to the
right and, hence \cite{burbank07}
\begin{equation}
F_{cl,par}(x) = \int_{x-L}^{x+L} F_{c} n_{c} \biggl[sign(y-x) + \frac{v(y)-v(x)}{V_c}\biggr] \rho(y) ~dy
\label{eq-Fslapar}
\end{equation}

We need separate equations for $\rho(y)$ and $\tilde{\rho}(z)$.
For simplicity, the following assumptions are made
\cite{burbank07}:
(i) that the MTs nucleate only at the center of the spindle
(corresponding to $x=0$) at the rate $R$, and (ii) that the MTs
have an average lifetime $\tau$.
Then, the conservation of the number of MTs is described by the
equation of continuity \cite{burbank07}
\begin{equation}
\frac{\partial \rho}{\partial t} + \frac{\partial(V\rho)}{\partial x}
= R \delta(x) - \frac{\rho}{\tau}
\label{eq-rhoburbank}
\end{equation}
where the ``source'' and ``sink'' terms are written on the right
hand side of the equation. In the steady state,
$\partial \rho/\partial t = 0$,
$F_{net}(x) = 0$ (force balance) and, by symmetry,
$\tilde{\rho}(z) = \rho(-z)$, $\tilde{V}(z) = - V(-z)$.
Then, the steady-state profiles $\rho(x)$ and $v(x)$ are determined
by the conservation law (\ref{eq-rhoburbank}) and the force balance
equation (\ref{eq-Fburbank}). Analytical calculation is possible in
the limiting condition when the MT length $L$ is much longer than
the region over which most of the minus ends are distributed.
This condition, in turn, implies almost complete overlap between the
parallel MTs and small overlap between antiparallel MTs. Consequently,
the steady-state distribution of the {\it minus} ends of the MTs is
determined completely by a single dimensionless parameter \cite{burbank07}
\begin{equation}
\phi = \biggl(\frac{V_c}{V_s}\biggr) \biggl[\frac{\{F_c n_c/V_c\}}{\{F_c n_c/V_c\} + \{F_s c_s L/(2V_s)\}}\biggr]
\end{equation}
which captures the fractional contribution of the clustering motors
in the combined effects of the two types of motors. There exists a
critical value $\phi=\phi_{c}$, such that the clustering forces
dominate for $\phi > \phi_c$. For all $\phi < \phi_c$, spindles do
not have sharply defined poles. On the other hand, for $\phi \geq \phi_c$,
sharp poles form. Moreover, the pole-pole separation increases with
increasing $\phi$ beyond $\phi_c$. In the large $\phi$ limit almost
all the minus ends are clustered at the poles and the pole-pole
separation saturates.

\subsubsection{\bf Amphitelic attachments: a determinant of fidelity of segregation} 

Because of the intrinsic randomness of the search-and-capture mechanism 
of the astral pathway, only rarely both the sister kinetochores attach 
with MTs simultaneously. Therefore, after MTs from one of the poles 
capture a kinetochore, the sister kinetochore remains in an unattached 
state for a further period of time. In other words, most of the chromosomes 
remain in a ``monooriented'' state for some time before the unattached 
sister kinetochore is finally captured by MTs growing from the opposite 
pole of the spindle thereby making the pair ``bioriented'' \cite{watanabe12}. 

The major source of mitotic error is wrong attachment 
of the sister chromatids to the MTs. 
Proper segregation of the chromosomes in the anaphase can take place 
only if, in the earlier phases, kinetochores of the two sisters are 
attached to the plus ends of MTs emanating from opposite poles; 
such attachments are called {\it amphitelic}. In contrast, in 
{\it syntelic} attachment both the sister kinetochores are attached 
to the same pole of the spindle; in this case, segregation of the two 
sisters towards opposite poles cannot take place. If a single 
kinetochore is attached to MTs coming from both the poles, and its 
sister kinetochore is attached to a single pole, such an 
attachment is called {\it merotelic}; the chromosome would not move 
towards either pole even after the two sister chromatids separate. 
Large number of syntelic attachments result if the two spindle poles 
are too close to search the entire cellular space \cite{silkworth12}.   
Efforts are on to unambiguously identify the molecular ``sensors'' 
that detect, and correct, the merotelic attachments \cite{cimini07}.

A comparison of the mechanisms of error detection and error correction 
during DNA replication with those during chromosome segregation brings 
out the stark contrasts vividly \cite{nicklas88b}. During replication,  
following error detection, the incorrect nucleotide is excised and the 
correct nucleotide is added by the replisome. In contrast, there is no 
macromolecular machine for detecting and correcting wrong attachments 
of sister chromatids to MTs. Interestingly, just as the MT-kinetochore 
attachment is the outcome of chance encounters, chance is exploited 
also in correcting errors that result from such a random process 
\cite{nicklas97}. Incorrect MT-kinetochore attachments are less stable 
than correct attachments and, therefore, more likely to break. Only the 
correct MT-kinetochore attachment has sufficient stability to drive the 
subsequent steps of chromosome segregation. 

Why does a cell rely solely on chance for correcting errors of MT-kinetochore 
attachments, instead of deploying a machine as it does for error correction 
during replication? A speculative answer is based on the difference of 
length scales involved in the two processes \cite{nicklas88b}. 
A machine as large as the replisome will have no difficulty in examining 
a single MT-kinetochore junction for any possible molecular defects. 
But, the syntelic or merotelic attachments involve MTs that are much 
farther apart. Detection of such wrong attachments of sister kinetochores 
to MTs would require machines much larger than all the known molecular 
machines within an eukaryotic cell. That is why, perhaps, cell has left 
the correction of wrong MT-kinetochore attachments to chance rather 
than to a machine. In fact, a Darwinian-like ``selection'' process has 
been invoked to emphasize the role of chance in this mode of error 
correction \cite{nicklas88b,nicklas97}: the ongoing random 
attachment-detachment (with the kinetochore) produces ``mutant'' MT arrays, 
the most stable MT array is ``selected''.

\subsubsection{\bf Chromosomal congression driven by poleward and anti-poleward forces} 

Bioriented chromosome undergoes some further translations and rotations 
so as to finally position themselves at the spindle equator completing 
the process of chromosome {\it congression} \cite{kapoor02,kops10}.
Monitoring spindle formation in 3-dimensional space of the cell 
\cite{magidson11}, 
recently it has been discovered that, surprisingly, there is an 
important stage of chromosome congression that was overlooked in 
all the earlier works which monitored only a 2-dimensional image of 
the process. Lateral interaction between the MTs and kinetochores 
arrange the chromosomes in a toroidal region that overlaps with 
the equatorial ring of the spindle. Such a toroidal distribution of 
the chromosomes facilitates more frequent interactions with the plus 
ends of the spindle MTs thereby speeding up the congression. 

For quite some time in the mid-twentieth century it was assumed that 
the chromosome congression was caused by a position-dependent force 
\cite{ostergren51}. In this scenario, it was hypothesized that 
chromosomes attached to two spindle poles experience forces directed 
to both the poles; however, the magnitude of each of these two forces 
is proportional to the length of the corresponding kinetochore fiber 
that connects it to that pole. The eventual alignment of the chromosomes 
at the spindle equator is the result of balancing the two opposing 
poleward forces on the chromosomes. However, this scenario is 
inconsistent with the experimental observation that mono-oriented 
chromosomes exhibit both poleward and away-from-the-pole movements. 
In the last two decades, experiments have established that the dominant 
effects of the {\it polar ejection force} can account for the observed 
shape and movements of the chromosomes leading to their congression 
\cite{kapoor02,kops10}. 

\noindent$\bullet${\bf Chromosome arm-MT interactions: Polar ejection force}

MT-kinetochore interactions are essential for segregation of chromosomes. 
It is not always well appreciated that interactions between the MTs 
and chromosome arms are equally important in the earlier stages before 
the segregation can begin.

It has been observed that a kinetochore often moves {\it away} from the 
pole to which it is coupled by kMTs even when its sister kinetochore 
is not attached to any MT emanating from either of the poles. In such 
situations it appears that the kinetochore is experiencing a {\it pull} 
towards the pole to which is it coupled by attached kMTs while the 
chromosome arms are experiencing a pushing force away from that pole. 
A ``polar ejection force'' \cite{rieder94} has been postulated to 
explain this phenomenon. This force is generated by a class of plus-end 
directed kinesin motors, called chromokinesins \cite{vanneste11,mazumdar05},  
which walk on the spindle MTs while their tails remain attached to the 
chromosome.

\noindent $\bullet${\bf Positioning chromosomes: congression by ``smart'' or ``dumb'' kinetochore?} 

It has been proposed \cite{murray94} that a ``dynamic, smart, 
tension-sensitive'' kinetochore (i) would be ``told'' by the polar 
ejection force where it is located on the spindle and, in turn,  
(ii) would pull or push the poles in the appropriate direction thereby 
providing a mechanism for determining the length of the spindle. 
In other words, a ``smart'' kinetochore can ``sense'' its position 
on the spindle and can use this information to control the forces 
not only for its own movements but also to push and pull the poles.  
In this model, chromosome congression results from the coordinated 
movements of the sister kinetochores which act as ``tensiometers''. 
This speculative idea gave rise to a debate on whether kinetochores 
are really ``smart'' or just appear to be ``smart''; an alternative 
scenario based on ``dumb'' kinetochores has also been proposed 
\cite{khodjakov99}.

\subsubsection{\bf Positioning and orienting spindle: role of MT-cortex coupling} 

So far we have discussed the mechanisms of positioning of the chromosomes 
at the equatorial plane of the spindle. Now we discuss the positioning of 
the spindle poles with respect to the equator (i.e., the pole-to-pole 
separation).

We have already reviewed the kinetic mechanisms of relative sliding of 
MT filaments {\it in-vitro} by tetrameric kinesin-5 (e.g., Eg5) and 
dimeric kinesin-14 \cite{peterman09,ferenz10,hentrich10}. 
By sliding the two antiparallel ipMTs a cross-linking 
kinesin-5 tends to increase the pole-to-pole separation whereas a 
crosslinking kinesin-14 tends to shorten the spindle \cite{dumont09,goshima10,peterman09,tanenbaum10}. Interplay of Eg5 and dynein in crosslinking and 
sliding of MTs during spindle morphogenesis has also been elucidated 
by experiments and mathematical modeling \cite{ferenz09}.
Inspired by the sliding filament model of muscle contraction, a sliding 
filament hypothesis for spindle contraction (or extension) was proposed 
already in the nineteen sixties \cite{mcintosh69} although the identity 
of the sliders emerged much later. Utilizing the contemporary knowledge 
on the various force generators, Cytrinbaum et al. 
\cite{cytrinbaum03,cytrinbaum05}
have developed a differential equation for the force balance in the system. 
Assuming physically justified distributions of the force generators, 
they solved the force balance equation to determine the resulting 
steady pole-to-pole separation.

The positioning of the spindle as a whole in the middle of the parent 
cell \cite{grill05b,manneville06} and its proper orientation 
\cite{thery07,moore10} may arise from a delicate balance between the 
(i) {\it pushing} of the cell cortex / plasma membrane by growing aMTs 
polymerized at (or near) the spindle poles, 
(ii) lateral {\it sliding} of the microtubules along the cell cortex 
without losing contact with the cortex, and 
(iii) {\it pulling} forces exerted by the aMTs that depolymerize at the 
plus end and are anchored on the cell cortex / plasma membrane.  
The anchoring of aMTs on the cell cortex may take place by a 
``search-and-capture'' mechanism that resembles anchoring of kMTs by the 
kinetochores \cite{schuyler01a}.

\subsection{\bf Pull to the poles} 

\noindent $\bullet${\bf Timing events: entry to mitosis and signaling anaphase} 

The mitotic spindle has a ``check-point'' mechanism that monitors 
the alignments of the chromosomes. The anaphase-promoting complex 
(APC) triggers chrosomosome segregation only if all the chromosomes 
are properly aligned. A fundamental question is: how do sister 
kinetochores sense misalignment and inhibit the APC till the error 
is corrected \cite{maresca10,murray11,pinsky05,kotwaliwale06}? 
Strong evidences have accumulated over decades to establish that 
mechanical forces are needed not only for pulling the chrosmosomes 
apart in the anaphase, but also to detect and correct errors; tension 
generated by the MT-kinetochore coupling triggers chemical signals 
of the checkpoint \cite{nicklas97,bloom10b,matsson09}. 
Since signaling is not the main focus of this review, we'll not 
delve deeper into the kinetics of the signaling processes underlying 
this checkpoint mechanism. Nevertheless, it is worth pointing out that 
such mechano-transduction processes, whereby mechanical 
force is transduced into a chemical signal, is just the opposite of 
the transduction of chemical energy into mechanical force by molecular 
motors. 

\subsubsection{\bf Kinetochore pulling by MT filaments: Brownian ratchet or power stroke?} 

The existence of bundle of MTs, called kinetochore fiber, has been known 
for a long time (see ref.\cite{rieder05} for a review from a historical 
perspective).
Next, we summarize the conceptual frameworks developed to account for 
the kinetochore pulling by depolymerizing kMT 
\cite{hill85b,joglekar02,joglekar10,santaguida09,molodtsov05a,molodtsov05b,efremov07,mcintosh10,vladimirou11} 
in terms of 
(I) a Brownian ratchet mechanism, and 
(II) an alternative power stroke mechanism.

\noindent$\bullet${\bf MT-kinetochore coupling device: sleeve / ring, grappling hook, connecting rods} 

The captured MTs are stabilized at the kinetochore. But this stabilization 
is not achieved by preventing catastrophes. The catastrophes can take place 
at the captured plus ends of the MTs; nevertheless, the MTs are stabilized 
because they cannot detach from the kinetochore \cite{hyman96,biggins03}. 
A remarkable feature of the MT-kinetochore coupling is that the kinetochore 
remains attached to the MT tips even when the same MT shortens because of 
depolymerization. The identity of the coupler and the nature of its kinetics 
remain controversial \cite{davis07,cheeseman08,foley13,welburn08,bader10,asbury11,lampert11}.

There are at least three different models which attempt to account
for the MT-kinetochore coupling. First, ATP-powered molecular motors
can link the tip of a MT with the kinetochore. But, by deletion
or depletion of motor population the MT-kinetochore coupling is
is not affected significantly. Therefore, although motors may
play the role of MT-kinetochore coupler, it is not believed to be
the dominant one.

\noindent$\bullet${\bf Biased diffusion of a ``sleeve'': Brownian ratchet mechanism} 

Long before the identities of the molecular components of the kinetochore 
and those of the MT-kinetochore coupling device were established, a 
mechanism for the chromosome pulling by depolymerizing MTs was proposed 
assuming the existence of a ``sleeve'' with some special properties 
\cite{hill85b}. 
About 40 nm of the plus end of the MT is assumed to be surrounded by 
a coaxial ``sleeve''. It also postulates that there are several 
(possibly, equispaced) binding sites for the MT on the inner surface 
of the sleeve. The position of the plus-end of the MT in the sleeve 
is labelled by the integer index $n$ ($n=1,2,...M$); $n=1$ denotes the 
state in which the MT is fully inserted into the sleeve whereas 
$n=M$ corresponds to the position in which the MT is almost unattached 
from the sleeve. 

The position of the tip of the MT inside the sleeve can change because 
of thermal fluctuations of the sleeve at the rate $k$. Suppose $w$ ($<0$) 
is the free energy of interaction of a single subunit of the MT with 
the inner wall of the sleeve. Therefore, the insertion of each additional 
subunit into the sleeve lowers the free energy of the system by an amount 
$w$. However, any repositioning of a MT within the sleeve involves 
breaking the prior interactions and reforming interactions in the new 
position. Therefore, a free energy barrier $b$ has to be overcome for 
such repositioning of a subunit. Obviously the barrier height increases 
with the increasing number of subunits in the sleeve. In other words, 
$b$ arises from the ``roughness'' of the interface between the outer 
surface of the MT and the inner surface of the sleeve whereas thermal 
fluctuation works effectively as a ``lubricant''.

Let $\alpha$ and $\beta$ denote the rates of polymerization and 
depolymerization of the MT and $F$ is the load force. 
Defining $r = exp(-b/k_BT)$, $s = exp(w/k_BT)$ and 
$f = exp(-F{\ell}/k_BT)$, the kinetic scheme can be depicted as 
\begin{equation}
1 \mathop{\rightleftharpoons}^{ksr^Mf^{-1}+\beta s}_{kr^Mf+\alpha c} 2 ...M-1 \mathop{\rightleftharpoons}^{ksr^2f^{-1}+\beta s}_{kr^2f+\alpha c} M \mathop{\rightarrow}^{ksrf^{-1}+\beta s}
\end{equation}
where $F$ is the load force and $c$ is the concentration of the MT 
subunits in the solution. Because of thermal fluctuations, the sleeve 
executes a one-dimensional Brownian motion along the axis of the MT. 
However, this Brownian motion is biased towards the MT than away from 
it because the larger the number of MT-sleeve bindings the lower is 
the total energy of the system. The poleward bias arises from the 
depolymerization of the MT. Since depolymerization is caused either 
by the loss of GTP cap or by a depolymerase motor (both of which 
involve hydrolysis of NTP), the biased diffusion of the sleeve is a 
physical realization of the Brownian ratchet mechanism. An additional 
bias can arise from the curling protofilaments at the plus end of the MT.
Dhtylla and Keener \cite{shtylla11} have extended this picture by 
modeling the kinetochore-MT interface in terms of kinetochore binders.

The discovery of the Dam1 and DASH complexes \cite{nogales09,buttrick11} 
and the Ndc80 complex \cite{tooley11} in recent years have triggered 
renewed interest in the ``sleeve'' model of MT-kinetochore coupler.
Note that, for generating forces strong enough to pull the chromosomes, 
the gap between the outer surface of the MT and the inner surface of 
the sleeve must be sufficiently small. Moreover, the mechanism would 
fail if the inner space of the sleeve is always fully occupied by the 
MT (i.e., if all the binding sites on the inner surface of the sleeve 
are always bound to the partner sites on the outer surface of MT). 
Very recent experiments \cite{dumont12b} indicate the possibility that 
the MT-kinetochore coupler is a hybrid of passive and active force 
generators.

\noindent$\bullet${\bf Ring pulled by curled tip of MT: Power stroke mechanism} 

Based on the images of the MT-kinetochore couplers revealed by 
electron microscopy, it is now widely believed that curling 
protofilaments at the plus end of a depolymerizing MT can pull 
a ring or sleeve (e.g., the Dam1 ring) which is connected to 
the kinetochore by linking rods; the Ndc80 complex on the 
kinetochore is a strong candidate for the rod-like structures.  
Different versions of this {\it power stroke} mechanism have been 
named ``conformational wave'' mechanism \cite{koshland88}, 
``forced walk'' mechanism, etc.

The {\it longitudinal} interaction between the subunits (each 
subunit being $\alpha-\beta$ hetero-dimer) of a protofilament 
can be described by a bending potential energy $U_{||}(\chi)$ 
where the spontaneous bending angle $\chi_0$ depends on whether the subunit 
is bound with GTP or GDP \cite{molodtsov05a,molodtsov05b}. 
Moreover, the {\it transverse} interactions between two neighboring 
protofilaments is described by another potential energy function 
$U_{\perp}(r)$ that depends on the distance $r$ between the points 
of interaction on the corresponding adjacent subunits 
\cite{molodtsov05a,molodtsov05b}. 
The net potential energy $U$ is a sum of $U_{||}(\chi)$ and $U_{\perp}(r)$. 
The mean force $<F>$ acting on the ring, as the MT is shortened by 
one subunit, is obtained from $<F> = - (U_{z'}-U_{z})/(z'-z)$ where 
$U_z$ and $U_{z'}$ are the potentials at the initial and final 
positions $z$ and $z'$, respectively \cite{molodtsov05a,molodtsov05b}.

If the inner surface of the sleeve is so close to the MT surface that 
it does not allow the curling protofilaments to bend sufficiently, 
the force generated will be much less than the maximum possible 
value achievable by unconstrained bending of the protofilaments. 
Therefore, for optimal design of the device for power stroke the inner 
radius of the sleeve must be at least 1-2 nm larger than that of the 
outer radius of the MT \cite{molodtsov05a,molodtsov05b}.

The original power stroke model (``conformational wave model'') 
\cite{koshland88} 
did not assume any interaction between the inner surface of the sleeve 
and the outer surface of MT; leaning of the curling protofilaments 
against the edge of the sleeve is adequate for poleward pulling of the 
sleeve. Therefore, in this version of the power stroke model, the 
length of the sleeve is irrelevant. However, some later versions 
\cite{efremov07,armond10} 
explored  effects of (i) electrostatic interactions of the 
charged ring with the oppositely charged subunits of MT, or (ii) 
contact interactions of rigid or flexible linkers of the ring with 
the MT surface. 
One of the questions that needs attention is: if several MTs are attached to the same 
kinetochore, why and how do all these kMTs synchronize their kinetics 
so as to depolymerize simultaneously \cite{hyman96}?

\subsubsection{\bf Chromosome oscillation}

Mono-oriented chromosomes in the pro-metaphase as well as bi-oriented 
chromosomes in the metaphase are known to exhibit an oscillatory 
movement; periodic switching between poleward movement and movement 
away from the pole takes place for durations as long as tens of minutes. 
Joglekar and Hunt \cite{joglekar02} extended the Hill-sleeve model 
\cite{hill85b} to explain this phenomenon as a ``directional instability'' 
arising from a competition between the ``poleward'' force exerted on 
the kinetochores and the ``polar ejection'' forces on the chromosome arms. 
They postulated an inverse-square law for the polar ejection force; 
the force is maximum at the poles (symmetrically) and vanishes at the 
equator. 

Campas and Sens \cite{campas06b} developed a model from an 
alternative perspective that treats the force at the kinetochore 
phenomenologically while the role of the chromokinesin motors 
\cite{vanneste11,mazumdar05} in generating the polar ejection 
was described explicitly. Suppose, $N$ is the total number of 
chromokinesins attached permanently to the chromosome arms. However, 
because of the possibility of attachment to and detachment from the 
MTs, the actual instantaneous number $n(t)$ of chromokinesins attached 
to the MTs keeps fluctuating with time $t$. A kinetic equation accounts 
for the time-dependence of $n$ arising from this process; the rates of 
attachment and detachment being $k_b$ and $k_u$, respectively. 
These motors exert polar ejection force. Campas and Sens 
\cite{campas06b} assumed that (a) the motors contribute equally to the 
polar ejection force, and (b) the force-velocity relation for the 
individual chromokinesins is linear. Starting from a force balance 
equation for poleward force, polar ejection force and viscous drag, 
and then identifying the chromosome velocity $dr/dt$ with the  
chromokinesin velocity on the MTs, they obtained the second dynamical 
equation. Unlike the Joglekar-Hunt model \cite{joglekar02}, no 
spatial-dependence of the polar ejection force is assumed directly; 
the spatial information enters dynamics only through the postulated 
form of $k_b(r)$ whose $r$-dependence arises from its proportionality 
to the MT density $\rho_{MT}(r)$. A stable fixed point of the coupled 
nonlinear dynamics of the two variables $n$ and $r$ indicates that the 
chromosome would stall at a fixed distance from the pole. On the other 
hand, an unstable fixed point corresponds to the periodic oscillation 
observed in the experiments.
 
\subsubsection{\bf Force-exit time relation: a first passage problem}

As we have seen repeatedly in this article, for a single molecular motor 
the force-velocity relation is one of its most fundamental characteristic 
properties. However, for the kMT-kinetochore coupling device, a more 
appropriate counterpart of the force-velocity relation has emerged from 
most recent experiments. In the context of mitotic spindle, ever since 
the pioneering experiments performed by Nicklas \cite{nicklas88a}, force 
measurement has become quite routine work. However, to my knowledge, 
none of the works reported till recently applied load force on a single 
kinetochore. Therefore, none of those experiments could be regarded as 
the counterpart of a {\it single-motor} experiment where the motor molecule 
is subjected to a load force exerted, for example, by an optical tweezer. 

Recently, for the first time, Akiyoshi et al.\cite{akiyoshi10} 
have achieved this goal. By subjecting a single kinetochore to a load 
force $F$, they have measured the mean life time $<T>$ of the 
kMT-kinetochore coupler. The plot $<T>-F$ characterizes the physical 
strength of this coupling. Contrary to expectations based on physical 
intuition and theoretical modeling \cite{armond10}, they observed a 
nonmonotonic variation of $<T>$ with $F$ and interpreted their 
observation in terms of a mechanism based on the concept of catch 
bond \cite{thomas08,sokurenko08,prezhdo09}. 

\subsubsection{\bf Chromosome segregation in the anaphase: separated sisters transported to opposite poles}

Once the proteins holding the two sister chromatids in a tight embrace 
falls apart, the final journey of the two sister chromatids towards 
the two opposite poles of the spindle begins. This phase of mitosis 
involves a few key kinetic processes which we have not discussed 
explicitly so far in this section.

\noindent $\bullet${\bf (I) Poleward flux: MT flux towards stationary centrosome}

Poleward flux \cite{margolis78,mitchison90,rogers05,kwok07}
can be visualized in optical fluorescence microscopy by
putting an appropriate fiduciary mark on the MTs. For example,
\cite{kwok07} marking all the MTs along a diameter in the equatorial
plane of the spindle one gets a single initial rectangle-like narrow
band which, with the passage of time, splits into two bands that move
towards the two opposite poles.

One of the models of poleward flux is based on the assumption
of depolymerization of spindle MTs from their {\it minus} ends at the
spindle poles. In this scenario, poleward flux has been defined as
follows, irrespective of the mitotic stage \cite{rogers05}: poleward flux
is the  poleward movement of the $\alpha-\beta$ dimeric subunits of a
MT, along the contour of the filament, that is coupled to the
depolymerization of the {\it minus} end of the same MT at the spindle
pole. Depolymerases,  which are members of the kinesin superfamily (and
which have been discussed earlier), are believed to drive the
depolymerization process at the spindle poles.

\noindent $\bullet$ {\bf (II) Pacman mechanism and chromosomeward MT flux}

This mechanism, named after ``Pacman'' in the famous video game, can
account for the poleward journey of the sister chromatids. Just like
the Pacman of the video game, a chromatid can ``chew'' its way poleward
by actively depolymerizing kMTs from their {\it plus} ends \cite{sharp04}.
This mechanism is, in a sense, ``mirror image'' of the poleward flux
process. Recall that poleward flux is actually MT flux towards a
stationary centrosome. Similarly, in the frame of reference of a
chromatid, Pacman mechanism is essentially MT flux towards a stationary
chromosome. 

\noindent$\bullet${\bf Anaphase A}

The recent reviews of the mechanisms of chromosome movements in the 
anaphase \cite{maiato10,rath11} 
are quite comprehensive. Therefore, we'll present here only some of 
the theoretical aspects which are not available in these reviews but 
are interesting from the perspective of physicists.   

A quantitative theory of anaphase A was developed by Scholey et al. 
\cite{scholey06} using force-balance equations. 
Both the Pacman mechanism and the poleward flux are believed to
contribute towards the journey of the sister chromatids towards
their respective spindle poles in the anaphase A. Combining these
two processes, a unified mechanism (named ``Pacman-flux''
\cite{sharp04,rath11}), has been proposed.

Most of the models of anaphase describe only the motion of the kinetochore, 
ignoring that of the chromosome arising from its flexibility. An extended 
model that describes the coupled motion of the kinetochore and the 
chromosome has been developed by Raj and Peskin \cite{raj06}. This model 
is based on a postulate of an ``imperfect'' Brownian ratchet, captured 
through a tilted saw-tooth-like potential. Moreover, energies of both 
elastic stretching and bending of the chromosome were incorporated. 
One of the main results \cite{raj06} is that the velocity of the 
chromosome becomes independent of its length in the limit of high 
flexibility (small stiffness) whereas in the opposite limit the longer 
the chromosome the slower it moves.

\noindent$\bullet${\bf Anaphase B}

We summarize a mathematical model of anaphase-B \cite{mascher04,mascher11} 
based on a variant of force-balance relations. 
Suppose $S(t)$ is the pole-to-pole separation at time $t$ and $L(t)$ 
is the length of the region of overlap between ipMTs at time $t$. 
We also define $V_{depoly}^{+}$ and $V_{depoly}^{-}$ as the average 
rates of depolymerization of the plus and minus ends of the ipMTs, 
respectively. Moreover, $V_{sliding}(t)$ is the average rate of 
sliding of ipMTs at time $t$. 
Let us begin with the simplified picture with {\it identical overlap}
region $L(t)$ for all pairs of ipMTs. For this simple situation, we 
have the kinematic equation 
\begin{equation}
\frac{dS}{dt} = 2 [V_{sliding}(t) - V_{depoly}^{-}] 
\label{eq-dsdt}
\end{equation}
If $V_{sliding}(t)$ is a time-independent constant, then pre-anaphase B 
situation $dS/dt = 0$ results from the condition $V_{sliding} = V_{depoly}$. 
However, at the onset of anaphase B, $V_{depoly}^{-} \simeq 0$ and, hence, 
$S(t)$ would increase linearly with time. But, $V_{sliding}(t)=$constant  
is oversimplification of reality. To be more realistic \cite{mascher04} 
the equation (\ref{eq-dsdt}) must be coupled with the kinematic equation 
\begin{equation}
\frac{dL}{dt} = 2 [V{poly}^{+}(t) - V_{sliding}(t)] 
\label{eq-dLdt}
\end{equation} 
and the dynamic equation 
\begin{equation}
\frac{\mu}{2}\biggl(\frac{dL}{dt}\biggr) = k N L(t) f  
\label{eq-mudSdt}
\end{equation} 
where $\mu$ is the effective drag coefficient, $k$ is the average number 
of motors per unit length of overlap, $N$ is the total number of 
crosslinked ipMT pairs and $f$ is the force produced by each motor. 
Assuming linear relation 
\begin{equation}
f = F_{m} \biggl[1 - \frac{V_{sliding}}{V_m}\biggr],
\end{equation}
where $F_{m}$ is the stall force and $V_{m}$ is the maximal unloaded 
motor velocity, one gets the complete set of equations for computing 
$S(t)$ and $L(t)$ 
as functions of time. In a real mitotic spindle different pairs of 
ipMTs have different extents of overlap and not all the pairs are 
necessarily antiparallel. However, the simple arguments presented 
above can be easily extended to capture the more general scenarios 
\cite{mascher04}. 

One of the important results \cite{mascher04}, which is consistent 
with experimental observations, is that $S(t)$ increases practically 
monotonically whereas $L(t)$ shows hardly any change with time. 
This implies that the two opposite poles of the bipolar spindle move 
away from each other while the overlap of the antiparallel ipMTs 
at the equator is maintained by continuous polymerization of the 
MTs at their plus ends.

\subsection{Section summary}

In contrast to the earlier sections where we studied force generation 
by either specific motors or filaments, in this section we have reviewed 
the kinetics of a system that consists of not only molecular motors of 
different types (porters, sliders, chippers, etc.), but also cytoskeletal 
filaments that generate force by their polymerization and depolymerization.
We have considered most of the main stages of mitosis and the kinetics of 
the key processes as well as the roles of the force generators in those 
stages. 

Not all motors are primarily force generators. For many motors, like 
polymerases, ribosome and mitotic spindle, accuracy of the assigned 
task is more important than power output. Just like muscle and eukaryotic 
flagella, the mitotic spindle is another complex machine where motors 
slide filaments relative to each other. However, its specific power 
output is about six orders of magnitude lower than that of the 
acto-myosin system. The structural and kinetic design of the spindle 
must have been tuned by nature so that the genetic blueprint of a cell 
are segregated accurately and passed onto the next generation; the speed 
of segregation or the force driving the segregation are less important 
than the accuracy of segregation.

The stability of the MT-kinetochore coupling and the strength of the force 
generated depend on the structure and kinetics of the coupler. 
Although significant progress has been made in the last few years, one 
of the fundamental open questions that needs serious attention is the 
following: in those cells where a bundle of MTs, rather than a single MT, 
is attached to a single kinetochore, how are the depolymerization of the 
MTs coordinated and how is the load on the kinetochore fiber shared by 
the individual MTs?

\section{\bf Macromolecule translocation through nano-pore by membrane-associated motors: specific exporters, importers and packers} 
\label{sec-specificexim}

In section \ref{sec-genericexim} we have discussed some generic models 
of export and import of a linear chain through a narrow pore in a thin 
surface. However, several complexities of the macromolecules and the 
nature of the membranes were ignored in section \ref{sec-genericexim}. 
For example, a linear chain is too simple to capture some of the distinct 
structural and conformational features of DNA, RNA and proteins which 
may have important implications for their translocation across a membrane. 
Similarly, instead of being a passive barrier, a membrane may actively 
participate in the export/import process exploiting some specific 
structural features that cannot be captured by a thin structureless surface. 
Furthermore, the molecular composition of the aqueous media on the two 
sides of the membrane, which were almost completely ignored in section 
\ref{sec-genericexim}, often have nontrivial effects on export/import of 
macromolecules across the membrane. In this section we review specific 
examples of the membrane-associated translocation motors and mechanisms 
of their operation. 
We also discuss the mechanisms of motor-driven import, and packaging, of 
viral genomes into pre-fabricated empty capsids. 
  
\subsection{\bf Properties of macromolecule, membrane and medium that affect translocation} 

In general, the speed of translocation is expected to depend on the
(i) properties of the macromolecule, (ii) those of the membrane,
(iii) the nature of the macromolecule-membrane (and macromolecule-pore)
interactions, and (iv) nature of the aqueous media on the cis and
trans sides of the membrane. We list some of the relevant properties 
in this subsection.

The properties of a typical macromolecule that can affect the speed 
of its translocation across a membrane are its (a) length, (b) elastic 
stiffness, (c) electric charge, (c) existence of binding sites on it 
where other molecules can bind and, if such sites exist, their distribution 
along the chain, (i.e., on its primary structure), and (d) its higher 
order structures (i.e., secondary and tertiary structures). 

Normally a membrane is a bilayer of lipids which also contains many proteins 
and other types of molecules. 
(for more details see appendix \ref{app-membranes}).
Some organelles, e.g., mitochondria, have two membranes, called inner 
membrane (IM) and outer membrane (OM). Gram-positive bacteria have 
an outer cell wall in addition to the inner membrane. 
Therefore, the unique features of a membrane can have distinct effects on 
the passage of macromolecules across it. In this context, some of the 
relevant properties of a membrane are its (a) thickness (b) spontaneous
curvature and (c) bending elastic stiffness, (d) variation of its 
composition within the plane and across its thickness, (e) the nature 
of interaction of the macromolecules with the surfaces of the membrane 
as well as with the pore in the membrane, 
etc. 

The macromolecule may adsorb preferentially on one side (cis or trans) 
of the membrane which, in turn, may influence its translocation across 
it through a pore. The size, shape and composition of the pore as well 
as the nature of its interaction with the macromolecule translocating 
through it affect the speed of translocation. At least one of the pores 
that we'll consider is large enough to allow unhindered passage of small 
molecules (and ions) whereas permits passage of macromolecules only if 
``chaperoned'' by a specific type of molecules which consume free energy 
for their operation. In contrast, most of the other pores are much narrower 
and only unfolded macromolecules can pass through such pores. 
The walls of the pore may form a hydrophilic conduit, thereby screening 
out the hydrophobic region of the membrane, making it easier for the 
passage of the macromolecule. Electrostatic charge on the macromolecule 
may give rise to additional interactions with the charged amino acid 
residues of the proteins which form the wall of the pore. 

Properties of the aqueous media that affect translocation of macromolecules 
are the (a) concentration, and (b) charge of the small ions. Concentration 
and binding affinity of ligands that can bind to the translocating 
macromolecule also influence the rate of translocation.

\subsection{\bf Export and import of proteins} 
\label{section-proteinexim}

Export and import of proteins by cells and intracellular organelles 
of eukaryotes are ubiquitous (see Mindell \cite{mindell98} for a 
nice biblical analogy). These processes are carried out by protein 
translocation motors (also called translocases and translocons)
\cite{eichler03,schnell03,wickner05,tomkiewicz07,cross09}
which use input energy to drive their operation \cite{alder03}. 

The protein translocase motors have to meet some essential requirements 
\cite{pohlschroder05}: (i) it must be able to distinguish between its 
correct substrate (i.e, the specific protein to be translocated) from 
incorrect substrates, (ii) it must be capable of discriminating between 
the proteins to be exported (or, imported) and those to be integrated 
into the membrane, and act accordingly, (iii) it should perform its 
function without compromising the integrity of the membrane and without 
allowing undesirable passage of small ions during protein translocation.

The general principles of protein translocation 
\cite{jungnickel94,schatz96a,agarraberes01} 
are as follows: 
(i) Normally, because of size constraints, a folded protein cannot be 
translocated across a membrane; it has to be unfolded before its 
translocation can begin and, usually, gets refolded after crossing 
the barrier.
(ii) Protein translocation can take place (a) during synthesis
({\it co-translation}, e.g., in ER), or (b) after completion of
synthesis ({\it post-translation}, e.g., in mitochondria) 
\cite{mitra06a,mitra06c,kramer09}.


\subsubsection{\bf Bacterial protein secretion machineries} 

Bacteria resort to three principal acts to mount a successful infection 
of an eukaryotic host \cite{lee01b,cerda06}: 
(i) they stick to the surface of the target host cell using specialized 
adhesion proteins or more sophisticated appendages called {\it pili}; 
(ii) they secret toxins in the extracellular environment of the target 
host to neutralize the attack of the host immune system by keeping the 
immune cells at bay (like laying ``mine fields'' in warfare), and 
(iii) secret proteins that get injected into the target host cell;  
some are shot into the host from a distance (like an ``intelligent missile'' 
in a high-tech war) whereas others are injected in large numbers through 
a conduit formed specifically for this purpose (like ``close combatants''); 
In this section we focus almost exclusively on (iii), namely protein 
secretion 
\cite{holland10,thanassi00,kostakioti05,stathopoulos08}.  

The bacterial secretion systems have been divided broadly into two classes 
depending on the number of stages involved \cite{waksman12}: 
(i) {\it one-stage} and (ii) {\it two-stage} systems. In the one-stage 
secretion systems of Gram-negative bacteria the protein is picked up 
for secretion from the cytoplasmic side and released directly in the 
extracellular milieu crossing both the IM and OM without being released 
in the periplasm. On the other hand, in a two-stage secretion system of a 
Gram-negative bacteria the protein is released in the periplasm after 
it is translocated across the IM in the first stage of its journey; 
then it is again captured by another transporter which translocates it 
across the OM into the extracellular space in the second stage of its 
journey. 

In contrast, Gram-positive bacteria have only one membrane and has a cell 
wall. The mechanism of protein secretion across the cell wall of 
Gram-positive bacteria is not well understood \cite{schneewind12,forster12}. 
The protein exporters operating in the membrane of Gram-positive bacteria 
are essentially identical to those operating in the IM of Gram-negative 
bacteria. Therefore, it is not surprising that in some of the pathways, 
the secretion across the inner membranes of both Gram-positive and 
Gram-negative bacteria take place following identical routes-either the 
general secretion (Sec) pathway or the twin-arginine translocation (Tat) 
pathway
\cite{dekeyzer03,pohlschroder05,lee06,economou02,papanikou07,driessen08,yuan10,dalbey12,park12,nijeholt12,frobel12,palmer12,feltcher12}. 
But, once the protein crosses the inner membrane along such a two-stage 
route, the second stage for the ultimate secretion in Gram-negative 
bacteria can be very different from those in crossing the cell wall of 
Gram-positive bacteria. 

Moreover, Gram-negative bacteria also have few other secretory one-stage 
pathways that integrate the machineries of IM and OM without co-opting 
the machines of the Sec-dependent pathway. In this pathway the protein is 
not released freely in the periplasm, but transported sequentially across 
the IM, the periplasmic space and the OM by a single machine that consists 
of several subunits (or modules) that perform distinct functions. Some of 
these integrated machines can even translocate the protein directly into 
a host cell by docking onto the host cell surface. In Gram-negative 
bacteria, at least six functionally independent pathways for protein 
secretion have been discovered; according to the latest convention, these 
are labelled as type-I, type-II,...,typeVI 
\cite{gerlach07,tseng09,delepelaire04,johnson06,douzi12,buttner02,ghosh04,cornelis06,galan06,galan08,erhard10,cascales03,christie01,christie04,christie05,ding03,fronzes09,martinez09,juhasz08,zechner12,henderson04,leo12,dautin07,hazes08,bingle08,filloux08,bonemann10,cascales12}.
These pathways utilize different types of cell surface structures for 
protein secretion \cite{hayes10}. For example, type-III system uses a 
tube to deliver protein effectors into host cells. Type-IV system uses 
a pilus of the type that we have discussed in section 
\ref{sec-specificpistonhookspring}. Type-VI system deploys a device that injects 
effector proteins by puncturing the host cell membrane. 
Some of these machines are also either similar or evolutionarily related 
to other bacterial machines, e.g., type-IV pili or bacterial flagella, etc.

So far as the energetics of the bacterial protein translocation is 
concerned, translocating motors powered by ATP hydrolysis and IMF have 
been discovered 
\cite{kusters11,plessis11}. 
For example, the nano-motor SecA, a key component of the SecA system, 
is fulled by ATP hydrolysis. In contrast, the Tat system is powered 
by protom-motive force (PMF). Moreover, Sec translocates unfolded 
proteins whereas Tat translocates folded proteins. Therefore, Tat 
requires a wider channel than what is adequate for Sec system. The 
conformational changes of Tat must be flexible so that the opening 
of the channel can adapt according to the size of the folded protein. 
At the same time, it must prevent the unwanted passage of smaller 
proteins and small molecules (or small ions). Thus, the task of the Tat 
system seems to be more challenging than that of the Sec system. 

To my knowledge, no quantitative kinetic model has been developed so far 
for any type of bacterial protein secretion system by integrating all 
parts of the corresponding machineries. However, only the basic process 
of for Sec- and Tat-pathways for translocation have been modeled. We'll 
discuss the Tat later in the next subsubsection in the context of 
protein translocation across organellar membranes in eukaryotic cells. 
Here we mention only the models reported for Sec-dependent pathway. 

So far four different mechanisms have been suggested for SecA-mediated 
protein translocation (see ref.\cite{kusters11,plessis11} for reviews); 
these are based on (i) power stroke, (ii) Brownian ratchet, (iii) 
peristalsis, and (iv) subunit recruitment. The peristalsis model is 
actually a combination of a power stroke and Brownian ratchet. 
The subunit recruitment model is analogous to ``active rolling'' of 
dimeric helicase motors, that we discuss in section 
\ref{sec-specificunzippers}.
A reciprocating piston model has been proposed as an attempt to have 
a unified model for protein translocation by SecA \cite{kusters11}.

\subsubsection{\bf Machines for protein translocation across membranes of organelles in eukaryotic cells} 

The organelles are not static objects; their dynamic shapes emerge from 
the interplay of several physical phenomena \cite{shibata06,shibata09}. 
There are two distinct major pathways of protein transport in eukaryotic 
cells: (i) the vesicular pathway, and (ii) non-vesiular pathway.  
We focus here on the non-vesicular pathway where proteins are translocated 
across membranes of organelles by protein-translocating machines.

\noindent$\bullet${\bf Machines for importing proteins by endoplasmic reticulum} 

ER is a very dynamic organelle (see appendix \ref{app-organelles}). 
One special feature of ER is that many ribosomes are located on its 
membrane. Therefore, co-translational translocation is a dominant 
pathway for protein import into ER in which the ribosome partners 
with the translocation machinery. Nevertheless, post-translation 
translocation of protein across ER membrane also takes place 
\cite{hegde11}. 

Sec protein complex, which we have mentioned earlier in the context of 
bacterial protein secretion, also plays a key role in protein translocation 
across ER membrane in eukaryotic cells. However, depending on the 
source and type of the cell used, this translocation process can be 
explained by either power stroke or Brownian ratchet 
\cite{matlack98,tsai02,osborne05,rapoport07,rapoport08}.  
Simon et al.'s Brownian ratchet mechanism of protein translocation 
\cite{simon92,peskin93a} has been applied to the concrete example 
of post-translational translocation across the ER membrane  
\cite{liebermeister01}. Without assuming steady-state condition, 
Liebermeister et al.\cite{liebermeister01} could account for the 
experimental data where a protein BiP plays the role of chaperonins.  
In their model the translocation of the protein was described in 
terms of discrete steps each occurring with a transition probability 
$s$ per unit time. In contrast, in a model developed almost simultaneously 
by Elston \cite{elston02}, the translocating protein was modeled as 
a one-dimensional continuous object which diffuses with an effective 
diffusion constant $D$. Elston \cite{elston02} also made a detailed 
comparison of the two models and the corresponding results. 
One of Elston's conclusions was that the theoretical analysis could 
not decisively and unambiguously establish whether the observed 
translocation is driven by a power stroke or caused by a Brownian 
ratchet.

\noindent$\bullet${\bf Machines for importing proteins by mitochondria and chloroplats} 

Ancestors of the two organelles mitochondria and chloroplasts are bacteria 
(see appendix \ref{app-organelles} for a summary on these organelles). 
However, because of evolutionary changes, now mitochondria and chloroplasts 
have only small genomes of their own. Most of their proteins are encoded by 
nuclear DNA and are synthesized in the cytosol. These proteins are then 
imported by mitochondria and chloroplasts by elaborate mechanisms of 
post-translational translocation. Some of the protein import machineries of 
mitochondria and chloroplasts are adapted from the import/export machineries 
that they inherited from their bacterial ancestor. However, some totally 
new machines were also added to their toolbox thereby opening novel pathways 
for import \cite{balsera09}.

Mitochondria have a translocase of the outer membrane (called TOM) and a 
translocase of the inner membrane (called TIM). Similarly, the corresponding 
translocases of chloroplasts are names as TOC and TIC, respectively.
Do the proteins cross the space between the two membranes of a mitochondrion 
without active assistance of the translocation machinery? If not, how do 
the TOM and TIM (and, similarly, TOC and TIC) cooperate to import proteins 
\cite{pfanner01,perry05,laan06,kutic07,li10b,schmidt10}? 
Moreover, how are the proteins translocated across the thylakoid membrane 
after entering a chloroplast?

The inventory of the components of the translocase complexes of mitochondria 
and chloroplasts have been prepared by extensive experimental investigations 
of several research groups over many years 
\cite{neupert97,neupert07,aronson09,gutensohn06,agne07}. 
In spite of many similarities between the protein import machines and 
mechanisms of these two organelles, there are also some crucial 
differences \cite{schleiff11}. 

Protein import into mitochondria has been analyzed separately using power 
stroke and Brownian ratchet mechanisms \cite{neupert02}. 
Initially, Neupert et al.\cite{neupert90} hypothesized that a Brownian 
ratchet mechanism which was quantified by Simon et al.\cite{simon92}. 
However, a re-examination of the phenomenon revealed \cite{chauwin98} 
that a power stroke mechanism is more plausible than a Brownian ratchet. 
A hybrid mechanism based on a combination of power stroke and Brownian 
ratchet cannot be ruled out \cite{neupert02}. 

Two alternative pathways for translocation across thylakoid membrane exist; 
one is based on ATPase Sec-dependent whereas the other is dependent on 
twin-arginine translocation (Tat) which uses the IMF across the thylakoid 
membrane as the fuel \cite{gutensohn06,albiniak12}. The Tat-pathway of the thylakoid membrane of 
chloroplasts and their prokaryotic counterparts share some common features.

\noindent$\bullet${\bf Machines for protein import by peroxisome} 

The main difference between the translocation of proteins across peroxisome 
and that across the membranes of other organelles is that fully folded 
proteins can be imported into peroxisome without unfolding for this purpose 
\cite{azevedo04}. 
Maintaining such a large pore permanently may not be desirable for the 
peroxisome. The possibility of assembling the machinery at a transient 
pore for protein import has been explored \cite{erdmann05}. 
The process certainly needs energy input from ATP hydrolysis. But, the 
existence and use of IMF across the peroxisomal membrane is, at present, 
uncertain. 

According to the currently accepted ``receptor recycling model'' 
special molecular receptors bind the protein cargo on the cytosolic 
side of the membrane forming a complex. These receptors ``guide'' the 
protein cargo to the peroxisomal protein translocation machinery on 
the peroxisomal membrane. Once the complex enters the channel on the 
peroxisomal membrane, the folded protein continues its onward journey 
on its own while its guide receptors return to the cytosol for being 
recycled or degraded 
\cite{azevedo04,grou09,rucktaschel11}. 
The removal of the receptor is driven by ATP hydrolysis whereas the actual 
translocation 
of the protein does not require any energy expenditure. Since the export 
of the receptor drives the import of the cargo, this ``piggyback ride'' 
of proteins across peroxisomal membrane is sometimes referred to as 
``export-driven import'' \cite{dejonge06,schliebs10}.

\subsection{\bf Export and import of macromolecules across eukaryotic nuclear envelope} 

In an eukaryotic cell, pre-mRNA is synthesized in the nucleus and has to 
be exported to the cytoplasm where an mRNA can serve as a template for 
protein synthesis 
\cite{carmody09,kohler07,komeili01,rougemaille08,kelly09,hocine10,grunwald11,stewart07b,stewart10,cole06,rodriguez04,vinciguerra04,cullen03a,oeffinger12}. 
Similarly, proteins, including polymerases and accessory proteins involved 
in the synthesis of RNA, have to be imported into the nucleus after these 
are synthesized in the cytoplasm 
\cite{gorlich99,koepp96,talcott99,nakielny99,quimby01,lei02,sorokin07,stewart07a}. 
Foreign DNA that enter a cell either during viral infection or during 
gene therapy \cite{marieke06} 
also enter the nucleus from the cytoplasm by crossing the nuclear envelope.

Nuclear pore allows small and medium-size molecules to pass through it by 
passive diffusion, but erects a free energy barrier against the passage 
of macromolecules. But, unlike the molecular motors that drive protein 
translocation across membranes of the other eukaryotic compartments, no 
ATP-driven translocator directly pulls or pushes macromolecules through 
nuclear pore. For the export and import of macromolecules through the 
nuclear pore complex (NPC) 
\cite{fried03,aitchison12},
every eukaryotic cell has an elaborate mechanism that requires expenditure 
of energy 
\cite{gorlich99,koepp96,talcott99,nakielny99,quimby01,lei02,sorokin07,stewart07a,stewart07b,stewart10,carmody09,kohler07,komeili01,rougemaille08,kelly09,hocine10,grunwald11,cole06,rodriguez04,vinciguerra04,cullen03a,zenklusen01}. 
The cargo protein or mRNA crosses the barrier in association with a 
``carrier molecule'' (e.g., importins $\alpha$ and $\beta$ in one pathway). 

Translocation of proteins and mRNA through the NPC involves essentially 
the following four steps \cite{stewart07a}: 
(i) formation of a cargo-carrier complex in the donor compartment, 
(ii) translocation of the carrier-cargo complex from the donor to the 
acceptor compartment by diffusion through the NPC, 
(iii) release of the cargo in the acceptor compartment by a energy-driven 
disassembly of the cargo-carried complex, and (iv) recycling of the 
carriers by their transfer from the acceptor compartment to the donor 
compartment and recharging for the next round of cargo transport.  
When the carrier dissociates from the cargo and binds with its new 
partner RanGTP after reaching the acceptor compartment, the cargo gets 
trapped there and cannot return to the donor compartment. The energy 
required for the step (iv) is supplied from hydrolysis of GTP by Ran 
GTPase. This scenario fits into the generic Brownian ratchet model 
of macromolecule translocation \cite{simon92} that we have already 
reviewed in section \ref{sec-genericexim}.  
Computer simulations of a model \cite{mincer11} based on the structural 
elements of the NPC has provided insight into the molecular level 
details of the Brownian ratchet mechanism of macromolecular transport 
through NPC. 
A Brownian ratchet mechanism has been proposed also for translocation 
of mRNA through NPC. The basic mechanism is very similar to that for 
translocation of protein through NPC except that the energy supplying 
GTP hydrolysis is replaced by ATP hydrolysis \cite{stewart07b}.

The Brownian ratchet model is based on the assumption of passive diffusion 
of the cargo-carried complex in step (ii) of the four-steps listed above. 
In this context one of the key questions in the following: how does 
the cargo-carrier complex pass through the pore by passive diffusion 
whereas the cargo alone cannot do that? To answer this question a 
theoretical model has to incorporate some of the main molecular components 
of the NPC and their interactions with the carrier and cargo. 
Several theoretical models of this type have been introduced in the last 
ten years to understand the molecular origin of the ``virtual gate'' and 
the mechanisms of the diffusive transport of the cargo-carrier complex 
through NPC 
\cite{becskei05,rout03,suntharalingam03,peters05,peters06,peters09,tran06,photos07,zilman07,miao09,terry09,wente10,walde10,colwell10,baygi11,tu11b}

How does the cell coordinate bidirectional traffic through a NPC? 
This question has been addressed by Kapon et al.\cite{kapon08} by 
analyzing a theoretical model inspired by asymmetric exclusion 
process. This model exhibits two distinct modes of transport 
depending on the concentration of the cargoes and the rates of 
dissociation of the carrier-cargo complex. At low cargo concentration 
and high rate of carrier-cargo dissociation transport in the two 
directions can proceed uninterruptedly. In the other mode, which occurs 
at high cargo concentration and slow dissociation of carrier-cargo 
complex, traffic continues in one direction for some time before switching 
to the other direction. The latter mode resembles controlled passage of 
oppositely moving traffic along a narrow passage by alternately allowing 
traffic from each direction (while the opposite traffic waits).

\subsection{\bf Export/import of DNA and RNA across membranes} 

\subsubsection{\bf Export/import of DNA across bacterial cell membranes} 

Three basic mechanisms of intercellular DNA transfer in bacteria are 
\cite{dreiseikelmann94,hellingwerf96,burton10,kruger11}:\\
(i) {\it Transformation}, i.e., uptake of naked DNA (DNA which is
not associated with proteins or other cells) from extracellular
environment;\\
(iii) {\it Transduction}, i.e., indirect transfer of bacterial DNA
into a new cell by a bacteriophage that has the ability to ``inject''
their genomic DNA directly into bacterial cells  
\cite{zinder92} \\
(ii) {\it Conjugation}, i.e., direct transfer of DNA between two
bacteria which are in physical contact with each other, e.g., 
through a nanotube connecting the two cells \cite{ficht11}. 
Since bacteria are not capable of sexual reproduction, conjugation  
may be regarded as a primitive substitutes \cite{redfield01,narra06}.
TrwB \cite{cabezon06,llosa02}, 
one of the major machines for bacterial conjugation, has strong 
similarity of architectural design with a more famous rotary motor 
called F1-ATPase which we discuss in detail in section 
\ref{sec-specificrotary1}. 
DNA translocation across a pore connecting two cells also take place 
for chromosome segregation during cell division and sporulation. 
However, this mechanism and the corresponding machines will be 
discussed later in the section \ref{sec-miscellaneous}.

\noindent$\bullet${\bf DNA transfer across bacterial cell membranes} 

Free DNA is abundantly available in the extracellular environment, 
the main source being the dead organisms. Bacteria take up free 
DNA from this pool for at least three purposes \cite{chen04}: 
(i) to acquire genetic diversity, and thereby, resistance to 
antibiotics; (ii) to utilize DNA originated from a closely related 
bacteria to repair its own damaged DNA; (iii) to use the chemical  
components of the incoming DNA simply as raw material for its 
own maintenance. 
Once inside the host bacterial cell, the fate of the incoming DNA 
depends, at least partly, on its own sequence. It can be assimilated 
into the genome of the bacterial host in a number of different ways 
\cite{ochman00,thomas05}. 

The natural process of genetic transformation in bacteria
\cite{dubnau99,chen04,chen05a,claverys09}  
by DNA uptake from extracellular environment is driven by a machinery 
whose main component is {\it pseudopilus}, which has several structural 
similarities with type IV pili that we have mentioned earlier in  
the section \ref{sec-specificpistonhookspring}. 
After the external dsDNA binds on the surface of the bacterial 
cell, a segment of it is cleaved by a nuclease and one of the strands 
of the resulting duplex is degraded. The remaining ssDNA segment binds 
with the pseudopilus and the latter begins to disassemble thereby 
pulling the ssDNA into the bacterial cell. The main components of the 
machinery and the mechanism of transfer are more or less same for both 
gram positive and gram negative bacteria except that the latter 
has some extra components that facilitates the crossing of the 
outer membrane by the incoming DNA \cite{burton10}.

\subsubsection{\bf Machines for injection of viral DNA into host: phage DNA transduction as example} 

There are two distinct major pathways of viral entry into eukaryotic 
cells \cite{dimitrov04,barocchi05,poranen02}: (i) the vesicular pathway, and 
(ii) non-vesicular pathway.  
In the vesicular pathways the entire capsid may be encapsulated in 
a coated vesicle and internalized by endocytosis;  most of the animal 
viruses follow this pathway. However, once internalized, depending on 
the nature of the virus and the physiological conditions inside the 
host cell, the uncoating of the virion can take place at different 
locations, e.g., in the cytosol or at the nuclear membrane or inside 
the nucleus \cite{poranen02}.

Plant viruses have limited or mostly indirect way of infecting new plant 
cells \cite{poranen02}. 
Viruses of fungi lack machinery to penetrate the strong barriers of these 
eukaryotic cells \cite{poranen02}. Instead they persist in  fungal cells; 
they infect new fungal cells when they get transferred from one cell to 
another during their mating through a pathway that may be called 
non-vesicular. 

The membrane (and cell wall) of bacteria are not permeable to the 
capsids of the bacteriophages. Therefore, the strategy adopted by 
most of the bacteriphages for invading host bacterial cell is to 
inject only the viral genome, leaving the empty capsid outside the 
invaded host \cite{poranen02}. 

The physics of viral genome is intimately related to the physics 
of packaging of DNA into the viral capsid. Therefore, in the next 
subsection we'll discuss the energetics and kinetics of packaging 
of viral genome. 

To my knowledge, the one of the earliest kinetic theories of injection 
of phage DNA \cite{ore56}, based on Brownian motion of the DNA, was 
developed four years after the classic experiment reported by Hershey 
and Chase \cite{molineux06,hershey52,valen12}. 
However, purely diffusive entry into the host is ruled out by the fact 
that it would be too slow to account for the observed speed of translocation. 
An alternative speculative idea, which was quite appealing at that 
time, is that the high internal pressure of the capsid drives the ejection 
of the DNA. In case of tailed bacteriophages the contractile tail was 
believed to work like a hypodermic syringe following the ``uncorcking'' 
of the capsid \cite{zarybnicky69}. Although the contractile tail has a 
definite role to play in the injection process 
\cite{rossmann04,leiman12}, 
the syringe-like pressure-based injection mechanism remains controversial
\cite{molineux01,molineux06}. 
Nevertheless, even if the high pressure of the capsid is not the sole 
driving force for this process, it can assist the operation of specialized 
translocation motors at least during the initial stages of translocation. 
An altogether different mechanism was postulated by Grinius \cite{grinius80}. 
According to this mechanism the IMF across the host cytoplasmic membrane 
should be able to drive phage DNA across it. Although it might appear 
plausible, because of the highly charged nature of DNA, it fails to account 
for the data from many experiments \cite{letellier03}. 
One of the earliest systematic papers on the DNA ejection in virues was 
based on reptation \cite{gabashvili92}.
Finally, a mechanism based on a chemo-mechanical molecular can generate 
the required push or pull for translocation of the DNA from the viral 
capsid to the host cell. Does it exert a power stroke or is it a Brownian 
ratchet?
A comparative study of all possible alternative scenarios \cite{grayson07b} 
clarified their limitations, in spite of partial success of some of 
these mechanisms.

Suppose $P(x,t)$ is the probability of finding a length $x$ of the phage 
DNA ejected at time $t$. Inamdar et al.\cite{inamdar06} wrote down a 
Fokker-Planck equation for $P(x,t)$ which incorporates both a diffusion 
and a drift term. The potential $U(x)$ is is the free energy calculated 
in theories of phage DNA packaging (which we'll discuss in the next 
section). The potential gets contribution from the self-repulsion of the 
charged DNA, its bending and confinement, etc. The key point is that this 
potential as well as its slope (i.e., the ejection force) decreases with 
increasing $x$. The average time taken for injection of the entire DNA 
of length $L$ into the host cell is the corresponding mean first-passage 
time. 

Can this internal pressure, which continues to decrease with increasing 
$x$, sustain the injection of the phage DNA against the osmotic pressure 
of the host cell? Inamdar et al.\cite{inamdar06} extended the Brownian 
ratchet model for polymer translocation, developed earlier by Simon et 
al.\cite{simon92} (and discussed in section \ref{sec-specificpistonhookspring}) 
to argue that the late stages of phage DNA injection is dominated by the 
effective inward pull generated by the DNA-binding ``particles'' of the 
host cell.

\subsection{\bf Machine-driven packaging of viral genome} 
\label{section-viralcapsid}

In the last part of the preceding section we have mentioned briefly 
about the high internal pressure generated by the packaging of the 
DNA in a viral capsid. In this section we review (a) the energetics of 
the packaged viral genome, and (b) the mechanism by which the nano-motor 
at the entrance of the capsid pushes the nucleic acid strand into 
the capsid during packaging 
\cite{gelbert08,knobler09,black89,hendrix98,catalano05,jardine06,nurmemmedov07,bustamante10,casjens11,rossmann12b}. 
The crucial constraint of ``confinement'' 
within the small volume of the capsid makes the viral genome packaging 
more challenging than most of the polymer translocation processes that we have 
discussed in the preceding section \cite{marenduzzo10}.

Two different strategies for packaging of genome has been discovered in 
viruses \cite{kainov06}: (i) the capsid can self-assemble around the viral 
genome, or (ii) the viral nucleic acid can be injected into a pre-synthesized 
empty viral capsid by a packaging motor. The first strategy is adopted 
mostly by the viruses with ssDNA or ssRNA genomes. However, in this 
section we discuss only the second strategy which is adopted by viruses 
with dsDNA and dsRNA genomes \cite{guo07}. 

One of the model systems, which has been very popular among the researchers 
for studying motor-driven packaging of genome, is the bacteriophage $\phi29$ 
\cite{anderson05,bustamante10}; its genome consists of a double-stranded DNA. 
Other dsDNA viruses which have also served as model systems for this purpose  
are the T-odd (e.g., T3, T7) bacteriophages \cite{serwer05}, T-even (e.g., T4) 
bacteriophage \cite{rao05}, the bacteriophage $\lambda$ \cite{feiss05}, and 
bacteriophage P22 \cite{casjens05}.  
$\phi 6$ virus, which has also received attention for genome packaging 
\cite{poranen05}, has 
a dsRNA as its genome.  
For understanding the mechanism of packaging double-stranded RNA into the 
viral capsids, the bacteriophage $\phi6$ has been used as model system. 

About 2 bp are translocated per ATP molecule hydrolyzed by the packaging 
motor. The rate of packaging could be as high as about 2000 bp/s. But 
the translocation gradually slows down as internal pressure builds up with 
the progress of packaging. These motors are also highly processive although 
these are not very efficient.  
The highest pressure generated inside the capsid of the $\phi29$ is about 
$60$ times the normal atmospheric pressure (i.e., about $10$ times 
the pressure in a typical champagne bottle!) and the corresponding force 
applied by the packaging motor is about $60$ pN. Thus, genome packaging 
motors of viral capsids are among the strongest discovered so far. 
What is the mechanism used by these motors to generate such a relatively 
large force (large compared to the forces generated by most of the other 
motors)?

\subsubsection{\bf Energetics of packaged genome in capsids}

As stated earlier, the viral genomes may consist of DNA or RNA. There 
are two alternative mechanisms for packaging of the genome. In case 
of some viruses, the genome is encapsulated by molecules that  
self-assemble around it. In contrast, the genome of other viruses are 
packaged into a pre-fabricated empty container, called {\it viral 
capsid}, by a powerful motor. As the capsid gets filled, The pressure 
inside the capsid increases which opposes further filling. The effective 
force, which opposes packaging, gets major contributions from three 
sources  
\cite{riemer78,odijk98,odijk04,purohit05,locker07,petrov07,petrov08,li08c}:
(a) bending of stiff DNA molecule inside the capsid; 
(b) strong electrostatic repulsion between the negatively charged  
strands of the DNA; 
(c) loss of entropy caused by the packaging.

\subsubsection{\bf Structure and mechanism of viral genome packaging motor}

Are the nucleic acids ``pulled'' or ``pushed'' into the capsid head by 
the motor? Is the packaging motor linear or rotary? 
Single-molecule studies of viral genome packaging motors \cite{chemla12} 
have provided deep insight into the force generated and the mechanism 
of packaging. 
The process of packaging can be divided into different stages 
\cite{earnshaw80}: {\it initiation}, 
{\it packaging} and {\it termination}. The packaging motor is a transient 
multi-component, multi-subunit complex that is assembled at the entrance 
of the capsid just before packaging begins. Initiation also requires 
recognition of the correct substrate (i.e., the nucleic acid).  
Once packaging is complete, 
the packaging motor is disassembled and portal system is ``sealed'' to 
prevent leakage of the viral genome \cite{tavares12}.

\begin{figure}[htbp]
\includegraphics[angle=90,width=0.8\columnwidth]{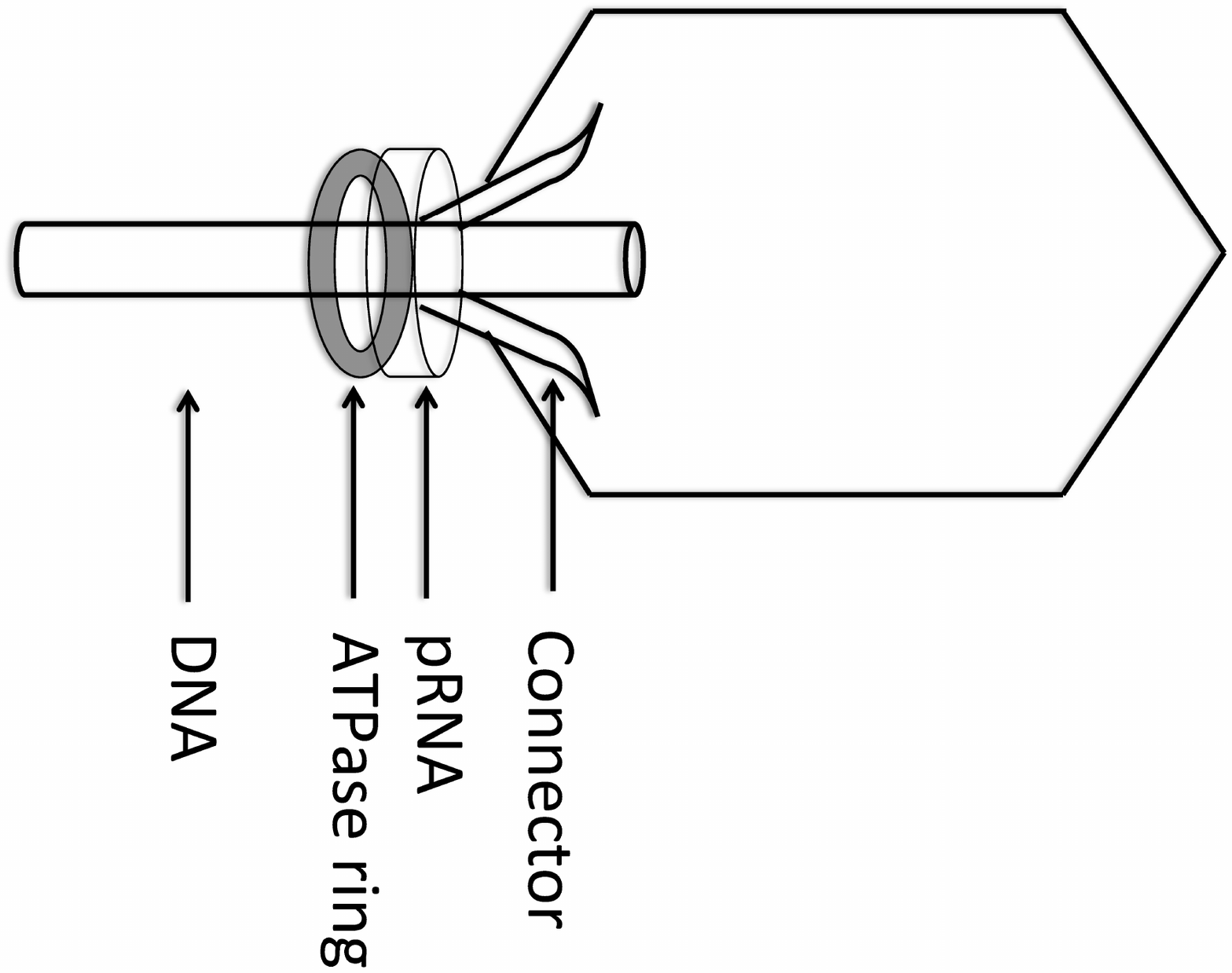}
\caption{A schemetic depiction of a viral capsid where the key components 
of the packaging motor are shown explicitly.
}
\label{fig-capsid}
\end{figure}

The three common structural units \cite{jardine06} of the packaging motor 
are as follows (see fig.\ref{fig-capsid}): 
(i) a ``portal ring'', made of proteins, that connects the capsid head 
to the phage tail, (ii) a ring of RNA molecules, called ``prohead RNA'' 
(pRNS) that is located at the narrow end of the connector, and (iii) 
an ATPase ring, which is known as the enzyme ``terminase''.

A minimal kinetic model of the viral DNA packaging machine has been 
suggested by Yang and Catalano \cite{yang04}. More detailed models,  
that capture structural features of the motor, have been broadly divided 
into two basic types \cite{rao08}: 
(I) {\it rotary} motors, and (II) {\it linear} 
motors. Rotary motors have also been termed as ``portal centric'' while 
the linear motors are ``terminase-centric'' \cite{chemla12}. 

(I) Rotary motor models: 

(1) ``Nut-and-bolt'' model \cite{hendrix78}: In this model the portal ring 
rotates like a ``nut'' causing the linear entry of the ``bolt-like'' DNA 
into the capsid head. The rotation of the portal ring is driven by 
the cyclic change in the conformation of the ATPase during its enzymatic 
cycle.

(2) Compression-relaxation model \cite{simpson00}: In this case, driven 
by ATP hydrolysis, the ATPase causes lengthwise expansion and contraction 
of the portal which, in turn is converted into a rotation of the portal 
that ultimately results in the linear translocation of the DNA into the 
capsid head. For this rotary motor, the portal is the rotor while the 
pRNA and the ATPase, together with the capsid, serve as the stator.

(3) DNA griping model \cite{guasch02}: In this rotary model, the 
packaging motor, as a whole, serves as the rotor while the capsid is 
the stator. The grip of the portal on the negatively charged phosphate 
backbone of DNA arises 
from the positively charged lysines on the portal. The ATPase rotates 
the portal thereby weakening its grip on the DNA and causing linear 
push of the DNA by about 2 nm before re-establishing the electrostatic 
grip with the next set of backbone phophates of the DNA.   

(4) Mexican wave model \cite{lebedev07}: 
There are several protein loops, each of which about 15 amino 
acid long, project into the portal channel. Moreover, each loop 
can take two different conformations. Therefore, in the molecular 
lever model the loops are postulated to behave as levers that can 
tightly grip the phage DNA. ATPase driven portal rotation gives 
rise to the conformational changes of the levers in a sequential 
manner that results in the translocation of the DNA into the 
capsid head. The sequential conformational changes of the loops 
has been described as a Mexican wave. 

(5) Push-and-roll model \cite{yu10}: In contrast to the models based 
on the rotation of the portal ring, this model postulates rotation of 
the DNA induced by a push from levers. These levers grip the DNA 
electrostatically and the release of the inorganic phosphate (P$_{i}$) 
is accompanied by the power stroke. In each power stroke, a lever from 
a single subunit of the motor pushes the DNA in such a way that in each 
complete cycle the DNA moves both into the capsid head as well as in 
a plane perpendicular to it. Upon release of the lever at the end of the 
power stroke the DNA rolls sequentially to the next subunit of the 
ring-shaped motor.   

(II) Linear motor models: \\
(i) Inchworm model \cite{draper07,sun07}: Inspired by structural 
similarities with helicases, terminase has been assumed to 
translocate DNA by an inchworm-like mechanism that successfully 
accounted for DNA translocation by monomeric helicase motors. 
In this speculative model, the opening and closing of the ATP-binding 
cleft drives the movement of the terminase domains which, in turn, 
is coupled to the translocation of the DNA. A plausible scenario is 
as follows \cite{rao08}: ATP-binding closes the cleft between the 
terminase domains and increases its affinity for the DNA; subsequent 
hydrolysis of ATP generates the force that not only opens the cleft 
and releases the products of hydrolysis but also translocates the DNA 
thereby completing one enzymatic cycle.

(ii) Supercoiling model \cite{black78}: This model is based on the 
assumption that upon entering the capsid the leading end of the 
DNA gets fixed onto the portal or the prohead in such a way that the 
terminase can introduce supercoils. This ATP-dependent supercoiling 
driven by the terminase is similar to the primary function of the 
ATP-dependent DNA gyrase. Supercoil of the DNA serves as a transient 
storage of  part of the energy released by ATP hydrolysis. Subsequent 
relaxation of the supercoil triggers DNA propulsion into the capsid.

Most of the models envisage execution of a power stroke by the genome 
packaging motor. However, alternative scenarios that envisage a 
Brownian ratchet mechanism for these motors have also been proposed 
\cite{fujisawa97,serwer03,serwer10}.

\subsection{\bf ATP-binding cassette (ABC) transporters: two-cylinder ATP-driven engines of cellular cleaning pumps}

An ATP-binding cassette (ABC) transporter is a membrane-bound
machine. These machines are found in all cells from bacteria to
humans. In prokaryotic cells, ABC transporters are located in
the plasma membrane. In eukaryotes, ABC transporters have been
found in the internal membranes of organelles like mitochondria,
peroxisomes, golgi and endoplasmic reticulum. These translocate
ions, nutrients like sugars and amino acids, drug molecules,
bile acids, steroids, phospholipids, small peptides as well as
full length proteins.

In spite of wide variations in their functions and substrates
translocated by them, they share some common features of structure
and dynamics 
\cite{altenberg03,higgins04,davidson04,davidson08,rees09}. 
Each ABC transporter consists of four core domains.
Out of this four, two transmembrane domains (TMDs) are needed
for binding the ligands which are to be transported while the
two nucleotide-binding domains (NBDs) bind, and hydrolyze, ATP.
Many ABC transporters are single four-domain proteins. In contrast,
``half-size'' ABC transporters consist of one TMD and one NBD;
many ABC transporters are actually homo-dimers or hetero-dimers
of ``half-size'' transporters.

Some of the fundamental questions specifically
related to the mechanisms of ABC transporters are as follows:
(i) why do these machines need two ATP-binding domains although
it consumes only one molecule of ATP for transporting one ligand?
(ii) Do the two NBDs act in alternating fashion, like a
two-cylinder engine where the cycles of the two cylinders are
coupled to each other? Or, do the two NBDs together form a
single ATP-switch \cite{higgins04}?

\subsection{Section summary}

In this section we have extensively reviewed machines and mechanisms 
of translocation of macromolecules of life, namely, nucleic acids 
and proteins, across the cell boundary and, in case of eukaryotic 
cells, membranes of organelles. While an importer or exporter 
translocates a macromolecule across a membrane the membrane must 
maintain barrier against the passage of small molecules and ions. 
This formidable task is performed by plug domains in the protein 
conducting channels that have been discovered, for example, in the 
Sec pathway. The role of the conformational kinetics of the plug 
domain in protein translocation has been elucidated by careful 
experiments. There is a need for inclusion of this conformational 
kinetics in the quantitative theoretical models of protein translocation. 

Packaging of the viral genome into a pre-fabricated empty capsid by 
the portal motor involves not only translocation of the nucleic acid 
strand through the narrow entrance but also its compact spatial 
organization inside the capsid. The large number of theoretical models 
for packaging motors that we have listed in this section indicates that 
their operational mechanism remains a hotly debated topic with many 
challenging open questions.

\section{\bf Motoring into nano-cage for degradation: specific examples of nano-scale mincers of macromolecules}
\label{sec-specificdegraders}

In this section we'll review the current status of understanding of 
the mechanism of operation of some machines which degrade nucleic 
acids and proteins. We'll also point out the structural and dynamic 
similarities between two machines one of which degrades RNA whereas 
the other degrades proteins. 

Nucleases are enzymes which function as ``scissors'' by cleaving the
phosphodiester bonds on nucleic acid molecules. Endonucleases cleave
the phosphodiester bond within the nucleic acid thereby cutting it 
into two strands whereas exonucleases remove the terminal nucleotide
either at the 3' end or at the 5' end.
Ribonucleases (whose commonly used abbreviation is RNase) are also
nucleases and function as ``scissors'' that cleave the phosphodiester
bonds on RNA molecules. Like all other nucleases, RNases are also
broadly classified into endoribonucleases and exoribonucleases.
Proteases are enzymes which perform functions that are analogous to
nucleases. Just as nucleases cleave the phosphodiester bonds on
nucleic acids (i.e., polynucleotides), proteases cleave peptide bonds
on polypeptides and, hence, sometimes also called peptidase.

\subsection{\bf Exosome: a RNA deagrading machine}

In eukaryotes, a barrel-shaped multi-protein complex, called exosome,
\cite{parker99,raijmakers04,schilders06,hopfner06,houseley06,vanacova07}
degrades RNA molecules. The bacterial counterpart of exosome is usually
referred to as the RNA degradosome. The fundamental questions on the
operational mechanism of these machines are of two types: 
(i) how is the RNA forced into the execution chamber of exosome; and
(ii) what is the size distribution of the RNA segments that are released 
by the exosome after shredding it into pieces?

\subsection{\bf Proteasome: a protein deagrading machine}

Earlier in the context of protein synthesis by ribosome, we have discussed 
the mechanisms of quality control that enhance the translational fidelity 
far beyond what would be achievable by discrimination among the aminoacyl 
tRNA molecules based purely on equilibrium thermodynamics. However, not all 
the polypeptides thus synthesized are competent to function properly because 
of, for example, misfolding. There are additional quality control mechanisms 
that monitor the nascent proteins during various stages of their maturation 
\cite{brenni12}. Defective products are detected and degraded by the cell.
Proteasome is a large and ATP-dependent complex machine for protein degradation 
\cite{rape02,groll03,thomas03,furlow04,pickart04,wolf04,heinemeyer04,bajorek04,sauer04,baker06,demartino07,hanna07,baumeister98,baumeister99,baumeister05,piwko06,smith06,sorokin09,furlow12}
This barrel-shaped machine has structural and functional similarities 
with exosome; what exosome does for RNA, proteasome does for proteins 
\cite{lorentzen06}. 
Obviously, the fundamental questions to be addressed are very similar to 
those in the case of exosomes.

The 26S proteasome consists of the 20S core proteasome and the 19S 
regulator which operates as an ATP-dependent chaperone. 
It has been claimed that the translocation of polypeptides into the 
proteasome and cleavage may be coordinated in such a way that the 
machine is a physical realization of the Brownian ratchet mechanism 
\cite{smith06}. Several possible modes of such coordination have 
been listed as plausible candidates \cite{smith06}. In addition to 
facilitating the biased diffusion of the protein substrate, ATP plays 
several other roles in the operation of a proteasome \cite{goldberg07}. 

Mathematical models have been developed for modeling the two above functions 
of proteasome 
\cite{luciani07,holzhutter00,peters02,hadeler04,luciani05,zaikin05,zaikin06,goldobin09}.
Zaikin and P\"oschel \cite{zaikin05} developed a Brownian ratchet 
mechanism for the translocation of a polypeptide into the proteasome 
and demonstrated a size-dependent rate of transport without explicitly 
incorporating cleavage. In more general versions of this model 
\cite{zaikin06,goldobin09}, 
a polypeptide that enters the channel of the proteasome encounters two 
cleavage centers. Goldobin and Zaikin \cite{goldobin09} also suggested 
how the translocation rate and cleavage strength functions can be 
reconstructed from the experimental data. Because of probabilistic 
cleavage, a polypeptide 
may either undergo cleavage or can escape cleave and continue getting 
translocated along the channel. The fragments generated by the cleavage 
also continue translocation along the channel till they leave the 
channel through the exit. Since smaller fragments move faster than the 
parent polypeptide, further cleavage of these segments before their 
exit from the chamber would be rare.

Holzhutter and collaborators \cite{holzhutter00,peters02} developed a 
kinetic model in terms of master equations for the probabilities of 
observing fragments of different sizes resulting from cleavage. Separate 
equations were written for fragments inside and outside the proteasome 
at time $t$. Illustration of these ideas with simpler situations was 
presented by Hadeler et al.\cite{hadeler04}. Luciani et al.\cite{luciani05} 
wrote down kinetic equations for $N_{k}(t)$ and $n_{k}(t)$ which denote 
the concentrations of fragments of length $k$ outside and inside the 
proteolytic chambers of the proteasomes. The general form of the equations 
are \cite{luciani05} 
\begin{eqnarray}
\frac{dN_{k}(t)}{dt} &=& - [{\rm Influx~ term}] + [{\rm Efflux~ term}] \nonumber \\
\frac{dn_{k}(t)}{dt} &=&  [{\rm Influx~ term}] - [{\rm Efflux~ term}] + [{\rm Cleavage~ terms}]\nonumber \\
\end{eqnarray}
The proteolytic action of the proteasome machine is captured by the form 
of the ``cleavage terms''. There are at least two possible choices for 
selecting the cleavage site: (a) a pre-defined cleavage site that depends 
on the amino acid sequence, or (b) a pre-determined approximate length of 
the fragments so that the most probable cleavage site is at a distance of, 
say, 9 amino acids from one end of the polypeptide, irrespective of the 
actual sequence. The latter choice was made by Luciani et al. 
\cite{luciani05}. 

\subsection{Section summary}

Proteasomes can not only cleave but also splice polypeptides; after cleaving 
it can catalyze a peptide bond between two distal fragments 
\cite{cresswell04,vigneron04,liepe10}. How does a proteasome decide whether 
to splice or merely degrade a protein that enters into its cage?
At present fairly reliable algorithms are available that predict potential 
proteasomal cleavage sites \cite{peters02}. However, to our knowledge, 
no reliable model-based prediction of proteolytic fragments, which are 
generated by two interrelated cleavages, is possible at present. In other 
words, the rules that govern the processing machinery inside the proteasome 
remain unclear.

\section{\bf Polymerases motoring along DNA and RNA templates: template-directed polymerization of DNA and RNA}
\label{sec-specificpoly}

Earlier, in section \ref{sec-genericpolyribo}, we have discussed
some generic aspects of template-directed polymerization. In this
section, we discuss the distinct features of a few specific
machines and their effects on the kinetics of the corresponding
polymerization process. Readers who are not familiar with the 
facts on template-directed polymerization can find a brief elementary 
introduction in the appendices. The tools for single molecule studies 
of these phenomena and typical applications have been reviewed recently 
\cite{dulin13}. Some aspects of template-directed polymerization that 
are covered in ref.\cite{sharma12bprl} will not be repeated here. 

{\it Transcription}, whereby a RNA is polymerized using a DNA template,
is carried out by a machine called RNA polymerase (RNAP). DNA is also 
replicated just before cell division so that genetic blueprint can be 
inherited by both the daughter cells. The molecular machine that 
polymerizes a DNA molecule using another DNA molecule as its template, 
is called a DNA polymerase (DNAP). However, in view of other type of 
polymerizing machines discovered in viruses, it is more appropriate 
to classify the polymerizing machines according to the nature of the 
template and product polymers, as presented in the table 
\ref{table-polymerase}.

There are several common architectural features of all polynucleotide
polymerases. The shape of the polymerase has some resemblance with the
``cupped right hand'' of a normal human being; the three major domains
of it are identified with ``fingers'', ``palm'' and ``thumb''. There
are, of course, some crucial differences in the details of the
architectural designs of these machines which are essential for their
specific functions.

\noindent$\bullet${Comparison between polynucleotide polymerases and 
cytoskeletal motors}

Let us compare these polymerase motors with the cytoskeletal motors.
(i) Polymerase motors generate forces which are about 3 to 6 times
stronger than that generated by cytoskeletal motors.
(ii) But, the step size of a polymerase is about 0.34 nm whereas that
of a kinesin is about 8 nm.
(iii) Moreover, the polymerase motors are slower than the cytoskeletal
motors by two orders of magnitude.
(iv) Furthermore, natural nucleic acid tracks are intrinsically
inhomogeneous because of the inhomogeneity of nucleotide sequences
\cite{kafri05a,kafri05b,kafri04} 
whereas, in the absence of MAPs and ARPs (see appendix \ref{app-cytoskeleton} 
for brief introduction to MAP and ARP), the cytoskeletal tracks are
homogeneous and exhibit perfect periodic order.

\subsection{\bf Transcription by RNAP: a DdRP}

RNAP is a molecular motor \cite{gelles98,buc09} that   
exhibits some distinct phenomena which are not exhibited by the 
other polymerases. We highlight particularly the kinetics of those 
phenomena.
Single subunit DdRP have been found in bacteriophages; one of the  
extensively studied examples being the T7 RNA polymerase. Single 
subunit RNA polymerases have been found also in mitochondria of 
eukaryotic cells \cite{clayton91,silva03,falkenberg07}. 
However, all the other DdRP in all kingdoms of 
life are multi-subunit enzymes. The
eukaryotic DdRP machines are not only larger in size than their
bacterial counterparts, but also consist of larger number of subunits.
There are three different types of DdRP in eukaryotic cells, namely,
RNAP-I, RNAP-II and RNAP-III. The mRNA, which serves as the template
for protein synthesis, is polymerized by RNAP-II whereas rRNA and
tRNA are synthesized by RNAP-I and RNAP-III, respectively.

Ever since the formulation of the central dogma and the discovery of 
RNA polymerase \cite{weiss76,hurwitz05}, many leading groups have 
put enormous efforts in determining the structure of these machines 
\cite{cramer08}.
For the success of the team led by Roger Kornberg, in determining a 
high-resolution structure of RNAP-II and for elucidating the mechanism 
of transcription, Kornberg was awarded the Nobel prize in chemistry 
in 2006 
\cite{landick06a,cramer06,svejstrup06}.
Single-molecule studies have provided insight into the mechanism of 
transcription 
\cite{bai06,herbert08}.

RNAP itself exhibits helicase activity and, unlike DNAP, does not 
require assistance of any separate helicase for unwinding dsDNA to 
expose its template. Moreover, unlike DNAP, it also does not require 
a separate clamp for its processivity and utilizes a segment of 
itself effectively as a clamp. However, if a RNAP somehow dissociates 
from its template, it does not have the capability to to re-associate 
and resume transcription to complete the interrupted process. In 
contrast, for DNAP dissociation and re-association is a routine part 
of its operation.
The RNAP, the dsDNA and the RNA transcript together form what is called 
a ``transciption-elongation complex'' (TEC).

\noindent$\bullet${\bf Transcription initiation: role of scrunching} 

\begin{figure}[htbp]
\includegraphics[angle=90,width=0.8\columnwidth]{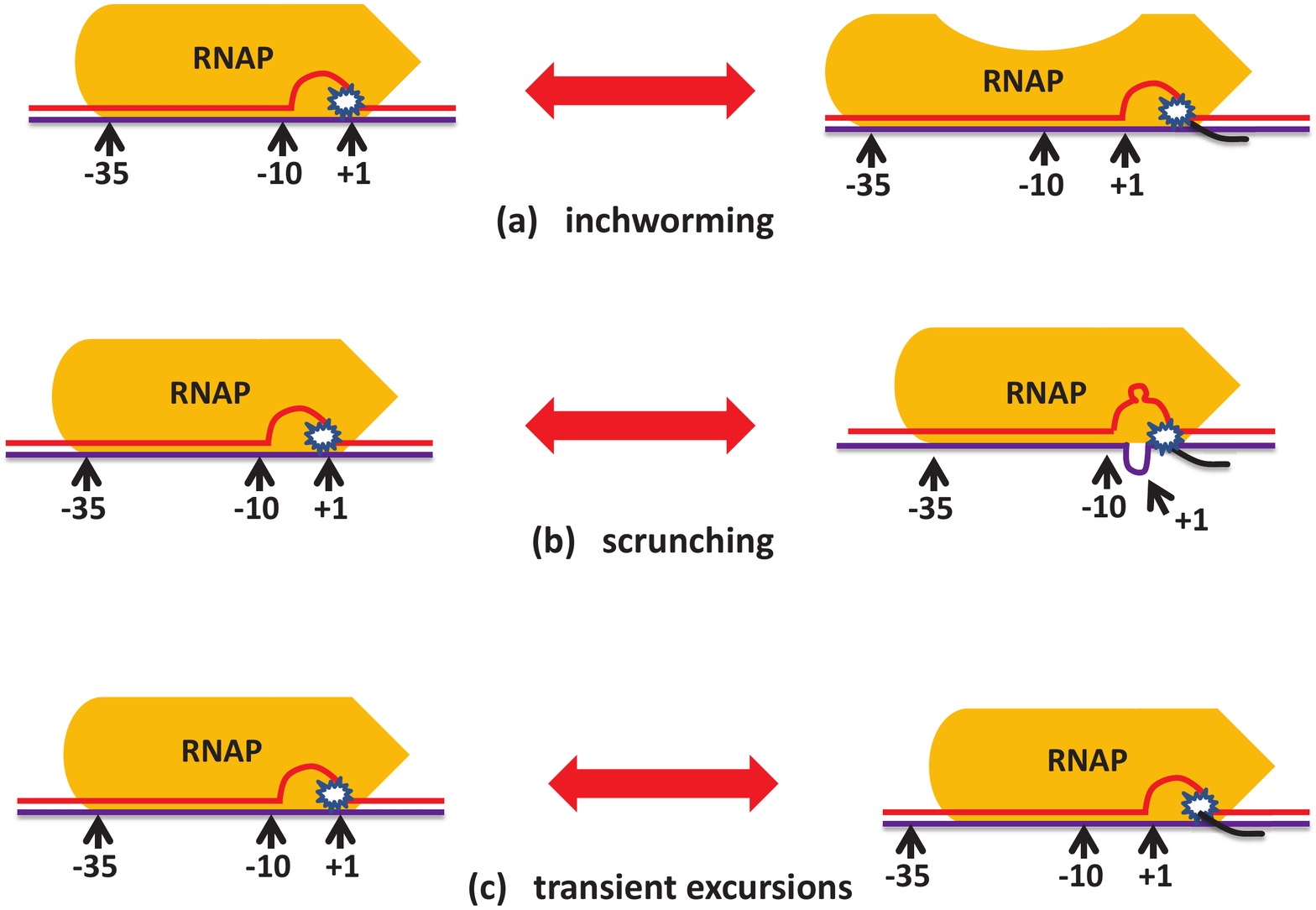}
\caption{Scrunching of RNAP compared with inchworming and transient 
excursions. (Courtesy of Ajeet K. Sharma).
}
\label{fig-scrunch}
\end{figure}

Initiation of transcription needs binding of the RNAP to the appropriate 
sites on its template. Although, in general, it would be expected to be 
a non-equilibrium phenomenon, it may be approximated as a process of 
binding in thermodynamic equilibrium under several realistic circumstances. 
Therefore, RNAP binding has been treated within the framework of equilibrium 
statistical mechanics \cite{bintu05a,bintu05b}.
In the initiation stage, a RNAP can {\it scrunch}  
(see fig.\ref{fig-scrunch} for an explanation of scrunching)
\cite{cheetham99,roberts06,revyakin06,kapanidis06,tang08,vahia11}. 
Some recent theoretical models of the kinetics of transcription initiation 
\cite{xue08}.
incorporate the scrunching-based pathways.

\noindent$\bullet${\bf Transcription termination: role of antitermination} 

Termination of transcription is also a complex phenomenon and it can 
be regulated by intrinsic or extrinsic signals 
\cite{kuehner11,santangelo11}. 
Assisted by anti-termination factors, a RNAP may ignore a stop signal and 
may continue transcribing beyond the stop signal sequence. 

\noindent$\bullet${\bf Elongation: Brownian ratchet or power stroke?} 

Theoretical modelers have paid most of the attention so far on the 
elongation phase. 
Each mechao-chemical cycle in 
the elongation stage involves two main sub-stages: polymerization and 
translocation. Polymerization elongates the RNA transcript by 1 nucleotide. 
During translocation, the RNAP moves forward by 1 nucleotide on the dsDNA 
which requires unzipping the downstream DNA by 1 nucleotide and simultaneous 
re-annealing of the upstream DNA by 1 nucleotide. Moreover, at the same time, 
the RNA-DNA hybrid also moves by 1 nucleotide so that the 3'end of the RNA 
vacates the active site of the RNAP and the RNA-DNA hybrid opens by 1 
nucleotide at the 5'-end of the RNA.

To our knowledge, one of the earliest attempts was the 
model developed by J\"ulicher and Bruinsma \cite{julicher98}. One special 
feature of the model is that is assumes an ``internal flexibility'' of 
the RNAP itself and captures this flexibility by an elastic element that 
allows compression of the separation between the catalytic site and front 
of the RNAP. Thus, the kinetics of transcription is described in terms of 
two stochastic variables: the position of the catalytic site along the 
DNA template and the internal deformation of the polymerase. 
Almost simultaneously Wang et al.\cite{wang98b} developed a stochastic 
model of transcription that included the detailed enzymatic cycle of the 
RNAP. It is essentially based on the Brownian ratchet mechanism of 
energy transduction where the binding of the NTP monomeric subunits 
rectify diffusion of the RNAP. 

In a Brownian ratchet model of RNAP 
\cite{gelles98,spirin02a,sousa05,guo06} 
the polymerase is assumed to fluctuate back and forth between the 
``pre-translocated'' and the ``post-translocated'' positions. In the 
absence of NTP, this motion is essentially an unbiased Brownian motion. 
However, when NTP binds, it rectifies fluctuations thereby biasing 
forward movement of the RNAP. Thus NTP would serve as a ``pawl''. 
High quality single-molecule data  
\cite{abbondanzieri05} 
as well as high-resolution structures obtained from X-ray crystallography 
\cite{brueckner08}
are consistent with this Brownian ratchet mechanism and inconsistent 
with the power stroke mechanism. The predictive power of the Brownian 
ratchet mechanism has been tested more stringently \cite{bai07}. 
Using the parameters extracted under chemical perturbation,predictions 
were made, and tested, for responses to mechanical perturbations. 
Chemical perturbation was achieved by varying the composition of the 
sequence-inhomogeneous template and the concentration of the NTPs. 
Transcription was perturbed mechanically by applying mechanical force. 
Because of the sequence inhomogeneity, the rates of transitions from 
the pre-translocated to the post-translocated states were also 
sequence dependent. These rates of transitions were computed from 
thermodynamic considerations quantifying \cite{bai04} earlier ideas 
formulated qualitatively by von Hippel and collaborators 
\cite{greive05}.         An even more elaborate Brownian ratchet 
mechanism has been proposed and quantified by Bar-Nahum et al. 
\cite{barnahum05}; it involves a static pawl with a asymmetric ratchet 
and a dynamic (reciprocating) asymmetric pawl with a symmetric ratchet. 
But, perhaps, not all RNAPs follow Brownian ratchet mechanism. Structural 
evidences have been presented in support of a power stroke mechanism for 
a class of RNAPs \cite{steitz09}. In order to reconcile these two 
apparently contradictory observations and settle the controversy on the 
mechanism, Yu and Oster \cite{yu12} have developed a model that allow two 
parallel paths- one of these is based essentially on a Brownian ratchet 
mechanism whereas the other utilizes a power stroke. The seven states 
and the kinetic pathways are shown in fig.\ref{fig-yuoster}. 
Approximate analytical expressions for the rates of elongation in 
the steady states of the process were derived separately for 
single-subunit RNAPs and multi-subunit RNAPs (see the supporting 
information of ref.\cite{yu12} for the details). 

\vspace{2cm}

\begin{figure}[htbp]
\begin{center}
{\bf Figure NOT displayed for copyright reasons}.
\end{center}
\caption{Parallel pathways based on facilitated translocation 
and Brownian ratchet within the framework of a single unified 
theoretical model for T7 RNAP. 
Reprinted from Biophysical Journal  
(ref.\cite{yu12}), 
with permission from Elsevier \copyright (2012) [Biophysical Society]. 
}
\label{fig-yuoster}
\end{figure}

An alternative 
formulation \cite{woo06}, based on the Fokker-Planck equation, also 
allows the possibilities of both the power stroke and Brownian rachet 
mechanisms by the appropriate shape of the free energy landscape.  
Not all kinetic models explicitly assume either the power stroke or the 
Brownian ratchet mechanism. Some purely kinetic models 
\cite{tripathi08,yamada09,yamada10} 
can be interpreted in either way because these assume movements of the 
RNAP without explicitly explaining how these movements are caused by 
the energy transduction mechanism. One of these is the ``look-ahead'' 
model \cite{yamada09,yamada10}, which assumes the existence of a 
``look-ahead'' window ahead of the catalytic site within the 
transcription bubble at which NTP can bind reversibly before they are 
covalently linked to the RNA transcript.

\noindent$\bullet${\bf Pause and backtracking of RNAP: kinetics and error correction} 

So far in the context of elongation stage, we have discussed, only the 
pathway that consists of the sub-stages of polymerization and translocation.
However, at each position on its template, multiple pathways are available 
to a real TEC. The pathways alternative to elongation include termination, 
pause and arrest (in backtracked positions), editing, etc. 
\cite{greive05}. 
The relative probabilities of entering into any of the available pathways 
depends not only on the sequence-specific interactions of the RNAP with 
its template DNA, but also on its interactions with the nascent RNA 
transcript as well as that with regulatory transcription factors. 

On the basis of the duration, pauses have been divided in two classes: 
Of all the pauses, about 95\% do not last longer than a few 
seconds and, therefore called ``short'' pause. The duration of the 
remaining 5\% of the pauses can be in the range 20 s - 30 min and 
these are called ``long pauses''. 

The pauses can be broadly divided into two categories also on the 
basis of their relation with backtracking \cite{landick06b}: 
(i) pausing without backtracking, in which only nucleotide addition is 
blocked, (ii) paused state following backtracking in which RNAP reverse 
translocates on its template by a few steps. 
Backtracking has been investigated by single-molecule techniques 
\cite{landick97,shaevitz03,galburt07}
as well as in structural studies 
\cite{wang09,cheung11}

The mechanism of pausing and backtracking as well as the effects of force 
on these phenomena remain controversial 
\cite{shaevitz03,zhou11,galburt07,bai04,landick09,dalal06}. 
According to some schools of thought, all pauses, irrespective of the 
duration, arise from backtracking. In contrast, adherents of alternative 
scenarios believe that the cause of short and long pauses are quite 
distinct; long pauses are associated with backtracking whereas short 
pauses do not necessarily need any backtracking. 
Stochastic models have been developed for the kinetics of backtracking 
\cite{voliotis08,depken09,xie08,xie09,xie12}.

Experimentally observed transcriptional fidelity is a consequence of 
two-stage selection of the nucleotide dictated by the template.

Some of these models explicitly incorporate steps for proofreading 
by either an isolated single RNAP \cite{voliotis09a} or by individual 
RNAP motors in a traffic of RNAPs \cite{sahoo11}.  
In the first stage, the complementarity of shape and Watson-Crick 
base pairing helps in the selection of the correct nucleotide. 
In spite of this selectivity, misincorporations occasionally do occur. 
If a misincorporation takes place, disruption of the catalytic site 
conformation slows down transcription thereby allowing sufficient time 
for activation of the exonuclease activity of the RNAP. Once the 
misincorporated nucleotide is cleaved a new elongation cycle begins 
to select the correct nucleotide. This mechanism for ensuring high 
transcriptional fidelity has been established by structural techniques 
\cite{sydow09} 
as well as single-molecule experiments.

Suppose $n$ ($n=0,1,...N$) denotes the position of the last transcribed 
nucleotide on the template. In other words, $n$ is also the length of 
the nascent transcript. Let $m$ ($m=0,1,...M$) denote the position of 
the TEC (more specifically, the position of the catalytic site of the 
RNAP) {\it relative to} $n$. In this notation, $m=0$ indicates that the 
TEC is an active state whereas backtracked states correspond to $m > 0$. 
The transcription process starts at $n=0,m=0$ and terminates at $n=N,m=M$.  
Voliotis et al.\cite{voliotis09a} defined ${\cal P}_{n}$ 
(and $\bar{{\cal P}}_{n}$) as the probabilities for reaching the 
termination site $n=N$, having incorporated the correct (and an incorrect) 
nucleotide at the position $n$. They obtained a site-wise detailed measure 
of the transcriptional error by calculating $\{{\varepsilon}_n\}$ where 
${\varepsilon}_n = \bar{{\cal P}}_{n}/{\cal P}_{n}$.

For a fixed $n$, $P_{m}(t)$ is the probability of finding the TEC at $m$ 
at time $t$, having started at $m=0$ at time $t=0$. Defining a column 
vector ${\bf{P}}$ whose $M+1$ elements are the probabilities 
$P_{m}(t)$ ($m=0,1,...,M$), the master equations for these probabilities 
can be expressed in a compact notation 
\begin{equation}
\frac{d{\bf{P}}(t)}{dt} = {\bf W}^{\{s\}} {\bf{P}}(t)
\label{eq-masterq}
\end{equation}
in which ${\bf W}$ is the $(M+1) \times (M+1)$ tridiagonal transition 
matrix. The superscript $\{s\}$ indicates that the matrix ${\bf W}$ 
depends on the sequence of nucleotides along the nascent transcript. 
Note that $s$ is a binary variable: $s=0$ for correct transcription 
whereas $s=1$ for incorrect transcription. Thus, for a transcript of 
length $n$, $\{s\}$ is a string of length $n$, each element of which 
can be either 0 or 1 depending on correct or incorrect incorporation 
at the respective positions on the RNA transcript. However, the matrix 
${\bf W}^{\{s\}}$ depends only on the $M$ elements of $\{s\}$, namely, 
on $s_n,s_{n-1},...,s_{s-(M-1)}$.
 
In this formulation, there are $M+1$ ``absorbing boundaries'' through 
which the TEC can make an exit from the template position $n$. 
The absorbing boundary at $m=0$ corresponds to polymerization so that 
the catalytic site moves from $n$ to $n+1$. In contrast, the absorbing 
boundaries at $m=1,...,M+1$ correspond to cleavage of the transcript 
that causes the catalytic site to move from $n$ to $n-m$ because of 
the cleavage occuring at the backtracked state $m$. Applying a technique 
based on Laplace transforms, Voliotis et al.\cite{voliotis09a,voliotis09b} 
calculated $p_{m}$, the probability of hitting the absorbing boundary at 
$m$, and the corresponding mean exit time $t_{m}$. On time scales much 
longer than those for elongation and cleavage of the transcript, the 
kinetics can be described on a coarse-grained temporal resolution in 
which the elongation and cleavage are captured only through the rates 
$p_{0}$ and $p_{m}$, respectively. Suppose $\Pi_{n}(t)$ denotes the 
probability of finding a transcript of length $n$, irrespective of the 
sequence $\{s\}$. Voliotis et al. \cite{voliotis09a} wrote down the 
master equation for $\Pi_{n}(t)$ where the expressions for probability 
flux involve $p_{0}$ and $p_{m}$ that were calculated for fixed $n$. 
From this master equation, they obtained estimates of $\{{\varepsilon}_n\}$ 
by analytical treatment.

\subsubsection{\bf Effects of RNAP-RNAP collision and RNAP traffic congestion}

Two RNAPs can collide while transcribing either (i) the same gene, or
(ii) two different genes. In the former case only co-directional
collision is possible whereas in the latter case both co-directional
and head-on collisions can take place depending on the relative
position of the two genes.

\noindent$\bullet${Two RNAPs transcribing the same gene}

While transcribing the same gene simultaneously, the two RNAPs would
move on the same DNA template strand and are co-directional. This
situation is analogous to that of two vehicles in the same lane of a
highway where both the vehicles are supposed to enter and exit the
traffic at the same entry and exit points on this highway.
In such a co-directional collision, does the trailing polymerase get
obstructed by the leading polymerase or does the leading polymerase
get pushed from behind?

The leading RNAP may stall either because of backtracking or because
of ``roadblocks'' created by other DNA-binding proteins. In both
these situations, the co-directional trailing RNAP can rescue the
stalled leading RNAP by ``pushing'' it forward from behind
\cite{epshtein03a,epshtein03b,jin10}.
Two distinct underlying mechanisms can manifest as ``pushing''
by the trailing RNAP \cite{galburt11}- in the first, the push exerted
by the trailing RNAP on the leading stalled RNAP is a ``power stroke'';
in the second, the leading stalled RNAP resumes transcription by
thermal fluctuation just when the trailing one reaches it from behind
thereby rectifying the backward movement of the leading RNAP by a
``Brownian ratchet'' mechanism.
The elasticity of the TECs may give rise to other possible outcomes of
RNAP-RNAP collisions. For example, if the leading RNAP is ``obstructed''
by a sufficiently strong barrier, the trailing RNAP can backtrack after
suffering collision with it \cite{saeki09}. 
Moreover, the stalled leading TEC must be at least 20 bp away from 
the promoter so that the trailing TEC can get stabilized before 
encountering the stalled TEC in front of it and restart transcription 
by the leading TEC \cite{zhou06}. 

\noindent$\bullet${Two RNAPs transcribing two different genes}

Next we consider the more complex situation where the two interacting
RNAPs transcribe two different genes. Their interaction can cause
transcriptional interference (TI). TI is defined as
\cite{shearwin05,mazo07,beiter09,faghihi09,georg11}
the ``suppressive influence of one transcriptional process, directly and
{\it in cis} on a second transcriptional process''.
TI can occur when both the genes are on the same DNA strand so that
both the RNAPs move from the 3' to 5' end of the template DNA strand.
Anti-sense RNA transcripts are synthesized by RNAPs that translocate
from the 3' to the 5' end of the (+)DNA strand (i.e., codingDNA strand).

\begin{figure}[htbp]
\begin{center}
{\bf Figure NOT displayed for copyright reasons}.
\end{center}
\caption{Three possible promoter arrangements that can give rise to 
transcriptional interference (TI); see the text for detailed explanations. 
Reprinted from Trends in Genetics 
(ref.\cite{shearwin05}),
with permission from Elsevier \copyright (2005).
}
\label{fig-promoters}
\end{figure}

The RNAP transcribing one gene can interfere with the {\it initiation},
or {\it elongation}, or {\it termination} of transcription of another
neighbouring gene. The possibility of a transcriptional interference
between two neighboring genes and the relative {\it direction} of
approach of the two RNAPs is decided by the arrangement of the
interfering promoters. The three alternative arrangements 
(see fig.\ref{fig-promoters}) are named as follows
\cite{shearwin05,mazo07,dinh09,palmer11}: \\
(i) {\it Convergent} promoters (fig.\ref{fig-promoters}(b)):
the RNAPs collide ``head-on''.
(ii) {\it Tandem} promoters (fig.\ref{fig-promoters}(a)):
the RNAPs are co-directional; the collision takes place between the
front edge of the trailing RNAP and the rear end of the leading RNAP.
(iii) {\it Divergent (overlapping)} promoters (fig.\ref{fig-promoters}(c)):
Except for the initial mutual hindrance in starting their journey on
their respective tracks, the RNAPs do not interact once they depart
from the sites of initiation and start elongating their own transcripts.

Based on the {\it stage} of transcription, one can envisage four
different types of interference and the corresponding possible outcomes:\\
(I) {\it Promoter competition}: Occupation of one promoter by a RNAP
hinders the occupation of the promoter by the other RNAP if the
promoters are {\it overlapping (divergent)}. Thus, initiation of
one transcription interferes with the initiation of the other.
Unless influenced by other regulatory molecules, only one of the
two RNAPs succeeds in initiating transcription at a time. The
winner of the competition could be decided randomly if their binding
affinities are comparable. The overall effect is either systematic or
random suppression of the expression of one or both of the genes. \\
(II) {\it Occlution}: A RNAP cannot even initiate transcription,
if the promoter is occluded, at least temporarily, because of the
occupation of the site by another RNAP that is already elongating
its transcript. This phenomenon can occur for both tandem and
convergent promoters. If the occupying RNAP transcribes at a normal
rate, the occlution effect is marginal because it vacates the
promoter site of the other RNAP within seconds. However, if the
occupying RNAP is in a stalled state because of backtracking, it can
delay initiation of transcription of the other gene for a long time 
\cite{palmer09}.
But, no real collision of two RNAPs takes place in this case.\\
(III) {\it Sitting duck mechanism}: If one of the two RNAPs in
still in the stage of transcription initiation (and yet to begin
transcript elongation), it is regarded as a ``sitting duck'' that
can get ``hit'' by another RNAP that is already in the stage of
elongation of its transcript. This mechanism of TI is significant
when the two promoters are sufficiently close but do not overlap.
The most likely outcome is that the sitting duck is dislodged from
the DNA template.\\
(IV) {\it Traffic collision}: If both the RNAPs are engaged in the
elongation of their respective transcripts, a ``head-on'' collision
between their TECs is possible if the promoters are convergent.
Alternatively, a co-directional collision of the two RNAPs is
also possible for tandem promoters if the trailing RNAP transcribes at
a faster rate than that of the leading RNAP. Strong TI can be caused
by such collision if the distance between the two promoters is
sufficiently large so that both the RNAPs are engaged in elongating
their respective transcripts when they collide. If one of the
RNAPs bind strongly to its template while the other binds weakly,
the strongly bound RNAP acts as a ``roadblock'' for the weakly
bound RNAP. The outcome of the collision could be any of the
following depending on the specific organism and the particular
gene:\\
(i) Either or both of the RNAPs can stall and/or backtrack.
(ii) Either or both of the RNAPs can get dislodged from the template
causing premature abortive termination of the corresponding
transcriptions.
(iii) In case of ``head-on'' collision, the two RNAPs can simply pass
each other just like two cars on two adjoining central lanes meant for
oppositely moving vehicular traffic \cite{ma09}.
(iv) In case of ``co-directional'' collision, the stalled leading RNAP
can restart transcription by the push of the trailing RNAP.

Based on the existing mathematical model \cite{sneppen05}, the
following predictions have been made:\\
(i) Occlusion is strong and dominating when the interfering promoter
is very strong.
(ii) Collision dominates TI if the interfering promoter is
strong or if the convergent promoters are sufficiently far apart
(typically $> 200$bp).
(iii) If the interfering promoter is not strong enough or if the
convergent promoters are not far enough from each other, the
sitting duck mechanism is expected to dominate TI.

\noindent$\bullet${Many RNAPs transcribing the same gene: traffic congestion}

Traffic-like collective movement of many RNAPs during transcription of 
a gene and the effects of traffic congestion on the rate of transcription 
has been studied theoretically \cite{tripathi08,klumpp08a,ohta11,klumpp11}. 
The effects of RNAP traffic congestions on the pausing and backtracking 
of RNAPs and the consequent implications for transcriptional proofreading 
have also been investigated theoretically by Sahoo and Klumpp 
\cite{sahoo11}. The theoretical framework of this work has many 
similarities with that formulated by Voliotis et al.\cite{voliotis09a} 
for the special case of nucleolytic proofreading by a single isolated 
RNAP via backtracking. 

\subsubsection{\bf Primase: a unique DdRP}

DdDP cannot begin polymerization of a polynucleotide from scratch.
First, a DNA primase
\cite{arezi00,frick01,siriex05,zenkin06,kuchta10}
polymerizes a short RNA primer using the DNA template. Just like a
RNAP, a primase identifies its binding site on the template DNA by
a particular sequence. This initiation is followed by the elongation
of the RNA primer. Thus, a primase in a special type of DdRP. 
One of the challenging questions is: what determines the length of 
the primer that the primase must polymerize? Once the primer reaches 
that length, the primase disengages and the DNAP takes over adding 
nucleotides to the primer thereby continuing DNA replication. On one of 
two template strands exposed by the unzipping of the dsDNA by a helicase,
the priming event occurs only once in the beginning of DNA replication.
But, on the other template strand priming occurs repeatedly because
on this template DNA replication takes place in discontinuous segments
each of which requires separate priming. The mechanism of the
coordination between the primase and DNAPs on the two template strands
will be discussed in the next subsection.

\subsection{\bf Replication by DNAP: a DdDP}

In this section we study the kinetics of DNAP, a DdDP, during replication 
of DNA \cite{kornberg92}.
The chromosome replication cycle can be broadly divided into a few
distinct stages \cite{diffley02} that are more complex than the stages 
of transcription. Therefore, we present the kinetic processes in the 
order of increasing complexity, rather than the actual sequence of the 
stages of replication of the whole genome.

\subsubsection{\bf Coordination of elongation and error correction by a single DdDP: speed and fidelity}

In their pioneering work on DNAP, Wuite et al.\cite{wuite00} a ssDNA 
molecule, that served as the template, was subjected to tension by 
holding it with a micro pipette at one end and an optical trap on the 
other. Almost simultaneously, Maier et al.\cite{maier00} carried out 
a similar experiment on a different DNAP where a magnetic trap was 
used, instead of an optical trap. In both these experiments, the DNAP 
converted the ssDNA into a dsDNA by polymerizing a strand that is 
complementary to the template. An interesting feature of the data is 
the nonmonotonic variation of the average rate of replication $k(F)$ 
with the tension $F$; $k(F)$ exhibits a maximum \cite{wuite00,maier00}.

The experimental data were interpreted quantitatively in terms of 
a phenomenological model \cite{wuite00,maier00} 
based on a thermally activated rate-limiting step. The rate constants 
$k(F)$ and $k(0)$, in the presence and absence, respectively, of the 
tension $F$ were assumed to be related by 
$k(F) = k(0) exp(-\Delta g/k_BT)$, where 
$\Delta g = \Delta G(F) - \Delta G(0)$ is the force-induced change in 
the free energy barrier $\Delta G$. The barrier $\Delta g$ was 
assumed to have the form 
\begin{equation}
\Delta g = n F[X_{ss}(F) - X_{ds}(F)] - T \Delta s(F) 
\end{equation}
where the first term on the right hand side is the enthalpic contribution 
while the second term is the entropic contribution to $\Delta g$. 
$X_{ss}(F)$ and $X_{ds}(F)$ are the measured (in different experiments) 
extensions per base of the ssDNA and dsDNA strands, respectively, at 
force $F$. Wuite et al.\cite{wuite00} evaluated the $\Delta s(F)$ from 
\begin{equation} 
T \Delta s(F) = n[\int_{0}^{X_{ss}(F)} F_{ss}(x) dx - \int_{0}^{X_{ds}(F)} F_{ds}(x) dx].
\end{equation}
A comparison of the force-extension curves of ssdNA and dsDNA 
explains the observed trend of variation of $k(F)$. For forces 
$F < F_{\ast}$, dsDNA contour length is longer than that of ssDNA 
whereas for $F > F_{\ast}$ the ssDNA is longer; the magnitude of 
the crossover length $F_{\ast}$ is about 5 pN. DNAP induces this 
change in length as it converts a ssDNA into a dsDNA. Therefore, 
for $F < F_{\ast}$ the DNAP is assisted by the external tension 
$F$ whereas for $F > F_{\ast}$ its operation is opposed by the 
applied tension $F$. 

Note that $X_{ss}(F)$ and $X_{ds}(F)$ are the contributions per base to the 
end-to-end net extension of the entire ssDNA and dsDNA chains, respectively. 
Each of these chains, typically, consists of thousands of nucleotides. Thus, 
this phenomenological model \cite{wuite00,maier00} implicitly assumes that 
the global force-extension behavior of ``bare'' long ssDNA and dsDNA chains, 
rather than their local interactions with the DNAP, is the dominant cause 
of the observed $F$-dependence of the rate of DNA replication. Therefore, 
this model was later named as the ``global'' model. The effect of the DNAP 
enters into the results of this model only through the parameter $n$, which 
denotes the number of ssDNA bases that get converted to dsDNA in each 
polymerization reaction catalyzed by the DNAP. It also ignores the fact 
that at the active site of DNAP the ss and ds segments of the DNA are far 
from collinear. The resulting torque generated by the tension $F$ has 
important energetic implications which the global model ignored.

As an alternative to these ``global'' models, a ``local'' model was 
developed by Goel et al. \cite{goel01,goel03} to calculate $\Delta g$ 
for a quantitative interpretation of the same experimental observations. 
This model focusses only on the nucleotides in the immediate vicinity of 
the active site of the DNAP. 

Both the global and local models described above assume that although 
the rate of the rate-limiting step is altered by the tension $F$, none 
of the non-rate-limiting steps become rate-limiting in this process. 
An atomistic model, that does not need to make any of these assumptions, 
was developed later by Andricioaei et al.  \cite{andricioaei04} to 
calculate the force-dependent barrier. This ``restricted-cone local 
model'' \cite{andricioaei04} correctly excludes many of the orientations 
of the DNA because of steric constraints imposed by the DNAP. All the 
models described above can account for the qualitative features, 
particularly the nonmonotonicity, of the variation of the rate of 
replication with increasing tension applied on the template DNA.

In an alternative approach, the force-dependence of the replication rate 
was interpreted by Venkataramani and Radhakrishnan \cite{venkataramani08} 
from a different dynamical perspective in terms of a subtle interplay of 
fast motions of the catalytic site and the slow delocalized modes of the 
DNA-DNAP complex. However, their analysis breaks down because of anharmonic 
effects at strong forces. Therefore, they could present results for applied 
tensions upto a maximum of $6$ pN.  

Application of external tension on the template DNA strand is just one of 
the many possible ways to control the rate of DNA replication. Secondary 
structures of the template, DNA-bound ligands and sequence inhomogeneities 
can also have significant effect on replication speed. Since replication 
is interrupted by pauses caused by the heterogeneous sequence of the 
template, the average replication rate extracted from ensemble measurements 
does not reflect the intrinsic ``speed limit'' of the DNAP motor. In a 
single-molecule study \cite{schwartz09}, that had sufficient spatial and 
temporal resolutions, the paused and burst phases of replication were 
separated to measure the true intrinsic ``speed limit'' of a DNAP.

A DdDP is a dual-purpose enzyme that plays two opposite roles in two
different circumstances during DNA replication. It plays its normal
role as a polymerase catalyzing the elongation of a new DNA molecule. 
However, it can switch its role to that of a exonuclease catalyzing 
the {\it shortening} of the nascent DNA by excision of the nucleotide 
at the growing tip of the elongating DNA 
\cite{kunkel00,kunkel09,krantz10,gill11}.
The two distinct sites where, respectively, polymerization (pol) and 
exonuclease (exo) reactions are catalyzed, are separated by 3-4 nm 
on the DNAP \cite{shevelev02,ibarra09}.
Once a misincorporation of a nucleotide takes place, the DNA is 
transferred from the pol-site to the exo-site. After excision of the 
wrong nucleotide from its growing tip, the trimmed DNA is returned to 
pol-site for resuming its elongation. The elongation and cleavage 
reactions are thus coupled by the forward and reverse transfers of 
the DNA between the pol- and exo-sites of the DNAP. 
``Exo-deficient'' mutants and ``transfer-deficient'' mutants have been
used in single-molecule experiments to understand the interplay of 
exonuclease and strand transfer processes on the platform of a single 
DdDP \cite{ibarra09,xie09m}. By varying the tension applied on the 
template strand, the kinetics of the pol-exo transfer has been probed.

Very recently Sharma and Chowdhury \cite{sharma13polexo} have extended 
the earlier kinetic models to develop a more detailed Markov model of 
DNA replication that captures, in addition to the pol and exo activities, 
also the palm closing and palm opening conformational transitions. 
This model accounts for the observed nonmonotonic variation of the 
average rate of replication with increasing tension on the template 
strand. Going beyond the earlier theoretical treatments of DNA replication, 
Sharma and Chowdhury \cite{sharma13polexo} have also defined, and 
calculated, 9 distinct conditional dwell times of the DNAP motor.

\subsubsection{\bf Replisome: coordination of machines within a machine}

Unlike RNAP, the DNAP is not capable of helicase activity. Therefore,
ahead of the DNAP, a helicase progressively unzips the dsDNA thereby
exposing the two single strands of DNA which serve as the templates
for DNA replication. For the processive translocation of a DNAP on
its template, it needs to be clamped with a ring-like ``DNA clamp'', 
which is assembled by a ``clamp loader'' in a ATP-dependent manner
\cite{bloom06,bloom09,davey02,indiani06,ellison01,kelch12}.
DNAP cannot initiate replication on its own and requires priming by
another enzyme called primase, which we have already described above.
Thus, DdDPs alone cannot replicate the genome; together with DNA clamp
and clamp loader, DNA helicase and primase, it forms a large multi-component
complex machinery which is often referred to as the {\it replisome}
\cite{baker98,benkovic01,johnson05,pomerantz07,egel06,frouin03,bell02,garg05,barry06,mcgeoch08,perumal10,oijen10,hamdan10,labib07,langston09,lee11a,patel11}.
How does the DNAP coordinate its motion with that of the helicase? 
How does a primase and a DNAP coordinate their operation so that when 
primase stops further elongation of the RNA primer and disengages 
itself, the DNAP takes charge and begins polymerization of a DNA strand?
The spatio-temporal coordination of the operation of the different
components of the replisome during DNA replication is the most
interesting aspect of its operational mechanism.

\subsubsection{\bf Coordination of two replisomes at a single fork}

\begin{figure}[htbp]
\begin{center}
\includegraphics[angle=90,width=0.45\columnwidth]{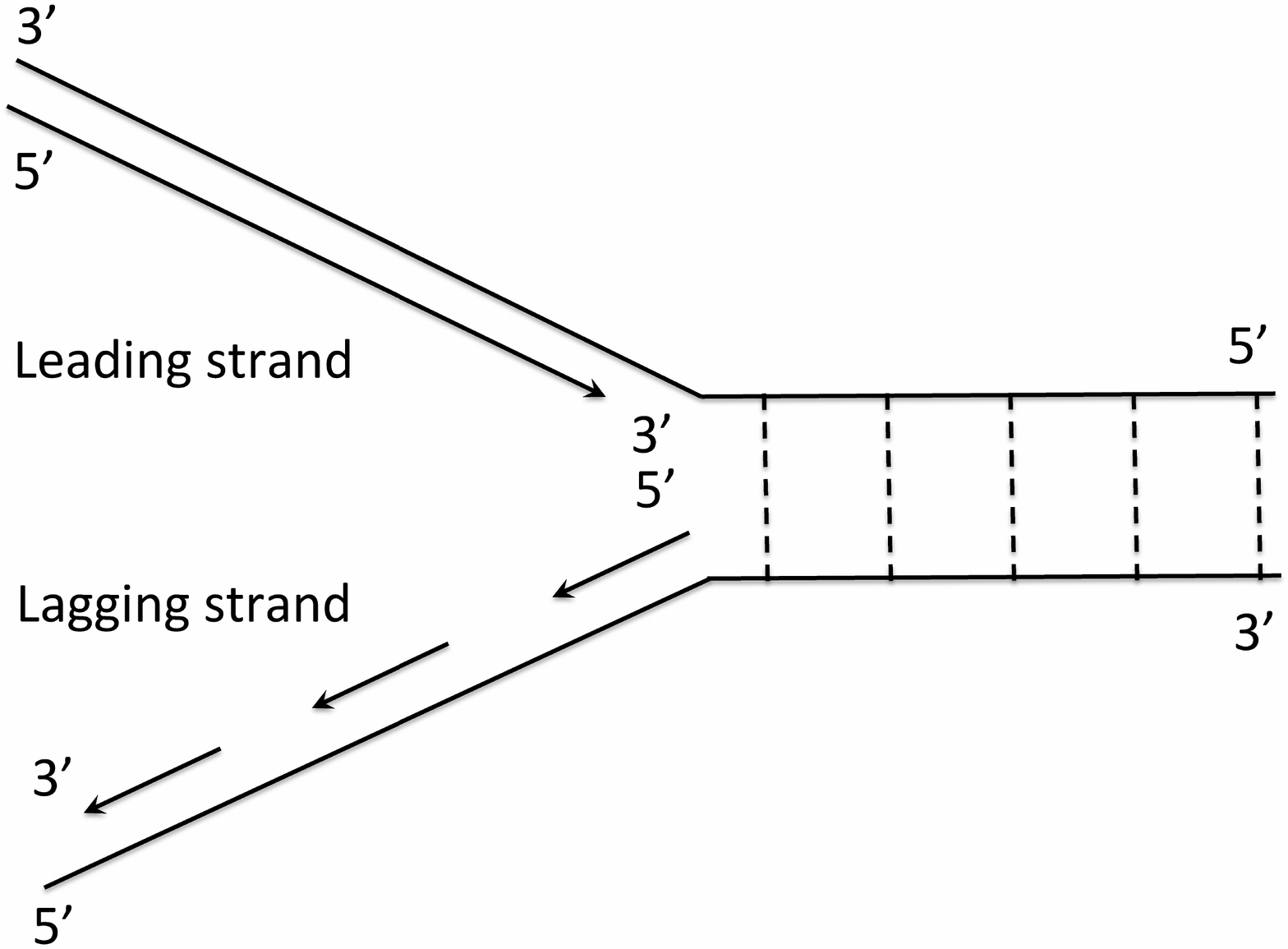}
\end{center}
\caption{A schemetic description of the continuous synthesis of the 
leading strand and discontinuous synthesis of the lagging strand 
during DNA replication.  
}
\label{fig-okazaki}
\end{figure}

Two DdDPs have to replicate two complementary strands of DNA. However, 
each DdDP is capable of translocating only unidirectionally($5' \to 3'$) 
for elongating the product strand. As a result, one of the strands 
(called the ``leading strand'') is synthesized processively, whereas 
the ``lagging strand'' is replicated discontinuously 
(see fig.\ref{fig-okazaki}); the ``Okazaki 
fragments'' synthesized by this discontinuous process are then joined 
together (ligated) by an enzyme called DNA ligase.  
Processing of the Okazaki fragments into a continuous DNA strand
takes place in three steps catalyzed by three enzymes which are
not part of the replisome: (a) removal of the RNA primer by a
separate 5'$\to$3' exonuclease (which is distinct from the 3'$\to$5'
exonuclease domain of the DNAP that is used for proofreading and
editing during elongation); (b) filling the gap, which is left
open by the removal of RNA primer, by a DNA polymerase; (c) sealing
the nick by a DNA ligase.\\

How are the replications of the leading and lagging strands maintain 
tight coordination as the replication fork moves forward? 
Do the DNAPs on the two strands polymerize at the same average rate? 
If so, does the DNAP on the leading strand pause at the replication 
fork till the DNAP on the lagging-strand catches up again? In such a  
scenario, can any component of the replisome, e.g., the primase, 
or helicase, operate effectively as a ``brake'' preventing the 
leading-strand synthesis from outpacing the lagging-strand synthesis 
of DNA \cite{lee11a}? 
Alternatively, does the DNAP on the lagging strand polymerize at a 
faster rate than that on the leading strand so as to make up for the 
time lost in the priming and in re-starting DNA elongation thereby
enabling it to catch up the DNAP on the leading strand \cite{patel11}?

\subsubsection{\bf Traffic rules for replication forks and TECs: DNAP-DNAP and DNAP-RNAP collisions}

In the preceding subsubsection we have studied the coordination of the 
two DNAPs at a single replication fork. Now we review coordination of 
multiple forks during genome-wide replication. 
Moreover, we review some of the common causes for a temporary pause or 
permanent stall of a replication fork by either damage of the track or 
by the present of ``blockage'' on its way \cite{branzei12}.

\noindent$\bullet${Coordination of replication and transcription: DNAP-RNAP collision}

Once replication of the genome begins, the replication forks may 
encounter TECs on the way. Therefore, for an orderly execution of 
replication and transcription, either (a) their progress must be 
coordinated in such a way as to avoid possibilities of collision, 
or (b) there must be a mechanism for resolving such collisions 
\cite{brewer88,rudolph07,pomerantz10b,poveda10,soultanas11,roa11}.
One could imagine that, in addition to the direct physical contact 
between a DNAP and a RNAP in a collision, topological changes in 
the track induced by one polymerase can affect the movement of the 
other. Moreover, a RNAP may find itself stalled in a backtracked 
state while a DNAP approaches it either from front or from behind. 
The outcome of the collision in such a situation need not be the 
same as in cases where the RNAP is actively transcribing a gene.  
Furthermore, if a gene is being transcribed by several RNAPs 
simultaneously while a replication fork approaches that segment of 
DNA, the replication fork may have to deal with multiple RNAPs 
sequentially.

Let us begin with the DNAP-RNAP encounter during their co-directional 
translocation. Suppose the DNAP, the faster of the two, approaches 
the RNAP from behind. The three alternative outcomes of such a collision 
are as follows: (a) the DNAP dislodges the RNAP from the template, 
thereby aborting transcription, and goes ahead with replication; 
(b) the DNAP slows down so as to avert collision; in this case 
the maximum rate of replication can equal that of transcription until 
RNAP detaches from the termination site after completing transcription; 
(c) the DNAP passes (or bypasses) the RNAP without dislodging it so 
that both the polymerases can come out of the encounter unscathed and 
continue their respective tasks. 

Since both DNAP and RNAP move from 3'-to-5' direction on the template 
DNA, they must move on the complementary strands while moving in the 
opposite directions on the same duplex DNA. Therefore, at first sight, 
a head-on collision between a DNAP and a RNAP, while approaching each 
other on the same duplex DNA, may seem unlikely. However, each DNAP 
on the leading strand is accompanied by the other members of the 
replisome and another DNAP on the lagging strand. Thus, although the 
DNAP on the leading strand would not collide with the RNAP approaching 
it from its front, the other proteins at the replication fork would 
certainly suffer a head-on collision with the RNAP. 

To our knowledge, the first systematic survey of the literature 
on the traffic rules of DNAP and RNAP motors was carried out by 
Brewer \cite{brewer88}. 
However, it was the experiments performed, a few years later, by Alberts 
and collaborators, both on co-directional \cite{liu93,liu94} and 
head-on collisions \cite{liu95} between DNAP and RNAP that opened up 
a new frontier of research. 
The general concensus now is that the head-on collisions affect 
replication more adversely than co-directional collision. Replication 
fork stalling \cite{rothstein00,mirkin07}, which can arise also from 
the interactions 
of the fork with non-RNAP proteins, can have severe consequences, e.g., 
genomic instability. However, several other proteins may play roles of 
regulators that either reduce the chances of fork stalling or restart 
stalled fork \cite{pomerantz10a,pomerantz10b,pomerantz10c}. 
To my knowledge, no serious effort has been made till now to develop 
kinetic models of heterogeneous traffic of DNAPs and RNAPs, incorporating 
possibilities of both co-directional and head-on collisions described above.    

\subsubsection{\bf Initiation and termination of replication: where, which, how and when?}

The mechanism of initiation \cite{aves09} and termination \cite{dalgaard09} 
of replication are quite different from those of transcription. 
In bacteria, there is a single location, called the origin of
replication (and denoted by OriC), from where replication fork
propagates {\it bidirectionally} (i.e., in both the clockwise and
counter-clockwise directions on the circular dsDNA)
\cite{kornberg92,mott07}.
The replication is completed when the two forks meet head-on.
Naively, one might think that in eukaryotes the replication of the 
mitochondrial DNA (mtDNA) and that of the chloroplast DNA (cpDNA) 
might be similar to that of bacterial DNA. However, the mechanism 
of replication of mtDNA remains controversial 
\cite{clayton91,clayton03,silva03,falkenberg07,holt09,pohjoismaki10} 
and that of cpDNA remain shrouded in mystery 
\cite{bendich04}. 
The most striking feature of replication of mammalian mtDNA is that 
the leading and lagging strands are replicated from separate origins 
designated as $O_H$ (for the heavy, or leading strand) and 
$O_L$ (for the light, or lagging strand). Moreover, the replication 
initiation of the two strands are also asynchronous; the synthesis of 
the lagging strand begins much later than that of the leading strand.  
The polymerase $\gamma$, the DdDP that drives replication of mtDNA, 
is assisted by a mitochondrial hexameric helicase called TWINKLE 
\cite{kaguni04}.

In comparison with DNA replication in bacteria, that in eukaryotes
seem to be much more complex
\cite{depamphilis93,mechali01,gilbert04,gilbert10,rhind06,rhind10,aladjem07,chagin10,hyrien03,barberis10,bryant11}.
First, the length of the DNA in a eukaryotic cell is so long that 
if it had a single origin of replication, a few years (typically, 
5 years for a human cell) would be needed to complete replication 
once. To circumvent this problem, most of the eukaryotic cells 
initiate replication at many sites (typically, thousands of sites 
in a human cell) so that replication can be completed in minutes 
(in embryos) to hours (in somatic cells). So, the fundamental 
questions are (i) {\it where} are these potential origins of 
replication located, (ii) {\it which} of these potential origins 
actually get activated to begin replication (i.e., ``fire''), 
(iii) {\it how}, i.e., by what kind of molecular signaling or 
interaction, does this ``firing'' take place, and (iv) {\it when}, 
i.e., in which type of temporal sequence, do the origins ``fire''?  
How does the spatio-temporal organization of replication initiation
and progress ensure that no segment is left unreplicated at the end 
and no segment is replicated more than once?
These fundamental questions have been addressed by studying the 
spatio-temporal pattern of firing of origins and the fork propagation 
with several ingeneous experimental techniques \cite{herrick09,hamlin10}.

\noindent$\bullet${\bf ``License'' to ``fire''}

The mechanism of genome wide DNA replication ensures that no segment is 
replicated more than once in a cell cycle. In other words, the 
replication origins get ``licence'' to ``fire'' once, and only 
once, in a cell cycle \cite{diffley96,blow05,machida05,truong11}.

\noindent$\bullet${\bf Where are the potential origins located: marked or unmarked?}

The first fundamental questions is: {\it where} are the potential origins
of replication located? Are these potential origins equispaced or 
distributed randomly along the DNA?  Are these chemically marked on the 
DNA by any specific sequence? Barring a few exceptions, most eukaryotic 
cells randomly select many sites, irrespective of the sequence, as the 
potential origins of replication \cite{hyrien03}. 
A pre-replication complex (pre-RC) of macromolecules, that includes
the origin recognition complex (ORC), is assembled at each selected
site. 

The next obvious question is: how are these potential replication 
sites selected, i.e., spontaneously or guided by any signal molecule(s)? 

\noindent$\bullet${\bf Which of the potential origins ``fire'' and how? }

To begin with, not all the selected potential origins need to get
activated (i.e., ``fire''). Drawing analogy with the biblical statement 
of St. Mathew, this principle has been articulated in the literature 
\cite{depamphilis93} as the statement ``many are called but few are chosen''. 
So, which of the selected origins fire- are they all equally likely to 
fire or some are more likely than others to fire? It is now generally 
believed that a random fraction of the pre-RC get activated for 
initiation of replication. 

\noindent$\bullet${\bf How do the origins fire: molecular communication and interactions?}

The major component of ORC is a helicase. A pre-RC is activated to a 
pre-initiation complex (pre-IC) by a special class of enzymes (kinase). 
Then, recruitment and loading of other components of the replisome, 
including DNAP, and the formation of the two replication forks are 
followed by priming.

\noindent$\bullet${\bf When do the origins fire: simultaneously, in ordered sequence or randomly?}

What is the {\it temporal sequence} in which the origins get activated,
(i.e., they ``fire'')-  simultaneously, or in any ordered sequence, or
random sequence \cite{prioleau10}?  There are strong evidences 
from experiments that not all the selected potential origins fire 
simultaneously. Instead, they fire in a random sequence; however, 
each origin that fires can fire once, and only once, in a cell cycle. 

But this random firing could give rise to another problem 
\cite{hyrien03,rhind06,rhind10}:
there is non-zero probability that a pair of activated sites may be
separated by a very large gap which would take enormously long time
to get replicated. To speed up the process, eukaryotic cells have a
smart strategy. The rate of firing itself keeps increasing with the
passage of time. Consequently, the longer a large gap persists the
higher is the probability that some other pre-RC located in this
gap would fire. 

It is obvious that not all the selected potential origins would get
an opportunity to fire in the cell cycle in which they were selected.
Why does an eukaryotic cell opt for such a redundancy? Perhaps, the
cell has its back-up plans- in case any running fork hits an unexpected
barrier and stalls, its pending job can be completed by one (or more)
of the back-up origins that fire later in the sequence.

Thus, during a specific cell cycle the potential origins of replication 
may have two alternative fates: (a) to ``fire'' and get replicated 
(called ``active'' replication), or (b) not to ``fire'' and get 
replicated when a fork initiated at some other origin passes through it 
(called ``passive'' replication). By monitoring this process over 
sufficiently large number of cell cycles, one can measure the ``firing 
{\it efficiency}'' of a specific origin \cite{luo10}. In general, the 
firing efficiency is not expected to be uniform across all origins.

\subsubsection{\bf Genome-wide replication: analogy with nucleation, growth and coalescence}

In order to develop a theoretical description of genome-wide replication, 
let us recall a classic problem in non-equilibrium statistical mechanics:
{\it nucleation} of ordered crystalline solid in a metastable supercooled 
liquid, followed by growth of the crystallites and their coalescence. 
``Crystals'' {\it nucleate} by thermally activated process. However, only 
those crystalline domains whose initial size is larger than a critical 
size {\it grow}, others simply shrink and disappear. If two (or more) 
growing crystalline domains impinge on one another, they {\it coalesce} 
to form a single crystal that can, then, continue to grow further. For 
nucleation, growth and coalescence of ordered crystalline domains, the 
Kolmogorov-Johnson-Mehl-Avrami (1D-KJMA) model provides many analytical 
results in one-dimension.

Stochastic models for the genome-wide DNA replication has been developed 
by several research groups (see, for example, ref.\cite{lygeros08}).
A formal mapping of the whole genome replication and the 1D-KJMA
was identified a few years ago and fully utilized for quantitative 
analysis \cite{herrick02,bechhoefer07,jun05a,jun05b,zhang06,yang08,yang10}
(see table \ref{table-genomedup}).

\begin{table}
\begin{tabular}{|c|c|} \hline
1D-KJMA  & Genome duplication \\\hline
Crystalling solid & Replicated region \\ \hline
Metastable liquid & Unreplicated region \\ \hline
Nucleation & Firing \\ \hline
Crystal growth & Replication progress \\ \hline
Coalescence of growing crystals & Meeting of replicated regions \\ \hline
\end{tabular}
\caption{One-to-one correspondence between nucleation, growth and
coalescence of crystals in the 1D-KJMA, on the one hand, and firing,
fork propagation and merging of replicated regions in genome duplication,
on the other.
 }
\label{table-genomedup}
\end{table}

The increasing rate of firing would correspond to a increasing rate of 
nucleation in the KJMA-type models \cite{goldar08,goldar09}. 
However, the original 1D-KJMA model 
assumed a stationary (i.e., time-independent) rate of nucleation $I$. 
Therefore, to adapt the nucleation-type theories for describing firing 
of the potential origins of replication, the 1D-KJMA theory had to be 
extended by allowing the rate $I$ to increase with time 
\cite{herrick02,bechhoefer07,jun05a,jun05b,zhang06,yang08,yang10}. 

The rate at which the total fraction of the replicated genome increases
with time quantifies the rapidity of the genome duplication process.
However, the evolving spatial pattern is characterized in more detail by
the distributions of (a) the sizes of the replicated segments, (b) the
spatial gap between the replicated segments, (c) the distance between the
centers of two neighboring replicated segments, etc.
The effects of defects on the kinetics has also been investigated within 
the framework of this formalism \cite{gauthier10}.

The main quantity calculated in the KJMA-based theory is the local 
initiation rate $I(x,t)$. On the other hand, experimental data on 
genome-wide replication provides information on the unreplicated 
fraction $s(x,t)$. In most of the early attempts in testing the 
predictions of the KJMA-based theory, curve-fitting strategies were 
used to estimate $I(x,t)$ from the genome-wide replication timing data. 
In a recent work Baker et al. \cite{baker12} have analytically inverted 
the KJMA-based model deriving the expression 
\begin{equation}
I(x,t) = - \frac{V}{2} \Box ln s(x,t) 
\label{eq-baker}
\end{equation}
where $\Box = (1/V^2)\partial_{t}^2 - \partial_{x}^2$ is the d'Alembertial 
operator. The equation (\ref{eq-baker}) can be used to extract $I(x,t)$ 
from the experimental data.

\section{\bf Ribosome motor translating mRNA track: template-directed polymerization of proteins}
\label{sec-specificribo}

Ribosome, one of the largest and most sophisticated macromolecular
machines within the cell, polymerizes polypeptides using a mRNA
as the corresponding template
\cite{frank11a,frank06,mitra06b,spirin99,spirin02b,spirin04,spirin09,rodnina11,moore12}. 
Although it works effectively as a polymerase, it differs from the 
polynucleotide polymerases in two important respects: 
(i) it is a ribonucleoprotein whereas polynucleotide polymerases 
are proteins, 
(ii) the track is a mRNA strand, rather than DNA, and the step size 
is 3 nucleotides, instead of a single-nuceotide step-size of polymerases, 
and  
(iiii) it ``translates'', instead of replicating or transcribing, the 
genetic message.

A critical analysis of the free energy cost of protein synthesis was 
initiated, to my knowledge, by Chetverin and Spirin \cite{chetverin82}.
In each successfully completed mechano-chemical cycle of a ribosome
two molecules of guanosine triphosphate (GTP) are hydrolyzed into
guanosine diphosphate (GDP). Moreover, one of the steps of this cycle 
needs the assistance of specifically prepared accessory molecular 
assembly (aa-tRNA) whose prior preparation also involves hydrolysis of 
a molecule of ATP. Because of these energy-consuming steps involved in 
the operation of a ribosome, it is regarded as a molecular motor
\cite{abel96}.
It has been argued \cite{spirin99} that the the energy of the chemical
bond between the amino acid and tRNA is later used by the ribosome for
forming a peptide bond between this amino acid and the nascent polypeptide.
Just like RNAP (and unlike DNAP), ribosome is capable of helicase activity. 
There are strong indications \cite{qu11} that it uses two active mechanisms 
for unwinding mRNA during translation.

A ribosome is not merely a ``protein-making motor protein''
\cite{cross97b} but it serves as a ``mobile workshop'' which provides
a platform where a coordinated action of many tools take place for
the selection of the appropriate subunits and for linking them to
synthesize each of the proteins. As this mobile workshop moves along
the ``assembly line'' (mRNA), new subunits (amino-acids) are brought
to it by the ``workers'' (tRNA molecules).

\subsection{\bf Composition and structure of a single ribosome and accessory devices}

The mechanisms of ribosomes in bacteria as well as those of the 
cytoplasmic and organellar ribosomes 
\cite{agrawal11,somanchi01,hauser06,navarro07,zehavi07} 
in eukaryotes have lot of similarities, in spite of differences in 
the details of their composition and kinetics. 
Numerical values of some of the parameters that characterize the physical 
properties of a typical ribosome have been listed by Moore (see table 2 
of ref.\cite{moore12}). 
 
Even in the simplest organisms like single-cell bacteria, a ribosome is
composed of few rRNA molecules as well as several varieties of protein
molecules. The reason for the complexity of its composition and structure 
can be understood by studying its possible origin and evolution over 
billions of years \cite{fox11}. 

The structure of both bacterial and eukaryotic ribosomes have been
revealed by extensive detailed investigation over several years by
a combination of X-ray diffraction, cryo-electron microscopy, etc.
\cite{westhof00,moore03,steitz08,steitz10,yonath02,yonath09,yonath10,ramakrishnan02,schmering09,ramakrishnan10,frank00,frank09,frank10,yusupov01,blaha04,dunkle10}. 
For this achievement, V. Ramakrishnan, T.A. Steitz and A. Yonath
shared the Nobel prize in chemistry in 2009
\cite{ramakrishnan10,steitz10,yonath10}.
For many years the mechano-chemical kinetics of ribosomes have been
investigated by studying bulk samples with biochemical analysis as
well as the structural probes mentioned above. Only in the last few
years, it has been possible to observe translation by single isolated
ribosome {\it in-vitro}
\cite{marshall08b,blanchard04,uemura07,munro08,munro09,blanchard09,uemura10,aitken10a,aitken10b,uemura11,petrov11,vanzi07,wang07c,wen08,tinoco09}.
Very recently Sanbonmatsu \cite{sanbonmatsu12} has presented an 
excellent review of the theoretical and computational studies of 
ribosome. 

\subsubsection{\bf Molecular composition and structural design of a ribosome}

Ribosomes found in nature can be broadly divided into two classes:
(i) prokaryotic $70$S ribosomes, and (ii) eukaryotic $80$S ribosomes;
the numbers $70$ and $80$ refer to their sedimentation rates in the
Svedberg (S) units.  There are separate channels in the 
ribosome for the passage of the template mRNA and the nascent polypeptide. 
In the earliest electron microscopy the prokaryotic
and eukaryotic ribosomes appeared to be approximately spherical
particles of typical diameters in the ranges $20-25$ nm and $20-30$nm,
respectively. 

\noindent$\bullet${\bf Large and small subunits of a ribosome} 

In the earliest electron micrographs of ribosome a visible groove was
found to divide each ribosome into two unequal parts; the larger and
the smaller parts are called, for obvious reasons, large and small
subunit, respectively. The sizes of the large and small subunits of
the $70$S ribosome are $50$S and $30$S respectively, whereas those of
the $80$S ribosome are $60$S and $40$S, respectively. 
The two subunits of a ribosome interact directly via ``intersubunit 
bridges'' \cite{gao06b}.

The small subunit assists in the decoding of the genetic message by 
mediating the base-pairing interaction between the tRNA and the 
template mRNA. But, the actual polymerization of the polypeptide 
takes place at a site, called peptidyl transferase center, that is 
located in the large subunit. These operations of the two subunits 
are coordinated by a L-shaped adapter molecule called tRNA. 

\noindent$\bullet${\bf tRNA and amino-acyl tRNA synthetase}

The tRNA molecules are sufficiently long so that their two ends can interact 
with the two subunits simultaneously. The intersubunit space is large 
enough to accomodate just three tRNA molecules which can bind, at a time, 
with the three binding sites designated as E, P and A. Moreover, the 
shape of the intersubunit space is such that it allows easy passage of the
L-shaped tRNA molecules. The end which carries the amino acid interacts 
with the large subunit while the other end, where the anticodon is located, 
interacts with the codon on the mRNA located on the small subunit. 
If the mismatch between the codon and anticodon is limited to just 
one codon, the tRNA is called near cognate whereas mismatch of larger 
number of codon-anticodon pairs occurs if the tRNA is non-cognate.
Aminoacyl tRNA synthetase (aa-tRNAsynth) ``charges'' a tRNA molecule
with an amino acid.  
\cite{ibba04,ling09,banerjee10,yadavalli12}.

\noindent$\bullet${\bf Elongation factors, initiation factors, release factors and NTP hydrolysis}

Elongation factors (EF), which are themselves proteins, play important
roles in the control of the major steps in each elongation cycle of a 
ribosome. Both EF-Tu and EF-G are GTPase. EF-Tu is a key player in the 
selection of cognate tRNA. Similarly, elongation factor G (EF-G) 
coordinates the orchestrated movements of the tRNA molecules within the 
ribosome with that of the ribosome along its mRNA track. 
As the name suggests, initiation factors (IF) are proteins that facilitate 
the steps involves in translation initiation \cite{sonenberg03}.
The ribosome recycling factor \cite{weixbaumer07}
plays an important role in the dissociation of the large and small 
subunits of a ribosome \cite{hirokawa05}.

\subsection{\bf Polypeptide elongation by a single ribosome: speed versus fidelity}

A ribosome covers about 30 nucleotides on the mRNA track. But, it is 
neither a hard sphere not a hard rod. It undergoes several functionally 
important conformational changes during each cycle 
\cite{fraser07,frank10,agirrezabala09,agirrezabala11} 
which can be described quantitatively as the conformational kinetics 
of the ribosome in an energy landscape \cite{whitford11}.

During the elongation stage, while translating a codon on the mRNA
template, the three major steps in the mechano-chemical cycle of a
ribosome are as follows \cite{cooperman11}:
In the first, based on matching the codon with the anticodon on the
incoming aa-tRNA, the ribosome {\it selects} the correct amino acid
monomer that, according to the genetic code, corresponds to this
codon \cite{banerjee10}. Next, it catalyzes the chemical reaction responsible for the
formation of the peptide bond between the nascent polypeptide and
the newly recruited amino acid resulting in the {\it elongation} of
the polypeptide. Final step of the cycle is {\it translocation} at
the end of which the ribosome finds itself at the next codon and is
ready to begin the next cycle.

\begin{figure}[htbp]
\begin{center}
\includegraphics[angle=90,width=0.45\columnwidth]{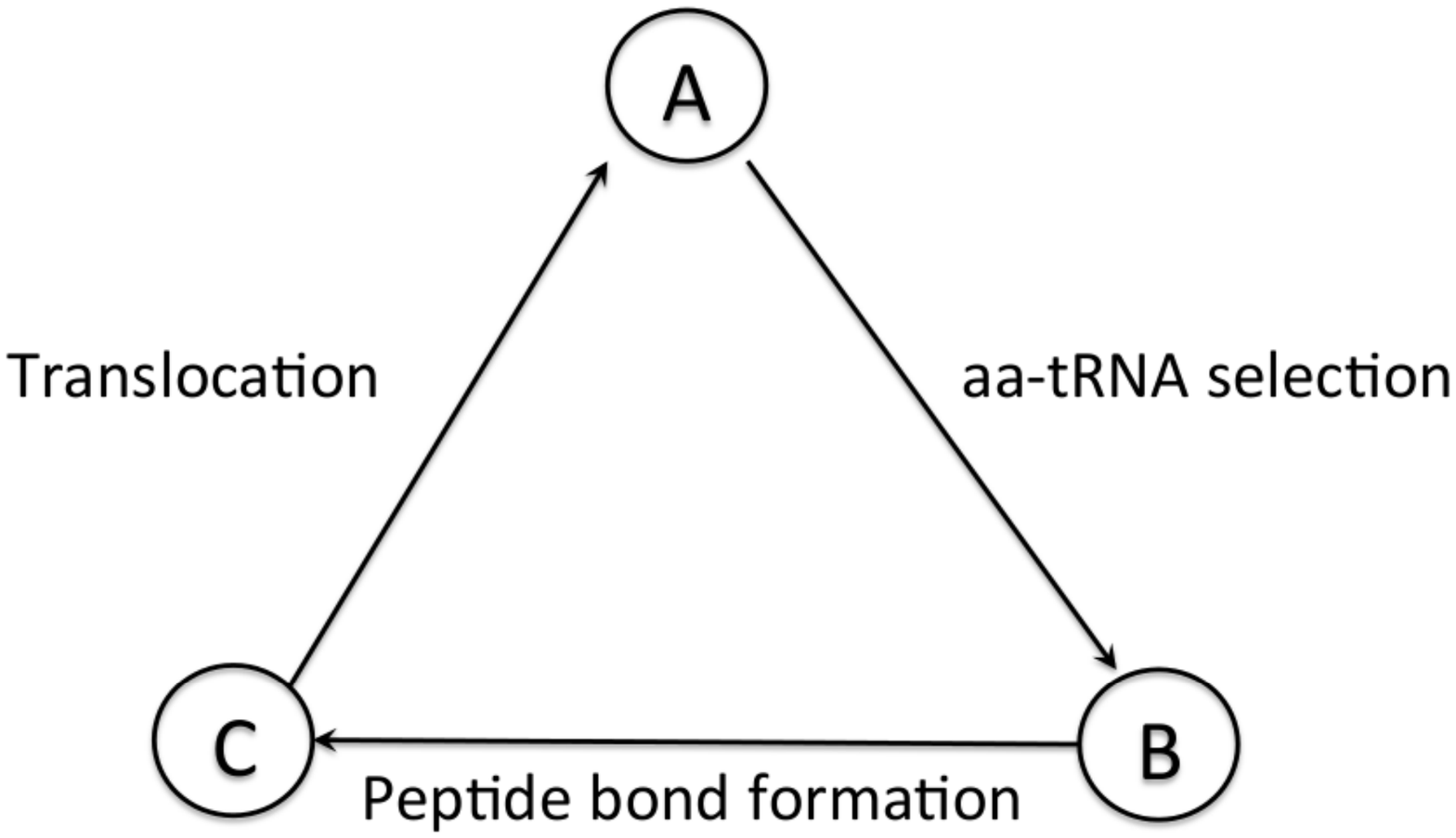}
\end{center}
\caption{A simplified 3-state markov model for the elongation 
cycle during translation. It captures only the three key steps 
of this cycle. 
}
\label{fig-3state}
\end{figure}

Clearly, the 3-state cycle sketched in fig.\ref{fig-3state} is an
oversimplified description of the mechano-chemical kinetics of a
ribosome during the elongation stage. It is inadequate to account
for most of the phenomena which answer the questions listed above.
We'll see in this section that at least two of the three steps in
fig.\ref{fig-3state} consist of important sub-steps. Moreover, the
aa-tRNA selected (erroneously) by the ribosome may not be the
correct (cognate) tRNA. Rejection of such non-cognate and near-cognate
tRNAs by the process of kinetic proofreading leads to an alternative
branch completion of which ends up in a futile cycle.

\subsubsection{\bf Selection of amino-acid: two steps and kinetic proofreading}

Selection of the amino-acid consists of a series of steps at least 
two of which have major implications  in optimization of translational 
speed and accuracy 
\cite{rodnina11c,rodnina12,daviter06,wohlgemuth11,lovmar06,johansson08,johansson11}. 
The ternary complex consisting of aa-tRNA, EF-Tu and GTP enters 
the ribosome and forms a labile complex with the ribosome. 
In the first major step non-cognate tRNA are ejected from the 
ribosome because of codon-anticodon mismatch. This step exploits 
mainly the difference between the free-energies of binding of 
cognate and non-cognate tRNAs. However, this difference is inadequate 
to discriminate between cognate and near-cognate tRNAs. 
In the next major step, EF-Tu gets activated and it hydrolyzes GTP. 
The release of inorganic phosphate induces a change of conformation 
of EF-Tu because of which EF-Tu loses its affinity for the 
aa-tRNA. The cognate tRNA, released from the grip of EF-Tu now 
moves to the binding site P; alternatively, at this stage, the 
tRNA gets rejected by the ribosome if it is near-cognate (or 
non-cognate). 
Kinetic proofreading takes place in the second step which involves GTP 
hydrolysis \cite{zaher09}. 
For a detailed discussion on the thermodynamics of the interaction 
between EF-Tu and aa-tRNA, see ref.\cite{schrader11}.  
How is information on the correctness/incorrectness of the codon-anticodon 
matching on the small subunit transmitted to the EF-Tu and to the large 
subunit where the aa-tRNA contributes the amino acid to the elongating 
protein? 
Models based on alternative pathways for signal transmission have been 
proposed \cite{powers94,pipenburg00,yarus03,sanbonmatsu06}. 
Codon-recognition by itself may not be sufficient for stimulating the 
GTPase activity of EF-Tu \cite{pipenburg00}; 
successful mechanical distortion of the tRNA by the ribosome seems to 
be necessary for sending the required signal on the codon-anticodom 
matching to the EF-Tu GTPase \cite{yarus03}. Action of tRNA as a 
``molecular spring'' in both decoding mRNA and translocation are 
well documented \cite{frank05,moran08}.

\subsubsection{\bf Peptide bond formation: peptidyl transfer}

The tRNA molecule that has donated the last amino acid monomer to
the nascent polypeptide and to which the growing end nascent
polypeptide remains bound is called the {\it peptidyl-tRNA}.
Throughout the elongation stage, the ribosome retains the peptidyl-RNA
in the intersubunit region. As long as the freshly arrived aa-tRNA
at the A site goes through all the identity checks by the ribosome's
quality control system, the peptidyl-tRNA remains bound to the P
site. Once the aa-tRNA at the A site is passed by the proofreading
system, the ribosome catalyzes the peptidyl transferase reaction
whereby the nascent polypeptide is transferred to the aa-tRNA by
the formation of a new peptide bond between its amino-acid cargo and
the nascent polypeptide.

\subsubsection{\bf Translocation: two steps of a Brownian ratchet?}

In the late 1960s, Bretscher \cite{bretscher68} and Spirin \cite{spirin69} 
independently proposed a concept of ``locking-unlocking'' 
\cite{spirin02b,spirin04,spirin09} 
of a macromolecular complex that should be applicable also to ribosomes.
The complex was assumed to oscillate between locked (closed) and 
unlocked (open) states. In the unlocked state it accepts or releases  
substrates (including other ligands), moves ligands inside the complex, 
and releases products (and other ligands). In the locked state ligands 
and subatrates remain practically immobile and chemical reactions take 
place. The system can have more than one locked and unlocked states.
The currently accepted mechano-chemical cycle of ribosomes can now be 
interpreted in terms of the locking-unlocking concept \cite{spirin02b}. 
The entry of the aa-tRNA and rejection of non-cognate tRNA are possible 
in the unlocked states whereas the peptidyl tranfer reaction requires 
a locked state \cite{valle03}. Completion of this reaction causes 
unlocking so that translocation process can begin. 

Immediately after the peptidyl transferase reaction is completed, 
the ribosome is in the {\it pre-translocational} state in which  
the P site remains occupied by the deacylated tRNA while the 
peptidyl-tRNA is located at the A site. Before translating the 
next codon, the following processes must take place so that 
the system finds itself in the initial state of the next elongation 
cycle: (i) the deacylated tRNA must move from the P site to 
the E site while the peptidyl-tRNA moves from the A site to the 
P site, and (ii) the ribosome moves forward, along its mRNA track, 
so that the next codon is exposed to its A site. 
Thus, the transition to this {\it post-translocational} state from the 
{\it pre-translational} state 
involves coupled movements of two species of RNA: forward movement 
of the tRNA molecules through the inter-subunit space to their next 
binding sites, and a coordinated movement of the mRNA template along 
a groove in the small subunit 
\cite{spirin85,wilson98,noller02,korostelev08,spirin02b,joseph03,frank07,shoji09,chen12}. 
As we explain below, translocation needs the action of the GTPase 
EF-G.
The dynamics of translocation has been investigated by several 
experimental techniques; the most important recent results have been 
obtained by a combination of the complementary techniques of smFRET 
and cryo-electron microscopy 
\cite{frank11b,noller11,rodnina11b}.

Interestingly, the two ends of a tRNA do not translocate simultaneously. 
Keeping in mind that tRNA molecules interact with both the large and 
small subunits whereas the mRNA interacts with only the small subunit, 
the actual process of translocation can be split into two steps: 
(i) translocation of the tRNAs with respect to the large subunit and 
(ii) translocation of the mRNA and tRNA with respect to the small subunit.

First let us consider the movement of the two tRNA molecules with 
respect to the large subunit of the ribosome.
The acceptor stems (i.e., the ends which can get aminoacylated) of 
the tRNAs located at the P and A sites translocate spontaneously 
to the E and P sites, respectively, on the large subunit while their 
opposite ends (anti-codon end) reside at the P and A sites on the 
small subunit thereby causing a transition from the ``classical'' 
P/P,A/A state to the ``hybrid'' E/P,P/A states, respectively 
\cite{moazed89,dunkle11}. 
The spontaneous fluctuation of the two tRNA molecules between the 
classical and the hybrid states is accompanied by the rotational 
Brownian motion of the small subunit with respect to the large subunit 
\cite{horan08,cornish08,marshall08c}; the angle of relative rotation 
of the two subunits is $\sim$6$^{0}$. 
The classical state of the tRNAs find themselves in the non-rotated 
state of the ribosome and this composite state, in modern terminology 
\cite{agirrezabala09}, is often referred to as macro-state I (MS-I). 
In contrast, in the rotated state of the ribosome the tRNAs are in the 
hybrid state and, in the most recent terminology \cite{agirrezabala09}, 
this composite state is denoted by MS-II. 
In the absence of EF-G, the ribosome fluctuates between MS-I and MS-II. 
 
The rotational dynamics around an axis normal to the plane separating 
the two subunits is possible because of some key structural features 
of the inter-subunit bridges \cite{frank07,agirrezabala09}. 
The stronger rRNA-rRNA intersubunit bridges are concentrated near the 
axis that passes through the center while the weaker elastic bridges, 
where at least one ribosomal protein is involved, are located near the 
periphery.
Why does the rotational Brownian motion of the ribosome begin only 
after peptidyl reaction is complete? Immediately after the peptidyl 
transfer the intersubunit interaction is reduced because of the 
deacylation of a tRNA which reduces resistance against the rotation.

The landscape scenario, that we presented in part I of this review to 
account for the interplay of conformational fluctuations and chemical 
reactions, has been adapted \cite{frank10} to develop a similar picture 
for the elongation cycle of translation.
GTP-bound elongation factor EF-G biases the forward transition 
MS-I $\to$ MS-II by altering the free energy landscape \cite{frank10}. 
This phenomenon (the completion of the step (i) of translocation) is one 
of the possible ways of physical realization of the Brownian ratchet 
mechanism.  No relative motion between the small subunit and the mRNA 
template or that between small subunit and the anticodon end of tRNA 
takes takes place during the transition MS-I $\to$ MS-II. 
Possibly, there are more states in 
between MS-I and MS-II and the existence of these states, discovered by 
cryo-electron microscopy \cite{agirrezabala11b}, need independent support 
also by other experimental techniques.
On the basis of strong evidences from experiments it has been claimed 
that not only the translocation of the acceptor ends of the tRNAs with 
respect to the large subunit (i.e., the step (i) of translocation), 
but also that of the tRNA anticodons and the mRNA with respect to small 
subunit (i.e., the step (ii) of translocation) are governed by Brownian 
ratchet mechanisms \cite{frank00b,frank09,frank10,frank11b,rodnina11b}.

The above scenario of translocation on the free energy landscape, which 
was painted qualitatively on the basis of experimental data, has been 
quantified later by Xie \cite{xie09y} in terms of a Langevin equation. 
One of the interesting quantities that Xie calculated using this 
theoretical model is the mean translocation time $T_{t}$ and predicted 
how $T_{t}$ would increase with the increasing magnitude of an externally 
applied load force. 

What could be the role of EF-G and the GTP hydrolysis that it 
catalyzes? In principle, there are at least three different ways in 
which EF-G can accelerate translocation \cite{rodnina11}: 
(i) Binding of GTP-bound EF-G to the ribosome can stabilize the 
hybrid state thereby supporting partial forward movement of the 
tRNAs (movement on the large subunit). (ii) At least part of the 
free energy released by the hydrolysis of GTP catalyzed by EF-G can 
also be utilized to drive conformational changes of the ribosome 
itself. (iii) The 
conformational change of EF-G, caused by the GTP hydrolysis, can 
further bias the rotational diffusion of the two subunits with respect 
to each other towards the translocated final state. 

\vspace{2cm}

\begin{figure}[htbp]
\begin{center}
{\bf Figure NOT displayed for copyright reasons}.
\end{center}
\caption{Detailed mechano-chemical transitions in the elongation cycle 
of a ribosome in the Sharma-Chowdhury model of translation. 
Reprinted from Journal of Theoretical Biology  
(ref.\cite{sharma11b}),
with permission from Elsevier \copyright (2011).
}
\label{fig-scribofull}
\end{figure}

\subsubsection{\bf Dwell time distribution and average speed of ribosome}

By the term ``dwell'' we mean the duration of stay of a ribosome at 
a codon while actively translating that codon. 
As observed in single molecule experiments
\cite{marshall08b,blanchard04,uemura07,munro08,munro09,blanchard09,uemura10,aitken10a,aitken10b,uemura11,petrov11,vanzi07,wang07c,wen08,tinoco09}.
the stochastic stepping of a ribosome is characterized by an alternating
sequence of pause and translocation. The sum of the durations of a pause
and the following translocation is defined as the time of a dwell of the
ribosome at the corresponding codon. The codon-to-codon fluctuation in
the dwell time of a ribosome arises from two different sources:
(i) {\it intrinsic} fluctuations caused by the Brownian forces as well
as the low of concentrations of the molecular species involved in the
chemical reactions, and (ii) {\it extrinsic} fluctuations arising from
the inhomogeneities of the sequence of nucleotides on the template mRNA
\cite{buchan07}. Because of the sequence inhomogeneity of the mRNA
templates used by Wen et al. \cite{wen08}, the dwell time distribution
(DTD) measured in their single-molecule experiment reflects a combined
effect of the intrinsic and extrinsic fluctuations on the dwell time.

The probability density $f_{dwell}(t)$ of the dwell times of a ribosome,
measured in single-molecule experiments \cite{wen08}, does not fit a
single exponential thereby indicating the existence of more than one
rate-limiting steps in the mechano-chemical cycle of each ribosome.
Best fit to the corresponding simulation data was achieved assuming
five different rate-determining steps \cite{tinoco09}.

We'll now sketch a theoretical framework \cite{garai09a,sharma11a}
which provides an exact analytical expression for $f_{dwell}(t)$
in terms of the rate constants for the individual transitions in
the mechano-chemical kinetics of a single ribosome. This scheme also
involves essentially five steps in each cycle during the elongation
stage of translation. For the kinetic model shown in fig.\ref{fig-scribofull}, 
the {\it exact} probability density of the dwell times is given by 
\cite{sharma11a}
\begin{eqnarray}
f(t)&=&\biggl[\dfrac{\omega_{h2}\omega_{bf}\omega_p\omega_{h1}\omega_a}{(\omega_1-\omega_2)(\omega_1-\omega_3)(\omega_1-\omega_4)(\omega_1-\omega_5)}\biggr]e^{-\omega_1 t}
+\biggl[\dfrac{\Omega_{h2}\Omega_{bf}\Omega_p\omega_{h1}\omega_a}{(\omega_1-\omega_2)(\omega_1-\omega_3)(\omega_1-\Omega_4)(\omega_1-\Omega_5)}\biggr]e^{-\omega_1 t}  \nonumber\\
&+&\biggl[\dfrac{\omega_{h2}\omega_{bf}\omega_p\omega_{h1}\omega_a}{(\omega_2-\omega_3)(\omega_2-\omega_4)(\omega_2-\omega_5)(\omega_2-\omega_1)}\biggr]e^{-\omega_2 t}
+\biggl[\dfrac{\Omega_{h2}\Omega_{bf}\Omega_p\omega_{h1}\omega_a}{(\omega_2-\omega_3)(\omega_2-\Omega_4)(\omega_2-\Omega_5)(\omega_2-\omega_1)}\biggr]e^{-\omega_2 t} \nonumber \\
&+&\biggl[\dfrac{\omega_{h2}\omega_{bf}\omega_p\omega_{h1}\omega_a}{(\omega_3-\omega_4)(\omega_3-\omega_5)(\omega_3-\omega_1)(\omega_3-\omega_2)}\biggr]e^{-\omega_3 t}
+\biggl[\dfrac{\Omega_{h2}\Omega_{bf}\Omega_p\omega_{h1}\omega_a}{(\omega_3-\Omega_4)(\omega_3-\Omega_5)(\omega_3-\omega_1)(\omega_3-\omega_2)}\biggr]e^{-\omega_3 t} \nonumber\\
&+&\biggl[\dfrac{\omega_{h2}\omega_{bf}\omega_p\omega_{h1}\omega_a}{(\omega_4-\omega_5)(\omega_4-\omega_1)(\omega_4-\omega_2)(\omega_4-\omega_3)}\biggr]e^{-\omega_4 t}
+\biggl[\dfrac{\Omega_{h2}\Omega_{bf}\Omega_p\omega_{h1}\omega_a}{(\Omega_4-\Omega_5)(\Omega_4-\omega_1)(\Omega_4-\omega_2)(\Omega_4-\omega_3)}\biggr]e^{-\Omega_4 t} \nonumber \\
&+&\biggl[\dfrac{\omega_{h2}\omega_{bf}\omega_p\omega_{h1}\omega_a}{(\omega_5-\omega_1)(\omega_5-\omega_2)(\omega_5-\omega_3)(\omega_5-\omega_4)}\biggr]e^{-\omega_5 t}
+\biggl[\dfrac{\Omega_{h2}\Omega_{bf}\Omega_p\omega_{h1}\omega_a}{(\Omega_5-\omega_1)(\Omega_5-\omega_2)(\Omega_5-\omega_3)(\Omega_5-\Omega_4)}\biggr]e^{-\Omega_5 t} \nonumber\\
\label{eq-ftfinal}
\end{eqnarray}
where $\omega_1$,$\omega_2$ and $\omega_3$ are solution of the cubic equation
\begin{eqnarray}
&&\omega^3 -\omega^2(\omega_{r1}+\omega_{h1}+\omega_a+\omega_{r2}+\omega_p+\Omega_p)+\omega(\omega_{h1}\omega_a+\omega_{r2}\omega_{r1}+\omega_{r2}\omega_{h1}+\omega_{r2}\omega_a\nonumber \\
&+&\omega_p\omega_{r1}+\omega_p\omega_{h1}+\omega_p\omega_a+\Omega_p\omega_{r1}+\Omega_p\omega_{h1}+\Omega_p\omega_a)-\Omega_p\omega_{h1}\omega_a+\omega_p\omega_{h1}\omega_a=0, \nonumber \\
\label{eq-w123}
\end{eqnarray}
$\omega_4$ and $\omega_5$ are the solution of the quadratic equation
\begin{equation}
\omega^2-\omega(\omega_{h2}+\omega_{br}+\omega_{bf})+\omega_{h2}\omega_{bf}=0
\label{eq-w45}
\end{equation}
and $\Omega_4$ and $\Omega_5$ are the solution of the quadratic equation
\begin{equation}
\Omega^2-\Omega(\Omega_{h2}+\Omega_{br}+\Omega_{bf})+\Omega_{h2}\Omega_{bf}=0
\label{eq-w45star}
\end{equation}

For the sake of simplicity of
analytical derivation of the exact expression for $f_{dwell}(t)$,
this theory assumed the template mRNA to have a {\it homogeneous}
sequence (i.e., all the codons of which are identical). Consequently,
the expression for $f_{dwell}(t)$ thus derived incorporates the
effects of fluctuations that are strictly {\it intrinsic}.
This model \cite{sharma11a} envisages a scenario that is very similar
to the protocol used in some single-ribosome experiments \cite{uemura10} 
that use a mRNA with homogeneous coding sequence.
The effects of mRNA degradation on the dwell time distribution (as well 
as the fluctuations in the protein copy number) has been reported 
\cite{gorissen12}.

\subsection{\bf Initiation and termination of translation: ribosome recycling}

Initiation of translation 
\cite{gualerzi90,spirin99,marintsev04,laursen05,boni06,myasnikov09,jackson10,pavlov11,simonetti11,malys11}
is a multi-step kinetic process and involves several initiation-factors 
\cite{desmit03,antoun06,grigoriadou07,na10}. 
In this multi-step process, the large subunit joins the small subunit 
after the latter already forms a multi-macromolecular complex at the 
start codon after locating it. However, the detailed molecular mechanism 
of translation initiation in eukaryotes differ in several respects from 
those observed in prokaryotes \cite{marintsev04,malys11}. 
For example, in prokaryotes a specific sequence on the mRNA, called 
Shine-Delgarno sequence, assists in finding the start codon, 
\cite{gualerzi90,laursen05,boni06,pavlov11}
whereas eukaryotes use a mRNA scanning mechanism \cite{sheinman12} 
to locate the start codon \cite{jackson10,hinnebusch11}.  

One of the difficulties faced by the small subunit in binding the mRNA 
template at the ribosome-docking site (RDS) is that, for stability RNA 
molecules form hairpins utilizing complementary base-pairing. In the 
``stand-by model'' \cite{desmit03}, one postulates that a ribosome can 
attach the mRNA and remains on a ``stand-by'' mode so that it can quickly 
occupy the RDS on the mRNA strand within the short time for which the 
hairpin can open spontaneously. Let us denote the small subunit and 
a folded (unfolded) hairpin by the symbols S and F (U), respectively. 
We also denote the complex formed by the small subunit with the folded 
and unfolded mRNA by SF and SU, respectively. The kinetics of the 
``stand-by model'' \cite{desmit03} is shown below 
\begin{eqnarray} 
S + F &\rightleftharpoons& SF \nonumber \\ 
~~~~~~~~~~\uparrow \downarrow ~~&~&~~ \uparrow \downarrow \nonumber \\
S + U &\rightleftharpoons& SU \nonumber \\
\end{eqnarray} 
The ribosome-binding kinetics has been described \cite{na10} in 
terms of three time-dependent variables, namely, number of RDS-exposed 
mRNAs, ribosome-bound mRNAs and free ribosomes.   
More detailed kinetic schemes of translation initiation, that take into 
account the influence of the initiation-factors explicitly, have also 
been developed to account for experimental data \cite{antoun06,grigoriadou07}. 

For releasing the nascent polypeptide after its complete synthesis, 
release factors (RF) catalyze the hydrolysis of the bond that links 
it with the tRNA at the P site. Following release of the peptide, 
the large and small subunits disaasemble and then recycled.

\subsection{\bf Translational error from sources other than wrong selection}

Translations errors are divided into two major categories: (a) nonsense 
error, and (b) missense error. Nonsense error occurs when a ribosome 
detaches from the mRNA template midway between start and stop codons 
thereby causing premature termination. In contrast, incorporation of 
a wrong (non-cognate or near-cognate) amino acid is a missense error. 

We have already discussed missense error that can arise from an 
erroneous selection of the aa-tRNA because of the failure of the 
quality control mechanisms, particularly kinetic proofreading.  
In this subsection we discuss some of the other sources of 
translational error. 

Potential sources of translational error keep lurking around during 
every stage of the process: 
(i) {\it Error in ``charging of tRNA by aminoacyl-tRNA synthetase}: 
During the charging of a tRNA, if a wrong amino acid is loaded on 
it by the aa-tRNAsynthetase, it would lead to translational error in 
the elongation stage in spite of correct codon-anticodon matching.  
In order to ensure high fidelity of translation, the aa-tRNAsynthetase 
must  have high specificity for its two substrates, namely, tRNA and amino 
acid. It has mechanism of proofreading for correction of possible errors 
\cite{ibba04,ling09,banerjee10,yadavalli12}. 
The error committed by by aa-tRNAsynthetase never exceeds once in $10^4$ 
enzymatic cycles.
Interestingly, aminoacyl-tRNA synthetase and DNAP share some common
mechanisms to ensure translational and replicational fidelities,
respectively \cite{francklyn08}.

(ii) {\it Frameshift error}: For polymerizing a specific protein, 
a ribosome initiates translation from a start codon and continues 
translation by reading successive adjacent codons. However, since 
there is no internal punctuation, the ribosome does not always 
succeed in faithfully maintaining the reading frame that recognizes 
adjacent triplets without any slippage of the reading frame.  
A slippage of the ribosome on its track by $3n+1$ and $3n+2$ 
nucleotides, where $n$ is an integer, is identified as +1 and -1 
frameshifts, respectively \cite{farabaugh96,farabaugh99,hansen03}. 
The consequence of frameshift can be regarded as translational 
``recoding''  \cite{dinman12} 
because the resulting polypeptide could be synthesized, in principle, 
by the translation of a recoded genetic message. 

(iii) {\it Stalled translation on aberrant mRNA and rescue of ribosome}: 
As we mentioned earlier, ribosomes pause for long durations at rare codons. 
However, these ``natural'' pauses, arising from stochastic fluctuations, 
are distinct from ``unnatural'' stalls that are normally more stable. 
Because of transcriptional error, often aberrant mRNAs are synthesized. 
On such defective tracks, the ribosome can stall either prematurely at a 
codon (which may be a erroneously placed stop codon or any other codon)  
or at the 3'-end of the mRNA that lacks the stop codon. Such stalled 
translational complexes, which cannot resume operation, can have 
detrimental effect on the overall production of proteins  
\cite{buchan07}. 
Cells have ``mRNA surveillance'' systems for monitoring mRNAs that are 
translated and degrade the troublesome ones \cite{shoemaker12}.
In order to rescue the ribosomes from stalled translational complexes, 
bacteria use a mechanism called {\it trans}-translation 
\cite{keiler08,keiler11,healey11,janssen12}. 
This mechanism provides a pathway for degrading the mRNA template as 
well as the nascent polypeptide, and releasing the ribosome from such 
stalled translational complexes. 
The main role in this process is played by transfer-messenger RNA 
(tmRNA), an RNA molecule that shares the properties of both tRNA and 
mRNA. It enters the translational complex in the guise of a tRNA, 
accepts the nascent polypeptide. Then, switching role, it replaces 
the original mRNA by one segment of itself and the original mRNA is 
destined for degrading. Providing an alternative to the original 
mRNA, the tmRNA allows the ribosome to resumes translation whose main  
purpose then is to incorporate a specific amino acid that tags the 
incomplete polypeptide for degrading upon its release by the ribosome.

\subsection{\bf Polysome: traffic-like collective phenomena}

\subsubsection{\bf Experimental studies: polysome profile and ribosome profile}

Often
many ribosomes move simultaneously on a single mRNA strand while
each synthesizes a separate copy of the same protein. Obviously,
at any instant the nascent polypeptides on different ribosomes have
different lengths because the ribosomes are at different stages of
their run from the start to the stop codon. Such a collective movement
of the ribosomes on a single mRNA strand has superficial similarities
with {\it single-lane uni-directional} vehicular traffic
\cite{chowdhury00,schadschneider11} and is, therefore, sometimes
referred to as ribosome traffic \cite{chowdhury05a}.
The ribosomes bound simultaneously to a single mRNA transcript are
the members of a polyribosome (or, simply, {\it polysome})
\cite{warner62,warner63,rich04,noll08}

The polysome profiling technique \cite{arava03,mikamo05} provides the
number of ribosomes bound to a mRNA, but not their individual positions
where they remained ``frozen'' when translation was stopped by the
experimental protocol. More detailed information on the numbers of
ribosomes associated with specified {\it segments} of a particular mRNA
can be obtained by using {\it ribosome density mapping} technique
\cite{arava05} which is based on site-specific cleavage of the mRNA
transcript. However, the ribosomes are not expected to be uniformly
distributed on the mRNA template. 

The detailed spatial distribution of the ribosomes on the mRNA template can 
be obtained by the most recent technique, called {\it ribosome profiling}
\cite{ingolia10,guo10,ingolia11}.
This technique effectively provides a ``snapshot'' of the ongoing
translation by the actively engaged ribosomes on the mRNA template.
There are three major steps in this method:
(i) The ribosomes are first ``frozen'' at their instantaneous positions;
(ii) the exposed parts of the mRNA transcripts (i.e., those segments
not covered by any ribosome) are digested by RNase enzymes and,
thereafter, the small ribosome ``footprints'' (segments protected
by the ribosomes against the RNases) are collected separately;
(iii) Finally, the ribosome-protected mRNA fragments thus collected
are converted into DNA which are then sequenced. ``Aligning'' these
footprints to the genome reveals the positions of the ribosomes at
the instant when they were suddenly frozen.

\subsubsection{\bf Modeling polysome: spatio-temporal organization of ribosomes}

Normally, collision between ribosomes and their queueing would reduce
the overall rate of protein synthesis when translation is initiation-
limited.
Computer simulations of ribosome traffic have been carried out on a
mRNA with a specially selected codon sequence near the start codon
and allowing mRNA to decay at an optimum rate \cite{mitarai08}.
In this case, the metabolic cost
of mRNA breakdown is more than compensated by the simultaneous
increase in translation efficiency because of reduced queueing of
the ribosomes.

To my knowledge, all the theoretical models of ribosome traffic
represent the mRNA as a one-dimensional lattice, where each site
corresponds to a single codon. Since an individual ribosome is
much larger than a single codon, the ribosomes are represented
by hard rod that can cover ${\ell}$ successive codons (${\ell} > 1$)
simultaneously. The inter-ribosome interactions are captured through 
hard-core mutual exclusion principle: none of codons can be covered 
simultaneously by more than one ribosome. Thus, these models of 
ribosome traffic are essentially TASEP for hard rods: a ribosome hops 
forward, by one codon, with probability $q$ per unit time, if an only 
if the hop does not lead to any violation of the mutual exclusion 
principle.  All the details of the mechano-chemical cycle of a ribosome 
during the elongation stage is captured by a single parameter $q$.

In most of the models the mRNA template  was assumed to remain stable 
throughout the period of observation. Effects of the degradation of 
the mRNA templates on the rate of translation have been modelled 
within the framework of TASEP-type models \cite{nagar11,valleriani11}. 
Since these models have been reviewed very recently both from the
perspective of statistical physics \cite{chou11} and mathematical 
modeling \cite{haar12}, we'll not discussed these here in detail.

\subsubsection{\bf Effects of sequence inhomogeneity: codon bias}

So far we have reviewed mostly those theoretical works which ignore 
many subtleties of translation that arise from the intrinsic sequence `
inhomogeneities of real mRNA templates. First, the codons do not appear 
in a random sequence along the contour of the mRNA template. second, 
degeneracy of the genetic code gives rise to further nontrivial effects. 
All the distinct codons which code for the same species of amino acid 
are called {\it synonymous} codons. Similarly, tRNA species whose 
anticodon match with different but synonymous codons are charged with 
the same amino acid species; these distinct species of tRNA are called 
isoacceptor tRNA. 

A change in a single nucleotide on the DNA, which is called point mutation, 
can alter a codon in such a way that the new codon codes for a wrong amino 
acid. Such a mutation causes a missense error. However, if the point 
mutation alters a nucleotide but the new codon is synonymous to the original 
codon, the mutation is called``silent'' \cite{chamary06,chamary09} because the 
corresponding amino acid species remains unchanged. Synonymous mutations 
are now found to be not so silent and have visible consequences 
\cite{chamary06,chamary09,plotkin11}, particular on the level of gene 
expression. Thus, a missense error 
is equivalent to a non-synonymous point mutation. On the other hand, if the 
new codon resulting from a point mutation happens to be a stop codon, it 
would give rise to a nonsense error. Moreover, a point mutation can alter 
a stop codon into a codon that encodes an amino acid; such a mutation, 
called ``sense'' mutation, results in a longer protein that the wild type 
gene.

Although naively one might expect statistically 
equiprobable usage of synonymous codons, real usage in living cells 
is far from this expectation.
Unequal frequency of usage of synonymous codons is called {\it codon 
bias}. 
In this article we'll not explore the evolutionary {\it causes} of codon 
bias \cite{herschberg08}. 
Instead, we review the {\it consequences} of codon bias only in 
the context of translation \cite{plotkin11}.
It is generally believed that there is strong positive correlation 
between the codon frequency bias and the abundance of the corresponding 
tRNA isoacceptors. However, a higher abundance of tRNA does not 
necessarily imply a lower missense error \cite{shah10}. 
Codon bias pattern varies from one gene to another of 
the same organism and may vary also from one species to another. 
Codon choice may affect not only the efficiency but also the fidelity 
of translation \cite{gingold11}. 

Protein production can be regulated by controlling the balance between 
the codon usage and abundance of isoacceptor tRNAs \cite{angov11}. 
It has also been discovered that the first 30-50 codons immediately 
after the start codon act as a ``ramp'' that slows down initiation of 
translation thereby reducing the possibility of queueing of the 
ribosomes and avoiding ribosome traffic jams \cite{frederick10}. 
Biased codon usage has important implications also for protein 
export \cite{zalucki09}.
Some synonymous 
codons are used very rarely and the corresponding isoacceptors tRNAs 
are also proportionately rare in the cell; such codons are called 
{\it hungry} codons because a ribosome has to wait longer at such 
codons for the arrival of the corresponding aatRNA from the surrounding 
medium. Thus, the most obvious consequence of codon usage bias is that 
rate of translation of codons can be controlled by regulating the bias 
in their usage. Kinetics models have been formulated for studying the 
effects of such hungry codons on the ``missense error'' \cite{gilchrist06}. 
Effects of codon distributions along the mRNA on the rate of protein 
synthesis has been modelled quantitatively \cite{zouridis08}.

\subsection{Summary of sections on machines and mechanisms for template-directed polymerization}

In this section we have reviewed the kinetics of several template-directed 
polymerization processes driven by molecular machines that utilize the 
respective templates as the track for their motor-like movements. We have 
also discussed many situations that require coordination of the operation 
of multiple motors either moving in the same direction or approaching 
one another head-on.  

Very little theoretical work has been done so far to study the traffic 
rules for co-directional and head-on approach of a DNA polymerase and 
a RNA polymerase during simultaneous transcription and DNA replication. 
Moreover, none of the published works has considered multiple co-directional 
RNA polymerases approaching the DNA polymerase although, as is well 
known, several RNA polymerases transcribe the same gene simultaneously. 

Several RNA-binding proteins are known to regulate the rate of protein 
synthesis at the initiation, elongation and termination stages 
\cite{kong12}. 
Incorporating these regulatory processes within a single unified kinetic 
model would bring us closer to a {\it in-silico} replica of in-vivo 
translation.

\section{\bf Helicase motors: unzipping of DNA and RNA}
\label{sec-specificunzippers}

Helicases use the free energy of ATP hydrolysis to catalyze the unzipping 
(or, more precisely, unwinding) of dsDNA or RNA and translocate along 
one of the strands. Therefore, these are molecular motors \cite{lohman98}. 
In this section we review models that take into account the structural 
and kinetic details of specific helicases motors 
\cite{levin03,delagoutte02,delagoutte03,pyle08,singleton07,yodh10,ha12} 
(for an historical account of the discoveries on helicases, particularly 
those found in plants, see ref.\cite{tuteja03}). 

\begin{figure}[htbp]
\begin{center}
\includegraphics[angle=90,width=0.65\columnwidth]{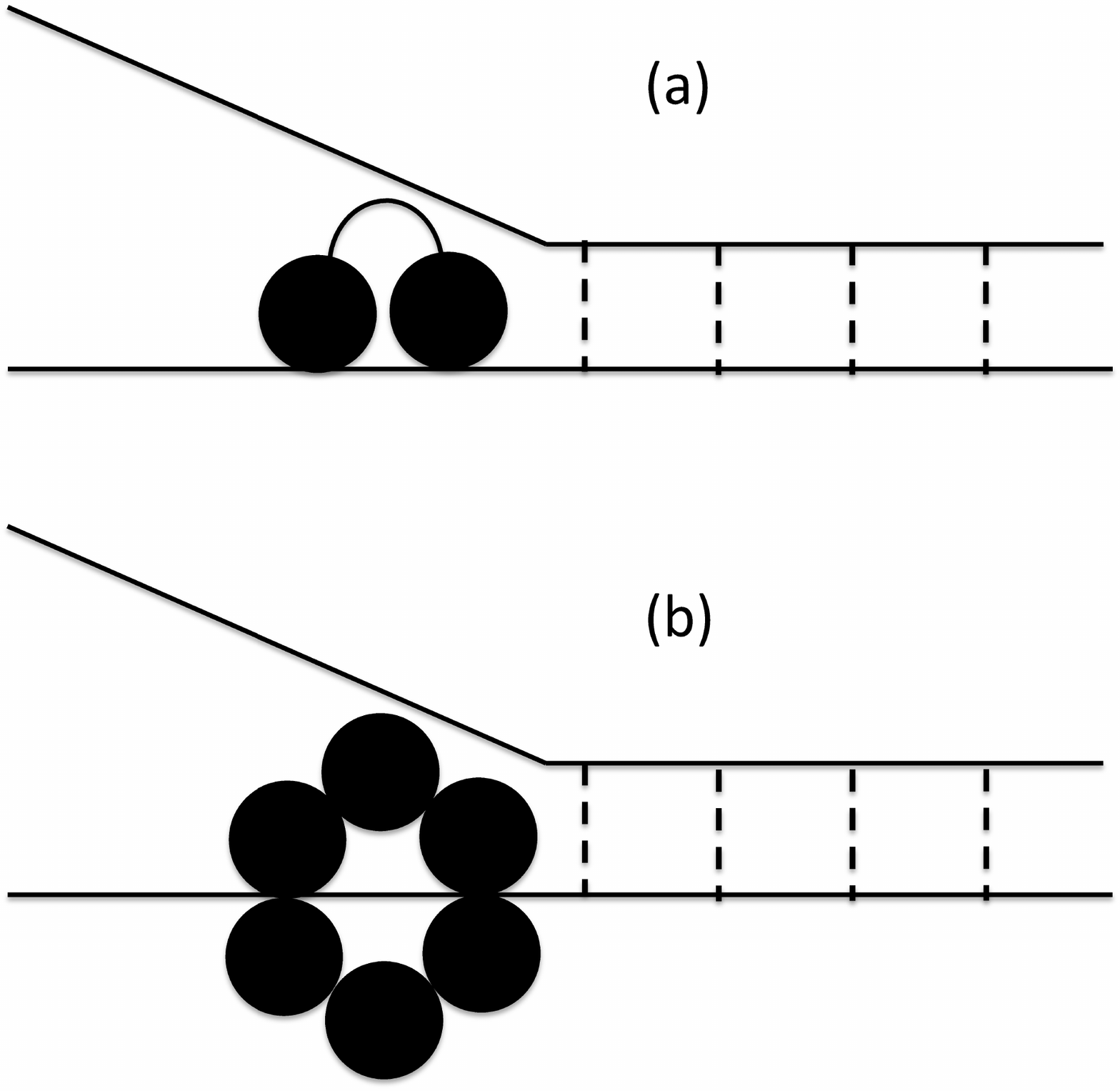}
\end{center}
\caption{A schematic representation of (a) dimeric, (b) hexameric DNA 
helicase. A monomeric helicase with two distinct domains would also be 
represented schematically by (a). Each of the solid lines represents a 
ssDNA strand whereas the dashed lines represent the base-pairing in the 
dsDNA.  (adapted from ref.\cite{delagoutte02}).}
\label{fig-helicase}
\end{figure}

Helicases have been classified in various ways using different criteria.
(i) Several conserved amino-acid sequences have been discovered in
helicases. On the basis of these ``helicase signature motifs'',
DNA helicases have been classified into superfamilies SF1, SF2, etc.\\
(ii) On the basis of the nature of the nucleic acid (DNA or RNA) track,
i.e., the nucleic acid which they unwind, helicases have been classified
into (a) DNA-helicases, (b) RNA-helicases and (c) hybrid helicases.
Some helicases are, however, hybrid in the sense that these can
unwind both DNA and RNA. \\
(iii) Some helicases move from 3' to 5' end of a ssDNA whereas others
move in the opposite direction. On the basis of directionality,
helicases have been classified into two groups: 3' to 5' helicases
and 5' to 3' helicases. \\
(iv) Helicases have also been grouped according to the the source of
these proteins, i.e., humans, plants, bacteria, viruses, etc. \\
(v) On the basis of the number of ATPase domains, helicases have been
classified into monomeric and multimeric types; dimeric and hexameric
being the most common multimeric helicases (see fig.\ref{fig-helicase}).
In the next two subsections, we study the mechanisms of helicases 
separately for hexameric \cite{patel00,donmez06} 
and non-hexameric helicases \cite{lohman96,lohman08}. \\

The concept of step size has been quite confusing in the helicase 
literature and sometimes gave rise to unnecessary controversies. 
A ``mechanical step-size'' should be defined as \cite{lohman08} the 
average distance moved by the center of mass of the helicase per 
ATP molecule hydrolyzed by it. In contrast a ``kinetic step-size'' 
is defined \cite{lohman08} is the average distance covered by the 
helicase in between two successive occurrences of the rate-limiting 
step of its mechano-chemical cycle. In general, these two step-sizes 
are not necessarily identical. Another quantitative characteristic 
of a helicase is its ATP-coupling stoichiometry \cite{lohman08} 
which is the average number of ATP molecules hydrolyzed per base 
pair unwound; it differs from step-size when, for example, futile 
hydrolysis of ATP takes place. 

One of the fundamental questions is the stepping pattern of a helicase 
on a single-stranded nucleic acid (ssNA)- is it analogous to hand-over-hand 
or inchworm pattern? Moreover, for this translocation on a ssNA, what 
is the mechanism of energy transduction- power stroke or Brownian ratchet? 
Furthermore, does it unzip the nucleic acid actively or passively? In the 
passive mode, it exploits (a) spontaneous opening of base pairs by thermal 
fluctuations, and (b) its own ability for directional translocation 
on ssNA to move forward, and stabilize the ssNA, before the base pair can 
close again. In contrast, in the active mode, it directly induces local 
destabilization of the dsNA instead of relying solely on thermal fluctuations 
for base-pair opening.

\subsection{\bf Non-hexameric helicases: monomeric and dimeric}

A few helicases are monomeric. Dimeric helicases are more common. 
Two different types of ssNA translocation patterns have been 
postulated for different types of non-hexameric helicase motors- 
(i) stepping, and (ii) Brownian ratchet. In the stepping model, 
the helicase much have at least two NA-binding sites on it. In 
the case of monomeric helicases, usually these binding sites are 
located on two different domains whereas each of the two monomeric 
constituents of a dimeric helicase can have a NA-binding site on it. 
Inchworm is the most common stepping pattern although sometimes 
the experimental observations are also consistent with a ``rolling'' 
pattern which is the analog of the hand-over-hand stepping pattern 
of cytoskeletal motors. 

The Brownian ratchet mechanism of non-hexameric helicase translocation 
on ssNA does not require more than one NA-binding site on the helicase. 
However, ATP-hydrolysis can cause a transition between two different 
conformational states in one of which the helicase has strong affinity 
for the NA whereas in the other it as weak affinity. 
Analog of the hand-over-hand mechanisms of the cytoskeletal
motors is called the ``rolling'' model. However, most of the dimeric
helicases are believed to follow the inchworm mechanism.

An oversimplified model for helicase motors was developed by Chen 
\cite{chen97}. 
A stochastic model for non-hexameric helicases was developed by 
Betterton and J\"ulicher \cite{betterton03,betterton05a,betterton05b}. 
This model has been extended by Garai et al.\cite{garai08a} to capture 
the effect of the ATPase activity of the helicase on its affinity for 
its nucleic acid track.
HCV helicase NS3 is a representative member of the non-hexameric 
helicases that have been studied extensively \cite{frick06a,frick06b}.  
A limiting case of the model studied by Garai et al.\cite{garai08a} 
corresponds to a Brownian ratchet mechanism for the NS3 helicase of 
HCV. 
Coarse-grained modeling of this helicase by elastic networks and 
NMA of the model has provided insight into its conformational 
kinetics in each cycle \cite{zheng07,flechsig10}.

\subsection{\bf Hexameric helicases}

A large number of helicases are hexameric and have an approximate
ring-like architecture \cite{donmez06,crampton03,doering95,liao05,johnson07}. 
For hexameric helicases, at least three
alternative mechanisms of enzymatic activities have been suggested;
these include, activities of all the ATP-binding domains in
(i) parallel, (ii) random, (iii) sequential manner. \\

(i) Parallel: In this mechanism all the subunits hydrolyze dTTP and
exert power stroke simultaneously. \\
(ii) Random: there are at least two possible different scenarios: \\
(a) random in time, where power stroke of each subunits starts and
finishes at random times independent of other units;
(b) random in space, where power strokes are sequential in time (i.e.,
each subunit can begin only after another finishes), but the order of
power strokes around the ring is random.\\
(iii) Sequential: there are at least two different sequences in which
the subunits can exert power stroke:\\
(a) paired sequential, i.e., sequentially around the ring, but with
diametrically opposite subunits in the same state;
(b) ordered sequential, i.e., sequential in the strict order 1,2,...6
around the ring.

Doering et al.\cite{doering95} developed a ``flashing-field model'' 
for DNA unwinding by hexameric ring-like helicases. This quantitative 
model is based on the following main assumption: ATP binding and 
hydrolysis induces conformational changes in the helicase that expose 
a pair of oppositely charged regions near the inner surface of the 
central channel of the ring. The negatively charged phosphates on 
the backbone of the DNA interact sequentially with this charge pair. 
The ``flashing'' charge pair gives a pulse of electrostatic push to 
the DNA before switching off in each cycle. Thus, in this model, 
the helicase is assumed to operate as a mechano-electrical transducer 
that transduces mechanical strains created by the conformational 
changes induced by ATP hydrolysis into an electrostatic force. It is 
this electrostatic force that, in turn, pushes charged DNA through the 
central channel of the helicase. It is not a Brownian ratchet; the 
Brownian force acts only as a ``lubricant'' reducing the stickiness 
caused by local minima in the free energy landscape. 

In the mathematical formulation of the model, Doering et al. 
\cite{doering95} placed the coordinate system on the DNA so that the 
helicase motor executes a helical motion on this DNA track. Using a 
cylindrical coordinate system, the motion of the helicase is described 
by the two coordinates, namely, the azimuthal angle $\theta$ and the 
axial coordinate $z$. The corresponding Langevin equations are 
\cite{doering95} 
\begin{eqnarray}
\zeta_{\theta} \frac{d\theta}{dt} &=& \underbrace{-\frac{dV(\theta,z,t)}{d\theta}}_\text{Torque derived from potentials} - \underbrace{\tau_{\theta}}_\text{Load torque} + \underbrace{\tau_{B}}_\text{Brownian torque} \nonumber \\ 
\zeta_{z} \frac{dz}{dt} &=& \underbrace{-\frac{dV(\theta,z,t)}{dz}}_\text{Axial force derived from potentials} - \underbrace{F_{z}}_\text{Load force} + \underbrace{F_{B}}_\text{Brownian force}  
\label{eq-ringHelicase}
\end{eqnarray}

The configuration of dipoles switches from one configuration to another; 
in the $n$-th configuration the full potential is given by \cite{doering95} 
\begin{equation}
V_{n} = V_{array}\biggl(x+2\pi r \frac{n-1}{N}\biggr); 
\end{equation}  
where $N$ is the total number of configurations. The potential $V_{array}(x)$ 
has the form 
\begin{equation}
V_{array}(x) = \sum_{m=1}^{M} V_{dip}\biggl(x-2\pi \frac{m}{M}r\biggr), 
\end{equation}  
with 
\begin{equation}
V_{dip}(x) \propto \frac{x}{b^2+x^2} e^{-\kappa \sqrt{b^2+x^2}}\biggl[\kappa+\frac{1}{\sqrt{b^2+x^2}}\biggr]
\end{equation}
where $x = r \theta$ is the coordinate along the DNA charges, $\kappa$ 
is the inverse screening length, and $b$ is the off-axis distance of 
the charges on the DNA backbone.

\subsection{Section summary} 

In this section we have reviewed the few models of helicase motors 
that have been reported so far in the literature. The mechanisms 
of coordination of the multiple ATPase domains in both hexameric 
and non-hexameric helicases are not well understood at this stage. 
There is a need for distinct theoretical models based on different  
plausible mechanisms of coordination of the ATPase domains of a 
single helicase motor. The predicted collective behaviour of these 
competing models can be compared with experimental data to rule out 
some of the speculated mechanims.

\section{\bf Rotary motors I: ATP synthase (F$_0$F$_1$-motor) and similar motors} 
\label{sec-specificrotary1}

In section \ref{sec-genericrotary} we have discussed generic models  
of rotary motors to highlight the common features of their operational 
mechanism. Now, in this and the next sections, we review the kinetics 
of two most important rotary motors in living cells 
\cite{berry00b,oster03,muench11},  
namely, ATP synthase and the bacterial flagellar motor, respectively, 
pointing out their main differences. 

The models of specific rotary motors are usually based on the construction 
of a structural model of the stator and rotor at an appropriate level of 
details. For modeling its stochastic kinetics one assumes 
(i) the trajectories of the ions through the model structure, and 
(ii) the nature of the interactions among the (a) stator, (b) rotor, 
(c) the mobile ions, and (d) the hydrophobic environment of the membrane 
in which the motor is embedded. These dynamic interactions give rise to 
the coupling between ionic movements and the directed (on the average) 
rotation of the rotor.  Significant progress have been made in 
understanding the mechanism of operation of these motors by 
a combination of structural studies and single-molecule experiments 
\cite{pilizota09}. 
We have already seen that Na$^{+}$ can substitute for H$^{+}$ as the 
coupling ion in secondary transporters. H$^{+}$ is not essential also 
for the operation of ATP synthases and in some ATP synthases Na$^{+}$ 
is used instead of H$^{+}$.  

In this section we review the kinetics of ATP sythase (also called 
F-ATPase, for reasons explained in the next subsection) as well as 
those of two other similar rotary motors, namely V-ATPase and A-ATPase. 
(which we describe in detail in the corresponding subsections) 
\cite{muench11}. 
These rotary motors in living cells are wonderful achievements of 
nature's evolutionary design. These can rotate at speeds exceeding 
even 100 revolutions per minute, generate torques as large as about 
50 pN.nm and transduce energy at efficiencies that are vey close 
to $100\%$.
There are also some interesting architectural similarities between 
the ATP synthase and the TrwB DNA translocase.

\subsection{\bf Rotary motor F$_0$F$_1$-ATPase} 

ATP synthase is the smallest rotary motor and is embedded in membranes. 
In bacteria it is located on the cytoplasmic membrane whereas in 
eukaryotes it is embedded in the membrane of specialized organalles 
called mitochondria (in animal cells) and chloroplasts (in plant cells) 
\cite{weber97,senior02,weber03,duncan04,nakamoto99,yoshida01,dimroth06,ballmoos08,ballmoos09,junge09,romanovsky10,bald11,spetzler12} 
(see appendix \ref{app-organelles} for a brief introduction to these 
organelles and ref.\cite{boyer02,junge04} for the history of the discovery 
of ATP synthase and its mechanism from the personal perspective of some of 
leading contributors). 

\vspace{4.5cm}

\begin{figure}[htbp]
\begin{center}
{\bf Figure NOT displayed for copyright reasons}.
\end{center}
\caption{Various modes of operation of the ATP synthase. (1) {\it Idling 
mode}: When no external energy source is available, the rotor fluctuates 
back-and-forth within a narrow range of angles with respect to the stator 
and exchanges Na$^{+}$ (or proton, depending on the species) between the 
two sides of the membrane. See the text for details of the 
(2) {\it synthesis mode}, and (3) the {\it hydrolysis mode}. 
Reprinted from Structure 
(ref.\cite{dimroth03}),
with permission from Elsevier \copyright (2003).
}
\label{fig-ATPsynthase}
\end{figure}

The free energy input for ATP-synthase is IMF and the final output is 
freshly synthesized ATP. Each ATP synthase consists of two coupled parts 
which are called F$_0$ and F$_1$. ATP synthase is also referred to as 
F$_0$F$_1$-ATPase. Individually, both F$_0$ and F$_1$ are rotary motors. 
ATP synthase motor is reversible 
\cite{vinogradov00,dimroth03} (see fig.\ref{fig-ATPsynthase}). 
In the ATP-synthesis mode, for F$_{0}$ motor, an IMF across the membrane 
is the input and the rotation of F$_0$, caused by the torque generated by 
the free energy transduction, is the corresponding mechanical output. 
Rotating F$_{0}$, drives the shaft that, in turn, rotates F$_1$ where 
ATP is synthesized from ADP and inorganic phosphate. 
In the reverse mode, free energy input from ATP hydrolysis is transduced 
by F$_1$ to power the rotation of the shaft in the reverse direction 
thereby rotating F$_0$ also in reverse while the latter operates effectively 
as a proton pump. 

The F$_1$ subunit has a threefold symmetry so that in a complete 
rotation by $360^{0}$ it synthesizes 3 molecules of ATP. However, the 
number of protons powering the corresponding $360^{0}$ rotation of the 
F$_0$ subunit depends on the organism; for some organisms it is 10 
whereas for some others it is 15 although even 12 and 13 are also 
quite common. Therefore, the ion-to-ATP ratio (the ``stoichiometry'') 
$n_s$ can vary between 3.3 to 5. Since flow of $n$ protons across a 
PMF of $\Delta \mu_{H+}$ would be utilized to synthesize a single 
molecule of ATP, one would expect $\Delta G(ATP) = - n \Delta \mu_{H+}$. 
Perhaps, the organisms have optimized the ``stoichiometry'' and the 
IMF by adapting to the environment of their habitat 
\cite{tomashek00,cross04b}.  
Therefore, understanding the operation of the ATP synthase motor requires 
addressing questions on the mechanisms of the rotary motors F$_0$ and F$_1$ 
separately and, then, their integration \cite{pedersen00}. 

The subunits of an ATP-synthase can also be clustered into two groups- 
those forming the parts of the ``rotor'' rotate with respect to a ``stator''
that consists of several other subunits.
Interesting comparison of the rotation of F$_0$ and F$_1$, with the 
stepping of linear motors has been presented by Kinosita et al. 
\cite{kinosita98} and Junge \cite{junge99}. 

\subsubsection{\bf F$_0$ motor: Brownian ratchet mechanism of energy transduction from PMF during ATP synthesis} 

\begin{figure}[htbp]
\begin{center}
\includegraphics[angle=-90,width=0.70\columnwidth]{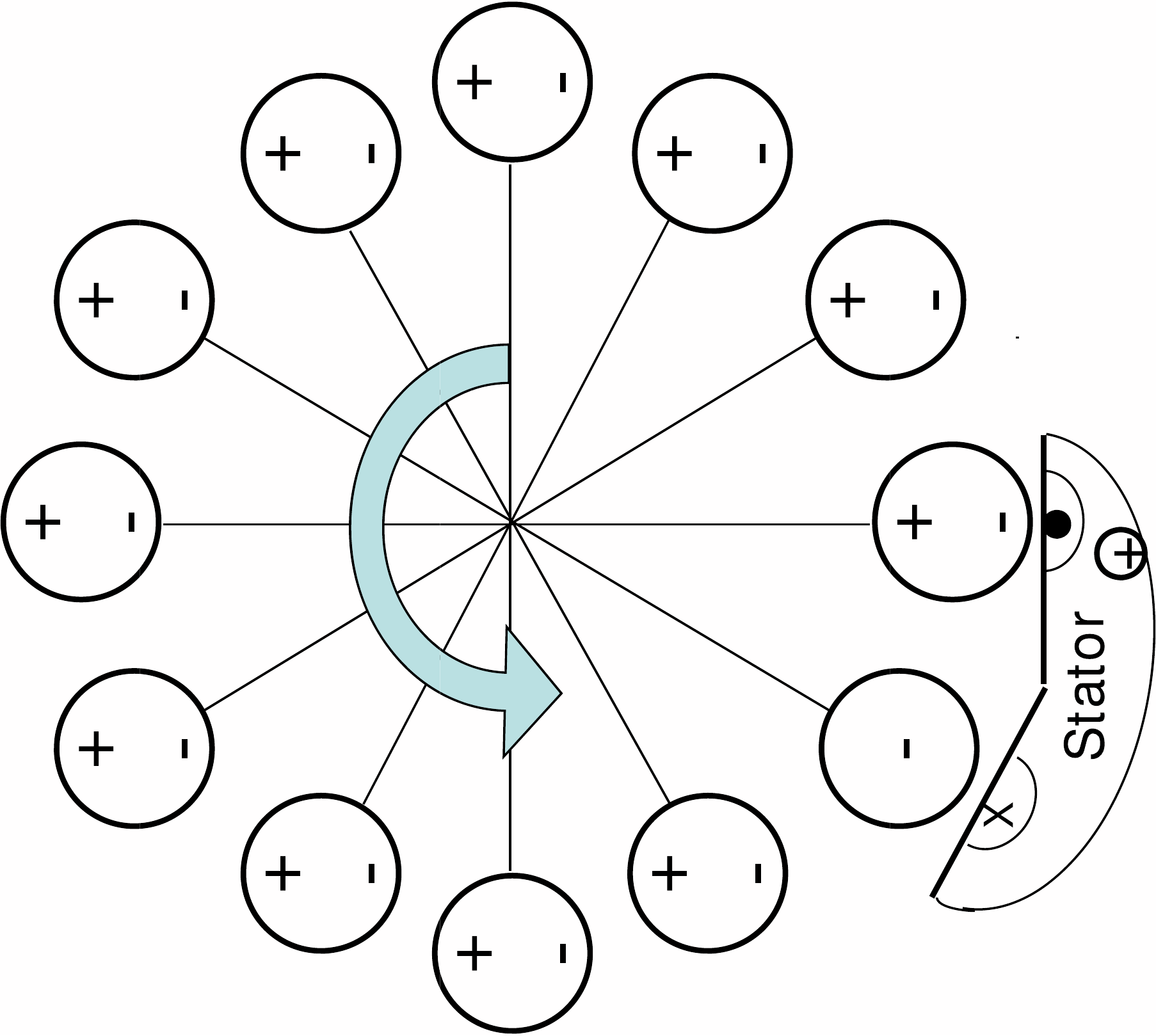}
\end{center}
\caption{A schematic representation of the model of rotary motor F$_0$: 
top view (adapted from ref.\cite{smirnov08}).}
\label{fig-F0model}
\end{figure}

\begin{figure}[htbp]
\begin{center}
\includegraphics[angle=-90,width=0.70\columnwidth]{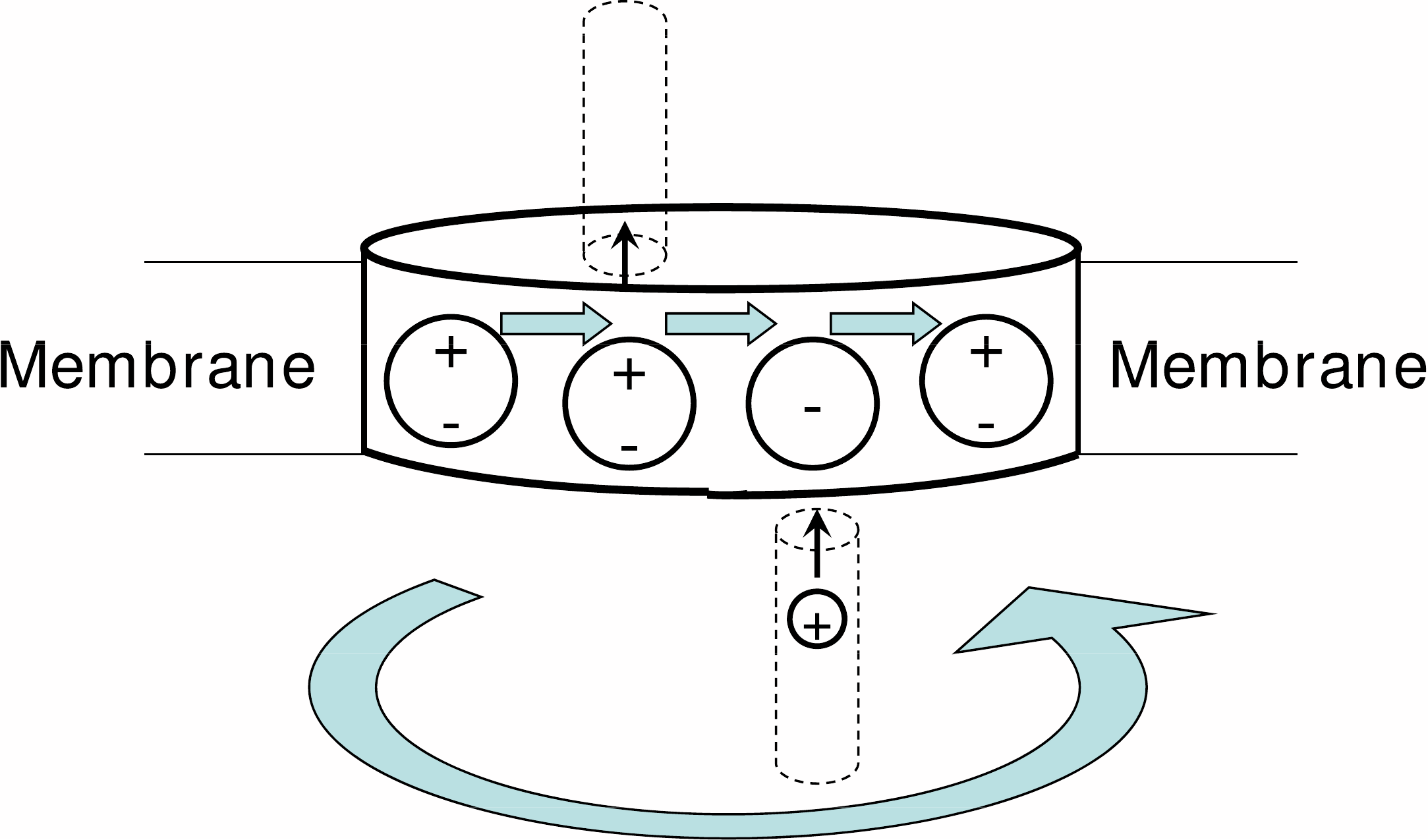}
\end{center}
\caption{A schematic representation of the model of rotary motor F$_0$: 
side view (adapted from ref.\cite{smirnov08}).}
\label{fig-F0model2}
\end{figure}

The two main components of F$_0$ motor are the {\it stator} and the 
cylindrical {\it rotor} which are separated by a narrow gap in between. 
Normally, the rotor consists of 12 identical segments each of which 
has a special negatively charged site on its surface that can be 
neutralized by adsorbing a single proton (protonation). Out of these 
twelve rotor subunits, two lie at the stator-rotor interface 
(see fig.\ref{fig-F0model}). 
There are two proton ``half-channels'' that are offset with respect 
to each other. Protons get access to one of the two rotor sites through 
the half channel that leads from the high-proton concentration side 
(acidic side) of the membrane to one of the two sites in front of 
the stator. Proton on the second site in front of stator can escape to 
the low-proton concentration site (basic side) of the membrane through 
the other half channel. This asymmetry of the two half-channels plays 
a crucial role in the rotational motion of the rotor 
(see figs.\ref{fig-F0model} and \ref{fig-F0model2}). Moreover, there 
is a positively charged site midway between the two proton channels. 
As we discuss below, the Coulomb repulsion between this proton and 
those on the rotor sites enhances the efficiency of the rotary motor 
F$_0$.

The operational mechanism of F$_0$ motor was proposed in the mid-1990s  
by several groups 
(see, for example, \cite{vik94,cross96,sabbert96,junge97,fillingame97}). 
But, all these 
provided mostly qualitative pictures rather than quantitative predictions. 
To our knowledge, the first mathematical model of F$_0$ motor was 
developed by Elston et al.\cite{elston98} (see also reviews in 
ref.\cite{oster99,oster00a,oster00b}). The original version of this model 
\cite{elston98} was formulated for PMF. Later similar treatments for SMF 
were also published \cite{dimroth99,xing04} although the details differ.
The mechanical state of the F$_0$ motor can be described by its angle of 
rotation around the central axis of the cylindrical rotor. The chemical 
state of the motor is denoted by the state of protonation of the two 
sites in front of the stator. The four possible chemical states are 
denoted by $E$ (both sites empty), $R$ (right site protonated), 
$F$ (fully, i.e., both sites protonated), and $L$ (left site protonated).
Elston et al. \cite{elston98} assumed that direct hopping of protons 
from one site to another is not possible. Therefore, transitions between 
the states $L$ and $R$ involved at least 2 steps. A direct transition 
between $L$ and $R$ indicates a mechanical rotation rather than a chemical 
transition. 
\begin{eqnarray}
&F&  \nonumber \\ 
k_{RF}\nearrow \swarrow k_{FR} &~& k_{LF}\nwarrow \searrow k_{FL}\nonumber \\
R~~~~~~~~~~&&~~~~~~~~~~L \nonumber \\
k_{RE}\searrow \nwarrow k_{ER} &~& k_{LE}\swarrow \nearrow k_{EL} \nonumber \\
&E& \nonumber \\
\end{eqnarray}

\begin{eqnarray}
\frac{d}{dt}\left(
\begin{array}{c}
dP_{E} \\
dP_{R} \\
dP_{F} \\
dP_{L}
\end{array}
\right) =
\begin{pmatrix}
-(k_{ER}+k_{EL}) & k_{RE} & 0 & k_{LE} \\
k_{ER} & -(k_{RE}+k_{RF}) & k_{FR} & 0 \\
0 & k_{RF} & -(k_{FR}+k_{FL}) & k_{LF} \\
k_{EL} & 0 & k_{FL} & -(k_{LE}+k_{LF}) 
\end{pmatrix}
\left(
\begin{array}{c}
P_{E} \\
P_{R} \\
P_{F} \\
P_{L}
\end{array}
\right) \nonumber \\
\label{eq-F0master}
\end{eqnarray}
The Langevin equation describing the rotational motion is essentially 
a torque-balance equation: 
\begin{equation}
\zeta (d\theta/dt) = \underbrace{\tau_{E}(\theta,\mu)}_\text{Electrostatic} + \underbrace{\tau_{H}(\theta,\mu)}_\text{Hydrophobic} - \underbrace{\tau_{L}}_\text{Load} + \underbrace{\tau_{B}}_\text{Brownian}
\label{eq-F0Langevin}
\end{equation}
where $\zeta$ is the viscous drag coefficient and the different sources 
of torque shown on the right hand side, in general, depend on the chemical 
state $\mu=E,R,F,L$. Alternatively, equations (\ref{eq-F0master}) and 
(\ref{eq-F0Langevin}) can be combined into a single hybrid equation which 
is a combination of Fokker-Planck and master equations \cite{elston98}. 
 
The effect of the membrane potential, which was not incorporated in the 
kinetic equations by Elston et al.\cite{elston98}, was taken into account 
in the corresponding equations for the Na$^+$-ion-driven F$_0$ motor 
formulated later by Dimroth et al.\cite{dimroth99}. In this equation `
\begin{equation}
\zeta (d\theta/dt) = \underbrace{\tau_{C}(\theta,\mu)}_\text{Rotor-Stator charge int.} + \underbrace{\tau_{M}(\theta,\mu)}_\text{Membrane pot.} + \underbrace{\tau_{H}(\theta,\mu)}_\text{Hydrophobic} + \underbrace{\tau_{P}(\theta,\mu)}_\text{Rot-Sta Passive int.}  - \underbrace{\tau_{L}}_\text{Load} + \underbrace{\tau_{B}}_\text{Brownian}
\label{eq-F0NaLangevin}
\end{equation}
$\mu$ labels the 16 distinct chemical states. 

Most of the results for these models of F$_{0}$ motor were obtained 
by solving the FP equations numerically. Bauer and Nadler \cite{bauer08} 
developed a simpler model for the F$_{0}$ motor that could be treated 
analytically. Instead of considering a pair of sites protected by the 
stator, this model focusses attention on a single site (more appropriately, 
a single ``protomer''). The movement of a protomer by one protomer is 
described asa cyclic process. A cyclic spatial variable $x$ denotes its 
spatial location while the subscripts $d$ and $p$ indicate the states of 
its protonation (i.e., deprotonated or protonated). 
Let $P_{d}(x,t)$ and $P_{p}(x,t)$ be the probability densities for a 
protomer unit of the F$_{0}$ motor to be located at $x$ and in the 
deprotonated and protonated states, respectively, at time $t$. 
The FP equations governing the time evolution of the system are 
\cite{bauer08}
\begin{eqnarray} 
\frac{\partial P_{d}(x,t)}{\partial t} = \frac{\partial}{\partial x}\biggl(D_{d}(x)\biggl[\frac{\partial}{\partial x}-F_{d}(x)\biggr] P_{d}(x,t)\biggr) - \Delta \noindent \\
\frac{\partial P_{p}(x,t)}{\partial t} = \frac{\partial}{\partial x}\biggl(D_{p}(x)\biggl[\frac{\partial}{\partial x}-F_{p}(x)\biggr] P_{p}(x,t)\biggr) + \Delta \noindent 
\end{eqnarray} 
where $\Delta$ accounts for the ``chemical'' transitions 
\begin{equation}
{\rm Deprotonated~ state} \rightleftharpoons {\rm Protonated~ state}
\end{equation}
In general, the diffusion constants depend not only on $x$, but also 
on the ``chemical'' state (i.e., the state of protonation). The force 
$F_{d}(x)$ and $F_{p}(x)$ account for the interactions of the rotor with 
its surroundings; it can be derived from an appropriate local potential.

A model of F$_0$-type motors was developed by Smirnov et al.\cite{smirnov08} 
using the theoretical formalisms that are usefully applied in condensed 
matter physics. The Coulomb interaction between the stator and rotor 
charges was shown to be dominant contributor to the torque that rotates 
the F$_0$ motor.

All the models of F$_0$ discussed so far in this review are essentially 
one-dimensional in the sense that the only mechanical movement allowed 
is the pure rotation (described by $\theta$) of the rotor unit that is 
effectively treated as a rigid body. A more detailed description of the 
possible mechanical movements of the stator-rotor combination emerged 
from MD simulation of an atomistic structural model \cite{fillingame00} 
of the system up to nanoseconds. 
Incorporating several of such different degrees of freedom in a 
coarse-grained model, a stochastic kinetic model of F$_0$ was developed 
by Aksimentiev et al. \cite{aksimentiev04}. Although this kinetic 
model also assumes a Brownian ratchet mechanism, this extended version 
is often referred to as a ``protein-roller bearing mechanism'' 
\cite{spetzler12}. Its numerical simulation over millisecond time scales 
provided a deeper insight into the underlying mechano-chemistry than that 
obtained from the older model of Elston et al.\cite{elston98}. 

\subsubsection{\bf F$_1$ motor: power stroke mechanism in reverse mode powered by ATP hydrolysis} 

The F$_1$ motor consists of a hexameric complex that looks somewhat 
like a skinned orange and is composed of alternating $\alpha$ and 
$\beta$ subunits; this structural organization is often denoted 
symbolically as $\alpha_3\beta_3$. The ATP-binding site is located 
on the $\beta$ subunits in the cleft between $\alpha$ and $\beta$. 
ATP hydrolysis drives a central ``shaft'', called $\gamma$-subunit, 
by executing a power stroke. 

\noindent$\bullet$ {\it Binding-change, cooperativity and rotational catalysis}

The three main ingredients in the mechanism of operation of F$_1$ are 
as follows \cite{cross81,senior88,boyer89,boyer93,boyer97,kresge06}:\\
(i) {\it binding-change}: the input energy is used not to form the ATP 
molecule, but to drive the release of an already formed molecule of ATP 
from the binding site.\\
(ii) {\it catalytic cooperativity}: each of the three catalytic sites of 
F$_1$ goes through three kinetic states in a cycle, but the cycles of 
the three are staggered. A catalytic site can be in one of the three states 
labelled by E (empty), T (tightly bound to substrate/product) and 
L (loosely bound to substrate/product). The three catalytic sites in F$_1$ 
operate cooperatively in such a way that ATP cannot be released from 
one site unless ADP and P$_i$ bind to another site while the third site 
is empty.  \\
(iii) {\it rotational catalysis} \cite{leslie00,matsui08}: the ion transport 
in the F$_0$ subunit is coupled to the ATP synthesis in the F$_1$ subunit 
through rotation of the $\alpha_3\beta_3$ hexamer.

\noindent$\bullet$ {\it Binding-zipper model}

In the binding-zipper model \cite{oster00a}, in the reverse mode of 
operation of the F$_{0}$F$_{1}$ motor, ATP binding to a catalytic site 
on F$_1$ takes place in a progressive manner by the sequential formation 
of bonds between the ATP molecule and the catalytic site. Conversely, 
in the ATP synthesizing mode, the binds bonds holding the freshly 
synthesized ATP molecule are broken (unzipped) sequentially. Thus, the 
binding-zipper model has been interpreted as an extension of the 
{\it induced fit model} of enzyme specificity (which we discussed in 
section \ref{sec-chemphys}); the single-step substrate-binding is replaced 
in this extended model by a multi-step binding process where bonds 
are formed between the substrate and the catalytic site sequentially 
\cite{oster00a}. Supporting evidences for the binding-zipper model 
have emerged from computer simulations of the {\it unbinding} and 
binding of ATP to and from the F$_{1}$ motor, respectively 
\cite{antes03,eide06}. 

\vspace{6cm}

\begin{figure}[htbp]
\begin{center}
{\bf Figure NOT displayed for copyright reasons}.
\end{center}
\caption{The crank-jack-like mechanism of chemo-mechanical
coupling in F$_1$ motor.
Reprinted from Biophysical Journal
(ref.\cite{sun04}),
with permission from Elsevier \copyright (2004) [Biophysical Society].
}
\label{fig-crankjack}
\end{figure}

The model for the ATP synthase operating in the ``ATP hydrolysis mode'', 
i.e., in the mode in which F$_1$ hydrolyzes ATP to function as a 
chemically-powered rotary motor, has been developed by Oster and 
coworkers \cite{wang98c,sun04}. The corresponding model for the ATP 
synthase running in the ``ATP synthesis mode'', also developed by 
Oster's group \cite{xing05}, utilizes essentially the same underlying 
structural features of the motor.  
In the ATP synthesis mode, rotary motion of the eccentric $\gamma$-shaft 
is converted into the ``hinge-bending'' motion of the $\beta$-subunits 
by a mechanism that is analogous to the vertical movement of automobile 
jack by turning the crank (see fig.\ref{fig-crankjack}). Conversely, 
in the ATP hydrolysis mode the hinge-bending motion of the 
$\beta$-subunits are converted into the rotation of the $\gamma$-shaft 
\cite{kinosita04}. 
The physical reasons for the surprisingly high mechanical efficiency 
of the F$_1$-ATPase have been explained by Oster and Wang in terms 
of underlying molecular mechanisms \cite{oster00c}. 


Gaspard and Gerritsma \cite{gaspard07} developed a mechano-chemical 
kinetic model. In this model the 6 chemical states are assumed to be 
adequate while for the mechanical state only a single angular variable 
$\theta$ is introduced.The dynamics of the system is formulated in 
terms of a hybrid set of equation that describes the changes in $\theta$ 
by the FP-like terms and changes in the chemical states by 
master-equation-like terms for discrete jump processes. The results 
are consistent with a tight mechano-chemical coupling for the F$_{1}$ 
motor. In a subsequent work, Gerritsma and Gaspard \cite{gerritsma10} 
replaced the continuous angle by a discrete 2-state model whose 
kinetics is governed by a set of master equations. Exact analytical 
expressions could be derived for the average angular speed of the 
$\gamma$-shaft and to show that it obeys a Michaelis-Menten-like 
equation with respect to the ATP concentration \cite{gerritsma10}. 
For sufficiently low external torque, the results are consistent with a 
tight mechano-chemical coupling of the F$_{1}$ motor.

\subsubsection{\bf F$_{0}$-F$_1$ coupling} 

In the two preceding subsubsections we have separately discussed the 
kinetics of the operations of the individual F$_0$ and F$_1$ subunits 
of the F$_0$F$_1$- ATPase. Now we briefly review the theoretical models 
of coupling the two subunits for the overall operation of the F$_{0}$F$_{1}$ 
motor.

For a long time, elastic  power-transmission has been a serious candidate 
for explaining the 
mechanism of coupling between F$_0$ and F$_1$ \cite{junge01}. 
Cherepanov et al.\cite{cherepanov99} developed a stochastic kinetic 
model of ATP synthase assuming an elastic coupling between F$_0$ and 
F$_1$. The 4 kinetic states are essentially same as the states 
E, L, F and R in the model developed by Elston et al.\cite{elston98}. 
Suppose the elastic coupler is in a relaxed at an instant of time 
when the angular position of the rotor is $\theta$. In the kinetic 
scheme formulated by Cherepanov et al.\cite{cherepanov99}, the four 
transition 
$\theta \to \theta+30^0 \to \theta+60^0 \to \theta+90^0 \to \theta+120^0$
increase the strain in the elastic coupler in a stepwise manner. 
Then the release of the elastic strain triggers the release of the 
ATP molecule synthesized in the F$_1$ subunit without causing any 
angular displacement of the rotor during this release of ATP. 
Using this kinetic model for the analysis of the channel conductance 
data it has been claimed that the F$_0$ is not voltage-gated 
\cite{feniouk04}.

A mesoscopic model of elastic coupling, developed by Czub et al. 
\cite{czub11} treats the F$_0$-F$_1$ coupling device as eight 
segments that are stacked layers each of which is harmonically 
coupled to its neighboring segments. This model may be viewed as an 
extension of the model developed earlier by Cherepanov et al. 
\cite{cherepanov99}.

\subsection{\bf Rotary motors similar to F$_0$F$_1$-ATPase} 

In this subsection we briefly describe two rotary motors that share 
many of the characteristic features of the structure and dynamics of 
F$_0$F$_1$-ATPase. 

\subsubsection{\bf Rotary motor V$_0$V$_1$-ATPase: a ``gear'' mechanism?} 

Vacuolar ATPases were initially identified in plant and fungal vacuoles
and hence the name. Later these were found also in plasma membrane and
organelle membranes of mammalian cells and plants. Therefore, it is
more appropriate to link the letter ``V'' in V-ATPase with
``{\it various}'' (various membranes) rather than ``vacuoles''. V-ATPases
operate {\it in-vivo} as ATP-dependent proton pumps that regulate the 
pH (acidify) intracellular compartments in eukaryotic cells 
\cite{nelson99,davies97,sze99,wagner04,nishi02,forgac07,mulkidjanian07,kane06,breton07,gaxiola07,hinton09,seidel09,toei10,schumacher11} 
(for a historical account, from the personal perspective of one of 
the leading contributors, see ref.\cite{nelson03a}).

Just like F$_0$F$_1$ motors, V-ATPases also consists of V$_0$ and V$_1$ 
domains \cite{saroussi09}. The V$_0$ domain is located on the cytoplasmic side of the 
membrane whereas V$_0$ is embedded in the membrane. V$_1$ hydrolyzes 
ATP to drive the proton pump V$_0$ whereby protons are translocated 
from the cytoplasm to the other side of the membrane. Unlike F$_0$F$_1$, 
V$_0$V$_1$ does not synthesize ATP from ADP. One key structural 
difference, which has functional implications, is that F$_0$ normally 
has 12 proton carriers whereas V$_0$ has only 6.

A mechano-chemical model of the V$_0$ion pump was developed by Grabe 
et al.\cite{grabe00} along the same line as followed by Elston et al. 
\cite{elston98} for formulating their model for the F$_0$ ion pump. 
However, unlike the two-channel (more appropriately, two half-channels) 
model of the F$_0$ ion pump, both two-channel and one-channel models 
are possible for V$_0$ \cite{grabe00}. The two-channel model of V$_0$ 
is very similar to the two-channel model of F$_0$. In contrast, in the 
one-channel model the proton-binding sites on the rotor are assumed to be 
just outside the level of the membrane and in equilibrium with the 
cytoplasm. Therefore, no channel is required for protonation of the 
proton-binding sites. Rotation bring a protonated site to the entrance 
to the channel where the stator charge forces its release from the 
binding site and passage through the channel to the other side of the 
membrane. The binding site that gets unprotonated in this process 
can continue its further rotation provided a narrow polar strip creates 
a corridor for it to make an exit to the cytoplasm where it again gets 
protonated. 

An interesting prediction of the model of Grabe et al.\cite{grabe00,grabe01} 
is the possibility of a ``gear'' mechanism. Suppose three of the 
six rotor sites of the V$_0$ pump withdraw from proton pumping. 
Then, the pumping rate would be slower but would be able to account 
for a much higher PMF. This is the analog of a gear changing that 
would generate larger force but smaller speed. Grabe and Oster 
\cite{grabe00} have developed a detailed quantitative kinetic theory 
of the regulation of organelle acidity incorporating the pumping 
of protons by V-ATPase.


\subsubsection{\bf Rotary motor A$_0$A$_1$-ATPase} 

A-ATPases can function as either ATP synthase mode or in the ion-pump 
mode.
Detailed comparison of the structures and functions of A-ATPases 
and V-ATPases have been reported \cite{arechaga01,lewalter06,gruber08}. 
The number of ion-binding sites on A$_0$ can vary from 6 to 13 
giving rise wide variability of the stoichiometry \cite{muller04}. 

The mechanisms of torque generation by the ATP synthase (or, 
F$_0$F$_1$-ATPase) in both the ATP-synthesizing and ATP-hydrolyzing modes 
have been investigated in great detail. The mechanism of the coupling 
of the two components, namely F$_0$ and F$_1$ is also fairly well 
understood.

\section{\bf Rotary motors II: Flagellar motor of bacteria} 
\label{sec-specificrotary2}

The flagellum is one of the most important prokaryotic motility structures 
\cite{bardy03,jarrell08}.
The diameter of the bacterial flagellar motor (BFM)
is approximately 50 nm and its angular speed can be as high as $10^{5}$ rpm.
Any satisfactory model of BFM 
\cite{schuster94,berg03,berg04,berry03,sowa08,baker10} 
has to be reversible because bacteria are known to switch the motor 
from CW to cCW and vice-versa.

A bacterial flagellum has three major parts: a rigid helical filament, 
a flexible hook and a basal body. The hook joins the filament with the 
basal body. The motor of the bacterial flagellum is located in the 
basal body which consists of a set of concentric rings mounted on a 
rod that passes through the central axis. The rings have been named 
according to their layer of the cell envelope in which they are located: 
L-ring (named after lipopolysaccharide), P-ring (named after peptidoglycan), 
S-ring, M-ring, etc. For our purpose here, we'll refer to these rings 
collectively as simply the ``ring''. Each BFM has several independent 
stators. The structure of the bacterial flagellum has been analyzed 
from an evolutionary perspective, i.e., how the flagellar structure 
in the living bacteria today might have resulted in the course of Darwinian 
evolution from its more primitive ancestors \cite{pallen06}.

The wealth of experimental data collected over the decades have imposed 
constraints on models of the BFM \cite{berg00}. 
Until about 40 years ago, it was 
commonly believed that each flagellum propagates a helical wave. 
But, Berg and Anderson \cite{berg73} emphatically argued that the 
experimental evidences indicate rotation of each of the flagella. 
Taking hint from the existing experimental results, they assumed 
that the flagellum is rigidly coupled to the ``M-ring'' which is 
free to rotate within the cytoplasmic membrane. However, they also 
imagined that the periphery of the M-ring would be linked to the 
cell wall by actomyosin-like ``cross-bridges'', whose real identity 
was unknown, that would exert the torque to rotate the M-ring. 
Thus, this scenario is based on a Brownian ratchet mechanism of 
free energy transduction.  
This idea was expanded and quantified by Khan and Berg \cite{khan83,khan88}. 
Several variants of this model as well as many other models 
\cite{kojima04,meister89,lauger77,lauger88,kleutsch90,wagenknecht86,mitchell84,walz05}
developed during the first twenty years of activity following Berg and 
Anderson's work \cite{berg73} have been reviewed 
\cite{khan88,macnab84,oosawa86,schuster94,berg03,berry03,sowa08,baker10}

\begin{figure}[htbp]
\begin{center}
\includegraphics[angle=90,width=0.6\columnwidth]{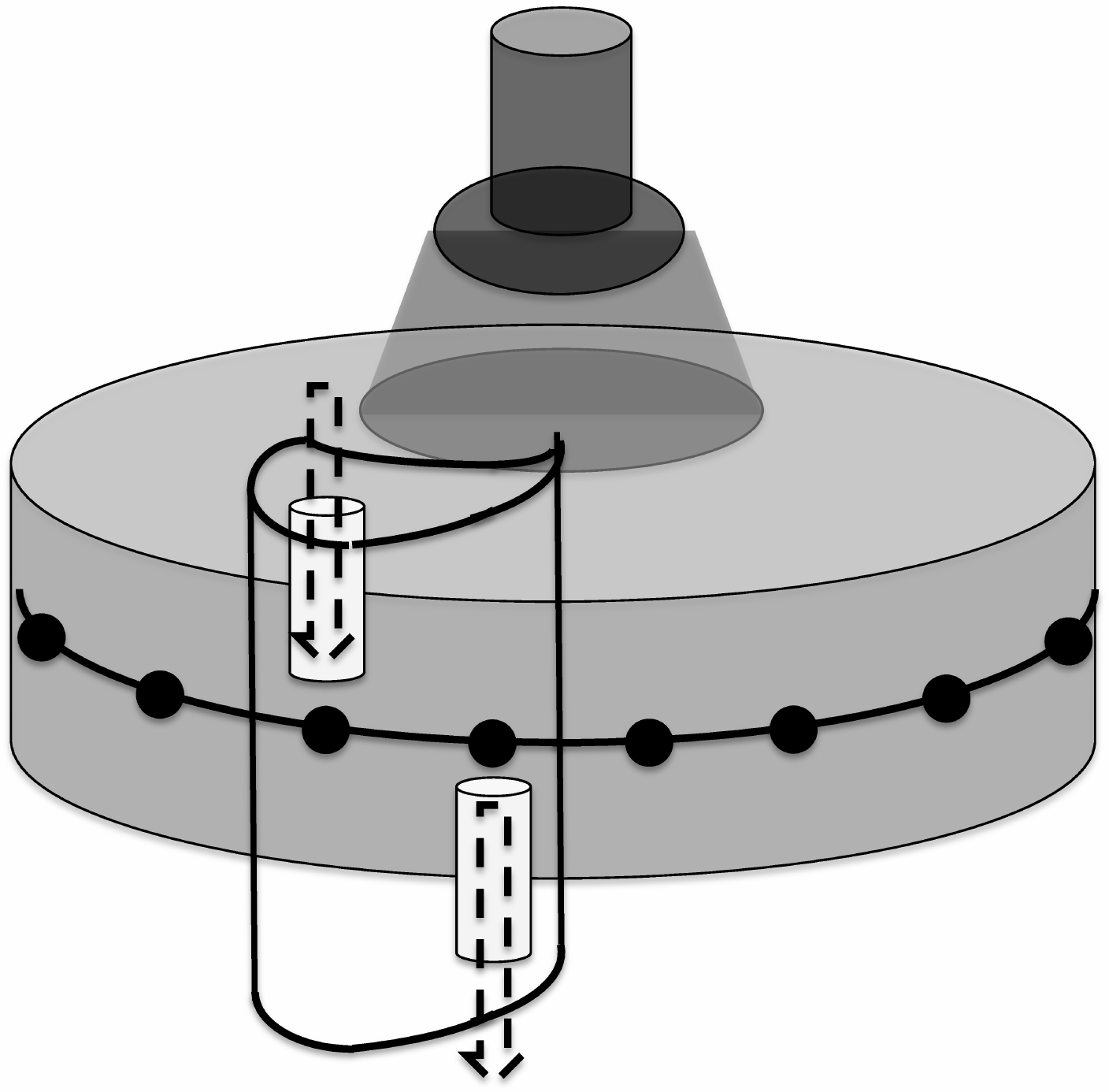}
\end{center}
\caption{The ``turnstile'' model of BFM (adapted from \cite{berry03})}
\label{fig-BFMturnstile}
\end{figure}

In the ``turnstile'' model \cite{meister89,berry00b,berry03}, 
(see fig.\ref{fig-BFMturnstile})
ions entering the bacterial cell from its external surroundings take a 
ride on the rotor. The rotor itself moves either because of electrostatic 
interactions among various charges on the rotor and stator or because of 
its own rotational Brownian motion. Each of the hitch-hiking ions disembark 
from the rotor after getting transported up to a certain distance by the 
rotor. Once the ion leaves, the rotor remains ``locked'' in its current 
position thereby completing one step of its rotational movement and waits 
for the arrival of the next ion. The key features of this mechanism are 
very similar to those of the F$_{0}$ motor that we discussed earlier in 
this section.

\begin{figure}[htbp]
\begin{center}
\includegraphics[angle=90,width=0.6\columnwidth]{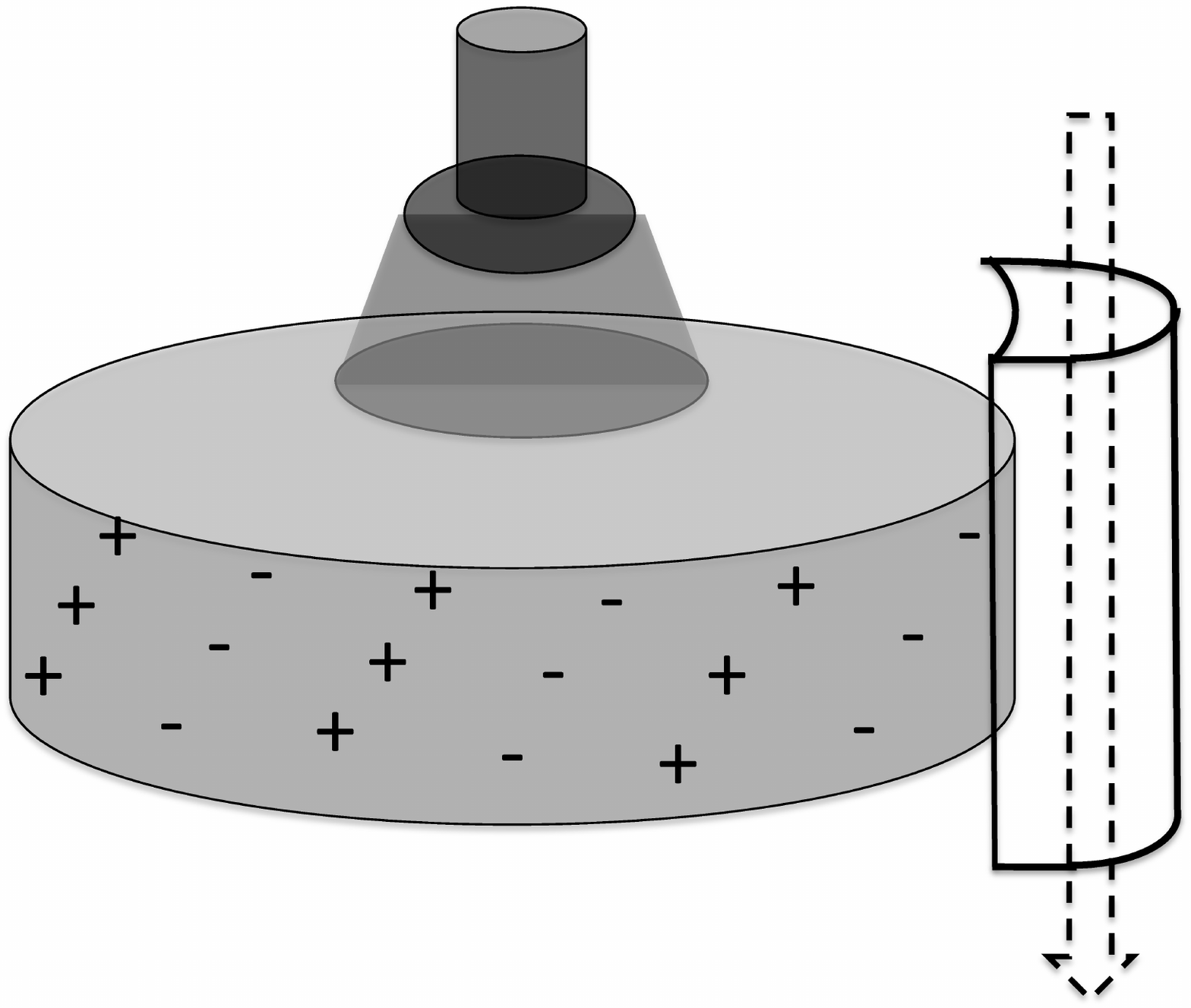}
\end{center}
\caption{The ``turbine'' model of BFM (adapted from \cite{berry03})}
\label{fig-BFMturbine}
\end{figure}

Among the earliest models of BFM that remain relevant even today, to my 
knowledge, the first one was proposed by L\"auger \cite{lauger77} and 
is based on the so-called ``turbine'' mechanism 
(see fig.\ref{fig-BFMturbine}).
In this model 
both the rotor and stator are assumed to be decorated with rows of 
chemical groups, called ``half-sites''. The special feature of these sites 
is that that individually each is incapable of binding to a proton, but 
become competent to bind a proton when paired up with another ``half-site''. 
The rows of the half-sites 
on the rotor are tilted with respect to the vertical row of half-sites on 
the stator. Therefore, at any arbitrary instant of time, the row on a 
given stator intersects a row on the rotor at a single point which forms 
the only proton-binding site at that instant of time. With the turning of 
the rotor, the point of intersection of the stator row and rotor row move 
across the vertical surface of the rotor. Therefore, if protons are 
constrained to move vertically hopping from one binding site to the next, 
along their concentration gradient, vertical proton flow would drive the 
rotation of the rotor in the horizontal plane. Moreover, a conformational 
change that reverses the title angle of the rows of half-sites on the rotor 
would account for the switching of the direction of rotation. 

The original version \cite{lauger77} of the ``turbine'' model of BFM was 
improved in its subsequent amendments \cite{lauger88,kleutsch90}. 
The essential feature of the ``turbine'' models is the rows charges 
that are tilted with respect to the ion channels in the stator. The 
tilted rows on the rotor need not be ``half-sites'', as assumed 
originally by L\"auger \cite{lauger77}. 
An alternative possibility of alternating rows of positive and negative 
charges has also been explored 
\cite{berry93,elston97}. 
In these ``ion turbine'' models, the positively charged ion moves 
exclusively along an ion channel in the stator and never moves onto the rotor.
As the positively charged ion moves along the channel, it tends to keep a 
row of negative charges on the rotor close to it. Consequently, the rotor 
rotates as the positively charged ion transits through the channel. 
This situation is analogous to hydro-electric turbines; the positively 
charged translocating ions are the analogs of water while the tilted rows 
of charges are the analogs of the turbine blades. Just as the water flowing 
through the turbine exerts torque on the rotor unit, the translocating ion 
in the BFM exerts torque on the corresponding rotor \cite{berry93}. 
Because of this analogy, this class of models are referred to as ``turbine'' 
\cite{berry93,elston97}.   
The CW-cCW switching can be explained by the change in the proton affinity 
of the proton-binding sites of the channel. When the proton-affinity is 
high the ionic current is carried by protons (+ve charge) into the cell 
whereas in the opposite case of low proton-affinity the effective ionic 
current is that of -vely charged ``holes'' flowing outward from the cell. 
The flipping of the sign of the effective carriers of ionic current leads 
to the CW-cCW switching. 

Let us assume that there are just 2 sites available in each proton channel 
for their protonation. The 4 possible states of the channel are then 
\cite{elston97}
denoted by the symbols E (both sites empty), T (top site occupied by a 
proton), B (bottom site occupied by a proton), and F (both sites full). 
The probabilities for the occurrence of these four states are the 
components of the column vector ${\bf P}(\theta,t)$ where $\theta$ 
denotes the angular position of the of the rotor at time $t$. The kinetic 
equation for ${\bf P}(\theta,t)$ can be expressed as \cite{elston97} 
\begin{equation}
\frac{\partial {\bf P}}{\partial t} = [{\bf L}(\theta) + {\bf W}(\theta)] {\bf P}
\label{eq-elstonturbine}
\end{equation}
where ${\bf L}(\theta)$ is the Fokker-Planck operator describing the 
mechanical rotation whereas ${\bf W}(\theta)$ is the master operator 
that accounts for the changes in the occupation of channel sites by 
protons. The operator ${\bf L}(\theta)$ incorporates not only rotational 
diffusion term but also rotational drift caused by electrostatic 
potentials $V_{\mu}(\theta)$ where the angular profile of the potential, 
in general, depends on the state of occupation of the designated sites 
by the protons \cite{elston97}. In this model the protons modulate the 
interactions of the negatively charged proton-binding sites with rotor 
charges by screening (or neutralizing) these negative charges. In the  
subsequent work of Walz and Caplan \cite{walz00} the protons are 
directly responsible for the torque generation. 

Kojima and Blair \cite{kojima01} proposed a power stroke mechanism in 
which a conformational change of the stator, caused by the inflow of 
a proton (or sodium ion) into the embedded channel, drives the rotation 
of the rotor.  Schmitt \cite{schmitt03} proposed a different mechanism 
of electrostatic transmission of force from the stator to the rotor. 
In this model the passage of proton or sodium ion induces a reversible 
rotation of the stator that, in turn, rotates the rotor because of their 
electrostatic interaction  at the stator-rotor interface. 
These ideas were extended even further, with minimum number of assumptions 
regarding the generic principles, by Xing et al. \cite{xing06} and 
Bai et al.\cite{bai09,bai12}. It was pointed out that there are two 
time scales in the system, namely, (a) the time scale of ion hopping on 
and off (the intrinsic dynamics), and (b) the time scale of relaxation 
of the load spring (an extrinsic dynamics). This time scale separation 
gives rise to a plateau in the force-velocity plot \cite{xing06}. 
The fact that the stators move approximately independent of each other 
allows a mean-field type approximation. Exploiting this simplification, 
Bai et al.\cite{bai12} have developed a unified theory that explains the 
coupling between the torque generation and the dynamics of switching 
between CW and CCW rotation.

\subsection{Summary of the sections on rotary motors} 

In this section and the preceeding section we have reviewed the models of 
two major classes of rotary molecular motors. 
Let us now draw analogy between the swimming of a single bacterium by 
multiple flagella, each of which is driven by a distinct rotary motor, 
and the transport of a single vesicular cargo by multiple cytoskeletal 
motors. In the latter case, as we have discussed in the subsection 
\ref{sec-collectiveporters}, significant progress has been made in 
recent years on the multi-motor cooperativity. However, in contrast to 
cytoskeletal motors, each BFM is a reversible rotary motor. Therefore, 
a kinetic model of the collective operation of the BFMs of a single 
bacterium should not only explain the mechanism of their cooperative 
rotation, but also their coordinated switching.

\section{\bf Some other motors}
\label{sec-miscellaneous}

\noindent $\bullet${\bf Myosin-I: membrane-cytoskeleton active crosslinker}

Among the unconventional myosins, myosin-V and myosin-VI have been 
discussed in detail in section \ref{sec-proc2myosins}. Here we 
briefly describe the special features of the members of another 
family of unconventional myosins, namely myosin-I.

Myosin-I can bind simultaneously to actin filaments and cell membranes 
thereby crosslinking the two. However, unlike other passive crosslining 
proteins, myosin-I crosslinks actively in the sense that it can bend, 
deform and move membrane by coupling these operations with ATP hydrolysis 
that it catalyzes. These operations of myosin-I are part of many biological 
functions that include, for example, exocytosis, endocytosis, vesicle 
shedding, blebbing and gating of ion channels \cite{mcconnell10}. 
All these biological processes have been modelled theoretically. But, 
to my knowledge, the mechanism of force generation by a single myosin-I 
motor in terms of its structure and mechano-chemical kinetics remains 
a challenging open problem.

\noindent $\bullet${\bf Chromatin remodellers and unwrapping of nucleic acids}

If nucleosomes
were static, segments of DNA buried in nucleosomes would not be
accessible for various functions involving the corresponding genes.
In order to get access to the relevant segments of DNA for various
processes in DNA metabolism, eukaryotic cells use
ATP-dependent chromatin-remodeling enzymes (CRE)
which alter the structure and/or position of the nucleosomes 
\cite{clapier09,eberharter04,tsukiyama02,flaus03,flaus11,halford04a,saha06,racki08,cairns07}.
In principle, there are at least four different ways in which a CRE
can affect the nucleosomes \cite{cairns07}:
(i) {\it sliding} the histone octamer, i.e., repositioning of the
entire histone spool, on the dsDNA; (ii) {\it exchange} of one
or more of the histone subunits of the spool with those in the
surrounding solution (also called {\it replacement} of histones)
(iii) {\it removal} of one or more of the histone subunits of the
spool, leaving the remaining subunits intact, and (iv) complete
{\it ejection} of the whole histone octamer without replacement.

Spontaneous thermal fluctuations can cause a transient unwrapping and
rewrapping of the nucleosomal DNA from one end of the nucleosome spool. 
On the nucleosomal DNA, the farther is a site from the entry and exit 
points, the longer one has to wait to access it by the rarer spontaneous 
fluctuation of sufficiently large size
\cite{li04,poirier08,poirier09,tims11}. 

Can a nucleosome slide {\it  spontaneously} by thermal fluctuations 
thereby exposing the nucleosomal DNA? 
If the DNA were to move unidirectionally along its own superhelical
contour on the surface of the histone, at every step it would have
to first {\it transiently} detach simultaneously from all the $14$
binding sites and then reattach at the same sites after its contour
gets shifted by $10$ bp (or multiples of $10$ bp). But, the energy
cost of the simultaneous detachment of the DNA from all the $14$
binding sites is prohibitively large 
\cite{blossey11,kulic03a,kulic03b}.

But, why can't the cylindrical spool simply roll on the wrapped
nucleosomal DNA thereby repositioning itself? If the nucleosome
rolls by detaching DNA from one end of the spool, cannot it compensate
this loss of binding energy by simultaneous attachment with a binding
at the other end? 
This rolling mechanism would
successfully lead to spontaneous sliding of the nucleosome only if
the histone spool were infinite with an infinite sequence of binding
sites for DNA on its surface; however, on a finite-size histone spool 
this would not be feasible \cite{blossey11}. 

An alternative possibility is to form a flap that can diffuse. 
In the process of normal ``breathing'', most often the spontaneously
unwrapped flap rewraps exactly to its original position on the
histone surface. However, if the rewrapping of a unwrapped flap takes
place at a slightly displaced location on the histone spool a small
bulge (or loop) of DNA forms on the surface of the histones. Since
the successive binding sites are separated by $10$bp, the length
of the loop is quantized in the multiples of $10$bp \cite{schiessel03}.
Such a spontaneously created DNA loop, can diffuse in an unbiased
manner on the surface of the histone spool. In the beginning of each
step DNA from one end of the loop detaches from the histone spool,
but the consequent energy loss is made up by the attachment of DNA
at the other end of the loop to the histone spool before the step
is completed. Consequently, by this diffusive dynamics, the DNA loop
can traverse the entire length of the $14$ binding sites on the
histone spool of a nucleosome which will manifest as sliding of the
nucleosome by a length that is exactly equal to the length of DNA in
the loop. The diffusing DNA bulge can be formed by a ``twist'',
rather than bending, of DNA  \cite{lusser03,becker02a,langst04}.
Spontaneous sliding of a nucleosome, however, is too slow to support
intranuclear processes which need access to nucleosomal DNA.
That is why ATP-dependent CRE is needed for active remodeling of the 
nucleosome. 

Various aspects of chromatin dynamics has received some attention of
theoretical modelers, including physicists, over the last few years
\cite{sakaue01,schiessel03,kulic03a,kulic03b,schiessel06,rafiee04,blossey11,mobius06,langowski06,lense06,vaillant06,ranjith07,lia06,chou07,garai12}.

\noindent $\bullet$ {\bf FtsK and SpoIIIE: Chromosome segregation motors in E-coli and Bacillus subtilis} 

So far there are no convincing direct evidence for the existence of any 
mitotic spindle-like machinery in bacteria for post-replication 
segregation of chromosomes before cell division. However, there are 
more primitive motors which carry out chromosome segregation in 
bacteria.  For example, in {\it E-coli} FtsK segregate chromosome in 
an ATP-dependent manner by translocates dsDNA during cell division 
\cite{strick06b,bigot06,bigot07,pease05,levy05,ptacin06,lowe06}.

Normally {\it Bacillus subtilis}, a rod shaped bacterium, divides to 
two similar daughter cells. However, under some special circumstances, 
which leads to spore formation, a {\it Bacillus subtilis} divides 
asymmetrically into a small prespore and a larger mother cell. The 
translocation of the chromosome into the small prespore compartment 
is carried out by the motor protein SpoIIIE. 

Most of the fundamental questions on the operational mechanism of 
FtsK and SpoIIIE are similar to those generic ones for helicases and 
translocases (including packaging motors for viral capsids). 
For example, how does SpoIIIE, which anchors itself at the septum 
between the two compartments, translocate the DNA in the desired 
direction, namely, from the larger to the smaller compartment?

\noindent $\bullet${\bf G-proteins}

G-proteins, which are believed to be the common evolutionary ancestors 
of myosins and kinesins \cite{vale96y}, also generate forces by 
hydrolyzing GTP. The force thus generated are comparable to that 
generated by myosin and kinesin both of which are powered by ATP 
hydrolysis \cite{kosztin02}.

\noindent $\bullet${\bf Topoisomerases and untangling of DNA}

As we mentioned in subsection \ref{sec-catalogue2}, topoisomerases can 
untangle DNA by passing one DNA through a transient cut in another. 
Topoisomerases are broadly 
divided into two classes, namely type I and type II, which cleave one 
or two strands of DNA, respectively 
\cite{wang96,wang02z,wang09z,charvin05,schoeffler08,liu09,neuman10,vos11}. 
In principle, mere strand pasage reaction catalyzed by a topoisomerase 
does not require input energy- the chemical energy of the cleaved 
phosphodiester bond (or bonds, in case of type-II topoiomerase) is 
stored within the DNA-enzyme complex and, therefore, can be utilized 
for the restoration of the bond(s) after the strand passage. Indeed, 
type-I topoisomerases perform the task of strand passage without the 
consumption of external energy. Then, why do type-II topoisomerases 
hydrolyze ATP for the same task? 

Classic experiment of Rybenkov et al.\cite{rybenkov97} established 
that the type-II topoisomerase are capable of suppresing the 
probability of self-entanglement (i.e., knots) and mutual entanglements 
(i.e., links) far below the levels expected from equilibrium statitical 
mechanics. In other words, type-II topoisomerases not marely catalyze 
the strand pasage reaction, but also control the overall entanglement 
of the DNA molecule. In its latter role, it essentially acts as a  
Maxwell's demon \cite{pulleyblank97,vologodski01} and does not 
violate the second law of thermodynamics becaue of the consumption of 
input energy supplied by ATP hydrolysis. The lower-than-equilibrium 
entanglement achieved by the type-II topoisomerases has been explained 
by a kinetic proofreading mechanim \cite{yan01} that, in spirit, is 
similar to the kinetic proofreading mechanism for lower-than-equilibrium 
error committed by a ribosome during translation.

For many years, it was not clear how a small machine like a topoisomerase 
can sense the DNA topology, which is a global property, through local 
DNA-protein interaction. In recent years, few alternative possible 
scenario have been proposed to explain this phenomenon \cite{vologodski09}; 
however, a detailed discussion of these models is beyond the scope of this 
review.

\noindent $\bullet${\bf Chaperones and folding of proteins}

Protein folding {\it in-vivo} is most often assisted by a group of 
molecular machines called chaperones 
\cite{richardson98,agashe00,walter02}. 
Members of many chaperone families are also known as heat shock 
proteins (HSP) and these families are classified according to their 
molecular weights, e.g., HSP60, HP70, HSP90, etc.\cite{taipale10,hartl11}. 
The operations of these machines are fuelled by ATP. Among the 
chaperones, the chaperonin proteins \cite{thirumalai01,mayer10,yebenes11} 
(GroEL in bacteria) have been characterized, both structurally and 
functionally, in great detail 
\cite{lin06,horwich06,horwich09}.
The heptameric ring-like structure of GroEL resembles a cage where a 
lid is formed by GroES. Encapulation of a protein within the cage 
protects it against aggregation or misfolding. ATP binding and hydrolysis 
modulates the affinities of the GroES caps for the GroEL ring thereby 
regulating the release of protein (partially-) folded inside the cage. 
The concept of allosterism, that we explained in section \ref{sec-allostery}, 
plays an important role in the kinetics of GroEL-GroES machinery. 
The positive cooperativity arising from the intra-ring interactions 
and the negative cooperativity caused by inter-ring interactions have 
been elucidated by molecular dynamic simulation \cite{ma00k}. 
Tehver and Thirumalai \cite{tehver08} developed a kinetic model that 
couples the allosteric transitions of the GroEL with the distinct 
stages of folding (or misfolding) of the protein substrate. 
The efficiency of the folding machinery is found to depend on the 
chaperonin concentration and the rate of binding of the protein 
substrate \cite{tehver08}.

\noindent $\bullet${\bf Synthetic molecular motors: biomimetics and nano-technology}

Initially, technology was synonymous with macro-technology. The first
tools applied by primitive humans were, perhaps, wooden sticks and
stone blades. Later, as early civilizations started using levers,
pulleys and wheels for erecting enormous structures like pyramids.
Until nineteenth century, watch makers were, perhaps, the only people
working with small machines. Using magnifying glasses, they worked
with machines as small as $0.1 mm$. Micro-technology, dealing with
machines at the length scale of micrometers, was driven, in the
second half of the twentieth century, largely by the computer
miniaturization.

In 1959, Richard Feynman delivered a talk \cite{feynman59} at a meeting
of the American Physical Society. In this talk, entitled ``{\it There's
Plenty of Room at the Bottom}'', Feynman drew attention of the scientific
community to the unlimited possibilities of manipulating and controlling
things on the scale of nano-meters. This famous talk is now regarded by
the majority of physicists as the defining moment of nano-technology
\cite{junk06}. In the same talk, in his characteristic style, Feynman
noted that ''many of the cells are very tiny, but they are very active,
they manufacture various substances, they walk around, they wiggle, and
they do all kinds of wonderful things- all on a very small scale''.

From the perspective of applied research, the natural molecular machines
opened up a new frontier of nano-technology
\cite{goodsell04,heuvel07,hess11,balzani03,barcohen05,sarikaya03,browne06,fulga09}. 
Even nano-robotics may no longer be a distant dream \cite{ummat05,ummat06}. 
A conventional 2-headed kinesin has strong resemblance, at least at a 
superficial level, with two-legged mobile robots (see ref.\cite{silva07} 
for a historical account of the development of legged robots, particularly 
the ``bipeds''). 
The miniaturization of components for the fabrication of useful devices,
which are essential for modern technology, is currently being pursued by
engineers following mostly a top-down (from larger to smaller) approach.
On the other hand, an alternative approach, pursued mostly by chemists,
is a bottom-up (from smaller to larger) approach. The bottom-up approach 
is also likely to enrich {\it synthetic biology}
\cite{serrano07,brent04,heinemann06,endy05,young10,andrianantoandro06,purnick09,mukherji09}.
The term {\it biomimetics} has already
become a popular buzzword \cite{barcohen05,sarikaya03}; this field deals
with the design of artificial systems utilizing the principles of natural
biological systems.
We can benefit from Nature's billion year experience in nano-technology.

\noindent$\bullet${\bf Importance of molecular motors in biomedical research- control and cure of disease}

Just as occasional disruption of work in any department of a factory
can bring entire operation factory to a standstill, defective molecular
machines can cause diseases. Moreover, viruses are known to hijack 
the motors to travel from the cell periphery to the cell nucleus.
If we understand how molecular machines work, we might be able to
devise ways to selectively either arrest those sub-cellular processes
that cause diseases like cancer or slow down metabolism of invading
organisms. The molecular motor transport system can be utilized even
for targeted drug delivery where molecular motors can be used as
vehicles for the drug. Thus, fundamental understanding of the mechanism
of biomolecular machines will help us to fix them when they malfunction
and, perhaps, to manipulate them to improve human health and fitness.

\section{\bf Summary and outlook}

In this article we have critically reviewed the stochastic kinetics 
of molecular motors and that of the processes that they drive. 
In part I, we have considered the fundamental principles, some essential 
background concepts and generic models for several different types of 
motors. Some of the common ``chemical'' proceses involved in their operation  
are listed below: \\
(i) chemical reactions are catalyzed either by the motor itself or by 
an accessory device; 
(ii) The motor forms non-covalent bonds with substrates and other 
ligands; 
(iii) many conformations of the motor are possible and the motor 
fluctuates between these conformations because the strengths of the 
non-covalent bonds are comparable to the thermal energy $k_BT$; 
(iv) binding or unbinding of a ligand can alter the relative stability 
of conformations thereby inducing transition from one to another.
Moreover, there are further formal similarities between the ``mechanical'' 
stepping of a single motor and chemical reaction catalyzed by a 
single enzyme molecule. Therefore, several aspects of the kinetics of 
chemical reactions, particularly those catalyzed by enzymes, have been 
discussed in significant detail in part I of this review. The strategies 
of modeling stochastic mechano-chemical kinetics at different levels of 
spatio-temporal resolution have also been explained before reviewing the 
generic models.

In part II we have presented applications of these basic concepts 
and techniques of chemical physics as well as those of nonequilibrium 
statistical mechanics to study the stochastic kinetics of specific 
molecular motors using theoretical models. Some of the key structural 
details or important features of the kinetics of the motors, which were 
ignored in part I in the context of generic models, have been incorporated 
in the specific models of these motors in part II of this review.
All the motor-driven movements reviewed here can be classified into 
four categories \cite{cozzarelli06a}: (i) rotation about an axis, (ii) 
translation along an axis, (iii) translation perpendicular to an axis, 
and (iv) lateral separation of two axes.

While summarizing each section, we have already listed some open questions 
in the context of specific motors. Next I list a few general open questions 
that can be raised for all the motors. Barring the exception of great 
visionaries, for anyone it is risky to speculate on the future directions 
of research on molecular motors. Neverthless, I point out some limitations 
of the current models so that the future course of action for theorists 
can be anticipated.

The atomistic structure of many motors are not known at present at 
sufficiently high-resolution (resolution of a few angtroms). 
Only when such high-resolution structures become available, fully 
atomistic {\it in-silico} modeling of the motor would be possible. 
However, even when such detailed structures are available, only 
coarse-grained models may be useful if the direct MD simulation of 
the atomistic models over the relevant time scales cannot be performed 
with the computational resources available. 

For modeling at the coarser Brownian level, based on Langevin or 
Fokker-Planck equations, the potential landscape is needed. However, 
for most of the motors, these potential landscapes are postulated, 
rather than derived from more microscopic considerations. For example, 
there are compelling evidences that electrostatic interactions play 
important roles in the operation of the individual members of many 
families of molecular motors. However, so far very little effort has 
been made to derive the effective interactions (and potential landscapes) 
from more microscopic considerations where the electric charges and 
their coulomb interactions would be treated explicitly. 

For solving the forward problem with process modeling, we expect significant 
progress over the next decade in two opposite directions:
(a) {\it in-silico} modeling of single individual motors with ever 
increasing structural and dynamical details of its coordinating parts 
in an aqueous medium;
(b) integration of nano-motors and motor-assemblies into a micro-factory- 
the living cell. 

Finally, let me emphasize the need of statistical inference drawn from 
analysis of empirical data for reverse-engineering of molecular motors.
In a rare example of model selection for a specific molecular motor, 
Bronson et al. \cite{bronson09} extracted the best model by optimizing 
the maximum {\it evidence}, rather than maximum likelihood, that treats, 
for example, the number of discrete states of the system as a variational 
parameter. This is a step in the right direction. However, I am not aware 
of any work that ranks alternative models of any molecular motor according 
to relative scores computed on the basis of the principle of strong 
inference 
\cite{chamberlin1890,platt64}.

The picture of the motor-driven intracellular processes in a living cell 
can be dramatized as follows \cite{chowdhury12col}: 
the cell is like an under water ``metro city'' which is, however, only 
about $10 \mu$m long in each direction! In this city, there are ``highways''
and ``railroad'' tracks on which motorized ``vehicles'' transport
cargo to various destinations. However, the highways and railroads are 
very dynamic- these are constructed when needed and often dismantled 
when not in use. It has a library for an efficienct storage of the 
chemically encoded blueprint of the construction and maintenance of 
the city. It has a system of machinery that provides rapid access to 
specific regions of this packaged blueprint. It has specialized machines 
that, utilizing the chemically encoded template stored in this library, 
synthesize materials that, then, form the components of various machine 
tools in this city. It has special ``waste-disposal plants'' which 
degrade waste into products that are recycled as raw materials for 
fresh synthesis. This eco-friendly city re-charges spent ``chemical fuel'' 
in uniquely designed ``power plants''. This city also uses a few 
``alternative energy'' sources, including ``electrical'' energy, directly 
in some operations. 

A complete understanding of the running of this fully automated  
``under water metro city'' in terms of the coordinated operation of 
the machineries would be the ultimate aim of scientific investigation 
on intracellular molecular motors. However, ``such triumphant victories 
come very rarely, and they are separated by the slow, plodding attack 
on a wide front'' \cite{adrian54}.



\appendix

\section{\bf Eukaryotic and prokaryotic cells: differences in the internal organization of the micro-factories}

From the evolutionary point of view, cells have been, traditionally, 
divided into two categories, viz., {\it prokaryotes} and {\it 
eukaryotes}. Most of the common bacteria like, for example, Escherichia 
Coli ({\it E-coli}) and Salmonella, are prokaryotes. Animals, plants 
and fungi are collectively called eukaryotes. Major difference 
between prokaryotic and eukaryotic cells lies in their internal 
architectures; the main distinct feature of eukaryotic cells is the 
cell nucleus where the genetic materials are stored. 
The prokaryotes are mainly uni-cellular organisms. The eukaryotes
which emerged first through Darwinian evolution of prokaryotes were
also uni-cellular; multi-cellular eukaryotes appeared much later. 
The plausible evolutionary routes leading to the birth of the first
eukaryotic cell from its prokaryotic ancestor(s) and its subsequent
evolution is an active area of scientific exploration
\cite{poole06,duve07,davidov09},

Moreover, the traditional division into prokaryotes and eukaryotes 
has been questioned and an alternative three-domain system, that 
classifies organisms into bacteria, archaea and eukaryota, has 
been proposed  \cite{sapp05,woese04,jarrell11}. We'll not get involved in 
the debate on the scheme for classification of organisms; for the 
purpose of this review, which is addressed mainly to physicists, 
we'll use both the schemes hoping that the context will make the 
usage unambiguous. 
 
Some of the molecular machines found in eukaryotes have been 
discovered also in the other domains in the kingdom of life.
But, some other machines seem to be present only in eukaryotes.
So far as the archaea are concerned, it is interesting to compare 
their machinery with those of bacteria and eukaryota. 
We'll consider concrete examples in the appropriate contexts.

\subsection{\bf Model eukaryotes and prokaryotes}

In biology, often the simplest among a family of objects is called
a model system for the purpose of experimental investigations 
\cite{muller10}. 
A short list of the model systems commonly used in molecular cell
biology, particularly those found useful for studying functions of
molecular machines, is now provided. 

The most popular model {\it animals} for biological studies are as
follows:
(i) the fruit fly {\it Drosophila melanogaster}, a model insect,
(ii) {\it Caenorhabditis elegans} (C-elegans), a transparent worm,
(iii) the zebra fish {\it danio rerio}, a model vertebrate;
(iv) the mouse, however, is more important for practical use of cell
biology in medical sciences.

{\it Arabidopsis thaliana} is the most popular model {\it plant}
while {\it Chlamydomonas reinhardtii} is a model of green algae. 
{\it Saccharomyces cerevisiae} (Baker's yeast) and
{\it Schizosaccharomyces pombe} (Fission yeast) are most widely
used models for fungi.
However, for studying filamentous fungi, {\it Neospora crassa} is
used most often as a model system. 

Bacteria are divided into two separate groups on the basis of their
response to a staining test invented by Hans Christian Gram. Those
which respond positively are called Gram-positive bacteria whereas
those whose response is negative are called Gram-negative. One of
the main differences between these two groups of bacteria is the
nature of the cell wall which we'll mention in an another appendix below. 

The commonly used models for Gram-positive bacteria are {\it Bacillus
subtilis}, {\it Listeria monocytogenes}, etc. The bacterium {\it
Escherichia coli} (E-coli), which is normally found in the colon of
humans and other mammals, and the bacterium {\it Salmonella} are the
most extensively used model for Gram-negative bacteria. Another prominent
member of the group of Gram-negative bacteria is {\it Proteus mirabilis}.\\

\section{\bf Molecules of a cell: motor components and raw materials}
\label{app-moleculesincell}

Water is a major abundant molecular species in a living cell. 
A molecular machine is either a macromolecular assembly or consists of
a single macromolecule. Moreover, these cross barriers created by 
medium-size (meso-molecules?) and are regulated by signals that are 
often carried by small molecules and ions. Therefore, for the convenience 
of the non-biologist readers, we list a few examples of these molecules 
(see table \ref{table-moleculesincell}). 
Somewhat longer introduction, intended for non-biologists, is available 
in ref.\cite{budd12a}.

\begin{table}
\begin{tabular}{|c|c|c|} \hline
Molecule type  & Group name   & Examples \\\hline
Macromolecule & Polynucleotide & DNA, RNA \\ \hline
Macromolecule & Polypeptide & Protein \\ \hline
Macromolecule & Polysaccharide & Starch, cellulose\\ \hline
Medium-size molecule & Lipid & Phospholipid, sphingolipid \\ \hline
Small molecule \& ions & & Na$^+$,K$^+$, etc. \\ \hline
\end{tabular}
\caption{Molecules in a cell classified by size.
 }
\label{table-moleculesincell}
\end{table}

\subsection{\bf Natural DNA, RNA, proteins}

The three main categories of macromolecules in a living cell are nucleic 
acids \cite{bloomfield00},
proteins \cite{whitford05} and polysaccharides \cite{poly}.
The individual {\it monomeric residues} that form nucleic acids and
proteins are {\it nucleotides} and {\it amino acids}, respectively.
Both these types of macromolecules are {\it unbranched} polymers.
The {\it complete covalent} structure is called the {\it primary}
structure of the macromolecule. It would be extremely time- (and space-)
consuming to write a chemical formula for the entire primary structure.
Therefore, it is customary to express primary structures in terms of
abbreviation using an alphabetic code. The most common convention uses
one-letter code for each nucleotide and three-letter code for each
amino acid. 

For proper biological function, these macromolecules form
appropriate {\it secondary} and {\it tertiary} structures. The term
{\it conformation} is synonymous with tertiary structure \cite{cantor1}
Different types of standard schemes followed for visual presentation of the
conformations of macromolecules \cite{goodsell05,richardson92,morange11} 
emphasize different aspects of the conformations. 
So far as the molecular motors are concerned, artificially synthesized 
motors need not be composed on natural polymers. Instead, ``{\it foldamers}'' 
\cite{hill01,goodman07},
a class of artificially synthesized polymers that can fold into a desirable 
conformation, may provide unlimited opportunities for wide range of 
applications.

\section{\bf Information transfer in biology: replication, gene expression and central dogma} 
\label{sec-infotransfer}

Like most of the literature in biology, we also make extensive use of 
the terms (genetic-) code, message, transcription, translation, 
proof-reading, editing, etc. However, most of these terms were originally 
coined in the context of information storage and transmission- at many 
levels starting from digital to linguistic. The 
\cite{yockey92}

DNA is {\it replicated} just before cell division so that exact copies 
of the genetic material of a cell can be inherited by both the daughter 
cells. {\it Gene expression}, on the other hand, is a process that 
consists of several steps, the main steps being {\it transcription} and 
{\it translation}. Initially, for many years, the ``central dogma''
\cite{crick58,crick70,morange08},
of molecular biology was interpreted loosely to imply the allowed 
pathway

$${\framebox{DNA}} \mathop{\rightarrow}^{\rm Transcription}~~~{\framebox{RNA}}~~~\mathop{\rightarrow}^{\rm Translation} ~~~{\framebox{Protein}}$$

for synthesizing a protein following the ``instructions'' encoded on 
the corresponding stretch of a DNA molecule. 

\begin{figure}[htbp]
\begin{center}
\includegraphics[angle=90,width=0.6\columnwidth]{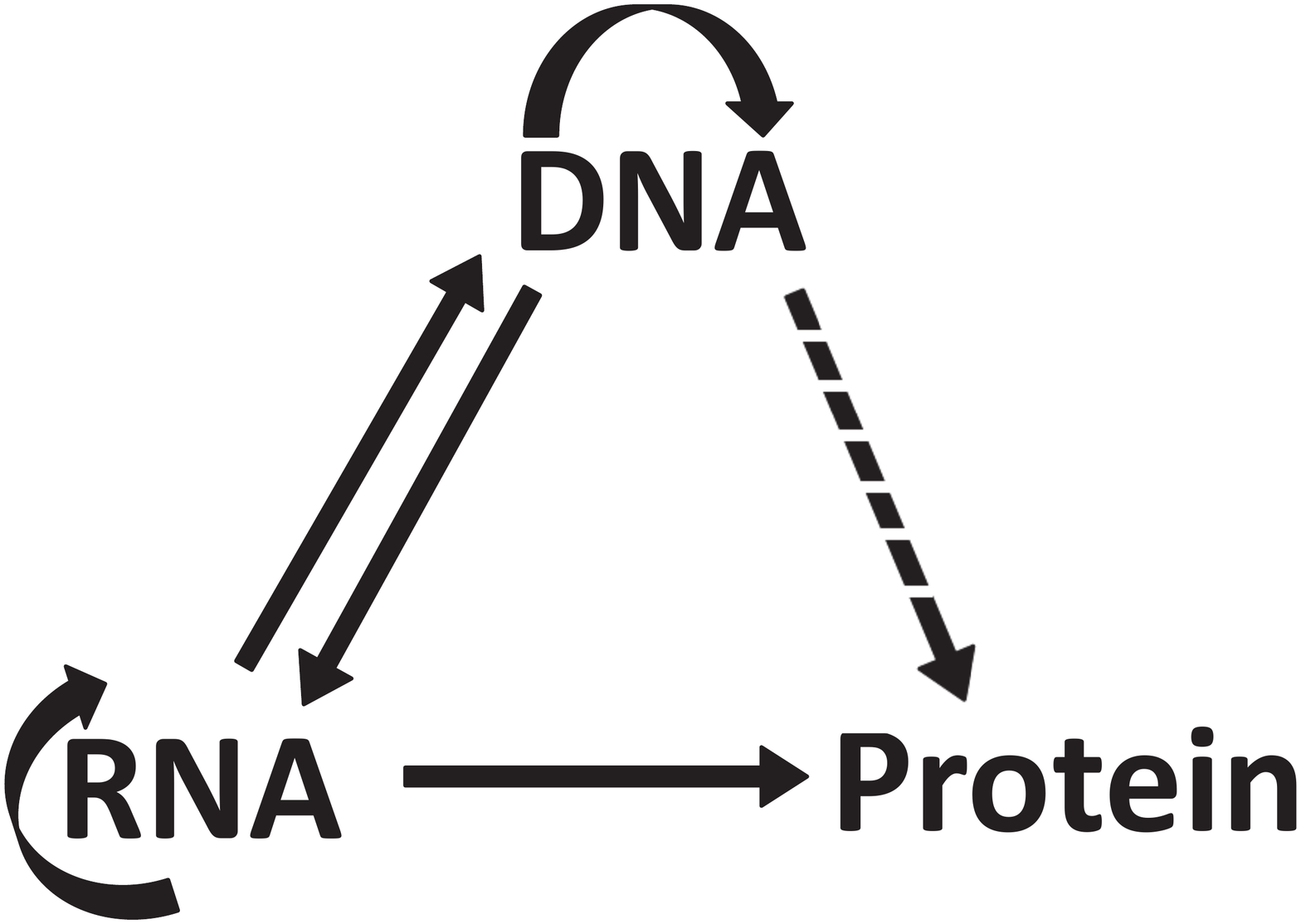}
\end{center}
\caption{The central dogma of molecular biology (adapted from \cite{crick70})}
\label{fig-dogma}
\end{figure}

However, it was re-emphasized by Crick \cite{crick70} in 1970 that
the central dogma is a negative statement- information transfer from
protein (i.e., from protein to nucleic acids and proteins to proteins)
does not take place. As we describe in the next subsection, genomes
of a large class of viruses consist of RNA, rather than DNA. However,
the possibility of information transfer from RNA to RNA and that from
RNA to DNA do not contradict the central dogma.
Moreover, the possibility of information transfer directly from DNA
to protein is not ruled out although such a process has never been
observed in nature so far. Thus, the current status of the central
dogma is represented schematically in fig.\ref{fig-dogma}.

\noindent$\bullet${The expanding world of RNA and its implications}

Among the macromolecules of life, RNA has a unique distinction-
on the one hand, just like DNA, it can serve as genetic material
and, on the other, like proteins, it can serve as a catalyst.
mRNA, rRNA and tRNA together form the group of ``core'' RNAs.
However, many other types of non-coding RNA (ncRNA) molecules have
been discovered and these may be just the ``tip of the iceberg''
\cite{darnell11,sharp09,mattick06,amaral08,mercer09,carninci10,collins11}.
Although some of their regulatory functions have been discovered,
at present, our understanding of the mechanisms of these processes
is far from clear.

\noindent$\bullet${Post-transcriptional processing of mRNA}

In bacteria freshly polymerized mRNA transcripts can be directly
translated into the corresponding proteins. But, in a eukaryotic cell
transcription takes place in the nucleus. The mRNA transcript undergoes
various types of post-transcriptional ``processing'' both inside nucleus
as well as in the cytoplasm after it is transported out of the nucleus
\cite{knoop11}.
One of the distinct features of eukaryotic DNA is the existence of
{\it introns}, patches of sequences which do not encode for proteins
and which, therefore, are removed during {\it splicing} of the pre-mRNA.
Why did evolution favor insertion of such apparently ``useless'' introns
into the DNA? Several plausible regulatory functions of introns have been
hypothesized.

\section{\bf Cytoplasmic and internal membranes of a cell}
\label{app-membranes}

Major molecular component of membranes are lipids. These amphiphilic 
molecules consist of a hydrophilic head and two hydrophobic tails. 
For energetic reasons, in aqueous medium, these molecules form 
bilayers where two oppositely oriented monolayers of lipids stick 
together with the heads outside and the tails inside the bilayer 
\cite{edidin03}. 
In animal cells, cholesterol is another important component of the 
cell membrane. 
From the perspective of molecular transport across the membrane, 
the membrane proteins are crucially important \cite{hedin11}.    
The four-decade old fluid mosaic model \cite{singer72,vereb03,singer04} 
is the best representation of the cytoplasmic membrane of animal 
cells; the molecular details of the membrane is not fairly well 
understood \cite{luckey08,lee11b}.  

Gram-positive bacteria have a single plasma membrane that consists of a 
inner membrane (IM) and a thick outer layer, called cell wall. In contrast, 
Gram-negative bacteria are enclosed by two membranes where the inner 
membrane (IM) and the outer membranes (OM) are separated by the periplasmic 
space that also contains a peptidoglycan layer \cite{beveridge99,dmitriev05}. 
Based on experimental investigations, it is now generally believed that 
the first ancestor of all eukaryotes on earth appeared more than 1.5 
billion years ago when a bacterial invader cell became captive of the host 
and became the precursor of present-day mitochondria \cite{dyall04}. 
Therefore, it is not surprising that mitochondrial membrane shares many 
features of Gram-negative bacteria; the inner and outer membranes of 
each mitochondrion is separated by an intermembrane space \cite{herrmann10}. 
Because of its important biological functions, which include protein 
translocation across mitochondrial membranes, it remained an indispensable 
feature of mitochondria in spite of many other evolutionary changes in 
the evolution of mitochondria from its bacterial ancestor.

\noindent$\bullet${\bf Nuclear envelope and nuclear pore complex (NPC)} 

The nuclear pore complex (NPC) \cite{hoelz11,jamali11,tu11} 
is a large assembly of more than two dozens of different types of 
proteins that are collectively called {\it nucleoporins} and denoted 
by the abbreviation Nups. Its shape resembles the upper half of an 
hourglass. The rim of the channel is sandwiched between the cytoplasmic 
ring and the nuclear ring. This assembly has a {\it eight-fold} symmetry
about an axis normal to the plane of the membrane. On the cytoplasmic
side of the membrane, {\it eight fibrils} extend from the eight lobes 
which are arranged in the form of a ring. On the nucleoplasmic side of the 
membrane these eight fibers join to form a {\it basket-like} structure. 
The diameter of the channel is about 60-70 nm at its widest point and 
about 25-45 nm at its narrowest point. The outer radius of the ring can 
be as large as 125 nm and the total length of the NPC in the direction 
perpendicular to the nuclear membrane can vary between 150-200 nm. 
In addition to the main central channel, the pore can also support some 
minor peripheral channels.

\section{\bf Internal compartments of a cell}
\label{app-organelles}

\begin{table}
\begin{tabular}{|c|c|} \hline
Compartment  & Special functional role \\\hline
Nucleus & Repository of genetic material \\ \hline
Mitochondria \& Chloroplast & Power plant \\ \hline
Endoplasmic reticulum &  Packaging center\\ \hline
Golgi apparatus & Post office \\ \hline 
Peroxisome &  \\hline
\end{tabular}
\caption{Internal compartments of eukaryotic cells and their special 
functional roles.
 }
\label{table-compartments}
\end{table}

Diatoms, an unicellular eukaryote, is an example of a special class 
of organisms that possess both mitochondria and chloroplasts 
\cite{prihoda12}.
In recent years, organelle-like compartments have been discovered 
also in prokaryotic cells \cite{murat10}. However, no counterpart 
of nucleus has been found so far in prokaryotic cells (see 
table \ref{table-compartments} for internal compartments of eukaryotic 
cells).

\section{\bf Viruses, bacteriophages and plasmids: hijackers or poor parasites?}
\label{sec-virusphageplasmid}

The concept of virus (including bacteriophages) has an interesting 
history and its proper definition was a matter of debate even half 
a century ago \cite{dherelle26,lwoff57}. 
A virus is composed of a genome encapsulated within a capsid that is 
made of proteins \cite{black12}. 
Different types of viruses differ in size, shape, structure of the 
capsid and the spatial organization of the genome within it 
\cite{prasad12}. 
Some have an envelope made of lipids whereas others and not enveloped. 
In contrast to living cells, where the genome is exclusively DNA, 
viruses can also have RNA as their genetic material. 
In fact, their genome can be either single-stranded 
or double-stranded RNA or DNA. Majority of the viruses are, in fact, 
RNA viruses. Among the RNA viruses, there is a special class with 
single-stranded RNA genome that are called {\it retrovirus} 
\cite{buchschacher02}.

The retroviruses not only exploit the machineries of their hosts, but 
also {\it integrate} dsDNA, polymerized from their RNA genome templates, 
into the genome of the host \cite{whitcomb92}. 
Such DNA sequences integrated into the host genome gets passed onto 
the daughter cells during division of the virus-infected host. It may 
become ``endogenous viral sequence'' 
\cite{blikstad08,feschotte12,nelson03b}
in the future generations of the host. 
The entire life cycle of a virus \cite{suzuki07}, 
including the mechanism of packaging, and the 
spatial organization of the packaged genome also varies depending 
on the nature of the genetic material \cite{johnson10}. 
Normally, the viral genome encodes 
the ``structural'' proteins which are constituents of its capsid. 
Moreover, it also encodes for some ``non-structural'' proteins which 
are essential for the replication of its genome. However, a virus 
cannot replicate using only the machines at materials at its disposal. 
Besides, unlike eukaryotic and prokaryotic cells, a virus does not 
``divide'' into ``daughter viruses''. Instead, a virus enters a living 
host cell not only to replicate its genome but also to assemble many 
copies of itself by exploiting the machineries of the host cell 
\cite{rossmann12b}. 
The capsid not only protects the viral genome \cite{roos07} but also 
participates in the process whereby the genome infects the host cell.
Contrary to the naive expectation, not all viruses are harmful; 
some viruses develop a symbiosis with their host and are ``good'' 
virus for the host \cite{roossinck11}.

Plasmids are extrachromosomal DNA in bacterial cells \cite{thomas00}. 
These can be replicated just like chromosomal DNA and can be passed 
to the daughter cells during cell division.

All the eukaryotes, namely, animals, plants (and algae) as well as 
fungi can be invaded by their respective viruses. 
Human immunodeficiency virus (HIV), which causes the disease acquired 
immunodeficiency syndrome (AIDS) 
\cite{karpas04}, 
is a retrovirus and is the most dreaded among the viruses that can 
infect {\it homo sapiens} (humans). 
Among the viruses which can infect plants, an well studied example 
is the {\it Tobacco mosaic virus }.

Bacteriophages are also viruses, but these infect prokaryotes.
T-odd (e.g., T7) and T-even (e.g., T4) bacteriophages, phage $\lambda$,
$\phi29$, etc. are some of the extensively used model bacteriophages. 
Not all phages are ``tailed'' \cite{letellier04}; 
filamentous bacteriophages are also widespread \cite{hemminga09,rakonjac12}. 
Viruses which infect archaea \cite{prangishvili06} are called archeovirus 
\cite{pina11}. 

\subsection{Baltimore classification of viruses according to their genome}

The viruses have been divided into six different classes each of
which has a characteristic distinct method of expressing its
genetic information \cite{baltimore71}.
An mRNA strand that can be translated directly into a protein is called
a (+)mRNA strand. A ssDNA that has the same polarity as the (+)mRNA is
called (+)DNA.But, a (+)DNA cannot be transcribed directly into a (+)mRNA.
Instead, a complementary DNA strand, called (-)DNA, is directly transcribed
into a (+)mRNA. Similarly, on direct replication, a (+)RNA produces a
complementary strand that is called a (-)RNA.

$${\rm Class~ I}: ~~{\framebox{dsDNA}} ~~~\mathop{\rightarrow} ~~~{\framebox{(+)mRNA}}$$
$${\rm Class~ II}: ~~{\framebox{ssDNA}} ~~~\mathop{\rightarrow} ~~~{\framebox{dsDNA}}~~~\mathop{\rightarrow} ~~~{\framebox{(+)mRNA}}$$
$${\rm Class~ III}: ~~{\framebox{($\pm$)RNA}} ~~~\mathop{\rightarrow} ~~~{\framebox{(+)mRNA}}$$
$${\rm Class~ IV}: ~~{\framebox{(+)RNA}} ~~~\mathop{\rightarrow} ~~~{\framebox{(-)RNA}}~~~\mathop{\rightarrow} ~~~{\framebox{(+)mRNA}}$$
$${\rm Class~ V}: ~~{\framebox{(-)RNA}} ~~~\mathop{\rightarrow} ~~~{\framebox{(+)mRNA}}$$
$${\rm Class~ VI}: ~~{\framebox{(+)RNA}} ~~~\mathop{\rightarrow}~~~{\framebox{ssDNA}} ~~~\mathop{\rightarrow} ~~~{\framebox{dsDNA}}~~~\mathop{\rightarrow} ~~~{\framebox{(+)mRNA}}$$
Similarities and differences between different classes of viruses
have been studied extensively \cite{ahlquist06,boon10}.

\section{\bf Organization of packaged genome: from virus and prokaryotes to eukaryotes}

The diverse ways in which the genome is organized in different systems 
\cite{budd12b}
shows the extraordinary richness of life.
In every cell the genetic information is encoded in the sequence of
the nucleotides. Thus, at some stage of biological evolution, Nature
chose an effectively {\it linear} device (namely, a NA strand) and a
{\it quaternary} code (i.e., four symbols, namely, A, T, C, G) for
storing genetic information. This was not the most efficient choice!
The fewer is the number of letters of the alphabet the longer is the
string of letters required to express a given message. 
Why does nature use exact 4 letters to write the genetic message on 
nucleic acids? Why does nature use 20 amino acids for making proteins? 
Are these numbers results of Darwinian evolution which initially could 
have been different?

One serious
consequence of nature's choice of the memory device and coding system
is that even for the most primitive organisms like an {\it E.coli}
bacterium, the total length of the DNA molecule is orders of magnitude
longer than the organism itself! The problem is more acute in case of
eukaryotic cells where an even longer DNA has to be accomodated within
a tiny nucleus! Moreover, random packaging of the DNA into the nucleus
would not be desirable because, for wide variety of biological processes
involving DNA, specific segments of the DNA molecules must be ``unpacked''
and made accessible to the corresponding cellular machineries.
Furthermore, at the end of the operation, the DNA must be re-packed.
Even in bacteria and viral capsids, the genome has to be packaged in
a manner which allows efficient access during various processes of
DNA and RNA metabolism. 

For detailed accounts of chromosome organization and function see, for 
example, refs.\cite{wolffe00,sumner03,olins03}.
A brief summary, appropriate as an introductory reading for uninitiated 
physicists is available in ref.\cite{budd12b}

As stated earlier, the viral genomes may consist of DNA or RNA. There
are two alternative mechanisms for packaging of the genome. In case
of some viruses, the genome is encapsulated by molecules that
self-assemble around it. In contrast, the genome of other viruses are
packaged into a pre-fabricated empty container, called {\it viral
capsid}, by a powerful motor.

Nature has solved the problem of packaging genetic materials in the
nucleus of eukaryotic cells by organizing the DNA strands in a
hierarchical manner and the final packaged product is usually referred
to as the {\it chromatin} \cite{lall07,travers10}
The primary repeating unit of chromatin at the lowest level of the
hierarchical structure is a nucleosome \cite{kornberg99b}. 
The cylindrically shaped core of each nucleosome consists of an octamer
of histone proteins around which 146 base pairs (i.e.,$\sim 50$ nm) of the
the double stranded DNA is wrapped about two turns (more precisely, 1.7
helical turns); the arrangement is reminiscent of wrapping of a thread
around a spool. There are $14$ equispaced sites, at intervals of $10$
base pairs (bp), on the surface of the cylindrical spool. Electrostatic
attraction between these binding sites on the histone spool and the
oppositely charged DNA seems to dominate the histone-DNA interactions
which stabilize the nucleosomes. The helical curve formed by the 
histone-DNA overlap is often referred to as the ``footprint'' of the DNA.

\section{\bf Experimental methods: introduction to the working principles}
\label{app-expttech}

Experiments play the most important roles in all natural sciences.
Throughout this review we use the term ''in-vitro'' to mean processes
occuring outside living cells whereas the term ``in-vivo'' is used
exclusively for processes occuring inside living cells. Naturally, more
controlled experiments are possible in-vitro than in-vivo. In analogy
with in-vitro and in-vivo experiments, computer simulation is often
referred to as in-silico experiments.

\subsection{\bf FRET: tool for monitoring conformational kinetics}

Fluorescence is the phenomenon in which a molecule gets excited when
illuminated with light of a specific wavelength and, then de-excites
by emitting light whose wavelength is usually longer than that of the
exciting light. A molecule that is capable of exhibiting the phenomenon
of fluorescence is called a {\it fluorophore}. Since the wavelength
of the emitted light is somewhat longer than that of the light absorbed,
the background can be darkened by rejecting the exciting light using
appropriate filters. In reality, both the excitation and emission
spectra of a molecule have characteristic shapes with a peak and a
non-zero width. Therefore, the positions of the peaks of the absorption
and emission spectra are identified as the corresponding characteristic
frequencies; the larger the difference in these two frequencies the
easier it is to record the emission from a fluorophore by filtering out
the exciting light.

In general, the desirable properties of fluorophores are as follows: \\
(i) it should be bright, \\
(ii) it should, preferably, emit light in the visible region of the spectrum,\\
(iii) fluctuations in its emission intensity, during the experiment, should be
as small as possible;\\
(iv) is should be sufficiently small and its interaction with the molecule 
under investigation should be weak so that it does not perturb the molecule 
under investigation,\\
(v) it should be available in a form which is suitable for ``attachment''
with the molecule under investigation. 

Extrinsic fluorescence reporters are usually organic dyes or quantum dots.
Intrinsic reporters are genetically engineered in the cell so that a 
fluorescent molecule forms a part of the protein under investigation; 
the green fluorescent protein (GFP) and its related cousins with other 
colors are now used routinely \cite{chalfie09}.

\begin{figure}[htbp]
\begin{center}
\includegraphics[angle=-90,width=0.65\columnwidth]{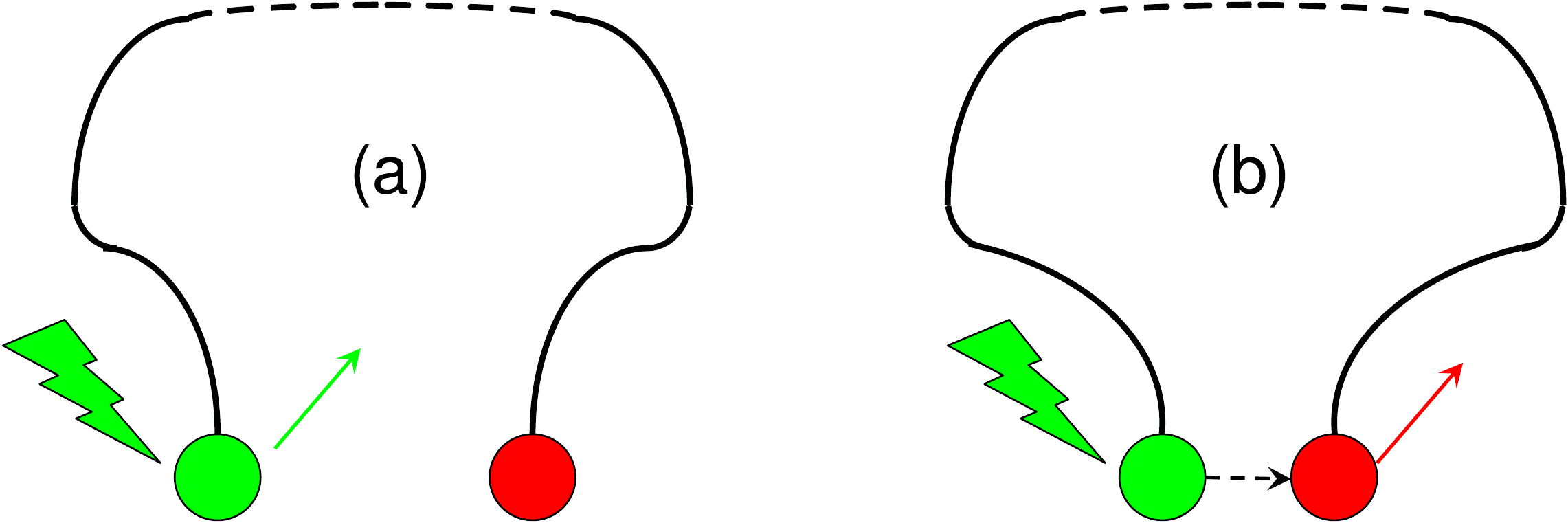}
\end{center}
\caption{Principle of FRET (see the text for the details).
  }
\label{fig-FRET}
\end{figure}

FRET (Fluorescence / F\"orster Resonance Energy Transfer) \cite{clegg06}
is a technique based on a quantum-mechanical phenomenon. It requires two
different species of fluorophores called a {\it donor} and an {\it acceptor}.
The donor absorbs laser light at a frequency higher than that of the
acceptor. Because of the electromagnetic interactions between the two,
non-radiative transfer of energy can take place from the donor to the
acceptor if the acceptor is sufficiently close to the donor. Since the
efficiency of the process falls with the sixth power of the separation
between the two fluorophores, it is significant only when the separation
is of the order of $\sim$ O(nm). Such a resonant transfer of excitation
energy of the acceptor to the donor excites fluorescence of the acceptor,
resulting in a decrease in the fluorescence of the donor. This is a
powerful technique for probing the separation between either two different
molecules or two different domains of a single macromolecule; the two
molecules or the two domains of interest are labelled by the two fluorophores
for this investigation (see fig.\ref{fig-FRET}).

\subsection{\bf Optical microscopy: diffraction-limited and beyond}

In the fluorescence-based techniques discussed above the kinetics of
the molecules are inferred from the pattern of temporal variation of
the intensity of the fluorescence. 
However, seeing is believing. Telescopes opened up the celestial world in front
of our eyes. Microscopes aid our vision to ``see'' the microscopic world
of both living and nonliving systems. Naturally, the most direct
approach to ``see'' the molecular machines in its natural environment
is to use laser based optical microscopy \cite{imhof03}.
Spectacular progress in optical microscopy over the last two decades
\cite{schmolze11}
has been possible because of important contributions from several
different disciplines; these include, for example, physics (principles
of optics), chemistry (synthesis of dyes), genetics (tagging molecules
with markers) and engineering (instrumentation and signal processing).
However, it is worth uttering a word of caution here: one has to be
very careful so as to avoid potential pitfalls in this apparently
straightforward approach \cite{north06}.

\centerline{\framebox{Optical microscopy}}

\centerline{$\swarrow$ ~~ $\searrow$}

\centerline{{\framebox{Diffraction-limited techniques}} ~~{\framebox{Sub-diffraction techniques}}}

\subsubsection{\bf Diffraction-limited microscopy}

The {\it contrast} between an object and its background can be enhanced 
by exploiting the variations in the refractive indices induced, for 
example, by appropriate {\it staining}; these techniques include the 
phase-contrast microscopy. Fluorescence provides an even better means of 
enhancing the contrast between the object and its background  
\cite{lichtman05,vonesh06,schliwa02,wang08b}. 
Although fluorescence microscopy is now done routinely, a beginner 
should be aware of the potential pitfalls \cite{brown07b}.

{\it Magnification} makes the observed object bigger, but does not improve 
spatial {\it resolution} which, for conventional optical microscope is 
limited by a natural constraint. 
Because of diffraction of light, which is a consequence of its wave nature, 
image of a point source of light is not a point, but a {\it volume}. The 
corresponding plot of the intensity profile in the image plane is called 
the {\it point spread function} (PSF). Because of the rotational symmetry 
in the image plane perpendicular to the optic axis, the {\it area} in the 
image plane illuminated by a point source is a circular pattern. The bright 
circular central patch, called the ``{\it Airy disc}'', is surrounded 
concentrically by increasingly darker circles.

According to Abbe's theory, the {\it spatial resolution} $r_m$ of
a microscope is half the radius of the first dark circular fringe.
Hence, $r_m = \lambda/2$(NA) where the numerical aperture (NA)
depends on the refractive index $n$ of the medium and the half
angular aperture $\alpha$: NA = n ~sin~$\alpha$. The largest value
of NA for a good quality microscope $\sim 1.5$ and, therefore,
using $\lambda = 500$nm, we get $r_m \sim 170 nm$. 

If the point objects are sufficiently far apart from one another, each 
forms its own distinct PSF without any overlap with that of another.
One can {\it locate} the center of the PSF by fitting, e.g., a Gaussian.
This center is expected to coincide with the position of the point object
that creates this PSF.

Interestingly, Abbe himself speculated in 1876 that ``future generations
will perhaps find other ways to circumvent the limits imposed by light
microscopy, which we consider unsurmountable, by making still unknown
processes and forces serve this purpose'' \cite{schliwa02}.

Recall that one of the assumptions made in the derivation of the Abbe
limit is that the two point sources emit light of {\it identical
spectral characteristics}. But, if the two sources emit lights of
different colors, they could be easily resolved by using appropriate
filters to reject light from one while localizing the other \cite{hell09}.
However,  success of this route to super-resolution hinges on the ability
to label different objects with different types of fluorophores which
would emit lights of different colors. For most real problems this is
an impractical approach.

\subsubsection{\bf Sub-difraction microscopy (or, super-resolution nanoscopy)}

The invention of the optical microscopes in the seventeenth century made
it possible to have a glimpse of the world of micro-organisms (bacteria,
etc.) \cite{zewail10}. But these microbes are typically micron-size
objects. For directly seeing nano-machines, it would be ideal if we
could have ``nanoscopes'' whose resolution surpasses the Abbe limit; 
however, such ``nanoscopes'' were not available until recent times 
\cite{garini05,pohl04}!

How do the sub-diffraction techniques developed in the last two
decades achieve ``{\it super-resolution}'' (i.e., spatial resolution
higher than the Abbe limit)
\cite{hell04,hell09,huang09,huang10,schermelleh10}?
We answer this question below by explaining the principles.

Recall that the derivation of the Abbe limit assumed that the two objects 
emitted light {\it simultaneously} leading to the overlap of the two PSFs. 
But, if only one of these two were fluorescent at a time, each could be 
{\it localized} separately and the full image of the pair could be 
constructed by superimposing their individual images recorded separately 
at two different instants of time.

Both the objects can be imaged separately if these are labelled by
fluorophores that are {\it photo-switchable}, i.e., switched ON
(fluorescent) and OFF (non-fluorescent) with appropriately selected
laser beams in a sequential manner \cite{hell09}.
Most of the nanoscopes achieve super-resolution exploiting essentially
this simple idea although the details of implementation vary from one
technique to another. 
Stimulated emission depletion (STED) microscopy is based targeted 
switching whereas stochastic switching is exploited in single-molecule 
localization microscopy (SMLM) \cite{mcevoy10}, photo-activated 
localization microscopy (PALM), stochastic optical reconstruction 
microscopy (STORM), etc.

\subsection{\bf Single-molecule imaging and single-molecule manipulation}

Broadly speaking, the single-molecule techniques
\cite{zlatanova06,ritort06,cornish07,deniz08,kapanidis09}
can be classified into two groups: (i) methods of imaging, and
(ii) methods of manipulation.

\centerline{\framebox{Single-molecule techniques}}

\centerline{$\swarrow$ ~~ $\searrow$}

\centerline{~~~~{\framebox{Single-molecule imaging}} ~$\longleftrightarrow$~ {\framebox{Single-molecule manipulation}}}

However, combining techniques of
manipulation with those of imaging makes it possible to achieve both
simultaneously \cite{weitzman03,wallace03,lang03a}.

Fluorescence, which we have already introduced above, has
been exploited extensively in imaging single molecules. 
\cite{moerner02,moerner03,peterman04,moerner07,lord10,michalet02,michalet03,hohlbein10}.
In addition to passive observation under a microscope, spectroscopic
analysis of the amplitude, frequency, polarization of the light emitted
by the single molecule and their variations in space and time provide
useful informations.

For observing
effects of chemical manipulations of molecular machines, imaging may
be adequate.
Several techniques have been developed for for mechanical manipulations of the nanomachines of life
\cite{strick01,bustamante00a,merkel01,knight01,greenleaf07,walter08,neuman08}.

\centerline{\framebox{Mechanical manipulation of a single biomolecule}}

\centerline{$\swarrow$ ~~ $\searrow$}

\centerline{{\framebox{Mechanical transducers}} ~~~~~~~ {\framebox{Field-based transducers}}}

\centerline{$\swarrow$ ~~ $\searrow$ ~~~~~~~~~~~~~~~~~~~~~~~~~~ $\swarrow$ ~~ $\searrow$ }

\centerline{~~~~~{\framebox{SFM}}~ {\framebox{Micro-needle}}~~~~~{\framebox{EM-field}}~ {\framebox{Flow-field}}   }
\centerline{~~~~~~~~~~~~~~~~~~~~~~~ $\swarrow$ ~~ $\searrow$ }

\centerline{~~~~~~~~~~~~~~~~~~~~~~~~~{\framebox{ Electric field}}~ {\framebox{Magnetic field}}   }


The existence of the {\it ``gradient''} force, which is exploited in setting 
up an optical trap, was discovered accidentally \cite{ashkin97,ashkin00}.
Consider a {\it dielectric} bead exposed to a laser light beam whose
intensity is spatially inhomogeneous. The physical origin of the
gradient force can be understood easily within the framework of
geometrical optics provided $R \gg \lambda$; however, the existence of 
the gradient force however, does not depend on the validity of this condition. 

Mathematical
expression for the gradient force has been derived \cite{ashkin92,neto00}.
For a guide to the scientific literature on optical tweezer (up to 2003)
see ref.\cite{lang03b}. Several elementary introductions to the basic
principles of optical tweezers and their design are available; we mention
here only a few of those which are presented in the context of molecular
machines \cite{appleyard07,rocha09,molloy02,grier03,gross03b,hormeno06,moffitt08}.

In magnetic tweezers \cite{hosu03,kim09}, the macromolecule is attached
between a surface and a {\it superparamagnetic} bead. Stretching force
can be applied on the macromolecule by controlled alterations of the
external magnetic field. A major advantage of the magnetic tweezer is
that the same set up can be used also to apply {\it torque} on the
molecule by merely rotating the magnetic field.

\subsection{\bf Determination of structure: X-ray crystallography and electrom microscopy} 

X-ray diffraction \cite{giege10}  and electron microscopy \cite{frank96}
are two of the most powerful techniques of determination of the structures 
of molecular machines. Both these techniques provide ensemble-averaged results.
Each sample of the molecular machine, when analyzed with X-ray diffraction 
or electron microscopy, provides essentially a ``static'' picture.
But, informations on time evolutions of structures can also be obtained 
from X-ray and electron microscopic studies by appropriate protocols 
for sample preparation and repetitions of the experiments.

\noindent$\bullet${\bf X-ray crystallography}

The basic principle of X-ray scattering for the determination of the
structure of macromolecules is as follows \cite{rhodes06}:
an atomic constituent of the macromolecule absorbs some energy of the
X-ray incident on it and then re-radiates the same in all directions.
A protein crystal has a periodic array of identical atoms. The X-rays
re-radiated by these atoms interfere constructively in some directions
whereas they interfere destructively in all the other directions.
Therefore, the detectors record a ``pattern'' in the intensity of X-ray
scattered by the protein crystal sample. But, such a  ``diffraction
pattern'' provides an indirect, and static, image of a molecular machine.
X-ray diffraction requires Fourier transform from momentum space to
get the structure in real space.

\noindent$\bullet${\bf Electron microscopy}

Electron microscopy is a powerful alternative for the determination of
structures of those macromolecules whose crystals are not available
\cite{frank96}.
The deBroglie wavelength associated with a material particle is given by
$\lambda = h/p$ where $p$ is the momentum of the particle. Therefore, a
desired short deBroglie wavelength can be attained by accelerating a
charged particle to the corresponding required momentum $p$ by applying
an external electric field. Electrons are ideal for this purpose because
an electron beam can also be easily bent and focussed using a suitable
magnetic field configuration. But, the generation and control of the
electro-magnetic fields makes the electron microscope costly. Moreover,
image obtained from an electron microscope requires special expertise to
interpret. Often, better results are obtained by a combination of X-ray
crystallography and electron microscopy \cite{rossmann05}.

Unlike optical microscopes, elaborate preparation of a sample is required
before observing under an electron microscope. Keeping the sample hydrated
(i.e., ``wet'') was a challenge which could be overcome by a technical
breakthrough. In this approach, the aqueous sample is cooled very rapidly
by plunging it into an appropriate liquid maintained at a sufficiently
low temperature. Because of the high rate of cooling, the water surrounding
the specimen does not get any opportunity to form ice crystal and, instead,
gets ``vitrified'' (i.e., amorphous). Moreover, because of the low
temperature of the vitrified water, nucleation of ice crystals in this
metastable medium is highly improbable. Because of the low-temperature
techniques used in sample preparation, this technique is called
{\it cryo-electron microscopy (cryo-EM)} \cite{glaeser08}.
The process of reconstruction of the full 3D structure of an object by
appropriately combining a series of 2D images recorded from different
angles is called cryo-EM {\it tomography}.
Modern electron microscopy has already provided deep insight
into the structure and function of many molecular machines
\cite{koster03,lucic05,jensen07,nogales01,chiu05,chiu06,hoenger07}.

\section{\bf Modeling of chemical reactions} 
\label{app-chemreaction}

\subsection{\bf Deterministic non-spatial models of chemical reactions: rate equations for bulk systems}

At this level, the problem of chemical kinetics can be formulated as follows:
suppose a macroscopically uniform mixture of $S$ chemical species is
confined in a fixed volume $V$ and can interact through $R$ reaction
channels. The main {\it assumption} of the rate equation approach is
that the population dynamics, formulated in terms of the {\it
concentrations} of the reactant and product species, is a {\it continuous}
and {\it deterministic} process. For a well stirred chemically reacting
bulk system, this is a reasonably good approximation. If the concentrations
of all the species are given at some initial instant of time, what will be
the corresponding concentrations at any later arbitrary instant of time
$t$? The traditional approach is based on ordinary differential equations,
called chemical reaction rate equations, for the concentrations of the
molecular species.

In chemical kinetics, the concentration of the $s$-th molecular species
is given by the molarity $c_{s} = n_{s}/V$, measured in the number of
moles per liter, where $n_{s}$ is the number of moles of the s-th molecular
species. Another related quantity is the molar fraction
$x_s = n_{s}/\sum_s n_{s} = n_{s}/n = c_{s}/\sum_s c_{s}$.
Thus, if $c_{s}$ of all the components in the solution are known, $x_{s}$
can be obtained. Conversely, if the $x_{s}$ are known, $c_{s}$ can be
obtained from $c_{s} = x_{s} n/V$.

Let us assume that the concentration of the $s$-th molecular species
is denoted by a continuous, single-valued function $c_s(t)$ of
time $t$. Then, the corresponding chemical rate equations can be
expressed as \cite{metiu06book,beard08book,phair01}
\begin{eqnarray}
\frac{dc_s}{dt} = f_s(c_1,c_2,...,c_s,...,c_S) ~(s=1,2,...,S) \nonumber \\
\end{eqnarray}
The specific forms of the functions $f_s$ are determined by the
actual nature of the reactions. For example, for the simple
{\it first order irreversible} reaction
\begin{equation}
{\cal E}_{1} \mathop{\rightarrow}^{k_f} {\cal E}_{2}
\label{eq-simplest}
\end{equation}
the ordinary differential equations governing the populations of
the two molecular species ${\cal E}_{1}$ and ${\cal E}_{2}$ are
\begin{equation}
d[{\cal E}_{1}]/dt = - d[{\cal E}]_{1}/dt = - k_{f} [{\cal E}_{1}]
\end{equation}
which are usually referred to as the rate equations and the square
brackets indicate the respective concentrations.
For the {\it first order reversible} reaction
\begin{equation}
{\cal E}_{1} \mathop{\rightleftharpoons}^{k_f}_{k_r} {\cal E}_{2}
\label{eq-simplerev}
\end{equation}
the corresponding reaction rate equations are
\begin{eqnarray}
d[{\cal E}_{1}]/dt &=& k_{r} [{\cal E}_{2}] - k_{f} [{\cal E}_{1}] \nonumber \\
d[{\cal E}_{2}]/dt &=& k_{f} [{\cal E}_{1}] - k_{r} [{\cal E}_{2}] \nonumber \\
\label{eq-rate12}
\end{eqnarray}
The coefficients $k_{f}$ and $k_{r}$ depend, in general, on the
temperature $T$ and pressure $p$, etc., but are independent of
the concentrations of the reactants and the products. At the
level of the chemical rate equations, the rate constants are
phenomenological parameters whose numerical values are to be
supplied from empirical data.
Rate constants usually depend strongly on temperature. An overwhelmingly
large number of rate constants are found to vary with temperature
according to the Arrhenius equation:
\begin{equation}
k(T) =  A~ exp[B/k_BT].
\label{eq-arrhenius}
\end{equation}
where $B$ is called the activation energy (barrier height).

Note that the rate equations (\ref{eq-rate12}) are linear. On the other
hand, for the {\it second order} reaction
\begin{equation}
{\cal E}_{1} + {\cal E}_{2} \mathop{\rightleftharpoons}^{k_f}_{k_r} {\cal E}_{3} 
\label{eq-2ndorder}
\end{equation}
the rate equations
\begin{eqnarray}
\frac{d[{\cal E}_{1}]}{dt} &=& \frac{d[{\cal E}_{2}]}{dt} = k_{r} [{\cal E}_{3}] - k_{f} [{\cal E}_{1}] [{\cal E}_{2}] \nonumber \\
\frac{d[{\cal E}_{3}]}{dt} &=& k_{f} [{\cal E}_{1}][{\cal E}_{2}] - k_{r} [{\cal E}_{3}]
\end{eqnarray}
are nonlinear.

\subsubsection{\bf Thermodynamic equilibrium, transient kinetics and non-equilibrium steady states}

With two examples of very simple reactions we introduce the concepts
of {\it transient} and {\it equilibrium} behavior in the context of
chemical reactions. We also demonstrate how {\it transient} kinetics
can be utilized to estimate the rate constants \cite{fierke95}.
Let us begin with the reaction (\ref{eq-simplest}). The concentration
of $[{\cal E}_{1}](t)$ varies with time $t$ exponentially:
\begin{equation}
[{\cal E}_{1}](t) = [{\cal E}_{1}](0) ~e^{-k_{f} t}.
\label{eq-expdec1}
\end{equation}
The slope of the plot of $ln [{\cal E}_{1}](t)$ versus $t$ yields
the rate constant $k_{f}$.

\noindent$\bullet${\bf Equilibrium constant}

Next consider the reaction (\ref{eq-simplerev}). Any arbitrary initial
concentrations $[{\cal E}_{1}]$ and $[{\cal E}_{2}]$ eventually reach
the corresponding equilibrium values
\begin{eqnarray}
[{\cal E}_{1}]^{eq} = \frac{k_r[{\cal E}]}{(k_f+k_r)}~~{\rm and}~~ 
[{\cal E}_{2}]^{eq} = \frac{k_f[{\cal E}]}{(k_f+k_r)}
\label{eq-eqconc}
\end{eqnarray}
In the equilibrium state, the {\it equilibrium constant} $K_{eq}$,
defined as
\begin{equation}
K_{eq} = \frac{k_{f}}{k_{r}} = \dfrac{[{\cal E}_{2}]^{eq}}{[{\cal E}_{1}]^{eq}}, 
\label{eq-kequilibrium}
\end{equation}
Note that
\begin{equation}
K_{eq} = e^{-\Delta G/(k_BT)}.
\end{equation}
Thus, the equilibrium constant is a thermodynamic parameter that
characterizes the {\it equilibrium} state of the reacting system.
For a given reaction, the equilibrium constant $K_{eq}$ can be
obtained by measuring the concentrations of the reactants and
products in equilibrium. The assay need not measure the concentrations
chemically; a common alternative is an optical assay where the
fluorescence intensities must be proportional to the respective
concentrations. In fact, any property, that is proportional to
the concentration, can be measured for estimating $K_{eq}$.
In order to avoid possible pitfalls, one has to design the
careful experiment and use appropriate protocols \cite{pollard10}.

\noindent$\bullet${\bf Transient kinetics}

Note that the fact
$d[{\cal E}_{1}]/dt + d[{\cal E}_{2}]/dt = 0$
in (\ref{eq-rate12}) reflects the conservation law:
\begin{equation}
[{\cal E}_{1}](t) + [{\cal E}_{2}](t) = [{\cal E}](0)
\end{equation}
where the total concentration $[{\cal E}](0)$ remains constant.
From equations (\ref{eq-rate12}) we get 
\begin{equation}
ln\biggl(\frac{\{[{\cal E}_{1}](t)-[{\cal E}_{1}^{eq}]\}}{\{[{\cal E}_{0}](t)-[{\cal E}_{1}^{eq}]\}}\biggr) = - k_{eff} t
\label{eq-keff}
\end{equation}
with $k_{eff} = k_f + k_r$. Thus, any fluctuation in the populations
of the two species decays exponentially with time with an effective
time constant $k_{eff}^{-1}$ to reach the thermodynamic equilibrium
values (\ref{eq-eqconc}). For a reacting system which can attain a
state of thermodynamic equilibrium, small deviations from such the
equilibrium state are merely {\it transients}. However, kinetics of
the decay of such transient states can be utilized to measure the
rate constants for both the forward and reverse transitions. In this
approach, one simply changes the condition of equilibrium of the
system and, then, monitors the time-dependent concentrations during
the process of re-equilibration. The slope of the plot of
$ln ([{\cal E}_{1}](t)-[{\cal E}_{1}^{eq}])$ yields the sum $k_f+k_r$.
Moreover, the $k_f/k_r$ can be obtained from (\ref{eq-kequilibrium})
provided the equilibrium concentrations $[{\cal E}_{1}]^{eq}$ and
$[{\cal E}_{2}]^{eq}$ are known.

\noindent$\bullet${\bf Non-equilibrium steady-states of chemically reacting systems}

The chemical system where the reaction (\ref{eq-simplerev}) takes place
exhibits transient behavior, described by the equation (\ref{eq-keff}),
till the system reaches the equilibrium state where the concentrations
attain the corresponding time-independent values (\ref{eq-eqconc}).
Thus, in the equilibrium state the concentrations of the reactants and
products remain steady (or, stationary). But, not every chemical steady
states is an equilibrium state of the system; equilibrium happens to be
just a special steady-state. Nonequilibrium steady states (NESS) can
exist only in open systems where the system exchanges matter with its
environment and consumes energy dissipating part of it as heat
\cite{hillbook,wyman75,qian05,qian06,qian07}.
Biochemical reactions within a living cell are typical examples where
such NESS can occur.

As an example, consider the reaction
\begin{equation}
\mathop{\rightarrow}^{J} {\cal E}_{1} \mathop{\rightleftharpoons}^{k_f}_{k_r} {\cal E}_{2} \mathop{\rightarrow}^{J}
\label{eq-simplest2}
\end{equation}
with a balanced input and output $J$. In this case the equations
(\ref{eq-rate12}) are modified to
\begin{eqnarray}
d[{\cal E}_{1}]/dt &=& k_{r} [{\cal E}_{2}] - k_{f} [{\cal E}_{1}] +J \nonumber \\
d[{\cal E}_{2}]/dt &=& k_{f} [{\cal E}_{1}] - k_{r} [{\cal E}_{2}] -J
\label{eq-rate12ne}
\end{eqnarray}
so that the system reaches a {\it non-equilibrium steady state}
where
\begin{eqnarray}
[{\cal E}_{1}]_{ss} = \frac{k_{r}[{\cal E}]+J}{(k_{f}+k_{r})}, ~~{\rm and}~~
[{\cal E}_{2}]_{ss} = \frac{k_{f}[{\cal E}]-J}{(k_{f}+k_{r})}
\label{eq-neeqconc}
\end{eqnarray}
Moreover, defining the forward and reverse fluxes $J_{f}$ and $J_{r}$
by the relations
\begin{equation}
J_{f} = k_{f} [{\cal E}_{1}], {\rm and} J_{f} = k_{r} [{\cal E}_{2}]
\end{equation}
$J = J_{f} - J_{r}$ in the non-equilibrium steady state of the system
whereas thermodynamic equilibrium demands that $J = 0$, i.e.,
$J_{f} = J_{r}$.

\noindent$\bullet${\bf Stoichiometry and reaction rates}

Next, let us consider more general complex reactions of the type
\begin{equation}
2A + 3B \rightarrow C + 4D
\label{eq-reac1}
\end{equation}
where A and B are the reactants while C and D are the products.
The stoichiometry of this reaction is such that the rate of
formation of one molecule of C (and that of four molecules of D)
is equal to the rate of consumption of two molecules of A (or,
that of 3 molecules of B). In general, for a reaction like
(\ref{eq-reac1}) the {\it stoichiometric coefficients}
$\nu_s > 0 (< 0)$ if $s$-th species is a product (reactant).
Thus for the reaction (\ref{eq-reac1}), the stoichiometric
coefficients of A,B,C,D are the integers $-2, -3, 1, 4$, respectively.

Often a more general notation is used for expressing the rate
equations. Suppose ${\vec c}(t)$ denotes is a column vector
whose three elements are the concentrations of ${\cal E}_{1}$,
${\cal E}_{2}$ and ${\cal E}_{3}$ at time $t$ for the reaction
(\ref{eq-2ndorder}). Then, the rate equations can be recast as
\begin{equation}
\frac{dc_i}{dt} = \nu_{fi} \tilde{\cal F}_{f}({\vec c}(t)) + \nu_{fr} \tilde{\cal F}_{r}({\vec c}(t))
\end{equation}
where $\nu_{f1} = \nu_{f2} = -1$, $\nu_{f3} = 1$ and the functions
$\tilde{\cal F}_{f}({\vec c}(t))$ and $\tilde{\cal F}_{r}({\vec c}(t))$
for the forward and reverse reactions are given by
\begin{eqnarray}
\tilde{\cal F}_{f}({\vec c}(t)) &=& k_{f} [{\cal E}_{1}][{\cal E}_{2}]\nonumber \\
\tilde{\cal F}_{r}({\vec c}(t)) &=& k_{r} [{\cal E}_{3}]
\end{eqnarray}

We define the {\it extent of the reaction} $\xi$ as follows:
when the chemical reaction advances by $d\xi$, the corresponding
changes in the amounts of the reactants and the products in the
above mentioned reaction are given by
\begin{equation}
dN_A = - 2 d\xi, dN_B = - 3 d\xi, dN_C = d\xi, dN_D = 4 d\xi
\end{equation}
i.e.,
\begin{equation}
\frac{dN_A(t)}{-2} = \frac{dN_B(t)}{-3} = \frac{dN_C(t)}{1} = \frac{dN_D(t)}{4} = d\xi(t)
\end{equation}
Any possible ambiguity in the definition of the rate of the reaction
is avoided by defining the reaction rate to be
$d\xi/dt = - (1/\nu_s) d[M_s]/dt$, so that, in general,
$dN_s = \nu_s d\xi$.

If the reaction (\ref{eq-reac1}) takes place in the opposite
direction, i.e.,
all the stoichiometric coefficients also reverse their sign and,
therefore, $d\xi$ flips its sign. Thus, the sign of $d\xi$
indicates whether the reaction is proceeding in the forward or
the reverse direction; $d\xi > 0$ for the forward reaction
whereas $d\xi < 0$ for the reverse reaction.

The actual extent of a reaction depends on the amount of substance
used in the reaction. A better definition of the rate of reaction
is $\eta(t) = \xi(t)/V$ where $V$ is the volume of the reaction
chamber. This definition allows one to associate a single rate with
the entire equation corresponding to a reaction. All practical
problems of chemical kinetics can be reduced to finding how $\eta(t)$
changes with time $t$. From $\eta(t)$ one can calculate the time
evolution of the concentrations of each chemical species involved
in the reaction \cite{metiu06book}.

\subsection{\bf Stochastic non-spatial models of reaction kinetics}

The rate equations conceal a great deal of detailed physical processes
involved in the reaction. In reality, the time evolution of the
populations cannot be {\it continuous} because the number of molecules
can change only by discrete integers. Moreover, the evolution is not
{\it deterministic} because it is impossible to predict the exact
molecular populations at an arbitrary time unless the positions and
velocities of all the molecules in the system, including those in the
solvent (i.e., reservoir or bath) are taken into account. Furthermore,
the smaller is the population of a reacting species, the stronger are
the fluctuations that makes a stochastic description unavoidable.
We now develop a theoretical formalism for stochastic chemical kinetics
that describes the {\it population} dynamics of the reacting species
as a {\it discrete}, {\it stochastic} process that is assumed to evolve
in a continuous time.

To our knowledge, one of the earliest studies of stochastic fluctuations
in chemical reactions was carried out by Max Delbr\"uck \cite{delbruck40}.
The literature on stochastic modeling of (bio-)chemical reactions,
including enzyme kinetics, is too vast to be covered in this appendix; 
only some representative original works of the successive decades
\cite{singer53,bartholamay62,smith71,aranyi77,gillespie77,hasstedt78}.
and a few useful reviews
\cite{mcquarrie67,erdi89,turner04,gillespie05,gillespie07,gadgil08,sun08,ullah10,qian12}
are listed in the references.
Authors of most of these works were fully aware of the fact that
their results on the fluctuations in chemical kinetics would be very
relevant in the limit of extremely low concentration of at least one
of the reactants. However, their ideas were far ahead of their time!
It took a few decades to develop the single molecule techniques
with which, for example, single-enzyme experiments are now carried
out routinely. Some of the old results can now be tested experimentally
while some of the subtle issues raised earlier may get resolved from
a modern perspective.

\subsubsection{\bf Chemical master equation}

Consider $S$ chemical species $(M_1, M_2, ..., M_s, ... M_S)$,
interacting through $R$ reaction channels. Let
$n_s(t) =$ Number of molecules of the s-th species at time $t$.
Our goal is to obtain the state vector
$\vec{n}(t) = (n_1(t), n_2(t),...,n_s(t),..., n_S(t))$, given the
state vector $\vec{n}(0) = (n_1(0), n_2(0),...,n_s(0), ..., n_S(0))$,
at time $t = 0$.

The stoichiometric coefficients form a matrix whose elements $\nu_{rs}$ 
is the stoichiometric coefficient for the s-th species in the r-th 
reaction. So, if the r-th reaction takes place,
the state vector $\vec{n}$ changes to $\vec{n} + \vec{\nu}_r$
where $\vec{\nu}_r = (\nu_{r1},\nu_{r2},...,\nu_{rs},...\nu_{rS})$.
We also define the {\it propensity function} (transition probabilities 
per unit time) $W_{r}(\vec{n})$ so that $W_{r}(\vec{n}) \Delta t$
is the probability that the r-th reaction takes place in time interval
between $t$ and $t+\Delta t$ thereby leading to a change of the
molecular population $\vec{n} \rightarrow \vec{n} + \vec{\nu}_r$.
Thus, a given reaction channel is characterized mathematically
by two quantities, namely, 
(i) the state change vector $\vec{\nu}_r$, and
(ii) The {\it propensity function} $W_r(\vec{n})$.

Let us assume that $dt$ is so small that no more than one reaction of
any kind can take place in the interval between $t$ and $t + dt$.
Then, we can write the ``chemical'' master equation
\cite{gillespie77,gillespie05,gillespie07,turner04}
\begin{eqnarray}
\frac{\partial P(\vec{n},t)}{\partial t} = \sum_{r=1}^{R} [W_r(\vec{n}-\vec{\nu}_r) P(\vec{n}-\vec{\nu}_r,t)] - \sum_{r'=1}^{R} [W_{r'}(\vec{n}) P(\vec{n},t)]
\label{eq-chemaster}
\end{eqnarray}
for the probability $P(\vec{n},t)$.

\noindent $\bullet${\bf Gillespie algorithm}

Except for a few special reactions
\cite{laurenzi00},
in general, it is very difficult
(practically impossible) to solve the CME either analytically or
numerically. Therefore, it is often investigated following an approach
which is essentially equivalent to Monte Carlo simulation of the
reaction. Gillespie algorithm \cite{gillespie77,gillespie05,gillespie07}
is the most popular technique for simulating the CME. There are
essentially three substeps in each step of this algorithm: \\
(i) it generates the time step $\Delta t$ till the next reaction; \\
(ii) it randomly picks up one of the reactions for its execution; \\
(iii) it implements the execution of the reaction by advancing the
time by $\Delta t$ and by updating the number of molecules to reflect
the occurrence of the reaction. Several different variants of this
algorithm and its extensions have been developed in the last three
decades.

\noindent $\bullet${\bf Deterministic limit: chemical rate equation from CME}

The steps in the systematic derivation of the deterministic chemical
reaction rate equations from the corresponding stochastic CME have
been clearly stated and the approximating assumptions have been
clarified \cite{widom65,gillespie09}.
The rate equations for a chemically reacting system can be derived
from the corresponding master equations. Let us define the average
population by
\begin{equation}
<\vec{n}(t)> = \sum_{\vec{n}} \vec{n}(t) P(\vec{n},t)
\end{equation}
It is straightforward to derive the rate equation
\begin{equation}
\frac{d<\vec{n}(t)>}{dt} = \sum_{r=1}^{R} \nu_r <W_r(\vec{n}(t))>
\label{eq-mtor}
\end{equation}
satisfied by $<\vec{n}(t)>$.

\subsubsection{\bf Chemical Langevin and Fokker-Planck equations}

Suppose the $s$-th component of the $S$-component column vector
$\vec{X}(t)$ denote the number of molecules of species $s$ at
time $t$. Then, the {\it chemical Langevin equation} (CLE)
\cite{gillespie00,zwanzig01} is given by \cite{andrews06,andrews09}
\begin{equation}
\frac{dX_{i}}{dt} = \sum_{j=1}^{S} \nu_{ij} {\cal F}_j({\vec X}(t))
+ \sum_{j=1}^{S} \Gamma_j(t) \nu_{ij} \sqrt{{\cal F}_j({\vec X}(t)}
\label{eq-CLE}
\end{equation}
The first term directly corresponds to the right hand side of the
rate equation except for the fact that $X_{i}$
denotes the numbers of molecules whereas $c_{i}$ is the concentration
of the molecules of $i$-th species. The second term on the right
hand side of (\ref{eq-CLE}) adds Gaussian noise of vanishing mean
and unit variance.

The CLE is formulated as the kinetics of an {\it individual based}
model. Just as in the case of Brownian mechanics, an alternative,
but equivalent, approach was developed in terms of the {\it chemical
Fokker-Planck equation} (CFPE) which described the kinetics in
terms of {\it populations} \cite{andrews09}. However, since this is hardly
ever used in the context of molecular motors we'll not discuss
it here.

\subsection{\bf Enzymatic reactions: regulation by physical and chemical means}

Enzymes are proteins and function as biological catalysts \cite{dixon79book}.
These are specific in the sense that a specific catalyst speeds up a
specific reaction by a factor of $10^6$ to $10^{20}$.
In spite of common essential features, there are also important
differences between non-biological catalysts and enzymes; the most important
differences arise from the macromolecular character of the proteins
\cite{keleti75}.
The evolution of the structural designs of enzymes from the primitive
(presumably inorganic) catalysts and optimization of their performance
during various stages of biological evolution are interesting subjects
of investigation \cite{heinrich91,maurel06} which, however, will not be discussed here.

Enzymatic reactions, which are of major interest here in the context of
molecular motors, can be regulated (i) by physical means, or (ii) by
chemical means. Physical means include temperature, force, etc. while
chemical means depend on interaction of the enzyme with other molecules.

Chemical regulation of an enzyme can be carried out following two
different approaches: (a) by regulating the {\it quantity} of the
enzyme through control on its production, or degradation (or, permanent
inactivation by irreversible covalent modification), and (b) controlling
the enzymatic activity (of a given amount of enzyme) by its binding and
dissociation with small molecules. Any small molecule which can bind
an enzyme reversibly is called a {\it ligand}. A ligand can be an
activator or inhibitor of the enzymatic activity \cite{keleti75}.
Many enzymes need a {\it cofactor} for enzymatic activity, the cofactor
can be a metal ion or an organic molecule.

\section{\bf Elastic stiffness of polymers}
\label{appendix-elasticpol}

DNA, RNA and proteins are linear polymers. These polymers not only 
introduced a new length scale (characterized by its size) and a time scale
(associated with its dynamics) but also brought in its ``flexibility''
which is not possible with only small molecules. This flexible nature
of macromolecules also gives rise to the importance of conformational
entropy. In fact, many biological processes are driven by entropic
elasticity. Apriori, it is not at all obvious that the phenomenological
concepts of classical theory of elasticity, which were developed for
macroscopic objects, should be applicable even for single molecules
of DNA, RNA, etc. Technological advances over the last two decades
made it possible to stretch, bend and twist a single macromolecule and
the corresponding moduli of elasticity have been measured 
\cite{marko05,benham05,bath07,bustamante00b,bustamante03,strick00a,lavery02,strick03,charvin04,lionnet06,zhuang05}.

The elasticity of nucleic acids is of particular significance because 
most often genome (DNA or RNA) are stored in bent conformation \cite{marko03}. 
For example, in eukaryotic cells, DNA is bent and wrapped around 
histones \cite{garcia06}. Similarly, in viral capsids, nucleic acids 
are strongly 
bent for efficient packaging. Furthermore, temporary bending of 
macromolecules take place in many biological processes driven by 
molecular motors. Stiff polymers can also play the role of a 
nano-piston. Therefore, the elasticity of the macromolecules of life 
is also interesting in the study of molecular machines which 
polymerize, manipulate and degrade these molecules.

\subsubsection{\bf Freely jointed chain model and entropic elasticity}

Let us model a linear polymer of $N$ monomers, each of length ${\ell}$. 
Freely jointed chain (FJC) model is based on the assumption that the 
relative orientation between the successive monomers is completely 
random and does not involve any energy change. Thus, the FJC model is 
essentially equivalent to a random walk. Therefore, for one-dimensional 
FJC with a given $N$ and end-to-end distance $x$, the entropy is given 
by 
\begin{equation}
S(N,x) = \frac{N!}{[N+(x/{\ell})]![N-(x/{\ell})]!} 
\end{equation}
For sufficiently large $N$ and $x$, using Stirling approximation, the 
force exerted by the FJC is found to be $-k_{eff} x$ where the effective 
spring constant $k_{eff}$ is 
\begin{equation}
k_{eff} = \frac{k_BT}{N{\ell}^2}
\end{equation} 
This spring-like behevior of the FJC is of entropic origin.

\subsubsection{\bf Worm-like chain and its relation with freely jointed chain: persistence length}

The polymer is represented by a flexible chain where an energy cost 
has to be paid for its bending. 
Suppose $\hat{t}(s)$ is the unit tangent to the chain at $s$. It turns 
out that the tangent-tangent correlation function 
$<\hat{t}(s)\bullet \hat{t}(0)>$ decays exponentially, i.e, 
\begin{equation}
<\hat{t}(s)\bullet \hat{t}(0)> \sim e^{-s/\xi_{P}} 
\end{equation}
with the arc length $s$. The {\it persistence length} 
$\xi_{P} \propto \kappa_B/(k_BT)$  of the polymer increases with the 
increasing bending stiffness $\kappa_B$ whereas it decreases with 
increasing temperature.

\section{\bf Cytoskeleton: beams, struts and cables}
\label{app-cytoskeleton}

The mechanical properties of the cell depends on its cytoskeleton
The cytoskeleton of an eukaryotic cell maintains its architecture
\cite{pollard03a,frixione00,schliwa02,fletcher10}.
The cytoskeleton is a complex dynamic network that can change in
response to external or internal signals. The cytoskeleton is also
responsible for intra-cellular transport of packaged molecular
cargoes as well as for the motility of the cell as a whole. The
cytoskeleton plays crucially important role also in cell division
and development of organisms.
The cytoskeleton of not only animals, but also those of plants and
algae \cite{kost02,westeneys02,westeneys04,mineyuki07,katsaros06}
as well as those of fungi
\cite{xiang03,steinberg00,geitman00}
have been investigated widely (see Table \ref{table-cytocomp}).
Counterparts of some molecular components of the eukaryotic
cytoskeleton have been discovered recently also in prokaryotic cells
\cite{lowe04,amos04,graumann04,pogliano08,kurner04,jensen05,gitai05,gitai07,watters06,lopez03,lopez06,ausmees03,erickson07}.

\subsection{\bf Cytoskeleton of eukaryotic cells}

The protein constituents of the cytoskeleton of eukaryotic cells
can be broadly divided into the following three categories:
(i) {\it Filamentous} proteins, (ii) {\it accessory} proteins, and
(iii) {\it motor} proteins.
The three classes of filamentous proteins, which form the main
scaffolding of the cytoskeleton, are:
(a) {\it actin}, (b) {\it microtubule}, and (c) {\it intermediate
filaments}.

The three superfamilies of motor proteins are:
(i) {\it myosin} superfamily,
(ii) {\it kinesin} superfamily, and
(iii) {\it dynein} superfamily.
Both kinesins and dyneins move on microtubules; in contrast, myosins
either move on actin tracks or pull the actin filaments.\\

\begin{table}
\begin{tabular}{|c|c|} \hline
Category  & Member \\\hline
Filamentous protein & Actin \\ \hline
Filamentous protein & Microtubule \\ \hline
Filamentous protein & Intermediate filaments \\ \hline
Accessory protein & Filament polymerization regulators \\ \hline
Accessory protein & Filament-filament linkers \\ \hline
Accessory protein &  Filament-plasma membrane linkers \\ \hline
Motor protein & Myosin \\ \hline
Motor protein & Kinesin \\ \hline
Motor protein & Dynein \\ \hline
\end{tabular}
\caption{Protein constituents of the cytoskeleton of an eukaryotic cell.
 }
\label{table-cytocomp}
\end{table}

On the basis of functions, accessory proteins can be categorized
as follows:
(i) regulators of filament polymerization,
(ii) filament-filament linkers,
(iii) filament-plasma membrane linkers.
Among the regulators of filament polymerization, some promote
nucleation of a filament, while some other species cap a filament
thereby terminating its growth. Some regulators enhance the rate
of filament growth whereas some others are involved in the
depolymerization and severing of filaments. Filament-filament
cross-linkers organize higher-order assemblies and networks of
the filaments.

Microtubules are cylindrical hollow tubes whose diameter is approximately
20 nm. The basic constituent of microtubules are globular proteins called
tubulin. Hetero-dimers, formed by $\alpha$ and $\beta$ tubulins, assemble
sequentially to form a protofilament. 13 such protofilaments form a
microtubule. The length of each $\alpha-\beta$ dimer is about 8 nm.
Since there is only one binding site for a motor on each dimeric subunit
of MT, the minimum step size for kinesins and dyneins is 8 nm.

Although the protofilaments are parallel to each other, there is
a small offset of about 0.92 nm between the dimers of the neighboring
protofilaments. Thus, total offset accumulated over a single looping of
the 13 protofilaments is $13 \times 0.92 \simeq 12 nm$ which is equal
to the length of three $\alpha-\beta$ dimers joined sequentially.
Therefore, the cylindrical shell of a microtubule can be viewed as
{\it three} helices of monomers. Moreover,
the asymmetry of the hetero-dimeric building block and their parallel
head-to-tail organization in all the protofilaments gives rise to the
polar nature of the microtubules. The polarity of a microtubule is such
an $\alpha$ tubulin is located at its - end and a $\beta$ tubulin is
located at its + end.

Filamentous actin are polymers of globular actin monomers. Each
actin filament can be viewed as a double-stranded, right handed
helix where each strand is a single protofilament consisting of
globular actin. The two constituent strands are half staggered
with respect to each other such that the repeat period is 72 nm.

\section{\bf Kinetics of nucleation, polymerization and depolymerization of polar filaments: treadmilling and dynamic instability}

Polymerizing and depolymerizing polar filaments generate force by 
mechanisms that we discuss in detail in several sections of this review. 
The role of $\gamma$-tubulin in the nucleation of MT filaments has been
known for quite some time
\cite{job03,wiese06,messina07}.
Theoretical models have been developed for the kinetics of MT nucleation
\cite{fygensen94,fygensen95,flyvbjerg96a,flyvbjerg97}.
Two classes of actin nucleating proteins are: 
(i) formin protein family; and
(ii) Arp2/3 complex
\cite{welch02,carlsson10,karalar11}.

The dynamics of polymerization and depolymerization of microtubules is quite 
different from those of most of the common proteins.  {\it Dynamic instability}
\cite{mitchison84,erickson92,desai97,gardner08a}
is now accepted as the dominant mechanism governing the dynamics of
microtubule polymerization. Each polymerizing microtubule persistently
grows for a prolonged duration and, then makes a sudden transition to a
depolymerizing phase; this phenomenon is known as ``catastrophe''.
However, the rapid shrinking of a depolymerizing microtubule can get
arrested when it makes a sudden reverse transition, called ``rescue'',
to a polymerizing phase. 

There are strong experimental evidences that the dynamic instability of
a MT is triggered by the loss of its {\it guanosine triphoshate} (GTP) cap
because of the hydrolysis of GTP into {guanosine diphosphate} (GDP).
Although the detailed mechanism, i.e., how the chemical process of cap loss
induces mechanical instability, remains far from clear, kinetic models have
been developed based on plausible mechanisms
\cite{flyvbjerg94,flyvbjerg96b,margolin06,antal07a,antal07b}.

Some small molecules can suppress the dynamic instability and infuence 
the rates of growth and/or shrinkage of the microbules when bound to the 
tubulins. These molecules are potential anti-cancer drugs because of the 
corresponding implications of the dynamic instability in cell division 
\cite{peterson02a,jordan04,honore05}.
Quantitative effects of such drug molecules on the kinetics of MT 
polymerization /depolymerization and the distribution of the microtubule 
lengths have been investigated \cite{mishra05}.

Pioneering theoretical model of MT polymerization
\cite{hill84a,hill84b,hill85a,rubin88}
and many of their more recent extensions
\cite{dogterom93,dogterom98,freed02,deymier05,govindan04,mishra05,margolin06}
treated each MT as an essentially one-dimensional object and and without
assuming any explicit scenario that causes catastrophe and rescue.
An one-dimensional model for the polymerization 
/ depolymerization kinetics of MT, incorporating catastrophe and rescue, 
was developed by Hill \cite{hill84a}. In this original version the 
kinetics is formulated in terms of, effectively, infinite number of 
coupled ordinary differential equations for $P_{\pm}(n,t)$, the 
probability of finding a filament consisting of $n$ subunits ($n$ being 
discrete) at time $t$ in the growing (+) and shrinking (-) phases. 
In a later work, Dogterom and Leibler \cite{dogterom93} described the 
kinetics in terms of two coupled partial differential equations for 
$P_{\pm}(x,t)$ where $x$, the length of a MT, is assumed to be a 
continuous variable. 
However, more recent works on the kinetics of dynamic instability capture 
the fact that a MT is a tubular object consisting of 13 protofilaments. 

Instead of {\it dynamic instability}, it is the {\it treadmilling}
\cite{bugyi10}
that dominates the kinetics of polymerization / depolymerization of 
actin. 
Kinetic models of actin polymerization should model it as double-stranded,
and capture the effects of ATP hydrolysis
\cite{stukalin05a,stukalin06,vavylonis05,hu07}.
For a deeper insight into the contrasting features of the polymerization 
kinetics of MT and F-actin, see \cite{mitchison92b}.

\noindent {\bf Acknowledgements}:

It is my great pleasure to thank all my students and collaborators for 
enjoyable collaborations on molecular motors. I sincerely thank 
John Bechhoefer, Zvonimir Dogic, Arnold Driessen, Pierre Gaspard, 
Manoj Gopalakrishnan, Steven P. Gross, Hermann-Georg Holzh\"utter, 
Mandar Inamdar, Ambarish Kunwar, Charles Lindemann, Roop Mallik, 
Amit Mitra, Alex Mogilner, Raja Paul, Hong Qian, Andreas Schadschneider, 
Sean Sun, Andrej Vilfan, Jianhua Xing, Alexey Zaikin and an anonymous 
referee for their comments and suggestions on various sections of the 
manuscript.  I am indebted to Joachim Frank for many enlightening 
correspondences and discussions as well as for his patient critical 
reading of an earlier preliminary shorter version of this manuscript. 
I thank Ashok Garai, Jeffrey Moffitt and Ajeet Sharma for their 
comments on earlier shorter drafts of this review and Ajeet Sharma for 
technical help in producing two figures. 
Over the last few years, I have also benefitted from many discussions 
and/or correspondences with Veronika Bierbaum, Stephan Grill, Joe Howard, 
Frank J\"ulicher, Stefan Klumpp, Reinhard Lipowsky, Sriram Ramaswamy and 
Ajay Sood.
This work has been supported at IIT Kanpur by the Dr. Jag Mohan Garg
Chair professorship and by Department of Biotechnology, government of 
India. Work on this review was in progress over the last 
five years during which I have enjoyed the hospitality of several 
research institutions. Parts of the literature survey for this review 
were carried out at those institutions, particularly, 
the Max-Planck Institute for the Physics of Complex Systems (MPI-PKS), 
Dresden, Germany, University of Cologne, Germany, and the Mathematical 
Biosciences Institute (MBI) at the Ohio State University, Columbus, USA. 
These visits have been supported by the Visitors Program of the MPI-PKS 
(in Dresden), Alexander von Humboldt Foundation (in Cologne) and 
by MBI and the National Science Foundation under grant DMS 0931642 
(in Columbus).




\newpage

\centerline{\bf Note added in proof}

Some interesting papers inadvertenty escaped my attention while writing 
this review. Moreover, some new relavent and important papers have 
appeared after the acceptance of the manuscript for pubication. These 
papers are listed beow.

\begin{enumerate}
\item{G.C. Lander, H.R. Saibil and E. Nogales, {\it Go hybrid: EM, 
crystallography, and beyond}, Curr. Opin. Struct. Biol. {\bf 22}, 
627-635 (2012).} 

\item{L.B. Oddershede, {\it Force probing of individual molecules inside 
the living cell is now a reality}, Nat. Chem. Biol. {\bf 8}, 879-886 (2013).}

\item{S. Forth, M.Y. Sheinin, J. Inman and M.D. Wang, {\it Torque 
measurement at the single-molecule level}, Annu. Rev. Biophys. {\bf 42}, 
(2013).}

\item{A. Ciudad and J.M. Sancho, {\it A unified phenomenological analysis 
of the experimental velocity curves in molecular motors}, J. Chem. Phys. 
{\bf 128}, 225107 (2008).}

\item{C.T. Friel and J. Howard, {\it Coupling of kinesin turnover to 
translocation and microtubule regulation: one engine, many machines}, 
J. Muscle Res. Cell Motil. {\bf 33}, 377-383 (2012).}

\item{A. Murugan, D.A. Huse and S. Leibler, {\it Speed, disipation, 
and error in kinetic proofreading}, PNAS {\bf 109}, 12034-12039 (2012).}

\item{Y. Sun and Y.E. Goldman, {\it Lever-arm mechanics of processive 
myosins}, Biophys. J. {\bf 101}, 1-11 (2011).} 

\item{R.E. McConnell and M.J. Tyska, {\it Leveraging the membrane- 
cytoskeleton interface with myosin-1}, Trends in Cell Biol. {\bf 20}, 
418-426 (2010).} 

\item{A. Ciudad and J.M. Sancho, {\it Analysis of the nucleotide-dependent 
conformations of kinesin-1 in the hydrolysis cycle}, J. Chem. Phys. 
{\bf 131}, 015104 (2009).}

\item{A. Rai, A. Rai, A.J. Ramaiya, R. Jha and R. Mallik, {\it Molecular 
adaptations allow dynein to generate large collective forces inside cells}, 
Cell {\bf 152}, 172-182 (2013).}

\item{D.G. Cole and W.J. Snell, {\it Snapshot: intraflagellar transport}, 
Cell {\bf 137}, 784-784.e1 (2009).}

\item{D.S. Gokhin and V.M. Fowler, {\it A two-segment model for thin 
filament architecture in skeletal muscle}, Nat. Rev. Mol. Cell Biol. 
{\bf 14}, 113-119 (2013).} 

\item{A.C. Martin, {\it Pulsation and stabilization: contractile forces 
that underlie morphogenesis}, Dev. Biol. {\bf 341}, 114-125 (2010).} 

\item{E.M. Craig, S. Dey and A. Mogilner, {\it The emergence of sarcomeric, 
graded-polarity and spindle-like patterns in bundles of short cytoskeletal 
polymers and two opposite molecular motors}, J. Phys. Condens. Matter 
{\bf 23}, 374102 (2011).} 

\item{R. Levayer and T. Lecuit, {\it Biomechanical regulation of 
contractility: spatial control and dynamics}, Trends in Cell Biol. {\bf 22}, 
61-81 (2012).}

\item{R. Lipowsky, S. Liepelt and A. Valleriani, {\it Energy conversion by 
molecular motors coupled to nuceotide hydrolyis}, J. Stat. Phys. {\bf 135}, 
951-975 (2009).}

\item{P. Ranjith, D. Lacoste, K. Mallick and J.F. Joanny, {\it Nonequiibrium 
self-assembly of a filament coupled to ATP/GTP hydrolysis}, Biophys. J. 
{\bf 96}, 2146-2159 (2009).} 

\item{P. Ranjith, K. Mallick, J.F. Joanny and D. Lacoste, {\it Role of 
ATP-hydrolysis in the dynamics of a single actin filament}, Biophys. J. 
{\bf 98}, 1418-1427 (2010).}

\item{K. Tsekouras, D. Lacoste, K. Mallick and J.F. Joanny, {\it Condensation 
of actin filaments pushing against a barrier}, New J. Phys. {\bf 13}, 
103032 (2011).} 

\item{R. Padinhateeri, A.B. Kolomeisky and D. Lacoste, {\it Random 
hydrolysis controls the dynamic instability of microtubules}, Biophys. J. 
{\bf 102}, 1274-1283 (2012).}

\item{F. Huber, J. Schnauss, S. R\"onicke, P. Rauch, K. M\"uller, 
C. F\"utterer and J. K\"as, {\it Emergent complexity of the cytoskeleton: 
from single filaments to tissue}, Adv. Phys. {\bf 62}, 1-112 (2013).}

\item{K.V. Korotkov, M. Sandkvist and W.G.J. Hol, {\it The type II secretion 
system: biogenesis, molecular architecture and mechanism}, Nat. Rev. 
Microbiol. {\bf 10}, 336-351 (2012).} 

\item{K. Struhl and E. Segal, {\it Determinants of nucleosome positioning}, 
Nat. Struct. Mol. Biol. {\bf 20}, 267-273 (2013).} 

\item{K.C. Neuman, T. Lionnet and J.F. Allemand, {\it
Single-molecule micromanipulation techniques}, Annu. Rev. Mater. Res.
{\bf 37}, 33-67 (2007).}

\item{N. Zuleger, A.R.W. Kerr and E.C. Schirmer, {\it Many mechanisms, one 
entrance: membrane protein translocation into the nucleus}, Cell. Mol. 
Life Sci. {\bf 69}, 2205-2216 (2012).}

\item{V. Bierbaum and R. Lipowsky, {\it Dwell time distributions of the 
molecular motor myosin V}, PLoS ONE {\bf 8}, e55366 (2013).} 

\item{J. Sparacino, P.W. Lamberti and C.M. Arizmendi, {\it Shock detection in 
dynamics of single-headed motor proteins KIF1A via Jensen-Shannon 
divergence}, Phys. Rev. E {\bf 84}, 041907 (2011).}`

\item{A.B. Kolomeisky, E.B. Stukain and A.A. Popkov, {\it Understanding 
mechanochemical couping in kinesins using first-passage-time processes}, 
Phys. Rev. E {\bf 71}, 031902 (2005).} 

\item{F. Berger, C. Keller, S. Klumpp and R. Lipowsky, {\it Distinct 
transport regimes for two elastically coupled molecular motors}, 
Phys. Rev. Lett. {\bf 108}, 208101 (2012).}

\item{J.G. Orlandini, C. Blanch-Mercader, J. Brugues and J. Casademunt, {\it 
Cooperativity of self-organized Brownian motors pulling on soft cargoes}, 
Phys. Rev. E {\bf 82}, 061903 (2010).} 

\item{J. Tailleur, M.R. Evans and Y. Kafri, {\it Nonequilibrium phase 
transitions in the extraction of membrane tubes by molecular motors}, 
Phys. Rev. Lett. {\bf 102}, 118109 (2009).} 

\item{J. Brugues and J. Casademunt, {\it Self-organization and cooperativity 
of weakly coupled molecular motors under unequal loading}, Phys. Rev. Lett. 
{\bf 102}, 118104 (2009).}

\item{S. Klumpp, Y. Chai and R. Lipowsky, {\it Effects of the chemomechanical 
stepping cycle on the traffic of molecular motors}, Phys. Rev. E {\bf 78}, 
041909 (2008).}

\item{V. Belitsky and G.M. Sch\"utz, {\it Cellular automaton model for 
molecular traffic jams}, JSTAT: theory and expt. P07007 (2011).} 

\item{H. Grzeschik, R.J. Harris and L. Santen, {\it Traffic of cytoskeletal 
motors with diordered attachment rates}, Phys. Rev. E {\bf 81}, 031929 (2010).} 

\item{Y. Chai, R. Lipowsky and S. Kumpp, {\it Transport by molecular motors 
in the presence of static defects}, J. Stat. Phys. {\bf 135}, 241-260 (2009).} 

\item{A. Ciudad, J.M. Sancho and G.P. Tsironis, {\it Kinesin as an electrostatic 
machine}, J. Biol. Phys. {\bf 32}, 455-463 (2006).} 

\item{J. Munarriz, J.J. Mazo and F. Falo, {\it Model for hand-over-hand 
motion of molecular motors}, Phys. Rev. E {\bf 77}, 031915 (2008).} 

\item{F. Slanina, {\it Interaction of molecular motors can enhance their 
efficiency}, EPL {\bf 84}, 50009 (2008).} 

\item{F. Slanina, {\it Efficiency of interacting Brownian motors: improved 
mean-fied treatment}, J. Stat. Phys. {\bf 135}, 935-950 (2009).} 

\item{F. Slanina, {\it Interacting molecular motors: efficiency and work 
fluctuations}, Phys. Rev. E {\bf 80}, 061135 (2009).}

\item{I. Pinkoviezky and N.S. Gov, {\it Modelling interacting molecular 
motors with an internal degree of freedom}, New J. Phys. {\bf 15}, 
025009 (2013).} 

\item{D.M. Suter and K.E. Miller, {\it The emerging role of forces in axonal 
elongation}, Prog. Neurobiol. {\bf 94}, 91-101 (2011).} 

\item{K. Szymanska and C.A. Johnson, {\it The transition zone: an essential 
functional compartment of cilia}, Cilia {\bf 1}, 10 (2012).}

\item{W.B. Ludington, K.A. Wemmer, K.F. Lechtreck, G.B. Witman and W.F. 
Marshall, {\it Avalanche-like behavior in ciliary transport}, PNAS 
{\bf 110}, 3925-3930 (2013).}

\item{C.T. Friel and J. Howard, {\it The kinesin-13 MCAK has an unconventional 
ATPase cycle adapted for microtubule depolymerization}, EMBO J. {\bf 30}, 
3928-3939 (2011).} 

\item{A. Melbinger, L. Reese and E. Frey, {\it Microtubule length regulation 
by molecular motors}, Phys. Rev. Lett. {\bf 108}, 258104 (2012).} 

\item{R. Loughlin, B. Riggs and R. Heald, {\it Snapshot: motor proteins in 
spindle assembly}, Cell {\bf 134}, 548.e1 (2008).}

\item{F.J. McNally, {\it Mechanisms of spindle positioning}, J. Cell Biol. 
{\bf 200}, 131-140 (2013).} 

\item{T.U. Tanaka, {\it Kinetochore-microtubule interactions: steps 
towards bi-orientation}, EMBO J. {\bf 29}, 4070-4082 (2010).}

\item{K. Tanaka, {\it Regulatory mechanisms of kinetochore-microtubule 
interaction in mitosis}, Cell. Mol. Life Sci. {\bf 70}, 559-579 (2013).} 

\item{J. Nilson, {\it Looping in on Ndc80- how does a protein loop at the 
kinetochore control chromosome segregation?}, Bioessays {\bf 34}, 1070-1077 
(2012).} 

\item{A. Khodjakov and J. Pines, {\it Centromere tension: a divisive issue}, 
Nat. Cell Biol. {\bf 12}, 919-923 (2010).}

\item{B. Akiyoshi and S. Biggins, {\it Reconstituting the 
kinetochore-microtubule interface: what, why, and how}, Chromosoma 
{\bf 121}, 235-250 (2012).} 

\item{N.T. Umbreit and T.N. Davis, {\it Mitosis puts sisters in a strained 
relationship: force generation at the kinetochore}, Expt. Cel Res. {\bf 318}, 
1361-1366 (2012).} 

\item{B. Shtylla and D. Chowdhury, {\it A theoretical model for attachment 
lifetimes of kinetochore-microtubules: Mechano-kinetic ``catch-bond' 
mechanism for error-correction}, arXiv.1301.5692 (2013).}

\item{A. Chaudhuri, B. Bhattacharya, K. Gowrishankar, S. Mayor and M. Rao, 
{\it Spatio-temporal regulation of chemical reactions by active cytoskeletal 
remodeling}, PNAS {\bf 108}, 14825-14830 (2011).}

\item{P. Rorth, {\it Fellow travellers: emergent properties of collective 
cell migration}, EMBO Rep. {\bf 13}, 984-991 (2012).} 

\item{C. Bustamante, W. Cheng and Y.X. Mejia, {\it Revisiting the central 
dogma one molecule at a time}, Cell {\bf 144}, 480-497 (2011).} 

\item{J. Zhou, V. Schweikhard and S.M. Block, {\it Single-molecule studies 
of RNAPII elongation}, Biochim. Biophys. Acta {\bf 1829}, 29-38 (2013).}

\item{Z. Bryant, F.C. Oberstrass and A. Basu, {\it Recent developments in 
single-molecule DNA mechanics}, Curr. Opin. Struct. Biol. {\bf 22}, 
304-312 (2012).} 

\item{J.F. Allemand, B. Maier and D.E. Smith, {\it Molecular motors for 
DNA translocation in prokaryotes}, Curr. Opin. Biotechnol. {\bf 23}, 
503-509 (2012).}

\item{I.J. Molineaux and D. Panja, {\it Popping the cork: mechanisms of 
phage genome ejection}, Nat. Rev. Microbiol. {\bf 11}, 194-204 (2013).} 

\item{H. Zhang, C. Schwarts, G.M. De Donatis and P. Guo, {\it ``Push through 
one-way valve'' mechanism of viral DNA packaging}, Adv. in Virus Res., 
{\bf 83}, 415-465 (2012).} 

\item{J. Telenius, A.E. Walin, M. Straka, H. Zhang, E.J. Mancini and R. Tuma, 
{\it RNA packaging motor: from structure to quantum mechanical modelling 
and sequential-stochastic mechanism}, Comp. and Math. Meth. in Med. 
{\bf 9}, 351-369 (2008).} 

\item{K.P. Santo and K.L. Sebastian, {\it Simple model for the kinetics of 
packaging of DNA into a capsid against an external force}, Phys. Rev. E 
{\bf 65}, 052902 (2002).}

\item{D. Michel, {\it How transcription factors can adjust the gene 
expression floodgates}, Prog. Biophys. Mol. Biol. {\bf 102}, 16-37 
(2010).}

\item{M. Djordjevic and R. Mundschuh, {\it Formation of the open complex 
by bacterial RNA poymerase- a quantitative model}, Biophys. J. {\bf 94}, 
4233-4248 (2008).}

\item{E. Nudler and M.E. Gottesman, {\it Transcription termination and 
anti-termination in E. coli}, Genes to Cells {\bf 7}, 755-768 (2002).}

\item{S. Borukhov and E. Nudler, {\it RNA polymerase: the vehicle of 
transcription}, Trends in Microbiol. {\bf 16}, 126-134 (2008).}

\item{E. Nudler, {\it RNA poymerase active center: the molecular engine 
of transcription}, Annu. Rev. Biochem. {\bf 78}, 335-361 (2009).}

\item{P. Rafael Costa, M. L. Acencio and N. Lemke, {\it Cooperative RNA 
polymerase molecules behavior on a stochastic sequence-dependent model 
for transcription elongation}, PLoS ONE {\bf 8}, e57328 (2013).}

\item{Y. Ohta, T. Kodama and S. Ihara, {\it Cellular-automaton model of the 
cooperative dynamics of RNA poymerase II during transcription in human cells}, 
Phys. Rev. E {\bf 84}, 041922 (2011).}

\item{J.R. Chubb and T.B. Lieverpool, {\it Bursts and pulses: insights from 
single cell studies into transcriptional mechanims}, Curr. Opin. Genet. 
Dev. {\bf 20}, 478-484 (2010).} 

\item{D.B. Wigley, {\it Bacterial DNA repair: recent insights into the 
mechanism of RecBCD, AddAB and AdnAB}, Nat. Rev. Microbiol. {\bf 
11}, 9-13 (2013).} 

\item{H. Merrikh, Y. Zhang, A.D. Grossman and J.D. Wang, {\it 
Replication-transcription conflicts in bacteria}, Nat. Rev. Microbiol. 
{\bf 10}, 449-458 (2012).} 

\item{A.M. Carr, A.L. Paek and T. Weinert, {\it DNA replication: failure and 
inverted fusions}, Sem. Cell \& Dev. Biol. {\bf 22}, 866-874 (2011). }

\item{M.D. Sutton, {\it Coordinating DNA polymerase traffic during high 
and low fidelity synthesis}, Biochim. et Biophys. Acta {\bf 1804}, 
1167-1179 (2010).} 

\item{S. Gokhale, D. Nyayanit and C. Gadgil, {\it A systems view of the 
protein expression process}, Syst. Synth. Biol. {\bf 5}, 139-150 (2011).} 

\item{J. Trylska, {\it Coarse-grained models to study dynamics of nanoscale 
biomolecules and their applications to the ribosome}, J. Phys. Condens. 
Matt. {\bf 22}, 453101 (2010). }

\item{P. Xie, {\it Dynamics of tRNA occupancy and dissociation during 
translation by the ribosome}, J. Theor. Biol. {\bf 316}, 49-60 (2013).} 

\item{P. Xie, {\it Model of ribosome translation and mRNA unwinding}, Eur. 
Biophys. J. (2013).} 

\item{E. N\"urenberg and R. Tampe, {\it Tying up loose ends: ribosome 
recycling in eukaryotes and archaea}, Trends in Biochem. Sci. (2012).}

\item{M.C. Romano, M. Thiel, I. Stansfield and C. Grebogi, {\it Queueing 
phase transition: theory of translation}, Phys. Rev. Lett. {\bf 102}, 
198104 (2009).}

\item{S. Melnikov, A. Ben-Shem, N. G. de Loubresse, L. Jenner, G. Yusupova 
and M. Yusupov, {\it One core, two shells: bacterial and eukaryotic 
ribosomes}, Nat. Struct. Mol. Biol. {\bf 19}, 560-567 (2012).} 

\item{S. Klumpp, J. Dong and T. Hwa, {\it On ribosome load, codon bias and 
protein abundance}, PLoS One {\bf 7}, e48542 (2012).} 

\item{E.M. Novoa and L. R. de Pouplana, {\it Speeding with control: 
codon usage, tRNAs, and ribosomes}, Trends in Genet. {\bf 28}, 574-581 
(2012).}

\item{Q. Liu and K. Frederick, {\it Contribution of intersubunit bridges 
to the energy barrier of ribosome translocation}, Nucleic Acids Res. 
{\bf 41}, 565-574 (2013).}

\item{L. Ciandrini, I Stansfield and M.C. Romano, {\it Ribosome traffic on 
mRNA maps to gene ontology: genome-wide quantification of translation 
initiation rates and polysome size regulation}, PLoS Comp. Biol. {\bf 9}, 
e1002866 (2013).} 

\item{A. Zinovyev, N. Morozova, N. Nonne, E. Barilot and A. Harel-Bellan, 
{\it Dynamical modeling of microRNA action on the protein translation 
process}, BMC Syst. Bio. {\bf 4}, 13 (2010).} 

\item{S. Reuveni, I. Meilijson, M. Kupiec, E. Ruppin and T. Tuller,  
{\it Genome-scale analysis of translation elongation with a ribosome 
flow model}, PLoS Comp. Biol. {\bf 7}, e1002127 (2011).}

\item{T. Tuller, I. Veksler-Lublinsky, N. Gazit, M. Kupiec, E. Ruppin and 
M. Ziv-Ukelson, {\it Composite effects of gene determinants on the 
translation speed and density of ribosomes}, Genome Biol. {\bf 12}, 
R110 (2011).} 

\item{I.J. Finkelstein and E.C. Greene, {\it Molecuar traffic jams on 
DNA}, Annu. Rev. Biophys. {\bf 42}, (2013).}

\item{J. Yu, T. Ha and K. Schulten, {\it Structure-based mode of the stepping 
motor of PcrA helicase}, Biophys. J. {\bf 91}, 2097-2114 (2006).} 

\item{V. Rajagopal, M. Gurjar, M.K. Levin and S.S. Patel, {\it The protease 
domain increases the translocation stepping efficiency of the Hepatitis 
C Virus NS3-4A helicase}, J. Biol. Chem. {\bf 285}, 17821-17832 (2010).} 

\item{J. Yu, W. Cheng, C. Bustamante and G. Oster, {\it Coupling translocation 
with nucleic acid unwinding by NS3 helicase}, J. Mol. Biol. {\bf 404}, 
439-455 (2010).} 

\item{R.A. Forties, J.A. North, S. Javaid, O.P. Tabba, R. Fishel, 
M.G. Poirier and R. Bundschuh, {\it A quantitative model of nucleosome 
dynamics}, Nucl. Acids Res. {\bf 39}, 8306-8313 (2011).}

\item{S. Mukherjee and A. Warshel, {\it Electrostatic origin of the 
mechanochemical rotary mechanism and the cataytic dwell of F1-ATPase}, 
PNAS {\bf 108}, 20550-20555 (2011).} 

\item{N. Nelson, A. Sacher and H. Nelson, {\it The significance of molecular 
slips in transport systems}, Nat. Rev. Mol. Cell Biol. {\bf 3}, 876-881 
(2002).}

\item{A. Basu, A.J. Schoeffler, J.M. Berger and Z. Bryant, {\it ATP binding 
controls distinct structural transitions of esherichia coli DNA gyrase 
in complex with DNA}, Nat. Struct. Mol. Biol. {\bf 19}, 538-546 (2012).} 

\item{S.H. Chen, N.L. Chan and T.S. Hsieh, {\it New mechanistic and 
functional insights into DNA topoisomerase}, Annu. Rev. Biochem. (2013).}

\end{enumerate}
\end{document}